\pdfoutput=1
\documentclass[11pt]{report}
\usepackage{upennstyle}
\usepackage[numbers,square]{natbib}
\usepackage{mathtools,mathabx}
\usepackage{mathpartir}
\DeclareSymbolFont{txsymbolsC}{U}{txsyc}{m}{n}
\DeclareSymbolFont{txlargesymbolsA}{U}{txexa}{m}{n}
\DeclareMathSymbol{\multimapinv}{\mathrel}{txsymbolsC}{18}
\DeclareMathDelimiter{\llbracket}{\mathopen}{txsymbolsC}{126}{txlargesymbolsA}{18}
\DeclareMathDelimiter{\rrbracket}{\mathclose}{txsymbolsC}{127}{txlargesymbolsA}{19}

\usepackage{tikz-cd}
\usetikzlibrary{decorations.markings}
\usepackage{adjustbox}

\usepackage{bussproofs}

\usepackage[most]{tcolorbox}

\usepackage{multicol}
\usepackage{enumitem}

\usepackage{soul}
\usepackage[makeroom]{cancel}

\usepackage{subfiles}

\usepackage{cleveref}
\usepackage{thmtools}
\declaretheorem[name=Theorem,     numberwithin=chapter, style=definition]{theorem}
\declaretheorem[name=Proposition, sibling=theorem,      style=definition]{proposition}
\declaretheorem[name=Lemma,       sibling=theorem,      style=definition]{lemma}
\declaretheorem[name=Corollary,   sibling=theorem,      style=definition]{corollary}
\declaretheorem[name=Definition,  sibling=theorem,      style=definition]{definition}

\declaretheorem[name=Example,     sibling=theorem,      style=remark]{example}

\crefname{theorem}{Theorem}{Theorems}             \Crefname{theorem}{Theorem}{Theorems}
\crefname{proposition}{Proposition}{Propositions} \Crefname{proposition}{Proposition}{Propositions}
\crefname{lemma}{Lemma}{Lemmas}                   \Crefname{lemma}{Lemma}{Lemmas}
\crefname{corollary}{Corollary}{Corollaries}      \Crefname{corollary}{Corollary}{Corollaries}
\crefname{definition}{Definition}{Definitions}    \Crefname{definition}{Definition}{Definitions}
\crefname{remark}{Remark}{Remarks}                \Crefname{remark}{Remark}{Remarks}
\crefname{example}{Example}{Examples}             \Crefname{example}{Example}{Examples}

\newcommand{\1}{{\mathbf 1}}
\newcommand{\C}{\mathcal C}
\newcommand{\mc}[1]{\mathcal #1}
\newcommand{\mf}[1]{\mathfrak #1}
\newcommand{\bs}[1]{\boldsymbol #1}

\newcommand{\fib}[2][]{\overset{\mc #1}{\underset{\mc #2}{\downarrow}}}

\newcommand{\xto}[2][]{\xrightarrow[#1]{#2}}
\newcommand{\nrightsquigarrow}{%
  \mathrel{\ooalign{%
    $\rightsquigarrow$\cr
    \hidewidth\raise0.2ex\hbox{\scriptsize/}\hidewidth\cr
  }}%
}

\newcommand{\parr}{\adjustbox{scale=0.8}{\rotatebox[origin=c]{180}{\&}}}
\newcommand{\quadt}[1]{\quad \text{#1} \quad}

\newcommand{\modec}{\ \mathsf{mode}^{\mathsf{c}}}
\newcommand{\model}{\ \mathsf{mode}^{\mathsf{l}}}

\newcommand{\subfilebibliography}{}

\title{A FIBRATIONAL FRAMEWORK FOR MULTI-CONCLUSION LOGICS}
\author{Oualid Merzouga}
\gradgroup{Mathematics}
\date{2026}
\supervisor{Henry Towsner}
\supervisortitle{Professor of Mathematics}
\cosupervisor{Paige Randall North}
\cosupervisortitle{Assistant Professor of Mathematics, Information and Computing Sciences}
\gradchair{Angela Gibney, Presidential Professor of Mathematics}
\committee{Henry Towsner, Professor of Mathematics}
\committee{Andre Scedrov, Professor of Mathematics, Computer and Information Science}
\committee{Paige Randall North, Assistant Professor of Mathematics, Information and Computing Sciences}

\dedication{To my mother and sister, for their unconditional love.}

\abstract{%
Sequent calculi are described by both their structural rules governing their hypotheses and the number of conclusions their judgments admit. The fibrational framework of Licata, Shulman, and Riley gives a uniform account of substructural and modal logics for single-conclusion judgments. This thesis extends that framework to a multiple-output setting in which cartesian resources may be used freely while linear resources occur exactly once on either side of the turnstile. The resulting judgments are indexed by descriptors: directed acyclic graphs that generalize the tree-shaped expressions of the original framework. Descriptor $2$-cells encode structural transformations, while explicit gates provide weakening, contraction, coweakening, and cocontraction at linear modes.

Over this mode theory we define a sequent calculus generated by two descriptor-indexed connectives, $F$ and $U$. We prove identity, the relevant invertibility properties, and cut admissibility; for signatures equipped with all four linear gates and their coherence data, a multicut theorem establishes the admissibility of cut for classical calculi. The mode and type theories present linear-non-linear $2$-polycategories. The syntactic projection is a local discrete fibration and a fibration, with opcartesian lifts for one-output base morphisms. Two adequacy results delimit the framework: its cartesian fragment recovers the Licata--Shulman--Riley calculus, and its pure linear fragment recovers a two-sided presentation of classical multiplicative linear logic.

The thesis also studies the relation between indexed and fibrational semantics. After developing the relevant representability correspondences for multicategories and polycategories, we establish an equivalence between multi-indexed categories over a fixed representable base and multicategorical opfibrations over that base, together with its biclosed bifibrational form. Finally, we give a one-way construction from planar poly-indexed data to a planar polycategory.
}

\acknowledgement{
  I would first like to thank my advisors, Henry Towsner and Paige Randall North. Thank you for your guidance and for standing by me throughout these years.

  Henry, I am continually in awe of the way you approach mathematics. Thank you for the time you have devoted to me over these seven years. Your ability to send me out of your office with renewed motivation and a smile on my face is a superpower, and for that I thank you.

  Paige, you set me on the path of type theory and category theory and have guided me along it ever since. Thank you for the invaluable resources you have shared and for explaining difficult ideas with such clarity. I really admire your razor-sharp intuition.

  Thank you to Philip Gressman and Ryan Hynd for making all of this possible. You believed in me and gave me the chance to pursue my dreams through your bridge program. Philip, your teaching has been a model for my own. Ryan, your optimism helped me believe I could overcome all my struggles.

  Scott Weinstein, thank you for welcoming me to Penn and inspiring my curiosity and enthusiasm for mathematics. Andre Scedrov, thank you for supporting me through the most important moments of my graduate studies.

  Thank you, Yumeng Ou, for your relentless support in my teaching journey. I attribute success in the classroom to you.

  Sasha Peterka, it is so wonderful to have been a TA for your class and have gotten to know you as more than a teacher. It feels incredible to witness the passion for mathematics shining through your eyes every time we speak.

  A special thank you to Noam Zeilberger from École Polytechnique for the discussions that guided me toward crucial answers in this thesis. The smooth wrap-up of the document is mostly due to your insights.

  I would like to thank both Ian Petrie and Karen Lagasse from the Center of Teaching and Learning for the opportunities you have given me and the trust you have placed in me. I grew as a teacher the most while being surrounded by your expertise, and you opened a new door into the world of teaching for me.

  My fellow graduate students played a decisive role in my studies.
  Marc, I owe you so much. Our discussions have shaped how I think about many aspects of life. I feel fortunate to know someone so caring and thoughtful. Without your insights, I would not have finished this thesis and would probably still be stuck in Plato's cave with my iPad.

  Alvaro, you are the most dedicated person I know. I will remember how you decided to spend days listening to me and supporting me through my lowest moments, despite how hard you work every day. Thank you for your friendship.

  Mauricio, thank you for turning back to introduce yourself that day in the classroom. Thank you for the board game nights, the math discussions over lunch, and all the laughter we shared. I dearly miss working with you in the library.

  Krishan, I admire the care with which you approach new material. Jin, your incredible intuition and ability to think on your feet always impress me. Thank you both for helping shape my graduate life.

  I would like to thank the undergraduate students that had a profound impact on my journey as well.
  Harith, thank you for taking my class. The best reward was to develop a friendship with you. I will miss your frequent visits to my office to discuss just about anything. Your sharpness and quick thoughts are inspiring.
  Andrew, thank you for being my student. You embody critical thinking and curiosity about knowledge. It is an absolute pleasure to entertain those deep questions with you.
  Leslie, thank you for trusting me during this formative year for you. I will always remember our discussions as being a turning point in my teaching philosophy and am blessed to see that you achieved your dreams.

  I am grateful to my family and friends, who often believed in me more than I believed in myself. Xavier, thank you for always being there for me. Your kindness and support mean a lot to me. Thank you both, Xavier and Karine, for showing me new trails to explore back in France and, of course, for the amazing vegan food.

  Jinyi, thank you for your continued care and support through our highs and lows. We grew together during the most formative period of my life.

  Hasna, thank you for sharing your life with me. I owe my resilience to you, and I cannot imagine growing up without you as my big sister. I miss fighting with you over who got to watch TV in the evening.

  Mom, thank you for making my happiness your priority. Your support through every life-changing decision has given me the strength to keep going. You have protected me, guided me, and made huge sacrifices for me. Thanking you is simply not enough.
}

\begin{document}
\maketitle
\setcounter{page}{2}

\makededication
\makeacknowledgement
\makeabstract
\tableofcontents

\begin{mainf}
\begingroup

\chapter{Introduction}\label{chap:introduction}

\section{Context and Motivation}

Sequent calculus makes the architecture of a logic visible. A judgment
\[
\Gamma \vdash \Delta
\]
records not only formulas but also two choices that substantially affect the resulting logic: the structural behavior of the contexts and the number of conclusions permitted on the right of the turnstile. Structural rules determine whether hypotheses may be reordered, discarded, or duplicated. Exchange, weakening, and contraction therefore distinguish ordinary structural reasoning from resource-sensitive systems such as linear logic \cite{Gir87}. Independently, the passage from a single conclusion to multiple conclusions is central to Gentzen's transition from the intuitionistic calculus $LJ$ to the classical calculus $LK$ \cite{Gen35}. A general account of sequent calculus must consequently control both axes rather than treating either as fixed.

This problem becomes more pronounced for modal and substructural logics. Modalities regulate not merely how often a hypothesis may be used, but also where it may be used and how information moves between contexts. Designing a separate calculus for every combination of structural and modal principles duplicates both syntax and metatheory. Licata, Shulman, and Riley address this issue by separating a logic into two layers \cite{LSR17}. A \emph{mode theory} describes the behavior of formulas in their context, and a type theory fibred over it supplies logical connectives and derivations. Changing the mode theory changes the judgmental structure without changing the metatheory.

The Licata--Shulman--Riley framework is intuitionistic: its context descriptors have one output, and its sequents have one conclusion. A descriptor can therefore be represented by a term, or equivalently by a tree whose root is the unique output. Multiple-conclusion calculi do not have this shape. A judgment with several outputs must record how resources contribute to all those outputs, and its composition must allow a selected output of one judgment to be cut against a selected input of another. Trees are no longer sufficient; the natural structure of a term becomes encoded in a graph. Semantically, this corresponds to the composition operation in a polycategory.

Polycategories generalize multicategories by allowing morphisms to have several outputs as well as several inputs \cite{Sza75}. They provide the categorical shape of two-sided calculi and, in particular, of classical multiplicative linear logic. At the same time, a framework intended to retain intuitionistic and modal examples must also accommodate reusable, cartesian data. Shulman's linear-non-linear (LNL) polycategories combine these cartesian and linear forms of judgment in a common structure \cite{Shu23}. They therefore supply the semantic setting for the extension developed in this thesis.

To produce a general account of multi-conclusion judgments while retaining the cartesian nature of single-conclusion logic, we propose a framework that preserves the two-layer design of \cite{LSR17} and replaces its tree-shaped context expressions by graph-shaped \emph{descriptors}.

\section{The Framework}

The framework distinguishes cartesian modes from linear modes. Cartesian variables may be weakened, contracted, and exchanged. The word ``may'' plays an important role here. The framework is still able to generate substructural logics and can, on demand, allow for weakening, contraction and exchange. On the other hand, linear variables are exact resources and occur once. Accordingly, the type theory has two forms of judgment:
\[
\Theta \vdash_{\alpha} A
\qquad\text{and}\qquad
\Theta\,;\Gamma \vdash_{\mf D} \Delta.
\]

The cartesian context $\Theta$ contains reusable resources. The linear contexts $\Gamma$ and $\Delta$ contain exact resources on the two sides of the turnstile. The index $\alpha$ in the first judgment is a single-output cartesian descriptor, while $\mf D$ in the second may have several linear inputs and outputs. The framework develops a unique general judgment of the form
$$\Theta\,;\Gamma \vdash_{\mf D} \Delta$$ that is restricted to the two forms above depending on the nature of the formulas.

A descriptor is presented as a finite directed acyclic graph. Its vertices record applications of generators from a mode signature, and its edges record how boundary and intermediate variables are connected. This representation generalizes the expression trees of \cite{LSR17}: when the output is cartesian, every well-formed descriptor corresponds to an expression of the original framework. In the linear case the graph records polycategorical behavior. Composition is defined by occurrence-indexed substitution, which distinguishes the possible repeated occurrences of a cartesian variable from the unique occurrence of a linear variable.

Structural transformations between descriptors are represented by $2$-cells. They allow a mode theory to declare equations and directed structural principles that are transported to the type theory as admissible operations. Cartesian weakening and contraction are available by construction. Their linear counterparts are not: a linear mode remains resource-sensitive unless the signature is equipped with explicit gates for weakening, contraction, coweakening, or cocontraction. This distinction separates structural properties intrinsic to cartesian contexts from additional structure imposed on linear ones.

The type theory is generated by two families of connectives. A positive connective $F$ packages the resources described by a descriptor into an output formula. A negative connective $U$ internalizes an input relative to its surrounding input and output contexts. These two schemes recover familiar connectives as instances: cartesian products and implications arise in the single-output fragment, while tensor and par arise in the multiple-output linear fragment. Mixed descriptors mediate between the cartesian and linear zones and provide the judgmental setting in which modalities can be expressed.

The categorical semantics reflects this syntax. The mode theory presents an LNL $2$-polycategory, and the type theory modulo its equations presents an LNL $2$-polycategory fibred over it. The projection from typed derivations to descriptors is a local discrete fibration and a fibration; the $F$-connectives provide opcartesian lifts when the base morphism has one output. Thus the construction extends the multicategorical fibrational semantics of \cite{LSR17} toward the polycategorical semantics used for classical linear logic \cite{Bla23}, without asserting unrestricted opfibrationality where the syntax does not provide it.

\section{Contributions}

The principal contributions of this thesis are as follows.
\begin{enumerate}
  \item We define a mode theory of graph-based descriptors with cartesian and linear boundaries. We establish their regularity, characterize their two possible sorts, define composition by cut, and prove its unit, associativity, and interchange properties. We also prove that cartesian descriptors are in bijection with the expressions of \cite{LSR17}.

  \item We develop descriptor $2$-cells and gates for structural rules. Weakening and contraction are admissible for cartesian inputs, whereas structural operations at linear modes are represented explicitly and are governed by coherence data.

  \item We define a two-zone sequent calculus generated by the descriptor-indexed connectives $F$ and $U$. Its metatheory includes well-formedness of sequent indexes, identity, left invertibility of $F$, right invertibility of $U$, admissibility of cartesian structural rules, and cut admissibility. For signatures carrying all four linear gates and the required coherence laws, we prove multicut and obtain classical cut harmony.

  \item We give a categorical account of both syntactic layers in terms of LNL $2$-polycategories. With a chosen transport cleavage, the syntactic projection is shown to be a local discrete fibration and a fibration, together with opcartesian lifts over one-output morphisms.

  \item We prove two logical adequacy results. The cartesian fragment recovers the judgmental structure of Licata--Shulman--Riley, so adequacy transfers to object logics already represented by their framework. The pure linear fragment recovers the specified two-sided calculus for classical multiplicative linear logic used by Blanco \cite{Bla23}.

  \item We connect the fibrational account with representability. Over the multicategory underlying a fixed unbiased monoidal category, we establish an equivalence between multi-indexed categories and multicategorical opfibrations, as well as the corresponding biclosed equivalence with bifibrations. We then define planar poly-indexed categories over an unbiased weakly distributive base and construct from them planar polycategories with canonical liftings. This last construction is one-way; its projection is a bifibration.
\end{enumerate}

These results give a common syntax and semantics for the single-output cartesian and multiple-output linear regimes. A particular object logic is encoded in the framework by designing a signature containing modes and structural data and establishing an adequacy argument.
\section{Organization of the Thesis}

\Cref{sec:prerequisites-logic,sec:prerequisites-category-theory} develop the logical and categorical prerequisites. The logical part reviews sequents, structural rules, cut, and proof terms. The categorical part introduces categories, monoidal categories, multicategories, polycategories, weakly distributive categories, fibrations, and LNL polycategories.

\Cref{chap:fibrational-framework} contains the main construction. It first develops descriptors, their composition, structural $2$-cells, gates, and coherence laws. It then defines the $F/U$ type theory, proves its structural and cut metatheory, equips derivations with an equational theory, and identifies the resulting categorical semantics.

\Cref{chap:examples} gives the two principal instances. The first translates the Licata--Shulman--Riley calculus into the cartesian fragment and proves equivalence of derivability. The second interprets a two-sided calculus for classical multiplicative linear logic in the pure linear fragment, including the tensor, par, unit, bottom, and negation constructions.

\Cref{chap:representability} turns from the syntax of logics to the relation between representable and fibred categorical structures. It reviews the relevant multicategorical and polycategorical representability results, proves the fixed-base multi-indexed equivalences, and develops the planar poly-indexed construction.

Finally, the remaining directions for future work include extending the indexed representation results to arbitrary multicategorical and polycategorical bases, replacing chosen tensor and cotensor objects by a factorization-system account, incorporating dependent types, and strengthening weak initiality results to initiality.

\endgroup{}
\begingroup

\chapter{Prerequisites}
In this chapter, we present the definitions that help make this thesis as self-contained as possible. We define all the concepts needed as well as set up the notation for the rest of the document. The two main sections concern \hyperref[sec:prerequisites-logic]{logic and the sequent calculus} and the required \hyperref[sec:prerequisites-category-theory]{category theory}, respectively. Most definitions in this chapter are standard and widely accessible in the literature.

\section{Logic}\label{sec:prerequisites-logic}
The study of logical systems has spanned thousands of years, since Aristotle's syllogisms. We will, however, be interested in a more modern and mathematical approach to logic: the sequent calculus. The fundamental concept of study is the notion of sequents defined in \cref{subsec:sequents}. The sequent calculus is the apparatus needed to reason about the sequents.

\textit{Note:} The sequent calculus was developed by Gerhard Gentzen \cite{Gen35} in the 1930s in an effort to prove the consistency of arithmetic. The presentation of logic through this lens relies on the expression of deductive rules manipulating mathematical statements to produce novel ones. A key idea is that those rules preserve truth, which means that if the premises are true, then the consequent is true as well.

A logic is a system to be used as a reasoning framework about mathematics. Thus, we encode mathematical statements in variables, e.g. $A, B, C, \dots$. Logic, as a field of study, is not interested in proving or analyzing a precise mathematical statement such as \emph{There exist infinitely many prime numbers}, and is rather preoccupied with studying the logical structure of those statements. This means that encoding the mathematical content in a propositional variable is not needed; only the relationships between statements are to be investigated. We therefore introduce the notion of propositional variables as well as connectives which are the means to encode the relationship between them.

\subsection{Propositional Variables}
Mathematics is interested in proving the truth of mathematical statements. A key insight consists of formalizing those very statements in order to reason about them. The first atomic notion we need is the \emph{propositional variable}. It will represent statements that can be assigned a truth value. Propositional variables will take the form of capital letters $A,B,C, \ldots$.

\begin{example}
To be able to reason about a statement such as \textit{The number zero has no predecessor}, we need to compactly denote and refer to it. We could set $A$ to denote this very proposition and refer to it by just using this letter.
\end{example}

\subsection{Connectives}
Combining statements is one way to gain complexity in our system and to render it interesting to study. The standard mathematical connectives mimic the ones in current language. Although their meaning is different in some cases, they serve as a nice approximation of our understanding. Technically, the sequent calculus we are setting up is a purely syntactic system. This means, among other things, that no meaning is attached to the symbols we are manipulating.
Mathematicians use the flavor of logic called \emph{classical logic}. The list of connectives for classical logic is
$\land, \lor, \Rightarrow, \neg $ and their intended meanings are ``and'', ``or'', ``implies'', ``not'', respectively. We recursively define the notion of a formula:

\begin{definition}[Formula]
The notion of \emph{formula} is generated inductively:
\begin{itemize}
  \item A propositional variable $A$ is a \emph{formula}.
  \item $\neg A$ is a \emph{formula} whenever $A$ is a formula.
  \item $A \land B$, $A \lor B$ and $A \Rightarrow B$ are \emph{formulas} whenever $A$ and $B$ are formulas.
\end{itemize}

\end{definition}

\begin{example}
The connective ``and'', denoted $\land$, is used to connect two statements together with the intended meaning of the conjunction. For instance, given two statements $A,B$, we write $A \land B$ to denote the statement $A$ and $B$. This can be understood as the formal internalization of the English notion of conjunction in the logical system. When first manipulating the notion of logic, one might feel confused about the word ``and'' as it stands for both the conjunction in natural language and the conjunction operator $\land$. For better or for worse, those coincide almost perfectly. The same cannot be said for the implication operator as the reader can check for themselves.
\end{example}

\subsection{Sequents}\label{subsec:sequents}
The notion of sequent internalizes the deductive processes of reasoning about mathematical statements in our object language. The sequents connect lists of formulas and individual formulas together and are characterized by the symbol $\vdash$. Deductive rules will manipulate sequents, informing us how to derive some sequents from others.

\begin{example}
Consider a syllogism:
\[
\begin{array}{ll}
   \textit{Aristotle is a human} \\
   \textit{All humans are mortal} \\
   \textit{Therefore Aristotle is a mortal}
\end{array}
\]
The form of this argument is such that it contains two premises and one conclusion. If we call the first premise \textit{Aristotle is a human} by the shorthand $A$, the second premise \textit{All humans are mortal} by $B$ and the conclusion \textit{Therefore Aristotle is a mortal} by $C$, then we can express it more concisely: if $A$ and $B$, then $C$. This encoding of statements into variables and connectives renders the study of those deductive arguments more amenable to mathematical analysis. The notation of a sequent captures this even more concisely as $$A \land B \vdash C.$$
Deduction rules will represent this syllogism more accurately. For the time being, we can think of this sequent as saying ``From the conjunction of the two premises, we can derive the conclusion'', leaving ``we can'' ambiguous for now.
\end{example}

Throughout this document, we call the formulas on the left of the turnstile $\vdash$ the \emph{antecedent} or the \emph{assumptions}, and the formula on the right of it the \emph{conclusion} or the \emph{consequent}; the word premise is reserved for the sequents standing above the line of an inference rule. For $\Gamma \vdash A$, where $\Gamma$ is a list of formulas $\Gamma := \Gamma_1, \ldots, \Gamma_n$, the intended meaning of this relation is that the conjunction of all the formulas in $\Gamma$ proves $A$.
The next point of interest is to describe the logical set of rules that encompass classical logic, arguably the most used logic in mathematics.

\subsection{The Sequent Calculus of Classical Logic}
The exposition of the classical sequent calculus used here is due to \cite{Gen35}. We build the system in two stages. The first stage presents the intuitionistic calculus $LJ$, where the right-hand side of the turnstile carries at most one formula. The second stage lifts this restriction and recovers the classical calculus $LK$. The number of formulas allowed on the right of the turnstile, in addition to some structure on the sequents, therefore decides the classicality of the system, and we will return to this observation throughout the thesis.

\textbf{The intuitionistic calculus.} A \emph{sequent} is of the form $\Gamma \vdash \Delta$ where $\Gamma$ is a list of formulas and $\Delta$ contains at most one formula and possibly none. The calculus starts from the identity axiom and the cut rule:
$$
\inferrule*[right=Id]{ }{A \vdash A}
\qquad
\inferrule*[right=Cut]{\Gamma \vdash A \\ \Gamma', A \vdash C}{\Gamma', \Gamma \vdash C}
$$
The identity axiom is the only rule without premises; it corresponds to the entry point of a derivation which consists of the sequential application of the derivation rules. The cut rule composes two derivations at the formula $A$. It corresponds to the informal notion of using a lemma when proving a larger mathematical statement. The formula $A$ corresponds to the lemma used to prove the final sequent. We note that the formula $A$ is not present in the conclusion $\Gamma', \Gamma \vdash C$, which means that it is difficult to backtrack a derivation in the presence of this rule. A class of rules, named \emph{structural rules}, manipulates the sequents irrespective of the nature of the formulas; exchange permutes two adjacent formulas, weakening adds an unused formula and contraction merges two copies of the same formula:
$$
\inferrule*[right=Ex]{\Gamma, A, B, \Gamma' \vdash C}{\Gamma, B, A, \Gamma' \vdash C}
\qquad
\inferrule*[right=Weak]{\Gamma \vdash C}{\Gamma, A \vdash C}
\qquad
\inferrule*[right=Contr]{\Gamma, A, A \vdash C}{\Gamma, A \vdash C}
$$
On the right-hand side of the turnstile, the restriction to at most one formula leaves no room for exchange and contraction. Weakening survives in a single instance, filling an empty right-hand side:
$$
\inferrule*{\Gamma \vdash \ }{\Gamma \vdash A}
$$
We note quickly that the flavor of logic without this rule is called \emph{Minimal logic} \cite{Joh37}. The connective rules are:

\begin{itemize}
  \item \textbf{Connective ``and'':} The rules for the connective $\land$ are:
  \[
  	\inferrule*[right=$\land$-intro]
  	{
    	\Gamma \vdash A \\
    	\Gamma \vdash B
  	}
  	{
    	\Gamma \vdash A \land B
  	}
  	\quad
  	\inferrule*[right=$\land$-elim]
  	{
    	\Gamma, A, B \vdash C \\
  	}
  	{
    	\Gamma, A \land B \vdash C
  	}
  \]       
  \item \textbf{Connective ``or'':} The rules for the connective $\lor$ are:
  \[
  	\inferrule*[right=$\lor$-intro${_1}$]
  	{
    	\Gamma \vdash A 
  	}
  	{
    	\Gamma \vdash A \lor B
  	}
  	\quad
  	\inferrule*[right=$\lor$-intro${_2}$]
  	{
    	\Gamma \vdash B \\
  	}
  	{
    	\Gamma \vdash A \lor B
  	}
  	\quad
  	\inferrule*[right=$\lor$-elim]
  	{
    	\Gamma, A \vdash C \\
    	\Gamma, B \vdash C
  	}
  	{
    	\Gamma, A \lor B \vdash C
  	}
  \]       
  \item \textbf{Connective ``implies'': } The rules for the connective $\Rightarrow$ are:
  \[
  	\inferrule*[right=$\Rightarrow$-intro]
  	{
    	\Gamma, A \vdash B 
  	}
  	{
    	\Gamma \vdash A \Rightarrow B
  	}
  	\quad
  	\inferrule*[right=$\Rightarrow$-elim]
  	{
    	\Gamma \vdash A \\
    	\Gamma, B \vdash C 
  	}
  	{
    	\Gamma, A \Rightarrow B \vdash C
  	}
  \]       
  \item \textbf{Connective ``not'':} The rules for the connective $\neg$ move a formula across the turnstile:
  \[
  	\inferrule*[right=$\neg$-intro]
  	{
    	\Gamma, A \vdash \
  	}
  	{
    	\Gamma \vdash \neg A
  	}
  	\quad
  	\inferrule*[right=$\neg$-elim]
  	{
    	\Gamma \vdash A
  	}
  	{
    	\Gamma, \neg A \vdash \
  	}
  \]
\end{itemize}
We usually refer to the rules for the connectives as \emph{introduction} and \emph{elimination} rules as they effectively introduce the connective on the conclusion side, i.e. the right-hand side of the turnstile, e.g. $\land$-right, and eliminate it from the assumptions side, i.e. the left-hand side of the turnstile, e.g. $\land$-left, in a given derivation. We sometimes call the introduction rules \emph{left introduction} rules and the elimination rules \emph{right introduction} as they display the location of the connective in the concluding sequent.

\textbf{The classical calculus.} Classical logic lifts the restriction of the sequents to at most one formula on the right-hand side. A sequent now takes the form $\Gamma \vdash \Delta$ where both $\Gamma$ and $\Delta$ are lists of formulas, read informally as saying that the collection of formulas in $\Gamma$ proves at least one formula in $\Delta$. The identity axiom is unchanged and the cut rule generalizes to lists in the conclusion:
$$
\inferrule*[right=Cut]{\Gamma \vdash \Delta, A \\ \Gamma', A \vdash \Delta'}{\Gamma', \Gamma \vdash \Delta, \Delta'}
$$
We feel free to reuse the same names for the sequent calculus rules and trust the reader to establish the context in which they occur.
The structural rules of exchange, weakening and contraction are now available for lists of formulas on the right-hand side of the sequents:
$$
\inferrule*[right=Ex-right]{\Gamma \vdash \Delta, A, B, \Delta'}{\Gamma \vdash \Delta, B, A, \Delta'}
\qquad
\inferrule*[right=Weak-right]{\Gamma \vdash \Delta}{\Gamma \vdash \Delta, A}
\qquad
\inferrule*[right=Contr-right]{\Gamma \vdash \Delta, A, A}{\Gamma \vdash \Delta, A}
$$
When the context is clear, we abbreviate the names of those rules to their equivalents on the left-hand side.
Each connective rule of the preceding section carries over with the replacement of the formula on the right-hand side of the turnstile by a list of formulas. For instance, the $\lor$-intro$_1$ rule derives $\Gamma \vdash A \lor B, \Delta$ from $\Gamma \vdash A, \Delta$. The rules for $\neg$ trade their empty right-hand sides for a list:
$$
\inferrule*[right=$\neg$-intro]{\Gamma, A \vdash \Delta}{\Gamma \vdash \neg A, \Delta}
\qquad
\inferrule*[right=$\neg$-elim]{\Gamma \vdash A, \Delta}{\Gamma, \neg A \vdash \Delta}
$$
Here we again decided to reuse the same names for the rules.
The gain in expressivity is exactly what is needed to recover the central feature of classical systems, i.e. the \emph{law of excluded middle}. It becomes derivable by using the fact that the conclusions of sequents accept a list of formulas and the availability of weakening, contraction and exchange on the right-hand side of the turnstile. It is derived by keeping two copies of the conclusion on the right and contracting them at the end, the right exchanges left implicit.
\begin{proposition}
  The sequent $\vdash A \lor \neg A$ is derivable in the classical calculus.
\end{proposition}
\textit{Proof.} The following derivation witnesses the sequent:
$$
\inferrule*[Right=ctr]
{
  \inferrule*[Right=$\lor$]
  {
    \inferrule*[Right=$\lor$]
    {
      \inferrule*[Right=$\neg$]
      {
        \inferrule*{ }{A \vdash A}
      }
      {\vdash \neg A, A}
    }
    {\vdash A \lor \neg A, A}
  }
  {\vdash A \lor \neg A, A \lor \neg A}
}
{\vdash A \lor \neg A}
$$
\hfill $\square$

Notice that the derivation uses the contraction on the right in an essential fashion that is not available to the intuitionistic system. 

\subsection{Context}
In a sequent $\Gamma \vdash A$, we usually call $\Gamma$ the assumptions. It is also called a \emph{context} in some instances. As a way to track which assumption is used where in a deductive step or in a full proof, it is useful to name those assumptions. We note that the letter used is not enough; indeed a context could contain duplicates such as $\Gamma := A,A,B,C$. In those cases, it will be useful to know which occurrence has been used. Thus we introduce new labels called \emph{variables}. From now on, we define \emph{contexts} to be lists of formula and variable pairs. The pairs are usually denoted as $a : A$ for a variable $a$ and a formula $A$.

\begin{definition}
  We define the notion of \emph{context} inductively:

\begin{itemize}
  \item The empty list $\cdot$ is a \emph{context}.
  \item Given a formula $A$ and a \emph{context} $\Gamma$, if the variable $x$ is not present in $\Gamma$ then $\Gamma, x : A$ is a \emph{context}.     
\end{itemize}
\end{definition}

\begin{example}
Expressing the presence of two assumptions $A, B$ to prove a formula $C$ is denoted using the notion of context as: $$x : A,y : B \vdash C.$$
\end{example}
\textit{Remark:} In the sequent-calculus presentation, contexts are bare lists of formulas; the annotated form carries in addition a variable naming each assumption. The two forms present the same data. 

\subsection{The Cut Rule}
Arguably, one of the most important operations on sequents is the cut rule. It is present in most logics of interest. Its manifestation in intuitionistic logic is
$$
\inferrule*
{
  \Gamma \vdash  A \\
  \Gamma', A \vdash  B \\
}
{
  \Gamma, \Gamma' \vdash  B \\
}
$$
The cut rule is sometimes referred to as the \emph{lemma rule} because it is akin to proving an intermediary result (here $A$) to show the desired conclusion. We note that the concluding sequent $\Gamma, \Gamma' \vdash B$ has no mention of the formula $A$. Thus, the cut rule introduces and eliminates a formula that is invisible to the concluding sequent. In a long derivation tree, we can easily imagine that many formulas would appear, get used and disappear before reaching the concluding sequent. Such a property is undesirable in certain aspects of proof theory. For instance, in proof search, trying to find a valid derivation of a sequent from the bottom up, i.e. starting with the conclusion and trying to apply the finitely many derivation rules to reach an axiom, is not possible in the presence of the cut rule. This is due to the fact that there are infinitely many formulas and it is not feasible to search them all as part of a potential cut rule instantiation.
However, if a calculus possesses the property of the conclusion containing every single formula that appears somewhere in the derivation tree, then we say that the calculus possesses the \emph{subformula property}. Again, the subformula property is desirable and the absence of cut is necessary for it to hold.

Why introduce the cut rule then if it is not desirable in the calculus? Traditionally, the cut rule is very convenient to work with as it makes derivations \emph{short} in a precise sense. In software engineering, this corresponds to writing a function that gets reused many times in the code base. We can inline those functions everywhere they are used, but doing so will grow the code base considerably, making it, in turn, less readable and understandable.
To get the best of both worlds, we would like the cut rule to be \emph{admissible}: cut is not among the defining rules, but adjoining it does not change which sequents are derivable. Equivalently, every derivation using cut can be replaced by a cut-free derivation. Gentzen proved cut elimination for sequent calculi in \cite{Gen35}. His consistency proof for first-order arithmetic uses the same principles \cite{Gen36}.

To infer consistency from cut elimination for a specified calculus, one must additionally show that its chosen contradiction judgment (for example, the empty sequent or $\vdash \bot$, depending on the calculus) has no cut-free derivation. That is a separate, calculus-specific analysis of the axioms and possible last rules. Cut elimination and this nonderivability fact together exclude a derivation of contradiction.

For our purposes, we seek cut admissibility because the admissible cut operation supplies composition in the categorical semantics. Any consistency conclusion still requires the separate cut-free nonderivability argument just described.

\subsection{Terms}
We have seen that a context consists of a pair of a variable and a type representing a formula in a specified logic. In this context, the variable plays the role of a placeholder that allows our calculus to record the position of the formula in the context. When defining mathematical structure, it is always beneficial to ensure that the extra structure imposed on some objects behaves nicely with the operations between them. For instance, in linear algebra, a linear map consists of a map between two spaces $A \xto{f} B$, together with conditions that allow the map to preserve the structure of the vector spaces $A$ and $B$. In particular, we require the map to satisfy $$f(x + y) = f(x) + f(y)$$ where $x,y \in A$ and the operation of addition belongs to the space $A$ on the left-hand side of the equality and to $B$ on the right-hand side. Similarly for us, we would like the variables in our context to behave nicely with the inference rules of our logic. To do this, we introduce the notion of terms for a calculus.

\begin{definition}
  A \emph{term} is defined by recursion using the following clauses:
  \begin{itemize}
    \item A variable is a \emph{term}
    \item For any function symbol $f$ of arity $n \in \mathbb{N}$, the expression $f(a_1, \ldots, a_n)$, where $a_1, \ldots, a_n$ are terms, is a \emph{term}
  \end{itemize}
\end{definition}
Terms allow us to annotate derivations so that every inference rule operates on the terms of its premises. We write
$$
\Gamma \vdash \alpha:A
$$
when $\alpha$ is a term witnessing a derivation of $\Gamma \vdash A$. We show how terms decorate sequents in the context of intuitionistic logic with the following function symbols and their intended arities $\mathsf{cut}_A, \mathsf{R_\land}, \mathsf{L}_\land,\mathsf{R}_{\lor,1}, \mathsf{R}_{\lor, 2}, \mathsf{L}_\lor, \mathsf{R}_{\Rightarrow}, \mathsf{L}_{\Rightarrow}$:
\begin{itemize}
  \item \textbf{Identity and cut:}
  \[
  \inferrule*[Right=Id]
  { }
  {\Gamma, x:A \vdash x:A}
  \qquad
  \inferrule*[Right=Cut]
  {\Gamma \vdash  \alpha:A \\
   \Gamma', x : A \vdash \beta:B}
  {\Gamma', \Gamma \vdash \mathsf{cut_A}(\alpha,\beta):B}
  \]

  \item \textbf{Conjunction:}
  \[
  \inferrule*[Right=$\land$R]
  {\Gamma \vdash \alpha:A \\
   \Gamma \vdash \beta:B}
  {\Gamma \vdash \mathsf{R}_{\land}(\alpha,\beta):A \land B}
  \qquad
  \inferrule*[Right=$\land$L]
  {\Gamma, x:A, y:B \vdash \alpha:C}
  {\Gamma, z:A \land B \vdash \mathsf{L}_{\land}(\alpha):C}
  \]

  \item \textbf{Disjunction:}
  \[
  \inferrule*[Right=$\lor$R$_1$]
  {\Gamma \vdash \alpha:A}
  {\Gamma \vdash \mathsf{R}_{\lor,1}(\alpha):A \lor B}
  \qquad
  \inferrule*[Right=$\lor$R$_2$]
  {\Gamma \vdash \beta:B}
  {\Gamma \vdash \mathsf{R}_{\lor,2}(\beta):A \lor B}
  \qquad
  \inferrule*[Right=$\lor$L]
  {\Gamma, x:A \vdash \alpha:C \\
   \Gamma, y:B \vdash \beta:C}
  {\Gamma, z:A \lor B \vdash \mathsf{L}_{\lor}(\alpha,\beta):C}
  \]

  \item \textbf{Implication:}
  \[
  \inferrule*[Right=$\Rightarrow$R]
  {\Gamma, x:A \vdash \alpha:B}
  {\Gamma \vdash \mathsf{R}_{\Rightarrow}(\alpha):A \Rightarrow B}
  \qquad
  \inferrule*[Right=$\Rightarrow$L]
  {\Gamma \vdash \alpha:A \\
   \Gamma, y:B \vdash \beta:C}
  {\Gamma, z:A \Rightarrow B \vdash \mathsf{L}_{\Rightarrow}(\alpha,\beta):C}
  \]
Adding terms to the calculus renders it proof-relevant. In some sense, we can keep track of the derivation used to produce a sequent. For instance, to derive the sequent $A \vdash A$ we can use the following two derivations:
\[
\begin{gathered}
\inferrule*
{ }
{
  x:A, y:B \vdash x:A
}
\\[1.5ex]
\inferrule*
{
  \inferrule*
  {
    x:A,y:B \vdash x:A \\
    x:A, y:B \vdash y:B \\
  }
  {
    x:A,y:B \vdash \mathsf{R}_\land(x,y) : A \land B
  }\\
  \inferrule*[Right=$\land$L]
  {x:A, y:B \vdash x:A}
  {z:A \land B \vdash \mathsf{L}_{\land}(x):A}
}
{
  x:A, y:B \vdash \mathsf{cut}_A(\mathsf{R}_\land(x,y), \mathsf{L}_{\land}(x)):A
}
\end{gathered}
\]
The calculus then differentiates between the two derivations even though the concluding sequent is exactly the same. Usually, we would want to identify them as the introduction and elimination of the connective $\land$ should not morally introduce any kind of new information compared to the axioms. The calculus then introduces equations between derivations. An example of such an equation is
$$\mathsf{cut}_A(\mathsf{R}_\land(x,y), \mathsf{L}_{\land}(x)) = x.$$ This specific rule is called a $\beta$\emph{-rule}, or a \emph{computation rule} in the lambda calculus language. To conclude this section, we summarize the utility of the introduction of the notion of terms as twofold:
\begin{itemize}
  \item Keep track of the derivation used to conclude a sequent.
  \item Unlock the ability to compare derivations and potentially introduce an equational theory to compare derivations.
\end{itemize}
The second point is an important one as it provides one of the main mechanisms for the calculus to take its semantics in category theory where the ability to compare morphisms is crucial.

\end{itemize}

\section{Category Theory}\label{sec:prerequisites-category-theory}
Category theory is a language used to capture the essence of mathematical objects. It rests on the foundational principle that not only are the objects of study important to study, but the relationships between them are as well. We start with the definition of a category, the most primitive concept of category theory.

\subsection{Categories}
\begin{definition}
  A \emph{category} $\mc M$ consists of:
  \begin{itemize}
    \item A collection of \emph{objects} denoted $Obj(\mc M)$.
    \item For any two objects $A, B \in Obj(\mc M)$ a collection of \emph{morphisms} denoted $\mc M(A, B)$.
    \item For any object $A$ an \emph{identity} morphism $\1_A \in \mc M(A,A)$.
    \item For any two morphisms $f \in \mc M(A,B),g \in \mc M(B,C)$ a \emph{composition} morphism $g \circ f \in \mc M(A,C)$. We say that $f,g$ are \emph{composable}.
    \item Equalities $f \circ \1_A = f$ and $\1_B \circ f = f$ for $f \in \mc M(A,B)$.
    \item Equalities $(f \circ g) \circ h = f \circ (g \circ h) $ for all composable morphisms $f,g$ and $g,h$. 
  \end{itemize}
\end{definition}

\phantomsection\label{anchor:category-size}
We can define a notion of morphisms between two categories in the hope of constructing a category of categories so as to be able to reason about them. We quickly notice that the standard set-theoretic paradoxes appear when facing the question of whether the category of categories exists. We redirect the reader to standard texts treating this concept as it is not the main focus of this document \cite[Chapter~I, \S~6]{Mac98}.

\begin{definition}
  A \emph{functor} $\mc M \xto{F} \mc P$ between two categories $\mc M, \mc P$ consists of:
  \begin{itemize}
    \item A function between the objects of $\mc M$ and $\mc P$, denoted $F_o : Obj(\mc M)\to Obj(\mc P)$.
    \item A function between the morphisms of $\mc M$ and $\mc P$, denoted $F_m : \mc M(A,B)\to \mc P(F_oA, F_oB)$.
    \item Coherence law for the identity: $F_m(\1_A) = \1_{F_oA}$.
    \item Coherence law for the composition: $F_m(f \circ g) = F_mf \circ F_m g$. 
  \end{itemize}  
\end{definition}
We can keep going up the abstraction ladder and consider morphisms between functors. These are called \emph{natural transformations}. They encapsulate what it should mean to have a morphism between functors by considering the different structures to preserve and respect.

\begin{definition}
  Given two functors with the same domain and codomain categories $F,G : \mc M \to \mc P$, we define a \emph{natural transformation} $\eta : F \to G$ to consist of:

\begin{itemize}
  \item  For each object $A \in Obj(\mc M)$, a morphism $\eta_A : FA \to GA$.
  \item  For any morphism $A \xto{f} B \in \mc M$, the naturality square commutes:
  \begin{center}
    \begin{tikzcd}[row sep=large, column sep=huge]
      FA \arrow[r, "\eta_A"] \arrow[d, "Ff"]   & GA \arrow[d, "Gf"] \\
      FB \arrow[r, "\eta_B"] & GB
    \end{tikzcd} 
      \quad i.e. \quad $Gf \circ \eta_A = \eta_B \circ Ff$  
  \end{center}
\end{itemize}
\end{definition}

\textit{Remark:} The collection of small categories with functors as morphisms itself forms a category. The adjective \emph{small} refers to the \hyperref[anchor:category-size]{size distinction}; we generally consider the categories that are contained in a category to be of a different ontology.

\begin{definition}
  Two functors $\mc M \xto{F}  \mc P$ and $\mc P \xto{G} \mc M$ form an \emph{adjunction} if $$\mc P(FA, B) \simeq \mc M(A, GB) \quad \text{Naturally in } A,B$$
  i.e.
  \begin{itemize}
  	\item There is a single family of isomorphisms $\varphi_{A,B} : \mc P(FA, B) \to \mc M(A, GB)$, one for every pair of objects, natural in $A$ and $B$ jointly
  	\item Naturality in each variable separately, the other held fixed, amounts to the joint naturality of the family
	\end{itemize}
	We call the functors $F,G$ left and right adjoints, respectively, and write $$F \dashv G.$$
\end{definition}
An equivalent definition of adjunction can be stated in terms of natural transformations relating the composition of the functors. 
\begin{definition}
  Two functors $\mc M \xto{F}  \mc P$ and $\mc P \xto{G} \mc M$ form an adjunction if there exist two natural transformations:
  \begin{itemize}
    \item $FG \xto{\epsilon} \1_{\mc P}$ called the counit and 
    \item $\1_{\mc M}  \xto{\eta} GF $  called the unit
	\end{itemize}
	such that $\1_F = \epsilon F \circ F\eta$ and $\1_G = G\epsilon \circ \eta G$. 
\end{definition}

There is a notion of composition of natural transformations sometimes called vertical composition:
\begin{definition}
  Let $F \xto{\eta} G$ and $G \xto{\epsilon}H$ be two natural transformations between the functors $F,G,H : \mc M \to \mc P$. We define the composition of $\eta$ and $\epsilon$ denoted $\epsilon \circ \eta$  by $$(\epsilon \circ \eta)_A = \epsilon_A \circ \eta_A$$     
\end{definition}

\begin{definition}
  Given two categories $\mc C, \mc D$, we define the \emph{product category} denoted $\mc C \times \mc D$ to be the category defined by:
	\begin{itemize}
    \item The objects are pairs $(A,B)$ where $A,B$ are objects of the categories $\mc C, \mc D$, respectively.
    \item A morphism between the objects $(A,B) \to (A', B')$ consists of a pair of morphisms $(f,g)$ where $f : A \to A'$, $g : B \to B'$ are morphisms in their respective categories.
  	\item The composition is defined in their respective categories as well:
  	$$(f,g) \circ (f',g') = (f \circ f', g \circ g')$$
    \item The identity morphism for the object $(A,B)$ is defined as $\1_{(A,B)} := (\1_A, \1_B)$, where $\1_A$ and $\1_B$ are the identity morphisms of the objects $A,B$ respectively.
	\end{itemize}
\end{definition}

\begin{definition}
  We define the \emph{Terminal category} denoted $\1$ to be the category with a single object denoted $\star$ and a single identity morphism $\1_\star : \star \to \star$.
\end{definition}

\subsection{Monoidal Categories}
\begin{definition}
A \emph{Monoidal category} $\mc C$ is a category equipped with a functor $\-- \otimes \-- : \mc C \times \mc C \to \mc C$ and data \cite{Mac98}:
\begin{itemize}
\item  A functor $\top : \1 \to \mc C$; we will abuse notation and use $\top$ for $\top(\star)$.
\item Associativity natural isomorphisms $a_\otimes : (\--_1 \otimes \--_2) \otimes \--_3 \Rightarrow \--_1 \otimes (\--_2 \otimes \--_3)$
\item Identity natural isomorphisms $u_\otimes : (\--) \otimes \top \Rightarrow (\--)$ and $u^\otimes : \top \otimes \-- \Rightarrow (\--)$  
\end{itemize}
subject to coherence laws \cite{Mac63}:
\begin{itemize}
\item \textbf{Invariance under associativity order:} Associativity can be applied in any order, i.e., the associativity pentagon commutes for all objects $A,B,C,D$:
\begin{center}
\begin{tikzcd}[row sep=large, column sep=huge]
((A \otimes B) \otimes C) \otimes D \arrow[r, "a_{\otimes(A \otimes B, C, D)}"] \arrow[d, "a_{\otimes(A,B,C)} \otimes D"] & (A \otimes B) \otimes (C \otimes D) \arrow[r, "a_{\otimes(A,B,C\otimes D)}"] & A \otimes (B \otimes (C \otimes D))\\
(A \otimes (B \otimes C)) \otimes D \arrow[r, "a_{\otimes(A, B \otimes C, D)}"] & A \otimes ((B \otimes C) \otimes D) \arrow[ru,swap, "A\otimes a_{\otimes(B,C,D)}"] 
\end{tikzcd}
\end{center}
\item \textbf{Invariance under identity simplification and associativity order:} Simplifying by the unit behaves well with associativity for all objects $A,B$:
\begin{center}
\begin{tikzcd}[row sep=large, column sep=huge]
(A \otimes \top) \otimes B \arrow[r, "u_{\otimes}(A) \otimes B "] \arrow[d, swap, "a_{\otimes(A, \top , B) }"]  & A \otimes B \\
A \otimes (\top \otimes B) \arrow[ru,swap, "A \otimes u^{\otimes}(B)"]   
\end{tikzcd}
\end{center}
\end{itemize}
\end{definition}

\textit{Remark:} Since $(\--)\otimes (\--) : \mc C \times \mc C \to \mc C$ is a functor, given two morphisms $f : A \to B$ and $g : C \to D$, we get a morphism $f \otimes g : A \otimes C \to B \otimes D$.\\
A related notion of \emph{unbiased monoidal category} will also be useful. A monoidal category is equipped with a bifunctor $\otimes : \mc C \times \mc C \to \mc C$ that can be iterated. Given $A,B,C$ we can form $A \otimes (B \otimes C)$ and $(A \otimes B) \otimes C$. Thanks to the natural isomorphism $a_\otimes$ those two objects are isomorphic. We can then think of them as a unique object potentially being written as $A \otimes B \otimes C$. Informally, this leads us to the notion that the bifunctor $\otimes$ defines a trifunctor $\otimes : \mc C \times \mc C \times \mc C \to \mc C$. Moving up, we can then define a functor $\otimes$ of every possible finite arity. This is part of the data that an unbiased monoidal category contains.

\textit{Remark:} The indices in the commutative diagrams for associativities are heavy. Their central idea is invariance under associativity, illustrated schematically by
\begin{center}
  \begin{tikzcd}
    & \otimes(\ldots \ldots \ldots \ldots \ldots) \arrow[dl, "\eta" ] \arrow[dr, "\eta"]  & \\
    \otimes(\ldots \ldots \otimes(\Gamma) \ldots \ldots) \arrow[dr, "\eta" ] & &  \otimes(\ldots \otimes(\ldots \Gamma \ldots) \ldots)\arrow[dl, "{\otimes(\ldots,\eta,\ldots)}" ]\\
    & \otimes(\ldots \otimes(\ldots \otimes(\Gamma) \ldots) \ldots) & 
  \end{tikzcd}
\end{center}
\begin{center}
  \begin{tikzcd}
    & \otimes(\ldots \ldots \ldots \ldots \ldots) \arrow[dl, "\eta" ] \arrow[dr, "\eta"]  & \\
    \otimes(\ldots \otimes(\Gamma) \ldots \ldots \ldots) \arrow[dr, "\eta" ] & &  \otimes(\ldots \ldots \ldots \otimes(\Delta) \ldots)\arrow[dl, "\eta" ]\\
    & \otimes(\ldots \otimes(\Gamma) \ldots \otimes(\Delta) \ldots) & 
  \end{tikzcd}
\end{center}
The nested-association diagram says that the order of association from inside out does not matter. The disjoint-association diagram says that the order of association on disjoint zones of $\otimes(\ldots)$ does not matter.

\begin{definition}\label{def:unbiased-monoidal-category}
  An \emph{Unbiased monoidal category} $(\mc C, (\otimes_n)_{n \in \mathbb{N}}, (\eta_{n,m})_{n,m \in \mathbb{N}}, \iota)$ consists of a category $\mc C$ equipped with the structure of \cite[\S 3.1]{Lei04}. We present a slight variation here:
  \begin{itemize}
    \item For every $n \in \mathbb{N}$ a functor: $$\otimes_n: \underbrace{\mc C \times \ldots \times \mc C}_{n-times}  \to \mc C$$
    \item For all $n_1, m, n_2 \in \mathbb{N}$, an associator natural isomorphism:
    \[
    \begin{aligned}
      \eta_{n_1,m,n_2} : {}& \otimes_{n_1 + m + n_2}(A_1, \ldots, A_{n_1}, B_1, \ldots, B_m, A_{n_1 + 1}, \ldots, A_{n_1 + n_2}) \\
      &\quad\to \otimes_{n_1 + n_2 + 1}(A_1, \ldots, A_{n_1}, \otimes_m(B_1, \ldots, B_m), A_{n_1 + 1}, \ldots, A_{n_1 + n_2})
    \end{aligned}
    \]
    \item A natural isomorphism $\iota$ with components: $$\iota_A : \otimes_1(A) \to A$$
    subject to coherence laws:
    \begin{itemize}
      \item \textbf{Invariance under associativity order: } The associativity diagrams commute.
      In the following two diagrams, we suppress the arity subscripts on $\otimes$; each arity is determined by the lengths of its arguments. We write $\1_\Gamma$ for the list of identity morphisms on the objects of $\Gamma$.

      \begin{samepage}
      For nested association, take lists $\Gamma_1,\Gamma_2,\Delta_1,\Delta_2,\Xi$ of lengths $n_1,n_2,m_1,m_2,k$, respectively:
      \begin{center}
        \begin{tikzcd}[row sep=huge, column sep=7em]
          \otimes(\Gamma_1,\Delta_1,\Xi,\Delta_2,\Gamma_2)
          \arrow[d, "{\eta_{n_1+m_1,k,m_2+n_2}}"]
          \arrow[r, "{\eta_{n_1,m_1+k+m_2,n_2}}"]
          & \otimes(\Gamma_1,\otimes(\Delta_1,\Xi,\Delta_2),\Gamma_2)
          \arrow[d, "{\otimes(\1_{\Gamma_1},\eta_{m_1,k,m_2},\1_{\Gamma_2})}"] \\
          \otimes(\Gamma_1,\Delta_1,\otimes(\Xi),\Delta_2,\Gamma_2)
          \arrow[r, "{\eta_{n_1,m_1+1+m_2,n_2}}"]
          & \otimes(\Gamma_1,\otimes(\Delta_1,\otimes(\Xi),\Delta_2),\Gamma_2)
        \end{tikzcd}
      \end{center}
      \end{samepage}
      \begin{samepage}
      For disjoint association, take lists $\Gamma_1,\Gamma_2,\Gamma_3,\Delta,\Xi$ of lengths $n_1,n_2,n_3,m,k$, respectively:
      \begin{center}
        \begin{tikzcd}[row sep=huge, column sep=7em]
          \otimes(\Gamma_1,\Delta,\Gamma_2,\Xi,\Gamma_3)
          \arrow[d, "{\eta_{n_1,m,n_2+k+n_3}}"]
          \arrow[r, "{\eta_{n_1+m+n_2,k,n_3}}"]
          & \otimes(\Gamma_1,\Delta,\Gamma_2,\otimes(\Xi),\Gamma_3)
          \arrow[d, "{\eta_{n_1,m,n_2+1+n_3}}"] \\
          \otimes(\Gamma_1,\otimes(\Delta),\Gamma_2,\Xi,\Gamma_3)
          \arrow[r, "{\eta_{n_1+1+n_2,k,n_3}}"]
          & \otimes(\Gamma_1,\otimes(\Delta),\Gamma_2,\otimes(\Xi),\Gamma_3)
        \end{tikzcd}
      \end{center}
      \end{samepage}
      \item \textbf{Invariance under associativity $\eta$  and $\iota$: } The unit-associativity diagram commutes for any list of objects $\Gamma := \Gamma_1, \ldots, \Gamma_n$:
      \begin{center}
        \begin{tikzcd}[row sep=large, column sep=huge]
          \otimes_1\otimes_n\Gamma   \arrow[r, "{\iota_{\otimes_n\Gamma}}"] & \otimes_n\Gamma \\
          \otimes_n\Gamma \arrow[ur, equals] \arrow[u, "\eta_{0,n,0}"] &
        \end{tikzcd}
        \quad
        \begin{tikzcd}[row sep=large, column sep=huge]
          \otimes_n(\Gamma_1, \ldots, \otimes_1(\Gamma_i), \ldots, \Gamma_n) \arrow[r, "{\otimes(\1_{\Gamma_1}, \ldots, \iota_{\Gamma_i}, \ldots, \1_{\Gamma_n} )}" ] & \otimes_n\Gamma\\
					\otimes_n\Gamma \arrow[ur, equals] \arrow[u, "{\eta_{i-1,1,n - i}}"]  &
        \end{tikzcd}
      \end{center}
    \end{itemize}
  \end{itemize}  

\end{definition}

\phantomsection\label{anchor:unbiased-monoidal-associator-notation}
\textit{Notation:} When no confusion arises, we will denote the associativities more concisely as follows:
\[
\begin{aligned}
  \eta_{n,m} : {}& \otimes_{n + m - 1}(A_1, \ldots, A_{i - 1}, B_1, \ldots, B_m, A_{i + 1}, \ldots, A_n) \\
  &\quad\to \otimes_{n}(A_1,\ldots, A_{i - 1}, \otimes_m(B_1, \ldots, B_m), A_{i + 1} \ldots, A_n)
\end{aligned}
\]
The insertion position $i$ is suppressed in the notation when clear from the context.

We hope to define a category of monoidal categories. To do this we need a notion of morphisms between monoidal categories. These will then be functors that will behave well with the structure of $\otimes$ and $\top$. There are numerous choices for the definition of those functors and it will be enough for us to consider \emph{(op)lax monoidal functors}.

\begin{definition}
A \emph{lax monoidal functor} between monoidal categories $(\mc C, \otimes)$  and $(\mc D, \oplus)$ is a functor $F : \mc C \to \mc D$ with two pieces of data:

\begin{itemize}
\item A morphism $F_\top : \top_\oplus \to F\top_\otimes$.
\item A natural transformation $\eta^{lax} : F(\--_1) \oplus F(\--_2) \to F(\--_1 \otimes \--_2)$ 
\end{itemize}
subject to coherence laws that describe the irrelevance of the order of operations between associativity and the functor application.
\begin{itemize}
\item  \textbf{Invariance under associativity order:}

\begin{center}
\begin{tikzcd}[row sep=large, column sep=huge]
FA \oplus (FB \oplus FC) \arrow[r, "FA \oplus \eta^{lax}_{B,C}"] \arrow[d, swap, "a_{\oplus(FA, FB, FC)}^{-1} "]  & FA \oplus F(B \otimes C) \arrow[r, "\eta^{lax}_{A, B \otimes C}"] & F(A \otimes (B \otimes C)) \arrow[d, "Fa_{\otimes(A,B,C)}^{-1} "] \\
(FA \oplus FB) \oplus FC \arrow[r, "\eta^{lax}_{A,B} \oplus FC"] & F(A \otimes B) \oplus FC \arrow[r, "\eta^{lax}_{A \otimes B, C}"] & F((A \otimes B) \otimes C)   
\end{tikzcd}
\end{center}

\item  \textbf{Invariance under identity simplification and application order:}

\begin{center}
\begin{tikzcd}[row sep=large, column sep=huge]
FA \oplus \top_\oplus \arrow[r, "FA \oplus F_\top "] \arrow[ddr, "u_\oplus"]  & FA \oplus F\top_\otimes \arrow[d, "\eta^{lax}_{A, \top_\otimes}"] \\
& F(A \otimes \top_\otimes) \arrow[d, "Fu_\otimes"] \\
& FA
\end{tikzcd} \quad
\begin{tikzcd}[row sep=large, column sep=huge]
\top_\oplus \oplus FA  \arrow[r, " F_\top \oplus FA"] \arrow[ddr, "u_\oplus"]  &  F\top_\otimes \oplus FA  \arrow[d, "\eta^{lax}_{\top_\otimes, A}"] \\
& F( \top_\otimes \otimes A ) \arrow[d, "Fu^\otimes"] \\
& FA
\end{tikzcd}
\end{center}
\end{itemize}
\end{definition}

\begin{definition}
An \emph{op-lax monoidal functor} between monoidal categories $(\mc C, \otimes)$  and $(\mc D, \oplus)$ is a functor $F : \mc C \to \mc D$ with two pieces of data:

\begin{itemize}
\item A morphism $F^\top : F\top_\otimes \to \top_\oplus$.
\item A natural transformation $\eta^{oplax} : F(\--_1 \otimes \--_2) \to F(\--_1) \oplus F(\--_2)$ 
\end{itemize}
subject to coherence laws that describe the irrelevance of the order of operations between associativity and the functor application.
\begin{itemize}
\item  \textbf{Invariance under associativity order:}

\begin{center}
\begin{tikzcd}[row sep=large, column sep=huge]
F(A \otimes (B \otimes C)) \arrow[r, "\eta^{oplax}_{A, B \otimes C}"] \arrow[d, swap, "Fa_{\otimes(A,B,C)}^{-1}"] & FA \oplus F(B \otimes C) \arrow[r, "FA \oplus \eta^{oplax}_{B,C}"] & FA \oplus (FB \oplus FC) \arrow[d, "a_{\oplus(FA, FB, FC)}^{-1}"] \\
F((A \otimes B) \otimes C) \arrow[r, "\eta^{oplax}_{A \otimes B, C}"] & F(A \otimes B) \oplus FC \arrow[r, "\eta^{oplax}_{A,B} \oplus FC"] & (FA \oplus FB) \oplus FC   
\end{tikzcd}
\end{center}

\item  \textbf{Invariance under identity simplification and application order:}

\begin{center}
\begin{tikzcd}[row sep=large, column sep=huge]
F(A \otimes \top_\otimes) \arrow[r, "\eta^{oplax}_{A, \top_\otimes}"] \arrow[ddr, swap, "Fu_\otimes"]  & FA \oplus F\top_\otimes \arrow[d, "FA \oplus F^\top"] \\
& FA \oplus \top_\oplus \arrow[d, "u_\oplus"] \\
& FA
\end{tikzcd} \quad
\begin{tikzcd}[row sep=large, column sep=huge]
F(\top_\otimes \otimes A) \arrow[r, "\eta^{oplax}_{\top_\otimes, A}"] \arrow[ddr, swap, "Fu^\otimes"]  &  F\top_\otimes \oplus FA  \arrow[d, "F^\top \oplus FA"] \\
& \top_\oplus \oplus FA \arrow[d, "u^\oplus"] \\
& FA
\end{tikzcd}
\end{center}
\end{itemize}
\end{definition}

\textbf{Remark:} The collection of monoidal categories and their lax monoidal functors forms a category itself.

\begin{definition}
  A monoidal category is \emph{left closed} if for every object $A$, the functor $A \otimes (\--)$ admits a right adjoint.
  \[A \otimes (\--) \dashv A \multimap (\--) \]
\end{definition}
\begin{definition}
  A monoidal category is \emph{right closed} if for every object $A$, the functor $(\--)\otimes A $ admits a right adjoint.
  \[(\--)\otimes A  \dashv (\--) \multimapinv A\]
\end{definition}

\begin{definition}
  A monoidal category is \emph{closed} if it is both left and right closed.
\end{definition}

\subsection{Multicategories}\label{subsec:multicategories}
\begin{definition}
  A \emph{multicategory} $\mc M$ consists of the following data: (\cite{Lam69})

  \begin{itemize}
    \item A collection of objects, denoted $\mc M$ again by abusing notation.
  	\item A collection of (multi)morphisms $\mc M(\Gamma;A)$ for each finite list of objects $\Gamma$ and object $A$.
  	\item An identity morphism $1_A \in \mc M(A;A)$ for every object $A$
  	\item For all multimorphisms $f \in \mc M(\Gamma;A)$ and $g \in \mc M(\Gamma', A, \Gamma'';B)$ a multimorphism $g \circ_A f \in \mc M(\Gamma',\Gamma, \Gamma'';B)$, the composition indexed by the object at which the composition happens
  	\item For every multimorphism $f \in \mc M(\Gamma, A, \Gamma';B)$ an equality $f \circ_A 1_A = f $
  	\item For every multimorphism $f \in \mc M(\Gamma;B)$ an equality $1_B \circ_B f = f $
  	\item Two associativity laws hold whenever the compositions make sense. For $h \in \mc M(\Gamma;A)$, $g \in \mc M(\Delta', A, \Delta'';B)$ and $f \in \mc M(\Lambda', B, \Lambda'';C)$, composing satisfies \[(f \circ_B g) \circ_A h = f \circ_B (g \circ_A h), \] both sides in $\mc M(\Lambda', \Delta', \Gamma, \Delta'', \Lambda'';C)$. For $f \in \mc M(\Gamma_1, A, \Gamma_2, B, \Gamma_3;C)$, $g \in \mc M(\Delta;A)$ and $h \in \mc M(\Lambda;B)$, composing at the two occurrences in either order gives \[(f \circ_A g) \circ_B h = (f \circ_B h) \circ_A g, \] both sides in $\mc M(\Gamma_1, \Delta, \Gamma_2, \Lambda, \Gamma_3;C)$, i.e. compositions at independent objects interchange.
		\end{itemize}
\end{definition}
\textit{Remark:} The index of the composition can be ambiguous when several copies of the same object $A$ appear in the domain of $g$; this can be resolved by explicitly indexing or labelling the input occurrences. Along the same lines as \cite[Remark~2.2.2]{Bla23}, we use the lighter notation aforementioned since no ambiguity will arise in this document.

\textit{Notation}: We will denote multimorphisms using the standard morphism convention, $f : \Gamma \to A$ where the arrow plays the role of the semicolon in the definition. We will also loosely call $\Gamma$ the domain of a morphism $f : \Gamma \to A$, and $A$ the codomain of the morphism.

\begin{definition}
A \emph{tensor multicategory} $\mc M$ is a multicategory equipped with a universal object and universal property:
For any list of objects $\Gamma := \Gamma_1, \ldots, \Gamma_n$ in $\mc M$, there exist an object $\otimes \Gamma$ and a morphism $u_{\Gamma} : \Gamma \to \otimes \Gamma$ such that for all lists of objects $\Gamma' := \Gamma'_1, \ldots, \Gamma'_{n'}$ and $\Gamma'' := \Gamma''_1, \ldots, \Gamma''_{n''}$ and any morphism $f : \Gamma', \Gamma, \Gamma'' \to A$  there exists a unique morphism $u_{\Gamma}^f : \Gamma', \otimes \Gamma, \Gamma'' \to A$ making the factorization triangle commute:
\begin{center}
\begin{tikzcd}
\underset{\Gamma', \Gamma, \Gamma''}{\Gamma} \arrow[dr, swap, "f"]  \arrow[rr, "u_\Gamma"] & & \underset{\Gamma', \otimes \Gamma, \Gamma''}{\otimes \Gamma} \arrow[dl, "u^f_\Gamma"]  \\
& A &
\end{tikzcd}
\end{center}
This is the first instance of an explicit definition of a universal object and a universal property. The definition as a whole is informally described as follows: an object is universal if it is the \emph{best} object that satisfies a criterion. The reader is of course encouraged to investigate as many universal objects and universal properties as possible to get a grasp on them and their relevance in mathematics. Informally, $\otimes \Gamma$ is the \emph{best} object representing the list of formulas $\Gamma := \Gamma_1, \dots, \Gamma_n$.
\end{definition}

\begin{definition}
  Let $\mc M \xto{F} \mc M'$ be a functor of the underlying multicategories of two tensor multicategories. For all objects $\Gamma_1, \ldots, \Gamma_n$ in $\mc M$, applying the universal property of $u_{F\Gamma_1, \ldots, F\Gamma_n}$ to the map $F(u_{\Gamma_1, \ldots, \Gamma_n}) : F\Gamma_1, \ldots, F\Gamma_n \to F(\otimes(\Gamma_1, \ldots, \Gamma_n))$ gives a unique \emph{canonical comparison map}
$$\otimes(F\Gamma_1, \ldots, F\Gamma_n) \to F(\otimes(\Gamma_1, \ldots, \Gamma_n)),$$
thus, every functor is in this sense a lax functor. We call $F$ a \emph{strong} functor whenever the canonical comparisons are isomorphisms. 
\end{definition}
There is a category of tensor multicategories that we will denote $\mathbf{MultiCat_{t}}$. The subscript indicates that it is the category of multicategories where all multicategories are tensor multicategories. Its morphisms are the functors of the underlying multicategories. Composition is functor composition. The canonical comparison of the identity functor is the identity, by the uniqueness in the universal property. Given two functors $\mc M \xto{F} \mc M'$ and $\mc M' \xto{G} \mc M''$, the composite $$\otimes(GF\Gamma_1, \ldots, GF\Gamma_n) \to G(\otimes(F\Gamma_1, \ldots, F\Gamma_n)) \to GF(\otimes(\Gamma_1, \ldots, \Gamma_n))$$ of the canonical comparison of $G$ at the objects $F\Gamma_i$ followed by the image under $G$ of the canonical comparison of $F$ factors $GF(u_{\Gamma_1, \ldots, \Gamma_n})$ through $u_{GF\Gamma_1, \ldots, GF\Gamma_n}$, and is therefore the canonical comparison of $GF$, by uniqueness again. The identity and associativity laws are inherited from the composition of functors.   

\begin{definition}
A \emph{bitensor multicategory} $\mc M$ is a tensor multicategory equipped with a universal object and universal property:
For any lists of objects $\Gamma_1, \Gamma_2$ and object $B$ in the multicategory, there exist an object $U(\Gamma_1 \mid B  \mid \Gamma_2)$ and a map $u_{\Gamma_1, B, \Gamma_2} : \Gamma_1, U(\Gamma_1 \mid B \mid \Gamma_2), \Gamma_2 \to  B$ such that for any map $f : \Gamma_1, \Gamma, \Gamma_2 \to B$  there exists a unique map $u_{\Gamma_1, B, \Gamma_2}^f : \Gamma \to U(\Gamma_1 \mid B \mid \Gamma_2)$ making the factorization triangle commute:
\begin{center}
\begin{tikzcd}
\underset{U(\Gamma_1 \mid B \mid \Gamma_2)}{\Gamma_1,U(\Gamma_1 \mid B \mid \Gamma_2), \Gamma_2 }   \arrow[rr, "u_{\Gamma_1, B, \Gamma_2}"] & & B   \\
\Gamma \arrow[u, "u_{\Gamma_1, B, \Gamma_2}^f"]& &  \Gamma_1, \Gamma, \Gamma_2 \arrow[u, "f", swap] 
\end{tikzcd}
\end{center}
\end{definition}
Finally, we conclude this section on multicategories with a notion that subsumes tensor and bitensor multicategories. A tensor multicategory is sometimes called \emph{representable} because \cref{thm:moncat-multicat-equivalence} identifies tensor multicategories with monoidal categories; see also \cite{Lam69,Her00}. A way to generate the tensors and the universal objects $U$ is to use a Grothendieck (op)fibration. This is useful for us for two reasons:
\begin{itemize}
  \item Grothendieck (op)fibrations offer the technical device necessary to express the universal tensors and homs of a multicategory.
  \item Grothendieck (op)fibrations are general enough to exhibit many universal objects under the same mechanism. It is an instance of ``one definition to rule them all''.
\end{itemize}

To be able to introduce the notion of \emph{op-fibration} we need to generalize the definition of the tensor as a universal object. To do this, we instead define a notion of \emph{cartesian} and \emph{op-cartesian} morphism in typical fashion. It will turn out that tensor objects are instances of objects that \emph{fit} nicely as the objects involved in the (op)-cartesian morphisms. The upshot of those definitions turns out to be useful when moving on to \cref{subsec:polycategories}.

\begin{definition}\label{def:cartesian-multimorphism}
  Given two multicategories $\mc T, \mc M$ and a functor $\mc T \xto{\pi} \mc M$, a multimorphism
  \[
    \Gamma_1, \ldots, \Gamma_i, \ldots, \Gamma_n \xto{f} A
  \]
  is \emph{cartesian} at $\Gamma_i$ if for any list of objects $\Gamma$ and multimorphisms
  \[
    \Gamma_1, \ldots, \Gamma, \ldots, \Gamma_n \xto{g} A
    \qquad\text{and}\qquad
    \pi \Gamma \xto{\beta} \pi \Gamma_i
  \]
  such that $\pi f \circ_{\pi \Gamma_i} \beta = \pi g$, there exists a unique multimorphism $\Gamma \xto{\phi} \Gamma_i$ such that $\pi \phi = \beta$ and $f \circ_{\Gamma_i} \phi = g$, as expressed by the diagrams:

\begin{center}
    \begin{tikzcd}[row sep=large, column sep=huge]
  \underset{\Gamma_i}{\Gamma_1, \ldots, \Gamma_i, \ldots, \Gamma_n} \arrow[rr, "f"]  & & A &    \\
  \Gamma \arrow[u, "\exists!\phi"] & & \Gamma_1, \ldots, \Gamma, \ldots, \Gamma_n  \arrow[u, "g"] & \mc T \arrow[d, "\pi" ] \\
  \underset{\pi \Gamma_i}{\pi \Gamma_1, \ldots, \pi \Gamma_i, \ldots, \pi \Gamma_n} \arrow[rr, "\pi f"]  & & A &   \mc M \\
  \pi \Gamma \arrow[u, "\beta"] & & \pi \Gamma_1, \ldots, \pi \Gamma, \ldots, \pi \Gamma_n  \arrow[u, "\pi g"]  & 
\end{tikzcd}
\end{center}
\end{definition}

\begin{definition}\label{def:opcartesian-multimorphism}
  Given two multicategories $\mc T, \mc M$ and a functor $\mc T \xto{\pi} \mc M$, a multimorphism $\Gamma \xto{f} A$ is \emph{opcartesian} at $A$ if for any lists of objects $\Gamma_1, \Gamma_2$ and multimorphisms $\Gamma_1, \Gamma, \Gamma_2 \xto{g} B$ and $\pi \Gamma_1, \pi A, \pi \Gamma_2 \xto{\beta} \pi B$ such that $\beta \circ_{\pi A} \pi f = \pi g$, there exists a unique multimorphism $\Gamma_1 , A, \Gamma_2 \xto{\phi} B$ such that $\pi \phi = \beta$ and $ \phi \circ_A f = g$, as expressed by the diagrams:

\begin{center}
    \begin{tikzcd}[row sep=large, column sep=huge]
  \underset{\Gamma_1, \Gamma, \Gamma_2}{\Gamma} \arrow[rr, "f"] \arrow[dr, "g"]   & & \underset{\Gamma_1, A, \Gamma_2}{A} \arrow[dl, "\exists!\phi"]  &    \\
 	  & B &  & \mc T \arrow[d, "\pi " ]  \\
  \underset{\pi \Gamma_1, \pi \Gamma, \pi \Gamma_2}{\pi \Gamma} \arrow[rr, "\pi f"] \arrow[dr, "\pi g"]   & & \underset{\pi \Gamma_1, \pi A, \pi \Gamma_2}{\pi A} \arrow[dl, "\beta"]  & \mc M    \\
 	  & \pi B &  &  \\
\end{tikzcd}
\end{center}
\end{definition}

\begin{definition}\label{def:multicategory-fibration}
  A functor of multicategories $\pi : \fib[T]{M}$ is a \textit{fibration} if given any lists of objects $\Gamma_1, \Gamma_2 \in \mc T$, any object $B \in \mc T$, any object $a \in \mc M$ and any multimorphism $\pi \Gamma_1, a,\pi \Gamma_2 \xto{\alpha} \pi B$ in the base multicategory $\mc M$, there exist an object $A \in \mc T$ over $a$, i.e. $\pi A = a$, and a multimorphism $\Gamma_1, A, \Gamma_2 \xto{f} B$, cartesian at $A$, such that $\pi f = \alpha$.             
\end{definition}

\begin{definition}\label{def:multicategory-opfibration}
  A functor of multicategories $\pi : \fib[T]{M}$ is an \textit{opfibration} if given any list of objects $\Gamma \in \mc T$, any object $a \in \mc M$ and any multimorphism $\pi \Gamma \xto{\alpha} a$ in the base multicategory $\mc M$, there exist an object $A \in \mc T$ over $a$ and an opcartesian multimorphism $\Gamma \xto{f} A$ such that $\pi f = \alpha$.             
\end{definition}

\textit{Remark:} We recover the standard definition of a Grothendieck (op)fibration if we consider multimorphisms with only one input. Notice that the lifted object is not asked to be unique; two lifts of the same multimorphism factor through each other by their universal properties and are canonically isomorphic. \\
\textit{Remark:} We can recover the tensor and bitensor multicategories from this definition by taking $\mc M$ to be the terminal multicategory; the opcartesian lifts provide the tensors and the cartesian lifts the universal objects $U(\Gamma_1 \mid B \mid \Gamma_2)$.  

\subsection{Polycategories}\label{subsec:polycategories}
The notion of a polycategory arises from the need to consider mappings with multiple inputs as well as multiple outputs \cite{Sza75}. Classical systems exhibit a duality between assumptions and conclusions; a proof of a conclusion from the assumptions corresponds to a refutation of an assumption from refutations of the conclusions.

\begin{definition}\label{def:polycategory}
  A \emph{polycategory} $\mc M$ consists of data:

  \begin{itemize}
    \item A collection of objects denoted $\mc M$ abusing notation.
    \item A collection of polymorphisms $\mc M(\Gamma;\Delta)$ for all finite lists of objects $\Gamma$ and $\Delta$.
  	\item An identity polymorphism $1_A : \mc M(A;A)$ for every object $A$
  	\item For all polymorphisms $f : \mc M(\Gamma;\Delta',A, \Delta'')$ and $g : \mc M(\Gamma', A, \Gamma'';\Delta)$ a polymorphism $g \circ_A f : \mc M(\Gamma',\Gamma, \Gamma'';\Delta', \Delta, \Delta'')$, subject to the planarity condition that either $\Delta'$ or $\Gamma'$ is empty and either $\Delta''$ or $\Gamma''$ is empty \cite[Definition~2.2.1]{Bla23}. 
  	\item For every polymorphism $f : \mc M(\Gamma, A, \Gamma';\Delta)$ an equality $f \circ_A 1_A = f $
  	\item For every polymorphism $f : \mc M(\Gamma;\Delta, A, \Delta')$ an equality $1_A \circ_A f = f $
  	\item For polymorphisms $h : \Gamma \to \Delta_1, B, \Delta_2$, $g : \Gamma'_1, B, \Gamma'_2 \to \Delta'_1, A, \Delta'_2$ and $f : \Gamma''_1, A, \Gamma''_2 \to \Delta''$, the associativity equality \[(f \circ_A g) \circ_B h =f \circ_A (g \circ_B h), \] 
  	\item For cuts at two objects $A, B$ possibly equal, the \emph{interchange laws} hold whenever the planarity conditions of both cuts do: \[(f \circ_A g) \circ_B h = (f \circ_B h) \circ_A g \quadt{and} f \circ_B (h \circ_A g) = h \circ_A (f \circ_B g)  \] 
		\end{itemize}
\end{definition}

\textit{Notation}: We will denote polymorphisms using the standard map convention, $f : \Gamma \to \Delta$ where the arrow plays the role of the semicolon in the definition. We will also loosely call $\Gamma$ the domain of a map $f : \Gamma \to \Delta$, and $\Delta$ the codomain of the map. For a list $X=(X_1,\ldots,X_r)$ and a function $\nu:[d]\to[r]$, write $X_\nu=(X_{\nu(1)},\ldots,X_{\nu(d)})$.

\begin{definition}[Structural classes]\label{def:structural-classes}
  Write $[n]=\{1,\ldots,n\}$, with $[0]=\varnothing$, and let $\mathbf{Fin}$ denote the category of finite sets and functions. A \emph{structural category} $\mathsf S$ is a subcategory of $\mathbf{Fin}$ that contains every bijection and is closed under finite coproducts and pullback along arbitrary maps. The structural classes used here are:
  \[
    \mathsf{Bij},\qquad \mathsf{Inj},\qquad
    \mathsf{Surj},\qquad \mathsf{Fin},
  \]
  consisting respectively of all bijections, injections, surjections, and functions.
\end{definition}

\begin{definition}[Structural cut]\label{def:structural-cut}
  Let $\lambda:[s]\to[r]$, let $a\in[r]$ with singleton fibre $\lambda^{-1}(a)=\{\widehat a\}$, and let $\mu:[n]\to[m]$. Cutting the map $\mu$ in $\lambda$ at $a$ is the map
  \[
    \lambda\star_a\mu:[s-1+n]\longrightarrow[r-1+m]
  \]
  defined by
  \[
    (\lambda\star_a\mu)(x)=
    \begin{cases}
      \iota_{a,m}(\lambda(x)),
        &1\leq x<\widehat a,\\
      \kappa_{a,m}(\mu(x-\widehat a+1)),
        &\widehat a\leq x<\widehat a+n,\\
      \iota_{a,m}(\lambda(x-n+1)),
        &\widehat a+n\leq x\leq s-1+n.
    \end{cases}
  \]
  where
  \[
    \iota_{a,m}:[r]\setminus\{a\}\longrightarrow[r-1+m],
    \qquad
    \iota_{a,m}(j)=
    \begin{cases}
      j,&j<a,\\
      j+m-1,&j>a,
    \end{cases}
  \]
  and
  \[
    \kappa_{a,m}:[m]\longrightarrow[r-1+m],
    \qquad
    \kappa_{a,m}(k)=a+k-1.
  \]
\end{definition}

\begin{example}
Let $\lambda:[3]\to[3]$ be the identity, take $a=\widehat a=2$, and let $\mu:[3]\to[2]$ be given by $\mu(1)=\mu(2)=1$ and $\mu(3)=2$. Then
\[
  \lambda\star_2\mu=(1,2,2,3,4):[5]\longrightarrow[4].
\]
\end{example}

\begin{definition}[Reindexed polycategory]\label{def:reindexed-polycategory}
  Let $\mathsf L$ and $\mathsf R$ be structural classes. An \emph{$(\mathsf L,\mathsf R)$-reindexed polycategory} $\mc P$ consists of a polycategory together with unrestricted cut operations
  \[
    \mc P(\Gamma;\Delta_1,A,\Delta_2)
    \times
    \mc P(\Gamma_1,A,\Gamma_2;\Delta)
    \longrightarrow
    \mc P(\Gamma_1,\Gamma,\Gamma_2;\Delta_1,\Delta,\Delta_2),
    \qquad
    (f,g)\longmapsto g\circ_A f,
  \]
  extending the planar cuts and satisfying the unit and nested associativity laws, together with reindexing structure.
  \begin{itemize}
    \item For lists $\Gamma=(A_1,\ldots,A_m)$ and $\Delta=(B_1,\ldots,B_r)$, and functions $\sigma:[n]\to[m]$ and $\tau:[p]\to[r]$, write
    \[
      \Gamma_\sigma=(A_{\sigma(1)},\ldots,A_{\sigma(n)}),
      \qquad
      \Delta_\tau=(B_{\tau(1)},\ldots,B_{\tau(p)}).
    \]
    For every $\sigma\in\mathsf L$ and $\tau\in\mathsf R$, there is a reindexing map
    \[
      (\--)^{\sigma\mid\tau}:
      \mc P(\Gamma_\sigma;\Delta_\tau)
      \longrightarrow \mc P(\Gamma;\Delta).
    \]
    If $\sigma':[k]\to[n]$ and $\tau':[q]\to[p]$ are also admissible, then
    \[
      f^{\1\mid\1}=f,
      \qquad
      \bigl(f^{\sigma'\mid\tau'}\bigr)^{\sigma\mid\tau}
      =f^{\sigma\circ\sigma'\mid\tau\circ\tau'}.
    \]
  \end{itemize}

  Reindexing is equivariant with respect to cut whenever the maps have singleton fibres at the cut positions. More precisely, suppose
  \[
    f\in\mc P\bigl(\Gamma_\sigma;(\Delta_1,A,\Delta_2)_\tau\bigr),
    \qquad
    g\in\mc P\bigl((\Gamma_1,A,\Gamma_2)_\rho;\Delta_\upsilon\bigr),
  \]
  where $\sigma,\rho\in\mathsf L$, $\tau,\upsilon\in\mathsf R$, and the fibre of $\tau$ over the output $A$ and the fibre of $\rho$ over the input $A$ are singletons. Set $a=|\Gamma_1|+1$ and $b=|\Delta_1|+1$ and define
  \[
    \rho\star_A\sigma:=\rho\star_a\sigma,
    \qquad
    \tau\star_A\upsilon:=\tau\star_b\upsilon.
  \]
  Then
  \[
    \bigl(g^{\rho\mid\upsilon}\bigr)\circ_A
    \bigl(f^{\sigma\mid\tau}\bigr)
    =
    (g\circ_A f)^{\rho\star_A\sigma\mid\tau\star_A\upsilon}.
  \]

  Finally, cuts at independent objects commute with permutations. Suppose
  \[
    f:\Gamma\to\Delta_1,A,\Delta_2,
    \qquad
    g:\Gamma'\to\Delta'_1,B,\Delta'_2,
    \qquad
    h:\Sigma_1,A,\Sigma_2,B,\Sigma_3\to\Theta,
  \]
  and 
    $(\Delta'_1,\Delta_1,\Theta,\Delta_2,\Delta'_2)_{\tau}
    =
    (\Delta_1,\Delta'_1,\Theta,\Delta'_2,\Delta_2).$
  Then
  \[
    (h\circ_A f)\circ_B g
    =
    \bigl((h\circ_B g)\circ_A f\bigr)^{\1\mid\tau}.
  \]
  Dually, suppose
  \[
    f:\Gamma\to\Delta_1,A,\Delta_2,B,\Delta_3,
    \qquad
    g:\Sigma_1,A,\Sigma_2\to\Theta,
    \qquad
    h:\Sigma'_1,B,\Sigma'_2\to\Theta',
  \]
  and
    $(\Sigma'_1,\Sigma_1,\Gamma,\Sigma_2,\Sigma'_2)_{\sigma}
    =
    (\Sigma_1,\Sigma'_1,\Gamma,\Sigma'_2,\Sigma_2).$
  Then
  \[
    h\circ_B(g\circ_A f)
    =
    \bigl(g\circ_A(h\circ_B f)\bigr)^{\sigma\mid\1}.
  \]
\end{definition}

\begin{definition}[Structural variants]\label{def:symmetric-polycategory}
  A $(\mathsf{Bij},\mathsf{Bij})$-reindexed polycategory is a \emph{symmetric polycategory}. More generally:
  \begin{itemize}
    \item an $(\mathsf{Inj},\mathsf{Inj})$-reindexed polycategory has \emph{affine} structural actions;
    \item a $(\mathsf{Surj},\mathsf{Surj})$-reindexed polycategory has \emph{relevant} structural actions;
    \item a $(\mathsf{Fin},\mathsf{Fin})$-reindexed polycategory has \emph{cartesian} structural actions and will be called a \emph{cartesian polycategory};
    \item an $(\mathsf S,\mathsf{Bij})$-reindexed polycategory is \emph{left-$\mathsf S$}, and a $(\mathsf{Bij},\mathsf S)$-reindexed polycategory is \emph{right-$\mathsf S$}. In particular, this gives left- and right-affine, left- and right-relevant, and left- and right-cartesian polycategories.
  \end{itemize}
\end{definition}

\begin{definition}
  A functor $F:\mc P\to\mc Q$ of $(\mathsf L,\mathsf R)$-reindexed polycategories consists of a function on objects and maps
  \[
    F:\mc P(\Gamma;\Delta)\longrightarrow\mc Q(F\Gamma;F\Delta)
  \]
  satisfying
  \[
    F(\1_A)=\1_{FA},
    \qquad
    F(g\circ_A f)=Fg\circ_{FA}Ff,
    \qquad
    F\bigl(f^{\sigma\mid\tau}\bigr)=(Ff)^{\sigma\mid\tau}.
  \]
  We write $\mathbf{SymPolyCat}$ for the category obtained in the $(\mathsf{Bij},\mathsf{Bij})$ case.
\end{definition}

\textit{Remark:} Restricting the cuts of a reindexed polycategory to the planar configurations of \cref{def:polycategory} and forgetting all reindexings gives an underlying planar polycategory.

\begin{example}[Categories]
A polycategory $\mc C$ such that every collection of morphisms $\mc C(\Gamma;\Delta)$ is empty unless $\Gamma$ and $\Delta$ each contain exactly one object is a category. Indeed, the objects of the category are the objects of the polycategory, and the morphisms $f : A \to B$ are the morphisms in the collection $\mc C(A;B)$. Identity maps are the same, cut reduces to composition in the category, and associativity and identity laws hold.
\end{example}
\begin{example}[Multicategories]
A polycategory $\mc C$ such that every collection of morphisms $\mc C(\Gamma;\Delta)$ is empty unless $\Delta$ contains exactly one object is a multicategory. The objects of the multicategory are the same objects as those of the polycategory and the morphisms $f : \Gamma \to A$ are the morphisms in the collection $\mc C(\Gamma;A)$. Identity maps are the same, cut reduces to composition in the multicategory, and associativity and identity laws hold.
\end{example}

\begin{definition}
  The \emph{terminal polycategory} denoted $\1$ is the polycategory with a unique object $\star$  such that every collection of polymaps $\1(\star^n; \star^m)$ contains a unique element for all $n,m \in \mathbb{N}$.
\end{definition}
\textit{Remark}: Besides being terminal in the category of polycategories, $\1$ converts bicartesian structure into a bifibration by \cref{thm:bicartesian-terminal-bifibration}. The terminal base is also used in the CMLL presentation of \cref{sec:examples-blanco-cmll}, whose proved comparison is the logical adequacy of \cref{prop:logical-adequacy-cmll} \cite{Bla23}.

\begin{definition}
  A polycategory such that every collection of polymaps $\mc M(\Gamma;\Delta)$ is a category, and every cut operation $\circ_A$ is a functor of two variables $\mc M(\Gamma;\Delta_1, A, \Delta_2) \times \mc M(\Gamma_1, A, \Gamma_2;\Delta) \to \mc M(\Gamma_1, \Gamma, \Gamma_2;\Delta_1, \Delta, \Delta_2)$, the identity, associativity and interchange laws holding as equalities of functors, will be called a $2$\emph{-polycategory}. 
\end{definition}

Up to this point, a polycategory can be thought of as a category such that the morphisms have multiple inputs and multiple outputs. This immediately prevents the standard specification of a category from being applied as no notion of composition is well defined. Instead, a notion of composition at a specific object is investigated. This closely matches the cut operation in logics if we interpret a sequent $\Gamma \vdash \Delta$ as a polymorphism $\Gamma \to \Delta$. The cut operations then match in both settings.  

\begin{definition}[Elementary structural operations]\label{def:elementary-polycategory-reindexings}
  The familiar exchange, weakening, and contraction operations are instances of context reindexing. Let each displayed source and target list determine its positional map.
  \begin{itemize}
    \item An adjacent transposition gives the exchange maps
    \[
      e_{A\leftrightarrow B}:
      \mc P(\Gamma_1,A,B,\Gamma_2;\Delta)
      \longrightarrow
      \mc P(\Gamma_1,B,A,\Gamma_2;\Delta),
    \]
    and dually $e^{A\leftrightarrow B}$ on outputs.
    \item An injection whose image omits $A$ gives the weakening maps
    \[
      w_A:\mc P(\Gamma_1,\Gamma_2;\Delta)
      \longrightarrow\mc P(\Gamma_1,A,\Gamma_2;\Delta),
    \]
    and dually $w^A$ on outputs.
    \item A surjection identifying two occurrences of $A$ gives the contraction maps
    \[
      c_{A_1}:\mc P(\Gamma,A,\Gamma_1,A,\Gamma_2;\Delta)
      \longrightarrow\mc P(\Gamma,A,\Gamma_1,\Gamma_2;\Delta),
    \]
    \[
      c_{A_2}:\mc P(\Gamma,A,\Gamma_1,A,\Gamma_2;\Delta)
      \longrightarrow\mc P(\Gamma,\Gamma_1,A,\Gamma_2;\Delta),
    \]
    and dually $c^{A_1},c^{A_2}$ on outputs. The subscript records which occurrence is retained.
  \end{itemize}
  Thus bijective actions give exchange, injective actions give exchange and weakening, surjective actions give exchange and contraction, and arbitrary finite-map actions give all three. Every finite function factors as a surjection followed by an injection, while the strict action laws make the resulting operation independent of such a factorization. 
\end{definition}

In a $2$-polycategory all the reindexing maps, and hence the elementary structural operations, are required to be functors.

\begin{definition}\label{def:left-biased-cartesian-polycategory}
  A \emph{strict left-biased cartesian polycategory} is a cartesian polycategory with additional laws for a non-bijective reindexing acting at the cut object.
  \begin{itemize}
    \item \textbf{weakening laws.} Write $w_{\Gamma}$ and $w^{\Delta}$ for block weakenings, i.e. the reindexings induced by the injections omitting the displayed block. For
    \[
      u:\Gamma\to A,
      \qquad
      f':\Gamma_1,\Gamma_2\to\Delta,
    \]
    require
    \[
      w_A(f')\circ_Au=w_{\Gamma}(f')
      \quad\in\quad
      \mc P(\Gamma_1,\Gamma,\Gamma_2;\Delta).
    \]
    Dually, for $n:A\to\Delta$ and $g':\Gamma\to\Delta_1,\Delta_2$, with $(\Delta_1,\Delta_2)$ nonempty, require
    \[
      n\circ_Aw^A(g')=w^{\Delta}(g')
      \quad\in\quad
      \mc P(\Gamma;\Delta_1,\Delta,\Delta_2).
    \]

    \item \textbf{contraction laws.} Write $c_{\Lambda}$ and $c^{\Delta}$ for the block contractions induced by surjections. Given two separately marked occurrences $A_1,A_2$ of the same object $A$ and maps
    \[
      f:\Gamma_1,A_1,\Gamma_2,A_2,\Gamma_3\to\Delta,
      \qquad
      g:\Lambda\to\Delta_1,A,\Delta_2,
    \]
    let $g^{(1)},g^{(2)}$ be two copies of $g$ with separately marked boundary occurrences. Write $c_{\Lambda_i}$ for the block retaining the position occupied by the $i$th copy of $\Lambda$. For $i\in\{1,2\}$ require
    \[
      c_{A_i}(f)\circ_Ag
      =
      (c_{\Lambda_i}\circ c^{\Delta_1}\circ c^{\Delta_2})
      \bigl((f\circ_{A_1}g^{(1)})\circ_{A_2}g^{(2)}\bigr).
    \]
    Dually, for two separately marked output occurrences $A_1,A_2$ of $A$ and maps
    \[
      f:\Gamma\to\Delta_1,A_1,\Delta_2,A_2,\Delta_3,
      \qquad
      g:\Gamma_1,A,\Gamma_2\to\Lambda,
    \]
    and separately marked copies $g^{(1)},g^{(2)}$, write $c^{\Lambda_i}$ for the output block retaining the position occupied by the $i$th copy of $\Lambda$, and require
    \[
      g\circ_Ac^{A_i}(f)
      =
      (c_{\Gamma_1}\circ c_{\Gamma_2}\circ c^{\Lambda_i})
      \bigl(g^{(2)}\circ_{A_2}(g^{(1)}\circ_{A_1}f)\bigr).
    \]
    Canonical block transports in these equations are implicit.
  \end{itemize}
\end{definition}

By definition, a functor between cartesian polycategories is a functor of $(\mathsf{Fin},\mathsf{Fin})$-reindexed polycategories. It therefore preserves exchange, weakening, and contraction automatically, since these are elementary reindexings.

\textit{Notation}: For $\Gamma = A_1, \ldots, A_n$ we will denote the list of images $FA_1, \ldots, FA_n$ by the shorthand $F\Gamma$.\\
A functor of planar polycategories is obtained by retaining only the object, identity, and cut-preservation clauses in the compatible planar configurations.

\begin{definition}
  A polycategory $\mc M$ has \emph{simple factorizations} if for every map $f : \Gamma \to \Delta$ there exist an object $A$ and two maps $g : \Gamma \to A$ and $h : A \to \Delta$ such that $h \circ_A g = f$.
\end{definition}

\begin{definition}
  A \emph{two-tensor polycategory} is a polycategory such that for all lists of objects $\Gamma, \Delta$, there exist objects $\otimes \Gamma$, $\oplus \Delta$ and maps $u_{\Gamma} : \Gamma \to \otimes \Gamma$, $u^{\Delta}: \oplus \Delta \to \Delta$ such that for any maps $f : \Gamma_1, \Gamma, \Gamma_2 \to \Delta$ and $g : \Gamma \to \Delta_1, \Delta, \Delta_2$ there exist unique maps $u_{\Gamma}^f : \Gamma_1, \otimes \Gamma, \Gamma_2 \to \Delta$ and $u^{\Delta}_g : \Gamma \to \Delta_1, \oplus \Delta, \Delta_2$ making the two factorization triangles commute:
\begin{center}
  \begin{tikzcd}
    \underset{\Gamma_1, \Gamma, \Gamma_2}{\Gamma} \arrow[dr, swap, "f"]  \arrow[rr, "u_\Gamma"] & & \underset{\Gamma_1, \otimes \Gamma, \Gamma_2}{\otimes \Gamma} \arrow[dl, "u^f_\Gamma"]  \\
    & \Delta &
   \end{tikzcd}
   \quad
  \begin{tikzcd}
    \underset{\Delta_1, \oplus \Delta, \Delta_2}{\oplus \Delta}  \arrow[rr, "u^\Delta"]   & & \underset{\Delta_1, \Delta, \Delta_2}{\Delta}    \\
    & \Gamma \arrow[ur, swap, "g"] \arrow[ul, "u_g^\Delta"] &
   \end{tikzcd}
\end{center}
  \[
		 \text{ie,} \quad u_\Gamma^f \circ_{\otimes \Gamma} u_\Gamma  = f  \quadt{and}  u^\Delta \circ_{\oplus \Delta} u^\Delta_g  = g
  \]

\end{definition}
In a two-tensor polycategory, the universal objects at the empty list deserve a special mention. The definition provides the two maps
\[u_{()} : () \to \otimes() \quad u^{()} : \oplus() \to ()\]
and the objects $\otimes()$ and $\oplus()$ play the roles of the units $\top$ and $\bot$ in the two-tensor construction of \cref{subsec:two-tensor-polycategories}. Notice that no map $\otimes() \to \oplus()$ is provided. The presence of this map corresponds to the nullary mix rule of linear logic \cite{Gir87}.

Promoting the operators $\otimes, \oplus$ to functors will require us to specify their action on morphisms when interpreting polycategories in weakly distributive categories. The universal objects provide the action one cut at a time. For $f : A \to B$ and $g : C \to D$, cutting $f$ and then $g$ into the universal map $u_{B,D} : B, D \to \otimes(B,D)$ produces the map \[(u_{B,D} \circ_B f) \circ_D g : A, C \to \otimes(B,D),\] and the $\oplus$ side is dual, cutting into $u^{A,C} : \oplus(A,C) \to A, C$ instead. Notice that it is not possible to merge two unrelated maps $\Gamma \to \Delta$ and $\Lambda \to \Theta$ into a single map $\Gamma, \Lambda \to \Delta, \Theta$ by way of cutting at the empty list. If that were possible, it would correspond to the mix rule in linear logic and we do not require our polycategory to possess it.

\begin{definition}
  A functor $F : \mc P \to \mc Q$ between two two-tensor polycategories is a functor of the underlying polycategories that respects the universal objects, i.e.
\[ F(\otimes_{\mc P} \Gamma) = \otimes_{\mc Q} F\Gamma \quadt{and} F(\oplus_{\mc P} \Delta) = \oplus_{\mc Q} F\Delta\]
and carries the universal maps to the universal maps, $F(u_{\Gamma}) = u_{F\Gamma}$ and $F(u^{\Delta}) = u^{F\Delta}$. The preservation is asked for on the nose, i.e. the functors are strict; a pseudo version relaxing the equalities to coherent isomorphisms exists and will not be needed here.

\end{definition}

\begin{definition}
  Given a cartesian polycategory $\mc P$ and lists of objects $\Gamma, \Delta_1, \Delta_2$, we say that a map $f : \Gamma \to \Delta_1, A, \Delta_2$ is \emph{opcartesian at the object $A$} if it has the universal property: \\
	For any other map $g : \Gamma_1, \Gamma, \Gamma_2 \to \Delta_1, \Delta, \Delta_2$ there exists a unique map $u_f : \Gamma_1, A, \Gamma_2 \to \Delta$ making the factorization triangle commute:
  \begin{center}
    \begin{tikzcd}
			\underset{\Gamma_1, \Gamma, \Gamma_2}{\Gamma} \arrow[r, "f"]\arrow[d, "g"]    &   \underset{\Gamma_1, A,  \Gamma_2}{\Delta_1, A, \Delta_2}\arrow[d, "u_f"] \\
			\Delta_1, \Delta, \Delta_2   &  \Delta 
    \end{tikzcd}
    \quad In other words $u_f \circ_A f = g$ 
  \end{center}  
\end{definition}

\textit{Notation: }Since the object $A$ is unique up to isomorphism, we write
\[
	FR : \Gamma \to \Delta_1, F(\Delta_1|\Gamma|\Delta_2), \Delta_2
\]
and the unique map generated by $g$ as $u_g^{FR}$ 

\begin{definition}
  Given a cartesian polycategory $\mc P$ and lists of objects $\Gamma_1, \Gamma_2, \Delta$, we say that a map $f : \Gamma_1, A, \Gamma_2 \to \Delta$ is \emph{cartesian} at the object $A$ if it has the universal property: \\
	For any other map $g : \Gamma_1, \Gamma, \Gamma_2 \to \Delta_1, \Delta, \Delta_2$ there exists a unique map $u^f : \Gamma \to \Delta_1, A, \Delta_2$ making the factorization triangle commute:
  \begin{center}
    \begin{tikzcd}
			\underset{\Delta_1, A, \Delta_2}{\Gamma_1, A, \Gamma_2} \arrow[r, "f"]    &   \underset{\Delta_1, \Delta,  \Delta_2}{\Delta} \\
			\Gamma \arrow[u, "u^f"]   &  \Gamma_1, \Gamma, \Gamma_2 \arrow[u, "g"]  
    \end{tikzcd}
    \quad In other words $f \circ_A u^f = g$ 
  \end{center}
\end{definition}

\textit{Notation: }Since the object $A$ is unique up to isomorphism, we write
\[
	UL : \Gamma_1, U(\Gamma_1|\Delta|\Gamma_2), \Gamma_2 \to \Delta
\] 
and the unique map generated by $g$ as $u_g^{UL}$ 

\begin{example}
Taking $\Delta_i$ empty in opcartesianness at $A$ recovers the tensor $\otimes \Gamma$. Taking $\Gamma_i$ empty in cartesianness at $A$ recovers the cotensor $\oplus \Delta$ in a two-tensor polycategory.
\end{example}

\begin{definition}
A functor $F : \mc D \to \mc M$ between two cartesian $2$-polycategories is a \emph{local fibration} if the induced functors $F_{\Gamma;\Delta} : \mc D(\Gamma;\Delta) \to \mc M(F\Gamma;F\Delta)$ are fibrations in the categorical sense.
\end{definition}

An alternative way of expressing cartesian and opcartesian morphisms in the context of polycategories is to reason about the collections of polymorphisms directly. It turns out that a polymorphism being cartesian or opcartesian corresponds to the existence of a pullback diagram.
\begin{definition}
If $\pi : \mc D \to \mc M$ is a functor between two cartesian polycategories, then a map $f : \Gamma_1, A, \Gamma_2 \to \Delta$ in $\mc D$ is cartesian at $A$ when composition with $f$ and projection along $\pi$ form a pullback of sets for all lists $\Gamma, \Delta_1, \Delta_2$:
\begin{center}
\begin{tikzcd}
    \mc D(\Gamma;\Delta_1, A, \Delta_2) \arrow[r, "f \circ (\--)"] \arrow[d, "\pi"] & \mc D(\Gamma_1, \Gamma, \Gamma_2;\Delta_1, \Delta, \Delta_2) \arrow[d, "\pi"]   \\
    \mc M(\pi\Gamma;\pi\Delta_1, \pi A, \pi\Delta_2) \arrow[r, "\pi f \circ (\--)"] & \mc M(\pi\Gamma_1, \pi\Gamma, \pi\Gamma_2;\pi\Delta_1, \pi\Delta, \pi\Delta_2)  \\
\end{tikzcd}
\end{center}
\end{definition}

\begin{definition}
If $\pi : \mc D \to \mc M$ is a functor between two cartesian polycategories, then a map $f : \Gamma \to \Delta_1, A, \Delta_2$ in $\mc D$ is opcartesian at $A$ when composition with $f$ and projection along $\pi$ form a pullback of sets for all lists $\Gamma_1, \Gamma_2, \Delta$:
\begin{center}
\begin{tikzcd}
    \mc D(\Gamma_1, A, \Gamma_2;\Delta) \arrow[r, "(\--) \circ f"] \arrow[d, "\pi"] & \mc D(\Gamma_1, \Gamma, \Gamma_2;\Delta_1, \Delta, \Delta_2) \arrow[d, "\pi"]   \\
    \mc M(\pi\Gamma_1, \pi A, \pi \Gamma_2;\pi \Delta) \arrow[r, "(\--) \circ \pi f"] & \mc M(\pi\Gamma_1, \pi\Gamma, \pi\Gamma_2;\pi\Delta_1, \pi\Delta, \pi\Delta_2)  \\
\end{tikzcd}
\end{center}
\end{definition}
In the context of $2$-polycategories the collections of polymaps are categories and the squares are asked to be pullbacks of categories. 
\textit{Remark:} Note that in the context of strict cartesian polycategories, the diagrams simplify a tiny amount, since order does not matter thanks to the exchange functors and their coherence laws. We can work with the two simplified diagrams:

\begin{center}
\begin{tikzcd}
    \mc D(\Gamma;\Delta, A) \arrow[r, "f \circ (\--)"] \arrow[d, "\pi"] & \mc D(\Gamma', \Gamma;\Delta, \Delta') \arrow[d, "\pi"]   \\
    \mc M(\pi\Gamma;\pi\Delta, \pi A) \arrow[r, "\pi f \circ (\--)"] & \mc M(\pi\Gamma', \pi\Gamma;\pi\Delta, \pi\Delta')  \\
\end{tikzcd}
\end{center}

\begin{center}
\begin{tikzcd}
    \mc D(\Gamma, A;\Delta) \arrow[r, "(\--) \circ f"] \arrow[d, "\pi"] & \mc D(\Gamma', \Gamma;\Delta, \Delta') \arrow[d, "\pi"]   \\
    \mc M(\pi\Gamma, \pi A;\pi \Delta) \arrow[r, "(\--) \circ \pi f"] & \mc M(\pi\Gamma', \pi\Gamma;\pi\Delta, \pi\Delta')  \\
\end{tikzcd}
\end{center}
for cartesian and opcartesian morphisms. The LNL framework uses the corresponding zone-sensitive notions of \cref{def:lnl-cartesian-morphism,def:lnl-opcartesian-morphism}, since its contexts may be cartesian, linear, or mixed.

\begin{definition}
A functor $\pi : \mc D \to \mc M$ between strict cartesian $2$-polycategories is a \emph{pull-fibration} if for every object $a \in \mc M$, all lists of objects $\Gamma, \Delta$ in $\mc D$ and every polymap $g : \pi\Gamma, a \to \pi\Delta$ in $\mc M$ there exist $g^\star(\Gamma \mid \Delta) \in \mc D$ over $a$, i.e. $\pi g^\star(\Gamma \mid \Delta) = a$, and a map $g_a^\star : \Gamma, g^\star(\Gamma \mid \Delta)  \to \Delta$ that is cartesian at $g^\star(\Gamma \mid \Delta)$ with $\pi g_a^\star = g$.
\end{definition}

\begin{definition}
A functor $\pi : \mc D \to \mc M$ between strict cartesian $2$-polycategories is a \emph{push-fibration}  if for every object $a \in \mc M$, all lists of objects $\Gamma, \Delta$ in $\mc D$ and every map $g : \pi\Gamma \to \pi\Delta, a$ there exist $g_\star(\Gamma \mid \Delta) \in \mc D$ over $a$, i.e. $\pi g_\star(\Gamma \mid \Delta) = a$, and a map $g_\star^a : \Gamma \to \Delta,g_\star(\Gamma \mid \Delta) $ that is opcartesian at $g_\star(\Gamma \mid \Delta)$ with $\pi g_\star^a = g$.
\end{definition}

\begin{definition}
  A functor $\pi : \mc D \to \mc M$ that is a pull-fibration and a push-fibration is called a \textit{Bifibration}. 
\end{definition}
The categorical semantics of the framework gives the fibration, together with one-output opcartesian lifts, of \cref{prop:type-theory-presents-lnl-polycategory}. This lifting apparatus provides the adjoint structure containing the information needed to encode the logical connectives.

\subsection{Weakly Distributive Categories}
The next step after monoidal categories is to equip our category of interest with a second monoidal structure. On its own, the addition is of limited interest, as the two monoidal operators could be treated in total isolation from each other. It becomes useful when the two operators interact with each other. The interaction will be very reminiscent of the relationship between the logical operators of conjunction and disjunction in any mainstream logic (e.g. classical logic). The formulation of a category with two monoidal structures and some interaction between them will be a little more general than the rigid specification of the logical operators $\land, \lor$ in classical logic with the De Morgan rules. This allows us to move towards a semantics for linear logic, which is the prime logic for accommodating substructural rules. Such categories are called \emph{weakly distributive categories} \cite{CS97,BCST96}. The qualifier weak indicates a weaker version of the distribution of the monoidal operators, and the qualifier distributive, as its name indicates, suggests the interaction between the monoidal operators is of a distributive nature.
\begin{definition}
  A \emph{weakly distributive category} is a category $\mc C$ equipped with two monoidal structures $\otimes, \oplus$ and two \emph{associative} natural transformations:
  \begin{align*}
	a_{\otimes\oplus} : A \otimes (B \oplus C) \to (A \otimes B) \oplus C \\
	a^{\otimes\oplus} : (A \oplus B) \otimes C \to A \oplus (B \otimes C) 
	\end{align*}
	as well as \emph{distributivity} natural transformations:
  \begin{align*}
	d_\otimes : A \otimes (B \oplus C) &\to B \oplus (A \otimes C)\\
	d^\otimes : (B \oplus C) \otimes A &\to (B \otimes A) \oplus C 
	\end{align*}
	natural in each of their arguments and satisfying coherence relations that make the monoidal operators behave well with each other.
\end{definition}

Before stating the coherence laws in full, let us lighten the notation and write $\top$ for the unit of $\otimes$ and $\bot$ for the unit of $\oplus$, i.e. shorthand for the objects $\top(\star)$ and $\bot(\star)$.\\
\textbf{Coherence laws:}
	\begin{itemize}
  	\item \textbf{Unit and Associativity laws:} We start by writing the laws specifying that the units $\top, \bot$ behave correctly with respect to associativity of $\otimes, \oplus$. The diagrams for $\otimes$ that are required to commute are:
  \begin{center}
    \begin{tikzcd}[row sep=large, column sep=huge]
      \top \otimes (A \oplus B)  \arrow[r, "u^{\otimes} "] \arrow[d, swap, "a_{\otimes\oplus(\top, A, B)}"]  & A \oplus B \\
			(\top \otimes A) \oplus B \arrow[ru,swap, "u^\otimes \oplus B"]   
    \end{tikzcd}
    \quad
    \begin{tikzcd}[row sep=large, column sep=huge]
      (A \oplus B)\otimes \top   \arrow[r, "u_{\otimes} "] \arrow[d, swap, "a^{\otimes\oplus(A,B,\top)}"]  & A \oplus B \\
			 A \oplus (B \otimes \top) \arrow[ru,swap, " A \oplus u_{\otimes} "]   
    \end{tikzcd}
  \end{center}
  The ones for $\oplus$ are:

  \begin{center}
    \begin{tikzcd}[row sep=large, column sep=huge]
      A \otimes (B \oplus \bot)  \arrow[r, "A \otimes u_{\oplus} "] \arrow[d, swap, "a_{\otimes\oplus( A, B, \bot)}"]  & A \otimes B \\
			(A \otimes B) \oplus \bot \arrow[ru,swap, "u_\oplus"]   
    \end{tikzcd}
    \quad
    \begin{tikzcd}[row sep=large, column sep=huge]
      (\bot \oplus A) \otimes B  \arrow[r, "u^{\oplus}\otimes B "] \arrow[d, swap, "a^{\otimes\oplus(\bot, A, B)}"]  & A \otimes B \\
			\bot \oplus (A \otimes B) \arrow[ru,swap, "u^\oplus"]   
    \end{tikzcd}
  \end{center}

  	\item \textbf{Unit and Distributive laws:} We turn to the laws for units and distributivity. They are similar to the associativity ones and in fact are derivable when the monoidal structures are symmetric. The ones for $\otimes$ are:

  \begin{center}
    \begin{tikzcd}[row sep=large, column sep=huge]
      \top \otimes (A \oplus B)  \arrow[r, "u^{\otimes} "] \arrow[d, swap, "d_{\otimes(\top, A, B)}"]  & A \oplus B \\
			A \oplus (\top \otimes B) \arrow[ru,swap, "A \oplus u^{\otimes}"]   
    \end{tikzcd}
    \quad
    \begin{tikzcd}[row sep=large, column sep=huge]
      (A \oplus B)\otimes \top   \arrow[r, "u_{\otimes} "] \arrow[d, swap, "d^{\otimes(A, B, \top)}"]  & A \oplus B \\
			 (A\otimes \top) \oplus B \arrow[ru,swap, " u_{\otimes}\oplus B "]   
    \end{tikzcd}
  \end{center}
  and for $\oplus$:

  \begin{center}
    \begin{tikzcd}[row sep=large, column sep=huge]
      A \otimes (\bot \oplus B)  \arrow[r, "A \otimes u^{\oplus} "] \arrow[d, swap, "d_{\otimes( A,\bot, B)}"]  & A \otimes B \\
			\bot \oplus (A \otimes B) \arrow[ru,swap, "u^{\oplus}"]   
    \end{tikzcd}
    \quad
    \begin{tikzcd}[row sep=large, column sep=huge]
      (A \oplus \bot) \otimes B  \arrow[r, "u_{\oplus} \otimes B "] \arrow[d, swap, "d^{\otimes( A,\bot, B)}"]  & A \otimes B \\
			(A \otimes B) \oplus \bot \arrow[ru,swap, "u_{\oplus}"]   
    \end{tikzcd}
  \end{center}
  \item \textbf{Associativity laws:} Associativity can be applied in any order for any monoidal operator. There is one pentagon for each mixed order of appearance of $\otimes,\oplus$ relating the objects $A,B,C,D$, which totals six. The six pentagons and four unit triangles are exactly the coherence conditions asked of the planar maps in \cite{CS97,BCST96}, in the count recalled by \cite[Definition~2.1.12]{Bla23}, beyond the monoidal structures. The pentagons are:
  \begin{itemize}

    \item For the order $\oplus, \otimes, \otimes$, require commutativity of
\begin{center}
  \begin{tikzcd}[row sep=large, column sep=huge]
    ((A \oplus B) \otimes C) \otimes D \arrow[r, "a_{\otimes(A \oplus B, C,D)}"] \arrow[d, "a^{\otimes\oplus(A,B,C)} \otimes D"] & (A \oplus B) \otimes (C \otimes D) \arrow[r, "a^{\otimes\oplus(A, B, C \otimes D)}"] & A \oplus (B \otimes (C \otimes D))\\
  	(A \oplus (B \otimes C)) \otimes D \arrow[r, "a^{\otimes\oplus(A, B \otimes C, D)}"] & A \oplus ((B \otimes C) \otimes D) \arrow[ru,swap, "A\oplus a_{\otimes(B,C,D)}"] 
  \end{tikzcd}
\end{center}

    \item For the order $\otimes, \otimes, \oplus$, require commutativity of
\begin{center}
  \begin{tikzcd}[row sep=large, column sep=huge]
    A \otimes (B \otimes (C \oplus D)) \arrow[r, "a_{\otimes(A, B, C\oplus D)}^{-1} "] \arrow[d, "A \otimes a_{\otimes\oplus(B,C,D)}"] & (A \otimes B) \otimes (C \oplus D) \arrow[r, "a_{\otimes\oplus(A \otimes B, C, D)}"] & ((A \otimes B) \otimes C) \oplus D \\
  	A \otimes ((B \otimes C) \oplus D) \arrow[r, "a_{\otimes\oplus(A, B \otimes C, D)}"] & (A \otimes (B \otimes C)) \oplus D \arrow[ru,swap, "a_{\otimes(A,B,C)}^{-1} \oplus D "] 
  \end{tikzcd}
\end{center}

    \item \begin{samepage}For the order $\otimes, \oplus, \oplus$, require commutativity of

\begin{center}
  \begin{tikzcd}[row sep=huge, column sep=9em]
    A \otimes (B \oplus (C \oplus D))
      \arrow[r, "A \otimes a_{\oplus(B,C,D)}^{-1}"]
      \arrow[dd, "a_{\otimes\oplus(A, B, C \oplus D)}"]
    & A \otimes ((B \oplus C) \oplus D)
      \arrow[d, "a_{\otimes\oplus(A, B \oplus C, D)}", swap] \\
    & (A \otimes (B \oplus C)) \oplus D
      \arrow[d, "a_{\otimes\oplus(A,B,C)}\oplus D", swap] \\
    (A \otimes B) \oplus (C \oplus D)
      \arrow[r, "a_{\oplus(A \otimes B, C, D)}^{-1}", swap]
    & ((A \otimes B) \oplus C) \oplus D
  \end{tikzcd}
\end{center}
\end{samepage}

    \item \begin{samepage}For the order $\oplus, \oplus, \otimes$, require commutativity of
\begin{center}
  \begin{tikzcd}[row sep=huge, column sep=9em]
    ((A \oplus B) \oplus C) \otimes D
      \arrow[r, "a_{\oplus(A,B,C)} \otimes D"]
      \arrow[dd, "a^{\otimes\oplus(A \oplus B,C,D)}"]
    & (A \oplus (B \oplus C)) \otimes D
      \arrow[d, "a^{\otimes\oplus(A, B \oplus C, D)}", swap] \\
    & A \oplus ((B \oplus C) \otimes D)
      \arrow[d, "A \oplus a^{\otimes\oplus(B, C, D)}", swap] \\
    (A \oplus B) \oplus (C \otimes D)
      \arrow[r, "a_{\oplus(A,B,C \otimes D)}", swap]
    & A \oplus (B \oplus (C \otimes D))
  \end{tikzcd}
\end{center}
\end{samepage}

    \item \begin{samepage}For the order $\otimes, \oplus, \otimes$, require commutativity of
\begin{center}
  \begin{tikzcd}[row sep=huge, column sep=9em]
    A \otimes ((B \oplus C) \otimes D)
      \arrow[r, "a_{\otimes(A, B \oplus C, D)}^{-1}"]
      \arrow[dd, "A \otimes a^{\otimes\oplus(B,C,D)}"]
    & (A \otimes (B \oplus C)) \otimes D
      \arrow[d, "a_{\otimes\oplus(A,B,C)} \otimes D", swap] \\
    & ((A \otimes B) \oplus C) \otimes D
      \arrow[d, "a^{\otimes\oplus(A \otimes B, C, D)}", swap] \\
    A \otimes (B \oplus (C \otimes D))
      \arrow[r, "a_{\otimes\oplus(A, B, C \otimes D)}", swap]
    & (A \otimes B) \oplus (C \otimes D)
  \end{tikzcd}
\end{center}
\end{samepage}
\item \begin{samepage}For the order $\oplus, \otimes, \oplus$, require commutativity of
\begin{center}
  \begin{tikzcd}[row sep=huge, column sep=9em]
    (A \oplus B) \otimes (C \oplus D)
      \arrow[r, "a_{\otimes\oplus(A \oplus B, C, D)}"]
      \arrow[dd, "a^{\otimes\oplus(A, B, C \oplus D)}"]
    & ((A \oplus B) \otimes C) \oplus D
      \arrow[d, "a^{\otimes\oplus(A,B,C)} \oplus D", swap] \\
    & (A \oplus (B \otimes C)) \oplus D
      \arrow[d, "a_{\oplus(A, B \otimes C, D)}", swap] \\
    A \oplus (B \otimes (C \oplus D))
      \arrow[r, "A \oplus a_{\otimes\oplus(B,C,D)}", swap]
    & A \oplus ((B \otimes C) \oplus D)
  \end{tikzcd}
\end{center}
\end{samepage}

\item The orders $\otimes, \otimes, \otimes$ and  $\oplus, \oplus, \oplus$ are covered by the ordinary monoidal coherence laws and will not be repeated here.
\end{itemize}

  \item \textbf{Associativity and distributivity laws:} Associativity and distributivity can be applied in any order for any monoidal operator. The only relevant cases are the orders $\otimes \otimes \oplus, \oplus\otimes\otimes, \otimes\oplus\oplus$ and $\oplus\oplus\otimes$, represented by four diagrams:
\begin{center}
  \begin{tikzcd}[row sep=large, column sep=huge]
    A \otimes ( B \otimes (C \oplus D)) \arrow[r, "A \otimes d_{\otimes(B,C,D)}"] \arrow[d, swap, "a_{\otimes(A,B,C \oplus D)}^{-1}"]  & A \otimes ( C \oplus (B \otimes D)) \arrow[r, "d_{\otimes(A,C, B \otimes D)}"] & C \oplus ( A \otimes (B \otimes D)) \\
    (A \otimes  B) \otimes (C \oplus D) \arrow[r, "d_{\otimes(A \otimes B, C, D)}"] & C \oplus  ((A \otimes B) \otimes D) \arrow[ru, swap, "C \oplus a_{\otimes(A,B,D)}"]  &
    \end{tikzcd}
\end{center}

\begin{center}
  \begin{tikzcd}[row sep=large, column sep=huge]
    ((A \oplus  B) \otimes C) \otimes D \arrow[r, "d^{\otimes(A,B,C)} \otimes D"] \arrow[d, swap, "a_{\otimes(A \oplus B,C, D)}"]  & ((A \otimes  C) \oplus B) \otimes D \arrow[r, "d^{\otimes(A \otimes C, B, D)}"] & ((A \otimes  C) \otimes D) \oplus B \\
    (A \oplus  B) \otimes (C \otimes D) \arrow[r, "d^{\otimes(A,B,C \otimes D)}"] & (A \otimes  (C \otimes D)) \oplus B \arrow[ru, swap, "a_{\otimes(A,C,D)}^{-1} \oplus B "]  &
    \end{tikzcd}
\end{center}

\begin{center}
  \begin{tikzcd}[row sep=large, column sep=huge]
    A \otimes ( B \oplus (C \oplus D)) \arrow[r, "d_{\otimes(A, B, C \oplus D)}"] \arrow[d, swap, "A \otimes a_{\oplus(B, C, D)}^{-1}"]  & B \oplus (A \otimes (C \oplus D)) \arrow[r, "B \oplus d_{\otimes(A, C, D)}"] & B \oplus (C \oplus (A \otimes D)) \\
    A \otimes  ((B \oplus C) \oplus D) \arrow[r, "d_{\otimes(A, B \oplus C, D)}"] & (B \oplus C) \oplus (A \otimes D) \arrow[ru, swap, "a_{\oplus(B, C, A \otimes D)}"]  &
    \end{tikzcd}
\end{center}

\begin{center}
  \begin{tikzcd}[row sep=large, column sep=huge]
     ((B \oplus C) \oplus D) \otimes A \arrow[r, "d^{\otimes(B \oplus C, D, A)}"] \arrow[d, swap, "a_{\oplus(B, C, D)} \otimes A"]  & ((B \oplus C) \otimes A) \oplus D \arrow[r, "d^{\otimes(B,C,A)} \oplus D"] & ((B \otimes A) \oplus C) \oplus D \\
    (B \oplus  (C \oplus D)) \otimes A \arrow[r, "d^{\otimes(B, C \oplus D, A)}"] & (B \otimes  A) \oplus (C \oplus D) \arrow[ru, swap, "a_{\oplus(B \otimes A, C, D)}^{-1} "]  &
    \end{tikzcd}
\end{center}

	\end{itemize}

\begingroup
\setlength{\emergencystretch}{1em}
\textit{Remark:} The distributivities $d_\otimes, d^\otimes$ are independent of the planar axiomatization of \cite{CS97,BCST96}. In the presence of symmetries for the monoidal operators, they are derivable. Our treatment keeps them as data without the presence of symmetries.
\par\endgroup

Just as monoidal categories admit useful unbiased versions, an unbiased structure for weakly distributive categories will be convenient in \cref{sec:polycategorical-representability}. We repeat the definition of an unbiased monoidal category as part of the data of an unbiased weakly distributive category in an effort to emphasize the direction of the arrows for the relevant isomorphisms. Although, since the morphisms are actually isomorphisms, the directions of the arrows do not play a meaningful role here, it is helpful to remember them as they map very well onto later concepts related to logical systems. It is also worth mentioning that weaker structures can be investigated where the arrows are only required to be natural transformations rather than natural isomorphisms. We do not repeat the coherence axioms for these structures, since they are those of \cref{def:unbiased-monoidal-category}.

\begin{definition}\label{def:unbiased-weakly-distributive-category}
  An \emph{unbiased weakly distributive category} is a category $\mc C$ equipped with two unbiased monoidal structures $(\otimes_n)_{n \in \mathbb{N}}$ and $(\oplus_n)_{n \in \mathbb{N}}$ described as follows:
	\begin{itemize}
  	\item \textbf{Unbiased monoidal structure $(\mc C, (\otimes_n)_{n \in \mathbb{N}}, (\eta^\otimes_{n_1, m, n_2})_{n_1,m, n_2 \in \mathbb{N}}, \iota_{(\--)} )$}
  	\begin{itemize}
        \item For every $n \in \mathbb{N}$ a functor: $$\otimes_n: \underbrace{\mc C \times \ldots \times \mc C}_{n-times}  \to \mc C$$
        \item For all $n, m, n' \in \mathbb{N}$, and lists of objects $\Gamma := \Gamma_1, \ldots, \Gamma_{n}$, $\Delta := \Delta_1, \ldots, \Delta_{m}$, and $\Gamma' := \Gamma'_1, \ldots, \Gamma'_{n'}$ an associator natural isomorphism: $$\eta^\otimes_{n,m,n'} : \otimes_{n + m + n'}(\Gamma, \Delta, \Gamma') \to \otimes_{n + 1 + n'}(\Gamma, \otimes_m\Delta, \Gamma')  $$
        \item A natural isomorphism: $$\iota_{(\--)} : \otimes_1(\--) \to (\--)$$
      \end{itemize}  
  	\item \textbf{Unbiased monoidal structure $(\mc C, (\oplus_n)_{n \in \mathbb{N}}, (\eta^\oplus_{n_1, m, n_2})_{n_1,m, n_2 \in \mathbb{N}}, \iota^{(\--)} )$ }
  	\begin{itemize}
        \item For every $n \in \mathbb{N}$ a functor: $$\oplus_n: \underbrace{\mc C \times \ldots \times \mc C}_{n-times}  \to \mc C$$
        \item For all $n, m, n' \in \mathbb{N}$, and lists of objects $\Gamma := \Gamma_1, \ldots, \Gamma_{n}$, $\Delta := \Delta_1, \ldots, \Delta_{m}$, and $\Gamma' := \Gamma'_1, \ldots, \Gamma'_{n'}$ an associator natural isomorphism: $$\eta^\oplus_{n,m,n'} : \oplus_{n + 1 + n'}(\Gamma, \oplus_m\Delta, \Gamma')\to \oplus_{n + m + n'}(\Gamma, \Delta, \Gamma')   $$
        \item A natural isomorphism: $$\iota^{(\--)} : (\--)\to \oplus_1(\--) $$
      \end{itemize}  
    	\item \textbf{Natural transformations}
        $$ a_i : \otimes(\Gamma, \oplus\Delta, \Gamma') \to \oplus(\otimes(\Gamma, \oplus\Delta_{\le i}), \otimes(\oplus\Delta_{>i}, \Gamma')) $$
      where $\Delta_{\le i}$ and $\Delta_{> i}$ denote the prefix and the suffix of the list $\Delta$ at the position $i$. The directions of the arrows, $\eta^\otimes$ nesting and $\eta^\oplus$ flattening, are the oplax and lax ones that the polycategorical reading requires.
    	\begin{align*}
        d^L &:  \otimes_{n + 1}(\Gamma, \oplus_{m + 1}(\Delta, A)) \to \oplus_{m + 1}(\Delta, \otimes_{n + 1}(\Gamma, A)) \\
       	d^R &:  \otimes_{n + 1}(\oplus_{m + 1}( A, \Delta),\Gamma) \to \oplus_{m + 1}(\otimes_{n + 1}(A, \Gamma), \Delta)\\ 
      \end{align*}
  	\end{itemize}
    \begin{samepage}
    such that for all lists of objects $$\Gamma := \Gamma_1, \ldots, \Gamma_n \quad \Gamma' := \Gamma'_1, \ldots, \Gamma'_{n'} \quad \Delta := \Delta_1, \ldots, \Delta_{m} \quad \Delta' := \Delta'_1, \ldots, \Delta'_{m'}$$ the coherence conditions are satisfied:
      \begin{center}
      \begin{tikzcd}[row sep=large, column sep=huge]
          \otimes(\Gamma, \oplus(\Delta, A, \Delta'), \Gamma')
          \arrow[r, "p"]
          \arrow[d, "q", swap]
          &
          \otimes(\oplus(\Delta, \otimes(\Gamma,\oplus(A, \Delta'))), \Gamma')
          \arrow[d, "r"] \\
          \otimes(\Gamma, \oplus(\Delta, \oplus(\otimes(A, \Gamma'), \Delta')))
          \arrow[r, "s", swap]
          &
          \oplus(\Delta, \otimes(\Gamma, A, \Gamma'), \Delta')
      \end{tikzcd}
      \end{center}
      Here the arrow labels denote the composites
      \begin{align*}
        p ={}& \otimes(d^L, \1_{\Gamma'})
          \circ \otimes(\otimes(\1_{\Gamma}, (\eta^\oplus_{m, m' + 1, 0})^{-1}), \1_{\Gamma'})
          \circ \eta^\otimes_{0,n + 1, n'}, \\
        q ={}& \otimes(\1_{\Gamma}, \oplus(\1_{\Delta}, d^R))
          \circ \otimes(\1_{\Gamma}, a_m)
          \circ \eta^\otimes_{n, n' + 1, 0}, \\
        r ={}& \oplus(\1_{\Delta}, \oplus((\eta^\otimes_{n, n' + 1, 0})^{-1}, \1_{\Delta'})) \\
          &\quad\circ \oplus(\1_{\Delta}, a_1 \circ \otimes(\1_{\Gamma}, d^R)
            \circ \eta^\otimes_{n,n' + 1, 0} \circ (\eta^\otimes_{0, n + 1, n'})^{-1}) \circ a_m, \\
        s ={}& \eta^\oplus_{m, 2, 0}
          \circ \oplus(\1_{\Delta}, \oplus((\eta^\otimes_{n, 1, n'})^{-1}, \iota_{\Delta'})) \\
          &\quad\circ \oplus(\1_{\Delta}, a_1) \circ d^L.
      \end{align*}
      \end{samepage}
      The choices of associativities are irrelevant according to the coherence laws as in \cite{BCST96}. We also omit the applications of $\iota$ to cancel $\otimes_1, \oplus_1$.

\end{definition}

\textit{Remark:} The distributivity natural transformations are the linear distributivities of \cite{CS97}. Another name for them is ``interchange laws'', which we reserve for the cut reordering laws of polycategories.
     
\phantomsection\label{anchor:two-sided-distributor-recovery}
\textit{Remark:} A standard presentation of unbiased weakly distributive categorical distributivity involves interchanging lists of objects on the left and the right at the same time: $$\otimes(\Gamma, \oplus(\Delta, A, \Delta'), \Gamma') \to \oplus(\Delta, \otimes(\Gamma, A, \Gamma'), \Delta').$$
It is possible to recover this map from our definition by chaining left and right distributivity with a suitable choice of associativities:

\begin{align*}
\otimes(\Gamma, \oplus(\Delta, A, \Delta'), \Gamma') &\xto{\eta^\otimes} \otimes(\otimes(\Gamma, \oplus(\Delta, A, \Delta')), \Gamma') \\
&\xto{\otimes(\otimes(\eta^\oplus), \1)} \otimes(\otimes(\Gamma, \oplus(\Delta, \oplus(A, \Delta'))), \Gamma') \\
&\xto{\otimes(d^L, \1)} \otimes(\oplus(\Delta, \otimes(\Gamma, \oplus(A, \Delta'))), \Gamma') \\
&\xto{a} \oplus(\Delta, \otimes(\Gamma, \oplus(A, \Delta'), \Gamma')) \\
&\xto{\oplus(\1, \eta^\otimes)} \oplus(\Delta, \otimes(\Gamma, \otimes(\oplus(A, \Delta'), \Gamma' ))) \\
&\xto{\oplus(\1, d^R)} \oplus(\Delta, \otimes(\Gamma, \oplus(\otimes(A, \Gamma'), \Delta'))) \\
&\xto{\oplus(\1, a)} \oplus(\Delta, \otimes(\Gamma, A, \Gamma'), \Delta') 
\end{align*}
where indices have been omitted.

\begin{definition}
  A functor $F: \mc C \to \mc D$ between two weakly distributive categories $\mc C, \mc D$ and their respective monoidal structures $(\otimes_1, \oplus_1)$ and $(\otimes_2, \oplus_2)$ is a lax functor between $\mc C$ and $\mc D$ with respect to their operators $\otimes_1, \otimes_2$ and oplax with respect to $\oplus_1, \oplus_2$.   

  \begin{itemize}
    \item \textbf{Distributivity coherence laws: } Distributing and applying the (op)lax functor is a well-behaved operation, i.e. the coherence diagrams commute:

		\begin{center}
    \begin{tikzcd}[row sep=large, column sep=huge]
  		FA \otimes_2 F(B \oplus_1 C) \arrow[r, "\eta^{lax}_{FA, F(B \oplus_1 C)}"] \arrow[d, "FA \otimes_2 \eta^{oplax}_{B,C}"] & F(A \otimes_1 (B \oplus_1 C)) \arrow[d, "Fd_{\otimes_1(A, B, C)}"] \\ 
  		FA \otimes_2 (FB \oplus_2 FC)  \arrow[d, "d_{\otimes_2(FA,FB,FC)}"] & F(B \oplus_1 (A \otimes_1 C)) \arrow[d, "\eta^{oplax}_{B, A \otimes_1 C}"] \\ 
  		FB \oplus_2 (FA \otimes_2 FC)  \arrow[r, "FB \oplus_2 \eta^{lax}_{A, C}"] & FB \oplus_2 F(A \otimes_1 C)  
		\end{tikzcd}
		\quad
   \begin{tikzcd}[row sep=large, column sep=huge]
  		 F(B \oplus_1 C)  \otimes_2 FA \arrow[r, "\eta^{lax}_{F(B \oplus_1 C), FA}"] \arrow[d, "\eta^{oplax}_{B,C} \otimes_2 FA"] & F( (B \oplus_1 C) \otimes_1 A) \arrow[d, "Fd^{\otimes_1(B,C,A)}"] \\ 
  		 (FB \oplus_2 FC)  \otimes_2 FA  \arrow[d, "d^{\otimes_2(FB,FC,FA)}"] & F( (B \otimes_1 A) \oplus_1 C) \arrow[d, "\eta^{oplax}_{B \otimes_1 A, C}"] \\ 
  		 (FB \otimes_2 FA)  \oplus_2 FC  \arrow[r, "\eta^{lax}_{B, A} \oplus_2 FC "] &  F(B \otimes_1 A) \oplus_2 FC
		\end{tikzcd}
		\end{center}

		\begin{center}
    \begin{tikzcd}[row sep=large, column sep=huge]
  		FA \otimes_2 F(B \oplus_1 C) \arrow[r, "\eta^{lax}_{A, B \oplus_1 C}"] \arrow[d, "FA \otimes_2 \eta^{oplax}_{B,C}"] & F(A \otimes_1 (B \oplus_1 C)) \arrow[d, "Fa_{\otimes_1\oplus_1(A, B, C)}"] \\
  		FA \otimes_2 (FB \oplus_2 FC)  \arrow[d, "a_{\otimes_2\oplus_2(FA,FB,FC)}"] & F((A \otimes_1 B) \oplus_1 C) \arrow[d, "\eta^{oplax}_{A \otimes_1 B, C}"] \\
  		(FA \otimes_2 FB) \oplus_2 FC  \arrow[r, "\eta^{lax}_{A, B} \oplus_2 FC"] & F(A \otimes_1 B) \oplus_2 FC
		\end{tikzcd}
		\quad
   \begin{tikzcd}[row sep=large, column sep=huge]
  		 F(A \oplus_1 B)  \otimes_2 FC \arrow[r, "\eta^{lax}_{A \oplus_1 B, C}"] \arrow[d, "\eta^{oplax}_{A,B} \otimes_2 FC"] & F( (A \oplus_1 B) \otimes_1 C) \arrow[d, "Fa^{\otimes_1\oplus_1(A,B,C)}"] \\
  		 (FA \oplus_2 FB)  \otimes_2 FC  \arrow[d, "a^{\otimes_2\oplus_2(FA,FB,FC)}"] & F( A \oplus_1 (B \otimes_1 C)) \arrow[d, "\eta^{oplax}_{A, B \otimes_1 C}"] \\
  		 FA \oplus_2 (FB \otimes_2 FC)  \arrow[r, "FA \oplus_2 \eta^{lax}_{B, C}"] &  FA \oplus_2 F(B \otimes_1 C)
		\end{tikzcd}
		\end{center}

  \end{itemize}

\end{definition}
\textit{Remark.} The subscripts for $\otimes, \oplus$ refer to the category rather than the arity of the operators, which have arity two.

For the symmetric weakly distributive categories of \cref{prop:wdc-to-polycat}, the relevant functors are Frobenius linear: besides these lax and oplax monoidal comparisons, they preserve the symmetries and satisfy the distributor coherence hexagons. The corresponding representability result is \cref{thm:wdc-sympolycat-equivalence}.

\begin{definition}
    For an object $A$ in a weakly distributive category, a \emph{left dual} $\star A$ is an object such that there exist morphisms:
  	\[ t_A : \top \to \star A \oplus A \quad b_A :  A \otimes \star A \to \bot\]
    subject to commutativity of

  	\begin{center}
      \begin{tikzcd}[row sep=large, column sep=4.4em]
        A \otimes \top \arrow[r, "A \otimes t_A"] \arrow[rrrr, bend left=20, "u_{\otimes} "]& A \otimes (\star A \oplus A) \arrow[r, "a_{\otimes\oplus(A,\star A, A)}"] & (A \otimes \star A) \oplus A \arrow[r, "b_A \oplus A"] & \bot \oplus A \arrow[r, "u^\oplus"] & A
     	\end{tikzcd}
  	\end{center}

  	\begin{center}
      \begin{tikzcd}[row sep=large, column sep=4.4em]
        \top \otimes \star A  \arrow[r, "t_A \otimes \star A"] \arrow[rrrr, bend right=20, "u^{\otimes} "]& (\star A \oplus A )\otimes \star A  \arrow[r, "a^{\otimes\oplus(\star A, A, \star A)}"] & \star A \oplus (A \otimes \star A) \arrow[r, "\star A \oplus b_A"] & \star A \oplus \bot  \arrow[r, "u_\oplus"] & \star A
     	\end{tikzcd}
  	\end{center}

\end{definition}

\begin{definition}
    For an object $A$ in a weakly distributive category, a \emph{right dual} $A\star$ is an object such that there exist morphisms:
  	\[ t^A : \top \to A \oplus A\star \quad  b^A : A\star \otimes A \to \bot\]
    subject to commutativity of

  	\begin{center}
      \begin{tikzcd}[row sep=large, column sep=4.4em]
        \top \otimes A  \arrow[r, "t^A \otimes A"] \arrow[rrrr, bend left=20, "u^{\otimes} "]& (A \oplus A \star)\otimes A  \arrow[r, "a^{\otimes\oplus(A, A \star, A)}"] & A \oplus (A \star \otimes A) \arrow[r, "A \oplus b^A"] & A \oplus \bot  \arrow[r, "u_\oplus"] & A
     	\end{tikzcd}
  	\end{center}

  	\begin{center}
      \begin{tikzcd}[row sep=large, column sep=4.4em]
         A \star \otimes \top \arrow[r, " A\star \otimes t^A"] \arrow[rrrr, bend right=20, "u_{\otimes} "]&  A\star \otimes ( A \oplus A \star) \arrow[r, "a_{\otimes\oplus(A \star, A, A \star)}"] & (A\star \otimes  A) \oplus A \star \arrow[r, "b^A \oplus A \star"] & \bot \oplus A \star \arrow[r, "u^\oplus"] & A \star
     	\end{tikzcd}
  	\end{center}

\end{definition}

\begin{definition}
  A weakly distributive category is called a $\star$\emph{-autonomous category} if it has all right and left duals.
\end{definition}
\textit{Remark:} Defining $\star$-autonomous categories through duals in a weakly distributive category is the characterization of \cite{CS97,BCST96}; the original definition of \cite{Bar79} proceeds through a dualizing object, and the two presentations agree.

\subsection{LNL Polycategories}\label{subsec:lnl-polycategories}
We conclude this chapter with a notion that combines the linear and the cartesian structures encountered so far. The reindexed polycategories of \cref{def:symmetric-polycategory} choose structural classes uniformly for all their objects. The logics of interest in this thesis instead require linear and cartesian objects in the same structure. The case of linear logic makes this need very salient. Indeed, this logic can be presented as containing two types of objects: linear and cartesian. They are characterized by the structural rules admissible for them. The modalities, in turn, relate linear and cartesian objects together. We are therefore inclined to define a category subsuming both types of objects at once. The notion that is adequate is the one of an \emph{LNL polycategory}, short for linear-non-linear polycategory, introduced by Shulman \cite{Shu23}. The name refers to the linear-non-linear adjunctions of Benton \cite{Ben95} that these polycategories subsume. An LNL polycategory carries two classes of objects: a \emph{nonlinear} class (that we sometimes call cartesian) subject to all the structural rules and a \emph{linear} class subject to exchange only.
\begin{definition}\label{def:lnl-polycategory}
  An \emph{LNL polycategory} $\mc P$ consists of the data:
  \begin{itemize}
    \item A collection of \emph{nonlinear objects} denoted by letters near the end of the alphabet $X, Y, Z$ and a collection of linear objects denoted by letters near the beginning of the alphabet $A, B, C$. We will denote finite lists of nonlinear objects by $\Theta, \Upsilon$ and finite lists of linear objects by $\Gamma, \Delta$
    \item Given a list of nonlinear objects $X_1, \ldots, X_m$, every function $\sigma : \{1, \ldots, n\} \to \{1, \ldots, m\}$ defines a structural map $$\sigma : X_1, \ldots, X_m \to X_{\sigma 1}, \ldots, X_{\sigma n}.$$ Similarly, every bijection $\tau : \{1, \ldots, n\} \to \{1, \ldots, n\}$ defines a structural permutation $\tau : A_1, \ldots, A_n \to A_{\tau 1}, \ldots, A_{\tau n}$ on lists of linear objects
    \item A collection of \emph{nonlinear morphisms} $\mc P(\Theta;X)$ for every list of nonlinear objects $\Theta$ and nonlinear object $X$, together with a reindexing map for every structural map $\sigma : \Theta \to \Upsilon$:
    \[(\--)^{\sigma} : \mc P(\Upsilon;X) \to \mc P(\Theta;X)\]
    \item A collection of \emph{linear morphisms} $\mc P(\Theta \mid \Gamma;\Delta)$ for every list of nonlinear objects $\Theta$ and lists of linear objects $\Gamma, \Delta$, together with a reindexing map for every structural map $\sigma : \Theta \to \Upsilon$ and structural permutations $\tau : \Gamma \to \Gamma'$ and $\rho : \Delta \to \Delta'$:
    \[(\--)^{\sigma \mid \tau \mid \rho} : \mc P(\Upsilon \mid \Gamma';\Delta) \to \mc P(\Theta \mid \Gamma;\Delta')\]
    \item An identity map $1_X \in \mc P(X;X)$ for every nonlinear object $X$ and an identity map $1_A \in \mc P(\cdot \mid A;A)$ for every linear object $A$, where $\cdot$ is the empty list
    \item For all nonlinear morphisms $f \in \mc P(\Upsilon;X)$ and $g \in \mc P(\Theta_1, X, \Theta_2;Y)$ a nonlinear morphism $g \circ_X f \in \mc P(\Theta_1, \Upsilon, \Theta_2;Y)$
    \item For every nonlinear morphism $f \in \mc P(\Upsilon;X)$ and linear morphism $g \in \mc P(\Theta_1, X, \Theta_2 \mid \Gamma;\Delta)$ a linear morphism $g \circ_X f \in \mc P(\Theta_1, \Upsilon, \Theta_2 \mid \Gamma;\Delta)$
    \item For all linear morphisms $f \in \mc P(\Theta \mid \Gamma;\Delta_1, A, \Delta_2)$ and $g \in \mc P(\Theta' \mid \Gamma_1, A, \Gamma_2;\Delta)$ a linear morphism $g \circ_A f \in \mc P(\Theta', \Theta \mid \Gamma_1, \Gamma, \Gamma_2;\Delta_1, \Delta, \Delta_2)$
    \item Identity, associativity and interchange laws for the three cut operations, analogous to the ones of multicategories and polycategories. The precise list can be found in \cite[\S 2.1]{Shu23}
    \item Functoriality and equivariance laws; reindexing along a composite of structural maps agrees with composing the reindexing maps, and reindexing commutes with the three cut operations.
  \end{itemize}
\end{definition}
\textit{Notation}: We will denote nonlinear morphisms using the standard map convention, $f : \Theta \to X$, and linear morphisms as $f : (\Theta \mid \Gamma) \to \Delta$ where the arrow plays the role of the semicolon in the definition.

\begin{example}[Symmetric polycategories]
An LNL polycategory with no nonlinear objects admits the empty list as its only nonlinear zone. Its linear morphisms form a symmetric polycategory in the sense of \cref{def:symmetric-polycategory}. These symmetric polycategories model the judgmental structure of classical multiplicative linear logic. To recover the judgmental structure of the intuitionistic fragment, we need symmetric multicategories which are obtained by restricting the linear morphisms to have exactly one output.
\end{example}
\begin{example}[Cartesian multicategories]
An LNL polycategory with no linear objects and no linear morphisms is a multicategory equipped with weakening, contraction and exchange functors, i.e. a cartesian multicategory. Notice that the first condition does not imply the second one since the collections $\mc P(\Theta \mid \cdot;\cdot)$ can be inhabited even in the absence of linear objects. These model the judgmental structure of intuitionistic nonlinear logic.
\end{example}
\begin{example}[LNL multicategories]
An LNL polycategory in which every linear morphism has exactly one output will be called an LNL multicategory. These model the judgmental structure of intuitionistic linear logic with exponentials.
\end{example}

\begin{definition}\label{def:lnl-2-polycategory}
  An LNL polycategory such that every collection of morphisms $\mc P(\Theta;X)$ and $\mc P(\Theta \mid \Gamma;\Delta)$ is a category, and every cut and reindexing map is a functor will be called an \emph{LNL $2$-polycategory}.
\end{definition}

The last ingredient of this chapter is the notion of bifibration for LNL polycategories. We follow the same path as for multicategories. We define the notion of functor between LNL polycategories, then a notion of cartesian and opcartesian morphisms, and finally a notion of fibrations. A functor of LNL polycategories transports each class of objects and morphisms to the corresponding class and preserves all the operations. We keep the shorthand $\pi\Theta$ for the list of images.

\begin{definition}
A functor $\pi : \mc D \to \mc M$ between two LNL polycategories consists of:
\begin{itemize}
  \item A nonlinear object $\pi X \in \mc M$ for each nonlinear object $X \in \mc D$ and a linear object $\pi A \in \mc M$ for each linear object $A \in \mc D$
  \item A nonlinear morphism $\pi f \in \mc M(\pi\Theta;\pi X)$ for each nonlinear morphism $f \in \mc D(\Theta;X)$ and a linear morphism $\pi f \in \mc M(\pi\Theta \mid \pi\Gamma;\pi\Delta)$ for each linear morphism $f \in \mc D(\Theta \mid \Gamma;\Delta)$
  \item Respect for the identity maps: $\pi(1_X) = 1_{\pi X}$ and $\pi(1_A) = 1_{\pi A}$
  \item Respect for the cuts: $\pi(g \circ_X f) = \pi g \circ_{\pi X} \pi f$ and $\pi(g \circ_A f) = \pi g \circ_{\pi A} \pi f$
\end{itemize}
such that the reindexing maps are preserved, i.e.: $\pi(f^{\sigma}) = (\pi f)^{\sigma}$ and $\pi(f^{\sigma \mid \tau \mid \rho}) = (\pi f)^{\sigma \mid \tau \mid \rho}$.
\end{definition}

Given a functor of LNL polycategories, a morphism will be cartesian or opcartesian at an object following the standard definitions already exposed in \cref{subsec:multicategories,subsec:polycategories}. Because now we are dealing with multiple zones and classes of objects, there are five possible configurations to expose. We record them explicitly following \cite{Shu23}. The structural actions let us only consider the universal object at the end of its containing list with no loss of generality.

\begin{definition}\label{def:lnl-cartesian-morphism}
  Let $\pi : \mc D \to \mc M$ be a functor of LNL polycategories.
  \begin{itemize}
    \item A nonlinear morphism $\psi \in \mc D(\Theta, X;Y)$ is \emph{cartesian at $X$} if for all nonlinear morphisms $h \in \mc D(\Theta, \Theta';Y)$ and $\beta \in \mc M(\pi\Theta';\pi X)$ such that $\pi\psi \circ_{\pi X} \beta = \pi h$, there exists a unique nonlinear morphism $\phi \in \mc D(\Theta';X)$ such that $\pi\phi = \beta$ and $\psi \circ_X \phi = h$
    \item A linear morphism $\psi \in \mc D(\Theta, X \mid \Gamma;\Delta)$ is \emph{cartesian at $X$} if for every linear morphism $h \in \mc D(\Theta, \Theta' \mid \Gamma;\Delta)$ and nonlinear morphism $\beta \in \mc M(\pi\Theta';\pi X)$ such that $\pi\psi \circ_{\pi X} \beta = \pi h$, there exists a unique nonlinear morphism $\phi \in \mc D(\Theta';X)$ such that $\pi\phi = \beta$ and $\psi \circ_X \phi = h$
    \item A linear morphism $\psi \in \mc D(\Theta \mid \Gamma, A;\Delta)$ is \emph{cartesian at $A$} if for all linear morphisms $h \in \mc D(\Theta, \Theta' \mid \Gamma, \Gamma';\Delta', \Delta)$ and $\beta \in \mc M(\pi\Theta' \mid \pi\Gamma';\pi\Delta', \pi A)$ such that $\pi\psi \circ_{\pi A} \beta = \pi h$, there exists a unique linear morphism $\phi \in \mc D(\Theta' \mid \Gamma';\Delta', A)$ such that $\pi\phi = \beta$ and $\psi \circ_A \phi = h$
  \end{itemize}
\end{definition}

\begin{definition}\label{def:lnl-opcartesian-morphism}
  Let $\pi : \mc D \to \mc M$ be a functor of LNL polycategories.
  \begin{itemize}
    \item A nonlinear morphism $\psi \in \mc D(\Theta;X)$ is \emph{opcartesian at $X$} if for all nonlinear morphisms $h \in \mc D(\Theta', \Theta;Y)$ and $\beta \in \mc M(\pi\Theta', \pi X;\pi Y)$ such that $\beta \circ_{\pi X} \pi\psi = \pi h$, there exists a unique nonlinear morphism $\phi \in \mc D(\Theta', X;Y)$ such that $\pi\phi = \beta$ and $\phi \circ_X \psi = h$, and if the same condition holds for all linear morphisms $h \in \mc D(\Theta', \Theta \mid \Gamma;\Delta)$ and $\beta \in \mc M(\pi\Theta', \pi X \mid \pi\Gamma;\pi\Delta)$ with a unique linear morphism $\phi \in \mc D(\Theta', X \mid \Gamma;\Delta)$
    \item A linear morphism $\psi \in \mc D(\Theta \mid \Gamma;\Delta, A)$ is \emph{opcartesian at $A$} if for all linear morphisms $h \in \mc D(\Theta', \Theta \mid \Gamma', \Gamma;\Delta, \Delta')$ and $\beta \in \mc M(\pi\Theta' \mid \pi\Gamma', \pi A;\pi\Delta')$ such that $\beta \circ_{\pi A} \pi\psi = \pi h$, there exists a unique linear morphism $\phi \in \mc D(\Theta' \mid \Gamma', A;\Delta')$ such that $\pi\phi = \beta$ and $\phi \circ_A \psi = h$
  \end{itemize}
\end{definition}

\textit{Remark.} The first two clauses of \cref{def:lnl-cartesian-morphism} both concern a nonlinear object $X$. They differ in the kind of morphism carrying the cartesian property: the first clause treats a nonlinear morphism $\psi \in \mc D(\Theta, X;Y)$, whereas the second treats a linear morphism $\psi \in \mc D(\Theta, X\mid\Gamma;\Delta)$. The remaining clauses distinguish the linear object $A$ and the input or output zone at which the universal property is imposed.

\phantomsection\label{anchor:lnl-opcartesian-pullback}
\textit{Remark:} As for polycategories, being cartesian or opcartesian at an object corresponds to the existence of a pullback diagram, here of sets. For instance, a linear morphism $\psi \in \mc D(\Theta \mid \Gamma;\Delta, A)$ is opcartesian at $A$ exactly when the displayed square is a pullback for all lists $\Theta', \Gamma', \Delta'$:
\begin{center}
\begin{tikzcd}
    \mc D(\Theta' \mid \Gamma', A;\Delta') \arrow[r, "(\--) \circ_A \psi"] \arrow[d, "\pi"] & \mc D(\Theta', \Theta \mid \Gamma', \Gamma;\Delta, \Delta') \arrow[d, "\pi"]   \\
    \mc M(\pi\Theta' \mid \pi\Gamma', \pi A;\pi\Delta') \arrow[r, "(\--) \circ_{\pi A} \pi\psi"] & \mc M(\pi\Theta', \pi\Theta \mid \pi\Gamma', \pi\Gamma;\pi\Delta, \pi\Delta')  \\
\end{tikzcd}
\end{center}
and similarly for the four other configurations. In the context of LNL $2$-polycategories, the same definitions apply with the collections of morphisms replaced by their categories and the squares asked to be pullbacks of categories, exactly as for the $2$-polycategories of \cref{subsec:polycategories}.

\begin{definition}
A functor $\pi : \mc D \to \mc M$ between two LNL $2$-polycategories is a functor of the underlying LNL polycategories whose actions on the collections of morphisms are functors of the hom-categories, commuting with the cut and reindexing functors. It is a \textit{local discrete fibration} if every hom-functor $\mc D(\Theta;X) \to \mc M(\pi\Theta;\pi X)$ and $\mc D(\Theta \mid \Gamma;\Delta) \to \mc M(\pi\Theta \mid \pi\Gamma;\pi\Delta)$ is a discrete fibration of categories, i.e. every $2$-cell of the base with a chosen lift of its codomain admits a unique lift.
\end{definition}

\phantomsection\label{anchor:lnl-local-discrete-elementwise}
\textit{Remark:} For a local discrete fibration the pullback-of-categories condition splits into two elementwise clauses, one for the objects of the hom-categories and one for their morphisms. In the \hyperref[anchor:lnl-opcartesian-pullback]{opcartesian pullback configuration}, $\psi \in \mc D(\Theta \mid \Gamma;\Delta, A)$ is opcartesian at $A$ exactly when:
\begin{itemize}
  \item for every $h \in \mc D(\Theta', \Theta \mid \Gamma', \Gamma;\Delta, \Delta')$ and $\beta \in \mc M(\pi\Theta' \mid \pi\Gamma', \pi A;\pi\Delta')$ with $\beta \circ_{\pi A} \pi\psi = \pi h$, there exists a unique $\phi \in \mc D(\Theta' \mid \Gamma', A;\Delta')$ with $\pi\phi = \beta$ and $\phi \circ_A \psi = h$, which is the object-level opcartesianness condition; and
  \item for every $2$-cell $b : \beta \Rightarrow \beta'$ of $\mc M(\pi\Theta' \mid \pi\Gamma', \pi A;\pi\Delta')$ and every $2$-cell $t : h \Rightarrow h'$ of $\mc D$ lying over $b \circ_{\pi A} 1_{\pi\psi}$, with factorizations $h = \phi \circ_A \psi$ and $h' = \phi' \circ_A \psi$ given by the first clause, there exists a unique $2$-cell $s : \phi \Rightarrow \phi'$ over $b$ with $s \circ_A 1_{\psi} = t$.
\end{itemize}
The other configurations are the same and the cartesian case is treated dually. When the hom-categories are discrete the second clause is vacuous and the notions collapse to the one-dimensional ones. For functors of LNL $2$-polycategories, the fibration, opfibration and bifibration conditions are read with these two-dimensional notions.

\begin{definition}
  A functor $\pi : \mc D \to \mc M$ of LNL polycategories is a \emph{fibration} if:
  \begin{itemize}
    \item for every list $\Theta$ and nonlinear object $Y$ in $\mc D$, every nonlinear object $x \in \mc M$ and every nonlinear morphism $g \in \mc M(\pi\Theta, x;\pi Y)$, there exist a nonlinear object $X$ with $\pi X = x$ and a nonlinear morphism $f \in \mc D(\Theta, X;Y)$ cartesian at $X$ with $\pi f = g$
    \item for all lists $\Theta, \Gamma, \Delta$ in $\mc D$, every nonlinear object $x \in \mc M$ and every linear morphism $g \in \mc M(\pi\Theta, x \mid \pi\Gamma;\pi\Delta)$, there exist a nonlinear object $X$ with $\pi X = x$ and a linear morphism $f \in \mc D(\Theta, X \mid \Gamma;\Delta)$ cartesian at $X$ with $\pi f = g$
    \item for all lists $\Theta, \Gamma, \Delta$ in $\mc D$, every linear object $a \in \mc M$ and every linear morphism $g \in \mc M(\pi\Theta \mid \pi\Gamma, a;\pi\Delta)$, there exist a linear object $A$ with $\pi A = a$ and a linear morphism $f \in \mc D(\Theta \mid \Gamma, A;\Delta)$ cartesian at $A$ with $\pi f = g$
  \end{itemize}
\end{definition}

\begin{definition}
  A functor $\pi : \mc D \to \mc M$ of LNL polycategories is an \emph{opfibration} if:
  \begin{itemize}
    \item for every list $\Theta$ in $\mc D$, every nonlinear object $x \in \mc M$ and every nonlinear morphism $g \in \mc M(\pi\Theta;x)$, there exist a nonlinear object $X$ with $\pi X = x$ and a nonlinear morphism $f \in \mc D(\Theta;X)$ opcartesian at $X$ with $\pi f = g$
    \item for all lists $\Theta, \Gamma, \Delta$ in $\mc D$, every linear object $a \in \mc M$ and every linear morphism $g \in \mc M(\pi\Theta \mid \pi\Gamma;\pi\Delta, a)$, there exist a linear object $A$ with $\pi A = a$ and a linear morphism $f \in \mc D(\Theta \mid \Gamma;\Delta, A)$ opcartesian at $A$ with $\pi f = g$
  \end{itemize}
\end{definition}

\begin{definition}
  A functor of LNL polycategories that is both a fibration and an opfibration is called a \textit{bifibration of LNL polycategories}.
\end{definition}

\subfilebibliography

\endgroup{}
\begingroup

\chapter{Fibrational Framework}\label{chap:fibrational-framework}
\setcounter{secnumdepth}{3}%
One of the features of substructural and modal logical systems is the ability of the sequent calculus to handle variables by permuting them, discarding them, or reusing them. Those actions are packaged as rules usually under the names of exchange, weakening and contraction respectively. Traditionally, a logic is set up with all, some or no substructural rules directly or indirectly in the calculus. There are effectively two ways of including those rules in the calculus. The direct method consists of including the specific rule in the calculus as an explicit rule. The indirect method consists of designing the calculus in such a fashion that the substructural rule is not explicitly present but adding it to the calculus does not, in some precise sense, change the underlying theory. The latter rules are called \emph{admissible}, and adding them explicitly to the calculus does not change the set of consequences that the rules entail.
Modal logics handle restrictions on which contexts certain variables can be used in (contrasting with the case where substructurality indicates how a variable can be used). For instance, a calculus with two modalities, say the modality of truths and the modality of validity, could require us to be able to prove that a formula is valid only from valid contexts of formulas, and proving a formula is true could lift this requirement to allow both valid and true formulas to be used.
We desire a general framework to model the various combinations of substructural and modal logics. Some work has been conducted in \cite{LSR17}, where a fibrational framework provides the judgmental structure for intuitionistic types of logics that are substructural and modal.

A classic move to recover standard mathematical reasoning is to construct classical logic from intuitionistic logic. Intuitionistic logic is constructive while the regular flavor of classical logic is not. This is ultimately encompassed by the fact that a provable disjunction in intuitionistic logic allows us to derive the relevant premise, i.e. if $A \lor B$ is derivable, then we can derive either $A$ or $B$. In classical logic, this is not always the case. This is called the disjunction property \cite{Avi22}. The judgmental structure is different between the two systems, and a hint on how to figure out a way to extend intuitionistic logic to classical logic comes from two facts. We recover classical logic by extending intuitionistic logic to multiple formulas on the right-hand side of the turnstile $\vdash$ while still retaining the substructural rules of weakening, contraction and exchange. Although this is a straightforward way to operate, it is somewhat hiding the central aspect that makes a system classical. A more precise insight comes from linear logic, which does away with all substructural rules but exchange and interprets classical logic through modalities.
This leads us to consider primarily two judgmental structures: the intuitionistic and the classical ones. To construct a general framework that will capture both, we need to account for both judgments. We somewhat deceptively call the judgmental structure of intuitionistic systems ``cartesian'' and the judgmental structure of classical systems ``linear'' to emphasize their reusable and exact resource behavior.
Shulman \cite{Shu23} investigated a mathematical structure that encompasses both judgmental structures (and more) at the same time. Such structures are called linear-non-linear polycategories \cite{Shu23}. The polycategorical aspect induces the judgmental structure of a sequent of the form $\Gamma \vdash \Delta$ necessary for expressing classicality (i.e. one of the formulas in $\Delta$ is true but we do not know which one) and the linear-non-linear part recovers the fundamental objects used in intuitionistic and classical systems. Linear objects are exact resources; they have to be used once and exactly once. Cartesian objects, on the other hand, have the capability of inheriting substructural properties.

The present chapter constructs such a framework. Following \cite{LSR17}, the construction is layered. The first layer, called the mode theory, describes the structure of the contexts and the targets of sequents together with the substructural and modal properties they are meant to obey. The second layer, the type theory, is a sequent calculus parametrized by the mode theory; its judgments, its connectives and its structural rules are generated from the declarations of the first layer. Finally, \cref{sec:categorical-semantics} returns to the semantical picture and identifies the two layers with linear-non-linear polycategories.

A mode theory begins with a signature which provides modes and generators. Each generator consumes a list of modes and produces another list of modes. Modes come in two types: the cartesian modes, whose variables can be discarded, permuted and reused freely, and linear modes, whose variables are exact resources. Every sequent of the framework carries two kinds of zones. The cartesian zone $\Theta$ collects the cartesian variables and behaves as an ordinary structural context; the linear zones $\Gamma$ and $\Delta$ sit on both sides of the turnstile and their formulas are used exactly once. The calculus splits therefore into two sequent judgments:
$$\Theta \vdash A \qquad \text{and} \qquad \Theta \ ; \ \Gamma \vdash \Delta$$
The cartesian judgment carries no linear zone and a single conclusion; it is the judgment of the intuitionistic world and will coincide with the sequent calculus of \cite{LSR17}. The linear judgment is multiple-conclusion and carries the classical structure. Notice that the latter collapses to the judgmental structure of a polycategory when $\Theta$ is empty; the cartesian zone acts as a glue between the two worlds. At this point, we can also notice that the presence of both zones at the same time defines new calculi not mentioned in this document.

\phantomsection\label{anchor:descriptor-overview}
Both judgments are indexed by the central objects of the mode theory: the descriptors. A descriptor is an algebraic object that indexes every sequent and records how the resources are used toward producing the targets. It encodes the substructural and modal properties of the variables contained in the resources and the targets of a logical sequent. In \cite{LSR17} this role is played by a term $\alpha$ built from the context variables and the generators of the signature, and the sequent $\Gamma \vdash_{\alpha} A$ is read ``$\Gamma$ proves $A$ using its resources as prescribed by $\alpha$''. A term describes a tree with a single root; a multiple-conclusion sequent produces several outputs at once and no term can describe it. Descriptors become in our setting directed acyclic graphs, presented by two sets: an assignment set, recording the applications of the generators, and a redirection set, connecting the produced values to targets. The linear sequent is then written $\Theta \ ; \ \Gamma \vdash_{\mf D} \Delta$ and read ``$\Gamma$ proves exactly $\Delta$ as prescribed in $\mf D$, using $\Theta$ freely''. By \cref{prop:cartesian-descriptors-are-terms}, a descriptor whose output is cartesian is a \cite{LSR17} term, so the cartesian judgment recovers the display $\Theta \vdash_{\alpha} A$ of \cite{LSR17}.

Substructural properties manifest in the two classes differently. Substructurality is a property at cartesian modes and a structure at linear modes. The cartesian mode possesses all the substructural rules by default. A particular logic specification (through the mode theory) declares in the signature transformations between descriptors of a common context. Those transformations are called $2$-cells. A $2$-cell $\mf D \Rightarrow \mf D'$ will transport any derivation indexed by $\mf D'$ to a derivation indexed by $\mf D$. Those transports become admissible in the subsequent logics. At a linear mode, weakening and contraction are not admissible operations on descriptors. A logic with substructural rules at linear modes must declare them as extra structure. We call this extra structure \emph{gates}. A gate declaration adjoins generators duplicating, discarding, merging or creating the resources of a chosen linear mode, together with coherence laws. With no gate declared, the linear layer remains purely multiplicative.

The type theory layer describes the behavior of two connectives. A positive connective $F$ internalizes the resources of a sequent into a single formula and generalizes the multiplicative products of linear logic and intuitionistic conjunction. A negative connective $U$ internalizes an input alongside the surrounding contexts and generalizes multiplicative disjunction and intuitionistic implication. Every connective of a desired logic arises as an instantiation of $F$ or $U$ indexed by a descriptor. The cartesian instances recover \cite{LSR17}, where descriptors collapse to trees, and the linear instances recover the polycategorical judgmental structure of \cite{Bla23}. Mixed instances connect the two zones; in particular, the right $U$ rule of a cartesian connective in a linear context is the promotion rule used in the exponential construction of \cref{ex:linear-logic-exponential-modality}. 

The metatheory is developed uniformly over the two judgments. Identity sequents exist at every type (\cref{thm:identity}), the $F$ and $U$ connectives are left and right invertible respectively (\cref{prop:left-invertibility-of-f,prop:right-invertibility-of-u}), and cut is admissible. \Cref{thm:cut-admissibility} covers both forms of the cut; the cartesian case behaves as a substitution; the linear case cuts a single formula between two sequents in the polycategorical fashion. Since the theorem is proven for the framework itself, it applies at once to every logic a mode theory describes. An equational theory identifies derivations along cuts and identity and establishes $\beta\eta$-laws for the two connectives.

The categorical semantics closes the chapter and returns to the structure that motivated the design. By \cref{prop:mode-theory-presents-lnl-polycategory}, the mode theory presents an LNL $2$-polycategory in the sense of \cref{def:lnl-2-polycategory}; the descriptors are its morphisms and the declared transformations its $2$-cells. The type theory, taken modulo its equational theory, presents another one. \Cref{prop:type-theory-presents-lnl-polycategory} shows that the assignment sending every sequent to its descriptor is a local discrete fibration and a fibration, with opcartesian lifts for base morphisms having a single output. The connectives $U$ and $F$ supply the cartesian lifts and these one-output opcartesian lifts, respectively. This exposes the multicategorical semantics of \cite{LSR17} and the polycategorical semantics of \cite{Bla23} as special cases of  the LNL polycategories of \cite{Shu23}.

\section{The Mode Theory}\label{sec:mode-theory}

\begingroup

\subsection{Descriptors}

As discussed in the \hyperref[anchor:descriptor-overview]{opening overview of this chapter}, the central objects of interest produced by the rules are called descriptors, which are, in essence, combinatorial objects capturing the precise use of the resources in the context of a sequent to produce the targets of a sequent.
The signature consists of a pre-signature and a collection of 2-cells (\cref{def:signature}); we let $\Sigma$ range over pre-signatures and write $\underline{\Lambda}$ for the pre-signature underlying $\Lambda$. The pre-signature defines the modes of the mode theory (modalities of the formulas) as well as generators which describe the relationship between the modes. For instance, in linear logic, two modes are present (cartesian and linear), and a map between the two generates the cartesian modality in a linear context. This mode-theoretic layer plays the role of a logic specification: using declarations of modes and generators, the framework produces connectives whose behavior is modulated by the generators. Structural rules such as weakening and contraction arise as $2$-cell declarations at cartesian modes and as gate declarations at linear modes, while exchange is admissible. This section develops the general theory of descriptors. We investigate them in their general form, keeping in mind that a restricted form is actually used in the framework itself. The hope is that future work could uncover more general settings in which the unrestricted form of the descriptors arises as a primitive object. 

Different kinds of descriptors are presented in this document. The most useful and interesting ones for our purposes are the \emph{well-formed descriptors}. They encode the substructural and modal properties of the variables contained in the resources and the targets of a logical sequent and are generated by the mode-theoretic rules of \cref{sec:well-formed-descriptors}. Thus, the well-formed part only refers to the guarantee of formation. Two kinds of descriptors are investigated:
\begin{itemize}
  \item Cartesian descriptors, which are the syntactic counterpart of nonlinear morphisms $\Theta \to A$ in the semantics, where $\Theta,A$ are nonlinear objects. In the semantics, this corresponds to imposing a nonlinear structure on the subcategory (more precisely, the submulticategory) containing those objects. In the type theory, it corresponds to the nonlinear nature of the context allowing all the substructural rules. This mimics the semantics very closely (and for good reasons). In the object logic generated by the type theory using the mode theory as a parametrization, it corresponds to allowing, but not enforcing, all the substructural rules: weakening, contraction, and exchange.
  \item Linear descriptors, which are the syntactic counterpart of linear morphisms denoted $\Theta \ ; \ \Gamma \to \Delta$, where $\Gamma,\Delta$ are linear list objects and $\Theta$ is a nonlinear set of objects. Again, this mirrors in the semantics the structure of a polycategory where the morphisms are indexed by nonlinear objects. In particular, if $\Theta = \varnothing$, the intended semantic environment collapses to a polycategory. The nonlinear formulas $\Theta$ act as glue in the total semantics containing both nonlinear and linear objects. Indeed, the mixed composition between purely nonlinear morphisms and linear morphisms occurs along the nonlinear boundary of the formulas contained in the nonlinear zone.    
\end{itemize}
We use the terminology of \emph{cartesian} to refer to nonlinear objects in this document so as to emphasize the substructural nature of the objects in question.
The computational interpretation of regular descriptors is a directed acyclic graph by \cref{prop:regular-descriptors-have-acyclic-dependency-graphs}, with the input, output, and internal variables of \cref{def:input-output-and-internal-variables} as its three kinds of nodes. The input and output nodes correspond to the boundary variables, each labelled with a mode of the mode theory, while the internal nodes represent (partial) computations. Usually, we will denote input variables with the letter ``x'' as in $x, x', x_i, \dots$, output variables with the letter ``y'' as in $y, y', y_i, \dots$, and internal variables with the letter ``w'' as in $w, w', w_i, \dots$. We will also use the letter ``a'' as in $a, a', a_i, \dots$ for input variables declared at cartesian modes, in particular for those of cartesian contexts. It is understood that variable kinds are pairwise disjoint, i.e., an input variable is not an output or internal one, and vice versa. Notice that the linear or cartesian character of a variable is not part of the variable data itself. It is determined by the mode attached to it by the ambient contexts and mode assignments.
The modes are denoted by the letters $p,q$, which will be chosen to be distinct from the variable labels.
We will also have generator symbols, denoted by the letters $f,g$, chosen, again, to be distinct from the variable and mode labels, with arities shown in a standard functional way such as $f : p_1, \dots, p_n \to q$ as well as $g : (p_1, \dots, p_k \mid l_1, \dots, l_m) \to l'_1, \dots, l'_n$. The former symbols are called \emph{c-generators} and the latter $\ell$\emph{-generators}.

The descriptor syntax uses this terminology. A raw descriptor is a pair $(\mc S \mid \mc T)$ consisting of an assignment set $\mc S$ and a redirection set $\mc T$. A descriptor class $\mf D$ is an $\alpha$-equivalence class of raw descriptors, denoted $\mf D := [\mc S \mid \mc T]$, where $(\mc S \mid \mc T)$ is a representative for the class; $\alpha$-equivalence renames internal variables only and is defined in \cref{def:internal-renaming-equivalence-alpha-equivalence}. A regular descriptor is a raw descriptor satisfying mode-validity alongside usage and graph conditions (\cref{def:regular-raw-descriptor}). A regular descriptor class is a descriptor class such that one, hence by \cref{prop:internal-renaming-preserves-regularity} any, representative is regular. Finally, the well-formed descriptor classes are those generated by the rules of the mode theory (\cref{sec:well-formed-descriptors}), and a well-formed descriptor is a raw representative of such a class. The relationship between all those objects is summarized by the diagram:

\begin{center}
  \begin{tikzcd}[row sep=large, column sep=huge]
    \mathrm{desc}^{\mathrm{raw}}_{\Sigma}(\gamma \ ; \ \delta) \arrow[r, "{(\--) \mapsto [(\--)]}"] & \mathrm{Desc}^{\mathrm{raw}}_{\Sigma}(\gamma \ ; \ \delta) \\
    \mathrm{desc}_{\Sigma}(\gamma \ ; \ \delta) \arrow[r, "{(\--) \mapsto [(\--)]}"] \arrow[u, hook] & \mathrm{Desc}_{\Sigma}(\gamma \ ; \ \delta) \arrow[u, hook]
  \end{tikzcd}
\end{center}
where the right column consists of the $\alpha$-equivalence classes of the representatives in the left column. The upward hooks are the injections from the well-formed descriptors defined in \cref{sec:well-formed-descriptors} to the raw descriptors.

\textit{Convention:}
We use the standard finite-set notation. For a set $A$, we write $\mc P_{\mathrm{fin}}(A)$ for the set of finite subsets of $A$. Finite sets are written with braces as in $\{a,b\}$. We also use the usual set-builder syntax $\{\,x\mid\dots\,\}$ and the union $\mc T\cup\mc T'$. Both the assignment component and the redirection component of a descriptor are finite sets. Repeated identical entries are therefore identified.

\subsubsection{Raw Descriptors}

\begin{definition}[Pre-Signature]\label{def:pre-signature}
  We define the \textit{Pre-Signature} judgment $\Sigma \ \mathrm{sig}$ by these five formation rules:
  $$
  \inferrule*
  { }
  {\cdot \ \mathrm{sig}}
  \quad
  \inferrule*
  {
    \Sigma \ \mathrm{sig} \\
    p \notin \Sigma
  }
  {(\Sigma, p \modec) \ \mathrm{sig}}
  \quad
  \inferrule*
  {
    \Sigma \ \mathrm{sig} \\
    p \notin \Sigma
  }
  {(\Sigma, p \model) \ \mathrm{sig}}
  $$
  $$
  \inferrule*
  {
    \Sigma \ \mathrm{sig} \\
    p_1 \modec \in \Sigma\\
    \dots\\
    p_n \modec \in \Sigma\\
    q \modec \in \Sigma\\
    f \notin \Sigma
  }
  {(\Sigma, f : p_1, \ldots, p_n \to q) \ \mathrm{sig}}
  $$
  $$\inferrule*
  {
    \Sigma \ \mathrm{sig} \\
    p_1 \modec \in \Sigma\\
    \dots\\
    p_k \modec \in \Sigma\\
    l_1 \model \in \Sigma\\
    \dots\\
    l_m \model \in \Sigma\\
    l'_1 \model \in \Sigma\\
    \dots\\
    l'_n \model \in \Sigma\\
    m+n \geq 1 \\
    g \notin \Sigma
  }
	  {(\Sigma, g : (p_1, \ldots, p_k \ \mid   l_1, \ldots, l_m) \to l'_1, \ldots, l'_n) \ \mathrm{sig}}
	 $$
	 For a pre-signature $\Sigma \ \mathrm{sig}$, we write
	\[
	\mathrm{Mode}(\Sigma):=\{p\mid p\modec\in\Sigma \ \text{or}\ p\model\in\Sigma\},
	\qquad
	\mathrm{Mode}^{\mathsf{l}}(\Sigma):=\{p\mid p\model\in\Sigma\}
\]
for the sets of modes and of linear modes declared in $\Sigma$.
\end{definition}

\textit{Remark:} In the formation rules, some of the lists may be empty. We only require $m + n \geq 1$ for the $\ell$-generators. When the cartesian argument list is empty, we abbreviate $g:(\cdot\, \mid \,l_1,\ldots,l_m)\to l'_1,\ldots,l'_n$ by $g:l_1,\ldots,l_m\to l'_1,\ldots,l'_n$.

A judgment $\Sigma \ \mathrm{sig}$ collects modes and generator symbols. The empty-signature rule creates the empty pre-signature, the two mode-extension rules append a mode of either class, and the two generator-extension rules add c-generators and $\ell$-generators.
\begin{example}[Linear--Nonlinear Logic]\label{ex:linear-logic-mode-signature}
Linear--nonlinear logic is a natural instance exhibiting multiple modes (linear and cartesian) as well as explicit substructural rules. We will use this logic as a
running example to illustrate the various parts of the syntactic system. As outlined in the \hyperref[anchor:descriptor-overview]{chapter overview},
linear logic can be seen as a logic of two modes. The mixed two-mode instance yields the exponential modality of \cref{ex:linear-logic-exponential-modality}, while its pure linear specialization is worked out in \cref{sec:examples-blanco-cmll}. We denote the cartesian mode by $c\modec$ and the
linear mode by $l\model$. The relationship between the two modes is captured by the existence of an $\ell$-generator $g : (c\, \mid \,\cdot)\to l$, which produces a linear resource from a cartesian one. This generator, as well as the apparatus of the type theory, will be expressive enough to encode the exponential modality which relates cartesian and linear resources \cite{Ben95,Mel09}.
In this mode-theoretic presentation, multiplicative positive connectives at the linear mode are represented by $\ell$-generators with a single target,
$$f:(\cdot\, \mid \,l_1,\ldots,l_m)\to l,$$
while multiplicative negative connectives are represented by $\ell$-generators with a single source,
$$g:(\cdot\, \mid \,l)\to l'_1,\ldots,l'_n.$$
Thus the pre-signature includes symbols such as
$$\otimes_l:l,l\to l \qquad\text{and}\qquad \parr_l:l\to l,l$$
in the abbreviated notation.
Here $\otimes_l$ represents the multiplicative conjunction, while $\parr_l$ is the multiplicative disjunction. Notice that this is not the ordinary binary arity of $\parr$ as a connective. It is an encoding in the descriptor calculus. The displayed arities describe how a descriptor consumes and produces variables and not the syntactic arity of formulas. An informal way to think about the arity of the connective the generators produce is to consider the domain to be the arity of the connective for the formulas on the left-hand side of the turnstile in the object logic, and the codomain to be the arity of the connective on the right-hand side of the turnstile. With this reading, $\otimes_l$ is a binary connective generated by two formulas on the left-hand side of the turnstile, so $x: A, y : B \vdash A \otimes B$ should be admissible. Similarly, $\parr_l$ is the binary connective generated by two formulas on the right-hand side of the turnstile, so $z : A \parr B \vdash x : A, y : B$ should be admissible. The nullary and co-nullary units are written in the same convention. For example, $I_l:\cdot\to l$ gives assignments of the form $\cdot\xmapsto{I_l}w$, while $\bot_l:l\to\cdot$ gives assignments of the form $x\xmapsto{\bot_l}\cdot$. Since assignments are set-based, duplicate identical co-nullary assignments are identified.
\end{example}

\begin{definition}[Contexts]\label{def:contexts}
For a pre-signature $\Sigma \ \mathrm{sig}$, we define the \textit{Cartesian Context}, \textit{Input Context}, and \textit{Output Context} judgments denoted $\theta \ \mathrm{ctx}^{\mathrm{c}}_{\Sigma}$, $\gamma \ \mathrm{ctx}^{\mathrm{in}}_{\Sigma}$, and $\delta \ \mathrm{ctx}^{\mathrm{out}}_{\Sigma}$ by these rules:
  \[
  \begin{aligned}
  \inferrule*{ }{\cdot \ \mathrm{ctx}^{\mathrm{c}}_{\Sigma}}
  &\quad
  \inferrule*
  {
    \theta \ \mathrm{ctx}^{\mathrm{c}}_{\Sigma} \\
    p \modec \in \Sigma \\
    a \ \text{an input variable} \\
    a \ \notin \theta
  }
  {(\theta , a : p ) \ \mathrm{ctx}^{\mathrm{c}}_{\Sigma}}
  \\[1.5em]
  \inferrule*{ }{\cdot \ \mathrm{ctx}^{\mathrm{in}}_{\Sigma}}
  &\quad
  \inferrule*
  {
    \gamma \ \mathrm{ctx}^{\mathrm{in}}_{\Sigma} \\
    p \in \mathrm{Mode}(\Sigma) \\
    x \ \text{an input variable} \\
    x \ \notin \gamma
  }
  {(\gamma , x : p ) \ \mathrm{ctx}^{\mathrm{in}}_{\Sigma}}
  \\[1.5em]
  \inferrule*{ }{\cdot \ \mathrm{ctx}^{\mathrm{out}}_{\Sigma}}
  &\quad
  \inferrule*
  {
    \delta \ \mathrm{ctx}^{\mathrm{out}}_{\Sigma} \\
    q \in \mathrm{Mode}(\Sigma) \\
    y \ \text{an output variable} \\
    y \ \notin \delta
  }
  {(\delta , y : q ) \ \mathrm{ctx}^{\mathrm{out}}_{\Sigma}}
  \end{aligned}
  \]
\end{definition}
These rules build the empty contexts and the lists of pairs of variables and modes. Cartesian contexts declare cartesian modes only, while input and output contexts may declare modes of both classes.

\begin{definition}[Context Permutations]\label{def:context-permutations}
Let $\gamma,\gamma'$ be input contexts. A context permutation $\pi:\gamma\cong \gamma'$ is a permutation of the declarations which preserves variable-mode labels.  We define output context permutations $\tau:\delta\cong \delta'$ and cartesian context permutations $\sigma:\theta\cong\theta'$ in the same way.
\end{definition}
\textit{Remark:} We note that the property ``$x:p \in \gamma$ if and only if $x:p \in \gamma'$'' determines the permutation.

\textit{Remark:} We note that a context permutation is not an equality of contexts. It is only a witness that the two displayed contexts differ by the order of their declarations.
\begin{example}[Linear--Nonlinear Logic]
A context may declare variables of both mode classes. For instance, in this two-mode presentation, a context could be $\gamma := a : c,\, x' : l$, where $a$ is declared at the cartesian mode and $x'$ at the linear mode. A cartesian context only declares variables at cartesian modes. For instance, the context $\theta := a : c$ is cartesian. The context $\gamma' := x' : l,\, a : c$ declares the same variables at the same modes, and the passage from $\gamma$ to $\gamma'$ is witnessed by a context permutation $\pi : \gamma \cong \gamma'$. The fragment of the framework in which every mode is of the cartesian class recovers the treatment of \cite{LSR17} exactly. There, the substructural rules allowed at a mode are controlled by the $2$-cells of \cref{def:descriptor-2-cells}.
\end{example}

\begin{definition}[Cartesian Expressions]\label{def:cartesian-expressions}
Let $\Sigma \ \mathrm{sig}$ be a pre-signature and $\theta \ \mathrm{ctx}^{\mathrm{c}}_{\Sigma}$ a cartesian context. The \textit{expression} judgment $\theta \vdash \alpha : q$, for $q$ a cartesian mode of $\Sigma$, is generated by these two rules:
$$
\inferrule*
{
  a : q \in \theta
}
{\theta \vdash a : q}
\quad
\inferrule*
{
  f : p_1, \ldots, p_n \to q \in \Sigma \\
  \theta \vdash \alpha_1 : p_1 \\
  \dots \\
  \theta \vdash \alpha_n : p_n
}
{\theta \vdash f(\alpha_1, \ldots, \alpha_n) : q}
$$
\end{definition}

\begin{example}[Linear--Nonlinear Logic]
We extend the linear--nonlinear pre-signature with a binary generator $\otimes_c : c, c \to c$ at the cartesian mode. For the cartesian context $\theta := a : c,\, a' : c$, the rules generate the judgments $\theta \vdash a : c$ and $\theta \vdash \otimes_c(a, a') : c$, as well as $\theta \vdash \otimes_c(a, a) : c$, in which the variable $a$ is used twice while $a'$ is not used at all. A cartesian variable may thus be used any number of times in an expression, including not at all.
\end{example}

Expressions are the ordinary terms in a standard treatment of intuitionistic type theories. In particular, they are the syntax of the \cite{LSR17} fragment. Here, they are generated by the c-generators. Our treatment of the syntactic nature of the classical and intuitionistic systems requires more expressive power. This added expressivity is encoded in the generation of terms by the other $\ell$-generators.
We are ready now to introduce the \emph{plat de r\'esistance} of this section. The descriptors are the combinatorial objects encoding the underlying information of resource management and target redirections.  
We start by defining the valid expressions that our descriptors will manipulate. The idea is that a generator performs some computation using its sources and redirects the results to its targets. A c-generator $f : p_1, \dots, p_n \to q$ computes a cartesian value from cartesian sources, while an $\ell$-generator $g : (p_1, \dots, p_k \ \mid \ l_1, \dots, l_m) \to l'_1, \dots, l'_n$ consumes cartesian and linear sources and produces linear results. Proof-theoretically, this is exactly what is needed to express classical systems. Intuitionistic systems, by contrast, perform a single computation and redirect it to the only output. The descriptors in the intuitionistic context are then just interpreted as (proof) trees with the computation result encoded in the root node and the inputs at the leaves. In the classical context, we work with directed acyclic (proof) graphs. Back to the formal definitions.

A descriptor consists of two sets: an assignment set and a redirection set. The assignments represent the storage of intermediate computations and ultimately encode the steps of a logical derivation tree. The redirections terminate the computations and make the composition (cut) operation well defined, as well as control the multiplicity of the outputs.
\begin{definition}[Assignments and Redirections]\label{def:assignments-and-redirections}
  The set $\mathrm{Asg}_{\Sigma}(\gamma)$ is generated by these clauses.
  \begin{itemize}
    \item If $f:p_1,\ldots,p_n\to q\in\Sigma$ is a c-generator, if $s_1,\ldots,s_n$ are either internal variables or input variables in $\gamma$, and if $w$ is an internal variable, then $$s_1,\ldots,s_n\xmapsto{f}w$$ is an assignment.
    \item If $g:(p_1,\ldots,p_k \ \mid  \ l_1,\ldots,l_m)\to l'_1,\ldots,l'_n\in\Sigma$ is an $\ell$-generator, if $s_1,\ldots,s_k$ and $s'_1,\ldots,s'_m$ are either internal variables or input variables in $\gamma$, and if $w_1,\ldots,w_n$ are internal variables, then $$(s_1,\ldots,s_k \ \mid  \ s'_1,\ldots,s'_m)\xmapsto{g}w_1,\ldots,w_n$$ is an assignment.
  \end{itemize}
  When $k=0$, we suppress the cartesian component from the notation and write $s'_1,\ldots,s'_m\xmapsto{g}w_1,\ldots,w_n$.

  The set $\mathrm{Rdr}_{\Sigma}(\gamma\ ;\ \delta)$ is generated by one clause: if $s$ is an internal variable or an input variable in $\gamma$, and $y$ is an output variable declared in $\delta$, then $$s\mapsto y$$ is a redirection.
\end{definition}

\begin{example}[Linear--Nonlinear Logic]\label{ex:linear-logic-assignments-redirections}
We work over the linear--nonlinear pre-signature extended with the binary c-generator $\otimes_c$. For the input context $\gamma := a : c,\, a' : c,\, x : l,\, x' : l$, these are assignments:
$$a, a' \xmapsto{\otimes_c} w, \qquad x, x' \xmapsto{\otimes_l} w', \qquad x \xmapsto{\parr_l} w_1, w_2, \qquad (a \ \mid \ \cdot) \xmapsto{g} w''.$$
The first stores in $w$ the result of applying the binary c-generator to the two cartesian inputs. The second stores in $w'$ the multiplicative conjunction of the two linear inputs. The third splits the input $x$ into the two targets $w_1, w_2$ of a multiplicative disjunction. The fourth produces the linear resource $w''$ from the cartesian input $a$. For an output context $\delta := y : l,\, y' : l$, examples of redirections include $w' \mapsto y$ and $x \mapsto y'$.
\end{example}

The assignment and redirection sets are the main tools we use to define the syntactic presentations of the (co)multimorphisms. Recall that the LSR treatment of intuitionistic type theories uses the syntactic presentation of multimorphisms as trees. This is adequate, as a multimorphism only possesses a single output, which in turn corresponds to the root of the given tree. For our purposes, the presence of multimorphisms and comultimorphisms requires a richer object for consideration. We elected to use directed acyclic graphs, and the reading is that the nodes correspond to assignments. The redirections are the technical tool needed to carve out the outputs of the graph, which was not required in the tree instances since the output is the unique root.

\textit{Global Freshness Convention:} Whenever a construction introduces a fresh input
variable, output variable, internal variable, or a fresh family of internal-variable copies, the
choice is made disjointly from the variables already in use. 

\begin{definition}[Targets and Sources of Assignments and Redirections]\label{def:targets-and-sources-of-assignments-and-redirections}
For an assignment $E$, define its \emph{target set} and \emph{source set} by
\begin{align*}
\mathrm{tgt}(s_1,\ldots,s_n\xmapsto{f}w)&:=\{w\}, \\
\mathrm{tgt}\big((s_1,\ldots,s_k \ \mid  \ s'_1,\ldots,s'_m)\xmapsto{g}w_1,\ldots,w_n\big)&:=\{w_1,\ldots,w_n\}. \\[0.5em]
\mathrm{src}(s_1,\ldots,s_n\xmapsto{f}w)&:=\{s_1, \dots, s_n\}, \\
\mathrm{src}\big((s_1,\ldots,s_k \ \mid  \ s'_1,\ldots,s'_m)\xmapsto{g}w_1,\ldots,w_n\big)&:=\{s_1,\ldots,s_k,\,s'_1,\ldots,s'_m\}.
\end{align*}
For a redirection $s \mapsto y$, define its \emph{target set} and \emph{source set} by
$$
\mathrm{tgt}(s \mapsto y):=\{y\}, \qquad \mathrm{src}(s \mapsto y):=\{s\}.
$$
Thus assignment targets are internal variables, while redirection targets are output variables.
For an assignment set or redirection set $\mc G$, define
$$\mathrm{tgt}(\mc G):=\bigcup_{E\in\mc G}\mathrm{tgt}(E), \qquad \mathrm{src}(\mc G):=\bigcup_{E\in\mc G}\mathrm{src}(E).$$
\end{definition}

\textit{Remark:} The maps $\mathrm{src}$ and $\mathrm{tgt}$ are set-valued. They forget the order and the multiplicity of the source list. A separate notion of  \emph{Mode-validity} (\cref{def:mode-assignment}) is used to check ordering and mode-assignment consistency with the morphism symbols. Two assignments with the same $\mathrm{src}$ may therefore impose distinct mode constraints.

\begin{example}[Linear--Nonlinear Logic]
For the assignments of the \hyperref[ex:linear-logic-assignments-redirections]{assignment-and-redirection example}, $\mathrm{src}(x, x' \xmapsto{\otimes_l} w') = \{x, x'\}$ and $\mathrm{tgt}(x, x' \xmapsto{\otimes_l} w') = \{w'\}$, while $\mathrm{src}(x \xmapsto{\parr_l} w_1, w_2) = \{x\}$ and $\mathrm{tgt}(x \xmapsto{\parr_l} w_1, w_2) = \{w_1, w_2\}$. For the redirection $w' \mapsto y$, we have $\mathrm{src}(w' \mapsto y) = \{w'\}$ and $\mathrm{tgt}(w' \mapsto y) = \{y\}$. The assignment $x, x \xmapsto{\otimes_l} w'$, whose source list repeats the variable $x$, has the singleton source set $\{x\}$.
\end{example}

\begin{definition}[Raw Descriptor]\label{def:raw-descriptor}
A \emph{raw descriptor} denoted by $(\mc S \mid \mc T)$ for input variables in $\gamma$ and output variables in $\delta$ consists of a finite assignment set $\mc S$ and a finite redirection set $\mc T$, i.e.,
$$(\mc S \mid \mc T) \in \mc P_{\mathrm{fin}}(\mathrm{Asg}_{\Sigma}(\gamma)) \times \mc P_{\mathrm{fin}}(\mathrm{Rdr}_{\Sigma}(\gamma \ ; \ \delta)).$$
We write $$\mathrm{desc}^{\mathrm{raw}}_\Sigma(\gamma \ ; \ \delta) := \mc P_{\mathrm{fin}}(\mathrm{Asg}_{\Sigma}(\gamma)) \times \mc P_{\mathrm{fin}}(\mathrm{Rdr}_{\Sigma}(\gamma \ ; \ \delta))$$
and call $(\gamma \ ; \ \delta)$ the \textit{boundary}.
\end{definition}

\textit{Remark:}
Note that assignments are finite sets; thus duplicate expressions are identified.
For example, two identical assignments $v\xmapsto{g}\cdot$ are represented by the same singleton assignment set $\{v\xmapsto{g}\cdot\}$. The same goes for redirections. Two identical redirections $s\mapsto y$ are represented by the same singleton $\{s\mapsto y\}$. In other words, the multiplicity of the occurrences of the assignments and redirections is not captured by the operations. A treatment that records multiplicity has been investigated by the author, and the descriptor theory can be adjusted to account for it. The main area where the recording of the multiplicity is beneficial is the ability to forgo the renaming requirements when composing descriptors together and in some sense make the treatment of $\alpha$-equivalence a little more tractable.

\begin{definition}[Permutation Transport of Raw Descriptors]\label{def:permutation-transport-of-raw-descriptors}
Let $\pi:\gamma\cong\gamma'$ and $\tau:\delta\cong\delta'$ be context permutations. We define the transport map
$$(\pi;\tau)_*:\mathrm{desc}^{\mathrm{raw}}_\Sigma(\gamma\ ;\ \delta)\to\mathrm{desc}^{\mathrm{raw}}_\Sigma(\gamma'\ ;\ \delta')$$
by
$$(\pi;\tau)_*(\mc S\mid\mc T):=(\mc S\mid\mc T).$$
\end{definition}

\textit{Remark:} This is well defined since assignments and redirections only use the declared variables of the contexts and not the positions of those declarations. The inverse of the transport is $((\pi;\tau)_*)^{-1} = (\pi^{-1};\tau^{-1})_*$.

\begin{example}[Linear--Nonlinear Logic]\label{ex:linear-logic-context-permutation}
The input contexts $\gamma := x : l,\, x' : l$ and $\gamma' := x' : l,\, x : l$ differ by the order of their declarations. Let $\pi : \gamma \cong \gamma'$ be the corresponding context permutation and $\tau$ the identity permutation on $\delta := y : l$. The raw descriptor $(\{x, x' \xmapsto{\otimes_l} w\} \mid \{w \mapsto y\})$ over the boundary $(\gamma \ ; \ \delta)$ is transported by $(\pi;\tau)_*$ to the same pair, now regarded as a raw descriptor over the boundary $(\gamma' \ ; \ \delta)$. The transport does not modify the assignment and redirection sets. It only changes the boundary.
\end{example}

Raw descriptors only record whether a variable is an input, output, or internal variable. They do not require the modes of those variables to match the arities of the generator symbols or the targets of redirections. A raw redirection $x \mapsto y$ is syntactically allowed even when $x$ and $y$ have different modes. We use the notion of \emph{Mode-validity} to enforce this requirement. It allows only the descriptors that possess internal variables consistently matching the symbols contained in the pre-signature.

Moreover, raw descriptors do not enforce the presence of a redirection to every output. They also do not require using all the inputs in some assignments. Regular descriptors (\cref{def:regular-raw-descriptor}) restrict these behaviors: they require internal variables and linear inputs to be used exactly once, and every output to be redirected to exactly once. Inputs at \emph{cartesian} modes may instead be used any number of times, including not at all. The substructural behavior therefore splits by class. At cartesian modes, descriptors may genuinely drop or reuse their inputs.

\begin{example}[Identity Descriptor]
The identity raw descriptor for a boundary $(x : p \ ; \ y : p)$ has an empty assignment set $\mc S = \varnothing$, and the redirection set is the singleton $\mc T = \{x \mapsto y\}$. This represents the identity map $p \to p$ where the input is sent to the output with no intermediate computations. At a cartesian mode, it corresponds to the identity expression $a : p \vdash a : p$ of \cref{def:cartesian-expressions}.
\end{example}

\begin{definition}[Componentwise Sum]\label{def:componentwise-sum}
For assignment sets $\mc S_1,\mc S_2$ and redirection sets $\mc T_1,\mc T_2$, define their componentwise sum by
$$
(\mc S_1 \mid \mc T_1)\cup(\mc S_2\mid\mc T_2)
:=
(\mc S_1\cup\mc S_2 \mid \mc T_1\cup\mc T_2).
$$
\end{definition}

\textit{Remark:}
If both pairs are raw descriptors in the same ambient context
$\mathrm{desc}^{\mathrm{raw}}_\Sigma(\gamma\ ;\ \delta)$, their componentwise sum is again a raw descriptor in that context. 

\begin{example}[Linear--Nonlinear Logic]\label{ex:linear-logic-componentwise-sum}
Over the boundary $(x : l \ ; \ y : l,\, y' : l)$, the componentwise sum
$$(\{x \xmapsto{\parr_l} w_1, w_2\} \mid \varnothing) \cup (\varnothing \mid \{w_1 \mapsto y,\ w_2 \mapsto y'\}) = (\{x \xmapsto{\parr_l} w_1, w_2\} \mid \{w_1 \mapsto y,\ w_2 \mapsto y'\})$$
builds the descriptor of a multiplicative disjunction from its assignment set and its redirection set.
\end{example}

\begin{definition}[Input, Output, and Internal Variables]\label{def:input-output-and-internal-variables}
  We define the sets of \emph{input, output}, and \emph{internal variables} of a raw descriptor as follows:
  \begin{itemize}
    \item For a variable $v$, we set
    \[
    \begin{aligned}
    \mathrm{Inp}(v) &=
    \begin{cases}
      \{v\} &\text{if $v$ is an input variable}\\
      \varnothing &\text{else}
    \end{cases} \\[0.5em]
    \mathrm{Int}(v) &=
    \begin{cases}
      \{v\} &\text{if $v$ is an internal variable}\\
      \varnothing &\text{else}
    \end{cases} \\[0.5em]
    \mathrm{Out}(v) &=
    \begin{cases}
      \{v\} &\text{if $v$ is an output variable}\\
      \varnothing &\text{else}
    \end{cases}
    \end{aligned}
    \]
		\item For a set of variables $W$ and any $\mathrm{V} \in \{\mathrm{Inp}, \mathrm{Int}, \mathrm{Out}\}$, we set
		$$\mathrm{V}(W) := \bigcup_{w \in W} \mathrm{V}(w),$$
    where the result is an ordinary set.
		\item For an assignment $ E := (s_1, \dots, s_n \xmapsto{f} w)$ and any $\mathrm{V} \in \{\mathrm{Inp}, \mathrm{Int}, \mathrm{Out}\}$, we set $$\mathrm{V}(E) := \mathrm{V}(\{s_1, \dots, s_n, w\}).$$
		\item For an assignment $ E := ((s_1, \dots, s_k \ \mid \ s'_1, \dots, s'_m) \xmapsto{g} w_1, \dots, w_n)$ and any $\mathrm{V} \in \{\mathrm{Inp}, \mathrm{Int}, \mathrm{Out}\}$, we set $$\mathrm{V}(E) := \mathrm{V}(\{s_1, \dots, s_k, s'_1, \dots, s'_m, w_1, \dots, w_n\}).$$
		\item For a redirection $R := (s \mapsto y)$ and any $\mathrm{V} \in \{\mathrm{Inp}, \mathrm{Int}, \mathrm{Out}\}$, we set $$\mathrm{V}(R) := \mathrm{V}(\{s, y\}).$$ 
		\item For an assignment set or redirection set $\mc G$ and $\mathrm{V} \in \{\mathrm{Inp}, \mathrm{Int}, \mathrm{Out}\}$, we set:
					$$\mathrm{V}(\mc G) := \bigcup_{E \in \mc G} \mathrm{V}(E)$$
		\item For a raw descriptor $(\mc S \mid \mc T)$ and $\mathrm{V} \in \{\mathrm{Inp}, \mathrm{Int}, \mathrm{Out}\}$, we set:
					$$\mathrm{V}((\mc S \mid \mc T)) :=\mathrm{V}(\mc S) \cup \mathrm{V}(\mc T) $$
  \end{itemize}
  We write $\mathrm{Var}(\--) := \mathrm{Inp}(\--) \cup \mathrm{Int}(\--) \cup \mathrm{Out}(\--).$
\end{definition}
\textit{Notation:} We write $\mathrm{V}(\mc S \mid \mc T)$ instead of the more cumbersome $\mathrm{V}((\mc S \mid \mc T))$.

\begin{example}[Linear--Nonlinear Logic]\label{ex:linear-logic-mixed-raw-descriptor}
We consider the raw descriptor
$$D := (\{(a \ \mid \ \cdot) \xmapsto{g} w,\ \ w, x \xmapsto{\otimes_l} w'\} \mid \{w' \mapsto y\})$$
over the boundary $(a : c,\, x : l \ ; \ y : l)$. The descriptor produces the linear resource $w$ from the cartesian input $a$ and stores in $w'$ the multiplicative conjunction of $w$ with the linear input $x$. We then have
$$\mathrm{Inp}(D) = \{a, x\}, \qquad \mathrm{Int}(D) = \{w, w'\}, \qquad \mathrm{Out}(D) = \{y\}, \qquad \mathrm{Var}(D) = \{a, x, w, w', y\}.$$
\end{example}

\begin{definition}[Mode Assignment]\label{def:mode-assignment}
Let $(\mc S\mid\mc T)\in \mathrm{desc}^{\mathrm{raw}}_\Sigma(\gamma;\delta)$.

A mode assignment for $(\mc S\mid\mc T)$ is a function
$$
\mu:\mathrm{Int}(\mc S\mid\mc T)\to \mathrm{Mode}(\Sigma).
$$

For each $v \in \mathrm{Var}(\mc S \mid \mc T)$, we write $\mathrm{mode}_\mu(v)$ for the mode of $v$ determined by $\mu$, $\gamma$, and $\delta$ as follows:
$$
\mathrm{mode}_\mu(v)=
\begin{cases}
\mu(v) & \text{if } v\in \mathrm{Int}(\mc S\mid\mc T),\\
p & \text{if } v:p\in\gamma,\\
q & \text{if } v:q\in\delta.
\end{cases}
$$

We say that $(\mc S\mid\mc T)$ is \textit{mode-valid} if there exists a mode assignment $\mu$ satisfying these conditions.
\begin{itemize}
\item If
$\big(s_1,\ldots,s_n\xmapsto{f}w\big)\in\mc S \qquad\text{and}\qquad f:p_1,\ldots,p_n\to q\in\Sigma,$
then
$$\mathrm{mode}_\mu(w)=q \qquad\text{and}\qquad \mathrm{mode}_\mu(s_i)=p_i \quad\text{for every }1\leq i\leq n.$$
\item If
$\big((s_1,\ldots,s_k \ \mid \ s'_1,\ldots,s'_m)\xmapsto{g}w_1,\ldots,w_n\big)\in\mc S \qquad\text{and}\qquad g:(p_1,\ldots,p_k \ \mid \ l_1,\ldots,l_m)\to l'_1,\ldots,l'_n \in\Sigma,$
then
\[
\begin{alignedat}{2}
\mathrm{mode}_\mu(s_i)&=p_i &\qquad &(1\leq i\leq k), \\
\mathrm{mode}_\mu(s'_i)&=l_i &\qquad &(1\leq i\leq m), \\
\mathrm{mode}_\mu(w_j)&=l'_j &\qquad &(1\leq j\leq n).
\end{alignedat}
\]
\item If a redirection $s\mapsto y$ belongs to $\mc T$ and $y:q\in\delta,$
then $\mathrm{mode}_\mu(s)=q.$
\end{itemize}
\end{definition}

\textit{Remark:} The mode assignment function is the extra data needed to test whether the untyped raw descriptor $(\mc S\mid\mc T)$ is compatible with the modes declared by $\Sigma$, $\gamma$, and $\delta$.
Together with the modes assigned by $\gamma$ and $\delta$ to the input and output variables, $\mu$ determines a mode for every variable occurring in $(\mc S\mid\mc T)$.

\begin{example}[Linear--Nonlinear Logic]
The raw descriptor $D$ of the \hyperref[ex:linear-logic-mixed-raw-descriptor]{mixed raw-descriptor example} is mode-valid. The mode assignment defined by $\mu(w) = \mu(w') = l$ satisfies every condition. The assignment $(a \ \mid \ \cdot) \xmapsto{g} w$ is compatible with the arity $g : (c\, \mid \,\cdot)\to l$ since $a : c \in \gamma$ and $\mu(w) = l$. The assignment $w, x \xmapsto{\otimes_l} w'$ is compatible with the arity $\otimes_l:l,l\to l$ since $\mu(w) = l$, $x : l \in \gamma$, and $\mu(w') = l$. The redirection $w' \mapsto y$ is compatible with the declaration $y : l \in \delta$ since $\mu(w') = l$. By contrast, the raw descriptor $(\varnothing \mid \{a \mapsto y\})$ over the same boundary is not mode-valid. The redirection forces $\mathrm{mode}_\mu(a) = l$, while the context declares $a : c$. Notice that in this pre-signature, a cartesian resource reaches a linear output only through the generator $g$.
\end{example}

\begin{lemma}[Uniqueness of Mode Assignments]\label{lem:uniqueness-of-mode-assignments}
A raw descriptor $(\mc S\mid\mc T)\in\mathrm{desc}^{\mathrm{raw}}_\Sigma(\gamma\ ;\ \delta)$ admits at most one mode assignment witnessing mode-validity.
\end{lemma}

\textit{Proof.} Suppose that there exist two mode assignment functions $\mu, \phi$ witnessing mode-validity. Then, by recursion on the assignments and on the redirections, the two mappings coincide. Indeed, suppose not. If they differ at an input or output mode, then one of the functions is not mode-valid (since there is only one mode per variable). If they differ at an internal variable, then the mode is enforced by a generator. Since the generator domain and codomain only refer to a unique mode, one of the mode assignment functions is not mode-valid, as it contains a mode different from the one provided by the generator. Thus, the two mode assignment functions must coincide at every input, output, and internal variable.
 \hfill $\square$

\textit{Remark:} Mode-validity is decidable, as there are only a finite number of constraints to check.
\endgroup{}
\begingroup

\subsubsection{Operations on Descriptors}

\paragraph{Reachability and the dependency graph.}
The goal of this section is to develop the tools necessary to arrive at a well-defined notion of composition of descriptors. To that end, we would like to understand and characterize paths in descriptors. By a path here, we mean the directed relationships between variables contained in assignment sets and redirection sets. For regular descriptors, the directed-acyclic-graph interpretation anticipated in the \hyperref[anchor:descriptor-overview]{chapter overview} is established by \cref{prop:regular-descriptors-have-acyclic-dependency-graphs}. We therefore investigate reachability and the associated dependency graph. \emph{Reachability} is a transitive relation that indicates whether two variables are connected by a sequence of assignments and redirections, while the dependency graph is a subgraph that contains all the variables reachable from a specific one. Alternatively, it is the set of all descendants of that variable.

\begin{definition}[Reachability Relation]\label{def:reachability-relation}
Let $(\mc S \mid \mc T) \in \mathrm{desc}^{\mathrm{raw}}_\Sigma(\gamma \ ; \ \delta)$ be a raw descriptor. We define the \textit{one-step reachability relation} $\leadsto$ by recursion. 
\begin{itemize}
  \item For an assignment $E := s_1,\ldots,s_n\xmapsto{f}w$, define $ \leadsto_E \ := \ \{(s_i,w)\}_{ 1 \leq i \leq n}. $
  \item For an assignment $E := (s_1,\ldots,s_k \ \mid \ s'_1,\ldots,s'_m)\xmapsto{g}w_1,\ldots,w_n$, define
  \[
  \leadsto_E \ := \ \{(s_i,w_j)\}_{1 \leq i \leq k,\ 1 \leq j \leq n} \cup \{(s'_i,w_j)\}_{1 \leq i \leq m,\ 1 \leq j \leq n}.
  \]
  \item For a redirection $R := s \mapsto y$, define $ \leadsto_R \ := \ \{(s,y)\}. $
  \item For an assignment set $\mc S$, define $ \leadsto_{\mc S} \ := \ \bigcup_{E \in \mc S} \leadsto_E. $
  \item For a redirection set $\mc T$, define $ \leadsto_{\mc T} \ := \ \bigcup_{R \in \mc T} \leadsto_R. $
  \item Finally, for a raw descriptor $(\mc S \mid \mc T)$, define $ \leadsto_{(\mc S \mid \mc T)} \ := \ \leadsto_{\mc S} \cup \leadsto_{\mc T}. $
\end{itemize}
The \textit{strict reachability relation} $\rightsquigarrow_{(\mc S \mid \mc T)}$ is the transitive closure of $\leadsto_{(\mc S \mid \mc T)}$. 

\end{definition}

We identify these relations with their infix equivalents, i.e., we write $a \leadsto_R b$ and $a \rightsquigarrow_R b$ for the statements $(a,b) \in \leadsto_R$ and $(a,b) \in \rightsquigarrow_R$, respectively. Notice that the relation is not reflexive.
When the raw descriptor is clear from context, we write $a \rightsquigarrow b$.

We also use $\rightsquigarrow_{(\mc S \mid \mc T)}$ as a partially applied operator. For a variable $v$, we write
$$v \rightsquigarrow_{(\mc S \mid \mc T)} \ := \ \{u \mid v \rightsquigarrow_{(\mc S \mid \mc T)} u\}$$
for the set of variables reachable from $v$ in $(\mc S \mid \mc T)$. When the descriptor is clear from context, we abbreviate this to $v \rightsquigarrow$.

\begin{example}[Linear--Nonlinear Logic]
We return to the raw descriptor of the \hyperref[ex:linear-logic-mixed-raw-descriptor]{mixed raw-descriptor example},
$$D := (\{(a \ \mid \ \cdot) \xmapsto{g} w,\ \ w, x \xmapsto{\otimes_l} w'\} \mid \{w' \mapsto y\})$$
over the boundary $(a : c,\, x : l \ ; \ y : l)$. The assignment $(a \ \mid \ \cdot) \xmapsto{g} w$ generates $a \leadsto w$, the assignment $w, x \xmapsto{\otimes_l} w'$ generates $w \leadsto w'$ and $x \leadsto w'$, and the redirection $w' \mapsto y$ generates $w' \leadsto y$. Taking the transitive closure, we obtain in addition the dependencies $a \rightsquigarrow w'$, $a \rightsquigarrow y$, $w \rightsquigarrow y$, and $x \rightsquigarrow y$. In the partially applied notation, we have $a \rightsquigarrow \ = \ \{w, w', y\}$ and $x \rightsquigarrow \ = \ \{w', y\}$. The internal variable $w$ is not reachable from the linear input $x$: it is produced from the cartesian input $a$ alone.
\end{example}

\begin{definition}[Underlying Dependency Graph]\label{def:underlying-dependency-graph}
Let $(\mc S\mid\mc T)$ be a raw descriptor. Its \textit{underlying dependency graph} is the directed graph
$$G(\mc S\mid\mc T):=(\mathrm{Var}(\mc S\mid\mc T),\leadsto_{(\mc S\mid\mc T)}).$$
\end{definition}

\textit{Remark:} The graph $G(\mc S\mid\mc T)$ is extracted from the descriptor. It records variable dependencies only and forgets the structure of assignments and redirections. In particular, an assignment with no targets ($n=0$), such as $v\xmapsto{g}\cdot$, contributes no edge to the graph $G(\mc S\mid\mc T)$.

\textit{Remark:}  Assignments with no sources and assignments with no targets behave differently. If a generator has no argument positions, i.e., if it is a nullary c-generator $f:\cdot\to q$ or an $\ell$-generator $g:\cdot\to l'$ with $k=m=0$, an assignment with no input $\cdot\xmapsto{g}w$ has an empty source set. But it creates the internal target $w$. It is therefore visible as an internal variable, although it contributes no assignment edge. For $g:l\to\cdot$ (that is, $n=0$), an assignment with no targets $v\xmapsto{g}\cdot$ has source $v$ and an empty target set. It creates no internal target and contributes no reachability edge. However, it is still present in the assignment set.

\begin{example}[Linear--Nonlinear Logic]
Over the boundary $(x : l \ ; \ y : l)$, we consider the raw descriptor
$$D' := (\{\cdot \xmapsto{I_l} w,\ \ x \xmapsto{\bot_l} \cdot\} \mid \{w \mapsto y\})$$
built from the units $I_l : \cdot \to l$ and $\bot_l : l \to \cdot$ of the linear--nonlinear pre-signature. Its underlying dependency graph has the nodes $\mathrm{Var}(D') = \{x, w, y\}$ and the single edge $w \leadsto y$, generated by the redirection. The assignment $\cdot \xmapsto{I_l} w$ contributes no edge, but it creates the internal target $w$, which the redirection then connects to the output $y$. The assignment $x \xmapsto{\bot_l} \cdot$ contributes no edge either. The input $x$ appears therefore in the graph as a node with no edges, although the assignment consuming it is present in the assignment set.
\end{example}

\textit{Remark:}
Explicitly, $a \rightsquigarrow_{(\mc S \mid \mc T)} b$ holds if and only if there exist variables $v_0,\ldots,v_k$ with $k \geq 1$, $v_0=a$, $v_k=b$, and
$$
v_0 \leadsto_{(\mc S \mid \mc T)} v_1 \leadsto_{(\mc S \mid \mc T)} \cdots \leadsto_{(\mc S \mid \mc T)} v_k.
$$
\phantomsection\label{anchor:input-output-reachability-remark}
\textit{Remark:} By definition of the reachability relation $\rightsquigarrow$, we immediately notice that no variables can reach an input $x$. This means that for any variable $s$, we have that $s \nrightsquigarrow x$. Similarly, for any output $y$, there are no variables $s$ such that $y \rightsquigarrow s$.

\textit{Remark:} We sometimes call $a \leadsto b$ an edge with a \emph{source} $a$ and a \emph{target} $b$. We reuse the same qualifiers for the reachability relation, i.e., $a \rightsquigarrow b$ has \emph{source} $a$ and \emph{target} $b$.

\begin{lemma}[One-step reachability distributes over componentwise sums]\label{lem:one-step-reachability-distributes-over-componentwise-sums}
Let $(D_i)_{i\in I}$ be a finite family of raw descriptors, with $D_i=(\mc S_i\mid\mc T_i)$, and suppose that their componentwise sum (\cref{def:componentwise-sum}) is formed as
$$D:=\bigcup_{i\in I}D_i=\left(\bigcup_{i\in I}\mc S_i\ \middle|\ \bigcup_{i\in I}\mc T_i\right).$$
Then
$$\leadsto_D=\bigcup_{i\in I}\leadsto_{D_i}.$$
\end{lemma}

\textit{Proof.}
By the definition of one-step reachability,
$$
\leadsto_D =\bigcup_{E\in\bigcup_i\mc S_i}\leadsto_E\cup\bigcup_{R\in\bigcup_i\mc T_i}\leadsto_R =\bigcup_i\left(\bigcup_{E\in\mc S_i}\leadsto_E\cup\bigcup_{R\in\mc T_i}\leadsto_R\right) =\bigcup_i\leadsto_{D_i}.
$$
\hfill $\square$

\begin{lemma}[Subrelations preserve irreflexive transitive closure]\label{lem:subrelations-preserve-irreflexive-transitive-closure}
Let $R\subseteq R'$ be binary relations on a set of variables, and let $\rightsquigarrow_R$ and $\rightsquigarrow_{R'}$ be their transitive closures. If $\rightsquigarrow_{R'}$ is irreflexive, then $\rightsquigarrow_R$ is irreflexive.
\end{lemma}

\textit{Proof.}
Notice that every non-empty $R$-path is also a non-empty $R'$-path. Hence $a\rightsquigarrow_R b$ implies $a\rightsquigarrow_{R'}b$. If $\rightsquigarrow_R$ had a reflexive instance $a\rightsquigarrow_R a$, then $\rightsquigarrow_{R'}$ would have the reflexive instance $a\rightsquigarrow_{R'}a$, a contradiction. \hfill $\square$

\begin{lemma}[Removing assignments or redirections cannot create cycles]\label{lem:removing-assignments-or-redirections-cannot-create-cycles}
Let $(\mc S\mid\mc T)$ be a raw descriptor whose strict reachability relation is irreflexive. If $\mc S_0\subseteq\mc S$ and $\mc T_0 \subseteq \mc T$, then the strict reachability relation of $(\mc S_0\mid\mc T_0)$ is irreflexive.
\end{lemma}

\textit{Proof.}
The definition of one-step reachability gives directly
$$\leadsto_{(\mc S_0\mid\mc T_0)}\subseteq\leadsto_{(\mc S\mid\mc T)}.$$
The claim follows then from \cref{lem:subrelations-preserve-irreflexive-transitive-closure}. \hfill $\square$

\begin{definition}[Redirection Filters]\label{def:redirection-filters}
Let $D=(\mc S\mid\mc T)$ be a raw descriptor and let $v$ be a variable. Write
$$
\mc T_{\mapsto v}
:= \{R\in\mc T\mid R=(s_R\mapsto v)\},
\qquad
\mc T_{\not\mapsto v}
:= \mc T\setminus\mc T_{\mapsto v},
$$
and
$$
\mc T^{\mathrm{src}}
:= \{s_R\mid R=(s_R\mapsto t_R)\in\mc T\}.
$$
\end{definition}

The subset $\mc T_{\mapsto v}$ captures all redirections to the variable $v$. These
redirections share the target $v$. They have therefore pairwise distinct sources. The source set $\mc
T_{\mapsto v}^{\mathrm{src}}$ collects those distinct source variables. The map $R=(s_R\mapsto
v)\mapsto s_R$ is then a bijection $\mc T_{\mapsto v}\to\mc T_{\mapsto v}^{\mathrm{src}}$.

\begin{example}
Consider this descriptor representing the multiplicative conjunction of linear logic $\otimes$ in the contexts $x : p, y : p \Vdash z : p$:
$$D:=\left( \{x,y \xmapsto{\otimes} w\} \mid \{w \mapsto z\} \right).$$
The subset of redirections targeting $z$ is $\{w \mapsto z\}$. The source set of redirections targeting $z$ is $\{w\}$.
\end{example}

We now prove the claim in the \hyperref[anchor:input-output-reachability-remark]{earlier reachability remark}: nothing can reach an input, and outputs cannot reach anything.
\begin{lemma}[Inputs have no incoming paths and outputs have no outgoing paths]\label{lem:inputs-have-no-incoming-paths-and-outputs-have-no-outgoing-paths}
Let $(\mc S \mid \mc T) \in \mathrm{desc}^{\mathrm{raw}}_\Sigma(\gamma \ ; \ \delta)$ be a raw descriptor. Then:
\begin{itemize}
  \item no one-step reachability edge targets an input variable;
  \item no one-step reachability edge starts from an output variable.
\end{itemize}
\end{lemma}
\textit{Proof.}
We proceed by case analysis on the generating clauses of the one-step reachability relation $\leadsto$.
For an assignment $s_1,\ldots,s_n\xmapsto{f}w$, the relation generates $s_i \leadsto w$. The target $w$ is an internal variable, and the $s_i$ are either input or internal variables, hence not output variables.
For an assignment $(s_1,\ldots,s_k \ \mid \ s'_1,\ldots,s'_m)\xmapsto{g}w_1,\ldots,w_n$, the relation generates $s_i \leadsto w_j$ and $s'_i \leadsto w_j$. Again, the targets $w_j$ are internal variables and all sources are either input or internal variables.
For a redirection $s \mapsto y$, the relation generates $s \leadsto y$. The target $y$ is an output variable, and the source $s$ is either input or internal, hence not an output variable.
In every case, $\leadsto$ has no edge with an input target or an output source. Since $\rightsquigarrow$ is the transitive closure of $\leadsto$ with $k \geq 1$, every dependence $a \rightsquigarrow b$ has a final edge $\_ \leadsto b$ and an initial edge $a \leadsto \_$. The variable $b$ cannot therefore be an input variable and $a$ cannot be an output variable.
\hfill $\square$

\textit{Remark:} In other words, the inputs are special nodes with no parents and the outputs are special nodes with no children. The entry points of a descriptor are therefore at the input variables and the exit points at the output variables. The composition constructions of \cref{def:raw-composition,def:composition-of-descriptors} connect descriptors along such output and input variables.

\Cref{lem:acyclicity-is-preserved-under-unions-of-components-sharing-only-boundary-variables} can be skipped on first reading. It is a technical lemma needed for the proofs of \cref{prop:c-comp-preserves-regularity,prop:pos-comp-preserves-regularity,prop:neg-comp-preserves-regularity}. It establishes irreflexivity of a relation when defined as the union of irreflexive subrelations that share potentially only boundary variables.
\begin{lemma}[Acyclicity is preserved under unions of components sharing only boundary variables]\label{lem:acyclicity-is-preserved-under-unions-of-components-sharing-only-boundary-variables}
Let $(\leadsto_i)_{i\in I}$ be a finite family of one-step reachability relations. Assume:
\begin{itemize}
  \item for each $i$, the transitive closure $\rightsquigarrow_i$ of $\leadsto_i$ is irreflexive;
  \item no edge in any $\leadsto_i$ targets an input variable, and no edge in any $\leadsto_i$ starts from an output variable;
  \item if $i\neq j$, then the variables shared by $\leadsto_i$ and $\leadsto_j$ are only input or output variables.
\end{itemize}
Then the transitive closure $\rightsquigarrow$ of
$$\leadsto:=\bigcup_{i\in I}\leadsto_i$$
is irreflexive.
\end{lemma}

\textit{Proof.}
Suppose for contradiction that $\rightsquigarrow$ is not irreflexive. Then there is a non-empty cycle
$$z_0 \leadsto z_1 \leadsto \cdots \leadsto z_n \leadsto z_0.$$
By the second hypothesis, no edge targets an input variable and no edge starts from an output variable. All $z_k$ are therefore internal variables for $0 \leq k \leq n$. 

Each edge $z_k \leadsto z_{k+1}$ belongs to some component $\leadsto_{i_k}$. We claim all the indices $i_k$ are equal. Indeed, if $i_k\neq i_{k+1}$, then the intermediate variable $z_{k+1}$ occurs in both $\leadsto_{i_k}$ and $\leadsto_{i_{k+1}}$. But $z_{k+1}$ is internal, contradicting the assumption that distinct components share only input or output variables. All edges of the cycle lie therefore in a single component $\leadsto_i$.

But then the same cycle witnesses that $\rightsquigarrow_i$ is not irreflexive, contradicting the first assumption. Hence $\rightsquigarrow$ is irreflexive. \hfill $\square$

\Cref{lem:fresh-output-cones-preserve-irreflexivity} states that adding variables $w, r \notin V_0$ to an irreflexive transitive closure of a relation $\leadsto_0$ on variables $V_0$, with $w$ one-step reachable from variables in $V_0$ and reaching $r$ in one step, preserves irreflexivity.
\begin{lemma}[Fresh Output Cones Preserve Irreflexivity]\label{lem:fresh-output-cones-preserve-irreflexivity}
Let $\leadsto_0$ be a relation on a finite set of variables $V_0$ whose transitive closure is irreflexive. Let $W$ be a finite set of variables disjoint from $V_0$, let $r$ be a variable not in $V_0\cup W$, and for each $w\in W$ let $P_w\subseteq V_0$. Define
$$\leadsto:=\leadsto_0\cup\{a\leadsto w\mid w\in W,\ a\in P_w\}\cup\{w\leadsto r\mid w\in W\}.$$
Then the transitive closure of $\leadsto$ is irreflexive.
\end{lemma}

\textit{Proof.}
Suppose for contradiction that there is a non-empty cycle
$$z_0\leadsto z_1\leadsto\cdots\leadsto z_n\leadsto z_0.$$
If every edge of the cycle lies in $\leadsto_0$, then this is a cycle in the transitive closure of $\leadsto_0$. This contradicts the irreflexivity hypothesis. Hence the cycle uses a new edge.

If the cycle uses an edge $a\leadsto w$ with $w\in W$, then the next edge out of $w$ must be $w\leadsto r$. Indeed, the only edges of $\leadsto$ with source $w$ are the edges $w\leadsto r$. But there is no edge of $\leadsto$ with source $r$, since $r\notin V_0\cup W$ and $r$ is not the source of any added edge. The cycle cannot therefore continue after reaching $r$, a contradiction.

If the cycle uses an edge $w\leadsto r$ with $w\in W$, the same argument applies. No edge starts at $r$. The cycle cannot close. No non-empty cycle exists therefore, and the transitive closure of $\leadsto$ is irreflexive. \hfill $\square$

\endgroup{}
\begingroup

\paragraph{Renaming and substitution.}

We consider two descriptors to be essentially the same if they differ only by the names given to their internal variables. Thus, to identify such descriptors, we set up a notion of $\alpha$-equivalence that is standard in the lambda calculus literature \cite{Bar84}. An associated notion is that of substitution. Since descriptors are the same up to renaming, it is possible to rename one into the other to collapse the equivalence to an actual equality. To perform these operations, we define what substitution means in our setting. First, we are interested in defining what it means to rename an input variable.

\begin{definition}[Input Substitution]\label{def:input-substitution}
  Let $x$ be an input variable. Let $u$ be an input or internal variable. We define the substitution of $x$ in:
  \begin{itemize}
    \item A variable $r$ by $u$ as:   
    $$
    	r[x:= u] =
    	\left\{
      	\begin{array}{ll}
        	u &\text{if } r = x \\
        	r &\text{if } r \neq x
      	\end{array}
      	\right.
    $$
		\item  An assignment $v_1, \dots, v_n \xmapsto{f} w$ by $u$ as:
	$$\big(v_1, \dots, v_n \xmapsto{f} w\big)[x := u] := \big(v_1[x := u], \dots, v_n[x := u] \xmapsto{f} w\big).$$  
		\item  An assignment $(v_1, \dots, v_k \ \mid \ v'_1, \dots, v'_m) \xmapsto{g} w_1, \dots, w_n$ by $u$ as:
    \[
    \begin{aligned}
    &\big((v_1, \dots, v_k \ \mid \ v'_1, \dots, v'_m) \xmapsto{g} w_1, \dots, w_n\big)[x := u] \\
    &\quad := \big((v_1[x := u], \dots, v_k[x := u] \ \mid \ v'_1[x := u], \dots, v'_m[x := u]) \xmapsto{g} w_1, \dots, w_n\big).
    \end{aligned}
    \]
		\item  A redirection $s \mapsto y$ by $u$ as:  
	$$\big(s \mapsto y \big)[x := u] := \big(s[x := u] \mapsto y \big).$$
		\item  A set of variables or assignments $\mc G$ by $u$ as:
		$$\mc G[x := u] := \{S[x := u] \mid S \in \mc G\}.$$
		\item A redirection set $\mc T$ by $u$ as:
		$$\mc T[x := u] := \{R[x := u] \mid R \in \mc T\}.$$
		\item A raw descriptor $(\mc S \mid \mc T)$ by $u$ as:
		$$(\mc S \mid \mc T)[x := u] := (\mc S[x := u] \mid \mc T[x := u]).$$
		\item A one-step reachability relation $\leadsto_D$ by $u$ as:   
$$(\leadsto_{D})[x := u] := \{(a[x := u], b[x := u]) \mid (a, b) \in \leadsto_{D}\}.$$
  \end{itemize}
\end{definition}

\textit{Remark:} Substitution does not change the ambient boundary. Let $(\mc S \mid \mc T)$ be a raw descriptor for input variables in $\gamma$ and output variables in $\delta$. If $u$ is an internal variable, or an input variable declared in $\gamma$, then $(\mc S \mid \mc T)[x := u]$ is again a raw descriptor over the same boundary. Indeed, the substitution replaces $x$ in source positions by $u$, which is again an internal variable or an input variable in $\gamma$, and every target is untouched.

\begin{example}[Linear--Nonlinear Logic]\label{ex:linear-logic-input-substitution}
We consider again the raw descriptor $(\{x, x' \xmapsto{\otimes_l} w\} \mid \{w \mapsto y\})$ of a multiplicative conjunction, over the boundary $(x : l,\, x' : l \ ; \ y : l)$. Substituting the input $x$ by an internal variable $w'$ gives
$$(\{x, x' \xmapsto{\otimes_l} w\} \mid \{w \mapsto y\})[x := w'] = (\{w', x' \xmapsto{\otimes_l} w\} \mid \{w \mapsto y\}).$$
The substitution replaces $x$ in the source position it occupies. The source $x'$, the target $w$, and the redirection $w \mapsto y$ remain untouched. The resulting descriptor is over the same boundary, and we note that the input $x$ is no longer present in it.
\end{example}

We now need an extended notion of bijection between variables. Since our variables come in three different types (input, output, and internal), we require our bijections to behave correctly with respect to them.
\begin{definition}[Typed Variable Bijection]\label{def:typed-variable-bijection}
  Let $\psi : U \to U'$ be a \emph{type-preserving} bijection between two finite sets of variables. This means that it carries input variables to input variables, output variables to output variables, and internal variables to internal variables. We call such a $\psi$ a \emph{typed variable bijection}. We extend $\psi$ to a total map $\overline{\psi}$ on all variables by fixing every variable outside $U$:
  $$\overline{\psi}(s)=\begin{cases}\psi(s)&\text{if }s\in U,\\ s&\text{if }s\notin U.\end{cases}$$
  The \emph{relabelling} operation $(\--)^\psi$ is defined using this total extension. We define the relabelling of
  \begin{itemize}
    \item A variable $s$ by $\psi$ as:
    $$s^\psi:=\overline{\psi}(s).$$
    \item A finite set of variables $M$ by $\psi$ as:
    $$M^\psi:=\{s^\psi\mid s\in M\}.$$
    \item An assignment $v_1, \dots, v_n \xmapsto{f} w$ by $\psi$ as:
    $$\big(v_1, \dots, v_n \xmapsto{f} w\big)^\psi := \big(v_1^\psi, \dots, v_n^\psi \xmapsto{f} w^\psi\big).$$
    \item An assignment $(v_1, \dots, v_k \ \mid \ v'_1, \dots, v'_m) \xmapsto{g} w_1, \dots, w_n$ by $\psi$ as:
    $$\big((v_1, \dots, v_k \ \mid \ v'_1, \dots, v'_m) \xmapsto{g} w_1, \dots, w_n\big)^\psi := \big((v_1^\psi, \dots, v_k^\psi \ \mid \ {v'_1}^\psi, \dots, {v'_m}^\psi) \xmapsto{g} w_1^\psi, \dots, w_n^\psi\big).$$
    \item A redirection $s \mapsto y$ by $\psi$ as:
    $$\big(s \mapsto y \big)^\psi := \big(s^\psi \mapsto y^\psi \big).$$
    \item A set of variables or assignments $\mc G$ by $\psi$ as:
    $$\mc G^\psi := \{G^\psi \mid G \in \mc G\}.$$
    \item A redirection set $\mc T$ by $\psi$ as:
    $$\mc T^\psi := \{R^\psi \mid R \in \mc T\},$$
    where $R$ ranges over redirections.
    \item A raw descriptor $(\mc S \mid \mc T)$ by $\psi$ as:
    $$(\mc S \mid \mc T)^\psi := (\mc S^\psi \mid \mc T^\psi).$$
    \item A one-step reachability relation $\leadsto_D$ by $\psi$ as:
    $$(\leadsto_D)^\psi := \{(a^\psi, b^\psi) \mid (a, b) \in \leadsto_D\}.$$
  \end{itemize}
\end{definition}

The relabelling operation preserves every operation symbol $f,g$ and every variable kind. Therefore, when the extension $\overline{\psi}$ is injective on the variables declared in the contexts, $(\mc S \mid \mc T)^\psi$ is again a raw descriptor. It lives over the relabelled contexts $\gamma^{\psi} := \{x^\psi : p \mid x : p \in \gamma\}$ and  $\delta^{\psi} := \{y^\psi : q \mid y : q \in \delta\}$. 

\begin{example}[Linear--Nonlinear Logic]\label{ex:linear-logic-typed-relabelling}
We relabel the descriptor of the \hyperref[ex:linear-logic-input-substitution]{input-substitution example}. Let $\psi : \{x, w, y\} \to \{x_1, w_1, y_1\}$ be the bijection defined by $\psi(x) := x_1$, $\psi(w) := w_1$, and $\psi(y) := y_1$. It carries the input $x$ to the input $x_1$, the internal variable $w$ to the internal variable $w_1$, and the output $y$ to the output $y_1$. It is therefore a typed variable bijection. The relabelling gives
$$(\{x, x' \xmapsto{\otimes_l} w\} \mid \{w \mapsto y\})^\psi = (\{x_1, x' \xmapsto{\otimes_l} w_1\} \mid \{w_1 \mapsto y_1\}).$$
The input $x'$ is not present in the domain of $\psi$; it is therefore fixed by the extension $\overline{\psi}$. The extension is injective on the declared variables, and the result is a raw descriptor over the relabelled contexts $x_1 : l,\, x' : l$ and $y_1 : l$.
\end{example}

\begin{definition}[Internal Renaming]\label{def:internal-renaming}
  An \emph{internal renaming} is a typed variable bijection that fixes every input and output variable.
\end{definition}

\textit{Remark:} Concretely, an internal renaming is determined by a bijection $\phi : V \to W$ between two sets of internal variables, extended to the total map $\overline{\phi}$ fixing every variable outside $V$:

  $$\overline{\phi}(s)=\begin{cases}\phi(s)&\text{if }s\in V,\\ s&\text{if }s\notin V.\end{cases}$$

\begin{example}[Linear--Nonlinear Logic]\label{ex:linear-logic-internal-renaming}
The bijection $\phi : \{w\} \to \{w_1\}$ between internal variables determines an internal renaming. The extension $\overline{\phi}$ renames $w$ to $w_1$ and fixes every other variable. Applying the bijection to the descriptor of the multiplicative conjunction gives
$$(\{x, x' \xmapsto{\otimes_l} w\} \mid \{w \mapsto y\})^\phi = (\{x, x' \xmapsto{\otimes_l} w_1\} \mid \{w_1 \mapsto y\}).$$
By contrast with the \hyperref[ex:linear-logic-typed-relabelling]{typed-relabelling example}, the boundary $(x : l,\, x' : l \ ; \ y : l)$ is unchanged. The two descriptors differ only by the name given to their internal variable. This is precisely the situation captured by $\alpha$-equivalence.
\end{example}

\begin{lemma}[Internal Renaming Preserves Redirections Targeting an Output]\label{lem:internal-renaming-preserves-redirections-targeting-an-output}
Let $(\mc S\mid\mc T)$ be a raw descriptor, let $y$ be an output variable, and let $\phi:V\to W$ be a bijection of internal variables whose extension is injective on the variables occurring in $\mc T$. Then
$$(\mc T_{\mapsto y})^\phi=(\mc T^\phi)_{\mapsto y}.$$
Moreover, taking source sets gives
$
(\mc T_{\mapsto y}^{\mathrm{src}})^\phi
=
((\mc T^\phi)_{\mapsto y})^{\mathrm{src}}.
$
Thus, $\phi$ induces an elementwise bijection of source sets and restricts to a bijection between their internal parts:
$$
(\--)^\phi:\mc T_{\mapsto y}^{\mathrm{src}}
\to
((\mc T^\phi)_{\mapsto y})^{\mathrm{src}},
\quad \text{and} \quad
(\--)^\phi:\mathrm{Int}(\mc T_{\mapsto y}^{\mathrm{src}})
\to
\mathrm{Int}(((\mc T^\phi)_{\mapsto y})^{\mathrm{src}}).
$$
\end{lemma}

\textit{Proof.}
For every redirection $R=(s\mapsto t)$ of $\mc T$, since $\phi$ fixes output variables, we have
$$
R^\phi=(s^\phi\mapsto t).
$$
Thus a redirection $R\in\mc T$ contributes to $(\mc T_{\mapsto y})^\phi$ if and only if $t=y$. Its internal renaming $R^\phi$ contributes to $(\mc T^\phi)_{\mapsto y}$ if and only if the target of $R^\phi$ is $y$, which again happens if and only if $t=y$. Both redirection sets are therefore represented by the same set builder
$$
\{R^\phi\mid R=(s\mapsto y)\in\mc T\}.
$$
This gives
$$
(\mc T_{\mapsto y})^\phi=(\mc T^\phi)_{\mapsto y}.
$$

Taking source sets gives
$$
(\mc T_{\mapsto y}^{\mathrm{src}})^\phi
=
\{s^\phi\mid (s\mapsto y)\in\mc T\}
=
((\mc T^\phi)_{\mapsto y})^{\mathrm{src}}.
$$
Each redirection $R=(s\mapsto y)$ contributes the source $s$ on the left and the source $s^\phi$ on the right. The map $s \mapsto s^\phi$ is surjective by construction and injective because the extension of $\phi$ is injective on the variables occurring in $\mc T$. Hence $\phi$ induces a bijection of sources
$$
(\--)^\phi:\mc T_{\mapsto y}^{\mathrm{src}}
\to
((\mc T^\phi)_{\mapsto y})^{\mathrm{src}}.
$$
Now $\phi$ maps internal variables to internal variables and fixes input variables. The bijection therefore restricts to the internal sources:
$$
(\--)^\phi:\mathrm{Int}(\mc T_{\mapsto y}^{\mathrm{src}})
\to
\mathrm{Int}(((\mc T^\phi)_{\mapsto y})^{\mathrm{src}}).
$$
\hfill $\square$

\begin{lemma}[Elementwise Operations Preservation]\label{lem:elementwise-operations-preservation}
Let $F$ be an operation defined on variables, assignments, redirections, or pairs of variables and extended elementwise to sets. Then for a finite family $(\mc G_i)_{i \in I}$ of sets of the same kind,
  $$F\Big(\bigcup_{i \in I}\mc G_i\Big)=\bigcup_{i \in I}F(\mc G_i),$$
  and if $\mc G\subseteq \mc H$, then
  $$F(\mc G)\subseteq F(\mc H).$$
\end{lemma}
The upshot of \cref{lem:elementwise-operations-preservation} is that it applies to internal renamings $(-)^\phi$ (\cref{def:internal-renaming}) and input substitutions $(-)[x:=u]$ (\cref{def:input-substitution}).

\textit{Proof.}
For ordinary sets, including the one-step reachability relations, which are sets of pairs of variables, these are the usual elementwise image identities.

For redirection sets, consider a finite family $(\mc T_i)_{i \in I}$. Applying $F$ elementwise gives
$$\bigcup_{i \in I} F(\mc T_i) = \bigcup_{i \in I} \{ F(R) \mid R \in \mc T_i \} = \Big\{ F(R) \ \Big| \ R \in \bigcup_{i \in I} \mc T_i \Big\} = F\Big(\bigcup_{i \in I} \mc T_i\Big).$$
If $\mc T_0$ is a subset of $\mc T$, then applying $F$ elementwise sends it into $F(\mc T)$. Hence $F(\mc T_0)$ is a subset of $F(\mc T)$.
\hfill $\square$

\textit{Remark:} As corollaries, for ordinary sets we have
$$(\mc G\cup \mc H)^\phi=\mc G^\phi \cup \mc H^\phi \quad \text{and} \quad (\mc G\cup \mc H)[x := u]=\mc G[x := u] \cup \mc H[x := u],$$
which also hold for redirection sets:
$$(\mc T\cup \mc U)^\phi=\mc T^\phi \cup \mc U^\phi \quad \text{and} \quad (\mc T\cup \mc U)[x := u]=\mc T[x := u] \cup \mc U[x := u].$$

\begin{lemma}[One-Step Reachability Is Natural]\label{lem:one-step-reachability-is-natural}
Let $F$ be either an input substitution $(-)[x:=u]$ or an internal renaming $(-)^\phi$. We extend $F$ to assignments, redirections, ordinary sets, redirection sets, descriptors, and one-step reachability relations by acting on every variable. Then, for every raw descriptor $(\mc S\mid\mc T)$,
$$
F\big(\leadsto_{(\mc S\mid\mc T)}\big)
=
\leadsto_{F(\mc S\mid\mc T)}.
$$
Equivalently,
$$
(\leadsto_{(\mc S\mid\mc T)})[x:=u]
=
\leadsto_{(\mc S\mid\mc T)[x:=u]}
\quad \text{and} \quad
(\leadsto_{(\mc S\mid\mc T)})^\phi
=
\leadsto_{(\mc S\mid\mc T)^\phi}.
$$
\end{lemma}

\textit{Proof.}
We prove the statement by cases on the generators of the one-step reachability relation. The cases for input substitution and internal renaming are identical in shape. We therefore write $F$ for either operation.

\begin{itemize}
	\item First, consider an assignment $E := (v_1,\ldots,v_n\xmapsto{f}w)$. Then
$
\leadsto_E=\{(v_i,w)\mid 1\leq i\leq n\}.
$
Applying $F$ to this relation gives
$
F(\leadsto_E)=\{(F(v_i),F(w))\mid 1\leq i\leq n\}.
$
On the other hand, the target $w$ is internal, so $F(w)=w$ when $F$ is a substitution, and in both cases
$
F(E)=(F(v_1),\ldots,F(v_n)\xmapsto{f}F(w)),
$
so
$
\leadsto_{F(E)}=\{(F(v_i),F(w))\mid 1\leq i\leq n\}.
$
Thus $F(\leadsto_E)=\leadsto_{F(E)}$.

\item Now consider an assignment $E := ((v_1,\ldots,v_k \ \mid \ v'_1,\ldots,v'_m)\xmapsto{g}w_1,\ldots,w_n)$. Then
$
\leadsto_E=\{(v_i,w_j)\mid 1\leq i\leq k,\ 1\leq j\leq n\}\cup\{(v'_i,w_j)\mid 1\leq i\leq m,\ 1\leq j\leq n\}.
$
Applying $F$ gives
$
F(\leadsto_E)=\{(F(v_i),F(w_j))\}_{i,j}\cup\{(F(v'_i),F(w_j))\}_{i,j}.
$
Since the targets $w_j$ are internal, the same reasoning gives
$
F(E)=((F(v_1),\ldots,F(v_k) \ \mid \ F(v'_1),\ldots,F(v'_m))\xmapsto{g}F(w_1),\ldots,F(w_n)),
$
so we also have
$
\leadsto_{F(E)}=\{(F(v_i),F(w_j))\}_{i,j}\cup\{(F(v'_i),F(w_j))\}_{i,j}.
$
Hence $F(\leadsto_E)=\leadsto_{F(E)}$.

\item Finally, consider a redirection $R := (s\mapsto y)$. Then
$
\leadsto_R=\{(s,y)\}.
$
Applying $F$ gives
$
F(\leadsto_R)=\{(F(s),F(y))\}.
$
But the target $y$ is an output variable, so $F(y)=y$ when $F$ is a substitution, and in both cases $F(R)=(F(s)\mapsto F(y))$. Therefore
$
\leadsto_{F(R)}=\{(F(s),F(y))\}.
$
Thus $F(\leadsto_R)=\leadsto_{F(R)}$.

\item Now take ordinary unions of reachability relations. By definition,
\[
\leadsto_{(\mc S\mid\mc T)}
=
\bigcup_{E\in\mc S}\leadsto_E
\cup
\bigcup_{R\in\mc T}\leadsto_R.
\]
Here $R\in\mc T$ ranges over redirections, while the outer union is an ordinary union of binary relations.
Using \cref{lem:elementwise-operations-preservation} together with the three established cases, we obtain
\begin{align*}
F\big(\leadsto_{(\mc S\mid\mc T)}\big)
&=
\bigcup_{E\in\mc S}F(\leadsto_E)
\cup
\bigcup_{R\in\mc T}F(\leadsto_R)\\
&=
\bigcup_{E\in\mc S}\leadsto_{F(E)}
\cup
\bigcup_{R\in\mc T}\leadsto_{F(R)}\\
&=
\leadsto_{F(\mc S\mid\mc T)}.
\end{align*}
\end{itemize}
\hfill $\square$

Recall that we consider two descriptors to be essentially the same if all their internal variables can be renamed to each other. This is formalized as $\alpha$-equivalence in \cref{def:internal-renaming-equivalence-alpha-equivalence}; first, we record two lemmas showing that internal renaming interacts well with input substitution and with reachability.

\begin{lemma}[Input Substitution and Internal Renaming Commute]\label{lem:input-substitution-and-internal-renaming-commute}
  Given a bijection $\phi : V \to W$ where $V,W$ are sets of internal variables, an input variable $x$, and an input or internal variable $w$, internal renaming and input substitution commute:
  $$((\--)[x := w])^{\phi} = (\--)^{\phi}[x := w^\phi].$$   
\end{lemma}

\textit{Proof.}
We prove the result by cases on the syntactic clauses.
\begin{itemize}
  \item First, let $r$ be a variable. We would like to show that
  $$r^{\phi}[x := w^\phi] = (r[x := w])^{\phi}.$$   
 	We proceed by cases on $r$:
  \begin{itemize}
  	\item If $r$ is an internal variable, then $r \neq x$. Its image $r^\phi$ is again internal, hence $r^\phi \neq x$, and we have $r^{\phi}[x := w^\phi]= r^\phi$. Commuting the operations, we evaluate $r[x := w]^\phi = r^\phi$, which shows equality.
		\item If $r$ is not an internal variable, then it is an input or output variable. Input and output variables are unaffected by internal renaming; thus $r^\phi = r$ and $r^\phi[x := w^\phi] = r[x := w^\phi]$. If $r = x$, then $x[x := w^\phi] = w^\phi$, and commuting the operations, we have $x[x := w]^{\phi} = w^\phi$.

		In the other case where $r \neq x$, we have $r^\phi[x := w^\phi] = r[x := w^\phi] = r$, and commuting gives $r[x := w]^{\phi} = r^\phi = r$.          
    \end{itemize}
    In all cases, for every variable $r$, we have $(r^\phi)[x := w^\phi] = (r[x := w])^\phi$.
  \item Now consider an assignment $v_1, \dots, v_n \xmapsto{f} t$. Applying internal renaming followed by substitution gives:
    \begin{align*}
    \big((v_1, \dots, v_n \xmapsto{f} t)^\phi\big)[x := w^\phi] &= \big(v_1^\phi, \dots, v_n^\phi \xmapsto{f} t^\phi\big)[x := w^\phi] \\
    &= (v_1^\phi)[x := w^\phi], \dots, (v_n^\phi)[x := w^\phi] \xmapsto{f} t^\phi
    \end{align*}
    By the variable case, we can commute the two operations:
    \begin{align*}
    &= (v_1[x := w])^\phi, \dots, (v_n[x := w])^\phi \xmapsto{f} t^\phi \\
    &= \big(v_1[x := w], \dots, v_n[x := w] \xmapsto{f} t\big)^\phi \\
    &= \big((v_1, \dots, v_n \xmapsto{f} t)[x := w]\big)^\phi.
    \end{align*}
  \item The same reasoning applies to assignments of the form $(v_1, \dots, v_k \ \mid \ v'_1, \dots, v'_m) \xmapsto{g} t_1, \dots, t_n$:
    \begin{align*}
    &\big(((v_1, \dots, v_k \ \mid \ v'_1, \dots, v'_m) \xmapsto{g} t_1, \dots, t_n)^\phi\big)[x := w^\phi] \\
    &\quad= \big((v_1^\phi, \dots, v_k^\phi \ \mid \ {v'_1}^\phi, \dots, {v'_m}^\phi) \xmapsto{g} t_1^\phi, \dots, t_n^\phi\big)[x := w^\phi] \\
    &\quad= \big(((v_1^\phi)[x := w^\phi], \dots \ \mid \ \dots, ({v'_m}^\phi)[x := w^\phi]) \xmapsto{g} t_1^\phi, \dots, t_n^\phi\big) \\
    &\quad= \big(((v_1[x := w])^\phi, \dots \ \mid \ \dots, (v'_m[x := w])^\phi) \xmapsto{g} t_1^\phi, \dots, t_n^\phi\big) \\
    &\quad= \big(((v_1, \dots, v_k \ \mid \ v'_1, \dots, v'_m) \xmapsto{g} t_1, \dots, t_n)[x := w]\big)^\phi.
    \end{align*}
    \item For a redirection $s \mapsto y$, since $\phi$ fixes the output variable $y$, we have
      \begin{align*}
      \big((s \mapsto y)^\phi\big)[x := w^\phi]
      &= (s^\phi \mapsto y)[x := w^\phi] \\
      &= ((s^\phi)[x := w^\phi] \mapsto y) \\
      &= ((s[x := w])^\phi \mapsto y) \\
      &= (s[x := w] \mapsto y)^\phi \\
      &= \big((s \mapsto y)[x := w]\big)^\phi.
      \end{align*}
  \item For the assignment set $\mc S$:
    \begin{align*}
    \mc S^\phi[x := w^\phi] &= \{ E[x := w^\phi] \mid E \in \mc S^\phi \} \\
    &= \{ (E^\phi)[x := w^\phi] \mid E \in \mc S \} \\
    &= \{ (E[x := w])^\phi \mid E \in \mc S \} \\
    &= \{ E'^\phi \mid E' \in \mc S[x := w] \} \\
    &= (\mc S[x := w])^\phi.
    \end{align*}
  \item For the redirection set $\mc T$, the reasoning is elementwise:
    \begin{align*}
    \mc T^\phi[x := w^\phi] &= \{ R[x := w^\phi] \mid R \in \mc T^\phi \} \\
    &= \{ (R^\phi)[x := w^\phi] \mid R \in \mc T \} \\
    &= \{ (R[x := w])^\phi \mid R \in \mc T \} \\
    &= \{ R'^\phi \mid R' \in \mc T[x := w] \} \\
    &= (\mc T[x := w])^\phi.
    \end{align*}
    \item Therefore, for a descriptor $(\mc S \mid \mc T)$,
      \begin{align*}
      \big((\mc S \mid \mc T)^\phi\big)[x := w^\phi]
      &= (\mc S^\phi \mid \mc T^\phi)[x := w^\phi] \\
      &= (\mc S^\phi[x := w^\phi] \mid \mc T^\phi[x := w^\phi]) \\
      &= ((\mc S[x := w])^\phi \mid (\mc T[x := w])^\phi) \\
      &= (\mc S[x := w] \mid \mc T[x := w])^\phi \\
      &= \big((\mc S \mid \mc T)[x := w]\big)^\phi.
      \end{align*}
\end{itemize}
This proves that input substitution and internal renaming commute as operations. \hfill $\square$

\begin{lemma}[Internal Renaming Preserves Strict Reachability]\label{lem:internal-renaming-preserves-strict-reachability}
Let $(\mc S \mid \mc T)$ be a raw descriptor, and let $\phi : V \to W$ be a bijection of internal variables. Assume that the extension of $\phi$ is injective on the variables occurring in $(\mc S \mid \mc T)$, i.e.,
$$a^\phi = b^\phi \quad \Longrightarrow \quad a=b$$
for all variables $a,b \in \mathrm{Var}(\mc S \mid \mc T)$. Then, for all variables $a,b$ either of input or output type, or occurring in $(\mc S \mid \mc T)$, we have   
$$a \rightsquigarrow_{(\mc S \mid \mc T)} b \quad \Longleftrightarrow \quad a^\phi \rightsquigarrow_{(\mc S \mid \mc T)^\phi} b^\phi.$$
\end{lemma}
\textit{Proof.}
If either $a$ or $b$ is an input or output variable not occurring in $(\mc S\mid\mc T)$, then it is fixed by $\phi$, and it does not occur in $(\mc S\mid\mc T)^\phi$ either, since $\phi$ only renames internal variables. Every one-step edge relates variables occurring in the ambient descriptor, so no path in either descriptor can have that variable as a source or target. Both sides are false in this case. We may therefore assume that $a,b\in\mathrm{Var}(\mc S\mid\mc T)$.

Suppose first that $a \rightsquigarrow_{(\mc S \mid \mc T)} b$. Then there is a path
$$a=v_0 \leadsto_{(\mc S \mid \mc T)} v_1 \leadsto_{(\mc S \mid \mc T)} \cdots \leadsto_{(\mc S \mid \mc T)} v_k=b$$
with $k\geq 1$. Applying $\phi$ to every variable gives
$$a^\phi=v_0^\phi \leadsto_{(\mc S \mid \mc T)^\phi} v_1^\phi \leadsto_{(\mc S \mid \mc T)^\phi} \cdots \leadsto_{(\mc S \mid \mc T)^\phi} v_k^\phi=b^\phi,$$
where each one-step edge is valid by \cref{lem:one-step-reachability-is-natural}. Hence $a^\phi \rightsquigarrow_{(\mc S \mid \mc T)^\phi} b^\phi$.

Conversely, suppose $a^\phi \rightsquigarrow_{(\mc S \mid \mc T)^\phi} b^\phi$. Then there is a path
$$a^\phi=r_0 \leadsto_{(\mc S \mid \mc T)^\phi} r_1 \leadsto_{(\mc S \mid \mc T)^\phi} \cdots \leadsto_{(\mc S \mid \mc T)^\phi} r_k=b^\phi$$
with $k\geq 1$. By \cref{lem:one-step-reachability-is-natural},
$$\leadsto_{(\mc S \mid \mc T)^\phi}=(\leadsto_{(\mc S \mid \mc T)})^\phi.$$
Hence for each edge $r_i\leadsto_{(\mc S \mid \mc T)^\phi}r_{i+1}$, there exists an edge $u_i\leadsto_{(\mc S \mid \mc T)}v_i$ such that $u_i^\phi=r_i$ and $v_i^\phi=r_{i+1}$. Since $u_0^\phi=r_0=a^\phi$ and both $u_0,a$ occur in $(\mc S\mid\mc T)$, injectivity gives $u_0=a$. Similarly, since $v_{k-1}^\phi=r_k=b^\phi$, injectivity gives $v_{k-1}=b$. For each $i<k-1$, we have
$$v_i^\phi=r_{i+1}=u_{i+1}^\phi,$$
and since $v_i,u_{i+1}$ occur in $(\mc S\mid\mc T)$, injectivity gives $v_i=u_{i+1}$. Therefore the lifted edges glue into a path
$$a=u_0 \leadsto_{(\mc S \mid \mc T)} v_0=u_1 \leadsto_{(\mc S \mid \mc T)} \cdots \leadsto_{(\mc S \mid \mc T)} v_{k-1}=b.$$
Hence $a \rightsquigarrow_{(\mc S \mid \mc T)} b$. \hfill $\square$
\endgroup{}
\begingroup

\subsubsection{Descriptor Classes}

We are now ready to formalize a notion of equivalence for descriptors that differ only by name. Indeed, we consider two raw descriptors sharing the same boundary to be essentially the same if all their internal variables can be renamed to each other. We therefore define an equivalence relation that will identify such raw descriptors.

\begin{definition}[Internal Renaming Equivalence ($\alpha$-equivalence)]\label{def:internal-renaming-equivalence-alpha-equivalence}
		Fix some input context $\gamma$ and output context $\delta$.\\
		We write $=_\alpha$ for the relation on the collection of raw descriptors $\mathrm{desc}^{\mathrm{raw}}_\Sigma(\gamma \ ; \ \delta)$ defined as
	\begin{align*}
  	(\mc S \mid \mc T) =_\alpha (\mc S' \mid \mc T') \quad &\text{if and only if} \\
  	&\text{there exists a bijection } \phi : \mathrm{Int}(\mc S \mid \mc T)   \to \mathrm{Int}(\mc S' \mid \mc T')  \quad \text{such that } \\
	  	& (\mc S \mid \mc T)^\phi = (\mc S' \mid \mc T')
	  	\end{align*}
\end{definition}

\begin{example}[Linear--Nonlinear Logic]\label{ex:linear-logic-alpha-equivalent-descriptors}
We consider again, over the boundary $(x : l,\, x' : l \ ; \ y : l)$, the two descriptors of a multiplicative conjunction from the \hyperref[ex:linear-logic-internal-renaming]{internal-renaming example},
$$(\{x, x' \xmapsto{\otimes_l} w\} \mid \{w \mapsto y\}) \quad \text{and} \quad (\{x, x' \xmapsto{\otimes_l} w_1\} \mid \{w_1 \mapsto y\}).$$
The bijection $\phi : \{w\} \to \{w_1\}$ is a bijection between their sets of internal variables, and renaming by $\phi$ sends the first descriptor to the second. The two descriptors are therefore $\alpha$-equivalent. Renaming $w$ to any other internal variable produces an $\alpha$-equivalent descriptor in the same way.
\end{example}

\begin{lemma}[Internal Renamings are Injective on Variables]\label{lem:internal-renamings-are-injective-on-variables}
Let $(\mc S\mid\mc T)=_\alpha(\mc S'\mid\mc T')$ be witnessed by a bijection
$$
\phi:\mathrm{Int}(\mc S\mid\mc T)\to \mathrm{Int}(\mc S'\mid\mc T')
$$
with $(\mc S\mid\mc T)^\phi=(\mc S'\mid\mc T')$. Then the extension of $\phi$ is injective on the variables occurring in $(\mc S\mid\mc T)$:
$$
a^\phi=b^\phi \quad\Longrightarrow\quad a=b
$$
for all $a,b\in\mathrm{Var}(\mc S\mid\mc T)$. The same statement holds for $\phi^{-1}$ on the variables occurring in $(\mc S'\mid\mc T')$.
\end{lemma}

\textit{Proof.}
Let $a,b\in\mathrm{Var}(\mc S\mid\mc T)$ and suppose $a^\phi=b^\phi$. If $a$ and $b$ are internal variables, then $a,b\in\mathrm{Int}(\mc S\mid\mc T)$. The injectivity of $\phi$ then gives $a=b$.

If exactly one of $a,b$ is internal, then exactly one of $a^\phi,b^\phi$ is internal, since $\phi$ sends internal variables to internal variables and fixes input and output variables. This is impossible, since input, output, and internal variables are pairwise disjoint.

Finally, if neither $a$ nor $b$ is internal, then both are fixed by $\phi$. Thus $a^\phi=b^\phi$ gives $a=b$. The extension of $\phi$ is therefore injective on variables occurring in $(\mc S\mid\mc T)$. The argument for $\phi^{-1}$ is identical, since $\phi^{-1}$ is again a bijection between internal variables and fixes input and output variables. \hfill $\square$

As expected, the operation of renaming the internal variables does not alter their modes. Therefore the operation is well behaved with respect to mode validity.

\begin{lemma}[Internal Renaming Preserves Mode Validity]\label{lem:internal-renaming-preserves-mode-validity}
Let $(\mc S\mid\mc T),(\mc S'\mid\mc T')\in
\mathrm{desc}^{\mathrm{raw}}_\Sigma(\gamma;\delta)$ be raw descriptors such that
$(\mc S\mid\mc T)=_\alpha(\mc S'\mid\mc T').$
Then $(\mc S\mid\mc T)$ is mode-valid if and only if $(\mc S'\mid\mc T')$ is mode-valid.
\end{lemma}

\textit{Proof.}
Suppose first that $(\mc S\mid\mc T)$ is mode-valid, witnessed by a mode assignment
$ \mu:\mathrm{Int}(\mc S\mid\mc T)\to \mathrm{Mode}(\Sigma). $ Let $ \phi:\mathrm{Int}(\mc S\mid\mc T)\to \mathrm{Int}(\mc S'\mid\mc T') $ witness the $\alpha$-equivalence, so that $ (\mc S\mid\mc T)^\phi=(\mc S'\mid\mc T') = (\mc S^\phi \mid \mc T^\phi)$. \\
Define $ \mu':\mathrm{Int}(\mc S^\phi\mid\mc T^\phi)\to \mathrm{Mode}(\Sigma) $ by $ \mu' :=\mu \circ \phi^{-1}. $ We now verify that $\mu'$ witnesses mode-validity of $(\mc S^\phi\mid\mc T^\phi)$.

Let
$ v'_1,\ldots,v'_n\xmapsto{f}w'\in \mc S^\phi $ with $f:p_1,\ldots,p_n\to q$. This assignment has the form $ (v_1,\ldots,v_n\xmapsto{f}w)^\phi =
v_1^\phi,\ldots,v_n^\phi\xmapsto{f}w^\phi
$ for some $v_1,\ldots,v_n\xmapsto{f}w\in\mc S$. Thus $$w'=w^\phi \quad \text{and} \quad v'_i=v_i^\phi.$$
Since $\mu$ is mode-valid, $\mathrm{mode}_\mu(w)=q$. Therefore $\mathrm{mode}_{\mu'}(w')=\mathrm{mode}_{\mu'}(w^\phi)=\mathrm{mode}_{\mu}(w)=q$. For each $i$, if $v_i:p_i\in\gamma$, then $v_i^\phi=v_i$, so $\mathrm{mode}_{\mu'}(v'_i)=p_i$. If $v_i$ is internal and $\mu(v_i)=p_i$, then $\mathrm{mode}_{\mu'}(v'_i)=\mu'(v_i^\phi)=\mu(v_i)=p_i$. This assignment is then mode-valid for $\mu'$. \\
The case of an $\ell$-generator assignment
$ (v'_1,\ldots,v'_k \ \mid \ v''_1,\ldots,v''_m)\xmapsto{g}w'_1,\ldots,w'_n\in\mc S^\phi $ with $g:(p_1,\ldots,p_k \ ; \ l_1,\ldots,l_m)\to l'_1,\ldots,l'_n$ is identical. We check the cartesian source positions against the $p_i$, the linear source positions against the $l_i$, and the targets against the $l'_j$.\\  
Finally, let
$$
s'\mapsto y\in\mc T^\phi
$$
with $y:q\in\delta$. This redirection is the internal renaming of some $s\mapsto y\in\mc T$. If $s:q\in\gamma$, then $s'=s$, since $\phi$ fixes input variables. If $s$ is internal, then $\mu(s)=q$, and hence
$$ \mathrm{mode}_{\mu'}(s')=\mu'(s^\phi)=\mu(s)=q. $$
Thus the redirection is mode-valid for $\mu'$.

So $(\mc S^\phi\mid\mc T^\phi)$ is mode-valid. For the converse, given a mode assignment $\mu'$ validating $(\mc S^\phi\mid\mc T^\phi)$, the assignment $\mu:=\mu'\circ\phi$ works for $(\mc S\mid\mc T)$. Each clause is checked against the corresponding image element $E^\phi$. Mode-validity is therefore invariant under internal renaming.
\hfill $\square$

We study the effect on the internal variables of the components of a descriptor (assignment and redirection sets) when the operations of internal renaming and input substitution are applied. In the case of internal variable renaming, the renamed internal variables are the internal variables of the renamings. For input substitution, we need to be careful about the nature of the substitution: whether the substituted element is an internal variable (then it will contribute to the final internal filter), or whether it is another input variable (in which case it will not contribute to the final internal filter).

\begin{lemma}[Internal Variables under Internal Renaming and Input Substitution]\label{lem:internal-variables-under-internal-renaming-and-input-substitution}
Let $\mc G$ be either a finite set of assignments or a finite set of redirections.
\begin{itemize}
  \item If $\phi : V \to W$ is a bijection of internal variables, then
  $$\mathrm{Int}(\mc G^\phi)=\mathrm{Int}(\mc G)^\phi.$$
  \item If $x$ is an input variable and $u$ is an input or internal variable, then
  $$\mathrm{Int}(\mc G[x:=u])=\mathrm{Int}(\mc G)\cup \mathrm{Int}(\mathrm{Inp}(\mc G)[x:=u]).$$
\end{itemize}
\end{lemma}

\textit{Remark.} Equivalently, the second equality says
$$
\mathrm{Int}(\mc G[x:=u])=
\begin{cases}
\mathrm{Int}(\mc G)\cup\{u\} & \text{if }u\text{ is internal and }x\in\mathrm{Inp}(\mc G),\\
\mathrm{Int}(\mc G) & \text{otherwise.}
\end{cases}
$$

\textit{Proof.}
For the internal renaming statement, a variable appears in $\mc G^\phi$ exactly when it has the form $v^\phi$ for some variable $v$ appearing in $\mc G$. The renaming $\phi$ sends internal variables to internal variables and fixes input and output variables. The image $v^\phi$ is therefore internal if and only if $v$ is internal. The internal variables of $\mc G^\phi$ are then exactly the images under $\phi$ of the internal variables of $\mc G$:
$$\mathrm{Int}(\mc G^\phi)=\mathrm{Int}(\mc G)^\phi.$$

For the substitution statement, let $z\in\mathrm{Int}(\mc G[x:=u])$. Then $z$ appears in some expression $E[x:=u]$ with $E\in\mc G$. If $z$ already appeared in $E$, then $z\in\mathrm{Int}(\mc G)$, since substitution only replaces the input variable $x$. If $z$ did not already appear in $E$, then it was introduced by replacing an occurrence of $x$ by $u$. Thus $x\in\mathrm{Inp}(E)\subseteq\mathrm{Inp}(\mc G)$ and $z\in\mathrm{Int}(u)\subseteq\mathrm{Int}(\mathrm{Inp}(\mc G)[x:=u])$. This proves
$$\mathrm{Int}(\mc G[x:=u])\subseteq \mathrm{Int}(\mc G)\cup \mathrm{Int}(\mathrm{Inp}(\mc G)[x:=u]).$$

Conversely, if $z\in\mathrm{Int}(\mc G)$, then $z$ appears in some $E\in\mc G$. Since $z$ is internal and $x$ is input, $z\neq x$, so $z$ still appears in $E[x:=u]$. Hence $z\in\mathrm{Int}(\mc G[x:=u])$. If $z\in\mathrm{Int}(\mathrm{Inp}(\mc G)[x:=u])$, then $x\in\mathrm{Inp}(\mc G)$ and $u$ is internal with $z=u$. Thus $x$ appears in some $E\in\mc G$, $u$ then appears in $E[x:=u]$, and therefore $z=u\in\mathrm{Int}(\mc G[x:=u])$. Hence
$$\mathrm{Int}(\mc G)\cup \mathrm{Int}(\mathrm{Inp}(\mc G)[x:=u])\subseteq \mathrm{Int}(\mc G[x:=u]).$$
The two inclusions prove the formula. We note that $\mathrm{Int}(\mathrm{Inp}(\mc G)[x:=u])$ is equal to the singleton $\{u\}$ when $u$ is internal and $x\in\mathrm{Inp}(\mc G)$. If this is not the case, the set is empty. \hfill $\square$

The operation of internal renaming behaves well with respect to composition. Renaming by $\phi$ and then by $\psi$ is the same as renaming by the composition $\psi \circ \phi$.   
\begin{proposition}[Internal Renaming Composition]\label{prop:internal-renaming-composition}
Internal renaming respects composition under a freshness condition. If
$$
\phi : V \to W
\qquad\text{and}\qquad
\psi : W \to U
$$
are bijections between sets of internal variables, then so is $\psi\circ\phi:V\to U$. For any assignment set or redirection set $\mc G$ such that
$$
(\mathrm{Int}(\mc G) \setminus V) \cap W = \varnothing,
$$
we have
$$
(\mc G^\phi)^\psi = \mc G^{\psi \circ \phi}.
$$
\end{proposition}

\textit{Proof.}
We prove the result by recursion on the structure of $\mc G$.
\begin{itemize}
  \item  Let $u$ be a variable. It suffices to consider variables $u$ appearing in $\mc G$, since the relabellings $(\--)^\phi$, $(\--)^\psi$, and $(\--)^{\psi\circ\phi}$ act non-trivially only on such variables.
    If $u$ is an input or output variable, then $u \notin V$ and $u \notin W$ since $V, W$ contain only internal variables. Therefore $u^\phi = u$, $u^\psi = u$, and $u^{\psi \circ \phi} = u$, so $(u^\phi)^\psi = u = u^{\psi \circ \phi}$.

    If $u$ is internal and $u \in V$, then $u^\phi = \phi(u) \in W$, so $(u^\phi)^\psi = \psi(\phi(u)) = (\psi \circ \phi)(u) = u^{\psi \circ \phi}$.

    If $u$ is internal and $u \notin V$, then $u^\phi = u$ and $u^{\psi \circ \phi} = u$, since $u$ is not in the domain of $\psi \circ \phi$. To obtain $(u^\phi)^\psi = u^\psi = u$, we need $u \notin W$. Since $u$ appears in $\mc G$, we have $u \in \mathrm{Int}(\mc G) \setminus V$. The freshness hypothesis $(\mathrm{Int}(\mc G) \setminus V) \cap W = \varnothing$ then gives $u \notin W$. Hence $(u^\phi)^\psi = u = u^{\psi \circ \phi}$.

	\item Now consider an assignment $v_1,\dots,v_n\xmapsto{f}s$. Then
    \begin{align*}
    \big((v_1,\dots,v_n\xmapsto{f}s)^\phi\big)^\psi &= \big(v_1^\phi,\dots,v_n^\phi\xmapsto{f}s^\phi\big)^\psi \\
    &= (v_1^\phi)^\psi,\dots,(v_n^\phi)^\psi\xmapsto{f}(s^\phi)^\psi \\
    &= v_1^{\psi\circ\phi},\dots,v_n^{\psi\circ\phi}\xmapsto{f}s^{\psi\circ\phi} \\
    &= (v_1,\dots,v_n\xmapsto{f}s)^{\psi\circ\phi}.
    \end{align*}

    The same computation applies to assignments $(v_1,\dots,v_k \ ; \ v'_1,\dots,v'_m)\xmapsto{g}w_1,\dots,w_n$:
    $$ \big(((v_1,\dots,v_k \ ; \ v'_1,\dots,v'_m)\xmapsto{g}w_1,\dots,w_n)^\phi\big)^\psi = ((v_1,\dots,v_k \ ; \ v'_1,\dots,v'_m)\xmapsto{g}w_1,\dots,w_n)^{\psi\circ\phi}. $$

    \item For a redirection $s\mapsto y$, we have
    \begin{align*}
    \big((s\mapsto y)^\phi\big)^\psi &= (s^\phi\mapsto y)^\psi \\
    &= ((s^\phi)^\psi \mapsto y) \\
    &= (s^{\psi\circ\phi}\mapsto y) \\
    &= (s\mapsto y)^{\psi\circ\phi}.
    \end{align*}

\item Finally, if $\mc G$ is an assignment set, then
    \begin{align*}
    (\mc G^\phi)^\psi &= \{S^\psi \mid S \in \mc G^\phi\} \\
    &= \{(S^\phi)^\psi \mid S \in \mc G\} \\
    &= \{S^{\psi\circ\phi} \mid S \in \mc G\} \\
    &= \mc G^{\psi\circ\phi}.
    \end{align*}

    If $\mc G$ is a redirection set, the same calculation applies elementwise:
    \begin{align*}
    (\mc G^\phi)^\psi &= \{R^\psi \mid R \in \mc G^\phi\} \\
    &= \{(R^\phi)^\psi \mid R \in \mc G\} \\
    &= \{R^{\psi\circ\phi} \mid R \in \mc G\} \\
    &= \mc G^{\psi\circ\phi}.
    \end{align*}
  \end{itemize}
This proves that internal renaming respects composition. \hfill $\square$

We now collect the fruits of our tree. We prove that the relation $=_\alpha$ is indeed an equivalence relation.

\begin{proposition}[Descriptor Equivalence Relation]\label{prop:descriptor-equivalence-relation}
  The relation $=_\alpha$ is an equivalence relation.
\end{proposition}
\textit{Proof.}
We must show that the relation $=_\alpha$ is reflexive, symmetric, and transitive.

\begin{itemize}
\item We first check reflexivity. Let $(\mc S \mid \mc T)$ be a raw descriptor. Consider the identity map $\mathrm{id} : \mathrm{Int}(\mc S \mid \mc T) \to \mathrm{Int}(\mc S \mid \mc T)$ defined by $\mathrm{id}(s) = s$ for all $s \in \mathrm{Int}(\mc S \mid \mc T)$. The identity map is trivially a bijection. By the inductive definition of renaming, extending this map to the entire descriptor replaces every internal variable $s$ with $s$. This leaves every assignment and redirection unchanged. Therefore $(\mc S \mid \mc T)^{\mathrm{id}} = (\mc S \mid \mc T)$. This proves $(\mc S \mid \mc T) =_\alpha (\mc S \mid \mc T)$.

\item For symmetry, assume $(\mc S \mid \mc T) =_\alpha (\mc S' \mid \mc T')$. By definition, there exists a bijective map $\phi : \mathrm{Int}(\mc S \mid \mc T) \to \mathrm{Int}(\mc S' \mid \mc T')$ such that $\mc S^\phi = \mc S'$ and $\mc T^\phi = \mc T'$. Since $\phi$ is a bijection, it has a well-defined inverse
$$\phi^{-1} : \mathrm{Int}(\mc S' \mid \mc T')\to \mathrm{Int}(\mc S \mid \mc T),$$
which is also a bijection. The side condition of \cref{prop:internal-renaming-composition} is satisfied since the domain of $\phi$ is $\mathrm{Int}(\mc S\mid\mc T)$. Hence it contains every internal variable occurring in $\mc S$ and $\mc T$. Applying the internal renaming $\phi^{-1}$ to $\mc S'$ and $\mc T'$ gives
$$
\mc S'^{\phi^{-1}} = \big(\mc S^\phi\big)^{\phi^{-1}} = \mc S^{\phi^{-1} \circ \phi} = \mc S
\quad \text{and} \quad
\mc T'^{\phi^{-1}} = \big(\mc T^\phi\big)^{\phi^{-1}} = \mc T^{\phi^{-1} \circ \phi} = \mc T.
$$
The composite $\phi^{-1} \circ \phi$ is exactly the identity map on $\mathrm{Int}(\mc S \mid \mc T)$. Renaming by the identity map leaves the descriptor unchanged by the reflexivity calculation. We therefore have $(\mc S' \mid \mc T')^{\phi^{-1}} = (\mc S \mid \mc T)$. Thus $(\mc S' \mid \mc T') =_\alpha (\mc S \mid \mc T)$.

\item Finally, for transitivity, assume $(\mc S \mid \mc T) =_\alpha (\mc S' \mid \mc T')$ and $(\mc S' \mid \mc T') =_\alpha (\mc S'' \mid \mc T'')$. There exist bijections
$$
\phi : \mathrm{Int}(\mc S \mid \mc T) \to \mathrm{Int}(\mc S' \mid \mc T')
\quad \text{and} \quad
\psi : \mathrm{Int}(\mc S' \mid \mc T') \to \mathrm{Int}(\mc S'' \mid \mc T'')
$$
such that $(\mc S \mid \mc T)^\phi=(\mc S'\mid\mc T')$ and $(\mc S'\mid\mc T')^\psi=(\mc S''\mid\mc T'')$. The composition
$$
\psi\circ\phi:\mathrm{Int}(\mc S \mid \mc T)\to\mathrm{Int}(\mc S'' \mid \mc T'')
$$
of two bijections is itself a bijection. The side condition of \cref{prop:internal-renaming-composition} is satisfied. Indeed, $\phi$ is defined on all internal variables occurring in $(\mc S\mid\mc T)$, and $\psi$ on all those occurring in $(\mc S'\mid\mc T')$. Applying this composed renaming to the initial descriptor amounts to successive applications of the individual renamings:
$$
\mc S^{\psi \circ \phi} = \big(\mc S^\phi\big)^\psi = \mc S'^\psi = \mc S''
\quad \text{and} \quad
\mc T^{\psi \circ \phi} = \big(\mc T^\phi\big)^\psi = \mc T'^\psi = \mc T''.
$$
This composed bijection explicitly witnesses that $(\mc S \mid \mc T) =_\alpha (\mc S'' \mid \mc T'')$.

\end{itemize}
Combining these three properties, we conclude that the relation $=_\alpha$ is an equivalence relation on $\mathrm{desc}^{\mathrm{raw}}_\Sigma(\gamma \ ; \ \delta)$. \hfill $\square$

We now define the set of equivalence classes of raw descriptors identified up to internal renaming of their internal variables.

\begin{definition}[Equivalence Classes of Descriptors]\label{def:equivalence-classes-of-descriptors}
The relation $=_\alpha$ partitions $\mathrm{desc}^{\mathrm{raw}}_\Sigma(\gamma\ ;\ \delta)$ into equivalence classes. We denote such a class by $[\mc S \mid \mc T]$, where $(\mc S \mid \mc T) \in \mathrm{desc}^{\mathrm{raw}}_\Sigma(\gamma\ ;\ \delta)$ is any representative. We write $$\mathrm{Desc}^{\mathrm{raw}}_\Sigma(\gamma\ ;\ \delta) := \mathrm{desc}^{\mathrm{raw}}_\Sigma(\gamma\ ;\ \delta)/_{=_\alpha}.$$
\end{definition}

\textit{Convention:} Descriptor classes are referenced by single letters (e.g., $\mf D, \mf E, \dots$). Representatives always expand to an assignment set and a redirection set (e.g., $(\mc S \mid \mc T),(\mc S' \mid \mc T'), (\mc S_i \mid \mc T_i), \dots $). We write $\mf D = [\mc S \mid \mc T]$ to express that $(\mc S \mid \mc T)$ is a representative of the descriptor class $\mf D$.

\textit{Convention:} Lowercase $\mathrm{desc}$ denotes collections of raw representatives, while uppercase $\mathrm{Desc}$ denotes $\alpha$-equivalence classes. 

\begin{example}[Linear--Nonlinear Logic]\label{ex:linear-logic-descriptor-class}
The descriptors of the \hyperref[ex:linear-logic-alpha-equivalent-descriptors]{$\alpha$-equivalence example} are representatives of a single descriptor class
$$\mf D := [\{x, x' \xmapsto{\otimes_l} w\} \mid \{w \mapsto y\}] \in \mathrm{Desc}^{\mathrm{raw}}_\Sigma(x : l,\, x' : l \ ; \ y : l).$$
The representatives of $\mf D$ are exactly the raw descriptors $(\{x, x' \xmapsto{\otimes_l} v\} \mid \{v \mapsto y\})$, where $v$ ranges over the internal variables. The class collects all the renamings of a single computation. The name of the internal variable is not part of the data of $\mf D$.
\end{example}

\begin{definition}[Permutation Transport of Descriptor Classes]\label{def:permutation-transport-of-descriptor-classes}
Let $\pi:\gamma\cong\gamma'$ and $\tau:\delta\cong\delta'$ be context permutations. We define the transport of descriptor classes by
$$(\pi;\tau)_*[\mc S\mid\mc T]:=[\mc S\mid\mc T].$$
\end{definition}

\textit{Remark:} Permutation transport does not change assignment sets, redirection sets, or internal variables. It therefore preserves $=_\alpha$. Hence it is well defined on descriptor classes.

\begin{example}[Linear--Nonlinear Logic]
We consider again the context permutation $\pi : \gamma \cong \gamma'$ of the \hyperref[ex:linear-logic-context-permutation]{context-permutation example}, between $\gamma := x : l,\, x' : l$ and $\gamma' := x' : l,\, x : l$, together with the identity permutation $\tau$ on $\delta := y : l$. The transport of the class of the multiplicative conjunction is
$$(\pi;\tau)_*[\{x, x' \xmapsto{\otimes_l} w\} \mid \{w \mapsto y\}] = [\{x, x' \xmapsto{\otimes_l} w\} \mid \{w \mapsto y\}],$$
now regarded as a descriptor class over the boundary $(\gamma' \ ; \ \delta)$. The representative with the internal variable $w_1$ gives the same transported class. As for raw descriptors, only the boundary changes.
\end{example}

\begin{proposition}[Redirections of Descriptors]\label{prop:redirections-of-descriptors}
Let $\mf D \in \mathrm{Desc}^{\mathrm{raw}}_\Sigma(\gamma\ ;\ \delta)$ and let $y \in \delta$. The redirection to $y$ of the descriptor, defined as
$$\mf D_{\mapsto y} := [(\mc S \mid \mc T_{\mapsto y})] \quad \text{for some representative} \quad (\mc S \mid \mc T),$$
is well defined, i.e., it does not depend on the representative.
\end{proposition}
\textit{Proof.}
Let $(\mc S \mid \mc T)$ and $(\mc S' \mid \mc T')$ be two representatives of $\mf D$. Since $(\mc S \mid \mc T) =_\alpha (\mc S' \mid \mc T')$, there exists a bijection 
$$\phi : \mathrm{Int}(\mc S \mid \mc T) \to \mathrm{Int}(\mc S' \mid \mc T')$$
such that $(\mc S \mid \mc T)^\phi = (\mc S' \mid \mc T')$. Equivalently, $\mc S^\phi = \mc S'$ and $\mc T^\phi = \mc T'$.

By \cref{lem:internal-renamings-are-injective-on-variables}, $\phi$ is injective on the variables occurring in $\mc T$. Hence, by \cref{lem:internal-renaming-preserves-redirections-targeting-an-output}, we have
$$(\mc T_{\mapsto y})^\phi = (\mc T^\phi)_{\mapsto y} = \mc T'_{\mapsto y}.$$
Combined with $\mc S^\phi = \mc S'$, this gives
$$(\mc S \mid \mc T_{\mapsto y})^\phi = (\mc S' \mid \mc T'_{\mapsto y}).$$

By \cref{lem:internal-variables-under-internal-renaming-and-input-substitution}, $\phi$ restricts to a bijection
$$\phi : \mathrm{Int}(\mc S \mid \mc T_{\mapsto y}) \to \mathrm{Int}(\mc S' \mid \mc T'_{\mapsto y}),$$
and the variables of $\mathrm{Int}(\mc S\mid\mc T)$ not in $\mathrm{Int}(\mc S\mid\mc T_{\mapsto y})$ do not occur in $(\mc S\mid\mc T_{\mapsto y})$. The restriction therefore acts as $\phi$ does and, together with this equality, witnesses
$$(\mc S \mid \mc T_{\mapsto y}) =_\alpha (\mc S' \mid \mc T'_{\mapsto y}).$$
Hence $[(\mc S \mid \mc T_{\mapsto y})] = [(\mc S' \mid \mc T'_{\mapsto y})]$, and $\mf D_{\mapsto y}$ does not depend on the choice of representative.
\hfill $\square$

\begin{lemma}[Input Occurrence is $\alpha$-Invariant]\label{lem:input-occurrence-is-alpha-invariant}
Let
$ (\mc S\mid\mc T)=_\alpha(\mc S'\mid\mc T') $ be witnessed by a bijection of internal variables $ \phi:\mathrm{Int}(\mc S\mid\mc T)\to \mathrm{Int}(\mc S'\mid\mc T') $ such that $ (\mc S\mid\mc T)^\phi=(\mc S'\mid\mc T'). $ Then $$ \mathrm{Inp}(\mc S\mid\mc T)=\mathrm{Inp}(\mc S'\mid\mc T'). $$
\end{lemma} 
\textit{Proof.}
The internal renaming $\phi$ acts only on internal variables and fixes input and output variables. A variable occurring in $(\mc S\mid\mc T)$ is therefore an input variable exactly when its image under $\phi$ is the same input variable. Hence
$$
\mathrm{Inp}\big((\mc S\mid\mc T)^\phi\big)=\mathrm{Inp}(\mc S\mid\mc T).
$$
Since $(\mc S\mid\mc T)^\phi=(\mc S'\mid\mc T')$, the claim follows.
\hfill $\square$

\begin{definition}[Input Occurrence of a Descriptor Class]\label{def:input-occurrence-of-a-descriptor-class}
Let $\mf D=[\mc S\mid\mc T]\in \mathrm{Desc}^{\mathrm{raw}}_\Sigma(\gamma;\delta)$. Define
$$
\mathrm{Inp}(\mf D):=\mathrm{Inp}(\mc S\mid\mc T).
$$
\end{definition}
\textit{Remark:} This is well defined by the $\alpha$-invariance of input occurrence.

\begin{example}[Linear--Nonlinear Logic]\label{ex:linear-logic-input-occurrences}
Consider the multiplicative conjunction descriptor $\mf D$ exposed in the \hyperref[ex:linear-logic-descriptor-class]{descriptor-class example}. Any representative of the class would give us $$\mathrm{Inp}(\mf D) = \{x, x'\}.$$
Now consider the descriptor class $[\varnothing \mid \{x \mapsto y\}]$ over the same boundary $(x : l,\, x' : l \ ; \ y : l)$. The input occurrence set is the singleton $\{x\}$. Even though the input $x'$ is present in the context, it does not occur anywhere in the descriptor and is therefore not present in the input set.
\end{example}
\endgroup{}
\begingroup

\subsubsection{Regular Descriptors}

Before giving the rules of generation for well-formed descriptors, we investigate five specific properties that will render the underlying graph of descriptors appropriate as an interpretation of morphisms in an LNL-polycategory. We remind the reader that LNL-polycategories are the semantics targeted by our syntax to unlock the possibility of embedding the two different judgment structures of cartesian and linear systems. The additional property of mode-validity is also required to guarantee the existence of a mode assignment for the descriptors. Such an assignment is automatically unique (\cref{lem:uniqueness-of-mode-assignments}). Mode-validity confirms that the descriptors respect the modes of the variables present in the context as well as the modes present in the application of the generators.
Regularity criteria are concerned with the structural properties of the underlying graph of the descriptors (not to be confused with the substructural properties of the logic those descriptors encode). Those properties turn out to be invariants of the well-formed descriptors.

\begin{definition}[Regular Raw Descriptor]\label{def:regular-raw-descriptor}
A raw descriptor $(\mc S\mid\mc T)$ is called \textit{regular} if it satisfies these six conditions:
\begin{itemize}
  \item \textbf{Mode validity.} The descriptor is mode-valid;
  \item \textbf{Irreflexive reachability.} The strict reachability relation $\rightsquigarrow$ of the descriptor is irreflexive, i.e., $s \nrightsquigarrow s$ for any variable $s$;
	\item \textbf{Unique internal target occurrence.} Every internal variable $w\in\mathrm{Int}(\mc S)$ occurs as the target of a unique assignment $E\in\mc S$, called the \emph{producer} of $w$, i.e.,
  $$
  \exists ! E\in\mc S \quad \text{such that} \quad w\in\mathrm{tgt}(E),
  $$
  and, for every assignment in $\mc S$, its targets are pairwise distinct;
  \item \textbf{Grounded redirections.} All the internal variables appearing in the redirection set $\mc T$ are present in $\mathrm{Int}(\mc S)$. Formally, $\mathrm{Int}(\mc T) \subseteq \mathrm{Int}(\mc S)$, where $\mathrm{Int}(\mc T)$ is the ordinary set of internal variables appearing in $\mc T$;
  \item \textbf{Unique internal source occurrence.} Every internal variable $w\in\mathrm{Int}(\mc S \mid \mc T)$ occurs in exactly one source position: there is exactly one pair $(E,i)$ of an element $E\in\mc S\cup\mc T$ and a source position $i$ of $E$ carrying $w$. In particular, no internal variable is left unconsumed, and no internal variable is shared between two consumers;
  \item \textbf{Boundary usage property.} Every input declared at a \emph{linear} mode in $\gamma$ occurs in exactly one source position of $\mc S\cup\mc T$. Inputs declared at cartesian modes are unrestricted. Every output $y\in\delta$ is the target of exactly one redirection of $\mc T$.
\end{itemize}
\end{definition}

\begin{example}[Linear--Nonlinear Logic]\label{ex:linear-logic-regular-tensor-descriptor}
The descriptor of the multiplicative conjunction $(\{x, x' \xmapsto{\otimes_l} w\} \mid \{w \mapsto y\})$ over the boundary $(x : l,\, x' : l \ ; \ y : l)$ is regular. It is mode-valid with the mode assignment $\mu(w) = l$, and its strict reachability relation is irreflexive. The internal variable $w$ is the target of a unique assignment, its producer, and occurs in exactly one source position, in the redirection $w \mapsto y$. The redirection set uses no internal variable beyond $w$, so the redirections are grounded. Finally, each of the linear inputs $x, x'$ occurs in exactly one source position, and the output $y$ is the target of exactly one redirection. By contrast, the raw descriptor $(\varnothing \mid \{x \mapsto y\})$ over the same boundary from the \hyperref[ex:linear-logic-input-occurrences]{input-occurrence example} fails the boundary usage property: the input $x'$ is declared at a linear mode and occurs in no source position. The raw descriptor $(\{x, x' \xmapsto{\otimes_l} w\} \mid \{w \mapsto y,\ w \mapsto y'\})$ over the boundary $(x : l,\, x' : l \ ; \ y : l,\, y' : l)$ fails unique internal source occurrence: the internal variable $w$ is shared between the two consumers $w \mapsto y$ and $w \mapsto y'$.
\end{example}

A substitution could in principle merge two elements of a set into one. The freshness of the substituted variable excludes this.

\begin{lemma}[No Collapse under Fresh Input Substitution]\label{lem:no-collapse-under-input-substitution}
Let $(\mc S\mid\mc T)\in\mathrm{desc}^{\mathrm{raw}}_\Sigma(\gamma\ ;\ \delta)$ be a raw descriptor, and let $x$ be an input variable and $u$ an input or internal variable not occurring in $(\mc S\mid\mc T)$. The map $E \mapsto E[x:=u]$ is a bijection from $\mc S$ to $\mc S[x:=u]$ and from $\mc T$ to $\mc T[x:=u]$; in particular, $|\mc S[x:=u]| = |\mc S|$ and $|\mc T[x:=u]| = |\mc T|$.
\end{lemma}

\textit{Proof.}
Since the substitution replaces every occurrence of $x$ with the fresh variable $u$, the descriptors remain unchanged and all assignments and redirections are mapped one to one by replacing $x$ with $u$. \hfill $\square$

\textit{Remark:} The freshness hypothesis cannot be dropped, even for regular descriptors. Take a co-nullary generator $g:l\to\cdot$ at a linear mode $l$ and the descriptor $(\{x\xmapsto{g}\cdot,\ u\xmapsto{g}\cdot\}\mid\varnothing)$ over the boundary $(x:l,\ u:l\ ;\ \cdot)$. The descriptor is mode-valid, each linear input occurs in exactly one source position, and the remaining conditions hold vacuously, so it is regular. The substitution $[x:=u]$ still merges the two assignments into the single assignment $u\xmapsto{g}\cdot$.

\textit{Remark:} The counting arguments of the composition section read the positions of assignments and redirections after a substitution of a fresh variable; \cref{lem:no-collapse-under-input-substitution} guarantees that the counts before and after agree.

The very first result we would like to show regarding regular descriptors is that the regularity conditions force mode-validity to be witnessed by a unique function.
\begin{proposition}[Mode Assignments Are Unique for Regular Descriptors]\label{prop:mode-assignments-are-unique-for-regular-descriptors}
Let $(\mc S\mid\mc T)\in\mathrm{desc}^{\mathrm{raw}}_\Sigma(\gamma\ ;\ \delta)$ be a regular raw descriptor. Then there is a unique mode assignment witnessing mode-validity of $(\mc S\mid\mc T)$.
\end{proposition}

\textit{Proof.}
Existence is part of regularity, since regularity includes mode-validity. Uniqueness holds for arbitrary raw descriptors, mode-valid or not, by \cref{lem:uniqueness-of-mode-assignments}. \hfill $\square$

The second desirable property for regular descriptors is the acyclicity of their underlying dependency graphs. This will ensure that the proof trees (which are encoded in the descriptors) have no cycles and are well-founded.
\begin{proposition}[Regular Descriptors Have Acyclic Dependency Graphs]\label{prop:regular-descriptors-have-acyclic-dependency-graphs}
If $(\mc S\mid\mc T)$ is a regular raw descriptor, then $G(\mc S\mid\mc T)$ is a finite directed acyclic graph.
\end{proposition}

\textit{Proof.}
The vertex set $\mathrm{Var}(\mc S\mid\mc T)$ is finite since $\mc S$ and $\mc T$ are finite and each assignment or redirection contains finitely many variables. The edge set $\leadsto_{(\mc S\mid\mc T)}$ is finite by the same argument. A cycle in $G(\mc S\mid\mc T)$ is a non-empty path from some variable $v$ to itself (\cref{def:reachability-relation}), i.e.,
$$v\rightsquigarrow_{(\mc S\mid\mc T)}v.$$
This contradicts the irreflexivity of strict reachability, which is part of regularity. The graph $G(\mc S\mid\mc T)$ is therefore acyclic. \hfill $\square$

\begin{lemma}[Fresh Input Substitution Preserves Irreflexivity]\label{lem:fresh-input-substitution-preserves-irreflexivity}
Let $(\mc S\mid\mc T)$ be a regular raw descriptor, let $x$ be an input variable, and let $w$ be a fresh internal variable for $(\mc S\mid\mc T)$, i.e., $w$ does not appear in $\mc S$ or in $\mc T$. Then the strict reachability relation $\rightsquigarrow_{(\mc S \mid \mc T)[x := w]}$ is irreflexive.
\end{lemma}

\textit{Proof.}
Suppose for contradiction that there is a non-empty cycle in the substituted descriptor (\cref{def:reachability-relation}):
$$
z_0 \leadsto z_1 \leadsto \cdots \leadsto z_n \leadsto z_0.
$$
Notice first that no one-step edge in $(\mc S[x:=w]\mid \mc T[x:=w])$ targets $w$. Indeed, by \cref{lem:one-step-reachability-is-natural}, every such edge is of the form
$$
a[x:=w]\leadsto b[x:=w]
$$
for some edge $a\leadsto b$ in $(\mc S\mid\mc T)$. If $b[x:=w]=w$, then either $b=x$ or $b=w$. The first case is impossible by \cref{lem:inputs-have-no-incoming-paths-and-outputs-have-no-outgoing-paths}: no one-step edge targets an input variable. The second is impossible since $w$ is fresh for $(\mc S\mid\mc T)$. No edge in the substituted descriptor therefore targets $w$.

The cycle cannot therefore contain $w$, since every variable on a cycle is the target of an edge that precedes it.
Now consider an edge $z_j\leadsto z_{j+1}$ of the cycle. By \cref{lem:one-step-reachability-is-natural}, the edge is of the form $$a[x := w] \leadsto b[x := w]$$ for some edge $a \leadsto b$ in $(\mc S \mid \mc T)$. Neither $z_j$ nor $z_{j+1}$ is $w$, and $w$ is fresh for $(\mc S\mid\mc T)$, so $a\neq x$ and $b\neq x$, hence $a=z_j$ and $b=z_{j+1}$. Every edge of the substituted cycle was therefore already present before substitution. The cycle is therefore already a cycle in $(\mc S\mid\mc T)$.

This contradicts the irreflexivity of the original descriptor; the substituted reachability relation is therefore irreflexive. \hfill $\square$

\begin{proposition}[Internal Renaming Preserves Regularity]\label{prop:internal-renaming-preserves-regularity}
Let $(\mc S\mid\mc T),(\mc S'\mid\mc T')\in\mathrm{desc}^{\mathrm{raw}}_\Sigma(\gamma\ ;\ \delta)$ be raw descriptors over a common boundary such that there exists a bijection
$\phi:\mathrm{Int}(\mc S\mid\mc T)\to \mathrm{Int}(\mc S'\mid\mc T')$
with $(\mc S\mid\mc T)^\phi=(\mc S'\mid\mc T')$. Then $(\mc S\mid\mc T)$ is regular if and only if $(\mc S'\mid\mc T')$ is regular.
\end{proposition}

\textit{Proof.}
It is enough to prove one direction, since the converse follows by applying the same argument to $\phi^{-1}$ (\cref{prop:descriptor-equivalence-relation}). Suppose $(\mc S\mid\mc T)$ is regular. Regularity includes mode-validity, so $(\mc S\mid\mc T)$ is mode-valid, and $(\mc S'\mid\mc T')$ is then mode-valid by \cref{lem:internal-renaming-preserves-mode-validity}. \\
It remains to prove the five remaining regularity properties for $(\mc S'\mid\mc T')$.

\begin{itemize}
	\item First, we prove the grounded-redirections property. We have $\mc T'=\mc T^\phi$, so \cref{lem:internal-variables-under-internal-renaming-and-input-substitution} gives:
  $$\mathrm{Int}(\mc T')=\mathrm{Int}(\mc T^\phi)=\mathrm{Int}(\mc T)^\phi.$$
  Since $(\mc S\mid\mc T)$ is regular, $\mathrm{Int}(\mc T)\subseteq\mathrm{Int}(\mc S)$. Hence
  $$\mathrm{Int}(\mc T')=\mathrm{Int}(\mc T)^\phi\subseteq \mathrm{Int}(\mc S)^\phi=\mathrm{Int}(\mc S^\phi)=\mathrm{Int}(\mc S'),$$
  again by \cref{lem:internal-variables-under-internal-renaming-and-input-substitution}. Thus $\mathrm{Int}(\mc T')\subseteq\mathrm{Int}(\mc S')$.

	\item Second, we prove the unique internal target occurrence property. Let $z\in\mathrm{Int}(\mc S')$. Since $\mc S'=\mc S^\phi$, \cref{lem:internal-variables-under-internal-renaming-and-input-substitution} provides a unique $u\in\mathrm{Int}(\mc S)$ such that $z=u^\phi$. Regularity of $(\mc S\mid\mc T)$ gives $u$ a unique producing assignment $E_u\in\mc S$, i.e., $u\in\mathrm{tgt}(E_u)$. Then $E_u^\phi\in\mc S'$ and $z=u^\phi\in\mathrm{tgt}(E_u^\phi)$, so $z$ is produced in $\mc S'$.

  If $F'\in\mc S'$ is another assignment with $z\in\mathrm{tgt}(F')$, then $F'=F^\phi$ for some $F\in\mc S$. Since $z\in\mathrm{tgt}(F^\phi)$, injectivity of the internal renaming (\cref{lem:internal-renamings-are-injective-on-variables}) gives $u\in\mathrm{tgt}(F)$, and uniqueness of the producer of $u$ in $\mc S$ forces $F=E_u$, hence $F'=E_u^\phi$. The producing assignment of $z$ in $\mc S'$ is therefore unique.

  Pairwise distinctness of targets inside an assignment is also preserved. An internal renaming carries the target tuple $w_1,\ldots,w_n$ of an assignment to $w_1^\phi,\ldots,w_n^\phi$, and if $w_i^\phi=w_j^\phi$, injectivity gives $w_i=w_j$, hence $i=j$ since the original targets were pairwise distinct.

	\item Third, we prove the unique internal source occurrence and boundary usage properties. The renaming $(\--)^\phi$ maps the elements of $\mc S\cup\mc T$ bijectively to the elements of $\mc S'\cup\mc T'$, preserving the shape of every assignment and redirection. It therefore maps the source positions of $\mc S\cup\mc T$ bijectively to those of $\mc S'\cup\mc T'$, carrying an occurrence of a variable $v$ to an occurrence of $v^\phi$. The extension of $\phi$ is injective on the variables of $(\mc S\mid\mc T)$ (\cref{lem:internal-renamings-are-injective-on-variables}) and fixes input and output variables; occurrence counts are therefore preserved. An internal variable $w$ has exactly one source occurrence if and only if $w^\phi$ does, an input has exactly one source occurrence if and only if it does after renaming, and a redirection targets $y$ if and only if its renaming does. Both properties transfer.

	\item Finally, we prove irreflexivity of strict reachability. Suppose for contradiction that for some variable $z$ we have the path
  $$z\rightsquigarrow_{(\mc S'\mid\mc T')} z.$$
  We know immediately that no input or output variable can lie on a cycle by \cref{lem:inputs-have-no-incoming-paths-and-outputs-have-no-outgoing-paths}. $z$ is therefore internal. Now $(\mc S'\mid\mc T')^{\phi^{-1}}=(\mc S\mid\mc T)$ by the symmetry of $\alpha$-equivalence (\cref{prop:descriptor-equivalence-relation}), and \cref{lem:internal-renamings-are-injective-on-variables} applies to $\phi^{-1}$, so \cref{lem:internal-renaming-preserves-strict-reachability} applied to $\phi^{-1}$ gives
  $$z^{\phi^{-1}}\rightsquigarrow_{(\mc S\mid\mc T)} z^{\phi^{-1}},$$
  contradicting the irreflexivity of $(\mc S\mid\mc T)$. Reachability in $(\mc S'\mid\mc T')$ is therefore irreflexive.
\end{itemize}

Therefore $(\mc S'\mid\mc T')$ is regular. \hfill $\square$

\textit{Remark:} It is therefore enough to check the defining properties on any convenient representative.

\textit{Remark:} Thanks to the grounded-redirections property giving $\mathrm{Int}(\mc S \mid \mc T) = \mathrm{Int}(\mc S)$, it suffices to consider only the internal variables in the assignment set $\mc S$ to cover all the internal variables of the descriptor.

\begin{definition}[Descriptor Class Regularity]\label{def:descriptor-class-regularity}
A descriptor class $\mf D=[\mc S\mid\mc T]$ is called \textit{regular} if one, equivalently every, representative $(\mc S\mid\mc T)$ is a regular raw descriptor (\cref{prop:internal-renaming-preserves-regularity}).

We write $$ \mathrm{Reg}(\mf D) $$ for the property of a descriptor class being regular.
\end{definition}

\begin{example}[Linear--Nonlinear Logic]
The descriptor class $\mf D = [\{x, x' \xmapsto{\otimes_l} w\} \mid \{w \mapsto y\}]$ of the multiplicative conjunction is regular: the representative in the \hyperref[ex:linear-logic-regular-tensor-descriptor]{regular-tensor example} satisfies the six conditions. We therefore write $\mathrm{Reg}(\mf D)$.
\end{example}

\begin{proposition}[Permutation Transport Preserves Regularity]\label{prop:permutation-transport-preserves-regularity}
Let $\pi:\gamma\cong\gamma'$ and $\tau:\delta\cong\delta'$ be context permutations. For any descriptor class $\mf D\in\mathrm{Desc}^{\mathrm{raw}}_\Sigma(\gamma\ ;\ \delta)$,
$$
\mathrm{Reg}(\mf D)
\quad\Longleftrightarrow\quad
\mathrm{Reg}((\pi;\tau)_*\mf D).
$$
\end{proposition}

\textit{Proof.}
Let $\mf D=[\mc S\mid\mc T]$. By definition of permutation transport (\cref{def:permutation-transport-of-raw-descriptors,def:permutation-transport-of-descriptor-classes}), $(\pi;\tau)_*\mf D=[\mc
S\mid\mc T]$ in the permuted ambient context. The graph-theoretic regularity conditions other than boundary usage depend only
on the assignment set and redirection set; they are therefore unchanged by transport. Mode-validity and the boundary usage property are
unchanged as well since context permutations preserve exactly the variable-mode declarations of the input
and output contexts. Hence $(\mc S\mid\mc T)$ is regular in the first ambient context if and only if
the transported representative is regular in the second. \hfill $\square$

\begin{definition}[c-Descriptors and $\ell$-Descriptors]\label{def:c-descriptors}
A \emph{c-descriptor} is a regular descriptor class over an all-cartesian input context with a single output at a cartesian mode, containing no variable at a linear mode under its unique mode assignment (\cref{prop:mode-assignments-are-unique-for-regular-descriptors}). An $\ell$\emph{-descriptor} is a regular descriptor class such that all the outputs are linear.
\end{definition}

\textit{Remark:} The two conditions do not depend on the chosen representative. The outputs are part of the boundary, which is fixed. The modes of the internal variables are given by the unique mode assignments, and those correspond along the internal renamings relating the representatives (\cref{lem:internal-renaming-preserves-mode-validity}).

\begin{example}[Linear--Nonlinear Logic]\label{ex:linear-logic-descriptor-sorts}
Over the pre-signature extended by the binary c-generator $\otimes_c : c, c \to c$, the class $[\{a, a' \xmapsto{\otimes_c} w\} \mid \{w \mapsto y\}]$ over the boundary $(a : c,\, a' : c \ ; \ y : c)$ is a c-descriptor. The input context is cartesian, the single output $y$ is declared at the cartesian mode, and the unique mode assignment gives $\mu(w) = c$, so no variable is at a linear mode. The class $\mf D$ of the \hyperref[ex:linear-logic-descriptor-class]{multiplicative conjunction} is an $\ell$-descriptor, since its single output is linear. The class of the raw descriptor from the \hyperref[ex:linear-logic-mixed-raw-descriptor]{mixed raw-descriptor example}, $(\{(a \ \mid \ \cdot) \xmapsto{g} w,\ \ w, x \xmapsto{\otimes_l} w'\} \mid \{w' \mapsto y\})$ over the boundary $(a : c,\, x : l \ ; \ y : l)$, is regular as well, and its single output is linear. It is therefore an $\ell$-descriptor, although its input context declares the cartesian variable $a$. The constraint only concerns the outputs.
\end{example}

\endgroup{}
\begingroup

\subsubsection{Dependent Trees}
For a raw descriptor, the property of regularity guarantees that the dependency graph of an output node is actually a tree. This corresponds to the fact that in the intuitionistic calculus, a derivation is a proof tree where the output is the root and the leaves are the axioms. For the classical calculus, this corresponds to the fact that a deduction tree is associated to each formula on the target side (formulas on the right-hand side of the turnstile). We use a procedure of starting from a given output in a raw descriptor and ``tracing back'' the directed edges until we reach the inputs. Every internal variable traversed corresponds to a node of the tree. The root is therefore the output variable, and the leaves are the inputs reached by this process. We investigate this procedure in this section.

\begin{definition}[Dependent Tree]\label{def:dependent-tree}
Let $(\mc S\mid\mc T)$ be a regular raw descriptor and $z$ one of its variables. The dependent tree of the variable $z$ in the descriptor $(\mc S \mid \mc T)$ is denoted $\llbracket z\rrbracket$. It is the finite expression over the input variables given by

\begin{itemize}
  \item If $x$ is an input variable, then $\llbracket x \rrbracket = x$.
  \item If $s \mapsto y$ is a redirection, then $\llbracket y \rrbracket = \llbracket s \rrbracket$.
  \item For any assignment of the form $s_1, \dots, s_n \xmapsto{f} w$, we have $\llbracket w \rrbracket = f(\llbracket s_1 \rrbracket, \dots, \llbracket s_n \rrbracket)$.
  \item For any assignment of the form $(s_1, \dots, s_k \mid s'_1, \dots, s'_m ) \xmapsto{g} w_1, \dots, w_n$, we have $\llbracket w_j \rrbracket = g_j(\llbracket s_1 \rrbracket, \dots, \llbracket s_k  \rrbracket\mid \llbracket s'_1 \rrbracket, \dots,  \llbracket s'_m \rrbracket)$,
\end{itemize}

where an $\ell$-generator $g$ with $n$ targets is counted as $n$ distinct operation symbols $g_{1},\ldots,g_{n}$. 
\end{definition}

\textit{Remark:} We note immediately that we took the liberty of not mentioning the associated descriptor the tree is being formed on in the notation $\llbracket \-- \rrbracket$. The reason is expository in nature. It will always be clear from the context which descriptor is being evaluated.

\textit{Remark:} Acyclicity makes the recursion terminate. The unique internal target occurrence, pairwise distinct targets, and boundary usage make each clause single-valued. The tree is therefore well defined and finite. When every assignment in the tree of the variable is a c-generator assignment, the dependent tree is an expression over the input variables. This is the technical device of \cref{prop:cartesian-descriptors-are-terms} through which the cartesian well-formed descriptors are fully determined by expressions.

\textit{Remark:} The dependent tree $\llbracket y\rrbracket$ of an output variable $y$ is well defined. Indeed, by the boundary usage property of \cref{def:regular-raw-descriptor}, exactly one redirection targets $y$.

\begin{example}[Linear--Nonlinear Logic]
We compute the dependent trees of three regular descriptors from the reusable examples. For the descriptor $(\{a, a' \xmapsto{\otimes_c} w\} \mid \{w \mapsto y\})$ of the binary c-generator $\otimes_c$ from the \hyperref[ex:linear-logic-descriptor-sorts]{descriptor-sorts example}, over the boundary $(a : c,\, a' : c \ ; \ y : c)$, the clauses give
$$\llbracket y \rrbracket = \llbracket w \rrbracket = \otimes_c(\llbracket a \rrbracket, \llbracket a' \rrbracket) = \otimes_c(a, a').$$
Every assignment traversed is a c-generator assignment, and the tree is an expression over the input variables. For the descriptor $(\{x \xmapsto{\parr_l} w_1, w_2\} \mid \{w_1 \mapsto y,\ w_2 \mapsto y'\})$ of the \hyperref[ex:linear-logic-componentwise-sum]{multiplicative-disjunction example}, over the boundary $(x : l \ ; \ y : l,\, y' : l)$, each output has its own tree,
$$\llbracket y \rrbracket = \llbracket w_1 \rrbracket = \parr_{l,1}(x) \qquad \text{and} \qquad \llbracket y' \rrbracket = \llbracket w_2 \rrbracket = \parr_{l,2}(x),$$
in which the generator $\parr_l$ is counted as the two operation symbols $\parr_{l,1}$ and $\parr_{l,2}$. The two trees share the input $x$. Finally, for the descriptor from the \hyperref[ex:linear-logic-mixed-raw-descriptor]{mixed raw-descriptor example}, $(\{(a \ \mid \ \cdot) \xmapsto{g} w,\ \ w, x \xmapsto{\otimes_l} w'\} \mid \{w' \mapsto y\})$ over the boundary $(a : c,\, x : l \ ; \ y : l)$, the trees nest:
$$\llbracket y \rrbracket = \llbracket w' \rrbracket = \otimes_{l,1}(\llbracket w \rrbracket, \llbracket x \rrbracket) = \otimes_{l,1}(g_1(a \ \mid \ \cdot),\, x).$$
\end{example}

\begin{lemma}[Fresh Representatives]\label{lem:fresh-representatives}
Let $\mf D_1,\ldots,\mf D_n$ be descriptor classes, and let $F$ be a finite set of internal variables. Then there exist representatives
\[
(\mc S_i\mid\mc T_i)\in \mf D_i
\]
such that the sets $\mathrm{Int}(\mc S_i\mid\mc T_i)$ are pairwise disjoint and disjoint from $F$.
\end{lemma}
\textit{Proof.}
Choose arbitrary representatives $(\mc S_i^0\mid\mc T_i^0)\in\mf D_i$. Each raw descriptor is finite. Each set $\mathrm{Int}(\mc S_i^0\mid\mc T_i^0)$ is therefore finite. The supply of internal variables is countably infinite. We may therefore choose finite sets $A_i$ of internal variables such that
\[
|A_i|=|\mathrm{Int}(\mc S_i^0\mid\mc T_i^0)|,
\]
the sets $A_i$ are pairwise disjoint, and each $A_i$ is disjoint from $F$.

For each $i$, choose a bijection
\[
\phi_i:\mathrm{Int}(\mc S_i^0\mid\mc T_i^0)\to A_i.
\]
Then $(\mc S_i^0\mid\mc T_i^0)^{\phi_i}$ is a representative of $\mf D_i$. Its internal variables are exactly $A_i$. Those representatives then have pairwise disjoint internal variables that are disjoint from $F$. \hfill $\square$
\endgroup{}
\begingroup

\subsubsection{Well-Formed Descriptors}\label{sec:well-formed-descriptors}
As is typical in any treatment of type theories, we generate the descriptors of interest to us using inductive rules. The constructors used in the rules manipulate raw descriptor classes and are defined by their actions on a representative. Thus, a big part of this subsection is to show that the rules are well defined for the classes they manipulate. Moreover, the rules are shown to preserve regularity, so not only is the property of preserving classes retained, but the property of preserving regularity is too. The upshot of the section is to conclude that the objects of interest are regular descriptor classes. Regularity provides graph-theoretic guarantees, and classes make the analysis and the manipulation of the objects $\alpha$-invariant, i.e., invariant under internal renamings.

Formally, we set $$ \mathrm{Desc}_\Sigma(\gamma \ ; \ \delta) := \{\mf D \in \mathrm{Desc}^{\mathrm{raw}}_\Sigma(\gamma \ ; \ \delta) \mid \gamma \Vdash_{\mf D} \delta \ \text{is derivable}\}. $$
We also write $$ \mathrm{desc}_\Sigma(\gamma;\delta) := \left\{ (\mc S\mid\mc T)\in \mathrm{desc}^{\mathrm{raw}}_\Sigma(\gamma;\delta) \ \middle|\ [\mc S\mid\mc T]\in \mathrm{Desc}_\Sigma(\gamma;\delta) \right\} $$
for the collection of raw representatives of well-formed descriptor classes.

Every judgment derivable by the generation rules of \cref{def:well-formed-descriptors-identity,def:well-formed-descriptor-cartesian,def:well-formed-descriptor-positive,def:well-formed-descriptor-negative} has one of two boundary shapes. A cartesian judgment, presented as $\theta \Vdash_{\mf D} b:q$, has all its inputs at cartesian modes and a single output, at a cartesian mode. A linear judgment, presented as $\theta \ ; \ \Gamma \Vdash_{\mf D} \delta$, has input context $\theta,\Gamma$ with $\theta$ its cartesian and $\Gamma$ its linear declarations, and all outputs at linear modes. It is shown in \cref{prop:sorts-of-well-formed-descriptors} that every derivable judgment in the mode theory has one of these two shapes.

\begin{definition}[Well-Formed Descriptors (Identity)]\label{def:well-formed-descriptors-identity}
Given a pre-signature $\Sigma \ sig$, a cartesian context $\theta \ \mathrm{ctx}^{\mathrm{c}}_{\Sigma}$, an input variable $x : p$ not present in $\theta$ and an output variable $y : p$, the descriptor $[\varnothing \mid \{x \mapsto y\}]$ is a \textit{well-formed descriptor} in the contexts $\theta,x:p$ and $y:p$. Its transports along context permutations are also well formed.

 We package this information in the Identity rule.
$$
\inferrule*[Right=Identity]
{
  \theta \ \mathrm{ctx}^{\mathrm{c}}_{\Sigma} \\
  \theta, x : p \ \mathrm{ctx}^{\mathrm{in}}_{\Sigma} \\
  y : p \ \mathrm{ctx}^{\mathrm{out}}_{\Sigma} \\
  \pi:\theta,x:p\cong\gamma' \\
  \tau:y:p\cong\delta'
}
{ \gamma' \Vdash_{(\pi;\tau)_*[\varnothing \mid \{x \mapsto y\}]} \delta' }
$$
\end{definition}

\textit{Remark:} Enforcing the condition that the descriptor is also well formed along context permutations is the mechanism by which exchange will be made admissible at the framework level. This is the technical device that allows the generation of descriptors irrespective of the order of the definitions of the variables. We emphasize again here that at the object logic level, the exchange substructural rule is enforced by the $2$-cell mechanism of \cref{def:descriptor-2-cells}.
The ambient input context $\theta$ is cartesian, and the output context is exactly $y:p$. The fact that the identity descriptor is contingent on only two variables $x$ and $y$, while living in an ambient cartesian context that may declare more variables, yields the weakening admissibility of \cref{thm:input-weakening-admissibility-for-well-formed-descriptors}. At linear modes no ambient side variables exist in the linear zone. The boundary usage property forbids unused linear inputs and unredirected outputs. Enabling substructural behavior at linear modes therefore requires the gates of \cref{def:gate-descriptor-generation}. When $p$ is cartesian, the conclusion is a cartesian judgment $\theta,x:p\Vdash y:p$. When $p$ is linear, it is the linear judgment $\theta \ ; \ x:p\Vdash y:p$.

\begin{example}[Linear--Nonlinear Logic]
Over the linear--nonlinear pre-signature of the \hyperref[ex:linear-logic-mode-signature]{running example}, take the cartesian context $\theta := a : c$. With $\pi$ and $\tau$ the identity permutations, the rule derives the linear judgment
$$a : c \ ; \ x : l \Vdash_{[\varnothing \mid \{x \mapsto y\}]} y : l.$$
The cartesian declaration $a$ does not occur in the descriptor. Taking instead an input $a' : c$ and an output $b : c$ gives the cartesian judgment $a : c,\, a' : c \Vdash_{[\varnothing \mid \{a' \mapsto b\}]} b : c$ over the same ambient context. No instance of the rule declares a second input at a linear mode, since every input declaration besides $x$ lives in the cartesian context $\theta$.
\end{example}

This identity generation rule is the entry point of the calculus. We now prove that the descriptor class produced by the rule is regular. This has the expected added benefit that all well-formed descriptors are regular as long as all the formation rules preserve regularity.
\begin{proposition}[Identity produces a regular descriptor]\label{prop:identity-produces-a-regular-descriptor}
  The identity rule is well defined and produces regular descriptors.
\end{proposition}

\textit{Proof.} We verify that the descriptor $(\varnothing \mid \{x \mapsto y\})$ satisfies the defining properties of regular descriptors over the inputs $\theta, x:p$ and the output $y:p$.\\
We first check that the raw descriptor is well defined. The assignment component is the empty set,
$\varnothing \in \mc P_{\mathrm{fin}}(\mathrm{Asg}_{\Sigma}(\theta, x:p)),$
and since $x$ is an input variable in the input context and $y$ is an output variable in the output context,
$\{x \mapsto y\} \in \mc P_{\mathrm{fin}}(\mathrm{Rdr}_{\Sigma}(\theta, x:p \ ;\ y:p)).$
Therefore
$$(\varnothing \mid \{x \mapsto y\}) \in \mathrm{desc}^{\mathrm{raw}}_{\Sigma}(\theta, x:p \ ;\ y:p).$$
We now check the regularity conditions.
\begin{itemize}
	\item \textbf{Mode Validity.} The descriptor has no internal variables. The only possible mode assignment is therefore the empty function
    $$\mu:\varnothing\to \mathrm{Mode}(\Sigma).$$
    The only redirection is $x\mapsto y$, and it is mode-valid since $x:p$ is in the input context and $y:p$ in the output context. Therefore $(\varnothing\mid\{x\mapsto y\})$ is mode-valid.
  \item \textbf{Irreflexive reachability.} The only generating edge is
  $$x \leadsto_{(\varnothing \mid \{x \mapsto y\})} y,$$
  coming from the redirection $x \mapsto y$. Since $y$ is an output variable, no edge originates from $y$, and no edges reach $x$. The reachability relation $\rightsquigarrow$ therefore contains only the pair $(x, y)$ and is irreflexive.

  \item \textbf{Unique internal target occurrence.} There are no assignments, so $\mathrm{Int}(\varnothing) = \varnothing$. The condition is vacuously satisfied.

  \item \textbf{Grounded redirections.} Since $x$ is an input variable and $y$ is an output variable, the redirection $x \mapsto y$ contains no internal variables, hence $\mathrm{Int}(\{x \mapsto y\}) = \varnothing \subseteq \varnothing = \mathrm{Int}(\varnothing)$.

  \item \textbf{Unique internal source occurrence.} There are no internal variables, so the condition is vacuously satisfied.

  \item \textbf{Boundary usage property.} If $p$ is linear, the input $x$ occurs in exactly one source position, namely in the redirection $x\mapsto y$. The remaining inputs are the cartesian declarations of $\theta$, which are unrestricted. If $p$ is cartesian, every input is cartesian and unrestricted. The unique output $y$ is the target of exactly one redirection.
\end{itemize}
Thus $(\varnothing \mid \{x \mapsto y\})$ is regular, and so is the class $[\varnothing \mid \{x \mapsto y\}]$ generated by the rule. The transports $(\pi;\tau)_*$ in the conclusion are well defined on classes by \cref{def:permutation-transport-of-descriptor-classes}, and any transported identity descriptor is regular by \cref{prop:permutation-transport-preserves-regularity}.
\hfill $\square$

\begin{definition}[Well-Formed Descriptor (Cartesian Composition)]\label{def:well-formed-descriptor-cartesian}
Suppose we are given a pre-signature $\Sigma \ sig$, a cartesian context $\theta \
\mathrm{ctx}^{\mathrm{c}}_{\Sigma}$, pairwise distinct output variables $(y_i : p_i)_{1 \leq i \leq n}$, an output variable $r : q$ distinct from the $y_i$, and a c-generator $f : p_1, \dots, p_n \to q \in \Sigma$. Let $(\mf D_i)_{1 \leq i \leq n}$ be descriptor classes appearing in premise judgments
$$
\theta \Vdash_{\mf D_i} y_i:p_i \quad \text{such that} \quad
|(\mc T_i)_{\mapsto y_i}| = 1 \quad \text{for every } i.$$
Fix representatives $(\mc S_i \mid \mc T_i) \in \mf D_i$ such that $\mathrm{Int}(\mc S_i\mid\mc T_i) \cap \mathrm{Int}(\mc S_j\mid\mc T_j) = \varnothing$ for all $i \neq j$ (by \cref{lem:fresh-representatives}). Each $\mc T_i$ is then the single redirection $\{s_i \mapsto y_i\}$. For a fresh internal variable $w_R$, the descriptor
$$
	\mf D := \left[\bigcup_{i=1}^n \mc S_i \cup \{s_1, \dots, s_n \xmapsto{f} w_R\} \mid
  \{ w_R \mapsto r  \} \right]
$$
is a \emph{well-formed descriptor} for the sequent $\theta \Vdash r : q$.
The corresponding inference rule is:
$$
\inferrule*[Right=c-Comp]
{
  f : p_1, \ldots, p_n \to q \in \Sigma \\
  \big(\theta \Vdash_{\mf D_i} y_i : p_i\big)_{i=1}^n \\
	|(\mc T_i)_{\mapsto y_i}| = 1 \quad \text{for every } i \\
  \pi:\theta\cong\theta' \\
}
{
  \theta' \Vdash_{(\pi;\1_{r : q})_*\mf D} r : q
}
$$
\end{definition}

\textit{Remark:} When $n=0$, there are no premises and $f:\cdot\to q$ is a constant. c-Comp creates one fresh internal variable $w_R$, one assignment $\cdot\xmapsto{f}w_R$, and one redirection $w_R\mapsto r$. The distinctness conditions on the $y_i$ and $r$ are hygiene only, as the $y_i$ do not occur in the produced descriptor. 

\begin{example}[Linear--Nonlinear Logic]
Consider the pre-signature including the binary c-generator $\otimes_c : c, c \to c$, and the cartesian context $\theta := a : c,\, a' : c$. The identity rule provides the premises
$$\theta \Vdash_{[\varnothing \mid \{a \mapsto y_1\}]} y_1 : c \qquad \text{and} \qquad \theta \Vdash_{[\varnothing \mid \{a' \mapsto y_2\}]} y_2 : c,$$
so that $s_1 = a$ and $s_2 = a'$. For a fresh internal variable $w$, the rule derives
$$\theta \Vdash_{[\{a, a' \xmapsto{\otimes_c} w\} \mid \{w \mapsto y\}]} y : c,$$
which is the precise result used for the binary c-generator examples throughout this chapter. If we instead consider the two premises $\theta' \Vdash_{[\varnothing \mid \{a \mapsto y_i\}]} y_i : c$ for $i = 1, 2$ over the context $\theta' := a : c$, we get the descriptor $[\{a, a \xmapsto{\otimes_c} w\} \mid \{w \mapsto y\}]$ with dependent tree $\llbracket y \rrbracket = \otimes_c(a, a)$. We note that here the cartesian input $a$ is used twice.
\end{example}

\begin{proposition}[c-Comp is well defined]\label{prop:c-comp-is-well-defined}
The descriptor class $\mf D$ produced by the c-Comp construction is independent of the chosen representatives $(\mc S_i \mid \mc T_i) \in \mf D_i$ and of the chosen fresh variable $w_R$.
\end{proposition}

\textit{Proof.}
We first verify that the representative $(\mc S \mid \mc T)$ produced by the rule is well defined. Each $(\mc S_i \mid \mc T_i)$ is a representative of a descriptor $\mf D_i \in \mathrm{Desc}_\Sigma(\theta \ ;\ y_i:p_i)$, so we have
$$
\mc S_i \in \mc P_{\mathrm{fin}}(\mathrm{Asg}_\Sigma(\theta))
\qquad\text{and}\qquad
\mc T_i \in \mc P_{\mathrm{fin}}(\mathrm{Rdr}_\Sigma(\theta \ ;\ y_i:p_i)).
$$
The sources $s_i$ of the new assignment are valid since each $s_i$ is the source of a redirection in $\mc T_i$, hence is either an input variable in $\theta$ or an internal variable. The hypothesis of the rule $|(\mc T_i)_{\mapsto y_i}|=1$, together with the fact that every redirection of $\mc T_i$ targets the unique output $y_i$, then makes each $\mc T_i$ the singleton $\{s_i\mapsto y_i\}$, so
$$
s_1,\ldots,s_n\xmapsto{f}w_R
$$
is a single assignment with fresh internal target $w_R$. Adjoining it to the finitely many old assignments keeps the collection finite, so $\mc S\in \mc P_{\mathrm{fin}}(\mathrm{Asg}_\Sigma(\theta))$. The redirection $w_R\mapsto r$ is valid since $w_R$ is internal and $r$ is an output variable, so $\mc T = \{w_R\mapsto r\}\in \mc P_{\mathrm{fin}}(\mathrm{Rdr}_\Sigma(\theta \ ;\ r:q))$, and $(\mc S\mid \mc T)\in \mathrm{desc}^{\mathrm{raw}}_\Sigma(\theta \ ;\ r:q)$.

The choice of representative is inconsequential. Given two representative families related by $\alpha$-renamings $\phi_i$ and two fresh variables $w_R, w_R'$, the map $\Theta$ consisting of the $\phi_i$ on old internal variables and sending $w_R\mapsto w_R'$ is a bijection on internal variables by freshness and disjointness. The condition $|(\mc T_i)_{\mapsto y_i}|=1$ is $\alpha$-invariant (\cref{lem:internal-renaming-preserves-redirections-targeting-an-output}). Each $\phi_i$ therefore carries the unique redirection $s_i\mapsto y_i$ to the unique redirection $s_i'\mapsto y_i$. The new assignment $s_1,\ldots,s_n\xmapsto{f}w_R$ is sent to $s_1',\ldots,s_n'\xmapsto{f}w_R'$ and the new redirection $w_R\mapsto r$ to $w_R'\mapsto r$. Hence the two constructed representatives are $\alpha$-equivalent and the descriptor class is the same. \hfill $\square$

\begin{proposition}[c-Comp Preserves Regularity]\label{prop:c-comp-preserves-regularity}
If all the descriptors $\mf D_i$ are regular, then the descriptor class produced by c-Comp is regular.
\end{proposition}

\textit{Proof.}
Choose regular representatives $(\mc S_i\mid\mc T_i)\in\mf D_i$ with pairwise disjoint internal variables and construct $(\mc S\mid\mc T)$ as in \cref{def:well-formed-descriptor-cartesian}. Regularity is $\alpha$-invariant. It therefore suffices to prove that $(\mc S\mid\mc T)$ is regular.
\begin{itemize}
  \item \textbf{Mode Validity.} Let $\mu_i$ witness mode-validity of $(\mc S_i\mid\mc T_i)$. The $\mu_i$ combine into one assignment on the old internal variables. Extend it by $\mu(w_R):=q$. Old assignments retain mode-validity. For the new assignment $s_1,\ldots,s_n\xmapsto{f}w_R$ with $f:p_1,\ldots,p_n\to q$, notice that each source $s_i$ occurs in the redirection $s_i\mapsto y_i$ of $\mc T_i$ with $y_i:p_i$. Mode-validity of the premise then gives $\mathrm{mode}_{\mu_i}(s_i)=p_i$, and $\mu(w_R)=q$ as required. The new redirection $w_R\mapsto r$ is mode-valid since $r:q$ and $\mu(w_R)=q$.
  \item \textbf{Irreflexive reachability.} For each $i$, set $\leadsto_i:=\leadsto_{(\mc S_i\mid\varnothing)}$, which has irreflexive transitive closure by \cref{lem:removing-assignments-or-redirections-cannot-create-cycles}. The components have pairwise disjoint internal variables and share only input variables of $\theta$. No $\leadsto_i$ targets an input or starts from an output (\cref{lem:inputs-have-no-incoming-paths-and-outputs-have-no-outgoing-paths}). $\leadsto_0:=\bigcup_i\leadsto_i$ then has irreflexive transitive closure by \cref{lem:acyclicity-is-preserved-under-unions-of-components-sharing-only-boundary-variables}. The full relation is
  $$\leadsto_0\cup\{s_i\leadsto w_R\mid 1\leq i\leq n\}\cup\{w_R\leadsto r\},$$
  and \cref{lem:fresh-output-cones-preserve-irreflexivity} with $W=\{w_R\}$ and $P_{w_R}=\{s_1,\ldots,s_n\}$ shows it has irreflexive transitive closure.
  \item \textbf{Unique internal target occurrence.} An old internal variable keeps its unique producer inside its own component by disjointness, and $w_R$ is produced only by the new assignment, which is the only new one. Pairwise distinctness of targets inside assignments is inherited, and the new assignment has the single target $w_R$.
  \item \textbf{Grounded redirections.} $\mathrm{Int}(\mc T)=\{w_R\}$, and $w_R\in\mathrm{Int}(\mc S)$ via the new assignment.
  \item \textbf{Unique internal source occurrence.} Let $u$ be an internal variable of the composite. If $u=s_i$ for some $i$ (a distinguished source that is internal), then $u$ loses its unique source occurrence $s_i\mapsto y_i$ (the redirection was removed) and gains exactly one source occurrence in the new assignment. Its count is again one. Every other old internal variable keeps its source occurrences unchanged, and they lie inside a single component by disjointness. The fresh variable $w_R$ occurs in exactly one source position, in the redirection $w_R\mapsto r$.
  \item \textbf{Boundary usage property.} All inputs of the conclusion are the cartesian declarations of $\theta$, which are unrestricted. The unique output $r$ is the target of exactly one redirection.
\end{itemize}
Therefore the c-Comp construction preserves regularity. \hfill $\square$

\begin{definition}[Well-Formed Descriptor (Positive)]\label{def:well-formed-descriptor-positive}
Suppose we are given a pre-signature $\Sigma \ sig$, a cartesian context $\theta \
\mathrm{ctx}^{\mathrm{c}}_{\Sigma}$, an $\ell$-generator $g : (p_1, \dots, p_k \ \mid \ l_1, \dots, l_m) \to l'_1, \dots, l'_n \in \Sigma$, and premise data as follows:
\begin{itemize}
\item for each cartesian position $1\leq i\leq k$, a cartesian premise judgment
$$\theta \Vdash_{\mf C_i} b_i:p_i \quad \text{with} \quad |(\mc T^{\mf C_i})_{\mapsto b_i}| = 1,$$
with the $b_i$ pairwise distinct output variables;
\item for each linear position $1\leq j\leq m$, a linear premise judgment
$$\theta \ ; \ \gamma_j \Vdash_{\mf D_j} \delta_j, y_j:l_j \quad \text{with} \quad |(\mc T^{\mf D_j})_{\mapsto y_j}| = 1,$$
where the linear contexts $\gamma_j$ are pairwise disjoint, the output contexts $\delta_j$ are pairwise disjoint, and the $y_j$ are pairwise distinct and fresh in $\delta:=\delta_1,\ldots,\delta_m$;
\item pairwise distinct output variables $r_1:l'_1,\ldots,r_n:l'_n$ fresh in $\delta$.
\end{itemize}
Fix representatives $(\mc S^{\mf C_i} \mid \mc T^{\mf C_i}) \in \mf C_i$ and $(\mc S^{\mf D_j} \mid \mc T^{\mf D_j}) \in \mf D_j$ with pairwise disjoint internal variables (by \cref{lem:fresh-representatives}). Write $c_i$ for the unique source with $(c_i\mapsto b_i)\in\mc T^{\mf C_i}$ and $s_j$ for the unique source with $(s_j\mapsto y_j)\in\mc T^{\mf D_j}$. For pairwise distinct fresh internal variables $w_1,\ldots,w_n$, the descriptor
$$
\mf D := \left[\begin{aligned}
  &\bigcup_{i=1}^k \mc S^{\mf C_i} \cup \bigcup_{j=1}^m \mc S^{\mf D_j} \cup \{(c_1, \dots, c_k \ \mid \ s_1, \dots, s_m) \xmapsto{g} w_1,\dots,w_n\} \\
  &{}\mid \bigcup_{j=1}^m (\mc T^{\mf D_j})_{\not\mapsto y_j} \cup \{ w_1 \mapsto r_1,\ldots,w_n\mapsto r_n  \}
\end{aligned}\right]
$$
is a \emph{well-formed descriptor} for the sequent $\theta \ ; \ \gamma_1,\ldots,\gamma_m \Vdash \delta, r_1:l'_1,\ldots,r_n:l'_n$.
The corresponding inference rule is:
$$
\inferrule*[right=l-Pos-Comp]
{
  g : (p_1, \ldots, p_k \ \mid \ l_1, \ldots, l_m) \to l'_1,\ldots,l'_n \in \Sigma \\
  \big(\theta \Vdash_{\mf C_i} b_i : p_i\big)_{i=1}^k \\
  \big(\theta \ ; \ \gamma_j \Vdash_{\mf D_j} \delta_j, y_j : l_j\big)_{j=1}^m \\
  \gamma_1,\ldots,\gamma_m \ \text{pairwise disjoint} \\
  \delta_1,\ldots,\delta_m \ \text{pairwise disjoint},\ \ \delta := \delta_1,\ldots,\delta_m \\
  y_1,\ldots,y_m \ \text{pairwise distinct and fresh in } \delta \\
  r_1,\ldots,r_n \ \text{pairwise distinct and fresh in } \delta \\
	|(\mc T^{\mf C_i})_{\mapsto b_i}| = 1, \quad |(\mc T^{\mf D_j})_{\mapsto y_j}| = 1 \\
  \pi:\theta,\gamma_1,\ldots,\gamma_m\cong\gamma' \\
  \tau:\delta,r_1:l'_1,\ldots,r_n:l'_n\cong\delta'
}
{
  \gamma' \Vdash_{(\pi;\tau)_*\mf D} \delta'
}
$$
\end{definition}

\textit{Remark:} Notice that the cartesian premises share $\theta$ with each other and with the linear premises, while the linear contexts $\gamma_j$ and the side outputs $\delta_j$ are multiplicative and disjoint among the premises. When $n=0$, the new assignment has an empty target set, no new redirections are created, and the outputs of the conclusion are exactly $\delta$. When $m=0$, there are no linear premises and the linear input context of the conclusion is empty.

\begin{example}[Linear--Nonlinear Logic]
Taking the premises generated by the identity rules
$$\cdot \ ; \ x : l \Vdash_{[\varnothing \mid \{x \mapsto y_1\}]} y_1 : l \qquad \text{and} \qquad \cdot \ ; \ x' : l \Vdash_{[\varnothing \mid \{x' \mapsto y_2\}]} y_2 : l,$$ and taking the $\ell$-generator $\otimes_l : l, l \to l$ in the rule with $k=0$, $m=2$, $n=1$, we get the sequent
$$\cdot \ ; \ x : l,\, x' : l \Vdash_{[\{x, x' \xmapsto{\otimes_l} w\} \mid \{w \mapsto y\}]} y : l$$
for a fresh internal variable $w$. The sources are $s_1 = x$ and $s_2 = x'$, and the side outputs are empty. For the $\ell$-generator $g : (c\, \mid \,\cdot)\to l$, the cartesian premise $a : c \Vdash_{[\varnothing \mid \{a \mapsto b\}]} b : c$ with $k=1$, $m=0$, $n=1$ gives the sequent
$$a : c \ ; \ \cdot \Vdash_{[\{(a \ \mid \ \cdot) \xmapsto{g} w\} \mid \{w \mapsto y_1\}]} y_1 : l,$$
whose linear input context is empty. Taking $\otimes_l$ in the rule a second time, with this sequent and the identity premise $a : c \ ; \ x : l \Vdash_{[\varnothing \mid \{x \mapsto y_2\}]} y_2 : l$ as the linear premises, we get
$$a : c \ ; \ x : l \Vdash_{[\{(a \ \mid \ \cdot) \xmapsto{g} w,\ \ w, x \xmapsto{\otimes_l} w'\} \mid \{w' \mapsto y\}]} y : l.$$
The two premises share the cartesian context $a : c$, while their linear contexts $\cdot$ and $x : l$ are disjoint.
\end{example}

\begin{proposition}[Pos-Comp is well defined]\label{prop:pos-comp-is-well-defined}
The descriptor class $\mf D$ produced by the $\ell$-Pos-Comp construction is independent of the chosen representatives and of the chosen fresh variables $w_1,\ldots,w_n$.
\end{proposition}

\textit{Proof.}
We first verify that the raw representative is well defined. Each $\mc S^{\mf C_i}$ lies in $\mc P_{\mathrm{fin}}(\mathrm{Asg}_\Sigma(\theta))$ and each $\mc S^{\mf D_j}$ in $\mc P_{\mathrm{fin}}(\mathrm{Asg}_\Sigma(\theta,\gamma_j))$, hence all lie in $\mc P_{\mathrm{fin}}(\mathrm{Asg}_\Sigma(\theta,\gamma_1,\ldots,\gamma_m))$. The sources $c_i$ and $s_j$ of the new assignment are sources of redirections of the premises, meaning that they are input or internal variables. The targets $w_1,\ldots,w_n$ are fresh pairwise distinct internal variables. The retained redirections in $(\mc T^{\mf D_j})_{\not\mapsto y_j}$ target outputs in $\delta_j\subseteq\delta$, and the new redirections $w_j\mapsto r_j$ are valid. Hence the constructed pair is a raw descriptor over $(\theta,\Gamma_1,\ldots,\Gamma_m \ ; \ \delta,r_1:l'_1,\ldots,r_n:l'_n)$.

We now prove independence of the choices. Given two representative families related by $\alpha$-renamings $\phi^{\mf C_i}, \phi^{\mf D_j}$ and fresh families $w_j$, $w'_j$, define $\Theta$ to consist of the $\phi$'s on the old internal variables and $w_j\mapsto w'_j$ on the fresh ones. By disjointness and freshness, $\Theta$ is a bijection on internal variables. The conditions $|(\mc T^{\mf C_i})_{\mapsto b_i}|=1$ and $|(\mc T^{\mf D_j})_{\mapsto y_j}|=1$ are $\alpha$-invariant (\cref{lem:internal-renaming-preserves-redirections-targeting-an-output}). Each renaming therefore carries the distinguished redirection (hence the distinguished source) of one representative to that of the other. The new assignment and the new redirections correspond under $\Theta$, and the two constructed representatives are $\alpha$-equivalent. \hfill $\square$

\begin{proposition}[Pos-Comp Preserves Regularity]\label{prop:pos-comp-preserves-regularity}
Assume the descriptor classes $\mf C_i$ and $\mf D_j$ are regular. Then the descriptor class produced by $\ell$-Pos-Comp is regular.
\end{proposition}

\textit{Proof.}
Choose regular representatives with pairwise disjoint internal variables and construct $(\mc S\mid\mc T)$ as in \cref{def:well-formed-descriptor-positive}.
\begin{itemize}
  \item \textbf{Mode Validity.} Combine the premise mode assignments and set $\mu(w_j):=l'_j$. The sources of the new assignment have the required modes. Indeed, $c_i$ occurs in the redirection $c_i\mapsto b_i$ with $b_i:p_i$, so $\mathrm{mode}(c_i)=p_i$. Likewise $\mathrm{mode}(s_j)=l_j$ from $s_j\mapsto y_j$ with $y_j:l_j$. The targets receive $\mathrm{mode}(w_j)=l'_j$ as required, and the new redirections $w_j\mapsto r_j$ match $r_j:l'_j$. Old assignments and retained redirections are mode-valid as in their premises.
  \item \textbf{Irreflexive reachability.} The components $(\mc S^{\mf C_i}\mid\varnothing)$ and $(\mc S^{\mf D_j}\mid(\mc T^{\mf D_j})_{\not\mapsto y_j})$ have irreflexive transitive closures by \cref{lem:removing-assignments-or-redirections-cannot-create-cycles}. They have pairwise disjoint internal variables and share only boundary variables. Their union $\leadsto_0$ therefore has irreflexive transitive closure by \cref{lem:acyclicity-is-preserved-under-unions-of-components-sharing-only-boundary-variables}. The new assignment and redirections contribute the edges $\{c_i\leadsto w_j\}\cup\{s_{j'}\leadsto w_j\}\cup\{w_j\leadsto r_j\}$. Applying \cref{lem:fresh-output-cones-preserve-irreflexivity} $n$ times, once for each $j$ with $W=\{w_j\}$, $P_{w_j}=\{c_1,\ldots,c_k,s_1,\ldots,s_m\}$, preserves irreflexivity at each step since each $w_j$ and $r_j$ is fresh for the relation built so far.
  \item \textbf{Unique internal target occurrence.} Old internal variables keep their unique producers inside their own components. Each $w_j$ is produced only by the new assignment, at a single position since the $w_j$ are pairwise distinct.
  \item \textbf{Grounded redirections.} Internal variables of retained redirections lie in their premise components. The internal variables of the new redirections are the $w_j$, which are targets of the new assignment.
  \item \textbf{Unique internal source occurrence.} The single source occurrence of each variable $c_i$ or $s_j$ disappears in the removed redirection set. The variable gets a single source position in the new assignment. Every other old internal variable keeps its status inside its own component. Each $w_j$ occurs in exactly one source position, in $w_j\mapsto r_j$.
  \item \textbf{Boundary usage property.} The linear inputs of the conclusion are those of the $\gamma_j$. Each occurs in exactly one source position inside its own premise component (premise regularity), and the construction preserves the count. If the distinguished source $s_j$ is itself a linear input, its occurrence moves from the removed redirection $s_j\mapsto y_j$ to the new assignment; otherwise, no occurrence is removed or added. Cartesian inputs are unrestricted. Each output of $\delta_j$ keeps its unique redirection (only the redirections to $y_j$ were removed, and $y_j\notin\delta_j$), and each new output $r_j$ is the target of exactly one redirection.
\end{itemize}
Therefore the $\ell$-Pos-Comp construction preserves regularity. \hfill $\square$

\begin{definition}[Well-Formed Descriptor (Negative)]\label{def:well-formed-descriptor-negative}
Suppose we are given a pre-signature $\Sigma \ sig$, a cartesian context $\theta \
\mathrm{ctx}^{\mathrm{c}}_{\Sigma}$, an $\ell$-generator $g : (p_1, \dots, p_k \ \mid \ l_1, \dots, l_m) \to l'_1, \dots, l'_n \in \Sigma$, and premise data as follows:
\begin{itemize}
\item for each cartesian position $1\leq i\leq k$, a cartesian premise judgment
$$\theta \Vdash_{\mf C_i} b_i:p_i \quad \text{with} \quad |(\mc T^{\mf C_i})_{\mapsto b_i}| = 1,$$
with the $b_i$ pairwise distinct;
\item for each target position $1\leq j\leq n$, a linear premise judgment
$$\theta \ ; \ \gamma_j, x_j:l'_j \Vdash_{\mf D_j} \delta_j \quad \text{with} \quad x_j \in \mathrm{Inp}(\mf D_j),$$
where the linear contexts $\gamma_j$ are pairwise disjoint, the $x_j$ are pairwise distinct and fresh in $\gamma_1,\ldots,\gamma_n$, and the output contexts $\delta_j$ are pairwise disjoint, with $\delta:=\delta_1,\ldots,\delta_n$;
\item pairwise distinct fresh linear input variables $z_1:l_1,\ldots,z_m:l_m$.
\end{itemize}
Fix representatives with pairwise disjoint internal variables (by \cref{lem:fresh-representatives}), write $c_i$ for the unique source with $(c_i\mapsto b_i)\in\mc T^{\mf C_i}$, and choose pairwise distinct fresh internal variables $w_1,\ldots,w_n$ disjoint from all internal variables of the representatives. The descriptor
	\begin{align*}
  	\mf D := \left[\bigcup_{i=1}^k \mc S^{\mf C_i} \cup \bigcup_{j = 1}^n \mc S^{\mf D_j}[x_j := w_j] \cup \{(c_1,\ldots,c_k \ \mid \ z_1,\ldots,z_m) \xmapsto{g} w_1, \dots, w_n\} \ \middle| \ \bigcup_{j = 1}^n \mc T^{\mf D_j}[x_j := w_j]\right]
	\end{align*}
	is a \emph{well-formed descriptor} for the sequent $\theta \ ; \ z_1:l_1,\ldots,z_m:l_m,\gamma_1,\ldots,\gamma_n \Vdash \delta$.
		The corresponding inference rule is:
$$
\inferrule*[right=l-Neg-Comp]
{
  g : (p_1,\ldots,p_k \ \mid \ l_1,\ldots,l_m) \to l'_1, \ldots, l'_n \in \Sigma \\
  \big(\theta \Vdash_{\mf C_i} b_i : p_i\big)_{i=1}^k \\
  \big(\theta \ ; \ \gamma_j, x_j : l'_j \Vdash_{\mf D_j} \delta_j\big)_{j=1}^n \\
  \gamma_1,\ldots,\gamma_n \ \text{pairwise disjoint} \\
  \delta_1,\ldots,\delta_n \ \text{pairwise disjoint},\ \ \delta := \delta_1,\ldots,\delta_n \\
  x_1,\ldots,x_n \ \text{pairwise distinct and fresh in } \gamma_1,\ldots,\gamma_n \\
  z_1,\ldots,z_m \ \text{pairwise distinct and fresh} \\
	|(\mc T^{\mf C_i})_{\mapsto b_i}| = 1, \quad x_j\in\mathrm{Inp}(\mf D_j) \ \text{for every } j \\
  \pi:\theta,z_1:l_1,\ldots,z_m:l_m,\gamma_1,\ldots,\gamma_n\cong\gamma' \\
  \tau:\delta\cong\delta'
}
{
  \gamma' \Vdash_{(\pi;\tau)_*\mf D} \delta'
}
$$
\end{definition}

\textit{Remark:} When $n=0$, there are no linear premises and the $\ell$-generator has no targets. $\ell$-Neg-Comp then creates one assignment $(c_1,\ldots,c_k \ \mid \ z_1,\ldots,z_m)\xmapsto{g}\cdot$ with an empty target set and no redirections. For $k=0$, $m=1$, this is the co-nullary case, e.g.,\ $\bot_l:l\to\cdot$. The hypotheses $x_j\in\mathrm{Inp}(\mf D_j)$ and $|(\mc T^{\mf C_i})_{\mapsto b_i}|=1$, like the counting hypotheses of c-Comp and $\ell$-Pos-Comp, are automatic by the regularity theorem \cref{thm:well-formed-descriptors-are-regular}. The boundary usage property gives each linear input exactly one occurrence and each output exactly one redirection. They are kept explicit so that the rules are self-contained.

\begin{example}[Multiplicative disjunction]
In the \hyperref[ex:linear-logic-mode-signature]{running example of linear logic}, the descriptor responsible for the multiplicative disjunction is generated by the $\ell$-generator $\parr : l \to l,l$ that takes a linear resource and splits it in two. Taking this generator in the rule with $k=0$, $m=1$, $n=2$, and the identity premises $\cdot \ ; \ x_1 : l \Vdash_{[\varnothing \mid \{x_1 \mapsto x\}]} x : l$ and $\cdot \ ; \ x_2 : l \Vdash_{[\varnothing \mid \{x_2 \mapsto y\}]} y : l$, we get the sequent
$$\cdot \ ; \ z : l \Vdash_{[\{z \xmapsto{\parr} w_1, w_2\} \mid \{w_1 \mapsto x, w_2 \mapsto y\}]} x : l, y : l$$
which can be read informally as the descriptor $x \ \parr \ y$.
\end{example}

\begin{proposition}[Neg-Comp is well defined]\label{prop:neg-comp-is-well-defined}
The descriptor class $\mf D$ produced by the $\ell$-Neg-Comp construction is independent of the chosen representatives and of the chosen fresh variables $w_j$.
\end{proposition}

\textit{Proof.}
We first verify that the raw representative is well defined. After substituting $x_j:=w_j$, the component $(\mc S^{\mf D_j}\mid\mc T^{\mf D_j})[x_j:=w_j]$ uses only input variables from $\theta,\gamma_j$ and internal variables, and its redirections target $\delta_j$. The new assignment $(c_1,\ldots,c_k \ \mid \ z_1,\ldots,z_m)\xmapsto{g}w_1,\ldots,w_n$ is valid since the $c_i$ are input or internal variables, the $z_j$ are input variables, and the $w_j$ are pairwise distinct fresh internal variables. Hence the constructed pair is a raw descriptor over $(\theta,z_1:l_1,\ldots,z_m:l_m,\gamma_1,\ldots,\gamma_n \ ; \ \delta)$.

The choice of the fresh variables $w_j$ is likewise inconsequential. The bijection sending $w_j\mapsto w_j'$ and fixing all other internal variables realizes the $\alpha$-equivalence, using \cref{lem:input-substitution-and-internal-renaming-commute} on each substituted component and the fact that internal renaming preserves unions (\cref{lem:elementwise-operations-preservation}).

It remains to prove independence with respect to the representatives. Given families related by $\alpha$-renamings $\phi^{\mf C_i}$, $\phi^{\mf D_j}$ and fresh families $w_j$, $\widehat w_j$, define
$$
\Phi(u)=
\begin{cases}
\phi^{\mf C_i}(u) & \text{if }u\in\mathrm{Int}(\mc S^{\mf C_i}\mid\mc T^{\mf C_i}),\\
\phi^{\mf D_j}(u) & \text{if }u\in\mathrm{Int}(\mc S^{\mf D_j}\mid\mc T^{\mf D_j}),\\
\widehat w_j & \text{if }u=w_j.
\end{cases}
$$
This is a well-defined bijection on internal variables by disjointness and freshness. The conditions $x_j\in\mathrm{Inp}(\mf D_j)$ hold for both families by \cref{lem:input-occurrence-is-alpha-invariant}. $\alpha$-invariance of $|(\mc T^{\mf C_i})_{\mapsto b_i}|=1$ (\cref{lem:internal-renaming-preserves-redirections-targeting-an-output}) then gives that each $\phi^{\mf C_i}$ carries the distinguished source $c_i$ to its counterpart. Input-substitution/internal-renaming commutation gives $\big((\mc S^{\mf D_j}\mid\mc T^{\mf D_j})[x_j:=w_j]\big)^\Phi=(\mc S'^{\mf D_j}\mid\mc T'^{\mf D_j})[x_j:=\widehat w_j]$, and the new assignment is carried to its counterpart. Hence the two constructed representatives are $\alpha$-equivalent. \hfill $\square$

\begin{proposition}[Neg-Comp Preserves Regularity]\label{prop:neg-comp-preserves-regularity}
If the descriptor classes $\mf C_i$ and $\mf D_j$ are regular, then the descriptor class produced by $\ell$-Neg-Comp is regular.
\end{proposition}

\textit{Proof.}
Choose regular representatives with pairwise disjoint internal variables and construct $(\mc S\mid\mc T)$ as in \cref{def:well-formed-descriptor-negative}.
\begin{itemize}
  \item \textbf{Mode Validity.} Combine the premise assignments and set $\mu(w_j):=l'_j$. In the $j$th substituted component, every occurrence of $x_j:l'_j$ was replaced by $w_j$ with $\mu(w_j)=l'_j$, so all old assignments and redirections remain mode-valid. The new assignment is mode-valid as well: $\mathrm{mode}(c_i)=p_i$ from the premise redirections $c_i\mapsto b_i$, the fresh inputs carry $z_j:l_j$ from the input context, and $\mu(w_j)=l'_j$ matches the targets of $g$.
  \item \textbf{Irreflexive reachability.} The substituted components have irreflexive transitive closure by \cref{lem:fresh-input-substitution-preserves-irreflexivity}, and the cartesian components $(\mc S^{\mf C_i}\mid\varnothing)$ by \cref{lem:removing-assignments-or-redirections-cannot-create-cycles}. All components have pairwise disjoint internal variables and share only boundary variables. The union $\leadsto_0$ therefore has irreflexive transitive closure by \cref{lem:acyclicity-is-preserved-under-unions-of-components-sharing-only-boundary-variables}. The new assignment contributes the edges $\{z_j\leadsto w_{j'}\}$ and $\{c_i\leadsto w_{j'}\}$. Suppose for contradiction that the full relation has a non-empty cycle. The $z_j$ are inputs and inputs have no incoming edges, so no $z_j$ lies on a cycle. A cycle avoiding new edges would lie in $\leadsto_0$, which is impossible. The cycle then uses an edge $c_i\leadsto w_{j'}$ and must return from $w_{j'}$ to $c_i$. Any path out of $w_{j'}$ proceeds through edges of the $j'$th substituted component. Its targets are internal variables of that component or outputs of $\delta_{j'}$; outputs have no outgoing edges, and no path can pass through an input variable since inputs have no incoming edges. The internal variables of distinct components are disjoint and $c_i$ is an internal variable of the $i$th cartesian component or an input. No path from $w_{j'}$ therefore reaches $c_i$. This contradiction shows the full relation has irreflexive transitive closure.
  \item \textbf{Unique internal target occurrence.} Old internal variables keep their unique producers inside their own components. The substitution $[x_j:=w_j]$ does not affect assignment targets since $x_j$ is an input. Each $w_j$ is produced only by the new assignment, at a single position. Pairwise distinctness of targets inside assignments is inherited and holds for the new assignment by the choice of the $w_j$.
  \item \textbf{Grounded redirections.} An internal variable of $\mc T$ lies in some substituted component. If it is not a $w_j$, it lies in $\mathrm{Int}(\mc S^{\mf D_j})$ by premise regularity, and if it is $w_j$, it is a target of the new assignment.
  \item \textbf{Unique internal source occurrence.} Old internal variables keep their source-occurrence counts inside their own components (substitution replaces only $x_j$, an input). Each variable $c_i$ disappears in the removed redirection $c_i\mapsto b_i$. It gains one source position in the new assignment. Each $w_j$ occurs in the source positions previously occupied by $x_j$. $x_j$ is a linear input of the regular premise. The boundary usage property then gives it exactly one source occurrence; hence $w_j$ has exactly one source occurrence.
  \item \textbf{Boundary usage property.} The fresh linear inputs $z_j$ occur in exactly one source position each, in the new assignment. The linear inputs of each $\Gamma_j$ keep their unique occurrences inside their own components. Cartesian inputs are unrestricted. Every output of $\delta_j$ keeps its unique redirection, since substitution does not change redirection targets.
\end{itemize}
Therefore the $\ell$-Neg-Comp construction preserves regularity.\hfill $\square$

\begin{proposition}[Sorts of well-formed descriptors]\label{prop:sorts-of-well-formed-descriptors}
Every derivable judgment $\gamma\Vdash_{\mf D}\delta$ has exactly one of two shapes:
\begin{itemize}
\item[(i)] \emph{cartesian}: every declaration of $\gamma$ is at a cartesian mode, $\delta$ is a single declaration $b:q$ at a cartesian mode, and $\mf D$ contains no variable at a linear mode at all;
\item[(ii)] \emph{linear}: every declaration of $\delta$ is at a linear mode.
\end{itemize}
\end{proposition}
\textit{Proof.}
By induction on the generating rules. For Identity, the ambient context $\theta$ is cartesian. If $p$ is cartesian, the conclusion is of shape (i) (the descriptor $(\varnothing\mid\{x\mapsto y\})$ has no linear variables), and if $p$ is linear, it is of shape (ii). For c-Comp, each premise has the single cartesian output $y_i:p_i$, hence is of shape (i) by the induction hypothesis. In particular, each premise input context is cartesian and each premise descriptor contains no linear variables. The conclusion has input context $\theta$, the single cartesian output $r:q$, and a descriptor which adds only the c-generator assignment and one redirection between cartesian variables. It is of shape (i). For $\ell$-Pos-Comp, the new outputs are at linear modes by the arities of $g$, and every linear premise has the linear output $y_j:l_j$, hence is of shape (ii) by the induction hypothesis. The retained side outputs are therefore linear and the conclusion is of shape (ii). For $\ell$-Neg-Comp, the outputs of the conclusion are the retained $\delta_j$, and the linear premises may have empty output contexts. Each declares the linear input $x_j:l'_j$, so it is not of shape (i), and the induction hypothesis then makes it of shape (ii). Every $\delta_j$ is therefore linear and the conclusion is of shape (ii). Finally, the two shapes are disjoint for every $\delta$; an empty $\delta$ fails the single-declaration demand of shape (i) and is of shape (ii) vacuously, and a nonempty $\delta$ of both shapes would declare an output whose mode is cartesian and linear at once, which is excluded since every mode belongs to exactly one class (\cref{def:pre-signature}). \hfill $\square$

\textit{Remark:} Shape (i) says exactly that the descriptor is a c-descriptor, and shape (ii) that it is an $\ell$-descriptor (\cref{def:c-descriptors}). The required regularity is supplied by \cref{prop:identity-produces-a-regular-descriptor,prop:c-comp-preserves-regularity,prop:pos-comp-preserves-regularity,prop:neg-comp-preserves-regularity}.

\begin{proposition}[Cartesian well-formed descriptors are expressions]\label{prop:cartesian-descriptors-are-terms}
Fix a cartesian context $\theta$ and a cartesian mode $q$. The dependent-tree reading of \cref{def:dependent-tree} induces a map
$$\mf D\longmapsto \llbracket b\rrbracket$$
from c-descriptors over $(\theta\ ;\ b:q)$ (\cref{def:c-descriptors}) to the set of \cite{LSR17}-style expressions $\alpha$ with $\theta\vdash\alpha:q$ (\cref{def:cartesian-expressions}), and this map is a bijection. Moreover, every c-descriptor is well formed, and the well-formed descriptor classes of cartesian shape $\theta\Vdash_{\mf D}b:q$ are exactly the c-descriptors. The generation rules are complete for the cartesian shape.
\end{proposition}
\textit{Proof.}
We prove that the map is well defined, that it is surjective and injective, and that the two collections of c-descriptors and cartesian well-formed descriptors coincide.

\begin{itemize}
\item \textbf{The map is well defined.} A c-descriptor is regular and, by \cref{def:c-descriptors} and mode validity, contains only c-generator assignments over variables at cartesian modes. The reading $\llbracket b\rrbracket$ therefore uses only the input, output, and c-generator clauses of \cref{def:dependent-tree} and produces an expression over the variables of $\theta$. Mode-validity types it as $\theta\vdash\llbracket b\rrbracket:q$ by induction along the recursion. Internal renaming fixes inputs and preserves generator symbols. $\llbracket b\rrbracket$ therefore depends only on the class $\mf D$.

\item \textbf{Surjectivity.} By induction on $\theta\vdash\alpha:q$. If $\alpha=a$ is a variable, the Identity rule gives $\theta',a:q\Vdash_{[\varnothing\mid\{a\mapsto b\}]}b:q$ where $\theta = \theta',a:q$ up to the permutation $\pi$ of the rule, and $\llbracket b\rrbracket=a$. If $\alpha=f(\alpha_1,\ldots,\alpha_n)$, the induction hypothesis provides cartesian well-formed $\theta\Vdash_{\mf D_i}y_i:p_i$ with $\llbracket y_i\rrbracket=\alpha_i$. The c-Comp rule applies (its counting hypothesis holds by regularity of the premises) and produces $\theta\Vdash_{\mf D}r:q$ with $\llbracket r\rrbracket=f(\llbracket s_1\rrbracket,\ldots,\llbracket s_n\rrbracket)=f(\alpha_1,\ldots,\alpha_n)$: the construction leaves the premise assignments untouched, so the subtree of $s_i$ in $\mf D$ is its subtree in $\mf D_i$, and $\llbracket s_i\rrbracket=\llbracket y_i\rrbracket$ by the output clause of the reading.

\item \textbf{Injectivity.} Let $\mf D$ be a c-descriptor over $(\theta\ ;\ b:q)$ with representative $(\mc S\mid\mc T)$, and consider its underlying graph. The argument uses regularity only. Every c-generator assignment has exactly one target, and by the unique internal source occurrence property every internal variable is consumed exactly once. The composite is finite and acyclic. Every maximal chain of consumptions therefore ends either at the unique redirection to $b$ or at a targetless assignment; c-generator assignments are never targetless. Hence every internal variable reaches $b$. The underlying graph is a tree with root redirected to $b$, whose leaves are the input-variable occurrences (an input of $\theta$ may label several leaves) and the nullary c-generator assignments. The assignment set is recovered from $\llbracket b\rrbracket$ by structural recursion, uniquely up to the choice of internal variable names. Two c-descriptors with the same reading therefore share a representative up to internal renaming, i.e.,\ they are equal.

\item \textbf{The two collections coincide.} A well-formed class of cartesian shape is regular by \cref{prop:identity-produces-a-regular-descriptor,prop:c-comp-preserves-regularity,prop:pos-comp-preserves-regularity,prop:neg-comp-preserves-regularity} and free of linear variables by \cref{prop:sorts-of-well-formed-descriptors}, hence a c-descriptor. Conversely, given a c-descriptor $\mf D$ with reading $\alpha$, surjectivity provides a \emph{well-formed} class with the same reading, and injectivity (which used regularity only) identifies it with $\mf D$. Every c-descriptor is therefore well formed. \hfill $\square$
\end{itemize}

\textit{Remark:} \cref{prop:cartesian-descriptors-are-terms} is the interface to the framework of \cite{LSR17}. The cartesian sort of the mode theory is exactly the \cite{LSR17} calculus, and we will move between the graph and expression presentations of a cartesian descriptor without comment. We use the expression when substitution is the operation in play, and the graph when the graph structure is.

\endgroup{}

\begingroup

\subsection{Composition and Cut}

\subsubsection{Raw Composition}
In \cref{sec:well-formed-descriptors}, we saw the identity, cartesian, and linear rules needed to construct well-formed descriptors. The descriptors are understood as being freely generated by the rules. The glue necessary for those descriptors to combine and generate more expressive ones is composition. We use the words composition and cut interchangeably in this document as they express almost exactly the same concept. In order to define a composition operation on the well-formed descriptor classes, we recall that since two types of descriptors exist, three types of compositions should exist as well.

\begin{itemize}
  \item Two cartesian descriptors can be composed together at a cartesian variable sharing the same mode.
  \item Two linear descriptors can be composed together at a linear variable sharing the same mode.
  \item A linear and a cartesian descriptor can be composed together along two cartesian variables sharing the same mode: the first an output variable of the cartesian descriptor, and the second an input variable in the cartesian zone of the linear descriptor.
\end{itemize}
When dealing with raw descriptor classes, the number of occurrences of the variable being cut can vary. The composition operation is therefore a kind of ``duplicated substitution''. All the resources associated to the descriptor being composed are duplicated to accommodate the multiplicity of the occurrences. The apparatus needed to even state the composition operation is developed in this section. We note quickly that the theory of composition is established very generally for raw descriptor classes. The only type of composition we will ever need is the composition defined on well-formed (in particular regular) descriptor classes. The text opts for more generality.

\begin{definition}[Fresh Family of Bijections]\label{def:fresh-family-of-bijections}
Let $W$ be a finite set of internal variables, let $I$ be a finite index set, and let $F$ be a finite set of internal variables. A \emph{fresh family of bijections} for $(W,I,F)$ is a family of bijections $\bs\beta=(\beta_i:W\to B_i)_{i\in I}$ whose codomains $B_i$ are finite sets of internal variables, pairwise disjoint and disjoint from $F$.
\end{definition}

\begin{lemma}[Existence of Fresh Families of Bijections]\label{lem:existence-of-fresh-families-of-bijections}
For every triple $(W,I,F)$ satisfying \cref{def:fresh-family-of-bijections}, a fresh family of bijections exists.
\end{lemma}
\textit{Proof.}
The supply of internal variables is countably infinite and $I$ is finite. We may therefore choose pairwise disjoint finite sets $B_i$ of internal variables, all disjoint from $F$, with $|B_i|=|W|$ for every $i\in I$. A bijection $\beta_i:W\to B_i$ for each $i$ then gives the desired family.
\hfill $\square$

\begin{example}[Linear--Nonlinear Logic]
For the internal variable set $W := \{w\}$, the index set $I := \{1, 2\}$, and $F := \{u\}$, the bijections $\beta_1 : \{w\} \to \{w_1\}$ and $\beta_2 : \{w\} \to \{w_2\}$ form a fresh family of bijections: the codomains are pairwise disjoint and disjoint from $F$. Sending both indices to $\{w_1\}$ instead does not give a fresh family.
\end{example}

\begin{definition}[Occurrences of an Input]\label{def:occurrences-of-an-input}
Let $(\mc S\mid\mc T)$ be a raw descriptor and $x$ an input variable. An \emph{occurrence} of $x$ in $(\mc S\mid\mc T)$ is a pair $o=(E,i)$ where $E\in\mc S\cup\mc T$ is an assignment or a redirection and $i$ is a source position of $x$ in $E$. We write $\mathrm{Occ}_x(\mc S\mid\mc T)$ for the finite set of occurrences of $x$.
\end{definition}

\begin{example}[Linear--Nonlinear Logic]\label{ex:cartesian-input-occurrences}
In the descriptor $(\{a, a \xmapsto{\otimes_c} w\} \mid \{w \mapsto y\})$ where the cartesian input $a$ is used twice, the assignment $E := (a, a \xmapsto{\otimes_c} w)$ possesses $a$ in its two source positions, and the occurrence set is $\mathrm{Occ}_a = \{(E, 1),\ (E, 2)\}$. For the identity descriptor $(\varnothing \mid \{x \mapsto y\})$, the input $x$ has the single occurrence $(x \mapsto y,\, 1)$, in a redirection. A linear input of a regular descriptor occurs in exactly one source position by the boundary usage property. Cartesian inputs are the ones that can have several occurrences, or none.
\end{example}

\begin{definition}[Occurrence-Indexed Substitution]\label{def:raw-duplicated-substitution}
Let $(\mc S\mid\mc T)$ be a raw descriptor and let $x$ be an input variable. Let $v=(v_o)_{o\in\mathrm{Occ}_x(\mc S\mid\mc T)}$ be a family of input or internal variables indexed by the occurrences of $x$ with $v_o\neq x$ for every occurrence $o$. The \emph{occurrence-indexed substitution}
$$(\mc S\mid\mc T)[x:=v]$$
replaces, in every element $E$ of $\mc S\cup\mc T$ and for every source position $i$ of $E$ carrying $x$, the variable at position $i$ by $v_{(E,i)}$. 
\end{definition}

\textit{Remark:} Notice that when $v$ is a constant family at a variable $u$, the operation agrees with the input substitution $(\--)[x:=u]$ of \cref{def:input-substitution}. A family indexed by a single occurrence is constant, so the case of a single occurrence is covered as well.

\begin{example}[Linear--Nonlinear Logic]
Continuing with the assignment $E = (a, a \xmapsto{\otimes_c} w)$ of the \hyperref[ex:cartesian-input-occurrences]{input-occurrence example}, take the family $v$ given by $v_{(E,1)} := u_1$ and $v_{(E,2)} := u_2$ for two distinct internal variables $u_1, u_2$. The occurrence-indexed substitution gives
$$(\{a, a \xmapsto{\otimes_c} w\} \mid \{w \mapsto y\})[a := v] = (\{u_1, u_2 \xmapsto{\otimes_c} w\} \mid \{w \mapsto y\}).$$
The two occurrences of the same input receive two different variables. The constant family at an internal variable $u$ gives instead $(\{u, u \xmapsto{\otimes_c} w\} \mid \{w \mapsto y\})$, which is the ordinary input substitution $[a := u]$ of \cref{def:input-substitution}.
\end{example}

\begin{lemma}[Occurrence-Indexed Substitution Removes the Input]\label{lem:raw-duplicated-substitution-removes-the-input}
Let $(\mc S\mid\mc T)$, $x$, and $v$ be as in \cref{def:raw-duplicated-substitution}. Then
$$x\notin\mathrm{Inp}\big((\mc S\mid\mc T)[x:=v]\big).$$
\end{lemma}
\textit{Proof.}
Every source position carrying $x$ is replaced by some $v_o\neq x$, and $x$ is an input variable. It therefore occurs at no target position. Hence $x$ does not occur in the substituted descriptor. \hfill $\square$

\begin{lemma}[Substitution and Renaming Preserve Redirection Targets]\label{lem:duplicated-substitution-preserves-redirection-targets}
Occurrence-indexed substitutions and internal renamings change only source positions. Hence, for every output variable $y$, if $\mc T_{\mapsto y}=\varnothing$ then the redirection set obtained from $\mc T$ by any combination of internal renamings and occurrence-indexed substitutions has no redirection targeting $y$. No new redirection targets are ever created.
\end{lemma}
\textit{Proof.}
A redirection $R=(s\mapsto t)$ has the output variable $t$ as its target. Internal renamings fix output variables, and occurrence-indexed substitutions replace source positions only. Hence every redirection obtained from $R$ has target $t$. \hfill $\square$

\begin{lemma}[Occurrence-Indexed Substitutions Commute]\label{lem:occurrence-substitutions-commute}
Let $\mc G$ be an assignment or redirection set and let $x\neq x'$ be input variables. Let $v=(v_o)_{o\in\mathrm{Occ}_x(\mc G)}$ and $v'=(v'_{o'})_{o'\in\mathrm{Occ}_{x'}(\mc G)}$ be families that do not contain $x$ or $x'$. Then
$$\mc G[x:=v][x':=v'] \;=\; \mc G[x':=v'][x:=v].$$
Moreover, for an internal renaming $\phi$ whose extension is injective on the variables occurring in $\mc G$,
$$\big(\mc G[x:=v]\big)^{\phi} \;=\; \mc G^{\phi}\big[x:=v^{\phi}\big],$$
where $v^{\phi}:=(v_o^{\phi})_o$ and the occurrences of $x$ in $\mc G^\phi$ are identified with those of $\mc G$ position by position.
\end{lemma}
\textit{Proof.}
Both operations act position by position on the finite source lists of the elements of $\mc G$. A position holding $x$ receives its $v$-entry, one holding $x'$ its $v'$-entry, and every other position is fixed. The position sets are disjoint ($x\neq x'$), and no substituted value creates a new occurrence of the other variable (the entry hypotheses). The two orders therefore produce the same entry at every position. This gives the first equality. For the second, $\phi$ fixes the input $x$, so the occurrences of $x$ correspond position by position. At every position both sides hold the $\phi$-image of the same entry, as in \cref{lem:input-substitution-and-internal-renaming-commute}. \hfill $\square$

We are now ready to define the first concept of composition. This is the general version operating on raw descriptors subject to two conditions.
\begin{definition}[Raw Composition]\label{def:raw-composition}
Let
$$(\mc S \mid \mc T) \in \mathrm{desc}^{\mathrm{raw}}_\Sigma(\gamma_1 \ ; \ \delta_1, y : p)
\qquad\text{and}\qquad
(\mc S' \mid \mc T') \in \mathrm{desc}^{\mathrm{raw}}_\Sigma(\gamma_2, x : p \ ; \ \delta_2)$$
be raw descriptors subject to these conditions:
\begin{itemize}
\item \textbf{definedness}: $|\mc T_{\mapsto y}|=1$. Write $\mc T_{\mapsto y}=\{s\mapsto y\}$ and call $s$ the \emph{cut source};
\item \textbf{compatibility}: $x$ occurs in neither $\gamma_1$ nor $\gamma_2$, $y$ occurs in neither $\delta_1$ nor $\delta_2$, the contexts $\gamma_1$ and $\gamma_2$ agree on their common declarations, and $\delta_1$, $\delta_2$ are disjoint.
\end{itemize}
Let $\bs\beta=(\beta_o:\mathrm{Int}(\mc S\mid\mc T)\to B_o)_{o\in\mathrm{Occ}_x(\mc S'\mid\mc T')}$ be a fresh family of bijections, its codomains $B_o$ pairwise disjoint and disjoint from $\mathrm{Int}(\mc S\mid\mc T)\cup\mathrm{Int}(\mc S'\mid\mc T')$, available by \cref{lem:existence-of-fresh-families-of-bijections}. Set $v:=(s^{\beta_o})_{o\in\mathrm{Occ}_x(\mc S'\mid\mc T')}$. The family satisfies the side condition $v_o\neq x$ of \cref{def:raw-duplicated-substitution}. The \emph{raw composition} of $(\mc S\mid\mc T)$ and $(\mc S'\mid\mc T')$ at $y,x$ using $\bs\beta$ is defined as:
$$(\mc S \mid \mc T) \circ_{y,x}^{\bs \beta} (\mc S' \mid \mc T') \ := \ \Bigl(\bigcup_{o\in\mathrm{Occ}_x(\mc S'\mid\mc T')} \bigl(\mc S\mid\mc T_{\not\mapsto y}\bigr)^{\beta_o}\Bigr) \ \cup\ \bigl(\mc S'\mid\mc T'\bigr)[x:=v].$$
Elementwise, this is the pair
$$\left(\bigcup_{o\in\mathrm{Occ}_x(\mc S'\mid\mc T')} \mc S^{\beta_o} \ \cup\ \mc S'[x:=v] \ \middle|\ \bigcup_{o\in\mathrm{Occ}_x(\mc S'\mid\mc T')} (\mc T_{\not\mapsto y})^{\beta_o} \ \cup\ \mc T'[x:=v]\right).$$
\end{definition}

\textit{Convention.}
 For $\mathrm{Occ}_x(\mc S'\mid\mc T')=\varnothing$ the sum of copies is empty and the composite is $(\mc S'\mid\mc T')$ itself.

This composition must seem mysterious at first glance. It nevertheless expresses the connection of an output variable $y$ on the graph attached to the descriptor $(\mc S \mid \mc T)$ to an input variable $x$ on the graph attached to the descriptor $(\mc S' \mid \mc T')$. Raw composition records all the occurrences of the cut variable appearing in the descriptor $(\mc S' \mid \mc T')$. It then duplicates the left descriptor with a fresh renaming of its internal variables, one duplication for each occurrence. The composition proceeds to connect the cut variable in the descriptor $(\mc S' \mid \mc T')$ to the source of the redirection of the cut variable in the descriptor $(\mc S \mid \mc T)$. This effectively grafts a fresh copy of the left descriptor at every occurrence of the right descriptor.

\textit{Remark:} If $s$ is internal, then $s^{\beta_o}=\beta_o(s)\in B_o$. That is, each occurrence of $x$ is fed by the cut source of its own fresh copy of the left descriptor. If $s$ is an input variable, then $s^{\beta_o}=s$ for every $o$. All occurrences therefore share the input leaf, exactly as an expression substitution of a variable shares that variable among the leaves. Since input variables at cartesian modes are unrestricted, this sharing is compatible with regularity.

\begin{example}[Linear--Nonlinear Logic]
Consider a cartesian mode $c$ as well as
$$(\mc S \mid \mc T) := (\{a', a'' \xmapsto{\otimes_c} u\} \mid \{u \mapsto b\})
\qquad\text{and}\qquad
(\mc S' \mid \mc T') := (\{a, a \xmapsto{\otimes_c} w\} \mid \{w \mapsto y\})$$
over the boundaries $(a' : c,\, a'' : c \ ; \ b : c)$ and $(a : c \ ; \ y : c)$. The cut source is $s = u$, and the occurrences of $a$ are $o_1 = (E, 1)$ and $o_2 = (E, 2)$, as computed in the \hyperref[ex:cartesian-input-occurrences]{input-occurrence example}. For the fresh family $\beta_{o_1} : \{u\} \to \{u_1\}$ and $\beta_{o_2} : \{u\} \to \{u_2\}$, the values are $v_{o_1} = u_1$ and $v_{o_2} = u_2$, and the raw composition $(\mc S \mid \mc T) \circ_{b,a}^{\bs\beta} (\mc S' \mid \mc T')$ is
$$(\{a', a'' \xmapsto{\otimes_c} u_1,\ \ a', a'' \xmapsto{\otimes_c} u_2,\ \ u_1, u_2 \xmapsto{\otimes_c} w\} \mid \{w \mapsto y\})$$
over the boundary $(a' : c,\, a'' : c \ ; \ y : c)$. The left factor redirects only to $b$, so the copies contribute no redirections. The disjoint codomains keep the two copies distinct in the assignment set. Each occurrence of $a$ is fed by its own copy of the left descriptor, and the dependent tree of the output is $\llbracket y \rrbracket = \otimes_c(\otimes_c(a', a''),\, \otimes_c(a', a''))$.
\end{example}

\begin{samepage}
\begin{lemma}[One Step Dependence Decomposition for Composition]\label{lem:one-step-dependence-decomposition-for-composition}
In the notation of \cref{def:raw-composition}, write $(\mc S^c\mid\mc T^c)$ for the raw composite. Then
\[
\begin{aligned}
\leadsto_{(\mc S^c \mid \mc T^c)}
&=
\bigcup_{o\in\mathrm{Occ}_x(\mc S'\mid\mc T')}\big(\leadsto_{(\mc S \mid \mc T_{\not\mapsto y})}\big)^{\beta_o}
\\
&\quad\cup
\{(a,b)\in\leadsto_{(\mc S'\mid\mc T')}\mid a\neq x\}
\\
&\quad\cup
\{(v_o,b)\mid o=(E,i)\in\mathrm{Occ}_x(\mc S'\mid\mc T'),\ b\in\mathrm{tgt}(E)\}.
\end{aligned}
\]
\end{lemma}
\end{samepage}
\textit{Proof.}
One-step reachability distributes over componentwise sums (\cref{lem:one-step-reachability-distributes-over-componentwise-sums}). This gives the union of the relations of the copies and of the substituted right descriptor. For each copy, naturality of one-step reachability under internal renaming (\cref{lem:one-step-reachability-is-natural}) gives $\leadsto_{(\mc S\mid\mc T_{\not\mapsto y})^{\beta_o}}=(\leadsto_{(\mc S\mid\mc T_{\not\mapsto y})})^{\beta_o}$. For the substituted right descriptor, consider an element $E$ and its generated edges. Every source position of $E$ not carrying $x$ contributes its original edges. Every position carrying $x$ at occurrence $o$ contributes the edges $(v_o,b)$ for $b\in\mathrm{tgt}(E)$. And since $x$ is an input variable, no edge of $(\mc S'\mid\mc T')$ targets $x$. No other edge is affected. \hfill $\square$

\begin{proposition}[Internal Variables of Raw Composition]\label{prop:internal-variables-of-raw-composition}
In the notation of \cref{def:raw-composition},
$$
\mathrm{Int}(\mc S^c\mid\mc T^c)
=
\mathrm{Int}(\mc S'\mid\mc T')\ \uplus \biguplus_{o\in\mathrm{Occ}_x(\mc S'\mid\mc T')} B_o.
$$
\end{proposition}
\textit{Proof.}
We have that $\mathrm{Int}\big((\mc S\mid\mc T_{\not\mapsto y})^{\beta_o}\big)\subseteq B_o$ by \cref{lem:internal-variables-under-internal-renaming-and-input-substitution}. The substituted right descriptor contributes $\mathrm{Int}(\mc S'\mid\mc T')$ together with the values $v_o$ which are internal. Internal $v_o=\beta_o(s)$ lie in $B_o$. No other internal variable occurs. Conversely, every variable of $B_o$ occurs in the composite. An internal variable $u\neq s$ of $(\mc S\mid\mc T)$ occurs in $\mc S\cup\mc T_{\not\mapsto y}$, so $\beta_o(u)$ occurs in the copy $o$. When $s$ is internal, $\beta_o(s)$ is the value $v_o$ itself. The unions are disjoint since the codomains $B_o$ are pairwise disjoint and disjoint from $\mathrm{Int}(\mc S'\mid\mc T')$. \hfill $\square$

\phantomsection\label{anchor:input-context-union-convention}
\textit{Remark (the union of input contexts):} The context $\gamma_1\cup\gamma_2$ denotes the list $\gamma_1$ followed by the declarations of $\gamma_2 \setminus \gamma_1$. The same convention applies wherever the union of two input contexts appears, in particular in the Cut rule (\cref{def:well-formed-composition-cut}) and in the $2$-Subst rule of \cref{def:descriptor-2-cells}.

\begin{proposition}[Raw Composition Lands in the Cut Context]\label{prop:raw-composition-lands-in-the-cut-context}
In the notation of \cref{def:raw-composition},
$$(\mc S\mid\mc T)\circ_{y,x}^{\bs\beta}(\mc S'\mid\mc T')\in\mathrm{desc}^{\mathrm{raw}}_\Sigma(\gamma_1\cup\gamma_2\ ;\ \delta_1,\delta_2).$$
\end{proposition}

\textit{Proof.}
We check the copies first. The sources of $\mc S^{\beta_o}$ and $(\mc T_{\not\mapsto y})^{\beta_o}$ are input variables of $\gamma_1$ (fixed by the renaming) or internal variables of $B_o$. Their assignment targets are internal and their redirection targets lie in $\delta_1$. On the substituted side, the sources of $\mc S'[x:=v]$ and $\mc T'[x:=v]$ are input variables of $\gamma_2$ other than $x$ (\cref{lem:raw-duplicated-substitution-removes-the-input}), internal variables of $(\mc S'\mid\mc T')$, or the values $v_o$. Each of these values is an input of $\gamma_1$ or internal. Their redirection targets lie in $\delta_2$ (\cref{lem:duplicated-substitution-preserves-redirection-targets}). All sets involved are finite. The composite is therefore a raw descriptor over $(\gamma_1\cup\gamma_2\ ;\ \delta_1,\delta_2)$. \hfill $\square$

We extend the notion of raw composition to an operation on raw descriptor classes. The idea is standard: we define the composition of two descriptor classes to be the resulting class of the composition of a choice of representatives.
\begin{definition}[Composition of Descriptors]\label{def:composition-of-descriptors}
  Consider descriptor classes $\mf D \in \mathrm{Desc}^{\mathrm{raw}}_\Sigma(\gamma_1 \ ; \ \delta_1, y : p)$ with $|\mc T_{\mapsto y}|=1$ for some, hence any, representative $(\mc S\mid\mc T)\in\mf D$, and $\mf D' \in \mathrm{Desc}^{\mathrm{raw}}_\Sigma(\gamma_2, x : p \ ; \ \delta_2)$, subject to the boundary compatibility conditions of \cref{def:raw-composition}. We define the \textit{composition} of $\mf D$ and $\mf D'$ at $y,x$ as the descriptor class
	$$\mf D \circ_{y,x} \mf D' := \big[(\mc S \mid \mc T) \circ_{y,x}^{\bs \beta}(\mc S' \mid \mc T') \big] \in \mathrm{Desc}^{\mathrm{raw}}_\Sigma(\gamma_1\cup\gamma_2\ ;\ \delta_1,\delta_2),$$
	where
	\begin{itemize}
  	\item $(\mc S \mid \mc T) \in \mf D$ and $(\mc S' \mid \mc T') \in \mf D'$ are representatives with $\mathrm{Int}(\mc S\mid\mc T)\cap \mathrm{Int}(\mc S'\mid\mc T')=\varnothing$ (\cref{lem:fresh-representatives}),
    \item $\bs\beta=(\beta_o:\mathrm{Int}(\mc S\mid\mc T)\to B_o)_{o\in\mathrm{Occ}_x(\mc S'\mid\mc T')}$ is a fresh family of bijections whose codomains are pairwise disjoint and disjoint from $\mathrm{Int}(\mc S\mid\mc T)\cup\mathrm{Int}(\mc S'\mid\mc T')$ (\cref{lem:existence-of-fresh-families-of-bijections}).
	\end{itemize}
\end{definition}

\textit{Remark:}
	Notice that the condition $|\mc T_{\mapsto y}|=1$ is $\alpha$-invariant (\cref{lem:internal-renaming-preserves-redirections-targeting-an-output}), and that the occurrence set $\mathrm{Occ}_x$ of the right representative is used as the index set.

\begin{example}[Linear--Nonlinear Logic]
At the linear mode $l$, take the class $\mf D = [\{x, x' \xmapsto{\otimes_l} w\} \mid \{w \mapsto y\}]$ of the multiplicative conjunction and the class $\mf D' := [\{x'' \xmapsto{\bot_l} \cdot\} \mid \varnothing]$ over $(x'' : l \ ; \ \cdot)$, built from the co-nullary unit $\bot_l$. The right representative has no internal variables, so the representatives are disjoint, and the input $x''$ has the single occurrence $o$ in the assignment. For the fresh bijection $\beta_o : \{w\} \to \{\widetilde{w}\}$, the raw composite is
$$(\{x, x' \xmapsto{\otimes_l} \widetilde{w},\ \ \widetilde{w} \xmapsto{\bot_l} \cdot\} \mid \varnothing) \in \mathrm{desc}^{\mathrm{raw}}_\Sigma(x : l,\, x' : l \ ; \ \cdot),$$
and the internal renaming $\widetilde{w} \mapsto w$ carries it to a representative written with $w$ itself. The composition of the classes is
$$\mf D \circ_{y, x''} \mf D' = [\{x, x' \xmapsto{\otimes_l} w,\ \ w \xmapsto{\bot_l} \cdot\} \mid \varnothing].$$
Passing to classes hides the choice of the fresh family.
\end{example}

Proving that the composition is well defined, i.e. that the descriptor class does not depend on the choices, proceeds in three steps. We show independence of the fresh family of bijections, independence of the right representative, and independence of the left representative.

\begin{lemma}[Fresh Families of Bijections Give Alpha-Equivalent Raw Composites]\label{lem:fresh-families-of-bijections-give-alpha-equivalent-duplicated-substitutions}
Let $(\mc S\mid\mc T)$ and $(\mc S'\mid\mc T')$ be as in \cref{def:composition-of-descriptors}, with disjoint internal variables. Let $\bs\beta=(\beta_o:\mathrm{Int}(\mc S\mid\mc T)\to B_o)_o$ and $\bs\beta'=(\beta'_o:\mathrm{Int}(\mc S\mid\mc T)\to B'_o)_o$ be two fresh families of bijections whose codomains are pairwise disjoint and disjoint from $\mathrm{Int}(\mc S\mid\mc T)\cup\mathrm{Int}(\mc S'\mid\mc T')$. Then
$$(\mc S\mid\mc T)\circ_{y,x}^{\bs\beta}(\mc S'\mid\mc T')=_\alpha(\mc S\mid\mc T)\circ_{y,x}^{\bs\beta'}(\mc S'\mid\mc T').$$
\end{lemma}

\textit{Proof.}
Define
$$
\Phi(z)=\begin{cases}
\beta'_o(\beta_o^{-1}(z))&\text{if }z\in B_o\text{ for some occurrence }o,\\
z&\text{otherwise.}
\end{cases}
$$
By \cref{prop:internal-variables-of-raw-composition} the internal variables of either composite are the right internal variables together with the fresh copy sets. The map $\Phi$ is therefore a well-defined bijection between them. It identifies $B_o$ with $B'_o$ and fixes $\mathrm{Int}(\mc S'\mid\mc T')$. For each copy, the restriction $\Phi|_{B_o}$ satisfies the freshness condition of \cref{prop:internal-renaming-composition}. The domain of $\beta_o$ is indeed all of $\mathrm{Int}(\mc S\mid\mc T)$. We therefore obtain
$$(\mc S^{\beta_o})^\Phi=\mc S^{(\Phi|_{B_o})\circ\beta_o}=\mc S^{\beta'_o},$$
and likewise for $(\mc T_{\not\mapsto y})^{\beta_o}$. On the substituted right descriptor, $\Phi$ fixes every variable except the internal values $v_o=\beta_o(s)$, which it carries to $v'_o=\beta'_o(s)$. Both substitutions act position-by-position, so $\big(\mc S'[x:=v]\big)^\Phi=\mc S'[x:=v']$ and likewise for the redirection component. Hence $\Phi$ carries the first composite to the second. \hfill $\square$

\begin{lemma}[Right Representative Preserves Composition]\label{lem:right-representative-preserves-composition}
Fix $(\mc S\mid\mc T)$ as in \cref{def:composition-of-descriptors}. Let $(\mc S'\mid\mc T')$ and $(\widehat{\mc S}'\mid\widehat{\mc T}')$ be representatives of $\mf D'$ with internal variables disjoint from $\mathrm{Int}(\mc S\mid\mc T)$, related by a bijection $\psi$ with $(\mc S'\mid\mc T')^\psi=(\widehat{\mc S}'\mid\widehat{\mc T}')$. Let $\bs\beta$, $\widehat{\bs\beta}$ be fresh families for the two composites. Then the two raw composites are $=_\alpha$.
\end{lemma}

\textit{Proof.}
The renaming $\psi$ maps the elements of $\mc S'\cup\mc T'$ bijectively to those of $\widehat{\mc S}'\cup\widehat{\mc T}'$, preserves shapes and positions, and fixes the input variable $x$. It therefore induces a bijection of occurrence sets $o=(E,i)\mapsto\widehat o=(E^\psi,i)$. Define
$$
\Psi(z)=\begin{cases}
\widehat\beta_{\widehat o}(\beta_o^{-1}(z))&\text{if }z\in B_o,\\
\psi(z)&\text{if }z\in\mathrm{Int}(\mc S'\mid\mc T').
\end{cases}
$$
The cases are disjoint and exhaust the internal variables of the first composite. The map $\Psi$ is a bijection onto those of the second. On each copy, $(\mc S^{\beta_o})^\Psi=\mc S^{\widehat\beta_{\widehat o}}$ by \cref{prop:internal-renaming-composition}, and likewise for the redirection copies. On the substituted right descriptor, $\Psi$ acts as $\psi$ on the old variables and carries $v_o$ to $\widehat v_{\widehat o}$ (both equal $s$ when $s$ is an input). Since $\psi$ maps the element carrying occurrence $o$ to the element carrying $\widehat o$ position-by-position, we get $\big(\mc S'[x:=v]\big)^\Psi=\widehat{\mc S}'[x:=\widehat v]$, and likewise for redirections. The renaming $\Psi$ therefore carries the first composite to the second. \hfill $\square$

\begin{lemma}[Left Representative Internal Renaming Preserves Raw Composition]\label{lem:left-representative-internal-renaming-preserves-raw-composition}
Fix $(\mc S'\mid\mc T')$ as in \cref{def:composition-of-descriptors}. Let $(\mc S\mid\mc T)$ and $(\widehat{\mc S}\mid\widehat{\mc T})$ be representatives of $\mf D$ with internal variables disjoint from $\mathrm{Int}(\mc S'\mid\mc T')$, related by a bijection $\phi$ with $(\mc S\mid\mc T)^\phi=(\widehat{\mc S}\mid\widehat{\mc T})$. Let $\bs\beta$, $\widehat{\bs\beta}$ be fresh families for the two composites. Then the two raw composites are $=_\alpha$.
\end{lemma}

\textit{Proof.}
By \cref{lem:internal-renaming-preserves-redirections-targeting-an-output}, $\phi$ carries the unique redirection $s\mapsto y$ of $\mc T$ to the unique redirection $\widehat s\mapsto y$ of $\widehat{\mc T}$. Here $\widehat s=\phi(s)$ if $s$ is internal and $\widehat s=s$ if $s$ is an input. Define
$$
\Theta(z)=\begin{cases}
\widehat\beta_o(\phi(\beta_o^{-1}(z)))&\text{if }z\in B_o,\\
z&\text{if }z\in\mathrm{Int}(\mc S'\mid\mc T').
\end{cases}
$$
On each copy, two applications of \cref{prop:internal-renaming-composition} (the freshness conditions hold since the codomains are fresh) give
$$(\mc S^{\beta_o})^\Theta=\mc S^{\widehat\beta_o\circ\phi}=(\mc S^{\phi})^{\widehat\beta_o}=\widehat{\mc S}^{\widehat\beta_o},$$
and likewise $((\mc T_{\not\mapsto y})^{\beta_o})^\Theta=(\widehat{\mc T}_{\not\mapsto y})^{\widehat\beta_o}$, using that $\phi$ carries $\mc T_{\not\mapsto y}$ to $\widehat{\mc T}_{\not\mapsto y}$. On the substituted right descriptor, $\Theta$ fixes the old variables and carries each value $v_o=s^{\beta_o}$ to $\widehat s^{\widehat\beta_o}$. The substituted components then correspond position-by-position, and $\Theta$ carries the first composite to the second. \hfill $\square$

Together, these three properties establish that raw composition is well defined.
\begin{proposition}[Composition is well defined]\label{prop:composition-is-well-defined}
The descriptor class
$$
\mf D \circ_{y,x} \mf D' := \big[(\mc S \mid \mc T)\circ^{\bs\beta}_{y,x}(\mc S'\mid\mc T')\big]
$$
is independent of the chosen representatives $(\mc S\mid\mc T)\in\mf D$, $(\mc S'\mid\mc T')\in\mf D'$, and of the chosen fresh family of bijections $\bs\beta$.
\end{proposition}

\textit{Proof.}
For fixed representatives, \cref{lem:fresh-families-of-bijections-give-alpha-equivalent-duplicated-substitutions} gives independence of the fresh family. Fixing the left representative, \cref{lem:right-representative-preserves-composition} then gives independence of the right representative. Fixing the right representative, \cref{lem:left-representative-internal-renaming-preserves-raw-composition} gives independence of the left representative. For two arbitrary simultaneous choices, we use transitivity of $=_\alpha$ to conclude. \hfill $\square$

We now record the behavior of the composition on the well-formed fragment. At a linear mode it is a single cut since there can only be one occurrence of the cut variable, and at a cartesian mode it is expression substitution through the reading of \cref{prop:cartesian-descriptors-are-terms} that maps exactly to the terms of \cite{LSR17}.

\begin{corollary}[Linear cut]\label{cor:linear-cut}
Let $\mf D$ and $\mf D'$ be regular descriptor classes as in \cref{def:composition-of-descriptors}, and let the cut mode $p$ be \emph{linear}. Then $x$ has exactly one occurrence in any representative of $\mf D'$, and for disjoint representatives $(\mc S\mid\mc T)\in\mf D$ and $(\mc S'\mid\mc T')\in\mf D'$ with cut source $s$,
$$
\mf D \circ_{y,x} \mf D' = \big[\,\mc S \cup \mc S'[x:=s]\ \big|\ \mc T_{\not\mapsto y} \cup \mc T'[x:=s]\,\big].
$$
\end{corollary}
\textit{Proof.}
By the boundary usage property of the regular right factor, the linear input $x$ occurs in exactly one source position. Hence $\mathrm{Occ}_x(\mc S'\mid\mc T')$ is a singleton $\{o\}$ and a single copy $\mc S^{\beta_o}$ is made. By \cref{prop:composition-is-well-defined} the class does not depend on $\bs\beta$. The renaming $\beta_o^{-1}$, extended by the identity on $\mathrm{Int}(\mc S'\mid\mc T')$, carries the composite to the displayed representative. It is a bijection since the representatives are disjoint. The occurrence-indexed substitution at a single occurrence is then the input substitution $[x:=s]$ as expected. \hfill $\square$

\begin{corollary}[Cartesian cut is expression substitution]\label{cor:cartesian-cut-is-term-substitution}
Let the cut mode $q$ be cartesian, let $\mf D$ be a c-descriptor over $(\theta_1\ ;\ y:q)$ (\cref{def:c-descriptors}), and let $\mf D'$ be a regular class over $(\gamma_2,x:q\ ;\ \delta_2)$, subject to the compatibility conditions of \cref{def:raw-composition}. Fix disjoint representatives $(\mc S\mid\mc T)\in\mf D$ and $(\mc S'\mid\mc T')\in\mf D'$ and a raw composite $(\mc S^c\mid\mc T^c)$ as in \cref{def:composition-of-descriptors}. Then the dependent trees satisfy, for every variable $z\neq x$ of $(\mc S'\mid\mc T')$,
$$
\llbracket z\rrbracket_{(\mc S^c\mid\mc T^c)}
=
\llbracket z\rrbracket_{(\mc S'\mid\mc T')}\big[x := \llbracket y\rrbracket_{(\mc S\mid\mc T)}\big].
$$
In particular, through \cref{prop:cartesian-descriptors-are-terms}, composition at a cartesian mode is the substitution of the expression $\llbracket y\rrbracket_{\mf D}$ for the variable $x$.
\end{corollary}
\textit{Proof.}
By \cref{def:c-descriptors}, $\mf D$ contains no variable at a linear mode and its output context is exactly $y:q$. Hence $\mc T=\{s\mapsto y\}$ and $\mc T_{\not\mapsto y}=\varnothing$, so the composite is $\big(\bigcup_o \mc S^{\beta_o}\cup\mc S'[x:=v]\mid\mc T'[x:=v]\big)$. It is regular by case (ii) of \cref{prop:cut-preserves-regularity}, so its reading is well defined.

We treat the copies first. We claim that $\llbracket\beta_o(u)\rrbracket_{(\mc S^c\mid\mc T^c)}=\llbracket u\rrbracket_{(\mc S\mid\mc T)}$ for every internal variable $u$ of $(\mc S\mid\mc T)$ and every occurrence $o$. The producer of $\beta_o(u)$ in the composite is the $\beta_o$-copy of the producer of $u$, with the same generator and the same positions. Its sources are inputs of $\theta_1$, fixed by the renaming, or variables of $B_o$. Inputs read as themselves, and the reading forgets internal names. The claim follows by induction along the acyclic structure of the left factor. In particular $\llbracket v_o\rrbracket_{(\mc S^c\mid\mc T^c)}=\llbracket y\rrbracket_{(\mc S\mid\mc T)}$: when $s$ is internal this is the claim at $u=s$ together with the output clause, and when $s$ is an input we have $v_o=s=\llbracket y\rrbracket_{(\mc S\mid\mc T)}$.

We now argue by induction along the acyclic structure of the right factor. If $z$ is an input variable other than $x$, both sides read $z$, and the substitution $[x:=\llbracket y\rrbracket]$ has no effect. If $z$ is internal, it is produced by a unique assignment $E$ of $\mc S'$, and the composite produces it by $E[x:=v]$. Both readings apply the generator clause of \cref{def:dependent-tree}, with the same generator and the same positions. A source $u\neq x$ contributes $\llbracket u\rrbracket_{(\mc S^c\mid\mc T^c)}=\llbracket u\rrbracket_{(\mc S'\mid\mc T')}[x:=\llbracket y\rrbracket]$ by the induction hypothesis. A position carrying $x$ at occurrence $o$ carries $v_o$ in the composite, and $\llbracket v_o\rrbracket=\llbracket y\rrbracket$ by the claim. This matches $x[x:=\llbracket y\rrbracket]=\llbracket y\rrbracket$. The two readings agree. If $z$ is an output of $\delta_2$, its unique redirection in $\mc T'[x:=v]$ keeps its target, and its source is treated by the input and internal cases, or by the claim when it is a value $v_o$. The output clause then gives the equality.
\hfill $\square$

\begin{proposition}[Cut Preserves Regularity]\label{prop:cut-preserves-regularity}
Let $(\mc S\mid\mc T)$ over $(\gamma_1\ ;\ \delta_1,y:p)$ and $(\mc S'\mid\mc T')$ over $(\gamma_2,x:p\ ;\ \delta_2)$ be \emph{regular} raw descriptors with disjoint internal variables, satisfying the boundary compatibility conditions of \cref{def:raw-composition}. Suppose additionally that every declaration shared by $\gamma_1$ and $\gamma_2$ is at a cartesian mode. Let $\bs\beta$ be a fresh family of bijections as in \cref{def:composition-of-descriptors}. Suppose that either
\begin{itemize}
\item[(i)] $p$ is a linear mode, or
\item[(ii)] $p$ is a cartesian mode, $(\mc S\mid\mc T)$ contains no variable at a linear mode, and $\delta_1=\varnothing$.
\end{itemize}
Then the raw composite $(\mc S^c\mid\mc T^c)$ is regular over $(\gamma_1\cup\gamma_2\ ;\ \delta_1,\delta_2)$. Consequently $\mf D\circ_{y,x}\mf D'$ is a regular descriptor class.
\end{proposition}

\textit{Proof.}
Write $r:=|\mathrm{Occ}_x(\mc S'\mid\mc T')|$. In case (i) the boundary usage property of the right factor gives $r=1$. In case (ii) the value of $r$ is arbitrary. Notice that in case (ii) we have $\mc T_{\not\mapsto y}=\varnothing$. The left factor redirects only to $y$ since $\delta_1=\varnothing$. In particular, in both cases no output of $\delta_1$ receives more than one copy of a redirection. In case (i) this is because there is a single copy; in case (ii) because there is nothing to copy.

By \cref{prop:raw-composition-lands-in-the-cut-context} the composite is a raw descriptor over the stated boundary. By \cref{prop:internal-variables-of-raw-composition} its internal variables decompose as the disjoint union
$$\mathrm{Int}(\mc S^c\mid\mc T^c)=\mathrm{Int}(\mc S'\mid\mc T')\ \uplus\ \biguplus_{o} B_o.$$
By \cref{prop:mode-assignments-are-unique-for-regular-descriptors} there exist unique mode assignments $\mu, \mu'$ for the descriptors $(\mc S \mid \mc T)$ and $(\mc S' \mid \mc T')$ respectively.

\begin{itemize}
\item \textbf{Mode Validity.} Define $\mu^c$ as $\mu'$ on $\mathrm{Int}(\mc S'\mid\mc T')$ and as $\mu\circ\beta_o^{-1}$ on each $B_o$. Every copied assignment and redirection is mode-valid since its original is and the renaming is mode-preserving by construction of $\mu^c$. Every substituted element of the right descriptor is mode-valid. The positions not involving $x:p$ keep their modes unchanged, and a position carrying $x:p$ now carries $v_o=s^{\beta_o}$, whose mode is $p$. Indeed, if $s$ is internal this is $\mu(s)=p$ from the redirection $s\mapsto y$ with $y:p$ and mode-validity of the left factor. If $s$ is an input it is the declaration of $s$ in $\gamma_1$.

\item \textbf{Irreflexive reachability.} Suppose for contradiction that the composite has a non-empty cycle. All its variables are internal (\cref{lem:inputs-have-no-incoming-paths-and-outputs-have-no-outgoing-paths}), hence lie in $\mathrm{Int}(\mc S'\mid\mc T')$ or in some $B_o$. By \cref{lem:one-step-dependence-decomposition-for-composition} the edges of the composite are the copy edges (inside a single $B_o$ together with $\gamma_1$-inputs) and the substituted right edges. The targets of the latter lie in $\mathrm{Int}(\mc S')$ or in the outputs. No edge of the composite targets a variable of $B_o$ except a copy-$o$ edge. Substituted right edges target right internal variables or outputs, and distinct copies are disjoint. Hence, if the cycle visits a variable of $B_o$, the edge arriving at it is a copy-$o$ edge. By backwards induction the entire cycle would have to lie inside the copy $o$. For the cycle to leave $B_o$ it must use a substituted right edge $(v_o,b)$ with $b$ a right internal variable. To return to $B_o$ it would need an edge targeting $B_o$ from outside. No such edge exists. A cycle inside the copy $o$ pulls back along $\beta_o$ to a cycle in $(\mc S\mid\mc T_{\not\mapsto y})$. This contradicts the regularity of the left factor via \cref{lem:removing-assignments-or-redirections-cannot-create-cycles} and \cref{lem:internal-renaming-preserves-strict-reachability}. A cycle avoiding every $B_o$ uses only edges $(a,b)$ of the substituted right descriptor with $a,b$ right internal variables. Every such edge is an original edge of $(\mc S'\mid\mc T')$, since the substituted positions carry the values $v_o$, which are not right internal variables. Such a cycle contradicts the regularity of the right factor. Hence no cycle exists.

\item \textbf{Unique internal target occurrence.} A right internal variable keeps its unique producer. The targets remain unchanged when performing the occurrence-indexed substitution, and the copies produce no right internal variable since their targets lie in the $B_o$. A variable $z=\beta_o(u)\in B_o$ is produced exactly by the $\beta_o$-copy of the unique producer of $u$ in $\mc S$. The copies are pairwise disjoint and fresh from the right. This producer is therefore unique in $\mc S^c$. The targets remain pairwise distinct inside assignments by the injectivity of the renamings.

\item \textbf{Grounded redirections.} An internal variable of $\mc T^c$ lies in a copy $(\mc T_{\not\mapsto y})^{\beta_o}$ or in $\mc T'[x:=v]$. In the first case it is $\beta_o(u)$ for $u\in\mathrm{Int}(\mc T)\subseteq\mathrm{Int}(\mc S)$, and the copy of its producer lies in $\mc S^{\beta_o}$. In the second case it is an old internal source of $\mc T'$, produced in $\mc S'$ by the grounded-redirections property of the right factor, or it is a value $v_o=\beta_o(s)$ with $s$ internal, produced in $\mc S^{\beta_o}$.

\item \textbf{Unique internal source occurrence.} A right internal variable keeps its source occurrences unchanged, since the substitution replaces only occurrences of $x$. For $z=\beta_o(u)\in B_o$ with $u\neq s$, the source occurrences of $z$ are exactly the $\beta_o$-copies of those of $u$. Of the latter there is exactly one, by regularity of the left factor. Now take $z=\beta_o(s)$ with $s$ internal. In the left factor the unique source occurrence of $s$ is the removed redirection $s\mapsto y$. It is a source occurrence, and regularity allows no other. In the composite the copies of it are therefore gone, and $z$ occurs in exactly one source position, namely the occurrence position $o$ itself. Hence every internal variable of the composite has exactly one source occurrence.

\item \textbf{Boundary usage property.} We treat the inputs first. Inputs at cartesian modes are unrestricted. An input at a linear mode belongs to exactly one of $\gamma_1$, $\gamma_2$, since shared declarations are cartesian. In case (ii) the left factor has no linear variables, so all linear inputs are right inputs. Their occurrences are unchanged by the substitution. In case (i) we have $r=1$. A linear input of $\gamma_1$ other than the cut source has its unique occurrence in the single copy. If the cut source $s$ is itself a linear input of $\gamma_1$, its unique occurrence in the left factor is the removed redirection $s\mapsto y$. In the composite its unique occurrence is the substituted position $o$ itself. Either way there is exactly one occurrence. A linear input of $\gamma_2$ keeps its unique occurrence. As for the outputs, each output of $\delta_2$ is the target of exactly one redirection of $\mc T'[x:=v]$ since substitution preserves targets. And each output of $\delta_1$ (case (i) only; in case (ii) $\delta_1=\varnothing$) is the target of exactly one redirection of the single copy $(\mc T_{\not\mapsto y})^{\beta_o}$.
\end{itemize}
Therefore $(\mc S^c\mid\mc T^c)$ is regular. \hfill $\square$

\endgroup{}
\begingroup

\subsubsection{The Cut Rule}

The key insight in our treatment of the mode theory is that composing descriptors is, in a sense, a free operation. In logic, the terminology used is that of \emph{admissibility}. We remind ourselves that an admissible rule is a rule that, when added to the calculus, does not prove more theorems than when not present. In our context of descriptors, this means that no new descriptors arise from the cut rule. 

\begin{definition}[The Cut Rule]\label{def:well-formed-composition-cut}
In the Cut rule, $\mc T$ denotes the redirection set of a representative of $\mf D$. The condition $|\mc T_{\mapsto y}| = 1$ does not depend on this choice (\cref{lem:internal-renaming-preserves-redirections-targeting-an-output}), and the composite $\mf D \circ_{y,x} \mf D'$ of the conclusion is the composition of descriptor classes of \cref{def:composition-of-descriptors}.
  $$
  \inferrule*[Right=Cut]
  {
    \gamma_1 \Vdash_{\mf D} \delta_1, y : p \\
    \gamma_2, x : p \Vdash_{\mf D'} \delta_2 \\
    |\mc T_{\mapsto y}| = 1 \\
    \gamma_1, \gamma_2 \ \text{agree on their common declarations} \\
    \text{every shared declaration is at a cartesian mode} \\
    x \notin \gamma_1, \quad y \notin \delta_2 \\
    \delta_1, \delta_2 \ \text{disjoint} \\
    \pi:\gamma_1\cup\gamma_2\cong\gamma' \\
    \tau:\delta_1,\delta_2\cong\delta'
  }
  {
    \gamma' \Vdash_{(\pi;\tau)_*(\mf D \circ_{y,x} \mf D')} \delta'
  }
  $$
\end{definition}

\textit{Remark (the union of input contexts):} By the \hyperref[anchor:input-context-union-convention]{input-context union convention}, $\gamma_1\cup\gamma_2$ denotes the list $\gamma_1$ followed by the declarations of $\gamma_2 \setminus \gamma_1$.

\textit{Remark:} Like the counting hypotheses of the other generation rules, the hypothesis $|\mc T_{\mapsto y}| = 1$ will be automatic once well-formed descriptors are proved regular (\cref{thm:well-formed-descriptors-are-regular}). The boundary usage property gives each output exactly one redirection. It is kept explicit so that the rule is self-contained.

\textit{Remark (the two instances):} In the sort displays of \cref{sec:well-formed-descriptors}, the rule has exactly two shapes (\cref{prop:sorts-of-well-formed-descriptors}). At a \emph{linear} mode $p$, the premises are linear judgments and
$$
\theta_1 \ ; \ \Gamma_1 \Vdash_{\mf D} \delta_1, y:p
\qquad
\theta_2 \ ; \ \Gamma_2, x:p \Vdash_{\mf D'} \delta_2
\qquad\Longrightarrow\qquad
\theta_1\cup\theta_2 \ ; \ \Gamma_1,\Gamma_2 \Vdash_{\mf D\circ_{y,x}\mf D'} \delta_1,\delta_2 :
$$
the cartesian zones are shared and the linear zones are multiplicative on both sides. The composition is computed as the linear cut (\cref{cor:linear-cut}). This is the polycategorical cut. At a \emph{cartesian} mode $q$, the left premise is a cartesian judgment $\theta_1\Vdash_{\mf D} y:q$ and the cut input $x:q$ lies among the cartesian inputs of the right premise. The conclusion has input context $\theta_1\cup\gamma_2$ and the sort of the right premise. The composition corresponds to the substitution of the expression $\llbracket y\rrbracket_{\mf D}$ for $x$ (\cref{cor:cartesian-cut-is-term-substitution}). This is the substitution form of cut.

\begin{example}[Linear--Nonlinear Logic, a linear cut]
We work over the linear--nonlinear pre-signature. The judgment $x : l,\, x' : l \Vdash_{\mf D} y : l$, where $\mf D := [\{x, x' \xmapsto{\otimes_l} w\} \mid \{w \mapsto y\}]$, is derivable by $\ell$-Pos-Comp from two identity premises. The judgment $x'' : l \Vdash_{\mf D'} y' : l,\, y'' : l$, where $\mf D' := [\{x'' \xmapsto{\parr_l} w_1, w_2\} \mid \{w_1 \mapsto y',\ w_2 \mapsto y''\}]$, is derivable by $\ell$-Neg-Comp from two identity premises. We cut the output $y$ against the input $x''$ at the linear mode $l$. Here $\gamma_1 = x : l,\, x' : l$, $\delta_1 = \cdot$, $\gamma_2 = \cdot$, and $\delta_2 = y' : l,\, y'' : l$. The premises of the rule hold: the set $\mc T_{\mapsto y} = \{w \mapsto y\}$ is a singleton, $x'' \notin \gamma_1$, $y \notin \delta_2$, the contexts $\gamma_1$ and $\gamma_2$ share no declaration, and $\delta_1$, $\delta_2$ are disjoint. Taking $\pi$ and $\tau$ to be identity permutations, the rule produces the judgment
$$x : l,\, x' : l \ \Vdash_{[\{x, x' \xmapsto{\otimes_l} w,\ \ w \xmapsto{\parr_l} w_1, w_2\} \mid \{w_1 \mapsto y',\ w_2 \mapsto y''\}]} \ y' : l,\, y'' : l,$$
computed by \cref{cor:linear-cut} with cut source $w$. The composite stores in $w$ the multiplicative conjunction of $x$ and $x'$ and splits it into the two outputs. The cartesian zones are empty and the linear zones are multiplicative. This is the linear instance.
\end{example}

\begin{example}[Linear--Nonlinear Logic, a cartesian cut]\label{ex:cartesian-cut-linear-logic}
We work over the pre-signature extended with the c-generator $\otimes_c : c, c \to c$. The judgment $a : c,\, a' : c \Vdash_{\mf D} y : c$, where $\mf D := [\{a, a' \xmapsto{\otimes_c} w\} \mid \{w \mapsto y\}]$, is derivable by c-Comp from two identity premises. It reads as the expression $\otimes_c(a, a')$ (\cref{prop:cartesian-descriptors-are-terms}). The judgment $a : c,\, x : c \Vdash_{\mf D'} y' : c$, where $\mf D' := [\{x, x \xmapsto{\otimes_c} w'\} \mid \{w' \mapsto y'\}]$, is derivable by c-Comp from two identity premises. It reads as the expression $\otimes_c(x, x)$, and the cartesian input $a$ is not used. We cut the output $y$ against the input $x$ at the cartesian mode $c$. Here the contexts $\gamma_1 = a : c,\, a' : c$ and $\gamma_2 = a : c$ share the declaration $a : c$ and agree on it. The union $\gamma_1 \cup \gamma_2$ is $a : c,\, a' : c$. The input $x$ has two occurrences in $\mf D'$. The composition therefore creates a fresh copy of $\mf D$ and places it at each occurrence, copying the internal variable $w$ to $w_1$ and $w_2$. Taking $\pi$ and $\tau$ to be identity permutations, the rule produces the judgment $a : c,\, a' : c \Vdash_{\mf D \circ_{y, x} \mf D'} y' : c$ with composite
$$\mf D \circ_{y, x} \mf D' = [\{a, a' \xmapsto{\otimes_c} w_1,\ \ a, a' \xmapsto{\otimes_c} w_2,\ \ w_1, w_2 \xmapsto{\otimes_c} w'\} \mid \{w' \mapsto y'\}].$$
The composite reads as the expression $\otimes_c(\otimes_c(a, a'), \otimes_c(a, a')) = \otimes_c(x, x)[x := \otimes_c(a, a')]$. This is the substitution of the expression $\llbracket y \rrbracket_{\mf D}$ for $x$ (\cref{cor:cartesian-cut-is-term-substitution}).
\end{example}

\begin{lemma}[Cut preserves the sorts]\label{lem:cut-preserves-the-sorts}
Every judgment derivable using the five generation rules of Identity, c-Comp, l-Pos-Comp, l-Neg-Comp, and Cut has one of the two shapes of \cref{prop:sorts-of-well-formed-descriptors}. Moreover, a cut at a linear mode has a conclusion of linear shape, and a cut at a cartesian mode has the shape of its right premise.
\end{lemma}
\textit{Proof.}
We proceed by induction on the derivation. All the rules but cut are treated by the proof of \cref{prop:sorts-of-well-formed-descriptors}, remembering that transports do not change the sets of modes. Suppose the last rule is Cut, with cut mode $p$.

Suppose first that $p$ is linear. This means that the sequent $\gamma_1 \Vdash_{\mf D} \delta_1, y : p$ is linear. Indeed, by the induction hypothesis it has one of the two shapes, and the linear output $y : p$ excludes the cartesian shape. Similarly, the other sequent $\gamma_2 , x : p \Vdash_{\mf D'} \delta_2$ is linear since $x : p$ is part of the input context. We conclude that the contexts $\delta_1, \delta_2$ are purely linear (contain only linear variables). The conclusion sequent, before the transport operation, is then linear.

If $p$ is cartesian, then $\delta_1$ is empty since the left premise is cartesian. Indeed, by the induction hypothesis it has one of the two shapes, and the cartesian output $y : p$ excludes the linear shape. The sequent attached to the descriptor $\mf D'$ can be either linear or cartesian. If it is cartesian, then the concluding sequent is also cartesian. It contains a cartesian context and a single cartesian output, and the composite descriptor contains no variable at a linear mode. Indeed, both premise descriptors contain no such variable by the cartesian shape. Every variable of the composite is a variable of $\mf D'$ or a fresh copy of a variable of $\mf D$ (\cref{prop:internal-variables-of-raw-composition}), and it has the mode of its original. If it is linear, then the concluding sequent is also linear as the composition happens at a cartesian variable and the linear contexts and conclusion are retained.\hfill $\square$

\Cref{thm:well-formed-descriptors-are-regular} is of crucial importance: it asserts that all well-formed descriptors are regular and so enjoy the properties necessary to interpret them as directed acyclic graphs. Most of its argument is supplied by \cref{prop:identity-produces-a-regular-descriptor,prop:c-comp-preserves-regularity,prop:pos-comp-preserves-regularity,prop:neg-comp-preserves-regularity,prop:cut-preserves-regularity}.
\begin{theorem}[Well-formed descriptors are regular]\label{thm:well-formed-descriptors-are-regular}
If $\gamma \Vdash_{\mf D} \delta$ is derivable, then $\mf D$ is regular.
\end{theorem}

\textit{Proof.}
We proceed by induction on the derivation of $\gamma\Vdash_{\mf D}\delta$.

\begin{itemize}
  \item \textbf{Identity.} If the last rule is Identity, then the identity descriptor is regular by \cref{prop:identity-produces-a-regular-descriptor}, and the transported conclusion descriptor is regular by \cref{prop:permutation-transport-preserves-regularity}.

	\item \textbf{Generators.} If the last rule is one of c-Comp, l-Pos-Comp, or l-Neg-Comp, then by the induction hypothesis, every premise descriptor is regular. The conclusion descriptor is regular by \cref{prop:c-comp-preserves-regularity}, \cref{prop:pos-comp-preserves-regularity}, or \cref{prop:neg-comp-preserves-regularity} respectively. Transport preserves this by \cref{prop:permutation-transport-preserves-regularity}.

	\item \textbf{Cut.} If the last rule is Cut, then by the induction hypothesis the premise descriptors $\mf D$ and $\mf D'$ are regular. In particular the definedness hypothesis $|\mc T_{\mapsto y}|=1$ holds by the boundary usage property of $\mf D$. The side conditions of the rule provide the boundary compatibility of \cref{def:raw-composition} and also the fact that every shared input declaration is at a cartesian mode. If the cut mode $p$ is linear, the first hypothesis (i) of \cref{prop:cut-preserves-regularity} holds. If $p$ is cartesian, the left premise is derivable with the cartesian output $y:p$. By \cref{lem:cut-preserves-the-sorts} it is of cartesian shape, i.e. $\mf D$ contains no variable at a linear mode and $\delta_1=\varnothing$. This is hypothesis (ii). In both cases \cref{prop:cut-preserves-regularity} makes the composite $\mf D\circ_{y,x}\mf D'$ regular, and the transported conclusion descriptor is regular by \cref{prop:permutation-transport-preserves-regularity}.
\end{itemize}
Thus every derivable descriptor is regular. \hfill $\square$

\begin{proposition}[Functoriality of Permutation Transport]\label{prop:functoriality-of-permutation-transport}
Let $\pi:\gamma\cong\gamma'$, $\pi':\gamma'\cong\gamma''$, $\tau:\delta\cong\delta'$, and $\tau':\delta'\cong\delta''$ be context permutations. Then, for every $\mf D\in\mathrm{Desc}^{\mathrm{raw}}_\Sigma(\gamma\ ;\ \delta)$,
$$
(\pi';\tau')_*((\pi;\tau)_*\mf D)=((\pi'\circ\pi);(\tau'\circ\tau))_*\mf D.
$$
\end{proposition}

\textit{Proof.}
Choose a representative $(\mc S\mid\mc T)\in\mf D$. The transport does not modify the assignment and redirection sets. It only changes the boundary (\cref{def:permutation-transport-of-raw-descriptors}), and it acts on classes through representatives (\cref{def:permutation-transport-of-descriptor-classes}). Both sides are therefore represented by the same assignment set and redirection set $(\mc S\mid\mc T)$, now regarded in the final ambient context $\gamma'';\delta''$. \hfill $\square$

\begin{lemma}[Permutation Transport Commutes with Cut]\label{lem:permutation-transport-commutes-with-cut}
Let $\pi_1:\gamma_1\cong\gamma_1'$ and $\pi_2:\gamma_2\cong\gamma_2'$ be context permutations, agreeing on the declarations shared by $\gamma_1$ and $\gamma_2$ (automatic, since context permutations preserve the variable-mode labels), and let $\tau_1:\delta_1\cong\delta_1'$ and $\tau_2:\delta_2\cong\delta_2'$ be context permutations. Write $\pi_x:\gamma_2,x:p\cong\gamma_2',x:p$ and $\tau_y:\delta_1,y:p\cong\delta_1',y:p$ for the extensions fixing $x:p$ and $y:p$, and $\pi_1\cup\pi_2:\gamma_1\cup\gamma_2\cong\gamma_1'\cup\gamma_2'$ for the common extension. Then, whenever the composition is defined,
$$(\pi_1\cup\pi_2;\tau_1,\tau_2)_*(\mf D\circ_{y,x}\mf E)=((\pi_1;\tau_y)_*\mf D)\circ_{y,x}((\pi_x;\tau_2)_*\mf E).$$
\end{lemma}

\textit{Proof.}
The side conditions of \cref{def:composition-of-descriptors} mention only the declarations of the contexts and the count $|\mc T_{\mapsto y}|$. Permutation transport changes neither, so one side is defined if and only if the other is.
Using \cref{lem:fresh-representatives} and \cref{lem:existence-of-fresh-families-of-bijections}, we can choose representatives with disjoint internal variables and a fresh family of bijections.
Notice that permutation transport does not change any of the assignment set, redirection set, internal variables, occurrence set, or the chosen family of bijections. It only changes the contexts. Both sides are therefore represented by the same raw composite. The two descriptor classes are equal. \hfill $\square$

\begin{theorem}[Exchange Admissibility for Descriptor Judgments]\label{thm:exchange-admissibility-for-descriptor-judgments}
If $\gamma \Vdash_{\mf D} \delta$ is derivable and $\pi:\gamma\cong\gamma'$, $\tau:\delta\cong\delta'$ are context permutations, then
$$
\gamma' \Vdash_{(\pi;\tau)_*\mf D} \delta'
$$
is derivable.
\end{theorem}

\textit{Proof.}
By induction on the derivation. In the Identity, c-Comp, l-Pos-Comp, and l-Neg-Comp cases, the rule already allows an arbitrary context permutation in the conclusion. In the c-Comp case the output permutation is fixed to $\1_{r : q}$. This is no restriction, as the output context $r : q$ admits no other permutation. If the given conclusion was first obtained with permutations $\pi_0$ and $\tau_0$, applying the same rule with the composite permutations $\pi\circ\pi_0$ and $\tau\circ\tau_0$ gives then the exchanged conclusion. By \cref{prop:functoriality-of-permutation-transport}, the resulting descriptor is $(\pi;\tau)_*\mf D$.

It remains to treat the cut rule. Suppose the last rule is Cut, with premise judgments $\gamma_1\Vdash_{\mf E}\delta_1,y:p$ and $\gamma_2,x:p\Vdash_{\mf E'}\delta_2$, and that the conclusion is transported by $\pi_0:\gamma_1\cup\gamma_2\cong\gamma$ and $\tau_0:\delta_1,\delta_2\cong\delta$. Thus $\mf D=(\pi_0;\tau_0)_*(\mf E\circ_{y,x}\mf E')$. Applying the same Cut rule to the same two premises, but using the composite permutations $\pi\circ\pi_0$ and $\tau\circ\tau_0$, gives $\gamma'\Vdash_{((\pi\circ\pi_0);(\tau\circ\tau_0))_*(\mf E\circ_{y,x}\mf E')}\delta'$. By \cref{prop:functoriality-of-permutation-transport}, this descriptor is exactly $(\pi;\tau)_*\mf D$. Exchange is therefore admissible for all the rules in the calculus. \hfill $\square$
\endgroup{}

\begingroup

\subsection{Signatures and \texorpdfstring{$2$}{2}-Cells}
This section introduces $2$-cells and the structural operations acting on them. Descriptor exchange is established in \cref{thm:exchange-admissibility-for-descriptor-judgments}, and the remaining cartesian structural rules are treated in \cref{subsec:substructural-rules}. A $2$-cell can be thought of as a way to relate two descriptors. The informal reading is that if a descriptor $\mf D$ gives permission to prove some targets $\Delta$ from the resources $\Theta ; \Gamma$, and there exists a way to relate $\mf D$ to $\mf D'$ (denoted $\mf D' \Rightarrow \mf D$ for now), then it should be possible to derive the targets from the resources using $\mf D'$. This is the main technical tool for using substructural rules at the logical level.

\subsubsection{Structural Operations on Descriptor Classes}

\Cref{thm:exchange-admissibility-for-descriptor-judgments} establishes that descriptors admit exchange. This means that the order of input and output variables does not matter for the purposes of forming a descriptor. Indeed, since every descriptor represents a directed acyclic graph, the inputs and outputs behave like multisets. More specifically, at cartesian modes, the descriptors benefit from the substructural rules of weakening and contraction as well. We now prove those properties.

\begin{definition}[Input Weakening of Descriptor Classes]\label{def:input-weakening-of-descriptor-classes}
Let \(a:p\) be an input variable at a \emph{cartesian} mode \(p\), fresh for an input context \(\gamma\). If
$
(\mc S\mid\mc T)\in \mathrm{desc}^{\mathrm{raw}}_\Sigma(\gamma\ ;\ \delta),
$
then the same pair \((\mc S\mid\mc T)\) may be regarded as a raw descriptor in the enlarged input context:
$
(\mc S\mid\mc T)\in \mathrm{desc}^{\mathrm{raw}}_\Sigma(\gamma,a:p\ ;\ \delta).
$

We write
$
\mathrm{wk}^{a:p}_{\mathrm{in}}(\mc S\mid\mc T):=(\mc S\mid\mc T)
$
for this same raw pair regarded in the enlarged input context. If
$
\mf D=[\mc S\mid\mc T]\in \mathrm{Desc}^{\mathrm{raw}}_\Sigma(\gamma\ ;\ \delta),
$
define
$
\mathrm{wk}^{a:p}_{\mathrm{in}}(\mf D)
:=
[\mc S\mid\mc T]
\in
\mathrm{Desc}^{\mathrm{raw}}_\Sigma(\gamma,a:p\ ;\ \delta).
$
\end{definition}

\textit{Remark:}
Indeed, every assignment and redirection that uses input variables from \(\gamma\) also uses input variables from \(\gamma,a:p\).

\begin{example}[Linear--Nonlinear Logic]
For the class $\mf D = [\{x, x' \xmapsto{\otimes_l} w\} \mid \{w \mapsto y\}]$ of the multiplicative conjunction and a fresh cartesian variable $a : c$, the weakening $\mathrm{wk}^{a:c}_{\mathrm{in}}(\mf D)$ is the same pair of sets, regarded over $(x : l,\, x' : l,\, a : c \ ; \ y : l)$. The new declaration does not occur in the descriptor. Weakening by a declaration at a linear mode would not preserve regularity: the pair $(\varnothing \mid \{x \mapsto y\})$ regarded over $(x : l,\, x' : l \ ; \ y : l)$ fails the boundary usage property, as shown in the \hyperref[ex:linear-logic-regular-tensor-descriptor]{regularity example}.
\end{example}

\begin{lemma}[Input Weakening is Well Defined]\label{lem:input-weakening-is-well-defined}
The operation \(\mathrm{wk}^{a:p}_{\mathrm{in}}\) is well defined on descriptor classes. Moreover, it preserves regularity.
\end{lemma}

\textit{Proof.}
Suppose
$
(\mc S\mid\mc T)=_\alpha(\mc S'\mid\mc T')
$
in \(\mathrm{desc}^{\mathrm{raw}}_\Sigma(\gamma\ ;\ \delta)\), witnessed by a bijection
$
\phi:\mathrm{Int}(\mc S\mid\mc T)\to
\mathrm{Int}(\mc S'\mid\mc T')
$
such that
$
(\mc S\mid\mc T)^\phi=(\mc S'\mid\mc T').
$
The bijection \(\phi\) acts only on internal variables, and input weakening changes only the ambient input context. The same \(\phi\) witnesses therefore
$
\mathrm{wk}^{a:p}_{\mathrm{in}}(\mc S\mid\mc T)
=_\alpha
\mathrm{wk}^{a:p}_{\mathrm{in}}(\mc S'\mid\mc T')
$
in the enlarged context \(\gamma,a:p\ ;\ \delta\). Input weakening is hence well defined on descriptor classes.

Since the assignment and redirection sets remain unchanged by the weakening operation, regularity is preserved as well. Since the new variable $a : p$ does not occur in the descriptor, mode validity is unchanged at the descriptor. All the other regularity conditions are not concerned with the addition of the phantom variable and so remain true of the descriptor. 
\hfill \(\square\)

\begin{lemma}[Input Substitution by an Input is well defined for Descriptor Classes]\label{lem:input-substitution-by-an-input-is-well-defined-for-descriptor-classes}
Let $x:p$ and $x':p$ be distinct input variables of a common mode $p$, of either class, in an input context
$\gamma,x:p,x':p$. If
$$
\mf D=[\mc S\mid\mc T]
\in
\mathrm{Desc}^{\mathrm{raw}}_\Sigma(\gamma,x:p,x':p\ ;\ \delta),
$$
define
$$
\mf D[x':=x]
:=
[\mc S[x':=x]\mid \mc T[x':=x]]
\in
\mathrm{Desc}^{\mathrm{raw}}_\Sigma(\gamma,x:p\ ;\ \delta).
$$
Then this definition is independent of the chosen representative.
\end{lemma}

\textit{Proof.}
Suppose $(\mc S\mid\mc T)=_\alpha(\mc S'\mid\mc T')$ is witnessed by a
bijection of internal variables
$\phi:\mathrm{Int}(\mc S\mid\mc T)\to\mathrm{Int}(\mc S'\mid\mc T')$
such that $(\mc S\mid\mc T)^\phi=(\mc S'\mid\mc T')$. Since $\phi$ fixes
input variables, we have $x^\phi=x$. By
\cref{lem:input-substitution-and-internal-renaming-commute},
$$
\big((\mc S\mid\mc T)[x':=x]\big)^\phi
=
\big((\mc S\mid\mc T)^\phi\big)[x':=x]
=
(\mc S'\mid\mc T')[x':=x].
$$
Moreover, substituting one input variable by another input variable does not
change the set of internal variables; the same $\phi$ witnesses then
$(\mc S\mid\mc T)[x':=x]=_\alpha(\mc S'\mid\mc T')[x':=x]$, and
$\mf D[x':=x]$ is well defined. \hfill $\square$

We turn to the study and notation for input contraction. We introduce a map between descriptors that contracts two variables into one.
\begin{definition}[Input Contraction Notation]\label{def:input-contraction-notation}
Let $p$ be a \emph{cartesian} mode and
$$
\mf D\in
\mathrm{Desc}^{\mathrm{raw}}_\Sigma(\gamma,x:p,x':p\ ;\ \delta).
$$
The contraction of $x'$ into $x$ is
$$
\mathrm{ctr}_{x'\to x}(\mf D)
:=
\mf D[x':=x]
\in
\mathrm{Desc}^{\mathrm{raw}}_\Sigma(\gamma,x:p\ ;\ \delta).
$$
When $\mf D$ is well formed, preservation of well-formedness for this
operation is supplied by \cref{thm:input-contraction-admissibility-for-well-formed-descriptors}.
\end{definition}

\begin{example}[Linear--Nonlinear Logic]
For the class of the binary c-generator over $(a : c,\, a' : c \ ; \ y : c)$, the contraction of $a'$ into $a$ is
$$\mathrm{ctr}_{a'\to a}\big([\{a, a' \xmapsto{\otimes_c} w\} \mid \{w \mapsto y\}]\big) = [\{a, a \xmapsto{\otimes_c} w\} \mid \{w \mapsto y\}],$$
the descriptor over $(a : c \ ; \ y : c)$ that uses its cartesian input twice. The classes read as the expressions $\otimes_c(a, a')$ and $\otimes_c(a, a)$.
\end{example}

We record finally the renaming of boundary variables. The $\alpha$-equivalence of \cref{def:internal-renaming-equivalence-alpha-equivalence} identifies representatives up to their \emph{internal} names only; a descriptor class is therefore attached to its named boundary. Boundary renaming, with its admissibility, makes the silent renaming of boundary variables used throughout the metatheory an operation of the calculus.

\begin{definition}[Boundary Renaming]\label{def:boundary-renaming}
Let $(\mc S\mid\mc T)\in\mathrm{desc}^{\mathrm{raw}}_\Sigma(\gamma\ ;\ \delta)$ and let $\sigma$ be a mode-preserving injection defined on a subset of the declarations of $\gamma;\delta$, its values pairwise distinct and fresh for $\gamma;\delta$, extended by the identity elsewhere. The \emph{boundary renaming} of $(\mc S\mid\mc T)$ along $\sigma$ is the raw descriptor
$$
\mathrm{ren}_{\sigma}(\mc S\mid\mc T)
\in
\mathrm{desc}^{\mathrm{raw}}_\Sigma(\gamma^{\sigma}\ ;\ \delta^{\sigma})
$$
obtained by replacing every occurrence of an input $x\in\gamma$ by $\sigma x$ and the target of every redirection $s\mapsto y$ with $y\in\delta$ by $\sigma y$, the contexts renamed declarationwise. We write $\mathrm{ren}_{x\mapsto a}$ for the renaming of the single declaration $x$ into $a$, and $\mathrm{ren}_{\sigma}(\mf D):=[\mathrm{ren}_{\sigma}(\mc S\mid\mc T)]$ on classes. On an input the operation is the input substitution of \cref{def:input-substitution} at a fresh variable value, $\mathrm{ren}_{x\mapsto a}(\mf D)=\mf D[x:=a]$. On an output it is the retargeting $\mf D[y:=b]$ of its redirection.
\end{definition}

\begin{example}[Linear--Nonlinear Logic]
For the class $\mf D$ of the multiplicative conjunction over $(x : l,\, x' : l \ ; \ y : l)$ and the renaming $\sigma$ sending the input $x$ to $x_1$ and the output $y$ to $y_1$, both values fresh, we get
$$\mathrm{ren}_{\sigma}(\mf D) = [\{x_1, x' \xmapsto{\otimes_l} w\} \mid \{w \mapsto y_1\}] \in \mathrm{Desc}^{\mathrm{raw}}_\Sigma(x_1 : l,\, x' : l \ ; \ y_1 : l).$$
On the input side the source positions carrying $x$ are replaced; on the output side only the target of the redirection changes. The declaration $x' : l$ and the internal variable $w$ keep their names: the renaming acts on the boundary, while the $\alpha$-equivalence identifies the internal names.
\end{example}

\begin{lemma}[Boundary Renaming is Well Defined]\label{lem:boundary-renaming-is-well-defined}
The operation $\mathrm{ren}_{\sigma}$ is well defined on descriptor classes and preserves regularity. Moreover:
\begin{itemize}
\item[(i)] (\emph{Functoriality.}) $\mathrm{ren}_{\mathrm{id}}(\mf D)=\mf D$, and $\mathrm{ren}_{\sigma'}(\mathrm{ren}_{\sigma}(\mf D))=\mathrm{ren}_{\sigma'\circ\sigma}(\mf D)$ whenever the composite renaming is defined.
\item[(ii)] (\emph{Transport and weakening.}) $\mathrm{ren}_{\sigma}((\pi;\tau)_*\mf D)=(\pi^{\sigma};\tau^{\sigma})_*(\mathrm{ren}_{\sigma}\mf D)$ for the conjugated permutations, and $\mathrm{ren}_{\sigma}(\mathrm{wk}^{a:p}_{\mathrm{in}}(\mf D))=\mathrm{wk}^{a:p}_{\mathrm{in}}(\mathrm{ren}_{\sigma}(\mf D))$ for a fresh $a$ untouched by $\sigma$.
\item[(iii)] (\emph{Composition.}) If $\mf D\circ_{y,x}\mf E$ is defined and $\sigma$ is a boundary renaming of the composite, extended by the identity on the variable, then
$$\mathrm{ren}_{\sigma}(\mf D\circ_{y,x}\mf E)=\mathrm{ren}_{\sigma_1}(\mf D)\circ_{y,x}\mathrm{ren}_{\sigma_2}(\mf E),$$
where $\sigma_1,\sigma_2$ are the restrictions of $\sigma$ to the two factor boundaries.
\end{itemize}
\end{lemma}

\textit{Proof.}
An internal renaming $\phi$ acts on internal variables and $\sigma$ on boundary occurrences; the two act therefore on disjoint variables and commute elementwise (\cref{lem:input-substitution-and-internal-renaming-commute}). The same $\phi$ witnesses the equivalence of the renamed representatives and $\mathrm{ren}_{\sigma}$ is well defined on classes. Regularity is preserved since a bijective relabelling of the boundary transports every occurrence count and every dependency edge, and $\sigma$ preserves the declared modes. Functoriality and the commutations of (ii) hold elementwise, permutation transport and weakening leaving the assignment and redirection sets untouched. For (iii), the composite is a function of the two representatives, the occurrence set of the cut input, the source of the removed redirection, and the fresh family alone (\cref{def:raw-composition}). The renaming fixes the variable, so the occurrence set and the removed redirection are those of the unrenamed factors, and every other boundary occurrence, in a factor, in its copies, or in a surviving redirection, is renamed identically on the two sides --- including the source of the removed redirection when it is an input of the left factor, renamed before or after the splice to the same effect. \hfill $\square$

\begin{theorem}[Renaming Admissibility for Descriptor Judgments]\label{thm:renaming-admissibility-for-descriptor-judgments}
If $\gamma\Vdash_{\mf D}\delta$ is derivable and $\sigma$ is a boundary renaming as in \cref{def:boundary-renaming}, then
$$\gamma^{\sigma}\Vdash_{\mathrm{ren}_{\sigma}(\mf D)}\delta^{\sigma}$$
is derivable.
\end{theorem}

\textit{Proof.}
By induction on the derivation, over the five generation rules. Every rule declares its boundary names as instance data and closes with a permutation transport.
For Identity, the instance at the renamed variables concludes $[\varnothing\mid\{\sigma x\mapsto\sigma y\}]=\mathrm{ren}_{\sigma}([\varnothing\mid\{x\mapsto y\}])$. For c-Comp, l-Pos-Comp and l-Neg-Comp, the premises are renamed by the induction hypothesis along the restrictions of $\sigma$. The same rules reapply on the renamed variables and transports. Each clause of the construction touches the boundary names elementwise; the canonical conclusion is therefore the renaming of the original one. For Cut, the story is similar. The induction hypothesis renames the two premises along the restrictions of $\sigma$. \cref{lem:boundary-renaming-is-well-defined}(ii),(iii) identify the conclusion of the re-applied rule.  \hfill $\square$

These definitions yield weakening, contraction, and exchange admissibility for well-formed descriptors in \cref{thm:input-weakening-admissibility-for-well-formed-descriptors,thm:input-contraction-admissibility-for-well-formed-descriptors,thm:exchange-admissibility-for-descriptor-judgments}. We re-emphasize that those substructural rules are not to be understood as the ones in the subsequent logics. In the cartesian case, the presence of 2-cells will encode the substructural information of the logics in question. In the linear case, the 2-cell mechanism cannot be used without collapsing the intended semantics to a cartesian structure. Instead, we define special gate operators that transform descriptors to their respective weakened or contracted contexts. Exchange itself is admissible in all cases.

\subsubsection{Signatures and Descriptor \texorpdfstring{$2$}{2}-Cells}

We define now signatures and the notion of $2$-cell for descriptors. The $2$-cells will specify the substructural rules available in the calculus at cartesian modes. The structural behaviour at linear modes will be declared through gates. A signature consists of an underlying pre-signature, which supplies the classified modes and generator symbols, together with a collection of generating $2$-cells between descriptor classes, an optional set of structural gates, and an optional collection of equations between parallel $2$-cell terms. The collection of generating $2$-cells is required to be closed under context permutations, i.e. if a generating $2$-cell is present in contexts $\gamma,\delta$, then its transport is also present in any permutation-equivalent contexts $\gamma',\delta'$. In the same vein, the generating 2-cells are also required to be closed under weakening, contraction and boundary renaming.

\begin{definition}[Signature]\label{def:signature}
  A signature $\Lambda \ \mathrm{Sig}$ consists of:
  \begin{itemize}
    \item a pre-signature $\underline{\Lambda} \ \mathrm{sig}$, called the underlying pre-signature of $\Lambda$;
    \item for every pair of contexts
    $\gamma \ \mathrm{ctx}^{\mathrm{in}}_{\underline{\Lambda}}$ and $
    \delta \ \mathrm{ctx}^{\mathrm{out}}_{\underline{\Lambda}},$ a collection
    $$\mathrm{Gen}_\Lambda(\gamma;\delta)$$
    of generating $2$-cell symbols in context $\gamma\Vdash\delta$, equipped with source and target functions
    $$
    s,t:\mathrm{Gen}_\Lambda(\gamma;\delta)
    \to
    \mathrm{Desc}_{\underline{\Lambda}}(\gamma\ ;\ \delta);
    $$
    \item an optional set of \emph{structural gates} at linear modes,
    $$\mathrm{Gate}_\Lambda \subseteq \{\mathsf{weak}(l),\ \mathsf{contr}(l),\ \mathsf{coweak}(l),\ \mathsf{cocontr}(l) \mid l \model \in \underline{\Lambda}\},$$
    where declaring a gate adjoins the indicated l-generator to the underlying pre-signature;
    $\mathsf{weak}(l)$ adjoins $w_l : l \to \cdot$,\ \ $\mathsf{contr}(l)$ adjoins $c_l : l \to l,l$,\ \ $\mathsf{coweak}(l)$ adjoins $w^\perp_l : \cdot \to l$,\ \ and $\mathsf{cocontr}(l)$ adjoins $c^\perp_l : l,l \to l$;
    \item an optional collection $E_\Lambda$ of \emph{equations between parallel $2$-cell terms}, pairs of terms of the calculus generated by the listed data (\cref{def:descriptor-2-cells}) sharing their contexts, their source and their target; the equations are required to be closed termwise under the four structural transports.
  \end{itemize}

  The generating $2$-cells are required to satisfy four closure conditions.

	\begin{itemize}
		\item Generating $2$-cells are closed under context permutations: if $\rho:\mf D \to_{\gamma \Vdash \delta} \mf D' \in \Lambda,$ and if $\pi:\gamma\cong\gamma'$, $\tau:\delta\cong\delta',$ then there is a transported generator $$(\pi;\tau)_*\rho:(\pi;\tau)_*\mf D \to_{\gamma' \Vdash \delta'}(\pi;\tau)_*\mf D' \in \Lambda.$$
		\item Generating $2$-cells are closed under cartesian input weakening: if $\rho:\mf D \to_{\gamma \Vdash \delta} \mf D' \in \Lambda$ and $a:p$ is an input variable at a cartesian mode $p$, fresh for $\gamma$, then there is a transported generator $$\mathrm{wk}^{a:p}_{\mathrm{in}}(\rho): \mathrm{wk}^{a:p}_{\mathrm{in}}(\mf D) \to_{\gamma, a : p \Vdash \delta}  \mathrm{wk}^{a:p}_{\mathrm{in}}(\mf D') \in \Lambda .$$
		\item Generating $2$-cells are closed under cartesian contraction: if
		$$
		\rho:\mf D\to_{\gamma,x:p,x':p\Vdash\delta}\mf D'\in\Lambda
		$$
		with $p$ a cartesian mode, then there is a transported contraction generator
		$$
		\mathrm{ctr}_{x'\to x}(\rho):
		\mathrm{ctr}_{x'\to x}(\mf D)
		\to_{\gamma,x:p\Vdash\delta}
		\mathrm{ctr}_{x'\to x}(\mf D')
		\in\Lambda.
		$$
		\item Generating $2$-cells are closed under boundary renaming: if $\rho:\mf D \to_{\gamma \Vdash \delta} \mf D' \in \Lambda$ and $\sigma$ is a boundary renaming as in \cref{def:boundary-renaming}, then there is a transported generator
		$$
		\mathrm{ren}_{\sigma}(\rho):
		\mathrm{ren}_{\sigma}(\mf D)
		\to_{\gamma^{\sigma}\Vdash\delta^{\sigma}}
		\mathrm{ren}_{\sigma}(\mf D')
		\in\Lambda.
		$$
		\end{itemize}
\end{definition}
We write
$  \rho:\mf D \to_{\gamma \Vdash \delta} \mf D' \in \Lambda
$  to mean
$  \rho\in\mathrm{Gen}_\Lambda(\gamma;\delta)$ with $s(\rho)=\mf D$ and $t(\rho)=\mf D'.$
Notice that the displayed typing of $\mathrm{Gen}_\Lambda(\gamma;\delta)$ requires $\mf D$ and $\mf D'$ to be well-formed descriptor classes over $\underline{\Lambda}$, in the same ambient contexts $\gamma$ and $\delta$. We continue to omit $\underline{\Lambda}$ from descriptor judgments when it is clear from the ambient signature.

\textit{Remark:} The weakening and contraction closures of \cref{def:signature} presuppose that cartesian input weakening and cartesian contraction preserve well-formedness of descriptor classes. This preservation is established in \cref{subsec:substructural-rules}. The weakening and contraction closures are stated at cartesian modes only. At linear modes the boundary is used exactly; no such closure is imposed there. Context permutations and boundary renamings act at every mode, neither operation changing which variables are consumed, the renaming closure resting on \cref{thm:renaming-admissibility-for-descriptor-judgments}. Outputs are never weakened or contracted; every output is produced exactly once by the boundary usage property. In addition, the Cut rule keeps the output contexts of its premises disjoint.

\phantomsection\label{anchor:ungated-four-generator-signature}
\textit{Remark (gates are declarations):} We are careful again to draw the distinction between the existence of the gates and the substructural properties the generators encode. A gate is data of the signature and not a property of its generators; a signature may contain generators of the displayed arities without declaring any gate. For instance, we could have one linear mode $l$ and the four l-generators $m:l,l\to l$, $u:\cdot\to l$, $w:l\to l,l$, $e:l\to\cdot$ without any gates on them, which would ensure the subsequent logic's deprivation of substructural rules. A declared gate additionally unlocks a structural sequent rule of the type theory and carries an axiom package of generating $2$-cells, specified in \cref{def:gated-structural-rules,def:gate-coherence-laws}.

\begin{example}[Weakening as a generating $2$-cell]\label{ex:weakening-generating-2-cell}
Suppose that the underlying pre-signature contains a cartesian mode $c$ and the binary c-generator $\otimes_c:c,c\to c$. Let $\gamma:=a:c,a':c$ and $\delta:=z:c$. In these contexts define
$$
\mf D_{\otimes}:=\left[\{a,a'\xmapsto{\otimes_c}w\}\mid\{w\mapsto z\}\right],
\qquad
\mf D_a:=\left[\varnothing\mid\{a\mapsto z\}\right].
$$
Both descriptor classes are well formed in the same ambient context $\gamma\Vdash\delta$. Indeed, $\mf D_{\otimes}$ is generated by c-Comp from the identity descriptors $[\varnothing\mid\{a\mapsto u\}]$ and $[\varnothing\mid\{a'\mapsto v\}]$ for fresh auxiliary outputs $u:c$ and $v:c$. The descriptor $\mf D_a$ is generated by Identity, with $a':c$ present as an unused ambient cartesian input. A signature may therefore include the generating $2$-cell
$$
\rho_{\otimes,a}:\mf D_{\otimes}\to_{\gamma\Vdash\delta}\mf D_a\in\Lambda.
$$
This is the descriptor-level content of weakening/projection  $a\otimes_c a'\Rightarrow a$. At a linear mode no such $2$-cell exists since a well-formed descriptor uses its linear boundary exactly; two descriptors over a common boundary cannot therefore differ in which linear inputs they consume. Substructural behaviour at linear modes enters through the gates instead. Note that without the presence of $\rho_{\otimes,a}$ in the signature, the mode $c$ would not possess weakening as a substructural rule in the subsequent logic, even though it is labelled ``cartesian'' at the framework level. This means in particular that cartesian modes have the ability to inherit substructural rules in their generated logic, and this ability is realized whenever the corresponding 2-cell is present in the signature. This unlocks the possibility of encoding substructural intuitionistic types of logic in the cartesian part of the framework (using cartesian descriptors).
\end{example}

\textit{Remark:} This is the main mechanism for adding substructural rules at cartesian modes in the consequent logic. The 2-cells dictate which sequents are derivable from which. Since the 2-cells are defined between descriptors, sequent manipulation remains purely structural.

\begin{definition}[Descriptor $2$-cells]\label{def:descriptor-2-cells}

  Given a signature $\Lambda \ \mathrm{Sig}$, a descriptor $2$-cell is a term judgment
  $$r :: \mf D \Rightarrow^{\Lambda}_{\gamma \Vdash \delta} \mf D'$$
  generated by four rules over the underlying pre-signature $\underline{\Lambda}$.

  $$
  \inferrule*[Right=2-Form]
  {
		\rho:\mf D \to_{\gamma \Vdash \delta} \mf D' \in \Lambda
  }
  {
    \mathrm{gen}(\rho) :: \mf D \Rightarrow^{\Lambda}_{\gamma \Vdash \delta} \mf D'
  }
  \quad\quad\quad\quad
  \inferrule*[Right=2-Identity]
  {
    \gamma \Vdash_{\mf D} \delta
  }
  {
    \mathbf{1}_{\mf D} :: \mf D \Rightarrow^{\Lambda}_{\gamma \Vdash \delta} \mf D
  }
  $$
  $$
  \inferrule*[Right=2-Comp]
  {
		r :: \mf D \Rightarrow^{\Lambda}_{\gamma \Vdash \delta} \mf D' \\
		s :: \mf D' \Rightarrow^{\Lambda}_{\gamma \Vdash \delta} \mf D''
  }
	  {
			s\circ r :: \mf D \Rightarrow^{\Lambda}_{\gamma \Vdash \delta} \mf D''
	  }
	  $$
	  $$
	  \inferrule*[Right=2-Subst]
	  {
			r :: \mf D \Rightarrow^{\Lambda}_{\gamma_1 \Vdash \delta_1, y : p} \mf D' \\
			s :: \mf E \Rightarrow^{\Lambda}_{\gamma_2, x : p \Vdash \delta_2} \mf E' \\
			\gamma_1, \gamma_2 \ \text{agree on their common declarations, all at cartesian modes} \\
			x \notin \gamma_1, \quad y \notin \delta_2 \\
			\delta_1, \delta_2 \ \text{disjoint} \\
			\pi:\gamma_1\cup\gamma_2\cong\gamma' \\
			\tau:\delta_1,\delta_2\cong\delta'
  }
  {
		(\pi;\tau)_*(r\circ_{y,x}s) :: (\pi;\tau)_*(\mf D \circ_{y,x} \mf E) \Rightarrow^{\Lambda}_{\gamma' \Vdash \delta'} (\pi;\tau)_*(\mf D' \circ_{y,x} \mf E')
  }
  $$

\end{definition}
\textit{Remark:}
The closing permutations of $2$-Subst mirror those of the Cut rule (\cref{def:well-formed-composition-cut}). They are instance data of the rule and not a term former, and we write $r\circ_{y,x}s$ for the instance closed by the identity permutations. When the particular term is irrelevant, we write
$\mf D \Rightarrow^{\Lambda}_{\gamma \Vdash \delta} \mf D'$
to mean that there exists a term $r :: \mf D \Rightarrow^{\Lambda}_{\gamma \Vdash \delta} \mf D'$.

\textit{Remark:} The mode theory generated by a signature $\Lambda$ consists therefore, in each pair
of contexts $\gamma;\delta$ over $\underline{\Lambda}$, of the well-formed descriptor classes
$\mathrm{Desc}_{\underline{\Lambda}}(\gamma;\delta)$ together with the generated descriptor $2$-cell
terms between them. The descriptor classes provide the structural shapes of sequents. The
$2$-cells specify the admissible structural transformations. The proof-irrelevant relation $\mf D
\Rightarrow^\Lambda_{\gamma\Vdash\delta}\mf D'$ is recovered by forgetting the term; only existence
is retained.

\textit{Remark:}	In the rule $2$-Subst, the composites in the conclusion are defined. Indeed, the sources and targets of the premises are well formed by \cref{prop:source-and-target-of-descriptor-2-cells-are-well-formed}, whose proof proceeds by the same induction. They are hence regular by \cref{thm:well-formed-descriptors-are-regular}, and the definedness hypothesis $|\mc T_{\mapsto y}|=1$ of \cref{def:composition-of-descriptors} follows from the boundary usage property. The term $r\circ_{y,x}s$ is the substitution of descriptor $2$-cell terms along this cut.

\begin{example}[Linear--Nonlinear Logic]\label{ex:linear-logic-descriptor-2-cell-terms}
Work over a signature $\Lambda$ declaring the generating $2$-cell $\rho_{\otimes,a} : \mf D_{\otimes} \to_{\gamma \Vdash \delta} \mf D_a$ of the \hyperref[ex:weakening-generating-2-cell]{weakening example}. The rule $2$-Form derives the term $\mathrm{gen}(\rho_{\otimes,a}) :: \mf D_{\otimes} \Rightarrow^{\Lambda}_{\gamma \Vdash \delta} \mf D_a$, and $2$-Identity provides $\mathbf{1}_{\mf D_{\otimes}}$ and $\mathbf{1}_{\mf D_a}$. The rule $2$-Comp forms the composite $\mathbf{1}_{\mf D_a} \circ \mathrm{gen}(\rho_{\otimes,a}) :: \mf D_{\otimes} \Rightarrow^{\Lambda}_{\gamma \Vdash \delta} \mf D_a$, a term distinct from $\mathrm{gen}(\rho_{\otimes,a})$ with the same source and target. For $2$-Subst, take the class $\mf D_c := [\{b, b' \xmapsto{\otimes_c} u\} \mid \{u \mapsto z'\}]$ of the conjunction of $b$ and $b'$ over $(b : c,\, b' : c \ ; \ z' : c)$ and cut its output $z'$ against the input $a'$. Closed by the identity permutations, the rule derives
$$\mathbf{1}_{\mf D_c} \circ_{z',a'} \mathrm{gen}(\rho_{\otimes,a}) :: \mf D_c \circ_{z',a'} \mf D_{\otimes} \Rightarrow^{\Lambda}_{b : c,\, b' : c,\, a : c \Vdash z : c} \mf D_c \circ_{z',a'} \mf D_a.$$
The source composite is $[\{b, b' \xmapsto{\otimes_c} u,\ \ a, u \xmapsto{\otimes_c} w\} \mid \{w \mapsto z\}]$ and reads as the expression $\otimes_c(a, \otimes_c(b, b'))$. In the target, the input $a'$ has no occurrence in $\mf D_a$, no copy of $\mf D_c$ is made, and the composite is $[\varnothing \mid \{a \mapsto z\}]$ itself, reading as $a$. The substituted $2$-cell discards the expression it substitutes.
\end{example}

\begin{proposition}[Source and Target of Descriptor $2$-cells are Well Formed]\label{prop:source-and-target-of-descriptor-2-cells-are-well-formed}
Let $\Lambda \ \mathrm{Sig}$. If $$r :: \mf D \Rightarrow^{\Lambda}_{\gamma \Vdash \delta} \mf D'$$ is derivable, then both descriptor judgments $$\gamma \Vdash_{\mf D} \delta \qquad\text{and}\qquad \gamma \Vdash_{\mf D'} \delta$$ are derivable.
\end{proposition}

\textit{Proof.}
We proceed by induction on the derivation of $r :: \mf D \Rightarrow^{\Lambda}_{\gamma \Vdash \delta} \mf D'.$

\begin{itemize}
	\item If the last rule is $2$-Form, then $\rho:\mf D \to_{\gamma \Vdash \delta} \mf D'\in\Lambda.$ Recall that generating $2$-cells in $\Lambda$ are only allowed between well-formed descriptors (\cref{def:signature}). Hence $\gamma \Vdash_{\mf D} \delta$ and $\gamma \Vdash_{\mf D'} \delta$ are derivable.

	\item If the last rule is $2$-Identity, then the premise is exactly $\gamma \Vdash_{\mf D} \delta,$ and the source and target descriptors are both $\mf D$.

	\item If the last rule is $2$-Comp, then we have premises $r_1 :: \mf D \Rightarrow^{\Lambda}_{\gamma \Vdash \delta} \mf D'$ and $r_2 :: \mf D' \Rightarrow^{\Lambda}_{\gamma \Vdash \delta} \mf D''.$ By the induction hypothesis applied to the first premise, $\gamma \Vdash_{\mf D} \delta.$ The second premise gives $\gamma \Vdash_{\mf D''} \delta.$ Thus the source and target descriptors of the composite $2$-cell are well formed.

	\item If the last rule is $2$-Subst, then the conclusion has the form $(\pi;\tau)_*(r_1\circ_{y,x}r_2) :: (\pi;\tau)_*(\mf D
\circ_{y,x} \mf E) \Rightarrow^{\Lambda}_{\gamma' \Vdash \delta'} (\pi;\tau)_*(\mf D' \circ_{y,x} \mf E').$ By
the induction hypothesis applied to $r_1$ and $r_2$ we obtain $\gamma_1 \Vdash_{\mf D} \delta_1,y:p $ and $
\gamma_2,x:p \Vdash_{\mf E} \delta_2,$ as well as $\gamma_1 \Vdash_{\mf D'} \delta_1,y:p $ and $
\gamma_2,x:p \Vdash_{\mf E'} \delta_2.$ These descriptors are regular by \cref{thm:well-formed-descriptors-are-regular}, so $|\mc T_{\mapsto y}|=1$ holds for $\mf D$ and $\mf D'$ by the boundary usage property. The side conditions of $2$-Subst supply the remaining premises of the Cut rule. Applying the Cut rule with the closing permutations $\pi$ and $\tau$ of the $2$-Subst instance to the
first pair gives $\gamma' \Vdash_{(\pi;\tau)_*(\mf D\circ_{y,x}\mf E)} \delta'$; the second pair gives $\gamma' \Vdash_{(\pi;\tau)_*(\mf
D'\circ_{y,x}\mf E')} \delta'.$ Thus both the source composite and the target composite are
well formed.
\end{itemize}\hfill $\square$

\begin{theorem}[Exchange Admissibility for Descriptor $2$-cells]\label{thm:exchange-admissibility-for-descriptor-2-cells}
Let $\Lambda \ \mathrm{Sig}$. If $$r :: \mf D \Rightarrow^{\Lambda}_{\gamma \Vdash \delta} \mf D'$$ is derivable and $\pi:\gamma\cong\gamma_1$, $\tau:\delta\cong\delta_1$ are context permutations, then there is a transported term, denoted
$$(\pi;\tau)_*r :: (\pi;\tau)_*\mf D \Rightarrow^{\Lambda}_{\gamma_1 \Vdash \delta_1}(\pi;\tau)_*\mf D'.$$
\end{theorem}

\textit{Proof.}
We proceed by induction on the derivation of $r :: \mf D \Rightarrow^{\Lambda}_{\gamma \Vdash \delta} \mf D'$.

\begin{itemize}
	\item If the last rule is $2$-Form, then $r=\mathrm{gen}(\rho)$ with $\rho:\mf D \to_{\gamma \Vdash \delta}\mf D'\in\Lambda$. By closure of generating $2$-cells under context permutations, there is a transported generator
  $$(\pi;\tau)_*\rho:(\pi;\tau)_*\mf D \to_{\gamma_1 \Vdash \delta_1}(\pi;\tau)_*\mf D'\in\Lambda,$$
  and the result follows by $2$-Form, with transported term $\mathrm{gen}((\pi;\tau)_*\rho)$.
	\item If the last rule is $2$-Identity, then $r=\mathbf{1}_{\mf D}$ and $\gamma\Vdash_{\mf D}\delta$. By \cref{thm:exchange-admissibility-for-descriptor-judgments}, $\gamma_1\Vdash_{(\pi;\tau)_*\mf D}\delta_1$. The result follows by $2$-Identity, with transported term $\mathbf{1}_{(\pi;\tau)_*\mf D}$.
	\item If the last rule is $2$-Comp, say $r=r_2\circ r_1$, the result follows by applying the induction hypothesis to $r_1$ and $r_2$ and then applying $2$-Comp.
	\item If the last rule is $2$-Subst, then $r=(\pi_0;\tau_0)_*(r_1\circ_{y,x}r_2)$ with premises
  $$r_1 :: \mf D \Rightarrow^{\Lambda}_{\gamma_1' \Vdash \delta_1',y:p} \mf D',
  \qquad
  r_2 :: \mf E \Rightarrow^{\Lambda}_{\gamma_2',x:p \Vdash \delta_2'} \mf E',
  $$
  and closing permutations $\pi_0:\gamma_1'\cup\gamma_2'\cong\gamma$ and $\tau_0:\delta_1',\delta_2'\cong\delta$, so that the source of $r$ is $(\pi_0;\tau_0)_*(\mf D\circ_{y,x}\mf E)$ and its target is $(\pi_0;\tau_0)_*(\mf D'\circ_{y,x}\mf E')$. Applying the same rule to the same two premises, with the composite closing permutations $\pi\circ\pi_0$ and $\tau\circ\tau_0$, yields a term over $\gamma_1\Vdash\delta_1$. By \cref{prop:functoriality-of-permutation-transport} its source is $((\pi\circ\pi_0);(\tau\circ\tau_0))_*(\mf D\circ_{y,x}\mf E)=(\pi;\tau)_*\bigl((\pi_0;\tau_0)_*(\mf D\circ_{y,x}\mf E)\bigr)$, and likewise for its target. It is therefore a transported term of the required type, exactly as in the Cut case of \cref{thm:exchange-admissibility-for-descriptor-judgments}.
\end{itemize}
\hfill $\square$

\begin{lemma}[Weakening commutes with composition]\label{lem:weakening-commutes-with-composition}
Let $(\mc S\mid\mc T)\in\mathrm{desc}^{\mathrm{raw}}_\Sigma(\gamma_1\ ;\ \delta_1,y:p)$ and
$(\mc S'\mid\mc T')\in\mathrm{desc}^{\mathrm{raw}}_\Sigma(\gamma_2,x:p\ ;\ \delta_2)$ be composable at
$(y,x)$, with fresh family $\bs\beta$ as in \cref{def:raw-composition}. If $a:q$ is an input
variable at a cartesian mode $q$, fresh for $\gamma_1\cup\gamma_2$, then
$$
\mathrm{wk}^{a:q}_{\mathrm{in}}\!\big((\mc S\mid\mc T)\circ^{\bs\beta}_{y,x}(\mc S'\mid\mc T')\big)
=\mathrm{wk}^{a:q}_{\mathrm{in}}(\mc S\mid\mc T)\ \circ^{\bs\beta}_{y,x}\ \mathrm{wk}^{a:q}_{\mathrm{in}}(\mc S'\mid\mc T').
$$
\end{lemma}

\textit{Proof.} By \cref{def:raw-composition} the composite is a function of the assignment and
redirection sets $\mc S,\mc T,\mc S',\mc T'$, of the occurrence set $\mathrm{Occ}_x(\mc S'\mid\mc
T')$, of the cut source $s$, and of the fresh family $\bs\beta$ alone. Input weakening leaves all
of these unchanged. It enlarges only the ambient input context by the fresh unused variable $a$.
Both sides are hence the same raw descriptor over the enlarged ambient
$(\gamma_1\cup\gamma_2,a:q\ ;\ \delta_1,\delta_2)$.
\hfill$\square$
\begin{lemma}[Ambient weakening of descriptor $2$-cells]\label{lem:ambient-weakening-of-descriptor-2-cells}
Let $r::\mf D\Rightarrow^{\Lambda}_{\gamma\Vdash\delta}\mf E$ be a descriptor $2$-cell. For every input variable $a:p$ at a cartesian mode $p$, fresh for $\gamma$, there is a descriptor $2$-cell
$$\mathrm{wk}^{a:p}_{\mathrm{in}}(r)::\mathrm{wk}^{a:p}_{\mathrm{in}}(\mf D)\Rightarrow^{\Lambda}_{\gamma,a:p\Vdash\delta}\mathrm{wk}^{a:p}_{\mathrm{in}}(\mf E).$$
The operation preserves identities, $2$-Comp, and $2$-Subst.
\end{lemma}

\textit{Proof.} We construct $\mathrm{wk}^{a:p}_{\mathrm{in}}(r)$ by recursion on the derivation of
$r$ (\cref{def:descriptor-2-cells}).

\begin{itemize}
\item \textbf{2-Form} If $r=\mathrm{gen}(\rho)$ with $\rho:\mf D\to_{\gamma\Vdash\delta}\mf D'\in\Lambda$, then by the cartesian input weakening closure of \cref{def:signature}, the transported generator
$\mathrm{wk}^{a:p}_{\mathrm{in}}(\rho):\mathrm{wk}^{a:p}_{\mathrm{in}}(\mf D)\to_{\gamma,a:p\Vdash\delta}\mathrm{wk}^{a:p}_{\mathrm{in}}(\mf D')$
is present in the signature $\Lambda$. We set $\mathrm{wk}^{a:p}_{\mathrm{in}}(\mathrm{gen}(\rho)):=\mathrm{gen}\big(\mathrm{wk}^{a:p}_{\mathrm{in}}(\rho)\big)$; this is a $2$-cell of the required type by 2-Form.

\item \textbf{2-Identity} If $r=\mathbf 1_{\mf D}$, then by
\cref{thm:input-weakening-admissibility-for-well-formed-descriptors},
$\gamma,a:p\Vdash_{\mathrm{wk}^{a:p}_{\mathrm{in}}(\mf D)}\delta$, and 2-Identity gives then
$\mathbf 1_{\mathrm{wk}^{a:p}_{\mathrm{in}}(\mf D)}$. Set
$\mathrm{wk}^{a:p}_{\mathrm{in}}(\mathbf 1_{\mf D}):=\mathbf 1_{\mathrm{wk}^{a:p}_{\mathrm{in}}(\mf D)}$.

\item \textbf{2-Comp} If $r=s\circ t$ with $t::\mf D\Rightarrow\mf D''$, $s::\mf D''\Rightarrow\mf E$ over
$\gamma\Vdash\delta$, then by the induction hypothesis $\mathrm{wk}^{a:p}_{\mathrm{in}}(t)$ and
$\mathrm{wk}^{a:p}_{\mathrm{in}}(s)$ are $2$-cells over $\gamma,a:p\Vdash\delta$ with matching
middle descriptor $\mathrm{wk}^{a:p}_{\mathrm{in}}(\mf D'')$. We set
$\mathrm{wk}^{a:p}_{\mathrm{in}}(s\circ t):=\mathrm{wk}^{a:p}_{\mathrm{in}}(s)\circ\mathrm{wk}^{a:p}_{\mathrm{in}}(t)$,
which is well typed by 2-Comp.

\item \textbf{2-Subst} If $r=(\pi;\tau)_*(t\circ_{y,x}s)$ with
$t::\mf D_1\Rightarrow_{\gamma_1\Vdash\delta_1,y:q'}\mf D_1'$ and
$s::\mf D_2\Rightarrow_{\gamma_2,x:q'\Vdash\delta_2}\mf D_2'$ and closing permutations
$\pi:\gamma_1\cup\gamma_2\cong\gamma$ and $\tau:\delta_1,\delta_2\cong\delta$, then with
$\mf D=(\pi;\tau)_*(\mf D_1\circ_{y,x}\mf D_2)$, $\mf E=(\pi;\tau)_*(\mf D_1'\circ_{y,x}\mf D_2')$,
and $a:p$ fresh for $\gamma_1\cup\gamma_2$, we apply the induction hypothesis to get
$$
\begin{aligned}
\mathrm{wk}^{a:p}_{\mathrm{in}}(t)&::\mathrm{wk}^{a:p}_{\mathrm{in}}(\mf D_1)\Rightarrow_{\gamma_1,a:p\Vdash\delta_1,y:q'}\mathrm{wk}^{a:p}_{\mathrm{in}}(\mf D_1'),
\\
\mathrm{wk}^{a:p}_{\mathrm{in}}(s)&::\mathrm{wk}^{a:p}_{\mathrm{in}}(\mf D_2)\Rightarrow_{\gamma_2,a:p,x:q'\Vdash\delta_2}\mathrm{wk}^{a:p}_{\mathrm{in}}(\mf D_2').
\end{aligned}
$$
The declarations shared by $\gamma_1,a:p$ and $\gamma_2,a:p$ are the old shared declarations together with $a:p$, all at cartesian modes. The output contexts $\delta_1,\delta_2$ remain disjoint since input
weakening does not touch outputs, so 2-Subst applies. Write $\pi_a:\gamma_1\cup\gamma_2,a:p\cong\gamma,a:p$ for the extension of $\pi$ fixing $a$. Closing the instance by $(\pi_a;\tau)$ yields
$$
\begin{aligned}
(\pi_a;\tau)_*\bigl(\mathrm{wk}^{a:p}_{\mathrm{in}}(t)\circ_{y,x}\mathrm{wk}^{a:p}_{\mathrm{in}}(s)\bigr)&::
(\pi_a;\tau)_*\bigl(\mathrm{wk}^{a:p}_{\mathrm{in}}(\mf D_1)\circ_{y,x}\mathrm{wk}^{a:p}_{\mathrm{in}}(\mf D_2)\bigr)
\\
&\quad\Rightarrow_{\gamma,a:p\Vdash\delta}
(\pi_a;\tau)_*\bigl(\mathrm{wk}^{a:p}_{\mathrm{in}}(\mf D_1')\circ_{y,x}\mathrm{wk}^{a:p}_{\mathrm{in}}(\mf D_2')\bigr).
\end{aligned}
$$
By \cref{lem:weakening-commutes-with-composition} the raw pair underneath the source is that of the weakened composite, and weakening commutes with permutation transport. Both operations leave the assignment and redirection sets untouched. The source equals therefore
$(\pi_a;\tau)_*\bigl(\mathrm{wk}^{a:p}_{\mathrm{in}}(\mf D_1\circ_{y,x}\mf D_2)\bigr)=\mathrm{wk}^{a:p}_{\mathrm{in}}(\mf D)$,
and the target equals $\mathrm{wk}^{a:p}_{\mathrm{in}}(\mf E)$. We set then
$\mathrm{wk}^{a:p}_{\mathrm{in}}\bigl((\pi;\tau)_*(t\circ_{y,x}s)\bigr):=(\pi_a;\tau)_*\bigl(\mathrm{wk}^{a:p}_{\mathrm{in}}(t)\circ_{y,x}\mathrm{wk}^{a:p}_{\mathrm{in}}(s)\bigr)$.
\end{itemize}

These exhaust the rules of \cref{def:descriptor-2-cells}; $\mathrm{wk}^{a:p}_{\mathrm{in}}(r)$ is
therefore defined for every $r$ with the stated type. The preservation of identities and of both compositions is
immediate from the 2-Identity, 2-Comp and 2-Subst clauses. \hfill$\square$

We write $\mathrm{wk}^{\theta}_{\mathrm{in}}(r)$ for the iterated ambient weakening, one declaration of $\theta$ at a time in the displayed order; two orders differ by the permutation transport of \cref{thm:exchange-admissibility-for-descriptor-2-cells}. 

\begin{lemma}[Boundary Renaming of Descriptor $2$-cells]\label{lem:boundary-renaming-of-descriptor-2-cells}
Let $r::\mf D\Rightarrow^{\Lambda}_{\gamma\Vdash\delta}\mf E$ be a descriptor $2$-cell and let $\sigma$ be a boundary renaming as in \cref{def:boundary-renaming}. There is a descriptor $2$-cell
$$\mathrm{ren}_{\sigma}(r)::\mathrm{ren}_{\sigma}(\mf D)\Rightarrow^{\Lambda}_{\gamma^{\sigma}\Vdash\delta^{\sigma}}\mathrm{ren}_{\sigma}(\mf E).$$
\end{lemma}

\textit{Proof.}
By recursion on the derivation of $r$, exactly as for
\cref{lem:ambient-weakening-of-descriptor-2-cells}. For $2$-Form, the
boundary-renaming closure of \cref{def:signature} supplies at least one
transported generator; choose one for this instance and use $2$-Form. For
$2$-Identity, \cref{thm:renaming-admissibility-for-descriptor-judgments}
gives the renamed descriptor judgment. For $2$-Comp the recursion is
componentwise. For $2$-Subst, rename the premises along the restrictions of
$\sigma$, extended by the identity on the variable, and close the re-applied rule
by the conjugated permutations $(\pi^{\sigma};\tau^{\sigma})$. The endpoints
match by \cref{lem:boundary-renaming-is-well-defined}(ii),(iii). Thus each rule
is sent to an instance of the same rule. The signature supplies no relation
between the generator chosen after two successive renamings and the generator
chosen after their composite; that coherence is imposed by
\cref{def:coherent-transport-completion}. \hfill $\square$

\endgroup{}

\begingroup

\subsection{Syntactic Properties of Mode Theory}

This section establishes the syntactic properties of the mode theory presented by a signature. We prove the coherence laws of composition, namely \cref{lem:associativity,lem:left-interchange-law,lem:right-interchange-law}, and the admissibility of the substructural rules at cartesian modes.

\subsubsection{Coherence Laws}

The composition of well-formed descriptor classes satisfies the laws of a polycategorical cut. In particular, it is associative and invariant under reordering of independent cuts (the \emph{left} and \emph{right interchange laws}). These are the \emph{strict coherence laws} applied throughout the metatheory. Associativity and left interchange allow arbitrary occurrence multiplicities, including zero. Right interchange has the single-occurrence hypotheses of \cref{lem:right-interchange-law}; these are automatic for well-formed descriptors because boundary usage makes each linear cut input occur exactly once. Moving forward, we choose representatives with pairwise disjoint internal variables (\cref{lem:fresh-representatives}). We choose also fresh families of bijections with pairwise disjoint codomains, disjoint from all representatives (\cref{lem:existence-of-fresh-families-of-bijections}). By \cref{prop:composition-is-well-defined} the resulting descriptor classes do not depend on these choices. It suffices therefore to exhibit, for each law, a bijection of internal variables carrying one raw composite to the other (\cref{def:internal-renaming-equivalence-alpha-equivalence}). We record also a boundary convention. The composition $\circ_{y,x}$ is insensitive to the positions of the declarations of $y$ and $x$ in the boundary lists, and the two sides of the associativity equation share their boundary lists on the nose. The two sides of an interchange equation differ instead by the order of the blocks in their boundary lists; they are compared along the canonical order permutation, whose transport is the identity on representatives (\cref{def:permutation-transport-of-raw-descriptors}). We will use the fact that internal renamings compose (\cref{prop:internal-renaming-composition}) and that they commute with input substitution (\cref{lem:input-substitution-and-internal-renaming-commute}). Recall also that occurrence-indexed substitutions at disjoint position sets commute with each other and with renamings position by position (\cref{lem:occurrence-substitutions-commute}).

\begin{lemma}[Associativity]\label{lem:associativity}
Let $\mf D_3, \mf D_2, \mf D_1$ be well-formed descriptor classes over the contexts
$$
(\gamma_3\ ;\ \delta_3, y_3:p_3),
\qquad
(\gamma_2, x_2:p_3\ ;\ \delta_2, y_2:p_2),
\qquad
(\gamma_1, x_1:p_2\ ;\ \delta_1),
$$
subject to these boundary conditions:

\begin{itemize}
  \item the input contexts pairwise agree on their common declarations, with all shared declarations at cartesian modes,
	\item the cut variables are distinct, $x_1\neq x_2$, with $x_2\notin\gamma_1\cup\gamma_3$ and $x_1\notin\gamma_2\cup\gamma_3$,
	\item the output contexts $\delta_1,\delta_2,\delta_3$ are pairwise disjoint, with $y_2\neq y_3$, $y_2\notin\delta_1\cup\delta_3$, and $y_3\notin\delta_1\cup\delta_2$.
\end{itemize}
Then all four displayed compositions are defined and
$$
\mf D_3 \circ_{y_3,x_2}\bigl(\mf D_2 \circ_{y_2,x_1}\mf D_1\bigr)
=
\bigl(\mf D_3 \circ_{y_3,x_2}\mf D_2\bigr)\circ_{y_2,x_1}\mf D_1.
$$
\end{lemma}

\textit{Proof.}
The descriptors are well formed, hence regular (\cref{thm:well-formed-descriptors-are-regular}). The boundary usage property gives then that each of $\mf D_3,\mf D_2$ produces its cut output by a unique redirection. Write $s_3\mapsto y_3\in\mc T_3$ and $s_2\mapsto y_2\in\mc T_2$ for these redirections, the representatives $(\mc S_i\mid\mc T_i)\in\mf D_i$ chosen with pairwise disjoint internal variables. Set
$$O_1:=\mathrm{Occ}_{x_1}(\mc S_1\mid\mc T_1),\qquad O_2:=\mathrm{Occ}_{x_2}(\mc S_2\mid\mc T_2).$$
All four compositions are defined. Three of the four definedness counts of \cref{def:composition-of-descriptors} are the uniqueness of the two redirections just recorded, and the case analysis identifies the fourth, the unique redirection targeting $y_2$ in $\mf D_3\circ_{y_3,x_2}\mf D_2$. The compatibility clauses of \cref{def:raw-composition} for the four cuts are the listed boundary conditions; in particular $y_2\neq y_3$ and $y_2\notin\delta_3$ make the output contexts of the inner cut of the right-hand side disjoint, and $x_2\notin\gamma_3$ makes the outer cut of the left-hand side compatible. Recall that the cut source $s_2$ is an internal variable or an input of $(\mc S_2\mid\mc T_2)$. We distinguish \emph{Case A}, where $s_2\neq x_2$, from \emph{Case B}, where $s_2=x_2$, treating Case A first. We compute now both compositions explicitly.

\begin{itemize} 
\item \textbf{Left-hand side.} Set $\mc C:=(\mc S_2\mid\mc T_2)\circ^{\bs\beta}_{y_2,x_1}(\mc S_1\mid\mc T_1)$ with $\bs\beta=(\beta_o)_{o\in O_1}$, so that
$$
\mc C=\Bigl(\bigcup_{o\in O_1}\mc S_2^{\beta_o}\ \cup\ \mc S_1[x_1:=v]\ \Bigm|\ \bigcup_{o\in O_1}(\mc T_2)_{\not\mapsto y_2}^{\beta_o}\ \cup\ \mc T_1[x_1:=v]\Bigr),
\qquad v_o=s_2^{\beta_o}.
$$
Write $\mc C=(\mc S_{\mc C}\mid\mc T_{\mc C})$ for its two components.

\emph{Case A ($s_2\neq x_2$):} Then $x_2$ occurs in $\mc C$ exactly at the $\beta_o$-copies of its occurrences in $(\mc S_2\mid\mc T_2)$. Since $s_2\neq x_2$, no occurrence of $x_2$ lies in the removed redirection, and we get
$\mathrm{Occ}_{x_2}(\mc C)=\{(o,e)\mid o\in O_1,\ e\in O_2\}$,
where $(o,e)$ denotes the $\beta_o$-copy of the occurrence $e$. The left-hand side of the associativity equation to be shown is then
$$
\Bigl(\bigcup_{(o,e)}\mc S_3^{\eta_{(o,e)}}\ \cup\ \mc S_{\mc C}[x_2:=u]\ \Bigm|\ \bigcup_{(o,e)}(\mc T_3)_{\not\mapsto y_3}^{\eta_{(o,e)}}\ \cup\ \mc T_{\mc C}[x_2:=u]\Bigr),
\qquad u_{(o,e)}=s_3^{\eta_{(o,e)}},
$$
for a fresh family $\bs\eta$ indexed by $\mathrm{Occ}_{x_2}(\mc C)$.

\item \textbf{Right-hand side.} Set $\mc C':=(\mc S_3\mid\mc T_3)\circ^{\bs\theta}_{y_3,x_2}(\mc S_2\mid\mc T_2)$ with $\bs\theta=(\theta_e)_{e\in O_2}$, so that
$$
\mc C'=\Bigl(\bigcup_{e\in O_2}\mc S_3^{\theta_e}\ \cup\ \mc S_2[x_2:=w]\ \Bigm|\ \bigcup_{e\in O_2}(\mc T_3)_{\not\mapsto y_3}^{\theta_e}\ \cup\ \mc T_2[x_2:=w]\Bigr),
\qquad w_e=s_3^{\theta_e}.
$$
Write $\mc C'=(\mc S_{\mc C'}\mid\mc T_{\mc C'})$ for its two components. In Case A, the unique redirection of $\mc C'$ targeting $y_2$ is $s_2\mapsto y_2$ with the same source $s_2$, since its source position is not an $x_2$-occurrence. The outer cut is therefore defined and the right-hand side becomes
$$
\Bigl(\bigcup_{o\in O_1}\mc S_{\mc C'}^{\zeta_o}\ \cup\ \mc S_1[x_1:=v']\ \Bigm|\ \bigcup_{o\in O_1}(\mc T_{\mc C'})_{\not\mapsto y_2}^{\zeta_o}\ \cup\ \mc T_1[x_1:=v']\Bigr),
\qquad v'_o=s_2^{\zeta_o},
$$
for a fresh family $\bs\zeta$ indexed by $O_1$, the value $v'_o$ being $s_2$ itself when $s_2$ is an input, since renamings fix inputs. Expanding $\mc S_{\mc C'}^{\zeta_o}$ with \cref{prop:internal-renaming-composition} and \cref{lem:input-substitution-and-internal-renaming-commute}:
$$
\mc S_{\mc C'}^{\zeta_o}=\bigcup_{e\in O_2}\mc S_3^{\zeta_o\theta_e}\ \cup\ \mc S_2^{\zeta_o}[x_2:=w^{\zeta_o}],
\qquad
(w^{\zeta_o})_e=s_3^{\zeta_o\theta_e},
$$
and likewise for the redirection component.
\end{itemize}

Define the bijection $\Phi$ of internal variables of the left composition onto those of the right by
$$
\Phi:=
\begin{cases}
(\zeta_o\theta_e)\circ\eta_{(o,e)}^{-1} & \text{on the copy set of } \eta_{(o,e)},\\
\zeta_o\circ\beta_o^{-1} & \text{on the copy set of } \beta_o,\\
\mathrm{id} & \text{on } \mathrm{Int}(\mc S_1\mid\mc T_1).
\end{cases}
$$
By the freshness and disjointness of all codomains this is a well-defined bijection, the domains and codomains decomposing by two applications of \cref{prop:internal-variables-of-raw-composition}. We check now that it carries the left composite to the right one elementwise. The $\mc S_3$-copy at index $(o,e)$ is carried to the $\mc S_3$-copy at $\zeta_o\theta_e$ by \cref{prop:internal-renaming-composition}. The $\mc S_2$-copy at $\beta_o$ is carried to the $\mc S_2$-copy at $\zeta_o$. Its $x_2$-position $e$ holds on the left the value $u_{(o,e)}=s_3^{\eta_{(o,e)}}$, and $\Phi$ sends this value to $s_3^{\zeta_o\theta_e}=(w^{\zeta_o})_e$, the value at that position on the right. The $\mc S_1$-part is fixed by $\Phi$, and its $x_1$-position $o$ holds, on the left, $v_o=s_2^{\beta_o}$. $\Phi$ sends it to $s_2^{\zeta_o}$ if $s_2$ is internal and fixes it if $s_2$ is an input, in both cases the value $v'_o$ on the right. The redirection components correspond by the same computation, using that substitutions and renamings change only sources (\cref{lem:duplicated-substitution-preserves-redirection-targets}). In particular the retained redirections of both sides are the $(o,e)$-indexed copies of $(\mc T_3)_{\not\mapsto y_3}$, the $o$-indexed copies of $(\mc T_2)_{\not\mapsto y_2}$ with substituted $x_2$-positions, and $\mc T_1$ with substituted $x_1$-positions. Hence the renaming of the left-hand side by $\Phi$ is equal to the right-hand side.

\emph{Case B ($s_2=x_2$):} Now the unique occurrence of $x_2$ in the $y_2$-redirection, say $o^*\in O_2$, is removed from each copy in $\mc C$. The $x_1$-positions of $(\mc S_1\mid\mc T_1)$ come then to hold $v_o=s_2^{\beta_o}=x_2$ since inputs are fixed by renamings. Hence
$$
\mathrm{Occ}_{x_2}(\mc C)=\{(o,e)\mid o\in O_1,\ e\in O_2\smallsetminus\{o^*\}\}\ \cup\ \{\widehat o\mid o\in O_1\},
$$
where $\widehat o$ is the former $x_1$-position $o$, and the left bracketing substitutes $u_{\widehat o}=s_3^{\eta_{\widehat o}}$ there. On the right, $\mc C'$ has its $y_2$-redirection sourced at $w_{o^*}=s_3^{\theta_{o^*}}$, so the outer cut substitutes $v'_o=s_3^{\zeta_o\theta_{o^*}}$ at the $x_1$-positions. The same $\Phi$ as in Case A, extended by
$$
\Phi:=(\zeta_o\theta_{o^*})\circ\eta_{\widehat o}^{-1}\quad\text{on the copy set of }\eta_{\widehat o},
$$
matches the two sides. The value at the $x_1$-position $o$ is $u_{\widehat o}=s_3^{\eta_{\widehat o}}$ on the left. It is carried to $s_3^{\zeta_o\theta_{o^*}}=v'_o$ on the right, and all other correspondences are as in Case A. Hence the renaming of the left-hand side by $\Phi$ is equal to the right-hand side.

In both cases the two sides have $\alpha$-equal raw representatives over the same boundary, and the descriptor classes agree. \hfill $\square$

\textit{Remark:} The copies of $\mf D_3$ on either side are indexed by pairs: one copy for each occurrence of $x_1$ in $\mf D_1$ and each occurrence of $x_2$ in $\mf D_2$. When the cut modes are linear, boundary usage makes both occurrence sets singletons and the law is the strict associativity of the cut. When a cut mode is cartesian, the law states that substituting-then-copying agrees with copying-then-substituting. Through \cref{cor:cartesian-cut-is-term-substitution} this is the associativity of expression substitution (\cref{def:cartesian-expressions}), the law $\alpha_1[x_1:=\alpha_2][x_2:=\alpha_3]=\alpha_1[x_1:=\alpha_2[x_2:=\alpha_3]]$ for $x_2$ not occurring in $\alpha_1$.

\begin{example}[Linear--Nonlinear Logic]
We work over the pre-signature extended with the binary c-generator $\otimes_c : c, c \to c$ and illustrate the cartesian case, continuing the \hyperref[ex:cartesian-cut-linear-logic]{cartesian-cut example}. Take
$$
\begin{aligned}
\mf D_3 &:= [\{a, a' \xmapsto{\otimes_c} w_3\} \mid \{w_3 \mapsto y_3\}],
\\
\mf D_2 &:= [\{x_2, x_2 \xmapsto{\otimes_c} w_2\} \mid \{w_2 \mapsto y_2\}],
\\
\mf D_1 &:= [\{x_1, x_1 \xmapsto{\otimes_c} w_1\} \mid \{w_1 \mapsto y'\}]
\end{aligned}
$$
over the boundaries $(a : c,\, a' : c \ ; \ y_3 : c)$, $(x_2 : c \ ; \ y_2 : c)$, and $(x_1 : c \ ; \ y' : c)$, reading as the expressions $\otimes_c(a, a')$, $\otimes_c(x_2, x_2)$, and $\otimes_c(x_1, x_1)$. Both occurrence sets have two elements. On the left, the inner cut computes $\otimes_c(x_1, x_1)[x_1 := \otimes_c(x_2, x_2)]$, with four occurrences of $x_2$, and the outer cut grafts one fresh copy of $\mf D_3$ at each of them. On the right, the inner cut computes $\otimes_c(x_2, x_2)[x_2 := \otimes_c(a, a')]$ with two copies of $\mf D_3$, and the outer cut duplicates the whole composite once per occurrence of $x_1$. Either way four copies of $\mf D_3$ appear, one per pair of occurrences, and both sides read as
$$\otimes_c(\otimes_c(\beta, \beta),\, \otimes_c(\beta, \beta)), \qquad \beta := \otimes_c(a, a'),$$
substituting-then-copying agreeing with copying-then-substituting.
\end{example}

\begin{lemma}[Left Interchange Law]\label{lem:left-interchange-law}
Let $\mf D_1,\mf D_2,\mf D_3$ be well-formed descriptor classes over the contexts
$$
(\gamma_1\ ;\ \delta_1, y_1:p_1),
\qquad
(\gamma_2\ ;\ \delta_2, y_2:p_2),
\qquad
(\gamma_3, x_1:p_1, x_2:p_2\ ;\ \delta_3),
$$
subject to these boundary conditions:

\begin{itemize}
  \item the input contexts pairwise agree on their common declarations, with all shared declarations at cartesian modes,
	\item the cut variables satisfy $x_1,x_2\notin\gamma_1\cup\gamma_2$,
	\item the output contexts $\delta_1,\delta_2,\delta_3$ are pairwise disjoint, with $y_1,y_2$ occurring in none of them.
\end{itemize} Then all four displayed compositions are defined and the interchange law holds at the level of the descriptor classes
$$
\mf D_1\circ_{y_1,x_1}\bigl(\mf D_2\circ_{y_2,x_2}\mf D_3\bigr)
=
\mf D_2\circ_{y_2,x_2}\bigl(\mf D_1\circ_{y_1,x_1}\mf D_3\bigr).
$$
\end{lemma}

\textit{Proof.}
Choose representatives with pairwise disjoint internal variables and write $s_1\mapsto y_1\in\mc T_1$ and $s_2\mapsto y_2\in\mc T_2$ for the unique cut redirections. Set then $O_1:=\mathrm{Occ}_{x_1}(\mc S_3\mid\mc T_3)$ and $O_2:=\mathrm{Occ}_{x_2}(\mc S_3\mid\mc T_3)$. These position sets are disjoint since $x_1\neq x_2$. All four compositions are defined, the definedness counts being the two uniquenesses just recorded and the compatibility clauses of \cref{def:raw-composition} the listed boundary conditions.

Consider the left side. The inner composite $\mc C:=(\mc S_2\mid\mc T_2)\circ^{\bs\beta}_{y_2,x_2}(\mc S_3\mid\mc T_3)$ replaces the positions of $O_2$ by the values $v_e=s_2^{\beta_e}$ ($e\in O_2$) and adjoins the copies of $\mf D_2$. Since $x_1\notin\gamma_2$, the variable $x_1$ occurs in $\mc C$ exactly at the positions of $O_1$, unchanged. The outer composite replaces those by $u_o=s_1^{\eta_o}$ ($o\in O_1$) and adjoins the copies of $\mf D_1$. The two substitutions act on disjoint position sets of $(\mc S_3\mid\mc T_3)$, and neither value family contains the other's cut variable, the values being fresh copies or inputs with $x_1,x_2\notin\gamma_1\cup\gamma_2$, which is the entry hypothesis of \cref{lem:occurrence-substitutions-commute}. The copies of $\mf D_1$ contain no $x_2$ and those of $\mf D_2$ no $x_1$. The left side is therefore
$$
\Bigl(\ \bigcup_{o\in O_1}\mc S_1^{\eta_o}\ \cup\ \bigcup_{e\in O_2}\mc S_2^{\beta_e}\ \cup\ \mc S_3[x_1:=u,\ x_2:=v]\ \Bigm|\ \bigcup_{o}(\mc T_1)^{\eta_o}_{\not\mapsto y_1}\ \cup\ \bigcup_{e}(\mc T_2)^{\beta_e}_{\not\mapsto y_2}\ \cup\ \mc T_3[x_1:=u,\ x_2:=v]\ \Bigr),
$$
where $[x_1:=u,\ x_2:=v]$ denotes the simultaneous occurrence-indexed substitution at the disjoint position sets. The right side unfolds to the same expression with the roles of the two cuts exchanged and fresh families $\bs\eta',\bs\beta'$. The bijection which is $\eta'_o\circ\eta_o^{-1}$ on each $\mf D_1$-copy, $\beta'_e\circ\beta_e^{-1}$ on each $\mf D_2$-copy, and the identity on $\mathrm{Int}(\mc S_3\mid\mc T_3)$ carries then one to the other, matching the substituted values ($s_1^{\eta_o}\mapsto s_1^{\eta'_o}$, or fixed when $s_1$ is an input, and likewise for $s_2$). Hence the two sides are $\alpha$-equal. \hfill $\square$

\begin{lemma}[Right Interchange Law]\label{lem:right-interchange-law}
Let $\mf D_3,\mf D_1,\mf D_2$ be regular descriptor classes over the contexts
$$
(\gamma_3\ ;\ \delta_3, y_1:p_1, y_2:p_2),
\qquad
(\gamma_1, x_1:p_1\ ;\ \delta_1),
\qquad
(\gamma_2, x_2:p_2\ ;\ \delta_2),
$$
subject to these boundary conditions:
\begin{itemize}
  \item the input contexts pairwise agree on their common declarations, with all shared declarations at cartesian modes,
	\item the cut variables $x_1\neq x_2$ satisfy $x_1\notin\gamma_2\cup\gamma_3$ and $x_2\notin\gamma_1\cup\gamma_3$,
	\item the output contexts $\delta_1,\delta_2,\delta_3$ are pairwise disjoint, with $y_1,y_2\notin\delta_1\cup\delta_2$,
	\item $x_1$ has exactly one occurrence in $\mf D_1$ and $x_2$ has exactly one occurrence in $\mf D_2$.
\end{itemize} Then all four displayed compositions are defined and the right interchange law holds at the level of the descriptor classes
$$
\bigl(\mf D_3\circ_{y_1,x_1}\mf D_1\bigr)\circ_{y_2,x_2}\mf D_2
=
\bigl(\mf D_3\circ_{y_2,x_2}\mf D_2\bigr)\circ_{y_1,x_1}\mf D_1.
$$
\end{lemma}

\textit{Proof.}
Choose representatives with pairwise disjoint internal variables and write $s_1\mapsto y_1$ and $s_2\mapsto y_2$ for the two cut redirections of $\mc T_3$ (unique by regularity, and distinct since $y_1\neq y_2$). The two inner compositions are defined by these counts and the listed boundary conditions. Consider the left side. The variable $x_1$ has the single occurrence $o_1$, so the inner composite makes a single copy of $\mf D_3$. By \cref{prop:composition-is-well-defined} the class does not depend on the fresh family, and as in the computation of \cref{cor:linear-cut}, whose single-occurrence argument applies verbatim, the renaming $\beta_{o_1}^{-1}$, extended by the identity, carries the raw composite to the representative
$$
\bigl(\mc S_3\cup\mc S_1[x_1:=s_1]\ \bigm|\ (\mc T_3)_{\not\mapsto y_1}\cup\mc T_1[x_1:=s_1]\bigr).
$$
The redirection $s_2\mapsto y_2$ is retained and is still the unique one targeting $y_2$, since $\mc T_1[x_1:=s_1]$ redirects only into $\delta_1$. Hence the outer cut is defined. Since $x_2$ has the single occurrence $o_2$, the outer cut makes a single copy again, carried by the same renaming argument to
$$
\bigl(\mc S_3\cup\mc S_1[x_1:=s_1]\cup\mc S_2[x_2:=s_2]\ \bigm|\ (\mc T_3)_{\not\mapsto y_1,\not\mapsto y_2}\cup\mc T_1[x_1:=s_1]\cup\mc T_2[x_2:=s_2]\bigr),
$$
using that the substitution $[x_2:=s_2]$ does not touch $\mc S_1[x_1:=s_1]$ or $\mc T_1[x_1:=s_1]$, since $x_2$ does not occur in $\mf D_1$ and $s_1\neq x_2$ by $x_2\notin\gamma_3$. The right side is defined symmetrically and evaluates to the same raw descriptor with the two splices performed in the other order. The two substitutions act at disjoint positions of disjoint components, and the results are therefore equal on the nose, each side carried to its specified representative. \hfill $\square$

\textit{Remark:} The single-occurrence hypotheses of \cref{lem:right-interchange-law} are automatic in the well-formed setting. A well-formed descriptor with two outputs is of linear sort (\cref{prop:sorts-of-well-formed-descriptors}), so the cut modes $p_1,p_2$ are linear, and the boundary usage property gives the cut inputs of regular consumers exactly one occurrence each. They are needed at this level of generality since a producer with a second output cannot be duplicated without duplicating that output. Indeed, copying $\mf D_3$ once per occurrence of $x_1$ would give $y_2$ one redirection per copy, and the second cut would then no longer be defined. This is the phenomenon behind the \emph{purity of the substituted producer} invoked in the metatheory: a c-descriptor has a single output and no side outputs to misroute. It surfaces as a definedness constraint and not as a failed theorem.

\endgroup{}
\begingroup

\subsubsection{Admissibility of Substructural Rules}\label{subsec:substructural-rules}

In this section we prove that the cartesian context enjoys all of the substructural rules of weakening, contraction and exchange. We recall again that this fact has no bearing on the admissibility of the substructural rules in the object logic. Informally, the admissibility of the substructural rules at the framework level grants us permission to use them at the object logic level. This permission is unlocked by the presence of $2$-cells in the mode theory. In a cartesian zone, weakening uses the notion of \emph{input weakening} $\mathrm{wk}^{a:p}_{\mathrm{in}}$ from \cref{def:input-weakening-of-descriptor-classes}, while contraction uses the notation $\mathrm{ctr}_{x'\to x}$ of \cref{def:input-contraction-notation}. The linear zone only enjoys the substructural rules of exchange using context permutations. For a regular descriptor, leaving a variable unused or using it multiple times is impossible by the boundary usage property. Weakening and contraction at linear modes are therefore not admissible operations on descriptors. To recover them, new structural data need to be introduced at the signature level.

\begin{theorem}[Input Weakening Admissibility for Well-Formed Descriptors]\label{thm:input-weakening-admissibility-for-well-formed-descriptors}
If $\gamma \Vdash_{\mf D} \delta$ is derivable, and if $a:p$ is an input variable at a cartesian mode $p$, fresh for $\gamma$, then $\gamma,a:p \Vdash_{\mathrm{wk}^{a:p}_{\mathrm{in}}(\mf D)} \delta$ is derivable.
\end{theorem}

\textit{Proof.}
We proceed by induction on the derivation of $\gamma \Vdash_{\mf D} \delta$. We use \cref{thm:exchange-admissibility-for-descriptor-judgments} freely to place the newly added input variable $a:p$ in the position required by the formation rules. We use also the usual freshness convention that the auxiliary input and output variables introduced by the rules are chosen fresh from the finite set of variables already under discussion, in particular fresh from $a$.

\begin{itemize}
\item \textbf{Identity.}
Suppose the derivation ends with the Identity rule. Then, before the final context permutation, the
descriptor has the form $[\varnothing\mid\{x\mapsto y\}]$ over the input context $\theta,x:q$ with $\theta$ cartesian, and the output context $y:q$. The mode $p$ is cartesian. The Identity rule applies therefore again with the enlarged ambient cartesian context $\theta,a:p$. It produces the same descriptor and hence $\gamma,a:p
\Vdash_{\mathrm{wk}^{a:p}_{\mathrm{in}}(\mf D)} \delta$. If the original Identity conclusion was
obtained after a context permutation, we extend that permutation by fixing $a:p$ and use
\cref{prop:functoriality-of-permutation-transport}.

\item \textbf{Cartesian case.} For the c-Comp rule,
suppose the last rule is c-Comp, with premises $\theta \Vdash_{\mf D_i} y_i:p_i$ sharing the cartesian context $\theta$, and conclusion $\theta\Vdash_{\mf D} r:q$ for a c-generator $f:p_1,\ldots,p_n\to q$. By the induction hypothesis, $\theta,a:p \Vdash_{\mathrm{wk}^{a:p}_{\mathrm{in}}(\mf D_i)} y_i:p_i$ for every $i$. Applying c-Comp in the enlarged shared context $\theta,a:p$ gives a conclusion in $\theta,a:p \Vdash r:q$. The raw construction is unchanged. The assignment sets, the redirection sets, the unique sources $s_i$, and the fresh internal variable $w_R$ are unchanged; only the ambient input context is enlarged. The conclusion descriptor is therefore exactly $\mathrm{wk}^{a:p}_{\mathrm{in}}(\mf D)$. The nullary case $n=0$ is included. Transported conclusions are handled by extending the permutation by $a:p$.

\item \textbf{Linear case.} Regarding the $\ell$-Pos-Comp and $\ell$-Neg-Comp rules,
we start by supposing that the last rule is $\ell$-Pos-Comp, with cartesian premises $\theta\Vdash_{\mf C_i} b_i:p_i$ and linear premises $\theta \ ; \ \Gamma_j \Vdash_{\mf D_j} \delta_j,y_j:l_j$. By the induction hypothesis each premise weakens by $a:p$. Using Exchange Admissibility we place $a$ in the cartesian zone and obtain $\theta,a:p\Vdash \mf C_i$-premises and $\theta,a:p \ ; \ \Gamma_j\Vdash \mf D_j$-premises. Reapplying $\ell$-Pos-Comp with the shared cartesian context $\theta,a:p$ produces the same raw descriptor since the variables $c_i$, the sources $s_j$, the fresh variables $w_1,\ldots,w_n$, and all assignment and redirection sets are unchanged, now regarded over the enlarged input context. Hence the conclusion descriptor is $\mathrm{wk}^{a:p}_{\mathrm{in}}(\mf D)$. The $\ell$-Neg-Comp case is identical, with the substitutions $x_j:=w_j$ and the new assignment unchanged and the fresh linear inputs $z_1,\ldots,z_m$ unaffected by the added cartesian variable.

\item \textbf{Cut case.}
Suppose the last rule is Cut, with premises $\gamma_1 \Vdash_{\mf E} \delta_1,y:q$
and $\gamma_2,x:q \Vdash_{\mf F} \delta_2$. By the induction hypothesis, the premises weaken to $\gamma_1,a:p
\Vdash_{\mathrm{wk}^{a:p}_{\mathrm{in}}(\mf E)} \delta_1,y:q$ and, after Exchange Admissibility, $\gamma_2,a:p,x:q
\Vdash_{\mathrm{wk}^{a:p}_{\mathrm{in}}(\mf F)} \delta_2$. Now apply Cut with input contexts
$\gamma_1,a:p$ and $\gamma_2,a:p$. The shared declarations are the old shared declarations together with $a:p$, all at cartesian modes, the premise $|\mc T_{\mapsto y}|=1$ is unchanged, and $\delta_1,\delta_2$ remain disjoint. This gives $\gamma_1\cup\gamma_2,a:p \Vdash_{\mathrm{wk}^{a:p}_{\mathrm{in}}(\mf
E)\circ_{y,x}\mathrm{wk}^{a:p}_{\mathrm{in}}(\mf F)} \delta_1,\delta_2$. By \cref{lem:weakening-commutes-with-composition},
$$
\mathrm{wk}^{a:p}_{\mathrm{in}}(\mf E)
\circ_{y,x}
\mathrm{wk}^{a:p}_{\mathrm{in}}(\mf F)
=
\mathrm{wk}^{a:p}_{\mathrm{in}}(\mf E\circ_{y,x}\mf F),
$$
so the conclusion is exactly $\gamma_1\cup\gamma_2,a:p \Vdash_{\mathrm{wk}^{a:p}_{\mathrm{in}}(\mf E\circ_{y,x}\mf F)} \delta_1,\delta_2$. If the original Cut conclusion was followed by a context permutation, we extend the input permutation by fixing $a:p$ and use \cref{prop:functoriality-of-permutation-transport}.
\end{itemize}

All possible last rules have been treated. Hence, whenever $\gamma \Vdash_{\mf D} \delta$ is derivable and $a:p$ is a cartesian-mode input variable fresh for $\gamma$, the weakened judgment $\gamma,a:p \Vdash_{\mathrm{wk}^{a:p}_{\mathrm{in}}(\mf D)} \delta$ is derivable. \hfill $\square$

\textit{Remark:} \Cref{thm:input-weakening-admissibility-for-well-formed-descriptors} does not extend to the output zone. An output variable added to the boundary would carry no redirection. This violates the boundary usage property and no regular descriptor inhabits the weakened boundary. The theorem fails likewise for $p$ linear, since the added input would be unused.

We turn to proving that contraction is admissible at cartesian modes. In this case, contraction is encoded by the composition with the descriptor that ``renames'' the variable to be contracted. For instance, contracting the cartesian variable $x$ in a descriptor $\mc D$ that contains another input cartesian variable $x'$ of the same mode is equivalent to composing the descriptor $\mc D$ with the descriptor $[\varnothing \mid \{x \mapsto x'\}]$. We use this fact now.

\begin{lemma}[Identity Cut Realizes Input Contraction]\label{lem:identity-cut-realizes-input-contraction}
Let $p$ be a cartesian mode, let $\mf D=[\mc S\mid\mc T]\in\mathrm{Desc}_\Sigma(\gamma,x:p,y:p\ ;\ \delta)$ be well formed with $x$ and $y$ distinct input variables, and let $a:p$ be an output variable. Let $[\varnothing\mid\{x\mapsto a\}]\in\mathrm{Desc}_\Sigma(x:p\ ;\ a:p)$ be the identity descriptor from $x$ to $a$. Then
$$
[\varnothing\mid\{x\mapsto a\}]\circ_{a,y}\mf D=\mf D[y:=x]
$$
as descriptor classes in $\mathrm{Desc}^{\mathrm{raw}}_\Sigma(\gamma,x:p\ ;\ \delta)$.
\end{lemma}

\textit{Proof.}
The left factor redirects to $a$ by the single redirection $x\mapsto a$. The definedness hypothesis of \cref{def:raw-composition} holds therefore, with cut source $s=x$. The boundary compatibility conditions hold as well. The input contexts $x:p$ and $\gamma,x:p$ share only the declaration $x:p$, at a cartesian mode.

In the raw composition, the copies of the left factor contribute nothing since its assignment set is empty and $(\{x\mapsto a\})_{\not\mapsto a}=\varnothing$. The cut source is an input variable. It is fixed therefore by every renaming of the fresh family. $s^{\beta_o}=x$ for every occurrence $o$ of $y$, and the occurrence-indexed substitution at this constant family is the ordinary input substitution $[y:=x]$ (\cref{def:raw-duplicated-substitution}). Hence, for any representative $(\mc S\mid\mc T)\in\mf D$,
$$
(\varnothing\mid\{x\mapsto a\})\circ^{\bs\beta}_{a,y}(\mc S\mid\mc T)
=
(\mc S[y:=x]\mid\mc T[y:=x]),
$$
a representative of $\mf D[y:=x]$, well defined on classes by \cref{lem:input-substitution-by-an-input-is-well-defined-for-descriptor-classes}.
\hfill $\square$

\textit{Remark (units for $\circ$):} \Cref{lem:identity-cut-realizes-input-contraction} exhibits the identity descriptors as units for composition, up to the name of the cut variable. Cutting an identity into an input renames that input. Dually, cutting the output of a descriptor against an identity relabels that output, at any mode. For regular $\mf D\in\mathrm{Desc}^{\mathrm{raw}}_\Sigma(\gamma\ ;\ \delta,y:p)$ and the identity $[\varnothing\mid\{x_0\mapsto b\}]$ over $(x_0:p\ ;\ b:p)$, the input $x_0$ has a single occurrence, so a single copy of $\mf D$ is kept, its redirection to $y$ now feeding $b$:
$$\mf D\circ_{y,x_0}[\varnothing\mid\{x_0\mapsto b\}]=\mf D[y:=b],$$
the descriptor $\mf D$ with its output $y$ renamed to $b$ (here $\mf D[y:=b]$ denotes the retargeting of the unique redirection $s\mapsto y$ to $s\mapsto b$).

\begin{theorem}[Input Contraction Admissibility for Well-Formed Descriptors]\label{thm:input-contraction-admissibility-for-well-formed-descriptors}
If $\gamma,x:p,y:p \Vdash_{\mf D} \delta$ is derivable, with $x$ and $y$ distinct input variables at a \emph{cartesian} mode $p$, then $\gamma,x:p \Vdash_{\mf D[y:=x]} \delta$ is derivable.
\end{theorem}

\textit{Proof.}
Choose an output variable $a:p$. By the Identity rule with empty ambient context, $x:p
\Vdash_{[\varnothing\mid\{x\mapsto a\}]} a:p$. By hypothesis, $\gamma,x:p,y:p \Vdash_{\mf D}
\delta$. The Cut rule applies, cutting the output $a:p$ of the identity descriptor against the
input $y:p$ of $\mf D$. Indeed, the input contexts $x:p$ and $\gamma,x:p$ share only the declaration $x:p$, at a cartesian mode. The identity produces $a$ by its single redirection, and the output contexts $\varnothing$ and $\delta$ are disjoint. This gives $\gamma,x:p \Vdash_{[\varnothing\mid\{x\mapsto a\}]\circ_{a,y}\mf D}
\delta$. By \cref{lem:identity-cut-realizes-input-contraction},
$$
[\varnothing\mid\{x\mapsto a\}]\circ_{a,y}\mf D
=
\mf D[y:=x].
$$
 \hfill $\square$
\endgroup{}

\section{The Type Theory}\label{sec:type-theory}

\begingroup

\subsection{Type Formation and Sequent Calculus}
We reach the second most important exposition section of the framework: \emph{the type theory}.
We recall that the mode theory parametrizes the logic of interest. The type theory, in addition, uses this parametrization to generate the logical rules that govern the logic. The mode theory specifies the definition of descriptors and the type theory specifies the types of logical formulas and their connectives. Every formula is assigned a mode, and there will be the understanding that whenever $x : p$ is a variable of mode $p$ such that $x : A$ is also declared in the type theory, then the formula $A$ is of mode $p$ as well.

The type theory specifies two connectives $F,U$. In the same way that the connective $A \land B$ internalizes the concatenation operator in the context $A,B$, the connectives respectively and intuitively represent the internalization of the resources and the targets of a sequent. One novelty (compared to \cite{LSR17}) is that the comma operator on the target side is now able to be internalized using the connective $U$. Thus, in the sequent $\Gamma \vdash C,D$, the introduction rule for the connective $U$ allows for the context $C,D$ to be internalized as the formula $C \parr D$, which corresponds to the multiplicative disjunction of linear logic.

Recall that the mode theory possesses two types of well-formed descriptors: cartesian and linear (\cref{def:c-descriptors}). Thus, our type theory produces rules sensitive to those two types. To streamline the exposition, whenever possible, we give a rule schema that can be specialized to the two cases. For completeness, we also give the specialized rules.
We note that the specialization at cartesian contexts recovers \cite{LSR17} and the specialization at linear contexts recovers the polycategorical semantics of \cite{Bla23}. The final mixed contexts recover a novel setting. Mixed instances express the classical-linear-logic modality ``!'' internally as the $F$-$U$ composite of \cref{ex:linear-logic-exponential-modality}. Its intended semantics is the comonad of the cartesian-linear adjunction. The main difference with the exposition of \cite{LSR17} is that  the comonad equations are not imposed as additional structure (in the sense of additional equations).

\phantomsection\label{anchor:well-formed-index-convention}
\textit{Convention (Well-formed indexes):} Every judgment of the type theory is indexed by a well-formed descriptor class or by a c-descriptor. The rules combine those indexes only by the operations of the mode theory: the Cut rule (\cref{def:well-formed-composition-cut}) on classes (\cref{def:composition-of-descriptors}), input substitution (\cref{def:input-substitution,cor:cartesian-cut-is-term-substitution}), boundary renaming (\cref{def:boundary-renaming}), and permutation transport (\cref{thm:exchange-admissibility-for-descriptor-judgments}). Every index appearing in a derivable judgment is therefore well formed by construction, hence regular (\cref{thm:well-formed-descriptors-are-regular}; in a gated signature, \cref{prop:gate-rules-preserve-regularity}). In particular, the definedness hypothesis $|\mc T_{\mapsto y}|=1$ of each displayed composition holds by the boundary usage property, and the coherence laws established in \cref{lem:associativity,lem:left-interchange-law,lem:right-interchange-law} apply to the indexes of a derivation. Notice finally that the sorts of the premises of each schema are determined by those of the conclusion and of the connective descriptor, via \cref{lem:cut-preserves-the-sorts}.

\subsubsection{Types and the Formers}

\begin{definition}[Type Contexts]\label{def:type-contexts}
Let $\Lambda \ \mathrm{Sig}$ be a signature. We recall that its underlying pre-signature is denoted $\underline{\Lambda}$. There are two types of contexts: cartesian and linear. We define the judgments that form those contexts now.   
$$
\inferrule*{ }{\cdot\ \mathrm{ctx}^{\mathrm{c}}_{\Lambda;\cdot}}
\qquad
\inferrule*
{
  \Theta\ \mathrm{ctx}^{\mathrm{c}}_{\Lambda;\theta}
  \\
  A\ \mathrm{type}_{\Lambda,p}
  \\
  p \modec
  \\
  (\theta,x:p)\ \mathrm{ctx}^{\mathrm{c}}_{\underline{\Lambda}}
}
{(\Theta,x:A)\ \mathrm{ctx}^{\mathrm{c}}_{\Lambda;\theta,x:p}}
$$
$$
\inferrule*{ }{\cdot\ \mathrm{ctx}^{\mathrm{in}}_{\Lambda;\cdot}}
\qquad
\inferrule*
{
  \Gamma\ \mathrm{ctx}^{\mathrm{in}}_{\Lambda;\gamma}
  \\
  A\ \mathrm{type}_{\Lambda,l}
  \\
  l \model
  \\
  (\gamma,x:l)\ \mathrm{ctx}^{\mathrm{in}}_{\underline{\Lambda}}
}
{(\Gamma,x:A)\ \mathrm{ctx}^{\mathrm{in}}_{\Lambda;\gamma,x:l}}
$$
$$
\inferrule*{ }{\cdot\ \mathrm{ctx}^{\mathrm{out}}_{\Lambda;\cdot}}
\qquad
\inferrule*
{
  \Delta\ \mathrm{ctx}^{\mathrm{out}}_{\Lambda;\delta}
  \\
  B\ \mathrm{type}_{\Lambda,p}
  \\
  (\delta,y:p)\ \mathrm{ctx}^{\mathrm{out}}_{\underline{\Lambda}}
}
{(\Delta,y:B)\ \mathrm{ctx}^{\mathrm{out}}_{\Lambda;\delta,y:p}}.
$$
\end{definition}
\textit{Remark:} The superscript $(-)^c$ indicates that the context is fully cartesian (contains only cartesian variables), while the other superscripts $(-)^{\mathrm{in}}$ and $(-)^{\mathrm{out}}$ indicate the location of a context (left or right with respect to the turnstile). The formulas of an input context are linear, the cartesian declarations living in the zone of the judgment $\mathrm{ctx}^{\mathrm{c}}$. The output judgment carries no mode-class premise on purpose: output contexts are linear in the linear sequents of \cref{def:sequent-judgments}, but the same judgment forms the single cartesian output of a cartesian sequent and the output context of a cartesian $U$ formation.

\begin{example}
Suppose $c \modec$ and $l \model$ are modes of the signature $\Lambda$, let $P$ be a type at the cartesian mode $c$, and let $A, B$ be types at the linear mode $l$. The context formation rules derive
$$
(x : P,\, x' : P)\ \mathrm{ctx}^{\mathrm{c}}_{\Lambda;x:c,x':c},
\qquad
(x_1 : A,\, x_2 : B)\ \mathrm{ctx}^{\mathrm{in}}_{\Lambda;x_1:l,x_2:l},
\qquad
(y : B)\ \mathrm{ctx}^{\mathrm{out}}_{\Lambda;y:l},
$$
We note that the declaration $x_1 : A$ cannot be present in a cartesian context because $x_1 : l$ forces a linear formula, and the declaration $x : P$ cannot be present in an input context for a dual reason. The pair $(x : P,\, x : P)$ is not a context either. Its underlying pair $(x : c,\, x : c)$ declares the variable $x$ twice and is therefore not a context of the pre-signature.
\end{example}

We suppose now that we are given a family of type symbols $\mathrm{Prop}_{\Lambda,p}$ for each mode $p$ present in the signature $\Lambda$. The type formation rules introduce atomic and non-atomic types. The atomic types are the ones present in this family. The non-atomic types are the $F$ and $U$ connectives of the framework.     
\begin{definition}[Type Formation Rules]\label{def:type-formation-rules}
$$
\inferrule*
{P\in \mathrm{Prop}_{\Lambda,p}}
{P\ \mathrm{type}_{\Lambda,p}}
\qquad\qquad
\inferrule*
{
  \Theta_F\ \mathrm{ctx}^{\mathrm{c}}_{\Lambda;\theta_F}
  \\
  \Gamma_F\ \mathrm{ctx}^{\mathrm{in}}_{\Lambda;\gamma_F}
  \\
  \theta_F \ ; \ \gamma_F\Vdash_{\mf D_F}y:q
}
{F_{\mf D_F}^{y}(\Theta_F;\Gamma_F)\ \mathrm{type}_{\Lambda,q}}
$$
$$
\inferrule*
{
  \Theta_U\ \mathrm{ctx}^{\mathrm{c}}_{\Lambda;\theta_U}
  \\
  \Gamma_U\ \mathrm{ctx}^{\mathrm{in}}_{\Lambda;\gamma_U^L,\gamma_U^R}
  \\
  \Delta_U\ \mathrm{ctx}^{\mathrm{out}}_{\Lambda;\delta_U}
  \\
  \theta_U \ ; \ \gamma_U^L,x:q,\gamma_U^R \Vdash_{\mf D_U}\delta_U
}
{U^{\mf D_U}_{x}(\Theta_U;\Gamma_U\mid\Delta_U)\ \mathrm{type}_{\Lambda,q}}
$$
\end{definition}

\phantomsection\label{rem:type-formation-input-zone}
\textit{Remark:}
The input $x:q$ stands in the cartesian or linear context depending on the specialization of the rule. If it is present in the linear context, then the rule can be read verbatim. If, instead, it is present in the cartesian context, then the rule can be read as $\theta_U,x:q \ ; \ \gamma_U \Vdash_{\mf D_U}\delta_U$. 

\textit{Remark:} In the semantics, $F$ represents the pushforward (or the opcartesian lift of the morphism in the mode theory represented by the descriptor). This operation effectively internalizes the input into a single formula (read object in the polycategory). The situation is similar for the connective $U$, which is the cartesian lift of the descriptor in the mode theory. It internalizes the input slot with respect to the rest of the contexts (input and output sides).
The sorts of \cref{prop:sorts-of-well-formed-descriptors} specialize the instances of the connectives. For the connective $F$, given a c-descriptor $\mf D_F$, the linear part $\gamma_F$ is empty and the output mode of $y$ is cartesian. The connective internalizes a cartesian computation, and we abbreviate its instances as $\mathsf F_{\mathsf c}:=F^{y}_{\mf D_F}(\Theta_F;\cdot)$. If, on the other hand, $\mf D_F$ is a linear descriptor, then the output $q$ is linear and the input linear context $\gamma_F$ could be empty or contain linear variables.

The situation for $U$ is similar. Suppose that the descriptor $\mf D_U$ is a c-descriptor. Then $x$ is necessarily cartesian and the input context $\Gamma_U$ is empty. Moreover, the context $\Delta_U$ contains only one variable $b : X$ at a cartesian mode, and we abbreviate $\mathsf U_{\mathsf c}:=U^{\mf D_U}_{x}(\Theta_U;\cdot\mid b:X)$. These cartesian instances present the connectives of \cite{LSR17}.

\textit{Convention (connective notation):} For the connective $F$ we write the descriptor as a subscript and the output as a superscript to reflect the position of the formula in the sequent:  $F_{\mf D_F}^{y}$. For the connective $U$, we invert the notation and place the input as a subscript and the descriptor as a superscript, again to reflect the position of the formula in the sequent:  $U_{x}^{\mf D_U}$.

\phantomsection\label{rem:context-variable-naming-invariance}
\textit{Remark (context variable naming invariance):} The contexts $\Theta_F;\Gamma_F$, the output $y$ and the descriptor $\mf D_F$ are bound in the type expression $F_{\mf D_F}^{y}(\Theta_F;\Gamma_F)$, and similarly for the data of $U$. Type expressions are identified up to boundary renaming of this bound data (\cref{thm:renaming-admissibility-for-descriptor-judgments}), so that the freshness conditions of \cref{def:rule-schemas} can always be met.

\begin{example}\label{ex:linear-logic-type-formers}
	Suppose $l$ is a linear mode, i.e., $l \model \in \mathrm{Mode}(\underline{\Lambda})$ and
	$\otimes_l:l,l\to l$ is an $\ell$-generator of the underlying
	pre-signature $\underline{\Lambda}$. Applying the linear positive composition rule to form the descriptor generated by $\otimes_l$ with two identity premises gives us the well-formed descriptor $	 \mf D_\otimes	 :=	 [x,y\xmapsto{\otimes_l}w\mid w\mapsto z]	 \in	 \mathrm{Desc}_{\underline{\Lambda}}(x:l,y:l\ ;\ z:l),$ as in  \cref{def:well-formed-descriptor-positive}. The multiplicative conjunction is then represented by the $F$ connective for this descriptor.
	$$ A\otimes B := F_{\mf D_\otimes}^{z}(\cdot\ ;\ x:A,y:B). $$
	The implication connective associated to this conjunction is then represented by the $U$ connective over the same descriptor. 
	$$ A\multimap B := U_y^{\mf D_\otimes}(\cdot\ ;\ x:A\mid z:B). $$
	This connective is characterized by the evaluation map $ A\otimes(A\multimap B)\vdash B$. Note that without the substructural rule of exchange (at the logic level), this implication connective records the order of the formulas and is then not symmetric. This means that it is different from the connective implementing the other evaluation map $(A\multimap' B)\otimes A\vdash B$.
\end{example}

\begin{example}\label{ex:linear-logic-multiplicative-units}
  Suppose we add to the signature of \cref{ex:linear-logic-type-formers} a unit $u: \cdot \to l$, a counit $e: l \to \cdot$, and a linear generator $\parr: l \to l,l$ (note that this generator cannot be added in a fully cartesian signature, i.e., \cite{LSR17}-style signatures). Applying the generator rules for those maps gives rise to these descriptors:
  $$
  \mf D_u:=[\,\cdot\xmapsto{u}v\mid v\mapsto z\,],
  \qquad
  \mf D_e:=[\,x\xmapsto{e}\cdot\mid\varnothing\,],
  \qquad
  \mf D_{\parr}:=[\,z\xmapsto{\parr}w_1,w_2\mid w_1\mapsto x,w_2\mapsto y\,].
  $$
  Composing $\mf D_\otimes$ of \cref{ex:linear-logic-type-formers} with the newly generated $\mf D_e$, its input renamed to $x_e$ to keep the boundaries disjoint, gives $\mf D_\otimes\circ_{z,x_e}\mf D_e = [\,x,y\xmapsto{\otimes_l}w,\ w\xmapsto{e}\cdot\mid\varnothing\,]$; renaming the input $y$ to $x'$ gives the descriptor $\mf D_{\mathrm{cap}}$ over $\cdot\ ;\ x:l,\,x':l\Vdash\cdot$. The type formers then give us these connectives:
  $$
  \begin{aligned}
  \mathbf 1&:=F^{z}_{\mf D_u}(\cdot\,;\cdot),
  & A\parr B&:=U^{\mf D_{\parr}}_{z}(\cdot\,;\cdot\mid x{:}A,y{:}B),
  \\
  \bot&:=U^{\mf D_e}_{x}(\cdot\,;\cdot\mid\cdot),
  & A^{\perp}&:=U^{\mf D_{\mathrm{cap}}}_{x}(\cdot\,;x'{:}A\mid\cdot),
  \end{aligned}
  $$
  which instantiate the multiplicative theory of classical linear logic.
\end{example}

\subsubsection{Judgments}

With the context and connective formation rules of \cref{def:type-contexts,def:type-formation-rules} established, we introduce the logical rules of the framework: the left and right introduction rules and the axiom rule. We first state the two types of sequents that arise from our descriptors, then introduce a \emph{uniform sequent display} that generalizes them.
\begin{definition}[Sequent Judgments]\label{def:sequent-judgments}
Let $\Lambda\ \mathrm{Sig}$ be a signature. Our type theory has two types of \emph{sequent judgments}, one for each of the cartesian and linear types.
\begin{itemize}
\item Given a cartesian context $\Theta\ \mathrm{ctx}^{\mathrm{c}}_{\Lambda;\theta}$,  the cartesian sequent $\Theta\vdash^{\Lambda}_{\mf D}A$ is such that the output of the sequent is cartesian as well. This is the \cite{LSR17} equivalent of the sequent $\theta\vdash\alpha:p$, where $\alpha$ is a \cite{LSR17} descriptor.
\item Given the contexts $\Theta\ \mathrm{ctx}^{\mathrm{c}}_{\Lambda;\theta}$, $\Gamma\ \mathrm{ctx}^{\mathrm{in}}_{\Lambda;\gamma}$, $\Delta\ \mathrm{ctx}^{\mathrm{out}}_{\Lambda;\delta}$, the \emph{linear sequent} $\Theta\ ;\ \Gamma\vdash^{\Lambda}_{\mf D}\Delta$ is such that the formulas in $\Theta$ are cartesian while the formulas in all other contexts are linear. 
\end{itemize}
\end{definition}

\textit{Convention (uniform sequent display):} We display both sequent judgments uniformly as
$$
\Theta\ ;\ \Gamma\vdash^{\Lambda}_{\mf D}\Delta .
$$
When $\mf D$ is an $\ell$-descriptor, this is the linear sequent. When $\mf D$ is a c-descriptor, $\Gamma=\cdot$ and $\Delta$ is a single declaration $b:A$ at a cartesian mode by \cref{def:c-descriptors}, and the sequent notation collapses to the cartesian sequent $\Theta\vdash^{\Lambda}_{\mf D}A$.

\begin{example}
Let $A, B$ be types at the linear mode $l$ and recall the tensor descriptor $\mf D_\otimes$ of \cref{ex:linear-logic-type-formers}, well formed over $(x:l,y:l\ ;\ z:l)$. Forming the contexts $\Gamma := x : A,\, y : B$ and $\Delta := z : A \otimes B$ gives the linear sequent
$$
\cdot\ ;\ x:A,\, y:B \vdash^{\Lambda}_{\mf D_\otimes} z : A \otimes B,
$$
with empty cartesian zone and linear formulas on both sides of the turnstile. For a cartesian mode $p \modec$, a type $X$ at $p$, and the identity c-descriptor $[\varnothing\mid\{a\mapsto b\}]$ over $(a:p\ ;\ b:p)$, the cartesian sequent reads $a:X \vdash^{\Lambda}_{[\varnothing\mid\{a\mapsto b\}]} b:X$. In the uniform display it is the sequent $a:X\ ;\ \cdot \vdash^{\Lambda}_{[\varnothing\mid\{a\mapsto b\}]} b:X$, whose linear zone is empty and whose single output $b:X$ stands at the cartesian mode $p$.
\end{example}

\subsubsection{The Rule Schemas}

We introduce first a notation for the cumbersome iterated cut:
\begin{align*}
  \bigl(\mf D_{\bar x_i}\bigr)_i
  \circ_{(\bar x_i,x_i)_{1 \leq i \leq n}}
  \mf U
  &:=
  \mf D_{\bar x_1}
  \circ_{\bar x_1,x_1}
  \bigl(
    \cdots
    (
      \mf D_{\bar x_n}
      \circ_{\bar x_n,x_n}
      \mf U
    )
    \cdots
  \bigr), \\
  \mf U
  \circ_{(y_j,\bar y_j)_{1 \leq j \leq m}}
  \bigl(\mf D_{\bar y_j}\bigr)_j
  &:=
  \bigl(
    \cdots
    \bigl(
      (
        \mf U
        \circ_{y_1,\bar y_1}
        \mf D_{\bar y_1}
      )
      \circ_{y_2,\bar y_2}
      \cdots
    \bigr)
    \circ_{y_m,\bar y_m}
    \mf D_{\bar y_m}
  \bigr).
\end{align*}
The first notation indicates a succession of left cuts, and the second pertains to a succession of right cuts. If $n=0$ or $m=0$, the notation returns $\mf U$ for the left iterated and the right iterated cut, respectively. When the context is clear, we will feel free to omit the indices at the composition and descriptor level. The mere sight of an index $i$ such as in $\mf D_i$, paired with a composition operation, should suggest to the reader that an iterated composition is meant. Thus the iterated composition displayed before can be written concisely as:
$$\mf D_{\bar x_i} \circ \mf U \quad \text{and} \quad \mf U \circ \mf D_{\bar y_j}.$$

\textit{Convention (principal zones):} The schemas display every principal declaration in the linear zone. When its mode is cartesian, the declaration stands in the cartesian zone instead, in accordance with the \hyperref[rem:type-formation-input-zone]{mode-class convention for distinguished inputs}: the variable rule then reads $x:P$ from $\Theta$, and the $U$-left rule reads $z:\mathsf U$ from $\Theta$. The specializations of \cref{prop:rule-specialization} display those readings explicitly.

\textit{Convention (boundary arrangement):} The composition of \cref{def:composition-of-descriptors} produces its boundary in a canonical arrangement: the cut asks for the distinguished output of its left factor to stand last, and the composite lives over $(\gamma_1\cup\gamma_2\ ;\ \delta_1,\delta_2)$. In the schemas, every composite index is therefore silently pre- and post-composed with the permutation transports of \cref{thm:exchange-admissibility-for-descriptor-judgments}: the premise indexes are transported to place the cut variables in the required positions, and the composite is transported to the arrangement displayed in the conclusion. Similarly, the renamings $[y:=z]$ and $[x:=z]$ in the schemas rename the distinguished port of the connective descriptor to the conclusion label (\cref{thm:renaming-admissibility-for-descriptor-judgments}).

\begin{definition}[Rule Schemas]\label{def:rule-schemas}
The two sequent judgments are generated by the five schemas displayed here. We state them first in the uniform notation of \cref{def:sequent-judgments} and specialize them in \cref{prop:rule-specialization}.

\begin{itemize}
  \item \textbf{Variable.} For $P\in\mathrm{Prop}_{\Lambda,q}$ at a mode $q$ of either class:
  $$
  \inferrule*[Right=$\mathsf{v}$]
  {
    P\in \mathrm{Prop}_{\Lambda,q}
    \\
    r::\mf D
    \Rightarrow^{\Lambda}
    [\varnothing\mid\{x\mapsto y\}]
  }
  {
    \Theta\ ;\ x:P
    \vdash_{\mf D}^{\Lambda}
    y:P
  }
  $$

  \item \textbf{F-left.} Let $\mathsf F := F_{\mf D_F}^{y}(\Theta_F;\Gamma_F)$ with $\theta_F \ ; \ \gamma_F\Vdash_{\mf D_F}y:q$ of either sort:
$$
\inferrule*[Right=$F$L]
{
  \Theta,\Theta_F\ ;\
  \Gamma_L,\Gamma_F,\Gamma_R
  \vdash_{\mf D_F\circ_{y,z}\mf D}^{\Lambda}
  \Delta
}
{
  \Theta\ ;\
  \Gamma_L,\,
  z:\mathsf F,\,
  \Gamma_R
  \vdash_{\mf D}^{\Lambda}
  \Delta
}
$$

  \item \textbf{F-right.} Let $\mathsf F$ have $\Gamma_F=x_1:A_1,\ldots,x_n:A_n$ over $\gamma_F$, let $\sigma=(\sigma_a)_{a:p\in\theta_F}$ be a family of c-descriptors over the ambient cartesian context $\Theta$, i.e., over its index $\theta$, let the $\bar x_i:p_i$ be fresh output labels, and let the side outputs be partitioned into disjoint slices $\Delta_L=\Delta_L^{(1)},\ldots,\Delta_L^{(n)}$ and $\Delta_R=\Delta_R^{(1)},\ldots,\Delta_R^{(n)}$ as well as the side input $\Gamma :=\Gamma^{(1)},\ldots,\Gamma^{(n)}$:
$$
\inferrule*[Right=$F$R]
{
  r::\mf D
  \Rightarrow^{\Lambda}
  \bigl(\mf D_{\bar x_i}\bigr)_i
  \circ_{(\bar x_i,x_i)_{1 \leq i \leq n}}
  \Bigl(\bigl(\sigma_a\bigr)_{a\in\theta_F}\circ_{(b_a,a)}\mf D_F\Bigr)[y:=z]
  \\
  \bigl(\Theta\vdash^{\Lambda}_{\sigma_a}A_a\bigr)_{a\in\theta_F}
  \\
  \bigl(
    \Theta\ ;\ \Gamma^{(i)}
    \vdash_{\mf D_{\bar x_i}}^{\Lambda}
    \Delta_L^{(i)},\,\bar x_i:A_i,\,\Delta_R^{(i)}
  \bigr)_{1\leq i\leq n}
}
{
  \Theta\ ;\ \Gamma
  \vdash_{\mf D}^{\Lambda}
  \Delta_L,\,z:\mathsf F,\,\Delta_R
}
$$

  \item \textbf{U-right.} Let $\mathsf U := U_{x}^{\mf D_U}(\Theta_U;\Gamma_U\mid\Delta_U)$ with connective descriptor of either sort and distinguished input $x:q$ in either zone:
$$
\inferrule*[Right=$U$R]
{
  \Theta,\Theta_U\ ;\ \Gamma_U^L,\, \Gamma,\, \Gamma_U^R
  \vdash_{\mf D\circ_{z,x}\mf D_U}^{\Lambda}
  \Delta_L,\Delta_U,\Delta_R
}
{
  \Theta\ ;\ \Gamma
  \vdash_{\mf D}^{\Lambda}
  \Delta_L,\,z:\mathsf U,\,\Delta_R
}
$$

  \item \textbf{U-left.} Let $\mathsf U$ have $\Gamma_U=\Gamma_U^L,\Gamma_U^R=x_1:A_1,\ldots,x_n:A_n$ and $\Delta_U=y_1:B_1,\ldots,y_m:B_m$, let $\sigma=(\sigma_a)_{a:p\in\theta_U}$ be a family of c-descriptors over the ambient cartesian context $\Theta$, i.e., over its index $\theta$, let the $\bar x_i:p_i$ be fresh output labels and the $\bar y_j$ fresh variables, each declared in the zone of its mode class, and let the side output $\Delta$ be partitioned into disjoint slices indexed by the $n+m$ components: each input component carries a two-part slice $\Delta_L^{(i)},\Delta_R^{(i)}$, each output component carries a single slice $\Delta^{(n+j)}$, no output being distinguished among the latter (its distinguished port $\bar y_j:B_j$ is an input), and let the side input be partitioned as $\Gamma_L,\Gamma_R=\Gamma^{(1)},\ldots,\Gamma^{(n+m)}$. We write the composite:
$$
\mf C:=\Bigl(\bigl(\mf D_{\bar x_i}\bigr)_i\circ_{(\bar x_i,x_i)_{1\le i\le n}}\bigl(\bigl(\sigma_a\bigr)_{a\in\theta_U}\circ_{(b_a,a)}\mf D_U\bigr)[x:=z]\Bigr)
   \circ_{(y_j,\bar y_j)_{1\le j\le m}}\bigl(\mf D_{\bar y_j}\bigr)_j\ ;
$$
then
$$
\inferrule*[Right=$U$L]
{
  r::\mf D
  \Rightarrow^{\Lambda}
  \mf C
  \\
  \bigl(\Theta\vdash^{\Lambda}_{\sigma_a}A_a\bigr)_{a\in\theta_U}
  \\
  \bigl(
    \Theta\ ;\ \Gamma^{(i)}
    \vdash_{\mf D_{\bar x_i}}^{\Lambda}
    \Delta_L^{(i)},\,\bar x_i:A_i,\,\Delta_R^{(i)}
  \bigr)_{1\leq i\leq n}
  \\
  \bigl(
    \Theta\ ;\ \Gamma^{(n+j)},\,\bar y_j:B_j
    \vdash_{\mf D_{\bar y_j}}^{\Lambda}
    \Delta^{(n+j)}
  \bigr)_{1\leq j\leq m}
}
{
  \Theta\ ;\ \Gamma_L,\,z:\mathsf U,\,\Gamma_R
  \vdash_{\mf D}^{\Lambda}
  \Delta
}
$$
\end{itemize}
\end{definition}

The schemas of \cref{def:rule-schemas} have been written in their full generality. Our framework makes use of only two types of sequents, the cartesian and the linear. The schemas then specialize to those two sequents. In the cartesian case, we recover a single-conclusion cartesian calculus; in the linear case, the rules with the intended polycategorical semantics.
\subsubsection{Specialization: the Two Layers}

We write down now in full the sequent rules. Notice that there are many cases to consider: the mode of the connective type (cartesian or linear) and the position of its formula (in the cartesian or the linear zone of the sequent). A formula at a linear mode cannot stand in a cartesian zone, so one of the four combinations is impossible and three remain. The case of a cartesian-mode type splits further according to the sort of its connective descriptor: a c-descriptor produces the cartesian layer, while an $\ell$-descriptor whose distinguished input is cartesian produces the shift layer. This is why \cref{prop:rule-specialization} displays a third layer next to the cartesian and the linear ones. We note also that a cartesian connective can stand in a linear sequent, since linear sequents possess cartesian contexts; the corresponding left rules are listed with the cartesian layer. Finally, the connective $F$ has no shift instance: by \cref{prop:sorts-of-well-formed-descriptors} the output of an $\ell$-descriptor is at a linear mode, so an $F$ type formed over an $\ell$-descriptor never sits at a cartesian mode.

\begin{proposition}[Rule Specialization]\label{prop:rule-specialization}
The instances of the five schemas, determined by the sort of the connective descriptor, the class of the principal mode, and the sort of the conclusion index, are exactly these rules.

\phantomsection\label{item:rule-specialization-cartesian-layer}
\emph{(i) The cartesian layer.} When all the data is cartesian, the uniform presentation collapses to cartesian sequents. With $\mathsf F_{\mathsf c}:=F^{y}_{\mf D_F}(\Theta_F;\cdot)$ and $\mathsf U_{\mathsf c}:=U^{\mf D_U}_{x}(\Theta_U;\cdot\mid b:X)$, we are left with the simplified rules:
$$
\inferrule*[right=$\mathsf{v}$]
{x{:}P\in\Theta \\ r::\mf D\Rightarrow^{\Lambda}[\varnothing\mid\{x\mapsto u\}]}
{\Theta\vdash^{\Lambda}_{\mf D}u:P}
\qquad\qquad
\inferrule*[right=$F_{\mathsf c}$L]
{\Theta,\Theta_F\vdash^{\Lambda}_{\mf D_F\circ_{y,z}\mf D}u:C}
{\Theta,\,z:\mathsf F_{\mathsf c}\vdash^{\Lambda}_{\mf D}u:C}
$$
$$
\inferrule*[right=$F_{\mathsf c}$R]
{r::\mf D\Rightarrow^{\Lambda}\bigl((\sigma_a)_{a \in \theta_F} \circ_{(b_a,a)} \mf D_F\bigr)[y:=z] \\ \bigl(\Theta\vdash^{\Lambda}_{\sigma_a}A_a\bigr)_{a \in \theta_F}}
{\Theta\vdash^{\Lambda}_{\mf D}z:\mathsf F_{\mathsf c}}
$$
$$
\inferrule*[right=$U_{\mathsf c}$R]
{\Theta,\Theta_U\vdash^{\Lambda}_{\mf D\circ_{z,x}\mf D_U}b:X}
{\Theta\vdash^{\Lambda}_{\mf D}z:\mathsf U_{\mathsf c}}
$$
$$
\inferrule*[right=$U_{\mathsf c}$L]
{z:\mathsf U_{\mathsf c}\in\Theta \\
 r::\mf D\Rightarrow^{\Lambda}\mf D'[x':=((\sigma_a)_{a\in\theta_U}\circ\mf D_U)[x:=z]] \\
 \bigl(\Theta\vdash^{\Lambda}_{\sigma_a}A_a\bigr)_{a\in\theta_U} \\
 \Theta,\,x':X\vdash^{\Lambda}_{\mf D'}u:C}
{\Theta\vdash^{\Lambda}_{\mf D}u:C}
$$
When the connective is cartesian but the conclusion sequent containing it is linear, the same instances connect the two judgments:
$$
\inferrule*[right=$F_{\mathsf c}$L-in-l]
{\Theta,\Theta_F\ ;\ \Gamma\vdash^{\Lambda}_{\mf D_F\circ_{y,z}\mf D}\Delta}
{\Theta,\,z:\mathsf F_{\mathsf c}\ ;\ \Gamma\vdash^{\Lambda}_{\mf D}\Delta}
$$
$$
\inferrule*[right=$U_{\mathsf c}$L-in-l]
{r::\mf D\Rightarrow^{\Lambda}\mf D'[x':=((\sigma_a)_{a\in\theta_U}\circ\mf D_U)[x:=z]] \\
 \bigl(\Theta,z:\mathsf U_{\mathsf c}\vdash^{\Lambda}_{\sigma_a}A_a\bigr)_{a\in\theta_U} \\
 \Theta,z:\mathsf U_{\mathsf c},\,x':X\ ;\ \Gamma\vdash^{\Lambda}_{\mf D'}\Delta}
{\Theta,\,  z:\mathsf U_{\mathsf c} \ ;\ \Gamma\vdash^{\Lambda}_{\mf D}\Delta}
$$

\begin{samepage}
\phantomsection\label{item:rule-specialization-linear-layer}
\emph{(ii) The linear layer.} When the connective descriptor is an $\ell$-descriptor present in a linear sequent, all judgments are linear and the schemas specialize to the polycategorical rules, with $\mathsf F=F^{y}_{\mf D_F}(\Theta_F;\Gamma_F)$ and $\mathsf U=U^{\mf D_U}_{x}(\Theta_U;\Gamma_U\mid\Delta_U)$:
$$
\inferrule*[right=$\mathsf{v}_{\mathsf l}$]
{P\in\mathrm{Prop}_{\Lambda,q} \\ q\model \\ r::\mf D\Rightarrow^{\Lambda}_{\theta;x:q\Vdash y:q}[\varnothing\mid\{x\mapsto y\}]}
{\Theta\ ;\ x{:}P\vdash^{\Lambda}_{\mf D}y{:}P}
$$
$$
\inferrule*[right=$F_{\mathsf l}$L]
{\Theta,\Theta_F\ ;\ \Gamma_L,\Gamma_F,\Gamma_R\vdash^{\Lambda}_{\mf D_F\circ_{y,z}\mf D}\Delta}
{\Theta\ ;\ \Gamma_L,\,z{:}\mathsf F,\,\Gamma_R\vdash^{\Lambda}_{\mf D}\Delta}
$$
\end{samepage}
$$
\inferrule*[right=$F_{\mathsf l}$R]
{
  r::\mf D
  \Rightarrow^{\Lambda}
  \bigl(\mf D_{\bar x_i}\bigr)_i
  \circ_{(\bar x_i,x_i)}
  \bigl(\bigl(\sigma_a\bigr)_{a\in\theta_F}\circ_{(b_a,a)}\mf D_F\bigr)[y:=z]
  \\
  \bigl(\Theta\vdash^{\Lambda}_{\sigma_a}A_a\bigr)_{a\in\theta_F}
  \\
  \bigl(
    \Theta\ ;\ \Gamma_i
    \vdash_{\mf D_{\bar x_i}}^{\Lambda}
    \Delta_L^{(i)},\,\bar x_i:A_i,\,\Delta_R^{(i)}
  \bigr)_{i}
}
{
  \Theta\ ;\ \Gamma
  \vdash_{\mf D}^{\Lambda}
  \Delta_L,\,z:\mathsf F,\,\Delta_R
}
$$
$$
\inferrule*[right=$U_{\mathsf l}$R]
{
  \Theta,\Theta_U\ ;\ \Gamma_U^L,\, \Gamma,\, \Gamma_U^R
  \vdash_{\mf D\circ_{z,x}\mf D_U}^{\Lambda}
  \Delta_L,\Delta_U,\Delta_R
}
{
  \Theta\ ;\ \Gamma
  \vdash_{\mf D}^{\Lambda}
  \Delta_L,\,z:\mathsf U,\,\Delta_R
}
$$
$$
\inferrule*[right=$U_{\mathsf l}$L]
{
  r::\mf D
  \Rightarrow^{\Lambda}
  \mf C
  \\
  \bigl(\Theta\vdash^{\Lambda}_{\sigma_a}A_a\bigr)_{a\in\theta_U}
  \\
  \bigl(
    \Theta\ ;\ \Gamma^{(i)}
    \vdash_{\mf D_{\bar x_i}}^{\Lambda}
    \Delta_L^{(i)},\,\bar x_i:A_i,\,\Delta_R^{(i)}
  \bigr)_{i}
  \\
  \bigl(
    \Theta\ ;\ \Gamma^{(n+j)},\,\bar y_j:B_j
    \vdash_{\mf D_{\bar y_j}}^{\Lambda}
    \Delta^{(n+j)}
  \bigr)_{j}
}
{
  \Theta\ ;\ \Gamma_L, \, z:\mathsf U,\,\Gamma_R
  \vdash_{\mf D}^{\Lambda}
  \Delta
}.
$$

\phantomsection\label{item:rule-specialization-linear-to-cartesian-layer}
\emph{(iii) The linear-to-cartesian layer.} When the connective descriptor is a linear $\ell$-descriptor in a cartesian slot, the type $\mathsf U=U^{\mf D_U}_{x}(\Theta_U;\Gamma_U\mid\Delta_U)$ sits at a cartesian mode and the $U$-schemas connect the two judgments:
$$
\inferrule*[right=$U_{\mathsf{shift}}$R]
{\Theta,\Theta_U\ ;\ \Gamma_U\vdash^{\Lambda}_{\mf D\circ_{z,x}\mf D_U}\Delta_U}
{\Theta\vdash^{\Lambda}_{\mf D}z:\mathsf U}
$$
$$
\inferrule*[right=$U_{\mathsf{shift}}$L]
{z:\mathsf U\in\Theta \\
 r::\mf D
  \Rightarrow^{\Lambda}
 \bigl(\bigl(\mf D_{\bar x_i}\bigr)_i\circ\bigl((\sigma_a)_a\circ\mf D_U\bigr)[x:=z]\bigr)\circ\bigl(\mf D_{\bar y_j}\bigr)_j \\
 \bigl(\Theta\vdash^{\Lambda}_{\sigma_a}A_a\bigr)_{a\in\theta_U} \\
 \bigl(
   \Theta\ ;\ \Gamma^{(i)}
   \vdash_{\mf D_{\bar x_i}}^{\Lambda}
   \Delta_L^{(i)},\,\bar x_i:A_i,\,\Delta_R^{(i)}
 \bigr)_{1\leq i\leq n} \\
 \bigl(
   \Theta\ ;\ \Gamma^{(n+j)},\,\bar y_j:B_j
   \vdash_{\mf D_{\bar y_j}}^{\Lambda}
   \Delta^{(n+j)}
 \bigr)_{1\leq j\leq m}}
{\Theta\ ;\ \Gamma^{(1)},\ldots,\Gamma^{(n+m)}\vdash^{\Lambda}_{\mf D}\Delta}
$$
\end{proposition}

\textit{Proof.}

The instantiation of the rules is done by enforcing the sorts of the three parameters. For a c-descriptor connective, \cref{prop:sorts-of-well-formed-descriptors} forces $\gamma_F=\cdot$ with a cartesian output for $F$, and $\Gamma_U=\cdot$, $\Delta_U=b:X$ with a cartesian distinguished input for $U$. The principal declaration then stands in the cartesian zone by the principal-zones convention, which yields the cartesian layer, in a cartesian or in a linear conclusion. For an $\ell$-descriptor connective whose type sits at a linear mode, all the judgments are linear and the schemas instantiate verbatim to the linear layer. For an $\ell$-descriptor $U$ whose distinguished input is cartesian, the type sits at a cartesian mode and the principal declaration again stands in the cartesian zone, which yields the shift layer; there is no such instance for $F$ since the output of an $\ell$-descriptor is linear. No other combination exists. In each rule, \cref{lem:cut-preserves-the-sorts} determines the sorts of the premise indexes, hence the types of sequents in the premises.\hfill $\square$

A judgment $\Theta;\Gamma\vdash_{\mf D}^{\Lambda}\Delta$ represents intuitively a proof of the targets contained in $\Delta$ using the resources in $\Theta,\Gamma$. The descriptor indexing the sequent records the information needed to construct the proof given the resources.

\begin{example}\label{ex:linear-logic-sequent-rules}
  We can now show how the framework derives the connectives of linear logic introduced in \cref{ex:linear-logic-type-formers,ex:linear-logic-multiplicative-units}.
   Let $P\in\mathrm{Prop}_{\Lambda,q}$ be an atomic formula at a linear mode $q$. The identity rule of \cref{def:well-formed-descriptors-identity} derives the descriptor $[\varnothing \mid \{x \mapsto y\}]$ in the linear context $x : q \Vdash y : q$. An application of the variable rule using this descriptor gives us
  $$
  \inferrule*
  {
    P\in\mathrm{Prop}_{\Lambda,q}
    \\
    [\varnothing\mid\{x\mapsto y\}]\Rightarrow^{\Lambda}_{x:q\Vdash y:q}[\varnothing\mid\{x\mapsto y\}]
  }
  {
    \cdot\ ;\ x:P\vdash_{[\varnothing\mid\{x\mapsto y\}]}^{\Lambda}y:P
  }
  $$
  where the second premise is the identity $2$-cell. This corresponds to the axiom in the logic. An interesting fact about the framework is that a cartesian version exists as well. By taking the cartesian context $\Theta$ to contain $x : P$ and using the identity 2-cell again, we recover the sequent $$\Theta \vdash^{\Lambda}_{[\varnothing \mid \{x \mapsto y\}]} y : P.$$

  In the tensor instance $\otimes$, the descriptor $\mf D_\otimes$ of \cref{ex:linear-logic-type-formers} generates the connective $A \otimes B$ using the $F$ introduction rule: $$
	\inferrule*
	{
	  \cdot\ ;\ x:A,y:B\vdash_{\mf D_\otimes^u\circ_{u,z}\mf E}^{\Lambda}r:C
	}
	{
	  \cdot\ ;\ z:A\otimes B\vdash_{\mf E}^{\Lambda}r:C
	}.
$$
Here $\mf D_\otimes^u$ is $\mf D_\otimes$ with its output renamed to $u$; the renaming avoids the clash with the input $z$ of $\mf E$. Now, consider the descriptor
	$\mf D'_\otimes := [x',y'\xmapsto{\otimes_l}w\mid w\mapsto z]$;
	the right introduction rule of the $F$ connective has as its conclusion:
	$$
	\cdot\ ;\ x':A,y':B\vdash_{\mf D'_\otimes}^{\Lambda}z:A\otimes B,
	$$
	and the full rule including premises becomes (forgetting about the descriptors)
	$$
	\inferrule*
	{
	  A\vdash A
	  \\
	  B\vdash B
	}
	{
	  A,B\vdash A\otimes B
	}.
	$$
\end{example}

\begin{example}\label{ex:linear-logic-exponential-modality} One of the interesting properties of the framework, already present in \cite{LSR17}, is the presence of the exponential modalities. Recall our signature possesses a cartesian mode $c \modec$ and a linear generator $g:c\to l$. The descriptor generated from this map is $\mf D_g:=[\,(\hat x\ ;\ )\xmapsto{g}v\mid v\mapsto y\,]$ over the sequent $\hat x:c \ ; \ \cdot\Vdash y:l$. The ``of course'' modality of linear logic is represented as the comonad of the cartesian-linear adjunction \cite{Ben95,Mel09}. It is defined as
$$
G(A):=U^{\mf D_g}_{\hat x}(\cdot\,;\cdot\mid y{:}A)\ \mathrm{type}_{c},
\qquad
{!A}:=F^{y}_{\mf D_g}\bigl(\hat x{:}G(A)\,;\cdot\bigr)\ \mathrm{type}_{l} .
$$
The rule of promotion is then $U_{\mathsf{shift}}$R, and the rule of dereliction $!A\vdash A$ is $F_{\mathsf l}$L followed by $U_{\mathsf{shift}}$L on $\hat x$ with a single output component given by the variable rule. Contraction and weakening of $!A$ are admissible by design by taking advantage of the cartesianness of $c$. Indeed, since $\hat x:G(A)$ lives in the cartesian zone, it can be reused in two $F_{\mathsf l}$R's and recombined by the linear cut of the metatheory, or simply ignored. The structural behaviour of $!$ is just the cartesian zone performing its role.
\end{example}

\subsubsection{Structural Rules at Linear Modes}

As established in \cref{subsec:substructural-rules}, the substructural rules at cartesian modes are handled using $2$-cells in the mode theory. This is possible due to the fact that, at the framework level, the substructural rules are admissible. The $2$-cell mechanism allows the logic to benefit from them. For the linear modes, the rules of contraction and weakening are not even admissible at the framework level since the resources are used exactly once. For instance, to construct a 2-cell between a tensor descriptor (abbreviated for the sake of the discussion as $x \otimes y$) and the identity descriptor $x$, we would need both of them to live over the same contexts. This is only possible if the contexts are cartesian. Thus, we
expand the type theory to accommodate the substructural rules at linear modes. 
The mechanism used here is novel to the knowledge of the author. We add the notion of \emph{gates} to the mode theory (\cref{def:signature}, \cref{def:gate-descriptor-generation}).   

\begin{definition}[Gate Rules for Descriptors]\label{def:gate-descriptor-generation}
Let $\Lambda$ declare gates at the linear mode $l$, with adjoined generators as in \cref{def:signature}. The well-formed descriptor judgment is extended by four gate rules. In each rule the premise class is an $\ell$-descriptor and all the variables displayed are chosen fresh. As for the generation rules of the ungated judgment (\cref{def:well-formed-descriptor-positive}), each gate rule closes with a permutation transport: its conclusion may be followed by any context permutation of the boundary, transported as in \cref{thm:exchange-admissibility-for-descriptor-judgments}. We leave this closing transport implicit in the displays. The operations themselves are indexed by the gate variables and read no positions in the contexts.
\begin{itemize}
\item \textbf{Gate-weak.} If the signature declares $\mathsf{weak}(l)$ with generator $w_l$, then we add this rule and descriptor:
$$
\inferrule*[Right=Gate-w]
{\theta \ ; \ \Gamma \Vdash_{\mf D} \delta}
{\theta \ ; \ \Gamma,x:l \Vdash_{\mf D\cup w_l(x)} \delta}
\qquad
\mf D\cup w_l(x):=\bigl[\,\mc S\cup\{(\ ;x)\xmapsto{w_l}\cdot\}\mid\mc T\,\bigr].
$$
\item \textbf{Gate-coweak.} If the signature declares $\mathsf{coweak}(l)$ and the generator $w^\perp_l$, then we add this rule and descriptor:
$$
\inferrule*[Right=Gate-cw]
{\theta \ ; \ \Gamma \Vdash_{\mf D} \delta}
{\theta \ ; \ \Gamma \Vdash_{\mf D\cup w^\perp_l(b)} \delta,b:l}
\qquad
\mf D\cup w^\perp_l(b):=\bigl[\,\mc S\cup\{(\ ;)\xmapsto{w^\perp_l}v\}\mid\mc T\cup\{v\mapsto b\}\,\bigr].
$$
\item \textbf{Gate-contr.} If the signature declares $\mathsf{contr}(l)$ and the generator $c_l$, then we add this rule and descriptor:
$$
\inferrule*[Right=Gate-c]
{\theta \ ; \ \Gamma,x:l,x':l \Vdash_{\mf D} \delta}
{\theta \ ; \ \Gamma,x_0:l \Vdash_{c_l(x_0;x,x')\cdot\mf D} \delta}
$$
$$
c_l(x_0;x,x')\cdot\mf D:=\bigl[\,\mc S[x:=v][x':=v']\cup\{(\ ;x_0)\xmapsto{c_l}v,v'\}\mid\mc T[x:=v][x':=v']\,\bigr].
$$
\item \textbf{Gate-cocontr.} If the signature declares $\mathsf{cocontr}(l)$ and the generator $c^\perp_l$, then we add this rule and descriptor:
$$
\inferrule*[Right=Gate-cc]
{\theta \ ; \ \Gamma \Vdash_{\mf D} \delta,b:l,b':l}
{\theta \ ; \ \Gamma \Vdash_{\mf D\cdot c^\perp_l(b,b';b_0)} \delta,b_0:l}
$$
$$
\mf D\cdot c^\perp_l(b,b';b_0):=\bigl[\,\mc S\cup\{(\ ;s,s')\xmapsto{c^\perp_l}v\}\mid(\mc T\smallsetminus\{s\mapsto b,\,s'\mapsto b'\})\cup\{v\mapsto b_0\}\,\bigr],
$$
where $s$ and $s'$ are the unique sources with $(s\mapsto b),(s'\mapsto b')\in\mc T$.
\end{itemize}
\end{definition}

\textit{Remark.} We note that in the case of allowing the substructural rules by way of $2$-cells at the cartesian modes, the descriptors remain the same. In the current linear situation, the addition of the gates changes the descriptors. This is an instantiation of the concept of property and structure. The substructural properties at cartesian modes are provided as properties, and they are provided as structure in the case of linear modes.

\begin{example}\label{ex:linear-logic-gate-descriptors}
Suppose the four gates are declared at the linear mode $l$, and recall the descriptors $\mf D_\otimes=[\,x,y\xmapsto{\otimes_l}w\mid w\mapsto z\,]$ and $\mf D_{\parr}=[\,z\xmapsto{\parr}w_1,w_2\mid w_1\mapsto x,w_2\mapsto y\,]$ of \cref{ex:linear-logic-type-formers,ex:linear-logic-multiplicative-units}. Applying Gate-w to $\mf D_\otimes$ at a fresh input $x'':l$ gives
$$
\mf D_\otimes\cup w_l(x'')=\bigl[\,x,y\xmapsto{\otimes_l}w,\ (\ ;x'')\xmapsto{w_l}\cdot\mid w\mapsto z\,\bigr]
$$
over the sequent $\cdot\ ;\ x:l,y:l,x'':l\Vdash z:l$. The new input is discarded in the descriptor and is present in the context. Applying Gate-c to the two inputs of $\mf D_\otimes$ gives
$$
c_l(x_0;x,y)\cdot\mf D_\otimes=\bigl[\,v,v'\xmapsto{\otimes_l}w,\ (\ ;x_0)\xmapsto{c_l}v,v'\mid w\mapsto z\,\bigr]
$$
over $\cdot\ ;\ x_0:l\Vdash z:l$; the single input $x_0$ produces the two nodes $v,v'$ to feed the tensor node. Dually, Gate-cw adjoins an output produced out of no sources, and Gate-cc applied to the two outputs of $\mf D_{\parr}$ gives
$$
\mf D_{\parr}\cdot c^\perp_l(x,y;b_0)=\bigl[\,z\xmapsto{\parr}w_1,w_2,\ (\ ;w_1,w_2)\xmapsto{c^\perp_l}u\mid u\mapsto b_0\,\bigr]
$$
over $\cdot\ ;\ z:l\Vdash b_0:l$, the two redirections $w_1\mapsto x$ and $w_2\mapsto y$ replaced by the cocontraction node at their sources.
\end{example}

\begin{proposition}[Gate rules preserve regularity]\label{prop:gate-rules-preserve-regularity}
The gate operations are well defined on descriptor classes, and every descriptor derivable in the judgment extended by the gate rules is regular. In other words, \cref{thm:well-formed-descriptors-are-regular} holds for gated signatures.
\end{proposition}

\textit{Proof.}
We first check well-definedness on classes. Each construction commutes with internal renamings of the premise representative since the adjoined nodes involve only boundary variables and fresh internals, and for Gate-cocontr the distinguished sources $s,s'$ are carried to the distinguished sources of the renamed representative, as in the well-definedness proofs of the generation rules. Different choices of the fresh internals $v,v'$ are matched by the evident renaming.

We check now regularity, rule by rule, for a regular premise representative:
\begin{itemize}
\item \emph{Gate-w.} The new node consumes the fresh linear input $x$ exactly once and has no targets, so it contributes no reachability edges and no internal variables. Every regularity condition is inherited, and boundary usage holds at $x$. 
\item \emph{Gate-cw.} The new component contributes the single edge $v\leadsto b$ on fresh variables. Acyclicity is preserved, $v$ is produced once (by the node) and consumed once (by the redirection), and $b$ is the target of exactly one redirection.
\item \emph{Gate-c.} By boundary usage each of $x,x'$ occurs in exactly one source position of the premise, so after the substitutions each of $v,v'$ occurs in exactly one source position, and each is produced exactly once by the new node. $x_0$ is consumed exactly once by it. The new edges run from the input $x_0$ to $v,v'$ and thence into the premise graph at the former positions of $x,x'$. Since no edge enters an input variable, no cycle can arise.
\item \emph{Gate-cc.} By boundary usage the sources $s,s'$ exist and are unique, and they are distinct. Each is at the linear mode $l$ by mode validity, so if internal it has a unique source occurrence and if an input it is consumed exactly once. Either way it cannot source two redirections. Replacing the two redirections by the node preserves the single consumption of $s$ and $s'$. The fresh $v$ is produced and consumed exactly once, $b_0$ is produced exactly once, and the new edges $s,s'\leadsto v\leadsto b_0$ end in a sink, so acyclicity is preserved.
\end{itemize}
The extension of \cref{thm:well-formed-descriptors-are-regular} then follows by induction on the extended derivation, the gate cases being the four clauses of \cref{def:gate-descriptor-generation}.\hfill $\square$

\begin{proposition}[The Descriptor Metatheory Extends to Gates]\label{prop:gated-descriptor-metatheory}
In a gated signature, the admissibility theorems for the well-formed descriptor judgment hold for the judgment extended by the gate rules: exchange (\cref{thm:exchange-admissibility-for-descriptor-judgments}), boundary renaming (\cref{thm:renaming-admissibility-for-descriptor-judgments}), and input weakening and contraction (\cref{thm:input-weakening-admissibility-for-well-formed-descriptors}, \cref{thm:input-contraction-admissibility-for-well-formed-descriptors}).
\end{proposition}

\textit{Proof.}
Each of those theorems proceeds by induction on the derivation, and the gate rules contribute one new case each, all of the same shape. For exchange, a transported conclusion is absorbed by the closing transport of the gate rule, composing the permutations (\cref{prop:functoriality-of-permutation-transport}). For boundary renaming, we extend the renaming by the identity on the consumed gate variables, rename the premise by the induction hypothesis, and re-apply the rule at the renamed gate variables; the conclusion is the renamed class since the operations touch the boundary names elementwise, as in the well-definedness clauses of \cref{prop:gate-rules-preserve-regularity}. For input weakening and contraction, the affected variables are cartesian while the gate rules read and modify only the linear boundary; the operations therefore commute, and re-applying the rule in the enlarged, respectively contracted, cartesian zone concludes.\hfill $\square$

Now that the gates have the ability to be added to the mode theory, and automatically make their way through the type theory, we finally add the mechanism to relate the descriptors to the ones which inherit the substructural rules. We use, as in the cartesian case, $2$-cells.

\begin{definition}[Gated $2$-cell formers]\label{def:gated-descriptor-2-cells}
Let $\Lambda$ declare gates. The $2$-cell calculus of \cref{def:descriptor-2-cells} is extended, for each declared gate and each choice of fresh gate variables, by a former applying the corresponding descriptor operation $G$ of \cref{def:gate-descriptor-generation} to a $2$-cell:
$$
\inferrule*[Right=2-Gate]
{r :: \mf D \Rightarrow^{\Lambda}_{\gamma \Vdash \delta} \mf E}
{G(r) :: G(\mf D) \Rightarrow^{\Lambda}_{\gamma' \Vdash \delta'} G(\mf E)}
$$
with $G$ among $(-)\cup w_l(x)$, $(-)\cup w^\perp_l(b)$, $c_l(x_0;x,x')\cdot(-)$ and $(-)\cdot c^\perp_l(b,b';b_0)$, and $\gamma'\Vdash\delta'$ the correspondingly gated boundary. The adjoined gate variables are fresh, the consumed ones are taken from the premise boundary, and the former closes with a permutation transport of the gated boundary, as the gate rules do.
\end{definition}

\begin{example}
Suppose the signature declares the gate $\mathsf{contr}(l)$ and a generating $2$-cell
$$
\rho\ ::\ \mf D_\otimes\ \Rightarrow^{\Lambda}_{x:l,\,y:l\,\Vdash\, z:l}\ [\,y,x\xmapsto{\otimes_l}w\mid w\mapsto z\,]
$$
exchanging the two inputs of the tensor. The gated former for contraction produces the $2$-cell
$$
c_l(x_0;x,y)\cdot\rho\ ::\ c_l(x_0;x,y)\cdot\mf D_\otimes\ \Rightarrow^{\Lambda}_{x_0:l\,\Vdash\, z:l}\ \bigl[\,v',v\xmapsto{\otimes_l}w,\ (\ ;x_0)\xmapsto{c_l}v,v'\mid w\mapsto z\,\bigr],
$$
whose source is the contracted tensor computed in \cref{ex:linear-logic-gate-descriptors} and whose target lists the two copies $v, v'$ in the exchanged order.
\end{example}

\begin{lemma}[The $2$-cell Metatheory Extends to Gates]\label{lem:gated-2-cell-metatheory}
In a gated signature, \cref{prop:source-and-target-of-descriptor-2-cells-are-well-formed}, \cref{thm:exchange-admissibility-for-descriptor-2-cells}, \cref{lem:ambient-weakening-of-descriptor-2-cells}, and \cref{lem:boundary-renaming-of-descriptor-2-cells} hold for the calculus extended by the gated formers.
\end{lemma}

\textit{Proof.}
The source and target $G(\mf D)$ and $G(\mf E)$ are well formed by the gate rules extending the judgment (\cref{def:gate-descriptor-generation}, \cref{prop:gate-rules-preserve-regularity}, \cref{prop:gated-descriptor-metatheory}). For exchange, the induction of \cref{thm:exchange-admissibility-for-descriptor-2-cells} gains the case of a gated former: a permutation $(\pi;\tau)$ of the gated boundary restricts to a permutation $(\tilde\pi;\tilde\tau)$ of the premise boundary fixing the consumed gate variables, and we set $(\pi;\tau)_*(G(r)):=G((\tilde\pi;\tilde\tau)_*r)$, the closing transports of the former identifying the two boundaries; the endpoints match since the gate operations read no positions. Weakening and renaming of the contexts only concern the variables that are distinct from the ones the gates are introducing. Therefore, the gate operations commute with them so that we set $\mathrm{wk}^{a:p}_{\mathrm{in}}(G(r)):=G(\mathrm{wk}^{a:p}_{\mathrm{in}}(r))$ and $\mathrm{ren}_{\sigma}(G(r)):=G(\mathrm{ren}_{\sigma}(r))$.\hfill $\square$

\begin{definition}[Gated Structural Rules]\label{def:gated-structural-rules}
For each gate declared at the linear mode $l$, with $A,B$ types at mode $l$ and descriptor operations as in \cref{def:gate-descriptor-generation}, the linear sequent judgment gains the corresponding rule:
$$
\inferrule*[right=$\mathsf{weak}(l)$]
{\Theta\ ;\ \Gamma\vdash^{\Lambda}_{\mf E}\Delta
 \\ r::\mf D\Rightarrow^{\Lambda}\mf E\cup w_l(x)}
{\Theta\ ;\ \Gamma,\,x{:}A\vdash^{\Lambda}_{\mf D}\Delta}
$$
$$
\inferrule*[right=$\mathsf{coweak}(l)$]
{\Theta\ ;\ \Gamma\vdash^{\Lambda}_{\mf E}\Delta
 \\ r::\mf D\Rightarrow^{\Lambda}\mf E\cup w^\perp_l(b)}
{\Theta\ ;\ \Gamma\vdash^{\Lambda}_{\mf D}\Delta,\,b{:}B}
$$
$$
\inferrule*[right=$\mathsf{contr}(l)$]
{\Theta\ ;\ \Gamma,\,x{:}A,\,x'{:}A\vdash^{\Lambda}_{\mf E}\Delta
 \\ r::\mf D\Rightarrow^{\Lambda}c_l(x_0;x,x')\cdot\mf E}
{\Theta\ ;\ \Gamma,\,x_0{:}A\vdash^{\Lambda}_{\mf D}\Delta}
$$
$$
\inferrule*[right=$\mathsf{cocontr}(l)$]
{\Theta\ ;\ \Gamma\vdash^{\Lambda}_{\mf E}\Delta,\,b{:}B,\,b'{:}B
 \\ r::\mf D\Rightarrow^{\Lambda}\mf E\cdot c^\perp_l(b,b';b_0)}
{\Theta\ ;\ \Gamma\vdash^{\Lambda}_{\mf D}\Delta,\,b_0{:}B}
$$
\end{definition}

\textit{Remark.} In all the rules, the 2-cells are over the sequent of the conclusion. Note that taking the 2-cells to be the identities recovers the substructural rules for the same descriptor; these are the versions that are usually needed the most. We provide the general version here so as to be consistent and closer to the admissible respect-for-transformations rule.

\begin{example}
Suppose the gate $\mathsf{contr}(l)$ is declared, and recall the tensor introduction of \cref{ex:linear-logic-sequent-rules}, instantiated at $B := A$:
$$
\cdot\ ;\ x':A,\,y':A\vdash^{\Lambda}_{\mf D'_\otimes}z:A\otimes A.
$$
Applying the rule $\mathsf{contr}(l)$ with the identity $2$-cell recovers the same-descriptor structural rule and gives the sequent
$$
\cdot\ ;\ x_0:A\vdash^{\Lambda}_{c_l(x_0;x',y')\cdot\mf D'_\otimes}z:A\otimes A,
$$
the conclusion index being the contracted tensor
$$
c_l(x_0;x',y')\cdot\mf D'_\otimes=\bigl[\,v,v'\xmapsto{\otimes_l}w,\ (\ ;x_0)\xmapsto{c_l}v,v'\mid w\mapsto z\,\bigr],
$$
and forgetting the descriptor index gives the sequent $A\vdash A\otimes A$. The single hypothesis $A$ feeds the two sources of the tensor through the diagonal node recorded in the index.
\end{example}

\begin{definition}[Gate coherence laws]\label{def:gate-coherence-laws}
In these coherence laws, $\mf T$ denotes a well-formed $\ell$-descriptor, $\mf D$ a well-formed descriptor, and every operation is the one of \cref{def:gate-descriptor-generation}, at fresh gate variables. For a linear context $\gamma$, we write $\mathrm{wk}^{w}_{\gamma}$ for the iterated Gate-weak, one declaration of $\gamma$ at a time, and $\mathrm{cowk}^{w^\perp}_{\delta}$ for the iterated Gate-coweak; $\mathrm{ctr}^{c}_{\gamma,\widetilde\gamma}$ merges each declaration of $\gamma$ with its copy in $\widetilde\gamma$ by Gate-contr, and $\mathrm{cctr}^{c^\perp}_{\delta,\widetilde\delta}$ does so dually by Gate-cocontr. Each law presupposes that the gates appearing in it are declared at every linear mode occurring in the displayed boundaries.

\begin{itemize}
  \item \textbf{Functoriality and localization equations.} For each gated former $G$ of \cref{def:gated-descriptor-2-cells}, these functoriality equations hold:
  $$G(\mathbf 1_{\mf D})\equiv\mathbf 1_{G(\mf D)},\qquad G(s\circ r)\equiv G(s)\circ G(r),$$
  together with the localization equations $G(t\circ_{y,x}s)\equiv G(t)\circ_{y,x}s$, respectively $t\circ_{y,x}G(s)$, whenever they make sense: a gate applied to a horizontal composition localizes to the factor carrying its gate variables.
  \item \textbf{Gate coherence and invertibility.} Contracting $c,d$ into $a$ and then $a,b$ into $x_0$ is equivalent to contracting $d,b$ into $e$ and then $c,e$ into $x_0$; contracting $x,x'$ into $x_0$ is equivalent to contracting $x',x$ into $x_0$; and weakening by $x$ and then contracting $x,x'$ into $x_0$ is equivalent to renaming $x'$ into $x_0$. We write $\Leftrightarrow$ for a pair of generating $2$-cells, one in each direction, adjoined together with the invertibility equations making both composites $\equiv$ identities. 
$$
c_l(x_0;a,b)\cdot\bigl(c_l(a;c,d)\cdot\mf T\bigr)\ \Leftrightarrow\ c_l(x_0;c,e)\cdot\bigl(c_l(e;d,b)\cdot\mf T\bigr),
\qquad
c_l(x_0;x,x')\cdot\mf T\ \Leftrightarrow\ c_l(x_0;x',x)\cdot\mf T,
$$
$$
c_l(x_0;x,x')\cdot\bigl(\mf T\cup w_l(x)\bigr)\ \Leftrightarrow\ \mathrm{ren}_{x'\mapsto x_0}(\mf T),
$$
and dually the coassociativity, cocommutativity and counit laws for $c^\perp_l$ and $w^\perp_l$.
  \item \textbf{Weakening and composition coherence.} Cutting a weakened variable discards the cut descriptor, up to the re-weakening of its boundary context. For $\mf D$ over $(\theta_{\mf D}\ ;\ \gamma_{\mf D}\Vdash\delta_L,y:l,\delta_R)$ we add the $2$-cell
$$
\mathrm{disc}_{\mf D,\mf T}\ ::\ \mf D\circ_{y,x}\bigl(\mf T\cup w_l(x)\bigr)\ \Rightarrow\ \mathrm{wk}^{w}_{\gamma_{\mf D}}\bigl(\mathrm{cowk}^{w^\perp}_{\delta_L,\delta_R}(\mf T)\bigr),
$$
and dually, for $\mf D$ over $(\theta_{\mf D}\ ;\ \gamma_L,x:l,\gamma_R\Vdash\delta_{\mf D})$,
$$
\mathrm{disc}^{\mathrm{op}}_{\mf D,\mf T}\ ::\ \bigl(\mf T\cup w^\perp_l(y)\bigr)\circ_{y,x}\mf D\ \Rightarrow\ \mathrm{wk}^{w}_{\gamma_L,\gamma_R}\bigl(\mathrm{cowk}^{w^\perp}_{\delta_{\mf D}}(\mf T)\bigr).
$$
  \item \textbf{Contraction and composition coherence.} Cutting two variables $x',x''$ against the same descriptor with its cut variable renamed duplicates the linear resources and targets. Contracting and cocontracting those afterwards leads to the same result as contracting $x',x''$ first into $x_0$ and cutting it.  For $\mf D$ as in $\mathrm{disc}$ and $\widetilde{\mf D}$ a copy of $\mf D$ with its linear boundary renamed and its cartesian zone shared (the cut allows shared cartesian declarations, \cref{def:well-formed-composition-cut}), we add the $2$-cell
$$
\mathrm{dup}_{\mf D,\mf T}\ ::\ \mf D\circ_{y,x_0}\bigl(c_l(x_0;x',x'')\cdot\mf T\bigr)\ \Rightarrow\
\mathrm{ctr}^{c}_{\gamma,\widetilde\gamma}\Bigl(\mathrm{cctr}^{c^\perp}_{\delta,\widetilde\delta}\bigl(\mf D\circ_{y,x'}(\widetilde{\mf D}\circ_{\widetilde y,x''}\mf T)\bigr)\Bigr).
$$
Dually, for a consumer $\mf D$ as in $\mathrm{disc}^{\mathrm{op}}$ and its copy $\widetilde{\mf D}$, 
$$
\mathrm{dup}^{\mathrm{op}}_{\mf D,\mf T}\ ::\ \bigl(\mf T\cdot c^\perp_l(b',b'';b_0)\bigr)\circ_{b_0,x}\mf D\ \Rightarrow\
\mathrm{ctr}^{c}_{\gamma,\widetilde\gamma}\Bigl(\mathrm{cctr}^{c^\perp}_{\delta,\widetilde\delta}\bigl((\mf T\circ_{b',x}\mf D)\circ_{b'',\widetilde x}\widetilde{\mf D}\bigr)\Bigr).
$$
\end{itemize}
Together, these functoriality, localization and invertibility equations, with the $2$-cells $\mathrm{disc}$ and $\mathrm{dup}$ and their duals, form the coherence laws of the gated signature. \end{definition}

\begin{example}
Suppose the gates $\mathsf{weak}(l)$ and $\mathsf{contr}(l)$ are declared, and take the identity descriptor $\mf T:=[\varnothing\mid\{x'\mapsto z\}]$ over $\cdot\ ;\ x':l\Vdash z:l$. Weakening by $x$ and contracting $x,x'$ into $x_0$ gives
$$
c_l(x_0;x,x')\cdot\bigl(\mf T\cup w_l(x)\bigr)=\bigl[\,(\ ;v)\xmapsto{w_l}\cdot,\ (\ ;x_0)\xmapsto{c_l}v,v'\mid v'\mapsto z\,\bigr],
$$
and the rule relates this descriptor to $\mathrm{ren}_{x'\mapsto x_0}(\mf T)=[\varnothing\mid\{x_0\mapsto z\}]$. Duplicating the input and discarding one copy is the identity. For the law $\mathrm{disc}$, let $\mf D':=[\,x_1,x_2\xmapsto{\otimes_l}w\mid w\mapsto y\,]$ be a boundary-renamed copy of the tensor descriptor. Cutting it into the weakened input $x$ gives
$$
\mf D'\circ_{y,x}\bigl(\mf T\cup w_l(x)\bigr)=\bigl[\,x_1,x_2\xmapsto{\otimes_l}w,\ (\ ;w)\xmapsto{w_l}\cdot\mid x'\mapsto z\,\bigr],
$$
the tensor computed and immediately consumed by the weakening node. The rule relates the two descriptors by the $2$-cell
$$
\mathrm{disc}_{\mf D',\mf T}\ ::\ \mf D'\circ_{y,x}\bigl(\mf T\cup w_l(x)\bigr)\ \Rightarrow\ \bigl[\,(\ ;x_1)\xmapsto{w_l}\cdot,\ (\ ;x_2)\xmapsto{w_l}\cdot\mid x'\mapsto z\,\bigr],
$$
which discards the tensor and re-weakens its two inputs.
\end{example}

Identity expansion, connective invertibility, and cut admissibility are the desirable metatheoretic properties that we now prove for every generated object logic.
\endgroup{}
\begingroup

\subsection{Metatheory}

This section establishes the metatheory of the two sequent judgments: respect for structural transformations (\cref{prop:respect-for-structural-transformation}), identity expansion (\cref{thm:identity}), invertibility of the hypothesis-exposing rules (\cref{prop:left-invertibility-of-f,prop:right-invertibility-of-u}), and the cut theorem (\cref{thm:cut-admissibility}), whose cartesian case is the substitution form of cut and whose linear case is the polycategorical cut. The gated harmony results are \cref{prop:harmony-affine,thm:multicut,prop:harmony-classical}. We treat the results in the general formulation of the schema of \cref{def:sequent-judgments}. The proofs separate the cartesian and the linear cases, but the treatment is general: every index is a descriptor class and every index rearrangement is an operation of the mode theory.

\textit{Convention (scope):} Except in \cref{subsec:harmony-for-gates}, the signature is assumed to declare \emph{no gates}. That subsection extends the cut theorem to gated signatures.

\subsubsection{Preliminaries}

\phantomsection\label{anchor:metatheory-conventions}
\textit{Conventions:}
We usually omit the input and output contexts for descriptor transformations. We write $\mf D\Rightarrow\mf E$ rather than $\mf D\Rightarrow^{\Lambda}_{\gamma\Vdash\delta}\mf E$ when it is clear from context. We also feel free to reorder the context, and we usually rename boundary variables implicitly. We also assume the freshness conditions on variables, which means that every time a new variable is introduced, it will be fresh with respect to every other variable already introduced.

\textit{Convention (connectives):} We write the cartesian connectives over their connective descriptors, $F^{y}_{\mf D_F}(\Theta_F;\cdot)$ and $U^{\mf D_U}_{x}(\Theta_U;\cdot\mid b:X)$ with $\mf D_F,\mf D_U$ c-descriptors, and we reason on all indexes by the operations of the mode theory alone. The input-substitution notation $(\--)[x:=\mf E]$ and its family form are the c-substitution of \cref{def:c-substitution}.

\begin{definition}[c-Substitution]\label{def:c-substitution}
Let $\mf E$ be a c-descriptor over $(\theta\ ;\ \cdot\Vdash u:p)$ and let $\mf D$ be a class declaring the cartesian input $x:p$, with $\theta$ compatible with the ambient contexts. The \emph{c-substitution} of $\mf E$ at $x$ is the cut
$$\mf D[x:=\mf E]:=\mf E\circ_{u,x}\mf D .$$
For the family form, let $\theta'=(a_i:p_i)_{1\leq i\leq n}$ be a cartesian context and suppose
$$
\theta_D,\theta'\ ;\ \gamma\Vdash_{\mf D}\delta,
\qquad
\bigl(\theta_\sigma\Vdash_{\sigma_{a_i}}b_{a_i}:p_i\bigr)_{1\leq i\leq n},
$$
where the latter are c-descriptor judgments over the shared cartesian context $\theta_\sigma$. Assume that $\theta_\sigma$ is disjoint from $\theta'$, that $\theta_\sigma$ and $\theta_D$ agree on common declarations, and that the outputs $b_{a_i}$ are pairwise distinct and otherwise fresh for $\theta_D,\theta',\theta_\sigma,\gamma,\delta$. The shared context $\theta_\sigma$ is part of the family data, including when $n=0$. If $n>0$, define the \emph{family c-substitution}
$$
\bigl(\sigma_a\bigr)_{a\in\theta'}\circ_{(b_a,a)_{a\in\theta'}}\mf D
:=
\sigma_{a_n}\circ_{b_{a_n},a_n}
\Bigl(\cdots\bigl(\sigma_{a_1}\circ_{b_{a_1},a_1}\mf D\bigr)\cdots\Bigr).
$$
It is independent of the order by \cref{lem:left-interchange-law} and is a descriptor over $((\theta_\sigma\cup\theta_D)\ ;\ \gamma\Vdash\delta)$, up to the input permutation. If $n=0$, the empty instance of the same notation is
$$
\bigl(\sigma_a\bigr)_{a\in\cdot}\circ_{(b_a,a)_{a\in\cdot}}\mf D
:=
()_{\theta_\sigma}\circ\mf D
:=
\mathrm{wk}^{\theta_\sigma\setminus\theta_D}_{\mathrm{in}}(\mf D),
$$
where the right-hand side is the iterated cartesian input weakening, followed by the canonical permutation to the chosen order on $\theta_\sigma\cup\theta_D$.
\end{definition}

\begin{example}[Linear--Nonlinear Logic]
We work over the linear--nonlinear pre-signature extended with the binary c-generator $\otimes_c : c, c \to c$, and we recall from \cref{ex:linear-logic-exponential-modality} the descriptor
$$\mf D_g := [\{(\hat x \ \mid \ \cdot) \xmapsto{g} v\} \mid \{v \mapsto y\}]$$
over $(\hat x : c \ ; \ \cdot \Vdash y : l)$, which produces a linear resource out of a cartesian input. The class
$$\mf E := [\{a, a' \xmapsto{\otimes_c} s\} \mid \{s \mapsto u\}]$$
over $(a : c,\, a' : c \ ; \ \cdot \Vdash u : c)$ is a c-descriptor, and $\mf D_g$ declares the cartesian input $\hat x : c$, so the c-substitution is defined and computes to
$$\mf D_g[\hat x := \mf E] = \mf E \circ_{u, \hat x} \mf D_g = [\{a, a' \xmapsto{\otimes_c} s,\ \ (s \ \mid \ \cdot) \xmapsto{g} v\} \mid \{v \mapsto y\}]$$
over $(a : c,\, a' : c \ ; \ \cdot \Vdash y : l)$. If $\mf E$ is the identity class descriptor $[\varnothing \mid \{a \mapsto u\}]$, then the substitution is a relabelling $\mf D_g[\hat x := a]$.
\end{example}

\Cref{lem:units-for-composition} analyzes the expected behaviour of identity descriptors under composition.
\begin{lemma}[Units for Composition]\label{lem:units-for-composition}
Let $p$ be any mode and $\mf D$ a regular descriptor class.
\begin{itemize}
\item[(i)] If $\mf D$ is a class over $(\gamma, x:p\ ;\ \delta)$ and $a:p$ is fresh, then
$[\varnothing\mid\{a\mapsto b\}]\circ_{b,x}\mf D=\mf D[x:=a]$
over $(\gamma,a:p\ ;\ \delta)$.
\item[(ii)] If $\mf D$ is a class over $(\gamma\ ;\ \delta,y:p)$ and $b:p$ is fresh, then
$\mf D\circ_{y,x_0}[\varnothing\mid\{x_0\mapsto b\}]=\mf D[y:=b]$
over $(\gamma\ ;\ \delta,b:p)$, where $\mf D[y:=b]$ retargets the unique redirection $s\mapsto y$ to $s\mapsto b$.
\end{itemize}
\end{lemma}

\textit{Proof.}
For the first point (i), choose a representative $(\mc S\mid\mc T)\in\mf D$. The left factor has no assignments, and its only redirection $a\mapsto b$ is removed by the cut. Its cut source is the input variable $a$, which is fixed by every internal renaming used in the raw composition. Consequently, the occurrence-indexed substitution at the occurrences of $x$ is the constant substitution by $a$, and
$$
(\varnothing\mid\{a\mapsto b\})
\circ^{\bs\beta}_{b,x}
(\mc S\mid\mc T)
=
(\mc S[x:=a]\mid\mc T[x:=a]).
$$
The right-hand side represents $\mf D[x:=a]$. This calculation includes the case where a cartesian input $x$ has no occurrences (or several occurrences); when $p$ is linear, regularity makes the occurrence unique.

For the second point (ii), there is only one occurrence of $x_0$ in the identity descriptor on the right-hand side of the composition; thus, the composition gives all the assignments present in the descriptor $\mf D$ and the redirections as well, with the exception of the one redirecting to the cut variable $y$. The source of the redirection $x_0 \mapsto b$ gets replaced by $s$ in the redirection $s \mapsto y$, where $s$ is the source of the only redirection to $y$ in the descriptor $\mf D$. This has the effect of obtaining the redirection $s \mapsto b$, which is exactly $(s \mapsto y)[y := b]$, which in turn leads to the conclusion. \hfill$\square$

\begin{lemma}[Cartesian Composition Commutes with Composition]\label{lem:c-substitution-commutes-with-composition}
Let $\mf E$ be a c-descriptor over $(\theta\ ;\ \cdot\Vdash u:p)$, and let $\mf D_1\circ_{y,z}\mf D_2$ be a defined composition of well-formed descriptor classes in the sense of \cref{def:well-formed-composition-cut}, such that their cartesian variables contain $x:p$, with $x$ distinct from the cut variable $z$. We assume that the cut variable $z$ is not present in the context $\theta$ of $\mf E$. Finally, we assume that $\theta$ is compatible with the ambient contexts. Then
$$
(\mf D_1\circ_{y,z}\mf D_2)[x:=\mf E]
=
\mf D_1[x:=\mf E]\circ_{y,z}\mf D_2[x:=\mf E],
$$
with the convention that $\mf D_i[x:=\mf E]:=\mf D_i$ when $x$ is not declared among the inputs of $\mf D_i$. Moreover, let $p'$ be a cartesian mode, let $\mf E'$ be a c-descriptor over $(\eta,x:p\ ;\ \cdot\Vdash u':p')$, and let $\mf D$ be a regular descriptor class over $(\gamma,x':p'\ ;\ \delta)$. Assume that $x\neq x'$, that $\gamma$ does not declare $x$, that neither $\theta$ nor $\eta$ declares $x'$, and that all the displayed boundaries and cuts are compatible. Then
$$
\mf D[x':=\mf E'][x:=\mf E]=\mf D[x':=\mf E'[x:=\mf E]].
$$
This is an equality of descriptor classes over $(\theta\cup\eta\cup\gamma\ ;\ \delta)$, up to the implicit boundary permutation transport. Consequently, composition with nonempty compatible families of c-descriptors is functorial,
$$
\bigl(\sigma_x\bigr)_{x\in\theta'}\circ_{(b_x,x)}\Bigl(\bigl(\tau_y\bigr)_{y\in\theta''}\circ_{(b_y,y)}\mf D\Bigr)=\bigl((\tau\circ\sigma)_y\bigr)_{y\in\theta''}\circ_{(b_y,y)}\mf D,
$$
where $(\tau\circ\sigma)_y:=\bigl(\sigma_x\bigr)_{x\in\theta'}\circ_{(b_x,x)}\tau_y$; the cartesian zones $\theta',\theta''$ are disjoint, $\mf D$ declares the inputs of $\theta''$ and no variable of $\theta'$, every $\tau_y$ declares the inputs of $\theta'$, and all the displayed boundaries and cuts are compatible.
\end{lemma}

It suffices to show the first equation, as the others follow.
Choose representatives with pairwise disjoint internal variables
$$
(\mc S_E\mid\{e\mapsto u\})\in\mf E,
\qquad
(\mc S_i\mid\mc T_i)\in\mf D_i\quad(i=1,2),
$$
and write $s_1\mapsto y$ for the unique $y$-redirection of $\mc T_1$. Put
$$
I_E:=\mathrm{Int}(\mc S_E\mid\{e\mapsto u\}),
\qquad I_i:=\mathrm{Int}(\mc S_i\mid\mc T_i),
$$
and
$$
O_i:=\mathrm{Occ}_x(\mc S_i\mid\mc T_i),
\qquad Q:=\mathrm{Occ}_z(\mc S_2\mid\mc T_2).
$$
If $s_1=x$, let $o_*\in O_1$ denote the occurrence of $x$ in the redirection $x\mapsto y$.

Choose a fresh family $\bs\beta=(\beta_q:I_1\to B_q)_{q\in Q}$ and form
$$
\mc C:=(\mc S_1\mid\mc T_1)\circ^{\bs\beta}_{y,z}(\mc S_2\mid\mc T_2).
$$
The occurrences of $x$ in the composite $\mc C$ are the occurrences in $\mf D_2$ and the occurrences in $\mf D_1$ that have been copied for every occurrence of $z$; thus, there is an identification
$$
\mathrm{Occ}_x(\mc C)\cong J:=O_2\uplus(Q\times O_1).
$$
The only edge case is if $s_1=x$; in this case, we use the occurrence $o_\star$, and the bijective correspondence still holds.

Choose a fresh family $\bs\eta=(\eta_j:I_E\to A_j)_{j\in J}$. The raw composite
$$
\mc L:=(\mc S_E\mid\{e\mapsto u\})\circ^{\bs\eta}_{u,x}\mc C
$$
is a representative of $(\mf D_1\circ_{y,z}\mf D_2)[x:=\mf E]$.

Now, for the right-hand side, we use the fresh families
$$
\bs\lambda=(\lambda_o:I_E\to A^1_o)_{o\in O_1},
\qquad
\bs\mu=(\mu_o:I_E\to A^2_o)_{o\in O_2}
$$
to form representatives
$$
\mc C_1:=(\mc S_E\mid\{e\mapsto u\})\circ^{\bs\lambda}_{u,x}(\mc S_1\mid\mc T_1),
\qquad
\mc C_2:=(\mc S_E\mid\{e\mapsto u\})\circ^{\bs\mu}_{u,x}(\mc S_2\mid\mc T_2).
$$
When $x$ is not declared in a factor, take that factor itself, as in the convention of the statement. Since $x\neq z$ and $z\notin\theta$, no copy of $\mf E$ contains $z$ and no $z$-position is changed; thus, $\mathrm{Occ}_z(\mc C_2)=Q$. Choose a fresh family
$$
\bs\kappa=(\kappa_q:\mathrm{Int}(\mc C_1)\to K_q)_{q\in Q}
$$
and let
$$
\mc R:=\mc C_1\circ^{\bs\kappa}_{y,z}\mc C_2.
$$
This is a representative of $\mf D_1[x:=\mf E]\circ_{y,z}\mf D_2[x:=\mf E]$.

\phantomsection\label{anchor:internal-variable-decompositions}
The analysis of the internal variables for the total left- and right-hand-side composites gives, using \cref{prop:internal-variables-of-raw-composition},
\begin{align*}
\mathrm{Int}(\mc L)
&=I_2
\uplus\biguplus_{q\in Q}\beta_q(I_1)
\uplus\biguplus_{o\in O_2}\eta_o(I_E)
\uplus\biguplus_{(q,o)\in Q\times O_1}\eta_{(q,o)}(I_E),\\
\mathrm{Int}(\mc R)
&=I_2
\uplus\biguplus_{o\in O_2}\mu_o(I_E)
\uplus\biguplus_{q\in Q}\kappa_q(I_1)
\uplus\biguplus_{(q,o)\in Q\times O_1}\kappa_q\lambda_o(I_E).
\end{align*}
Define $\Phi:\mathrm{Int}(\mc L)\to\mathrm{Int}(\mc R)$ by
$$
\Phi(v):=
\begin{cases}
v & v\in I_2,\\
(\kappa_q\circ\beta_q^{-1})(v) & v\in\beta_q(I_1),\\
(\mu_o\circ\eta_o^{-1})(v) & v\in\eta_o(I_E),\quad o\in O_2,\\
(\kappa_q\circ\lambda_o\circ\eta_{(q,o)}^{-1})(v)
& v\in\eta_{(q,o)}(I_E),\quad(q,o)\in Q\times O_1.
\end{cases}.
$$
We can see immediately that the map $\Phi$ is well-defined since the piecewise domains are disjoint. Moreover, it is defined on the whole $\mathrm{Int}$ because of the \hyperref[anchor:internal-variable-decompositions]{disjoint decompositions above}. It remains to show that this map is a bijection to conclude that the representative belongs to the same class, establishing the required equality.

First, we extend the map to be the identity on the other boundary variables. We verify now that the codomain coincides with $\mc R$.
The map matches the $\mf E$-copy at $o\in O_2$ with the $\bs\mu$-copy at $o$, the $\mf E$-copy at $(q,o)$ with the $\kappa_q\lambda_o$-copy, and the $\beta_q$-copy of the $\mf D_1$-part with the $\kappa_q$-copy of that part. At an $x$-position $o$ in the retained part of the $q$-copy of $\mf D_1$ (that is, at every $o\in O_1$ except $o_*$ when $s_1=x$), the source $e^{\eta_{(q,o)}}$ on the left is sent to $e^{\kappa_q\lambda_o}$, which is the source at the corresponding position on the right. At an $x$-position $o\in O_2$ of $\mf D_2$, the source $e^{\eta_o}$ is similarly sent to $e^{\mu_o}$.

It remains to check the source inserted at a $z$-position $q$. If $s_1\neq x$, the left representative has $s_1^{\beta_q}$ there and $\Phi$ sends it to $s_1^{\kappa_q}$, the source of the copied $y$-redirection of $\mc C_1$. If $s_1=x$, the left representative has $e^{\eta_{(q,o_*)}}$ there, whereas the $y$-redirection of $\mc C_1$ has source $e^{\lambda_{o_*}}$; after the $q$-copy, the right representative therefore has $e^{\kappa_q\lambda_{o_*}}$, again the $\Phi$-image of the left source. All remaining sources, assignment targets, and redirection targets correspond immediately, since internal renamings fix the boundary and the redirections to $u$ and $y$ are precisely the ones removed by the two cuts. Hence $\mc L^\Phi=\mc R$. The representatives are $\alpha$-equivalent, and the first equation follows. \hfill $\square$

An iterated composition with a linear variable allows a certain kind of interchange and associativity, packaged in \cref{lem:localization}.

\begin{lemma}[Localization of Linear Cuts]\label{lem:localization}
Let
$$
\mf C=
\left(
  (\mf D_{\bar x_i})_{i=1}^n
  \circ_{(\bar x_i,x_i)_{i=1}^n}
  \mf G
\right)
\circ_{(y_j,\bar y_j)_{j=1}^m}
(\mf D_{\bar y_j})_{j=1}^m
$$
be a defined iterated cut of well-formed descriptor classes, with every displayed cut at a linear mode. We call
$\mf G$, the classes $\mf D_{\bar x_i}$, and the classes
$\mf D_{\bar y_j}$ the \emph{factors} of $\mf C$.

For a factor $\mf F$, write
$$
\mf C[\mf F\leftrightarrow\mf F']
$$
for the same iterated cut with $\mf F$ replaced by $\mf F'$, leaving
all other factors and cut ports unchanged. Let $\mf E$ be a well-formed
class whose boundary is compatible with the displayed cuts.

\begin{itemize}
\item[(i)]
Suppose that the linear input $z:q$ of $\mf C$ comes from a unique
factor $\mf F$. If $\mf E$ has an output $y:q$, then
$$
\mf E\circ_{y,z}\mf C
=
\mf C[
  \mf F\leftrightarrow
  \mf E\circ_{y,z}\mf F
].
$$

\item[(ii)]
Suppose that the output $w:q$ of $\mf C$ comes from a unique factor
$\mf F$. If $\mf E$ has a linear input $x:q$, then
$$
\mf C\circ_{w,x}\mf E
=
\mf C[
  \mf F\leftrightarrow
  \mf F\circ_{w,x}\mf E
].
$$
\end{itemize}
\end{lemma}

\textit{Proof.}

We prove both statements simultaneously by induction on the number $n+m$ of cuts in the iterated composite. The base case $n=m=0$, where $\mf C=\mf G$, is immediate. All applications of \cref{lem:associativity,lem:left-interchange-law,lem:right-interchange-law} in this proof satisfy their boundary conditions because the displayed iterated cut is defined and $\mf E$ has compatible boundary. Whenever we use \cref{lem:right-interchange-law}, its two cut inputs are linear inputs of well-formed classes and therefore have exactly one occurrence by the boundary usage property.

Suppose first that $m>0$, and write the outermost cut as
$$
\mf C=\mf H\circ_{y_m,\bar y_m}\mf D_{\bar y_m},
$$
where $\mf H$ is the same iterated composite with the last factor $\mf D_{\bar y_m}$ removed. For (i), the linear input $z$ can be present in $\mf H$ or in the outermost descriptor $\mf D_{\bar y_m}$. If it is present in $\mf H$, \cref{lem:associativity} and the induction hypothesis give
\begin{align*}
\mf E\circ_{y,z}\mf C
&=\bigl(\mf E\circ_{y,z}\mf H\bigr)
  \circ_{y_m,\bar y_m}\mf D_{\bar y_m}\\
&=\mf H[\mf F\leftrightarrow\mf E\circ_{y,z}\mf F]
  \circ_{y_m,\bar y_m}\mf D_{\bar y_m}\\
&=\mf C[\mf F\leftrightarrow\mf E\circ_{y,z}\mf F].
\end{align*}
On the flip side, if $z$ is present in $\mf D_{\bar y_m}$, then $\mf F=\mf D_{\bar y_m}$ and \cref{lem:left-interchange-law} gives directly
$$
\mf E\circ_{y,z}
\bigl(\mf H\circ_{y_m,\bar y_m}\mf D_{\bar y_m}\bigr)
=
\mf H\circ_{y_m,\bar y_m}
\bigl(\mf E\circ_{y,z}\mf D_{\bar y_m}\bigr),
$$
which is the desired replacement.

For (ii), the output $w$ can likewise be present in the outermost descriptor or in $\mf H$. In the first case, \cref{lem:associativity} gives
$$
\bigl(\mf H\circ_{y_m,\bar y_m}\mf D_{\bar y_m}\bigr)
\circ_{w,x}\mf E
=
\mf H\circ_{y_m,\bar y_m}
\bigl(\mf D_{\bar y_m}\circ_{w,x}\mf E\bigr),
$$
which is again the desired replacement. If $w$ is present in $\mf H$, \cref{lem:right-interchange-law} followed by the induction hypothesis gives
\begin{align*}
\mf C\circ_{w,x}\mf E
&=\bigl(\mf H\circ_{w,x}\mf E\bigr)
  \circ_{y_m,\bar y_m}\mf D_{\bar y_m}\\
&=\mf H[\mf F\leftrightarrow\mf F\circ_{w,x}\mf E]
  \circ_{y_m,\bar y_m}\mf D_{\bar y_m}\\
&=\mf C[\mf F\leftrightarrow\mf F\circ_{w,x}\mf E].
\end{align*}

It remains to treat the case $m=0<n$. Write now the outermost input cut as
$$
\mf C=\mf D_{\bar x_n}\circ_{\bar x_n,x_n}\mf H,
$$
where $\mf H$ is the input iterated composite with $\mf D_{\bar x_n}$ removed. For (i), if $z$ is present in $\mf D_{\bar x_n}$, \cref{lem:associativity} gives
$$
\mf E\circ_{y,z}
\bigl(\mf D_{\bar x_n}\circ_{\bar x_n,x_n}\mf H\bigr)
=
\bigl(\mf E\circ_{y,z}\mf D_{\bar x_n}\bigr)
\circ_{\bar x_n,x_n}\mf H,
$$
which replaces that factor directly. If $z$ is present in $\mf H$, \cref{lem:left-interchange-law} and the induction hypothesis give
\begin{align*}
\mf E\circ_{y,z}\mf C
&=\mf D_{\bar x_n}\circ_{\bar x_n,x_n}
  \bigl(\mf E\circ_{y,z}\mf H\bigr)\\
&=\mf D_{\bar x_n}\circ_{\bar x_n,x_n}
  \mf H[\mf F\leftrightarrow\mf E\circ_{y,z}\mf F]\\
&=\mf C[\mf F\leftrightarrow\mf E\circ_{y,z}\mf F].
\end{align*}

For (ii), if $w$ is present in $\mf H$, \cref{lem:associativity} and the induction hypothesis give
\begin{align*}
\mf C\circ_{w,x}\mf E
&=\mf D_{\bar x_n}\circ_{\bar x_n,x_n}
  \bigl(\mf H\circ_{w,x}\mf E\bigr)\\
&=\mf D_{\bar x_n}\circ_{\bar x_n,x_n}
  \mf H[\mf F\leftrightarrow\mf F\circ_{w,x}\mf E]\\
&=\mf C[\mf F\leftrightarrow\mf F\circ_{w,x}\mf E].
\end{align*}
Finally, if $w$ is present in $\mf D_{\bar x_n}$, \cref{lem:right-interchange-law} gives
$$
\bigl(\mf D_{\bar x_n}\circ_{\bar x_n,x_n}\mf H\bigr)
\circ_{w,x}\mf E
=
\bigl(\mf D_{\bar x_n}\circ_{w,x}\mf E\bigr)
\circ_{\bar x_n,x_n}\mf H,
$$
which replaces the remaining possible outermost factor. This covers all the cases and completes the induction. \hfill $\square$

We define the \emph{complexity} $\mathrm{cx}(A)$ of a type as follows: propositions have complexity $0$, and each former, in either sort instance, has complexity one more than the maximum of the complexities of its argument types (the entries of $\Theta_F,\Gamma_F$, respectively $\Theta_U,\Gamma_U,\Delta_U,X$), the empty maximum being $0$. Every argument type of a former is a strict subformula, of strictly smaller complexity.

\begin{lemma}[Cartesian Weakening for Sequents]\label{lem:cartesian-weakening-for-sequents}
Let $x':A$ be defined at a cartesian mode $p$. If $\Theta\ ;\ \Gamma\vdash^{\Lambda}_{\mf D}\Delta$ is derivable and $x'$ is fresh, then $\Theta,x':A\ ;\ \Gamma\vdash^{\Lambda}_{\mathrm{wk}^{x':p}_{\mathrm{in}}(\mf D)}\Delta$ is derivable by a derivation of the same height.
\end{lemma}

\textit{Proof.}
We proceed by induction on the derivation $\Theta\ ;\ \Gamma\vdash^{\Lambda}_{\mf D}\Delta$. We treat all the cases separately.

	\begin{itemize}
    \item \textbf{Var rule.} If the derivation terminates with the variable rule, then we are faced with
 
   	$$
    	\inferrule*
      {
        P\in \mathrm{Prop}_{\Lambda,q}
        \\
        r::\mf D
        \Rightarrow^{\Lambda}
        [\varnothing\mid\{x\mapsto y\}]
      }
      {
        \Theta\ ;\ x:P
        \vdash_{\mf D}^{\Lambda}
        y:P
      }
    ,$$
    where we remember that the general schema reduces to the cartesian and linear versions. Since $\mathrm{wk}^{x':p}_{\mathrm{in}}(\--)$ behaves well with respect to 2-cells, we get $$\mathrm{wk}^{x':p}_{\mathrm{in}}(r) :: \mathrm{wk}^{x':p}_{\mathrm{in}}(\mf D) \Rightarrow \mathrm{wk}^{x':p}_{\mathrm{in}}([\varnothing\mid\{x\mapsto y\}]).$$ The weakened target is the identity class regarded over the enlarged ambient zone, and we reapply the variable rule to this new transformation to obtain
    $$
    	\inferrule*
      {
        P\in \mathrm{Prop}_{\Lambda,q}
        \\
        \mathrm{wk}^{x':p}_{\mathrm{in}}(r) :: \mathrm{wk}^{x':p}_{\mathrm{in}}(\mf D) \Rightarrow [\varnothing\mid\{x\mapsto y\}]
      }
      {
        \Theta, x':A\ ;\ x:P
        \vdash_{\mathrm{wk}^{x':p}_{\mathrm{in}}(\mf D)}^{\Lambda}
        y:P
      }
    .$$

    \item \textbf{F-left, U-right}. If the derivation terminates with the left F rule, then we are looking at
    $$
    \inferrule*
      {
      \Theta,\Theta_F\ ;\
      \Gamma_L,\Gamma_F,\Gamma_R
      \vdash_{\mf D_F\circ_{y,z}\mf D}^{\Lambda}
      \Delta
      }
      {
      \Theta\ ;\
      \Gamma_L,\,
      z:\mathsf F,\,
      \Gamma_R
      \vdash_{\mf D}^{\Lambda}
      \Delta
      }
    ;$$
      
    by the induction hypothesis, we can weaken the premise to end up with $$\Theta,\Theta_F, x' : A\ ;\
      \Gamma_L,\Gamma_F,\Gamma_R
      \vdash_{\mathrm{wk}^{x':p}_{\mathrm{in}}(\mf D_F\circ_{y,z}\mf D)}^{\Lambda}
      \Delta;$$
      again, the well-behavedness of the weakening operation allows us to extract the descriptor associated to the connective $\mf D_F$ to get  $\Theta,\Theta_F, x' : A\ ;\
      \Gamma_L,\Gamma_F,\Gamma_R
      \vdash_{\mf D_F\circ_{y,z}\mathrm{wk}^{x':p}_{\mathrm{in}}(\mf D)}^{\Lambda}
      \Delta$,
      and applying the rule once more gets us the result:
      $$
        \inferrule*
        {
        \Theta,\Theta_F,x':A\ ;\
        \Gamma_L,\Gamma_F,\Gamma_R
        \vdash_{\mf D_F\circ_{y,z}\mathrm{wk}^{x':p}_{\mathrm{in}}(\mf D)}^{\Lambda}
        \Delta
        }
        {
        \Theta,x':A\ ;\
        \Gamma_L,\,
        z:\mathsf F,\,
        \Gamma_R
        \vdash_{\mathrm{wk}^{x':p}_{\mathrm{in}}(\mf D)}^{\Lambda}
        \Delta
        }.
      $$
      The case for the U-right rule is symmetric: instead of pulling the connective descriptor $\mf D_U$ out on the left, we extract it on the right of the composition.

      \item \textbf{F-right, U-left}. In this case, we treat the connective $U$ as the other one is similar.
      $$\inferrule*
        {
          r::\mf D
          \Rightarrow^{\Lambda}
          \mf C
          \\
          \bigl(\Theta\vdash^{\Lambda}_{\sigma_a}A_a\bigr)_{a\in\theta_U}
          \\
          \bigl(
            \Theta\ ;\ \Gamma^{(i)}
            \vdash_{\mf D_{\bar x_i}}^{\Lambda}
            \Delta_L^{(i)},\,\bar x_i:A_i,\,\Delta_R^{(i)}
          \bigr)_{1\leq i\leq n}
          \\
          \bigl(
            \Theta\ ;\ \Gamma^{(n+j)},\,\bar y_j:B_j
            \vdash_{\mf D_{\bar y_j}}^{\Lambda}
            \Delta^{(n+j)}
          \bigr)_{1\leq j\leq m}
        }
        {
          \Theta\ ;\ \Gamma,\,z:\mathsf U
          \vdash_{\mf D}^{\Lambda}
          \Delta
        }
      ,$$
      with $\mf C:=\Bigl(\bigl(\mf D_{\bar x_i}\bigr)_i\circ_{(\bar x_i,x_i)_{1\le i\le n}}\bigl(\bigl(\sigma_a\bigr)_{a\in\theta_U}\circ_{(b_a,a)}\mf D_U\bigr)[x:=z]\Bigr)
   \circ_{(y_j,\bar y_j)_{1\le j\le m}}\bigl(\mf D_{\bar y_j}\bigr)_j$. 
      We apply weakening to the transformation as in the variable case and pull the iterated cut descriptor out of the composition to get the transformation: $$\mathrm{wk}^{x':p}_{\mathrm{in}}(r)::\mathrm{wk}^{x':p}_{\mathrm{in}}(\mf D) \Rightarrow \mathrm{wk}^{x':p}_{\mathrm{in}}(\mf C).$$ The weakening operation distributes to all the subdescriptors in the iterated cut, and applying the induction hypothesis to the premises gives

      $$
        \inferrule*
        {
				\mathrm{wk}^{x':p}_{\mathrm{in}}(r) :: \mathrm{wk}^{x':p}_{\mathrm{in}}(\mf D) \Rightarrow \mathrm{wk}^{x':p}_{\mathrm{in}}(\mf C)        \\
        \bigl(\Theta, x' : A\vdash^{\Lambda}_{\mathrm{wk}^{x':p}_{\mathrm{in}}(\sigma_a)}A_a\bigr)_{a\in\theta_U}
        \\
        \bigl(
        \Theta, x' : A\ ;\ \Gamma^{(i)}
        \vdash_{\mathrm{wk}^{x':p}_{\mathrm{in}}(\mf D_{\bar x_i})}^{\Lambda}
        \Delta_L^{(i)},\,\bar x_i:A_i,\,\Delta_R^{(i)}
        \bigr)_{1\leq i\leq n}
        \\
        \bigl(
        \Theta, x' : A\ ;\ \Gamma^{(n+j)},\,\bar y_j:B_j
        \vdash_{\mathrm{wk}^{x':p}_{\mathrm{in}}(\mf D_{\bar y_j})}^{\Lambda}
        \Delta^{(n+j)}
        \bigr)_{1\leq j\leq m}
        }
        {
        \Theta, x' : A\ ;\ \Gamma,\,z:\mathsf U
        \vdash_{\mathrm{wk}^{x':p}_{\mathrm{in}}(\mf D)}^{\Lambda}
        \Delta
        }
      .$$
      The F-right case is symmetric.
      \end{itemize}

\hfill$\square$

\subsubsection{Well-Formed Indexes}

The \hyperref[anchor:well-formed-index-convention]{well-formed-index convention} asks every judgment to be indexed by a well-formed class, and the rules combine their indexes by the operations of the mode theory alone. \Cref{prop:sequent-indexes-are-well-formed} records that the convention sustains itself across a derivation and fixes the sort of the index.

\begin{proposition}[Sequent Indexes Are Well-Formed]\label{prop:sequent-indexes-are-well-formed}
Let $\Lambda\ \mathrm{Sig}$. Every derivable sequent $\Theta;\Gamma\vdash^{\Lambda}_{\mf D}\Delta$ over the mode boundary $\theta;\gamma\Vdash\delta$ carries a derivable descriptor judgment $\theta;\gamma\Vdash_{\mf D}\delta$, of the sort read off the boundary by \cref{prop:sorts-of-well-formed-descriptors}: a c-descriptor for the cartesian sequent and an $\ell$-descriptor for the linear one.
\end{proposition}

\textit{Proof.}
By induction on the derivation, over the schemas of \cref{def:rule-schemas} and the gated rules of \cref{def:gated-structural-rules}. For the rules carrying a $2$-cell premise whose source is the conclusion index, i.e. the variable rule, the $F$-right and $U$-left instances in all their sort instances, and the four gated rules, the index is well-formed by \cref{prop:source-and-target-of-descriptor-2-cells-are-well-formed}, extended to the gated calculus by \cref{lem:gated-2-cell-metatheory}. For $F$-left and $U$-right, the instance is given by the connective descriptor, well-formed by the formation rule of the type (\cref{def:type-formation-rules}), and by the conclusion index, whose judgment is presupposed by the \hyperref[anchor:well-formed-index-convention]{well-formed-index convention} as data of the instance. The premise index is their cut, well-formed by \cref{def:well-formed-composition-cut}; the well-formedness flows therefore from the conclusion to the premise, and there is nothing to prove for the conclusion. The sorts follow. A cartesian sequent has an all-cartesian boundary with a single output, and its well-formed index is then a c-descriptor; a linear sequent declares zones of both classes, and its index is an l-descriptor, the sorts of the premises being determined by those of the conclusion and of the connective descriptor through \cref{lem:cut-preserves-the-sorts}. \hfill $\square$

\subsubsection{Respect for Structural Transformations}

\begin{proposition}[Respect for Structural Transformation]\label{prop:respect-for-structural-transformation}
Let $\Lambda\ \mathrm{Sig}$. In the general notation of a sequent, if $r::\mf D\Rightarrow^{\Lambda}_{\theta;\gamma\Vdash\delta}\mf E$ and $\Theta;\Gamma\vdash^{\Lambda}_{\mf E}\Delta$ are derivable, then $\Theta;\Gamma\vdash^{\Lambda}_{\mf D}\Delta$ is derivable.
\end{proposition}
The general notation of a sequent refers to the fact that it expresses the result for both cartesian and linear sequents by making the appropriate restrictions.

\textit{Proof.}
We proceed by induction on the derivation. For the rules that contain a 2-cell as a premise, i.e. the variable rules $\mathsf v$ and $\mathsf v_{\mathsf l}$, the $F$-right instances $F_{\mathsf c}$R and $F_{\mathsf l}$R, the $U$-left instances $U_{\mathsf c}$L, $U_{\mathsf c}$L-in-l, $U_{\mathsf l}$L, and $U_{\mathsf{shift}}$L, and the gated rules, the descriptor attached to the conclusion is the source of the rule's $2$-cell premise. We compose the 2-cell premise with $r$, and we re-apply the rule with all other premises unchanged. For illustration purposes, let us treat the variable case; all the other aforementioned cases are similar. Suppose we are given
      $$
    	\inferrule*
      {
        P\in \mathrm{Prop}_{\Lambda,q}
        \\
        s::\mf E
        \Rightarrow^{\Lambda}
        [\varnothing\mid\{w\mapsto y\}]
      }
      {
        \Theta,\ w:P
        \vdash_{\mf E}^{\Lambda}
        y:P
      }
    $$
    where the rule is cartesian (cartesian instantiation of the variable rule schema). Composing its $2$-cell premise with $r$ gives $$s \circ r :: \mf D \Rightarrow [\varnothing\mid\{w\mapsto y\}],$$ and the target is still the identity class. Thus an application of the variable rule again gives:
    $$
      \inferrule*
      {
      P\in \mathrm{Prop}_{\Lambda,q}
      \\
      s \circ r::\mf D
      \Rightarrow^{\Lambda}
      [\varnothing\mid\{w\mapsto y\}]
      }
      {
      \Theta, w:P
      \vdash_{\mf D}^{\Lambda}
      y:P
      }
    $$
  That is the cartesian or linear conclusion we desired.

	For rules with a single sequent premise, i.e. the $F$-left and $U$-right instances, the premise descriptor is a cut of the conclusion descriptor against the connective descriptor. The cut is performed in a different order depending on the connective, and the argument is symmetric. Using $2$-Subst with the identity descriptor, we can apply the induction hypothesis to the premise and then re-apply the corresponding connective rule. Since both cases are similar, we only treat the F case. Suppose the derivation ends with this rule:
      $$
        \inferrule*
        {
        \Theta,\Theta_F\ ;\
        \Gamma
        \vdash_{\mf D_F\circ_{y,x}\mf E}^{\Lambda}
        \Delta
        }
        {
        \Theta;
        \Gamma
        \vdash_{\mf E}^{\Lambda}
        \Delta
        }
      $$
      This is the case where the connective lives in the cartesian zone. For the cartesian or linear sequent cases, the argument is the same, and so we treat the sequent as linear here without loss of generality (for the cartesian case, just take $\Gamma = \varnothing$ and $\Delta$ to be a singleton). We apply the 2-Subst rule with the identity transformation on the descriptor $\mf D_F$ to get
			$$\1_{\mf D_F} \circ_{y,x} r :: \mf D_F \circ_{y,x} \mf D \Rightarrow \mf D_F \circ{y,x} \mf E.$$ We then apply the induction hypothesis to the premise to get the sequent decorated by the introduced descriptor composition: $$\Theta,\Theta_F\ ;\
        \Gamma
        \vdash_{\mf D_F\circ_{y,x}\mf D}^{\Lambda}
        \Delta$$ and use it for the left introduction of the connective F: 
        $$
          \inferrule*
          {
          \Theta,\Theta_F\ ;\
          \Gamma
          \vdash_{\mf D_F\circ_{y,x}\mf D}^{\Lambda}
          \Delta
          }
          {
          \Theta;
          \Gamma
          \vdash_{\mf D}^{\Lambda}
          \Delta
          }
        $$
      This argument effectively covers the cases $F_{\mathsf c}$L, $F_{\mathsf c}$L-in-l, and $F_{\mathsf l}$L.
      For $U_{\mathsf c}$R, $U_{\mathsf{shift}}$R, and $U_{\mathsf l}$R, the descriptor in the premise is $\mf E\circ_{z,x}\mf D_U$, and we use the composition $r\circ_{z,x}\mathbf 1_{\mf D_U}$ instead.

All rules are covered, and in each case exactly one rule is re-applied. \hfill $\square$

\subsubsection{Identity}

\begin{theorem}[Identity]\label{thm:identity}
Let $\Lambda\ \mathrm{Sig}$. For every type $A$ at a mode $q$ of either class and every cartesian $\Theta$, the identity sequent
$$
\Theta\ ;\ x:A\vdash^{\Lambda}_{[\varnothing\mid\{x\mapsto y\}]}y:A
$$
is derivable. 
\end{theorem}

Again, we remember that the result is stated in the general case where the sequent can be specialized to its cartesian or linear version.

\textit{Proof.}
We proceed by induction on the complexity of $A$. In all the cases (cartesian and linear) the descriptor is the identity class $[\varnothing\mid\{x\mapsto y\}]$, and any cut with a descriptor resorts to either input substitution or redirection relabelling. We split the cases by the complexity of the formula $A$.

\begin{itemize}
\item \textbf{Proposition.} Suppose $A=P$ is a proposition (not an F or U composite) at a mode $q$ that is cartesian or linear. We apply the variable rule directly with the descriptor $[\varnothing\mid\{x\mapsto y\}]$. We take the identity 2-cell as a transformation:
$$
\inferrule*
{
  P\in \mathrm{Prop}_{\Lambda,q}
  \\
  \mathbf 1::[\varnothing\mid\{x\mapsto y\}]
  \Rightarrow^{\Lambda}
  [\varnothing\mid\{x\mapsto y\}]
}
{
  \Theta\ ;\ x:P
  \vdash_{[\varnothing\mid\{x\mapsto y\}]}^{\Lambda}
  y:P
}
$$
This is just the rule $\mathsf v$ with $x:P\in\Theta$ when $q$ is cartesian, and $\mathsf v_{\mathsf l}$ with $x:P$ in the linear zone when $q$ is linear.

\item \textbf{F-Connective}. If the formula is $A=F^{y'}_{\mf D_F}(\Theta_F;\Gamma_F)$, with the descriptor $\mf D_F$ being either linear or cartesian, the standard procedure is to understand that the last rule that we need to apply is the left introduction rule. This means that the last rule of the derivation will be  
$$
\inferrule*
{
  \Theta',\Theta_F\ ;\
  \Gamma_F
  \vdash_{\mf D_F\circ_{y',x}[\varnothing\mid\{x\mapsto y\}]}^{\Lambda}
  y:\mathsf F
}
{
  \Theta\ ;\
  x:\mathsf F
  \vdash_{[\varnothing\mid\{x\mapsto y\}]}^{\Lambda}
  y:\mathsf F
}
$$
where $\Theta'$ is $\Theta$ without the variable $x$ in the cartesian case and $\Theta$ in the linear case. Using \cref{lem:units-for-composition}(ii), we compute the composition of descriptors in the premise to be $\mf D_F\circ_{y',x}[\varnothing\mid\{x\mapsto y\}]=\mf D_F[y':=y]$. Therefore, it suffices to show a derivation of
$$\Theta',\Theta_F\ ;\ \Gamma_F\vdash_{\mf D_F[y':=y]}^{\Lambda}y:\mathsf F.$$
The remaining operation at our disposal is to apply the right F-rule
 ($F_{\mathsf c}$R or  $F_{\mathsf l}$R depending on the cartesian nature of the connective) paired with the identity descriptors. Before applying it, \hyperref[rem:context-variable-naming-invariance]{context variable naming invariance} allows us to choose a fresh boundary-renamed copy $\widetilde{\mf D}_F$ of $\mf D_F$, with $\widetilde\Theta_F=(\tilde a:A_a)_{a\in\theta_F}$, $\widetilde\Gamma_F=(\tilde x_i:A_i)_{1\leq i\leq n}$, and output $\tilde y'$. Put
$$
\sigma_{\tilde a}:=[\varnothing\mid\{a\mapsto b_{\tilde a}\}],
\qquad
\mf D_{\bar x_i}:=[\varnothing\mid\{x_i\mapsto\bar x_i\}],
$$
and define
$$
\mf C_F:=
\left(
  \bigl(\mf D_{\bar x_i}\bigr)_i
  \circ_{(\bar x_i,\tilde x_i)_{1\leq i\leq n}}
  \left(
    \bigl(\sigma_{\tilde a}\bigr)_{a\in\theta_F}
    \circ_{(b_{\tilde a},\tilde a)_{a\in\theta_F}}
    \widetilde{\mf D}_F
  \right)
\right)[\tilde y':=y].
$$
By \cref{lem:units-for-composition} and the inverse boundary renaming, $\mf C_F=\mf D_F[y':=y]$. The right F-rule is therefore
$$
\inferrule*
{
  \mathbf 1::\mf D_F[y':=y]
  \Rightarrow^{\Lambda}
  \mf C_F
  \\
  \bigl(\Theta',\Theta_F\vdash^{\Lambda}_{\sigma_{\tilde a}}b_{\tilde a}:A_a\bigr)_{a\in\theta_F}
  \\
  \bigl(
    \Theta',\Theta_F\ ;\ x_i:A_i
    \vdash_{[\varnothing\mid\{x_i\mapsto\bar x_i\}]}^{\Lambda}
    \bar x_i:A_i
  \bigr)_{1\leq i\leq n}
}
{
  \Theta',\Theta_F\ ;\ \Gamma_F
  \vdash_{\mf D_F[y':=y]}^{\Lambda}
  y:\mathsf F
}
$$
The family $\sigma$ is the identity family from the exposed context $\Theta_F$ to the fresh bound copy $\widetilde\Theta_F$. The other premises are formed using the induction hypothesis on the strictly smaller formulas $A_a$ and $A_i$ (smaller in complexity with respect to $F$). In the event $\mf D_F$ is a cartesian descriptor, $\Gamma_F=\cdot$, so the linear premises disappear and the rule reduces to its $2$-cell and family premises.

\item \textbf{U-Connective.} Suppose now that the formula is $A=U^{\mf D_U}_{x_U}(\Theta_U;\Gamma_U\mid\Delta_U)$, with $\mf D_U$, again, of either cartesian or linear type. The argument is symmetric to the $F$-connective case. The last rule should be an instance of the $U$-right rule (i.e. $U_{\mathsf c}$R when $\mf D_U$ is cartesian, $U_{\mathsf{shift}}$R when $\mf D_U$ is linear with $x_U$ cartesian, and $U_{\mathsf l}$R otherwise):
$$
\inferrule*
{
  \Theta,\Theta_U\ ;\ \Gamma_U^L,\,x:\mathsf U,\,\Gamma_U^R
  \vdash_{[\varnothing\mid\{x\mapsto y\}]\circ_{y,x_U}\mf D_U}^{\Lambda}
  \Delta_U
}
{
  \Theta\ ;\ x:\mathsf U
  \vdash_{[\varnothing\mid\{x\mapsto y\}]}^{\Lambda}
  y:\mathsf U
}
$$
with $x:\mathsf U$ present in the cartesian or the linear zone depending on its type, retained in $\Theta$ in the two cartesian instances. Similarly, by \cref{lem:units-for-composition}(i), the premise descriptor computes to $[\varnothing\mid\{x\mapsto y\}]\circ_{y,x_U}\mf D_U=\mf D_U[x_U:=x]$; it suffices therefore to derive
$$\Theta,\Theta_U\ ;\ \Gamma_U^L,x:\mathsf U,\Gamma_U^R\vdash_{\mf D_U[x_U:=x]}^{\Lambda}\Delta_U.$$
Using the same technique as in the F case, we gather the premise using the induction hypothesis to get sequents indexed by their identity descriptors and take the identity transformation as the first premise of the U-left rule. We are careful to use the right U-left instance depending on the position of $x : \mathsf U$ in the context ($U_{\mathsf c}$L or $U_{\mathsf{shift}}$L for the cartesian case and $U_{\mathsf l}$L for the linear case). By \hyperref[rem:context-variable-naming-invariance]{context variable naming invariance}, choose a fresh boundary-renamed copy $\widetilde{\mf D}_U$ of $\mf D_U$, with $\widetilde\Theta_U=(\tilde a:A_a)_{a\in\theta_U}$, non-distinguished inputs $\tilde x_i$, distinguished input $\tilde x_U$, and outputs $\tilde y_j$. Put
$$
\sigma_{\tilde a}:=[\varnothing\mid\{a\mapsto b_{\tilde a}\}],
\qquad
\mf D_{\bar x_i}:=[\varnothing\mid\{x_i\mapsto\bar x_i\}],
\qquad
\mf D_{\bar y_j}:=[\varnothing\mid\{\bar y_j\mapsto y_j\}],
$$
and define
$$
\mf G_U:=
\left(
  \bigl(\sigma_{\tilde a}\bigr)_{a\in\theta_U}
  \circ_{(b_{\tilde a},\tilde a)_{a\in\theta_U}}
  \widetilde{\mf D}_U
\right)[\tilde x_U:=x],
$$
$$
\mf C_U:=
\left(
  \bigl(\mf D_{\bar x_i}\bigr)_i
  \circ_{(\bar x_i,\tilde x_i)_{1\leq i\leq n}}
  \mf G_U
\right)
\circ_{(\tilde y_j,\bar y_j)_{1\leq j\leq m}}
\bigl(\mf D_{\bar y_j}\bigr)_j.
$$
By \cref{lem:units-for-composition} and the inverse boundary renaming, $\mf C_U=\mf D_U[x_U:=x]$. The U-left rule is therefore
$$
\inferrule*
{
  \mathbf 1::\mf D_U[x_U:=x]
  \Rightarrow^{\Lambda}
  \mf C_U
  \\
  \bigl(\Theta,\Theta_U\vdash^{\Lambda}_{\sigma_{\tilde a}}b_{\tilde a}:A_a\bigr)_{a\in\theta_U}
  \\
  \bigl(
    \Theta,\Theta_U\ ;\ x_i:A_i
    \vdash_{[\varnothing\mid\{x_i\mapsto\bar x_i\}]}^{\Lambda}
    \bar x_i:A_i
  \bigr)_{1\leq i\leq n}
  \\
  \bigl(
    \Theta,\Theta_U\ ;\ \bar y_j:B_j
    \vdash_{[\varnothing\mid\{\bar y_j\mapsto y_j\}]}^{\Lambda}
    y_j:B_j
  \bigr)_{1\leq j\leq m}
}
{
  \Theta,\Theta_U\ ;\ \Gamma_U^L,\,x:\mathsf U,\,\Gamma_U^R
  \vdash_{\mf D_U[x_U:=x]}^{\Lambda}
  \Delta_U
}
$$
The premises are formed using the induction hypothesis on the strictly smaller formulas $A_a$, $A_i$, and $B_j$.
\end{itemize}
We covered all the cases and this finishes the proof.\hfill$\square$

\textit{Remark:} Identifying in the cartesian descriptor $[\varnothing \mid \{x \mapsto y\}]$ the variable $x$ with $y$, we recover the identity sequent $\Theta\vdash^{\Lambda}_{x}A$ of \cite{LSR17} where $x : A \in \Theta$.

The invertibility results \cref{prop:left-invertibility-of-f,prop:right-invertibility-of-u} are important for proving cut admissibility: the standard proof uses the left invertibility of positive connectives and the right invertibility of negative connectives. Semantically, these results validate the universal property of the connectives as (op)cartesian lifts of mode morphisms.

\subsubsection{Invertibility}

In the proofs of \cref{prop:left-invertibility-of-f,prop:right-invertibility-of-u}, whenever an induction hypothesis enlarges the shared cartesian zone of one component of a branching rule, we use \cref{lem:cartesian-weakening-for-sequents} to move the unchanged components and family premises to the same zone before reapplying the rule.

\begin{proposition}[Left-invertibility of F]\label{prop:left-invertibility-of-f}
Let $\mathsf F=F^{y}_{\mf D_F}(\Theta_F;\Gamma_F)$ with the connective descriptor being cartesian or linear. If $\Theta\ ;\ \Gamma_L,z:\mathsf F,\Gamma_R\vdash^{\Lambda}_{\mf D}\Delta$ is derivable, then so is the $F$-left premise
$$
\Theta,\Theta_F\ ;\ \Gamma_L,\Gamma_F,\Gamma_R\vdash^{\Lambda}_{\mf D_F\circ_{y,z}\mf D}\Delta,
$$
by a derivation of no greater height. 
\end{proposition}

\textit{Proof.}
We distinguish according to whether the connective descriptor $\mf D_F$ is cartesian or linear. Each case proceeds by induction on the derivation, re-applying at each step at most one rule to induction hypotheses of no greater height.
\begin{itemize}
\item \textbf{Cartesian connective.} In this case, we have $\Gamma_F=\cdot$ and the connective $z:\mathsf F$ lies in $\Theta$. The composition $\mf D_F\circ_{y,z}\mf D$ is the cartesian substitution $\mf D[z:=\mf D_F]$. Every rule commutes with this substitution. First, we consider the variable rule. We note immediately that the principal variable $x'$ of the rule satisfies $x'\neq z$, since its type is a proposition. Therefore, the connective is not principal in the derivation:
$$
\inferrule*
{
  P\in\mathrm{Prop}_{\Lambda,q}
  \\
  s::\mf D\Rightarrow^{\Lambda}[\varnothing\mid\{x'\mapsto u\}]
}
{
  \Theta,\,z:\mathsf F\ ;\ x':P
  \vdash_{\mf D}^{\Lambda}
  u:P
}
$$
Using 2-Subst with the identity descriptor on $\mf D_F$, we get $\mathbf 1_{\mf D_F}\circ_{y,z}s::\mf D[z:=\mf D_F]\Rightarrow[\varnothing\mid\{x'\mapsto u\}][z:=\mf D_F]$, whose target is the identity class over the enlarged zone, the cut landing at the unused cartesian input $z$. It remains to use the variable rule, taking advantage of the fact that the composition eliminated the variable $z : \mathsf F$:

$$
\inferrule*
{
  P\in\mathrm{Prop}_{\Lambda,q}
  \\
	\mathbf 1_{\mf D_F}\circ_{y,z}s::\mf D[z:=\mf D_F]\Rightarrow[\varnothing\mid\{x'\mapsto u\}]
}
{
  \Theta,\,\Theta_F ;\ x':P
  \vdash_{\mf D[z:=\mf D_F]}^{\Lambda}
  u:P
}
$$ which is exactly the conclusion desired for both cartesian and linear sequents (the displayed case is the linear one; for the cartesian case, $x'$ is present in $\Theta$).

For the F-left rules, if the connective is principal, then the premise is already what we need:
$$
\inferrule*
{
  \Theta,\Theta_F\ ;\ \Gamma
  \vdash_{\mf D[z:=\mf D_F]}^{\Lambda}
  \Delta
}
{
  \Theta,\,z:\mathsf F\ ;\ \Gamma
  \vdash_{\mf D}^{\Lambda}
  \Delta
}
$$
In the case where the connective is not principal, a second F-connective $\mathsf F'=F^{y''}_{\mf D_{F'}}(\Theta_{F'};\Gamma_{F'})$ is principal in the derivation, and we apply the induction hypothesis to the premise. We distinguish the cartesian and linear instances of its left rule.

If $\mathsf F'$ is cartesian, then $z':\mathsf F'$ lies in the cartesian zone, $\Gamma_{F'}=\cdot$, and the last rule is
$$
\inferrule*
{
  \Theta,z:\mathsf F,\Theta_{F'}\ ;\ \Gamma
  \vdash_{\mf D[z':=\mf D_{F'}]}^{\Lambda}
  \Delta
}
{
  \Theta,z:\mathsf F,z':\mathsf F'\ ;\ \Gamma
  \vdash_{\mf D}^{\Lambda}
  \Delta
}.
$$
The induction hypothesis on the premise gives the index $(\mf D[z':=\mf D_{F'}])[z:=\mf D_F]$. Since the variables $z,z'$ do not interact, \cref{lem:c-substitution-commutes-with-composition} permutes the two substitutions into $(\mf D[z:=\mf D_F])[z':=\mf D_{F'}]$, and we reapply the cartesian F-left rule.

If $\mathsf F'$ is linear, then $z':\mathsf F'$ lies in the linear context and the last rule is instead
$$
\inferrule*
{
  \Theta,z:\mathsf F,\Theta_{F'}\ ;\
  \Gamma_L,\Gamma_{F'},\Gamma_R
  \vdash_{\mf D_{F'}\circ_{y'',z'}\mf D}^{\Lambda}
  \Delta
}
{
  \Theta,z:\mathsf F\ ;\
  \Gamma_L,z':\mathsf F',\Gamma_R
  \vdash_{\mf D}^{\Lambda}
  \Delta
}.
$$
The induction hypothesis gives the premise index $(\mf D_{F'}\circ_{y'',z'}\mf D)[z:=\mf D_F]$. The boundary of $\mf D_{F'}$ is fresh for $z$, so \cref{lem:c-substitution-commutes-with-composition} gives
$$
(\mf D_{F'}\circ_{y'',z'}\mf D)[z:=\mf D_F]
=
\mf D_{F'}\circ_{y'',z'}(\mf D[z:=\mf D_F]).
$$
We can therefore reapply $F_{\mathsf l}$L at $z'$ and obtain the desired result.

The right rule of $U$ is handled just like the F-left rule, having a single premise as well. The connective $z:\mathsf F$ stays in the cartesian zone, and the premise cuts the conclusion descriptor into the descriptor $\mf D_U$:
$$
\inferrule*
{
  \Theta,z:\mathsf F,\Theta_U\ ;\ \Gamma_U^L,\Gamma,\Gamma_U^R
  \vdash_{\mf D\circ_{w,x_U}\mf D_U}^{\Lambda}
  \Delta_L,\Delta_U,\Delta_R
}
{
  \Theta,z:\mathsf F\ ;\ \Gamma
  \vdash_{\mf D}^{\Lambda}
  \Delta_L,\,w:\mathsf U,\,\Delta_R
}
$$
We use the induction hypothesis on the premise and commute substitution with composition using \cref{lem:c-substitution-commutes-with-composition} to get $(\mf D\circ_{w,x_U}\mf D_U)[z:=\mf D_F]=(\mf D[z:=\mf D_F])\circ_{w,x_U}\mf D_U$. An application of the U rule terminates the case.

We are left with the rules that carry a $2$-cell premise: the right rule of $F$ and the left rule of $U$. In both, the cartesian zone $\Theta,z:\mathsf F$ is shared by all the premises, so we use the induction hypothesis on the premises and apply the 2-Subst rule on the transformation just as we did for the variable rule. We show the right F rule:
$$
\inferrule*
{
  r::\mf D
  \Rightarrow^{\Lambda}
  \bigl(\mf D_{\bar x_i}\bigr)_i\circ_{(\bar x_i,x_i)}\bigl((\sigma_a)_{a\in\theta_{F'}}\circ_{(b_a,a)}\mf D_{F'}\bigr)
  \\
  \bigl(\Theta,z:\mathsf F\vdash^{\Lambda}_{\sigma_a}A_a\bigr)_{a\in\theta_{F'}}
  \\
  \bigl(\Theta,z:\mathsf F\ ;\ \Gamma^{(i)}\vdash_{\mf D_{\bar x_i}}^{\Lambda}\Delta_L^{(i)},\bar x_i:A_i,\Delta_R^{(i)}\bigr)_{1\le i\le n}
}
{
  \Theta,z:\mathsf F\ ;\ \Gamma^{(1)},\ldots,\Gamma^{(n)}
  \vdash_{\mf D}^{\Lambda}
  \Delta_L,\,w:\mathsf F',\,\Delta_R
}
$$
The composition on the transformation $\mathbf 1_{\mf D_F}\circ_{y,z}r$ has a target that commutes with the substitution again by \cref{lem:c-substitution-commutes-with-composition}, and the conclusion is produced over $\Theta,\Theta_F$. The cartesian and linear versions of the left rule of $U$ are similar. Its principal variable is distinct from $z$. Every rule of the calculus therefore commutes with the substitution $[z:=\mf D_F]$, which finishes the analysis of the cartesian connective.
\item \textbf{Linear connective.} In this case $z:\mathsf F$ lies in the linear zone, and the derivation is a linear judgment. We use induction on the derivation and consider the last rule.

For the variable rule $\mathsf v_{\mathsf l}$, since the connective is not a proposition and the linear context contains only one element (the proposition), the rule does not type check.
$$
\inferrule*
{
  P\in\mathrm{Prop}_{\Lambda,q}
  \\
  r::\mf D\Rightarrow^{\Lambda}[\varnothing\mid\{x\mapsto u\}]
}
{
  \Theta\ ;\ x:P
  \vdash_{\mf D}^{\Lambda}
  u:P
}
$$
immediately shows that $x \neq z$ and the case is impossible. 

If the last rule is the left F rule with $z$ principal, then the premise is the desired sequent:
$$
\inferrule*
{
  \Theta,\Theta_F\ ;\ \Gamma_L,\Gamma_F,\Gamma_R
  \vdash_{\mf D_F\circ_{y,z}\mf D}^{\Lambda}
  \Delta
}
{
  \Theta\ ;\ \Gamma_L,\,z:\mathsf F,\,\Gamma_R
  \vdash_{\mf D}^{\Lambda}
  \Delta
}
$$

If the last rule is a left F rule at another variable $z'\neq z$, with $z':\mathsf F'=F^{y''}_{\mf D_{F'}}(\Theta_{F'};\Gamma_{F'})$, its premise still carries $z$:
$$
\inferrule*
{
  \Theta,\Theta_{F'}\ ;\ \Gamma_L,\Gamma_{F'},\Gamma_R
  \vdash_{\mf D_{F'}\circ_{y'',z'}\mf D}^{\Lambda}
  \Delta
}
{
  \Theta\ ;\ \Gamma_L,\,z':\mathsf F',\,\Gamma_R
  \vdash_{\mf D}^{\Lambda}
  \Delta
}
$$
where $z$ occurs in $\Gamma_L,\Gamma_R$. The induction hypothesis on the premise gives the descriptor $\mf D_F\circ_{y,z}(\mf D_{F'}\circ_{y'',z'}\mf D)$. Note that the descriptors $\mf D_F$ and $\mf D_{F'}$ cut at different inputs $z,z'$ of $\mf D$, so \cref{lem:left-interchange-law} permutes them into $\mf D_{F'}\circ_{y'',z'}(\mf D_F\circ_{y,z}\mf D)$, and we reapply the left F rule at $z'$ to get the result.

If the last rule is the right F rule, it introduces some $w:\mathsf F'$ and, the linear contexts being split, $z$ sits in exactly one component context $\Gamma^{(k)}$:
$$
\inferrule*
{
  r::\mf D
  \Rightarrow^{\Lambda}
  \bigl(\mf D_{\bar x_i}\bigr)_i\circ_{(\bar x_i,x_i)}\bigl((\sigma_a)_{a\in\theta_{F'}}\circ_{(b_a,a)}\mf D_{F'}\bigr)
  \\
  \bigl(\Theta\vdash^{\Lambda}_{\sigma_a}A_a\bigr)_{a\in\theta_{F'}}
  \\
  \bigl(\Theta\ ;\ \Gamma^{(i)} \vdash_{\mf D_{\bar x_i}}^{\Lambda} \Delta_L^{(i)},\,\bar x_i:A_i,\,\Delta_R^{(i)}\bigr)_{1\le i\le n}
}
{
  \Theta\ ;\ \Gamma
  \vdash_{\mf D}^{\Lambda}
  \Delta_L',\,w:\mathsf F',\,\Delta_R'
}
$$
where $\Gamma$ is the list $\Gamma_1, \dots, \Gamma_n$.  
We use the induction hypothesis on the $k$-th component premise that contains the connective to get $\Theta,\Theta_F;\Gamma^{(k)}_L,\Gamma_F, \Gamma^{(k)}_R\vdash_{\mf D_F\circ_{y,z}\mf D_{\bar x_k}}\Delta_L^{(k)},\bar x_k:A_k,\Delta_R^{(k)}$. Using 2-Subst with the identity descriptor on $\mf D_F$, we get the composition on the transformation
$$\mathbf 1_{\mf D_F}\circ_{y,z}r::\mf D_F\circ_{y,z}\mf D\Rightarrow\mf D_F\circ_{y,z}\Bigl(\bigl(\mf D_{\bar x_i}\bigr)_i\circ\bigl((\sigma_a)_a\circ\mf D_{F'}\bigr)\Bigr).$$
Since $z$ is an input of the factor $\mf D_{\bar x_k}$ alone, the other components not declaring it and the family cut $(\sigma_a)_a\circ\mf D_{F'}$ are permuted using \cref{lem:localization}(i) to replace $\mf D_{\bar x_k}$ with $\mf D_F\circ_{y,z}\mf D_{\bar x_k}$. We reapply the right F rule with the modified $k$-th component to get the result.

The right U rule introduces $w:\mathsf U$ from a single premise which carries $z$:
$$
\inferrule*
{
  \Theta,\Theta_U\ ;\ \Gamma_U^L,\Gamma,\Gamma_U^R
  \vdash_{\mf D\circ_{w,x_U}\mf D_U}^{\Lambda}
  \Delta_L',\Delta_U,\Delta_R'
}
{
  \Theta\ ;\ \Gamma
  \vdash_{\mf D}^{\Lambda}
  \Delta_L',\,w:\mathsf U,\,\Delta_R'
}
$$
with $z$ in $\Gamma$. We apply the induction hypothesis to the premise and use \cref{lem:associativity} to transform $\mf D_F\circ_{y,z}(\mf D\circ_{w,x_U}\mf D_U)$ into $(\mf D_F\circ_{y,z}\mf D)\circ_{w,x_U}\mf D_U$, so we can reapply the right U rule to this new descriptor.

The left U rules $U_{\mathsf l}$L and $U_{\mathsf{shift}}$L go exactly like the right F rule: their principal variable carries a $U$-type, so $z$ again sits in one component context, and \cref{lem:localization}(i) localizes the composition on the transformation $\mathbf 1_{\mf D_F}\circ_{y,z}r$ to that component.

There remain the two rules that act on a cartesian connective inside the linear sequent. In $F_{\mathsf c}$L-in-l the premise descriptor is the substitution $\mf D[z':=\mf D_{F'}]$:
$$
\inferrule*
{
  \Theta,\Theta_{F'}\ ;\ \Gamma
  \vdash_{\mf D[z':=\mf D_{F'}]}^{\Lambda}
  \Delta
}
{
  \Theta,\,z':\mathsf F'\ ;\ \Gamma
  \vdash_{\mf D}^{\Lambda}
  \Delta
}
$$
where $z$ occurs in $\Gamma$. We use the induction hypothesis on the premise and commute substitution with composition using \cref{lem:c-substitution-commutes-with-composition}, the substitution $[z':=\mf D_{F'}]$ being vacuous on $\mf D_F$ whose boundary is fresh for $z'$, so that the index $\mf D_F\circ_{y,z}(\mf D[z':=\mf D_{F'}])$ becomes $(\mf D_F\circ_{y,z}\mf D)[z':=\mf D_{F'}]$, and we reapply the rule.

In $U_{\mathsf c}$L-in-l the last rule carries a $2$-cell, a family and a linear premise:
$$
\inferrule*
{
  r::\mf D\Rightarrow^{\Lambda}\mf D'[x'':=\tau]
  \\
  \bigl(\Theta,z':\mathsf U'\vdash^{\Lambda}_{\sigma_a}A_a\bigr)_{a\in\theta_{U'}}
  \\
  \Theta,z':\mathsf U',\,x'':X\ ;\ \Gamma
  \vdash_{\mf D'}^{\Lambda}
  \Delta
}
{
  \Theta,\,z':\mathsf U'\ ;\ \Gamma
  \vdash_{\mf D}^{\Lambda}
  \Delta
}
$$
with $\tau=((\sigma_a)_{a\in\theta_{U'}}\circ\mf D_{U'})[x:=z']$ and $z$ in $\Gamma$. We use the induction hypothesis on the linear premise, and using 2-Subst with the identity descriptor on $\mf D_F$, we get the composition on the transformation
$$\mathbf 1_{\mf D_F}\circ_{y,z}r::\mf D_F\circ_{y,z}\mf D\Rightarrow\mf D_F\circ_{y,z}\bigl(\mf D'[x'':=\tau]\bigr)=\bigl(\mf D_F\circ_{y,z}\mf D'\bigr)[x'':=\tau],$$
the variable $x''$ being fresh for $\mf D_F$, and we reapply the rule.
\end{itemize}
All cases re-apply at most one rule to induction hypotheses of smaller height; the resulting derivation has therefore no greater height. \hfill $\square$

\begin{proposition}[Right-invertibility of U]\label{prop:right-invertibility-of-u}
Let $\mathsf U=U^{\mf D_U}_{x_U}(\Theta_U;\Gamma_U\mid\Delta_U)$ with the connective descriptor being linear or cartesian, with variable $x_U$ in the cartesian or linear context. If $\Theta\ ;\ \Gamma\vdash^{\Lambda}_{\mf D}\Delta_L,z:\mathsf U,\Delta_R$ is derivable, then so is the $U$-right premise
$$
\Theta,\Theta_U\ ;\ \Gamma_U^L,\Gamma,\Gamma_U^R\vdash^{\Lambda}_{\mf D\circ_{z,x_U}\mf D_U}\Delta_L,\Delta_U,\Delta_R,
$$
by a derivation of no greater height.
\end{proposition}

\textit{Proof.}
We distinguish according to the nature of the connective descriptor $\mf D_U$ and of its input $x_U$. Each case proceeds by induction on the derivation, re-applying at each step at most one rule to induction hypotheses of no greater height.
\begin{itemize}
\item \textbf{Cartesian connective.} In this case $\mf D_U$ is cartesian, so $z:\mathsf U$ lies in the cartesian zone, and the derivation concludes the cartesian sequent $\Theta\vdash_{\mf D}z:\mathsf U$. We have $\Gamma_U=\cdot$ and $\Delta_U$ a single output $b:X$, and the composition $\mf D\circ_{z,x_U}\mf D_U$ is the cartesian substitution $\mf D_U[x_U:=\mf D]$. A cartesian sequent carries a single output, so either the last rule produces $z:\mathsf U$ or $z:\mathsf U$ is present in the premise. First, if the last rule is the right rule of $U$ with $z$ principal, then the premise is already what we need:
$$
\inferrule*
{
  \Theta,\Theta_U
  \vdash_{\mf D\circ_{z,x_U}\mf D_U}^{\Lambda}
  b:X
}
{
  \Theta
  \vdash_{\mf D}^{\Lambda}
  z:\mathsf U
}
$$
The variable rule is impossible, its conclusion being a proposition, and $U_{\mathsf{shift}}$R is excluded, its conclusion type carrying a linear connective descriptor. In every remaining case the last rule is a cartesian left rule, $F_{\mathsf c}$L or $U_{\mathsf c}$L, acting at a variable $z'$ and producing $z:\mathsf U$ in its premise. We show $F_{\mathsf c}$L:
$$
\inferrule*
{
  \Theta,\Theta_{F'}
  \vdash_{\mf D[z':=\mf D_{F'}]}^{\Lambda}
  z:\mathsf U
}
{
  \Theta,\,z':\mathsf F'
  \vdash_{\mf D}^{\Lambda}
  z:\mathsf U
}
$$
We use the induction hypothesis on the premise and commute substitution with composition using \cref{lem:c-substitution-commutes-with-composition}, so that $(\mf D[z':=\mf D_{F'}])\circ_{z,x_U}\mf D_U=(\mf D\circ_{z,x_U}\mf D_U)[z':=\mf D_{F'}]$, and we reapply the rule. The rule $U_{\mathsf c}$L is similar, its $2$-cell premise composed with $\mathbf 1_{\mf D_U}$ by 2-Subst. No other rule concludes a cartesian sequent with a $U$-type, which finishes the cartesian case.

\item \textbf{Linear connective, and $x_U$ cartesian.} Now $\mf D_U$ is linear and its input $x_U$ is cartesian, so $z:\mathsf U$ still lies in the cartesian zone, and the derivation again concludes the cartesian sequent $\Theta\vdash_{\mf D}z:\mathsf U$. This time the desired sequent is the linear premise of $U_{\mathsf{shift}}$R:
$$
\inferrule*
{
  \Theta,\Theta_U\ ;\ \Gamma_U
  \vdash_{\mf D\circ_{z,x_U}\mf D_U}^{\Lambda}
  \Delta_U
}
{
  \Theta
  \vdash_{\mf D}^{\Lambda}
  z:\mathsf U
}
$$
The same induction applies. If the last rule is $U_{\mathsf{shift}}$R with $z$ principal, its premise is already what we need. A last rule $\mathsf v$ or $U_{\mathsf c}$R is impossible by type mismatch. In the remaining cases the last rule is a cartesian left rule $F_{\mathsf c}$L or $U_{\mathsf c}$L acting at a variable $z'$. We use the induction hypothesis on the premise, now inverted into a linear sequent, and reapply the rule in its in-l form ($F_{\mathsf c}$L-in-l, respectively $U_{\mathsf c}$L-in-l), the substitutions commuting by \cref{lem:c-substitution-commutes-with-composition} and the $2$-cell premise $s$ of $U_{\mathsf c}$L composed with $\mathbf 1_{\mf D_U}$ into $s\circ_{z,x_U}\mathbf 1_{\mf D_U}$ by 2-Subst.

\item \textbf{Linear connective, and $x_U$ linear.} We spend more time in this case as this is the one that is different from the left invertibility of $F$. In this case $z:\mathsf U$ lies in the linear zone, and the derivation is a linear judgment. We use induction on the derivation and consider the last rule, dually to the linear connective of \cref{prop:left-invertibility-of-f}.

The variable rule $\mathsf v_{\mathsf l}$ cannot be the last rule, its output being a single propositional variable:
$$
\inferrule*
{
  P\in\mathrm{Prop}_{\Lambda,q}
  \\
  r::\mf D\Rightarrow^{\Lambda}[\varnothing\mid\{x\mapsto u\}]
}
{
  \Theta\ ;\ x:P
  \vdash_{\mf D}^{\Lambda}
  u:P
}
,$$
and $\mathsf U$ is not a proposition, so $z:\mathsf U$ can never be this $u:P$.

If the last rule is the right rule of $U$ with $z$ principal, then the premise is the sequent we need:
$$
\inferrule*
{
  \Theta,\Theta_U\ ;\ \Gamma_U^L,\Gamma,\Gamma_U^R
  \vdash_{\mf D\circ_{z,x_U}\mf D_U}^{\Lambda}
  \Delta_L,\Delta_U,\Delta_R
}
{
  \Theta\ ;\ \Gamma
  \vdash_{\mf D}^{\Lambda}
  \Delta_L,\,z:\mathsf U,\,\Delta_R
}
$$
If the last rule is a right rule of $U$ at another output $w\neq z$, with $w:\mathsf U'=U^{\mf D_{U'}}_{x_{U'}}(\Theta_{U'};\Gamma_{U'}\mid\Delta_{U'})$, its premise still carries $z$:
$$
\inferrule*
{
  \Theta,\Theta_{U'}\ ;\ \Gamma_{U'}^L,\Gamma,\Gamma_{U'}^R
  \vdash_{\mf D\circ_{w,x_{U'}}\mf D_{U'}}^{\Lambda}
  \Delta_L,\Delta_{U'},\Delta_R
}
{
  \Theta\ ;\ \Gamma
  \vdash_{\mf D}^{\Lambda}
  \Delta_L,\,w:\mathsf U',\,\Delta_R
}
$$
where $z$ occurs in $\Delta_L,\Delta_R$. The induction hypothesis on the premise gives the composition $(\mf D\circ_{w,x_{U'}}\mf D_{U'})\circ_{z,x_U}\mf D_U$. We apply \cref{lem:right-interchange-law} to permute the composition into $(\mf D\circ_{z,x_U}\mf D_U)\circ_{w,x_{U'}}\mf D_{U'}$, and we reapply the right U rule at $w$ to get the result.

If the last rule is the left rule of $F$, it has a single premise carrying $z$:
$$
\inferrule*
{
  \Theta,\Theta_{F'}\ ;\ \Gamma_L,\Gamma_{F'},\Gamma_R
  \vdash_{\mf D_{F'}\circ_{y'',z'}\mf D}^{\Lambda}
  \Delta_L,\,z:\mathsf U,\,\Delta_R
}
{
  \Theta\ ;\ \Gamma_L,\,z':\mathsf F',\,\Gamma_R
  \vdash_{\mf D}^{\Lambda}
  \Delta_L,\,z:\mathsf U,\,\Delta_R
}
$$
We apply the induction hypothesis to the premise and use \cref{lem:associativity} on the composition $(\mf D_{F'}\circ_{y'',z'}\mf D)\circ_{z,x_U}\mf D_U$ to get $\mf D_{F'}\circ_{y'',z'}(\mf D\circ_{z,x_U}\mf D_U)$. From this, we reapply the left F rule.

The rules $F_{\mathsf l}$R, $U_{\mathsf l}$L and $U_{\mathsf{shift}}$L split the linear contexts, so the output $z$ sits in only one component. We treat $F_{\mathsf l}$R:
$$
\inferrule*
{
  r::\mf D
  \Rightarrow^{\Lambda}
  \bigl(\mf D_{\bar x_i}\bigr)_i\circ_{(\bar x_i,x_i)}\bigl((\sigma_a)_{a\in\theta_{F'}}\circ_{(b_a,a)}\mf D_{F'}\bigr)
  \\
  \bigl(\Theta\vdash^{\Lambda}_{\sigma_a}A_a\bigr)_{a\in\theta_{F'}}
  \\
  \bigl(\Theta\ ;\ \Gamma^{(i)} \vdash_{\mf D_{\bar x_i}}^{\Lambda} \Delta_L^{(i)},\,\bar x_i:A_i,\,\Delta_R^{(i)}\bigr)_{1\le i\le n}
}
{
  \Theta\ ;\ \Gamma
  \vdash_{\mf D}^{\Lambda}
  \Delta_L,\,w:\mathsf F',\,\Delta_R
}
$$
where $\Gamma := \Gamma^{(1)},\ldots,\Gamma^{(n)}$ and $z$ lies in the $k$-th component. We use the induction hypothesis on that component premise to get the descriptor $\mf D_{\bar x_k}\circ_{z,x_U}\mf D_U$, and weaken the remaining components together with the family premise to the enlarged cartesian zone $\Theta,\Theta_U$ by \cref{lem:cartesian-weakening-for-sequents}. Using 2-Subst with the identity descriptor on $\mf D_U$, we get the composition on the transformation
$$r\circ_{z,x_U}\mathbf 1_{\mf D_U}::\mf D\circ_{z,x_U}\mf D_U\Rightarrow\Bigl(\bigl(\mf D_{\bar x_i}\bigr)_i\circ\bigl((\sigma_a)_a\circ\mf D_{F'}\bigr)\Bigr)\circ_{z,x_U}\mf D_U.$$
Since $z$ is a side output of the factor $\mf D_{\bar x_k}$ alone and the other components don't redirect to it, we can use \cref{lem:localization}(ii), whose right-interchange hypotheses hold automatically for well-formed classes. We rewrite the target as the same iterated cut with $\mf D_{\bar x_k}$ replaced by $\mf D_{\bar x_k}\circ_{z,x_U}\mf D_U$. We reapply the right F rule with the modified $k$-th component to get the result. The left U rules $U_{\mathsf l}$L and $U_{\mathsf{shift}}$L go the same way.

There remain the two rules that act on a cartesian connective inside the linear sequent. In $F_{\mathsf c}$L-in-l the output $z$ is produced in the premise, whose descriptor is the substitution $\mf D[z':=\mf D_{F'}]$:
$$
\inferrule*
{
  \Theta,\Theta_{F'}\ ;\ \Gamma
  \vdash_{\mf D[z':=\mf D_{F'}]}^{\Lambda}
  \Delta_L,\,z:\mathsf U,\,\Delta_R
}
{
  \Theta,\,z':\mathsf F'\ ;\ \Gamma
  \vdash_{\mf D}^{\Lambda}
  \Delta_L,\,z:\mathsf U,\,\Delta_R
}
$$
We use the induction hypothesis on the premise and commute substitution with composition using \cref{lem:c-substitution-commutes-with-composition}, the substitution $[z':=\mf D_{F'}]$ being vacuous on $\mf D_U$ whose boundary is fresh for $z'$, so that $(\mf D[z':=\mf D_{F'}])\circ_{z,x_U}\mf D_U$ becomes $(\mf D\circ_{z,x_U}\mf D_U)[z':=\mf D_{F'}]$, and we reapply the rule. In $U_{\mathsf c}$L-in-l the last rule carries a $2$-cell, a family and a linear premise, in which $z$ is produced:
$$
\inferrule*
{
  r::\mf D\Rightarrow^{\Lambda}\mf D'[x'':=\tau]
  \\
  \bigl(\Theta,z':\mathsf U'\vdash^{\Lambda}_{\sigma_a}A_a\bigr)_{a\in\theta_{U'}}
  \\
  \Theta,z':\mathsf U',\,x'':X\ ;\ \Gamma
  \vdash_{\mf D'}^{\Lambda}
  \Delta_L,\,z:\mathsf U,\,\Delta_R
}
{
  \Theta,\,z':\mathsf U'\ ;\ \Gamma
  \vdash_{\mf D}^{\Lambda}
  \Delta_L,\,z:\mathsf U,\,\Delta_R
}
$$
with $\tau=((\sigma_a)_{a\in\theta_{U'}}\circ\mf D_{U'})[x:=z']$. We use the induction hypothesis on the linear premise, and using 2-Subst with the identity descriptor on $\mf D_U$, we get the composition on the transformation
$$r\circ_{z,x_U}\mathbf 1_{\mf D_U}::\mf D\circ_{z,x_U}\mf D_U\Rightarrow\bigl(\mf D'[x'':=\tau]\bigr)\circ_{z,x_U}\mf D_U=\bigl(\mf D'\circ_{z,x_U}\mf D_U\bigr)[x'':=\tau],$$
the variable $x''$ being fresh for $\mf D_U$, and we reapply the rule.
\end{itemize}
All cases re-apply at most one rule to induction hypotheses of smaller height; the resulting derivation has therefore no greater height. \hfill $\square$

\subsubsection{Cartesian Contraction}

\begin{lemma}[Cartesian Contraction for Sequents]\label{lem:cartesian-contraction-for-sequents}
Let $x:A$ and $x':A$ be defined at a cartesian mode. If $\Theta,x:A,x':A\ ;\ \Gamma\vdash^{\Lambda}_{\mf D}\Delta$ is derivable, then $\Theta,x:A\ ;\ \Gamma\vdash^{\Lambda}_{\mf D[x':=x]}\Delta$ is derivable by a derivation of no greater height.
\end{lemma}

\textit{Proof.}
We proceed by lexicographic induction on the complexity of $A$ and the height of the derivation $\Theta,x:A,x':A\ ;\ \Gamma\vdash^{\Lambda}_{\mf D}\Delta$. The operation is the c-substitution $\mf D\mapsto\mf D[x':=x]$, i.e. the cut $[\varnothing\mid\{x\mapsto b\}]\circ_{b,x'}\mf D$ against the identity class of the retained variable $x$. It commutes with every rule exactly as weakening does, except for a last $F_{\mathsf c}$L or $F_{\mathsf c}$L-in-l rule principal at one of the contracted copies. If such a rule is principal at the retained copy $x$, we first exchange the boundary names $x$ and $x'$. Writing $\rho$ for this transposition, the rule is then principal at $x'$, while the contracted index is unchanged, since $\mathrm{ren}_{\rho}(\mf D)[x':=x]=\mf D[x':=x]$. Thus we may take the principal contracted copy to be $x'$. We treat all the cases separately.

\begin{itemize}
\item \textbf{Var rule.} If the derivation terminates with the variable rule, then we are faced with
$$
\inferrule*
{
  P\in \mathrm{Prop}_{\Lambda,q}
  \\
  r::\mf D
  \Rightarrow^{\Lambda}
  [\varnothing\mid\{w\mapsto y\}]
}
{
  \Theta,x:A,x':A\ ;\ w:P
  \vdash_{\mf D}^{\Lambda}
  y:P
}
,$$
where we remember once more that the general schema reduces to the cartesian and linear versions. Since the operation $(\--)[x':=x]$ behaves well with respect to $2$-cells, we get
$$r[x':=x] :: \mf D[x':=x] \Rightarrow [\varnothing\mid\{w\mapsto y\}][x':=x].$$
The target computes to the identity class $[\varnothing\mid\{w[x':=x]\mapsto y\}]$. The substitution leaves the redirection untouched, unless the rule is the cartesian $\mathsf v$ concluding at the very variable $x'$, in which case the substituted target names $x$ instead, available since $x:A\in\Theta$. We can thus reapply the variable rule to this new transformation to obtain
$$
\inferrule*
{
  P\in \mathrm{Prop}_{\Lambda,q}
  \\
  r[x':=x] :: \mf D[x':=x] \Rightarrow [\varnothing\mid\{w[x':=x]\mapsto y\}]
}
{
  \Theta,x:A\ ;\ w:P
  \vdash_{\mf D[x':=x]}^{\Lambda}
  y:P
}
.$$

\item \textbf{F-left, U-right}. If the derivation terminates with the left F rule at a principal $z\notin\{x,x'\}$, then we are looking at
$$
\inferrule*
{
  \Theta,x:A,x':A,\Theta_F\ ;\
  \Gamma_L,\Gamma_F,\Gamma_R
  \vdash_{\mf D_F\circ_{y,z}\mf D}^{\Lambda}
  \Delta
}
{
  \Theta,x:A,x':A\ ;\
  \Gamma_L,\,z:\mathsf F,\,\Gamma_R
  \vdash_{\mf D}^{\Lambda}
  \Delta
}
;$$
by the induction hypothesis, we can contract the premise to end up with
$$
\Theta,x:A,\Theta_F\ ;\ \Gamma_L,\Gamma_F,\Gamma_R
\vdash_{(\mf D_F\circ_{y,z}\mf D)[x':=x]}^{\Lambda}\Delta.
$$
Again, the well-behavedness of the contraction operation allows us to extract the descriptor associated to the connective $\mf D_F$, whose boundary is fresh for $x'$ and is therefore untouched by the substitution. By \cref{lem:c-substitution-commutes-with-composition}, we get
$$
\Theta,x:A,\Theta_F\ ;\ \Gamma_L,\Gamma_F,\Gamma_R
\vdash_{\mf D_F\circ_{y,z}(\mf D[x':=x])}^{\Lambda}\Delta,
$$
and applying the rule once more gets us the result:
$$
\inferrule*
{
  \Theta,x:A,\Theta_F\ ;\
  \Gamma_L,\Gamma_F,\Gamma_R
  \vdash_{\mf D_F\circ_{y,z}(\mf D[x':=x])}^{\Lambda}
  \Delta
}
{
  \Theta,x:A\ ;\
  \Gamma_L,\,z:\mathsf F,\,\Gamma_R
  \vdash_{\mf D[x':=x]}^{\Lambda}
  \Delta
}.
$$
The case for the U-right rule is symmetric: instead of pulling the connective descriptor $\mf D_U$ out on the left, we extract it on the right of the composition.

\item \textbf{F-right, U-left}. In this case, we treat the connective $U$ as the other one is similar. The last rule is $U$-left, with conclusion index $\mf D$ and iterated cut $\mf C$. We apply the substitution to the transformation as in the variable case and pull the iterated cut descriptor out of the composition to get the transformation
$$r[x':=x]::\mf D[x':=x] \Rightarrow \mf C[x':=x],$$
the substitution distributing to all the subdescriptors in the iterated cut by \cref{lem:c-substitution-commutes-with-composition}. Applying the induction hypothesis to the premises gives
$$
\inferrule*
{
  r[x':=x] :: \mf D[x':=x] \Rightarrow \mf C[x':=x]
  \\
  \bigl(\Theta,x:A\vdash^{\Lambda}_{\sigma_a[x':=x]}A_a\bigr)_{a\in\theta_U}
  \\
  \bigl(
    \Theta,x:A\ ;\ \Gamma^{(i)}
    \vdash_{\mf D_{\bar x_i}[x':=x]}^{\Lambda}
    \Delta_L^{(i)},\,\bar x_i:A_i,\,\Delta_R^{(i)}
  \bigr)_{1\leq i\leq n}
  \\
  \bigl(
    \Theta,x:A\ ;\ \Gamma^{(n+j)},\,\bar y_j:B_j
    \vdash_{\mf D_{\bar y_j}[x':=x]}^{\Lambda}
    \Delta^{(n+j)}
  \bigr)_{1\leq j\leq m}
}
{
  \Theta,x:A\ ;\ \Gamma,\,z:\mathsf U
  \vdash_{\mf D[x':=x]}^{\Lambda}
  \Delta
}
,$$
the displayed linear principal $z$ being untouched by the cartesian substitution. The cartesian left rules $U_{\mathsf c}$L, $U_{\mathsf c}$L-in-l and $U_{\mathsf{shift}}$L instead name a cartesian variable $z\in\Theta$, which they retain, named $x$ in the substituted premises if it was $x'$ itself, available since $x:A\in\Theta$. The F-right rule is symmetric.

\item \textbf{F-left, principal at $x'$.} This is the single exception, including the case originally principal at $x$ after exchanging the boundary names. Here the last rule is $F_{\mathsf c}$L consuming the contracted variable itself, so $A=F^{y}_{\mf D_F}(\Theta_F;\cdot)$. By \hyperref[rem:context-variable-naming-invariance]{context variable naming invariance}, write $\Theta_F=(a:A_a)_{a\in\theta_F}$ and choose a boundary-renamed copy $\widetilde\Theta_F=(\tilde a:A_a)_{a\in\theta_F}$ with descriptor $\widetilde{\mf D}_F$ and output $\tilde y$. The last rule is
$$
\inferrule*
{
  \Theta,x:A,\Theta_F\ ;\ \Gamma
  \vdash_{\mf D_F\circ_{y,x'}\mf D}^{\Lambda}
  \Delta
}
{
  \Theta,x:A,x':A\ ;\ \Gamma
  \vdash_{\mf D}^{\Lambda}
  \Delta
}
,$$
the premise carrying $\Theta_F$ in place of $x'$, so that commuting the substitution past the rule is unavailable. Instead, we invert the premise at the retained copy $x:A=x:\mathsf F$ by \cref{prop:left-invertibility-of-f}, obtaining a derivation of no greater height that consumes $x$ and exposes the fresh copy $\widetilde\Theta_F$:
$$
\Theta,\Theta_F,\widetilde\Theta_F\ ;\ \Gamma
\vdash_{\widetilde{\mf D}_F\circ_{\tilde y,x}(\mf D_F\circ_{y,x'}\mf D)}^{\Lambda}
\Delta.
$$
The two copies $\Theta_F$ and $\widetilde\Theta_F$ are now contracted componentwise by the induction hypothesis, retaining each $a:A_a$ and contracting the corresponding $\tilde a:A_a$. The formulas $A_a$ are strict subformulas of $A$ and so have strictly smaller complexity. This is where the lexicographic measure is spent. After the componentwise contractions, the inverse boundary renaming identifies the fresh connective descriptor with $\mf D_F$, and the resulting index is
$$
\mf D_F\circ_{y,x}(\mf D_F\circ_{y,x'}\mf D)
=
\mf D_F\circ_{y,x}(\mf D[x':=x]).
$$
Indeed, both sides compute to $\mf D_F\circ_{y,x'}(\mf D[x:=\mf D_F])$ by \cref{lem:c-substitution-commutes-with-composition} and \cref{lem:units-for-composition}. There remains a single copy of $\Theta_F$, over the context $\Theta,\Theta_F$, and re-applying $F_{\mathsf c}$L at $x$ yields the conclusion:
$$
\inferrule*
{
  \Theta,\Theta_F\ ;\ \Gamma
  \vdash_{\mf D_F\circ_{y,x}(\mf D[x':=x])}^{\Lambda}
  \Delta
}
{
  \Theta,x:A\ ;\ \Gamma
  \vdash_{\mf D[x':=x]}^{\Lambda}
  \Delta
}.
$$
This is exactly the $F_{\mathsf c}$L premise for the conclusion index $\mf D[x':=x]$. The $F_{\mathsf c}$L-in-l rule, which consumes its principal likewise, is treated verbatim. The inversion being of no greater height and the componentwise contractions being at strictly smaller complexity, the resulting derivation is of no greater height.
\end{itemize}
\hfill$\square$

\begin{corollary}[Cartesian Structural Rules for Sequents]\label{cor:cartesian-structural-rules-for-sequents}
Cartesian weakening and contraction are admissible in the calculus, with the height bounds of \cref{lem:cartesian-weakening-for-sequents,lem:cartesian-contraction-for-sequents}.
\end{corollary}

We will use \cref{cor:cartesian-structural-rules-for-sequents} silently in the remainder of the chapter, under the names \emph{cartesian weakening} and \emph{cartesian contraction}, in gated signatures through \cref{lem:gated-metatheory-extension}. In particular, when a rule is re-applied in an enlarged cartesian zone, its unchanged premises are weakened without comment, via \cref{lem:weakening-commutes-with-composition} on their indexes.

\subsubsection{Cut Admissibility}

\begin{theorem}[Cut Admissibility]\label{thm:cut-admissibility}
Let $\Lambda\ \mathrm{Sig}$ be a signature with no gates. Let $\Theta$ be a cartesian context, and $\Gamma, \Gamma', \Delta_L, \Delta_R$ be linear contexts. Cut is admissible for both cartesian and linear sequents.
\begin{itemize}
\item[(i)] Let $A,B$ be types at a cartesian mode. These statements hold:
$$
\begin{aligned}
&\text{If} \quad \Theta\vdash^{\Lambda}_{\mf D_1}y:A \quad \text{and} \quad \Theta,x:A\vdash^{\Lambda}_{\mf D_2}u:B
\\
&\text{are derivable, then} \quad
\Theta\vdash^{\Lambda}_{\mf D_1\circ_{y,x}\mf D_2}u:B \quad \text{is also derivable.}
\end{aligned}
$$
$$
\begin{aligned}
&\text{If} \quad \Theta\vdash^{\Lambda}_{\mf D_1}y:A \quad \text{and} \quad \Theta,x:A \; ; \; \Gamma'\vdash^{\Lambda}_{\mf D_2}\Delta'
\\
&\text{are derivable, then} \quad
\Theta \; ; \; \Gamma'\vdash^{\Lambda}_{\mf D_1\circ_{y,x}\mf D_2}\Delta' \quad \text{is derivable.}
\end{aligned}
$$

\item[(ii)] Let $A$ be a type at a linear mode.
$$
\begin{aligned}
&\text{If} \quad \Theta\ ;\ \Gamma\vdash^{\Lambda}_{\mf D_1}\Delta_L,y:A,\Delta_R
\quad\text{and}\quad
\Theta\ ;\ \Gamma'_L,x:A,\Gamma'_R\vdash^{\Lambda}_{\mf D_2}\Delta'
\\
&\text{are derivable, then} \quad
\Theta\ ;\ \Gamma'_L,\Gamma,\Gamma'_R
\vdash^{\Lambda}_{\mf D_1\circ_{y,x}\mf D_2}
\Delta_L,\Delta',\Delta_R \quad \text{is derivable.}
\end{aligned}
$$
\end{itemize}
\end{theorem}

\textit{Proof.}
The boundary hypotheses make every displayed cut well formed. We suppress the closing permutation transports according to the convention in the statement. The proof is by induction on the height of the derivation and the complexity of the formula $A$, denoted $(\mathrm{cx}(A),\ h_L+h_R)$, where $h_L,h_R$ are the heights of the two given derivations:

\begin{itemize}
\item \textbf{Cartesian cut formula, right sequent cartesian.} Since the cut formula is cartesian, the left premise must be cartesian; a sequent whose output stands at a cartesian mode is the cartesian sequent of the uniform display (\cref{prop:sorts-of-well-formed-descriptors}). Thus the left derivation decorated by the descriptor $\mf D_1$ is of the form $\Theta \vdash_{\mf D_1} y : A$. We use induction on the right derivation.

If the last rule of the right derivation is an instance of the variable rule, there are two cases. Either the variable introduced is the cut variable, or it is not. In the case it is not, let $x' \neq x$ and suppose the right derivation has the form:
$$
  \inferrule*
  {
  P\in\mathrm{Prop}_{\Lambda,q}
  \\
  r::\mf D_2\Rightarrow^{\Lambda}[\varnothing \mid \{x' \mapsto u\}]
  }
  {
  \Theta,\,x:A\vdash_{\mf D_2}^{\Lambda}u:P
  }
$$
Using 2-Subst with the identity descriptor on $\mf D_1$, we get the transformation $\1_{\mf D_1} \circ_{y,x} r :: \mf D_2[x := \mf D_1] \Rightarrow [\varnothing \mid \{x' \mapsto u\}][x := \mf D_1]$. We note that the cut lands at the unused cartesian input $x$ of the target, and the codomain collapses to $[\varnothing \mid \{x' \mapsto u\}]$. We reapply the variable rule to get:

$$
  \inferrule*
  {
  P\in\mathrm{Prop}_{\Lambda,q}
  \\
  r::\mf D_2[x := \mf D_1] \Rightarrow^{\Lambda}[\varnothing \mid \{x' \mapsto u\}] 
  }
  {
  \Theta\vdash_{\mf D_2[x := \mf D_1]}^{\Lambda}u:P
  }
$$
If it is the variable rule at $x$, then we are faced with the right derivation
$$
\inferrule*
{
  P\in\mathrm{Prop}_{\Lambda,q}
  \\
  r::\mf D_2\Rightarrow^{\Lambda}[\varnothing\mid\{x\mapsto u\}]
}
{
  \Theta,\,x:A\vdash_{\mf D_2}^{\Lambda}u:P
}
$$
Using 2-Subst with the identity descriptor on $\mf D_1$, we get the composition on the transformation
$$\mathbf 1_{\mf D_1}\circ_{y,x}r::\mf D_2[x:=\mf D_1]\Rightarrow[\varnothing\mid\{x\mapsto u\}][x:=\mf D_1]=\mf D_1[y:=u].$$
Now, by \cref{lem:units-for-composition}(ii), the left derivation renames its output to $u$, its index becoming $\mf D_1[y:=u]$. The displayed equality identifies this index with $\mf D_2[x:=\mf D_1]$, and \cref{prop:respect-for-structural-transformation} applied to the renamed sequent gives the final desired sequent $\Theta\vdash_{\mf D_2[x:=\mf D_1]}u:A$.

If the right derivation is $F_{\mathsf c}$L at a variable $z'\neq x$:
$$
\inferrule*
{
  \Theta,\Theta_{F'},x:A\vdash_{\mf D_2[z':=\mf D_{F'}]}^{\Lambda}u:B
}
{
  \Theta,\,z':\mathsf F',\,x:A\vdash_{\mf D_2}^{\Lambda}u:B
}
$$
We use the induction hypothesis on the premise, cutting it with the left derivation inverted at $z'$ by \cref{prop:left-invertibility-of-f} so that both sequents live over $\Theta,\Theta_{F'}$. We also use \cref{lem:c-substitution-commutes-with-composition} to get the equality of descriptors:
$$\bigl(\mf D_2[z':=\mf D_{F'}]\bigr)\bigl[x:=\mf D_1[z':=\mf D_{F'}]\bigr]=\bigl(\mf D_2[x:=\mf D_1]\bigr)[z':=\mf D_{F'}].$$
Using this new descriptor, we reapply the connective rule for the final derivation:

$$
\inferrule*
{
  \Theta,\Theta_{F'}\vdash_{\bigl(\mf D_2[x:=\mf D_1]\bigr)[z':=\mf D_{F'}]}^{\Lambda}u:B
}
{
  \Theta,\,z':\mathsf F',\vdash_{\mf D_2[x:=\mf D_1]}^{\Lambda}u:B
}
$$

If the right derivation is $F_{\mathsf c}$L at the cut variable $x$, then $A=F^{y_A}_{\mf D_{F_A}}(\Theta_{F_A};\cdot)$ and the derivation is:
$$
\inferrule*
{
  \Theta,\Theta_{F_A}
  \vdash_{\mf D_{F_A}\circ_{y_A,x}\mf D_2}^{\Lambda}
  u:B
}
{
  \Theta,\,x:A
  \vdash_{\mf D_2}^{\Lambda}
  u:B
}
$$

By induction on the left derivation, consider first the case of $F_{\mathsf c}$R with $r_L::\mf D_1\Rightarrow(\sigma_a)_a\circ\mf D_{F_A}$. We use the induction hypothesis to iteratively cut the premises together at the cut variables contained in $\Theta_{F_A}$. The resulting sequent is $\Theta\vdash_{\mf D_2[x:=(\sigma_a)_a\circ\mf D_{F_A}]}u:B$ by \cref{lem:c-substitution-commutes-with-composition}. Applying 2-Subst with the identity descriptor on $\mf D_2$ to the transformation premise of the left derivation, we get the composition $r_L\circ_{y,x}\mathbf 1_{\mf D_2}::\mf D_2[x:=\mf D_1]\Rightarrow\mf D_2[x:=(\sigma_a)_a\circ\mf D_{F_A}]$, and finally, applying respect for transformation via \cref{prop:respect-for-structural-transformation}, we conclude:

$$
	\inferrule*
	{
  	\Theta\vdash_{\mf D_2[x:=(\sigma_a)_a\circ\mf D_{F_A}]}u:B \\
		r_L\circ_{y,x}\mathbf 1_{\mf D_2}::\mf D_2[x:=\mf D_1]\Rightarrow\mf D_2[x:=(\sigma_a)_a\circ\mf D_{F_A}]
	}
	{
  	\Theta\vdash_{\mf D_2[x := \mf D_1]}u:B \\
	}
$$

The next left rule to consider is the left variable rule. Since the conclusion is a proposition while $A$ is a composite, the case is impossible.

Next, suppose the left derivation ends with $F_{\mathsf c}$L at some $z'':\mathsf F''\in\Theta$, $\mathsf F''=F^{y''}_{\mf D_{F''}}(\Theta_{F''};\cdot)$, writing $\Theta=\Theta',z'':\mathsf F''$. The rule consumes its principal: its premise lives over $\Theta',\Theta_{F''}$ while the right derivation lives over $\Theta',z'':\mathsf F''$; the two do not share a cartesian zone, and cutting them directly is not an instance of the induction hypothesis. The index it would produce, $\mf D_2[x:=\mf D_1[z'':=\mf D_{F''}]]$, retains the consumed $z''$ in its outer factor. We invert the right derivation at $z''$ first by \cref{prop:left-invertibility-of-f}, of no greater height, so that both sequents live over $\Theta',\Theta_{F''}$, and cut it against the rule's premise, of height one less, by the induction hypothesis. The resulting index is exactly the $F_{\mathsf c}$L premise index for the conclusion index $\mf D_2[x:=\mf D_1]$, by the first law of \cref{lem:c-substitution-commutes-with-composition}:
$$
\bigl(\mf D_2[z'':=\mf D_{F''}]\bigr)\bigl[x:=\mf D_1[z'':=\mf D_{F''}]\bigr]
=\bigl(\mf D_2[x:=\mf D_1]\bigr)[z'':=\mf D_{F''}],
$$
and re-applying the rule at $z''$ concludes. If the left derivation ends with $U_{\mathsf c}$L at $z''$ instead, the rule retains its principal and its sequent premise only adjoins the continuation hypothesis $x'':X$ to the zone. We weaken the right derivation by $x'':X$ (\cref{lem:cartesian-weakening-for-sequents}), cut it against the premise by the induction hypothesis, and re-apply the rule, its $2$-cell whiskered with $\mathbf 1_{\mf D_2}$ by 2-Subst, the second law of \cref{lem:c-substitution-commutes-with-composition} absorbing the continuation substitution, $x''$ being fresh for $\mf D_2$. In both branches the inverted or weakened right derivation is of no greater height and the left premise is of height one less, so the measure drops.
And that concludes the cases for the left rules.

If the right derivation is $F_{\mathsf c}$R, then $B=F^{y_B}_{\mf D_{F_B}}(\Theta_{F_B};\cdot)$ and the rule carries a $2$-cell and a family premise:
$$
\inferrule*
{
  r::\mf D_2\Rightarrow^{\Lambda}(\sigma_a)_a\circ\mf D_{F_B}
  \\
  \bigl(\Theta,x:A\vdash^{\Lambda}_{\sigma_a}A_a\bigr)_{a\in\theta_{F_B}}
}
{
  \Theta,\,x:A\vdash_{\mf D_2}^{\Lambda}u:B
}
$$
Using 2-Subst with the identity descriptor on $\mf D_1$, we get the composition on the transformation $\mathbf 1_{\mf D_1}\circ_{y,x}r::\mf D_2[x:=\mf D_1]\Rightarrow\bigl((\sigma_a)_a\circ\mf D_{F_B}\bigr)[x:=\mf D_1]=(\sigma_a[x:=\mf D_1])_a\circ\mf D_{F_B}$, and apply the induction hypothesis to cut the variable at all the premise sequents. The final family of sequents obtained has the shape $(\Theta \vdash_{\sigma_a[x := \mf D_1]}  A_a)_{a \in \theta_{F_B}}$. We can then reapply the $F_{\mathsf c}$  rule using the transformation to obtain the final required sequent.

If the right derivation is $U_{\mathsf c}$R, then $B=U^{\mf D_{U_B}}_{x_B}(\Theta_{U_B};\cdot\mid b:X)$ and the derivation is:
$$
\inferrule*
{
  \Theta,x:A,\Theta_{U_B}\vdash_{\mf D_2\circ_{u,x_B}\mf D_{U_B}}^{\Lambda}b:X
}
{
  \Theta,\,x:A\vdash_{\mf D_2}^{\Lambda}u:B
}
,$$
and its premise descriptor is $\mf D_2\circ_{u,x_B}\mf D_{U_B}$ with $\mf D_{U_B}$. We apply the induction hypothesis and \cref{lem:c-substitution-commutes-with-composition} to get the descriptor $(\mf D_2\circ_{u,x_B}\mf D_{U_B})[x:=\mf D_1]=\bigl(\mf D_2[x:=\mf D_1]\bigr)\circ_{u,x_B}\mf D_{U_B}$, and the rule reapplies.

If the right derivation is $U_{\mathsf c}$L at a variable $z'\neq x$, we proceed as in the $F_{\mathsf c}$L case at $z'\neq x$, the left derivation now weakened by the continuation hypothesis $x'':X$ in place of being inverted, taking all premises by the induction hypothesis and composing the $2$-cell with $\mathbf 1_{\mf D_1}$ by 2-Subst, and reapply the rule.

If the right derivation is $U_{\mathsf c}$L at the cut variable $x$, then $A=U^{\mf D_{U_A}}_{x'}(\Theta_{U_A};\cdot\mid b:X)$ and the rule carries a $2$-cell, a family and its last premise $\Theta,x:A,x'':X\vdash_{\mf D'}u:B$ retaining $x:A$:
$$
\inferrule*
{
  s::\mf D_2\Rightarrow^{\Lambda}\mf D'[x'':=\tau_A]
  \\
  \bigl(\Theta,x:A\vdash^{\Lambda}_{\sigma_a}A_a\bigr)_{a\in\theta_{U_A}}
  \\
  \Theta,x:A,\,x'':X\vdash^{\Lambda}_{\mf D'}u:B
}
{
  \Theta,\,x:A\vdash^{\Lambda}_{\mf D_2}u:B
}
$$
with $\tau_A=((\sigma_a)_a\circ\mf D_{U_A})[x':=x]$. We case on the last rule of the left derivation.
If the left derivation is $U_{\mathsf c}$R, with premise $\Theta,\Theta_{U_A}\vdash_{\mf D_1\circ_{y,x'}\mf D_{U_A}}b:X$, we first apply the induction hypothesis to every proper premise of the retained $U_{\mathsf c}$L rule. If $h_a$ is the height of the $a$-th family premise and $h'$ the height of the continuation premise, then $h_a,h'<h_R$, so the calls at
$$
(\mathrm{cx}(A),h_L+h_a),
\qquad
(\mathrm{cx}(A),h_L+h')
$$
are strictly smaller than the original measure. They give
$$
\bigl(\Theta\vdash^{\Lambda}_{\sigma_a[x:=\mf D_1]}A_a\bigr)_{a\in\theta_{U_A}},
\qquad
\Theta,x'':X\vdash^{\Lambda}_{\mf D'[x:=\mf D_1]}u:B.
$$
We then substitute the transformed family into the left premise by iterated cuts at the strictly smaller types of $\Theta_{U_A}$, obtaining $\Theta\vdash_{((\sigma_a[x:=\mf D_1])_a\circ\mf D_{U_A})[x':=\mf D_1]}b:X$, and cut this into $x'':X$ by the induction hypothesis at the strictly smaller type $X$. Put
$$\mf S_A:=(\sigma_a)_a\circ\mf D_{U_A},\qquad \mf S_A':=(\sigma_a[x:=\mf D_1])_a\circ\mf D_{U_A}=\mf S_A[x:=\mf D_1],$$
and let $\mf I_x=[\varnothing\mid\{x\mapsto v\}]$ for a fresh $v$. Then $\tau_A=\mf I_x\circ_{v,x'}\mf S_A$ by \cref{lem:identity-cut-realizes-input-contraction}.
Composing the right rule's $2$-cell with $\mathbf 1_{\mf D_1}$ by 2-Subst gives
$$
\mathbf 1_{\mf D_1}\circ_{y,x}s::
\mf D_2[x:=\mf D_1]
\Rightarrow
\bigl(\mf D'[x'':=\tau_A]\bigr)[x:=\mf D_1].
$$
The target is the descriptor just derived. Indeed, cartesian substitution distributes over the contraction cut, and \cref{lem:units-for-composition}(ii) gives
\begin{align*}
\tau_A[x:=\mf D_1]
&=\bigl(\mf I_x\circ_{v,x'}\mf S_A\bigr)[x:=\mf D_1]\\
&=\mf I_x[x:=\mf D_1]\circ_{v,x'}\mf S_A[x:=\mf D_1]\\
&=\mf D_1[y:=v]\circ_{v,x'}\mf S_A'
=\mf S_A'[x':=\mf D_1],\\
\bigl(\mf D'[x'':=\tau_A]\bigr)[x:=\mf D_1]
&=\mf D'[x:=\mf D_1]\bigl[x'':=\tau_A[x:=\mf D_1]\bigr]\\
&=\mf D'[x:=\mf D_1]\bigl[x'':=\mf S_A'[x':=\mf D_1]\bigr].
\end{align*}
The equalities include the canonical closing transports, so \cref{prop:respect-for-structural-transformation} concludes. The remaining left cases $F_{\mathsf c}$L and $U_{\mathsf c}$L are commuted as in the principal $F_{\mathsf c}$L case, the right derivation inverted at $z''$ or weakened by the continuation hypothesis, and a left variable rule is again impossible.

If the right derivation is $U_{\mathsf{shift}}$R, then $B$ is $U^{\mf D_U}_{x_s}(\Theta_U;\Gamma_U\mid\Delta_U)$ and its premise is the linear sequent:
$$
\inferrule*
{
  \Theta,x:A,\Theta_U\ ;\ \Gamma_U\vdash_{\mf D_2\circ_{u,x_s}\mf D_U}^{\Lambda}\Delta_U
}
{
  \Theta,\,x:A\vdash_{\mf D_2}^{\Lambda}u:B
}
$$
The induction hypothesis, the right sequent now linear and the premise of smaller height, gives $\Theta,\Theta_U;\Gamma_U\vdash_{(\mf D_2\circ_{u,x_s}\mf D_U)[x:=\mf D_1]}\Delta_U$, and $(\mf D_2\circ_{u,x_s}\mf D_U)[x:=\mf D_1]=\bigl(\mf D_2[x:=\mf D_1]\bigr)\circ_{u,x_s}\mf D_U$ by \cref{lem:c-substitution-commutes-with-composition}, so $U_{\mathsf{shift}}$R reapplies.

\item \textbf{Cartesian cut formula, right sequent linear.} If the cut formula is cartesian and the right sequent is linear, the standard procedure is to start the case analysis on the right derivation since the left being cartesian reduces the number of cases to treat. Suppose the right derivation is the linear sequent $\Theta,x:A\ ;\ \Gamma'\vdash_{\mf D_2}\Delta'$, its cut variable $x:A$ sitting in the cartesian zone that the whole derivation shares, and recall that the result descriptor is the cartesian substitution $\mf D_1\circ_{y,x}\mf D_2=\mf D_2[x:=\mf D_1]$. Since $x$ is cartesian, it is never the variable consumed by a linear rule, so the cut commutes past every linear rule and past every cartesian left rule at a variable $z'\neq x$, and only the cartesian left rules acting on $x$ itself stay principal. We treat the commuting rules first.

If the last rule is the linear variable rule $\mathsf v_{\mathsf l}$, then $\Gamma'$ is a single hypothesis $w:P$ and we are faced with:
$$
\inferrule*
{
  P\in\mathrm{Prop}_{\Lambda,q}
  \\
  r::\mf D_2\Rightarrow^{\Lambda}\mf [\varnothing \mid \{w \mapsto d\}]
}
{
  \Theta,x:A\ ;\ w:P\vdash_{\mf D_2}^{\Lambda}d:P
}
$$
Using 2-Subst with the identity descriptor on $\mf D_1$, we get the composition on the transformation $\mathbf 1_{\mf D_1}\circ_{y,x}r::\mf D_2[x:=\mf D_1]\Rightarrow[\varnothing \mid \{w \mapsto d\}]$. The redirection $(w\mapsto d)$ carries the consumed linear variable of the rule, and a cartesian substitution at $x$ leaves it untouched, so we can reapply the variable rule:
$$
\inferrule*
{
  P\in\mathrm{Prop}_{\Lambda,q}
  \\
  \mathbf 1_{\mf D_1}\circ_{y,x}r::\mf D_2[x:=\mf D_1]\Rightarrow^{\Lambda}[\varnothing \mid \{w \mapsto d\}]
  \\
}
{
  \Theta\ ;\ w:P\vdash_{\mf D_2[x:=\mf D_1]}^{\Lambda}d:P
}
$$

If the last rule is $F_{\mathsf l}$L at a linear connective $z:\mathsf F$, then we are looking at:
$$
\inferrule*
{
  \Theta,x:A,\Theta_F\ ;\ \Gamma'_L,\Gamma_F,\Gamma'_R\vdash_{\mf D_F\circ_{y_F,z}\mf D_2}^{\Lambda}\Delta'
}
{
  \Theta,x:A\ ;\ \Gamma'_L,\,z:\mathsf F,\,\Gamma'_R\vdash_{\mf D_2}^{\Lambda}\Delta'
}
$$
We use the induction hypothesis on the premise, cutting it with the left derivation, the right sequent still linear and the premise of smaller height. We also use \cref{lem:c-substitution-commutes-with-composition} to get the equality of descriptors:
$$(\mf D_F\circ_{y_F,z}\mf D_2)[x:=\mf D_1]=\mf D_F\circ_{y_F,z}(\mf D_2[x:=\mf D_1]),$$
the connective descriptor $\mf D_F$ untouched by the substitution, its boundary fresh for $x$, so that the cut reaches only the ambient factor $\mf D_2$. Using this new descriptor, we reapply the connective rule for the final derivation:
$$
\inferrule*
{
  \Theta,\Theta_F\ ;\ \Gamma'_L,\Gamma_F,\Gamma'_R\vdash_{\mf D_F\circ_{y_F,z}(\mf D_2[x:=\mf D_1])}^{\Lambda}\Delta'
}
{
  \Theta\ ;\ \Gamma'_L,\,z:\mathsf F,\,\Gamma'_R\vdash_{\mf D_2[x:=\mf D_1]}^{\Lambda}\Delta'
}
$$

The branching linear rules are where the cartesian zone does its work. If the last rule is $F_{\mathsf l}$R, then it carries a $2$-cell, a family premise and $n$ input components, every one of them over the full cartesian zone $\Theta,x:A$:
$$
\inferrule*
{
  r::\mf D_2\Rightarrow^{\Lambda}\bigl(\mf D_{\bar x_i}\bigr)_i\circ_{(\bar x_i,x_i)}\bigl((\sigma_a)_{a\in\theta_F}\circ_{(b_a,a)}\mf D_F\bigr)
  \\
  \bigl(\Theta,x:A\vdash^{\Lambda}_{\sigma_a}A_a\bigr)_{a\in\theta_F}
  \\
  \bigl(\Theta,x:A\ ;\ \Gamma^{(i)}\vdash_{\mf D_{\bar x_i}}^{\Lambda}\Delta_L^{(i)},\bar x_i:A_i,\Delta_R^{(i)}\bigr)_{1\le i\le n}
}
{
  \Theta,x:A\ ;\ \Gamma^{(1)},\ldots,\Gamma^{(n)}\vdash_{\mf D_2}^{\Lambda}\Delta_L,z:\mathsf F,\Delta_R
}
$$
The cut therefore reaches each premise. Using 2-Subst with the identity descriptor on $\mf D_1$, we get the composition on the transformation
$$\mathbf 1_{\mf D_1}\circ_{y,x}r::\mf D_2[x:=\mf D_1]\Rightarrow\Bigl(\bigl(\mf D_{\bar x_i}\bigr)_i\circ_{(\bar x_i,x_i)}\bigl((\sigma_a)_{a\in\theta_F}\circ_{(b_a,a)}\mf D_F\bigr)\Bigr)[x:=\mf D_1],$$
whose target distributes over every cut of the composite by \cref{lem:c-substitution-commutes-with-composition}, the connective descriptor $\mf D_F$ untouched because its boundary is fresh for $x$, to become
$$\bigl(\mf D_{\bar x_i}[x:=\mf D_1]\bigr)_i\circ_{(\bar x_i,x_i)}\bigl((\sigma_a[x:=\mf D_1])_{a\in\theta_F}\circ_{(b_a,a)}\mf D_F\bigr).$$
We then apply the induction hypothesis to cut the left derivation into the family premises, in the cartesian case, and into the $n$ input components, in the linear case, both at $h_R-1$. The family obtained has the shape $(\Theta\vdash_{\sigma_a[x:=\mf D_1]}A_a)_{a\in\theta_F}$ and the components carry the descriptors $\mf D_{\bar x_i}[x:=\mf D_1]$, exactly matching the target just computed. We can then reapply the $F_{\mathsf l}$R rule using the transformation to obtain the final required sequent $\Theta\ ;\ \Gamma^{(1)},\ldots,\Gamma^{(n)}\vdash_{\mf D_2[x:=\mf D_1]}\Delta_L,z:\mathsf F,\Delta_R$.

The remaining commuting rules are variations on these two shapes. For $U_{\mathsf l}$R, whose single premise carries the descriptor $\mf D_2\circ_{z,x_U}\mf D_U$, the induction hypothesis on that premise and \cref{lem:c-substitution-commutes-with-composition} give $(\mf D_2\circ_{z,x_U}\mf D_U)[x:=\mf D_1]=(\mf D_2[x:=\mf D_1])\circ_{z,x_U}\mf D_U$, the connective descriptor $\mf D_U$ untouched, and the rule reapplies, dually to $F_{\mathsf l}$L. The branching rules $U_{\mathsf l}$L and $U_{\mathsf{shift}}$L at a variable $z'\neq x$ are settled as $F_{\mathsf l}$R was, taking every component and family premise by the induction hypothesis at $h_R-1$, composing the $2$-cell with $\mathbf 1_{\mf D_1}$ by 2-Subst, and distributing the substitution over the composite by \cref{lem:c-substitution-commutes-with-composition}, the $m$ output components cut in as well. Finally the cartesian left rules at a variable $z'\neq x$, i.e. $F_{\mathsf c}$L-in-l and $U_{\mathsf c}$L-in-l at $z'$, commute past the cut by the same argument as their cartesian-sequent counterparts, taking all premises by the induction hypothesis and composing any $2$-cell with $\mathbf 1_{\mf D_1}$ by 2-Subst, the connective descriptor exposed at $z'$ having its boundary fresh for $x$.

Three cases are principal. If the last rule is $F_{\mathsf c}$L-in-l at $x$, then $A=F^{y_A}_{\mf D_{F_A}}(\Theta_{F_A};\cdot)$; we have:
$$
\inferrule*
{
  \Theta,\Theta_{F_A}\ ;\ \Gamma'\vdash_{\mf D_{F_A}\circ_{y_A,x}\mf D_2}^{\Lambda}\Delta'
}
{
  \Theta,\,x:A\ ;\ \Gamma'\vdash_{\mf D_2}^{\Lambda}\Delta'
}
$$
The construction used for $F_{\mathsf c}$L applies here. Suppose the left derivation ends with $F_{\mathsf c}$R, with premise transformation $r_L::\mf D_1\Rightarrow(\sigma_a)_a\circ\mf D_{F_A}$ and $\sigma$ being the family of descriptors of the other premises. We cut the premise $\Theta,\Theta_{F_A};\Gamma'\vdash_{\mf D_{F_A}\circ_{y_A,x}\mf D_2}\Delta'$ iteratively with the premises of $F_{\mathsf c}$L using the induction hypothesis to get the sequent $\Theta;\Gamma'\vdash_{\mf D_2[x:=(\sigma_a)_a\circ\mf D_{F_A}]}\Delta'$. Using 2-Subst with the identity descriptor on $\mf D_2$, we get the composition on the transformation $r_L\circ_{y,x}\mathbf 1_{\mf D_2}::\mf D_2[x:=\mf D_1]\Rightarrow\mf D_2[x:=(\sigma_a)_a\circ\mf D_{F_A}]$, and \cref{prop:respect-for-structural-transformation} concludes. If the left derivation ends with $F_{\mathsf c}$L or $U_{\mathsf c}$L, then the same commuting argument applies.

If the last rule is $U_{\mathsf c}$L-in-l at $x$, then $A=U^{\mf D_{U_A}}_{x'}(\Theta_{U_A};\cdot\mid b:X)$ and the rule carries a $2$-cell, a family and a linear premise:
$$
\inferrule*
{
  s::\mf D_2\Rightarrow^{\Lambda}\mf D'[x'':=\tau_A]
  \\
  \bigl(\Theta,x:A\vdash^{\Lambda}_{\sigma_a}A_a\bigr)_{a\in\theta_{U_A}}
  \\
  \Theta,x:A,x'':X\ ;\ \Gamma'\vdash^{\Lambda}_{\mf D'}\Delta'
}
{
  \Theta,\,x:A\ ;\ \Gamma'\vdash^{\Lambda}_{\mf D_2}\Delta'
}
$$
with $\tau_A=((\sigma_a)_a\circ\mf D_{U_A})[x':=x]$. As in the cartesian $U_{\mathsf c}$L case, we first apply the induction hypothesis at the same complexity $\mathrm{cx}(A)$ to every retained proper premise. If $h_a$ and $h'$ are the heights of a family premise and the linear continuation, respectively, then the calls have measures $(\mathrm{cx}(A),h_L+h_a)$ and $(\mathrm{cx}(A),h_L+h')$, both strictly smaller because $h_a,h'<h_R$. They give
$$
\bigl(\Theta\vdash^{\Lambda}_{\sigma_a[x:=\mf D_1]}A_a\bigr)_{a\in\theta_{U_A}},
\qquad
\Theta,x'':X\ ;\ \Gamma'
\vdash^{\Lambda}_{\mf D'[x:=\mf D_1]}\Delta'.
$$
If the left derivation ends with $U_{\mathsf c}$R, we use the transformed family in its premise by cuts at the strictly smaller types of $\Theta_{U_A}$ and then cut the resulting $X$-formula into the transformed linear continuation at the strictly smaller type $X$. Respect for structural transformation concludes as in the cartesian case. The remaining left rules use the same commuting argument.

If the last rule is $U_{\mathsf{shift}}$L at $x$, then $A=U^{\mf D_{U}}_{x_s}(\Theta_U;\Gamma_U\mid\Delta_U)$ with $x_s$ cartesian. This is the promotion reduction of Benton. The rule carries a $2$-cell, a family and the $n+m$ components:
$$
\inferrule*
{
  r::\mf D_2\Rightarrow^{\Lambda}\mf C
  \\
  \bigl(\Theta,x:A\vdash^{\Lambda}_{\sigma_a}A_a\bigr)_{a\in\theta_U}
  \\
  \bigl(\Theta,x:A\ ;\ \Gamma^{(i)}\vdash_{\mf D_{\bar x_i}}^{\Lambda}\Delta_L^{(i)},\bar x_i:A_i,\Delta_R^{(i)}\bigr)_{1\le i\le n}
  \\
  \bigl(\Theta,x:A\ ;\ \Gamma^{(n+j)},\bar y_j:B_j\vdash_{\mf D_{\bar y_j}}^{\Lambda}\Delta^{(n+j)}\bigr)_{1\le j\le m}
}
{
  \Theta,\,x:A\ ;\ \Gamma'\vdash^{\Lambda}_{\mf D_2}\Delta'
}
$$
with $\mf C$ the composite of the $U_{\mathsf{shift}}$L schema. We case on the left derivation of $\Theta\vdash_{\mf D_1}y:A$. If it ends with $U_{\mathsf{shift}}$R, its premise is $\Theta,\Theta_U;\Gamma_U\vdash_{\mf D_U[x_s:=\mf D_1]}\Delta_U$. Every family and component premise of the retained $U_{\mathsf{shift}}$L rule has height strictly less than $h_R$. Applying the induction hypothesis at the same complexity $\mathrm{cx}(A)$ to each of them gives
$$
\bigl(\Theta\vdash^{\Lambda}_{\sigma'_a}A_a\bigr)_{a\in\theta_U},
\qquad
\sigma'_a:=\sigma_a[x:=\mf D_1],
$$
$$
\bigl(
  \Theta\ ;\ \Gamma^{(i)}
  \vdash^{\Lambda}_{\mf D_{\bar x_i}[x:=\mf D_1]}
  \Delta_L^{(i)},\bar x_i:A_i,\Delta_R^{(i)}
\bigr)_{1\leq i\leq n},
$$
$$
\bigl(
  \Theta\ ;\ \Gamma^{(n+j)},\bar y_j:B_j
  \vdash^{\Lambda}_{\mf D_{\bar y_j}[x:=\mf D_1]}
  \Delta^{(n+j)}
\bigr)_{1\leq j\leq m}.
$$
Each call has measure $(\mathrm{cx}(A),h_L+h_P)$ for the height $h_P<h_R$ of the corresponding proper premise, and is therefore strictly smaller than the original measure. We substitute $\sigma'$ into the left premise by iterated applications of the induction hypothesis at the strictly smaller types of $\Theta_U$, obtaining $\Theta;\Gamma_U\vdash_{((\sigma'_a)_a\circ\mf D_U)[x_s:=\mf D_1]}\Delta_U$. Now we cut, in the linear case of the induction hypothesis at the strictly smaller complexities $\mathrm{cx}(A_i),\mathrm{cx}(B_j)$, each input component's output $\bar x_i:A_i$ into the input $x_i:A_i$ of this sequent, then each of its outputs $y_j:B_j$ into the corresponding output component's input $\bar y_j:B_j$. The resulting sequent is $\Theta;\Gamma^{(1)},\ldots,\Gamma^{(n+m)}\vdash_{\mf C'}\Delta$ with, after regrouping the iterated cuts by \cref{lem:localization,lem:associativity},
\begin{align*}
\mf C'&=\Bigl(\bigl(\mf D_{\bar x_i}[x:=\mf D_1]\bigr)_i\circ\bigl((\sigma'_a)_a\circ\mf D_U\bigr)[x_s:=\mf D_1]\Bigr)\circ\bigl(\mf D_{\bar y_j}[x:=\mf D_1]\bigr)_j\\
&=\biggl(\Bigl(\bigl(\mf D_{\bar x_i}\bigr)_i\circ\bigl((\sigma_a)_a\circ\mf D_U\bigr)[x_s:=x]\Bigr)\circ\bigl(\mf D_{\bar y_j}\bigr)_j\biggr)[x:=\mf D_1],
\end{align*}
the second equality by \cref{lem:c-substitution-commutes-with-composition} applied to every cut of the bracketed composite, whose gate factor transforms as $((\sigma_a)_a\circ\mf D_U)[x_s:=x][x:=\mf D_1]=((\sigma'_a)_a\circ\mf D_U)[x_s:=\mf D_1]$ by the same lemma and \cref{lem:units-for-composition}. The bracketed composite is exactly the target of the right rule's $2$-cell. Composing that $2$-cell with $\mathbf 1_{\mf D_1}$ by 2-Subst gives $\mf D_2[x:=\mf D_1]\Rightarrow\mf C'$, and \cref{prop:respect-for-structural-transformation} concludes. If the left derivation ends with $F_{\mathsf c}$L or $U_{\mathsf c}$L instead, it is commuted as in the principal $F_{\mathsf c}$L case of the cartesian block, the right derivation inverted at $z''$ or weakened by the continuation hypothesis, and a left variable rule is impossible since $A$ is a composite.

\item \textbf{Linear cut formula.} Both premises are now linear sequents, and we case on the last rules of the two derivations. Notice that the conclusion of $\mathsf v_{\mathsf l}$ contains only one formula, so the cut formula is principal in every variable rule. We organize the cases by considering whether the cut formula is principal in the derivations.

\begin{itemize}
\item \textit{The cut formula is not principal on the left.} In this case, the cut formula is present in the premise(s) of the last left derivation, so we can reapply the rule after cutting its premises.

If the last rule is $F_{\mathsf l}$L with a linear input $z:\mathsf F'$, then $y:A$ stays among the outputs of its single premise:
$$
\inferrule*
{
  \Theta,\Theta_{F'}\ ;\ \Gamma_L,\Gamma_{F'},\Gamma_R\vdash_{\mf D_{F'}\circ_{y_{F'},z}\mf D_1}^{\Lambda}\Delta_L,y:A,\Delta_R
}
{
  \Theta\ ;\ \Gamma_L,\,z:\mathsf F',\,\Gamma_R\vdash_{\mf D_1}^{\Lambda}\Delta_L,y:A,\Delta_R
}
$$
The induction hypothesis cuts this premise against the right derivation, and \cref{lem:associativity} gives:
$$(\mf D_{F'}\circ_{y_{F'},z}\mf D_1)\circ_{y,x}\mf D_2=\mf D_{F'}\circ_{y_{F'},z}(\mf D_1\circ_{y,x}\mf D_2),$$
so that $F_{\mathsf l}$L reapplies with conclusion descriptor $\mf D_1\circ_{y,x}\mf D_2$.

If the last rule is $U_{\mathsf l}$R at an output $w\neq y$, its single premise has the descriptor $\mf D_1\circ_{w,x_U}\mf D_U$ and carries $y:A$ still. The same argument applies with \cref{lem:right-interchange-law} in place of \cref{lem:associativity}, the connective leaving the output $w$ and the cut the distinct output $y$:
$$(\mf D_1\circ_{w,x_U}\mf D_U)\circ_{y,x}\mf D_2=(\mf D_1\circ_{y,x}\mf D_2)\circ_{w,x_U}\mf D_U,$$
and $U_{\mathsf l}$R reapplies.

If the last rule is among $F_{\mathsf l}$R, $U_{\mathsf l}$L or $U_{\mathsf{shift}}$L, then the linear contexts split and the cut output $y:A$ sits in exactly one component indexed by, say, $k$. For the sake of exposition, we treat the case $F_{\mathsf l}$R as the others are similar:
$$
\inferrule*
{
  r::\mf D_1\Rightarrow^{\Lambda}\bigl(\mf D_{\bar x_i}\bigr)_i\circ_{(\bar x_i,x_i)}\bigl((\sigma_a)_{a\in\theta_{F'}}\circ_{(b_a,a)}\mf D_{F'}\bigr)
  \\
  \bigl(\Theta\vdash^{\Lambda}_{\sigma_a}A_a\bigr)_{a\in\theta_{F'}}
  \\
  \bigl(\Theta\ ;\ \Gamma^{(i)}\vdash_{\mf D_{\bar x_i}}^{\Lambda}\Delta_L^{(i)},\bar x_i:A_i,\Delta_R^{(i)}\bigr)_{1\le i\le n}
}
{
  \Theta\ ;\ \Gamma \vdash_{\mf D_1}^{\Lambda}\Delta_L,w:\mathsf F',\Delta_R
}
$$
with $y:A$ in the component carrying $\Delta_L^{(k)},\Delta_R^{(k)}$. The induction hypothesis cuts that component against the right derivation, and using 2-Subst with the identity descriptor on $\mf D_2$, we get the composition on the transformation $r\circ_{y,x}\mathbf 1_{\mf D_2}$, which \cref{lem:localization}(ii) localizes to component $k$, so that the rule reapplies with the $k$-th component now carrying $\Delta'$ in place of $y:A$, the components remaining pairwise disjoint.

If the last rule is $F_{\mathsf c}$L-in-l or $U_{\mathsf c}$L-in-l at a cartesian variable $z'\in\Theta$, the shared cartesian zone must be matched on the right before the cut. Taking $F_{\mathsf c}$L-in-l, we expose $\Theta_{F'}$, and its premise is
$$\Theta,\Theta_{F'}\ ;\ \Gamma\vdash_{\mf D_{F'}\circ_{y',z'}\mf D_1}^{\Lambda}\Delta_L,y:A,\Delta_R ;$$
we use invertibility of the connective on the right derivation along the same $z'$ by \cref{prop:left-invertibility-of-f} giving the sequent $\Theta,\Theta_{F'}\ ;\ \Gamma'_L,x:A,\Gamma'_R\vdash_{\mf D_{F'}\circ_{y',z'}\mf D_2}\Delta'$ so that both sides carry $\Theta,\Theta_{F'}$. Using the induction hypothesis, we cut the two premises together, and \cref{lem:c-substitution-commutes-with-composition} pulls the substitution out of the cut:
$$(\mf D_{F'}\circ_{y',z'}\mf D_1)\circ_{y,x}(\mf D_{F'}\circ_{y',z'}\mf D_2)=\mf D_{F'}\circ_{y',z'}(\mf D_1\circ_{y,x}\mf D_2).$$
A reapplication of $F_{\mathsf c}$L-in-l grants the result. The rule $U_{\mathsf c}$L-in-l is the same with the right derivation weakened by the fresh hypothesis $x'':X$ in place of being inverted. Write $s_L::\mf D_1\Rightarrow\mf E[x'':=\tau]$ for the transformation premise of the left rule and $\widetilde{\mf D}_2$ for the index of the weakened right derivation, so that $\widetilde{\mf D}_2[x'':=\tau]=\mf D_2$ because the new input is unused. The induction hypothesis gives the continuation index $\mf E\circ_{y,x}\widetilde{\mf D}_2$, while 2-Subst gives the correctly typed whiskering
$$s_L\circ_{y,x}\mathbf 1_{\mf D_2}::\mf D_1\circ_{y,x}\mf D_2\Rightarrow\mf E[x'':=\tau]\circ_{y,x}\mf D_2.$$
These endpoints match for the reapplication because
$$
\bigl(\mf E\circ_{y,x}\widetilde{\mf D}_2\bigr)[x'':=\tau]
=\mf E[x'':=\tau]\circ_{y,x}\widetilde{\mf D}_2[x'':=\tau]
=\mf E[x'':=\tau]\circ_{y,x}\mf D_2,
$$
the first equality being cartesian substitution functoriality and the second the vacuous cut at the weakened input.

\item \textit{The cut formula is principal on the left but not on the right.} These are the mirror images, with the roles of the two derivations exchanged: now the right derivation's last rule is reapplied after cutting its premises, the cut input $x:A$ lying in a premise since it is not principal there. We show the branching representative, the right derivation ending with $F_{\mathsf l}$R at a side output $w:\mathsf F'$ and $x:A$ in the input slice of exactly one component $k$:
$$
\inferrule*
{
  r_R::\mf D_2\Rightarrow^{\Lambda}\bigl(\mf D_{\bar x_i}\bigr)_i\circ_{(\bar x_i,x_i)}\bigl((\sigma_a)_{a\in\theta_{F'}}\circ_{(b_a,a)}\mf D_{F'}\bigr)
  \\
  \bigl(\Theta\vdash^{\Lambda}_{\sigma_a}A_a\bigr)_{a\in\theta_{F'}}
  \\
  \bigl(\Theta\ ;\ \Gamma'^{(i)}\vdash_{\mf D_{\bar x_i}}^{\Lambda}\Delta_L'^{(i)},\bar x_i:A_i,\Delta_R'^{(i)}\bigr)_{1\le i\le n}
}
{
  \Theta\ ;\ \Gamma'^{(1)},\ldots,\Gamma'^{(n)}\vdash_{\mf D_2}^{\Lambda}\Delta_L',w:\mathsf F',\Delta_R'
}
$$
with $x:A$ in the input zone $\Gamma'^{(k)}$ of component $k$. Using the induction hypothesis and using 2-Subst with the identity descriptor on $\mf D_1$, we get the composition on the transformation $\mathbf 1_{\mf D_1}\circ_{y,x}r_R$, which \cref{lem:localization}(i) localizes to component $k$, so that $F_{\mathsf l}$R reapplies with the $k$-th component now carrying $\Gamma$ in place of $x:A$. The remaining right rules are the duals of the not-principal-on-the-left cases: $U_{\mathsf l}$L and $U_{\mathsf{shift}}$L localize $x:A$ to one input component by \cref{lem:localization}(i) as well. $F_{\mathsf l}$L at a variable $z'\neq x$ keeps $x$ in its single premise and regroups by \cref{lem:left-interchange-law}, the two inputs $z',x$ distinct. $U_{\mathsf l}$R keeps $x$ in its single premise and regroups by \cref{lem:associativity}, and a cartesian left rule at $z'\in\Theta$ passes by inverting or weakening the left derivation along $z'$.

\item \textit{The cut formula is principal on both sides.} There are three shapes. First, both derivations end with $\mathsf v_{\mathsf l}$:
$$
\inferrule*
{
  P\in\mathrm{Prop}_{\Lambda,q} \\ r_L::\mf D_1\Rightarrow^{\Lambda}[\varnothing\mid\{a\mapsto y\}]
}
{
  \Theta\ ;\ a:P\vdash_{\mf D_1}^{\Lambda}y:P
}
$$
$$
\inferrule*
{
  P\in\mathrm{Prop}_{\Lambda,q} \\ r_R::\mf D_2\Rightarrow^{\Lambda}[\varnothing\mid\{x\mapsto d\}]
}
{
  \Theta\ ;\ x:P\vdash_{\mf D_2}^{\Lambda}d:P
}
$$
By 2-Subst, $r_L\circ_{y,x}r_R::\mf D_1\circ_{y,x}\mf D_2\Rightarrow[\varnothing\mid\{a\mapsto y\}]\circ_{y,x}[\varnothing\mid\{x\mapsto d\}]=[\varnothing\mid\{a\mapsto d\}]$, the unit composition of \cref{lem:units-for-composition}(i), and the variable rule concludes $\Theta;a:P\vdash_{\mf D_1\circ_{y,x}\mf D_2}d:P$.

Second, $A=F^{y'}_{\mf D_F}(\Theta_F;\Gamma_F)$ and the left derivation ends with $F_{\mathsf l}$R at $y$:
$$
\inferrule*
{
  r_L::\mf D_1\Rightarrow^{\Lambda}\bigl(\mf D_{\bar x_i}\bigr)_i\circ_{(\bar x_i,x_i)}\bigl((\sigma_a)_{a\in\theta_F}\circ_{(b_a,a)}\mf D_F\bigr)
  \\
  \bigl(\Theta\vdash^{\Lambda}_{\sigma_a}A_a\bigr)_{a\in\theta_F}
  \\
  \bigl(\Theta\ ;\ \Gamma^{(i)}\vdash_{\mf D_{\bar x_i}}^{\Lambda}\Delta_L^{(i)},\bar x_i:A_i,\Delta_R^{(i)}\bigr)_{1\le i\le n}
}
{
  \Theta\ ;\ \Gamma^{(1)},\ldots,\Gamma^{(n)}\vdash_{\mf D_1}^{\Lambda}\Delta_L,y:\mathsf F,\Delta_R
}
$$
We invert the right derivation by \cref{prop:left-invertibility-of-f}:
$$\Theta,\Theta_F;\Gamma'_L,\Gamma_F,\Gamma'_R\vdash_{\mf D_F\circ_{y',x}\mf D_2}\Delta'.$$
We substitute the left rule's $\sigma$ by iterated applications of the cartesian case at the strictly smaller types of $\Theta_F$, turning the composition into $\bigl((\sigma_a)_a\circ\mf D_F\bigr)\circ_{y',x}\mf D_2$. Then we cut the left rule's $n$ components into the exposed inputs $x_i:A_i$ by the induction hypothesis at the strictly smaller complexities $\mathrm{cx}(A_i)$, and regroup by \cref{lem:localization}(i) and \cref{lem:associativity} into $\bigl((\mf D_{\bar x_i})_i\circ((\sigma_a)_a\circ\mf D_F)\bigr)\circ_{y,x}\mf D_2$, the bound label $y'$ $\alpha$-converted to $y$. Using 2-Subst with the identity descriptor on $\mf D_2$, we get the composition on the transformation $r_L\circ_{y,x}\mathbf 1_{\mf D_2}$, and \cref{prop:respect-for-structural-transformation} concludes.

Third, $A=U^{\mf D_U}_{x_U}(\Theta_U;\Gamma_U\mid\Delta_U)$ with $x_U$ linear; the left derivation ends with $U_{\mathsf l}$R at $y$ and the right with $U_{\mathsf l}$L at $x$. The left rule introduces $y:\mathsf U$:
$$
\inferrule*
{
  \Theta,\Theta_U\ ;\ \Gamma_U^L,\Gamma,\Gamma_U^R\vdash_{\mf D_1\circ_{y,x_U}\mf D_U}^{\Lambda}\Delta_L,\Delta_U,\Delta_R
}
{
  \Theta\ ;\ \Gamma\vdash_{\mf D_1}^{\Lambda}\Delta_L,y:\mathsf U,\Delta_R
}
$$
The $U_{\mathsf l}$R/$U_{\mathsf l}$L shape is dual to the $F_{\mathsf l}$R/$F_{\mathsf l}$L shape and reduces through the left premise. Put
$$
\mf S_U:=(\sigma_a)_a\circ\mf D_U,
\qquad
\mf C:=\Bigl((\mf D_{\bar x_i})_i\circ\mf S_U[x_U:=x]\Bigr)\circ(\mf D_{\bar y_j})_j,
$$
so that the right rule has $r_R::\mf D_2\Rightarrow\mf C$. We substitute the right rule's $\sigma$ into the left premise by iterated applications of the cartesian case, obtaining the index $\mf D_1\circ_{y,x_U}\mf S_U$. We then cut the right rule's $n$ input components into the inputs $x_i:A_i$ and the resulting outputs $y_j:B_j$ into its $m$ output components by the induction hypothesis at strictly smaller complexities. To identify the resulting endpoint, let $\mf I_x=[\varnothing\mid\{x\mapsto v\}]$ for a fresh $v$. The two unit laws and associativity from \cref{lem:units-for-composition,lem:associativity} compute the central factor as
\begin{align*}
\mf D_1\circ_{y,x}\bigl(\mf S_U[x_U:=x]\bigr)
&=\mf D_1\circ_{y,x}\bigl(\mf I_x\circ_{v,x_U}\mf S_U\bigr)\\
&=\bigl(\mf D_1\circ_{y,x}\mf I_x\bigr)\circ_{v,x_U}\mf S_U\\
&=\mf D_1[y:=v]\circ_{v,x_U}\mf S_U
=\mf D_1\circ_{y,x_U}\mf S_U.
\end{align*}
The last equality only renames the internal cut wire. Consequently, \cref{lem:localization,lem:associativity}, with the canonical closing transports, give
$$
\mf D_1\circ_{y,x}\mf C
=\Bigl((\mf D_{\bar x_i})_i\circ(\mf D_1\circ_{y,x_U}\mf S_U)\Bigr)\circ(\mf D_{\bar y_j})_j,
$$
which is exactly the descriptor obtained from the component cuts. Finally, 2-Subst gives $\mathbf 1_{\mf D_1}\circ_{y,x}r_R::\mf D_1\circ_{y,x}\mf D_2\Rightarrow\mf D_1\circ_{y,x}\mf C$, and \cref{prop:respect-for-structural-transformation} concludes.
\end{itemize}
\end{itemize}
In every case where the cut formula is not principal on both sides, one height decreases at fixed complexity. The principal retained-$U$ cases first apply the induction hypothesis to proper rule premises at the same complexity and lower height; their remaining cuts, like the other principal composite cases, are at strict subformulas and therefore lower the complexity. The proposition case terminates outright. This exhausts all cases and completes the simultaneous induction. \hfill $\square$

\textit{Remark:} Notice that no duplication occurs anywhere in the proof. Since the linear contexts split, the cut formula always lies in exactly one component. No derivation is ever copied, no descriptor is ever distributed over another, and the only rearrangements are the strict coherence laws in \cref{lem:associativity,lem:left-interchange-law,lem:right-interchange-law}. The two events which refuted the one-zone cut theorem, a junk side output at a variable leaf and a discarded producer orphaning its side outputs, are unstatable here. The zones of the variable rule are exact, and a producer is discarded only when it is pure.

\subsubsection{Harmony for Gates}\label{subsec:harmony-for-gates}

We extend now the cut theorem to gated signatures. Because of the contraction gate, \cref{thm:multicut} requires a stronger induction hypothesis. Indeed, in the presence of contraction, cutting a contracted formula leads to applying the induction hypothesis on premises which contain the formula duplicated, but neither the formula complexity nor the height of the derivation decreases. We therefore prove cut admissible in its \emph{multicut} form, one sequent with several designated outputs against one sequent with several designated inputs. We use an induction on a rank that measures the ancestry of the designated variables instead of the heights.

\begin{lemma}[Gate Operations Commute with Composition]\label{lem:gate-operations-commute}
Let the displayed operations be defined as in \cref{def:gate-descriptor-generation}, with fresh gate variables, and let $G$ range over the four operations $(-)\cup w_l(x)$, $(-)\cup w^\perp_l(b)$, $c_l(x_0;x,x')\cdot(-)$ and $(-)\cdot c^\perp_l(b,b';b_0)$.
\begin{itemize}
  \item[(i)] (\emph{Cuts at disjoint variables.}) If the acted variables of $G$ lie in the factor $\mf E$ and are distinct from the cut variable, then, at a variable of either mode,
  $$G(\mf E)\circ_{y,z}\mf F=G(\mf E\circ_{y,z}\mf F),\qquad \mf F\circ_{y,z}G(\mf E)=G(\mf F\circ_{y,z}\mf E).$$
  In particular, the gate operations pass c-substitution, $G(\mf E)[a:=\sigma]=G(\mf E[a:=\sigma])$ for a cartesian input $a$ and a c-descriptor $\sigma$.
  \item[(ii)] (\emph{Cuts at a weakening variable.}) These equalities hold:
  $$\mf G\circ_{y,x}\bigl(\mf E\cup w_l(x)\bigr)=\mf E\cup(\mf G\!\cdot\! w_l),\qquad
  \bigl(\mf E\cup w^\perp_l(b)\bigr)\circ_{b,x}\mf F=\mf E\cup(w^\perp_l\!\cdot\!\mf F),$$
  where $\mf G\cdot w_l$ places the source of the redirection to $y$ in $\mf G$ in the $w_l$-node, and $w^\perp_l\cdot\mf F$ places the $w^\perp_l$-node's output in the consumed input position of $\mf F$.
  \item[(iii)] (\emph{Disjoint gate operations.}) $G(G'(\mf E))=G'(G(\mf E))$ whenever the acted variables of $G$ and $G'$ are disjoint.
  \item[(iv)] (\emph{Structural operations.}) The gate operations commute with permutation transport, with cartesian input weakening, with c-substitution as in (i), and with boundary renaming at variables distinct from the gate variables.
\end{itemize}
\end{lemma}

\textit{Proof.}

All the equalities are equalities of descriptor classes, up to the implicit closing context permutations. We choose representatives of the factors with pairwise disjoint internal variables, and choose the internal variables introduced by the gates and the fresh families used by composition jointly fresh. By the well-definedness of composition and of the gate operations, it is enough to compare the resulting raw representatives up to an internal renaming.

We start with (i). Suppose first that the factor carrying the gate is on the left of the cut. Since a gate acts on an $\ell$-descriptor, this cut is necessarily linear. Let
$$
(\mc S_E\mid\mc T_E^0\cup\{s\mapsto y\})\in\mf E,
\qquad
(\mc S_F\mid\mc T_F)\in\mf F.
$$
The cut input $z$ has exactly one occurrence by the boundary usage property, and \cref{cor:linear-cut} computes the cut by removing $s\mapsto y$ and substituting $s$ at that occurrence. Gate-weak and Gate-coweak only adjoin their displayed components, which are disjoint from this composition. Gate-contr performs the two source substitutions at its consumed inputs and then adjoins its contraction node. These substitutions commute positionwise with the composition. If the cut source $s$ is itself one of the consumed inputs, applying Gate-contr before the cut replaces $s$ by the corresponding fresh target and inserts that target at the $z$-position, while applying it after the cut replaces the inserted occurrence of $s$ by the same target. Gate-cocontr replaces the two redirections at its acted outputs, which are distinct from $y$, and hence commutes with the removal of $s\mapsto y$. Thus $G(\mf E)\circ_{y,z}\mf F=G(\mf E\circ_{y,z}\mf F)$.

Suppose now that the factor carrying the gate is on the right. At a linear cut, the same one-occurrence cut computation applies. The only additional point is Gate-cocontr when the occurrence of $z$ is the source of one of the two redirections being replaced. Applying the gate first moves this occurrence to the corresponding source position of the $c^\perp_l$-node and the cut then replaces it by $s$; cutting first replaces the source of the redirection by $s$, which Gate-cocontr then places in that same source position. At a cartesian cut, the left factor $\mf F$ is a c-descriptor and every gate variable is linear. None of the four gate operations therefore creates, removes or changes an occurrence of the cartesian cut input $z$. We use the same fresh copy of $\mf F$ at each occurrence of $z$ on both sides of the equation. The gate component and all the copies then coincide, up to the names of their fresh internal variables. This proves $\mf F\circ_{y,z}G(\mf E)=G(\mf F\circ_{y,z}\mf E)$ and, taking $\mf F=\sigma$, proves the c-substitution equation as well.

For (ii), choose disjoint representatives
$$
(\mc S_G\mid\mc T_G^0\cup\{s\mapsto y\})\in\mf G,
\qquad
(\mc S_E\mid\mc T_E)\in\mf E.
$$
The variable $x$ occurs exactly once in $\mf E\cup w_l(x)$, as the source of the adjoined weakening node. The linear cut formula gives
$$
\mf G\circ_{y,x}\bigl(\mf E\cup w_l(x)\bigr)
=
\bigl[\,\mc S_G\cup\mc S_E\cup\{(\ ;s)\xmapsto{w_l}\cdot\}
\mid\mc T_G^0\cup\mc T_E\,\bigr].
$$
This is precisely $\mf E\cup(\mf G\!\cdot\!w_l)$: the redirection to $y$ has been removed and its source $s$ has been placed in the weakening node. Dually, for a representative $(\mc S_F\mid\mc T_F)\in\mf F$ and a fresh internal variable $v$, the redirection $v\mapsto b$ of the coweakening component is removed by the cut and
$$
\bigl(\mf E\cup w^\perp_l(b)\bigr)\circ_{b,x}\mf F
=
\bigl[\,\mc S_E\cup\{(\ ;)\xmapsto{w^\perp_l}v\}\cup\mc S_F[x:=v]
\mid\mc T_E\cup\mc T_F[x:=v] \,\bigr].
$$
This is $\mf E\cup(w^\perp_l\!\cdot\!\mf F)$, since the output $v$ of the $w^\perp_l$-node has been placed at the unique occurrence of the consumed input $x$ of $\mf F$.

For (iii), choose corresponding fresh internal variables for the two possible orders. Weakening and coweakening adjoin independent components, so their disjoint instances commute by commutativity of union. Two contractions perform source substitutions at disjoint variables, which commute positionwise by \cref{lem:occurrence-substitutions-commute}, and then adjoin their two independent nodes. Two cocontractions remove disjoint pairs of redirections and adjoin independent nodes. In the only mixed case requiring attention, a source of a redirection acted on by Gate-cocontr may be an input acted on by Gate-contr. Contracting first substitutes that source and the cocontraction then uses its substituted value; cocontracting first places the old source in the new $c^\perp_l$-node and the contraction performs the same substitution there. Thus the resulting assignment and redirection sets agree in either order. Matching the fresh internal variables of the two gate instances gives the required internal renaming, and hence $G(G'(\mf E))=G'(G(\mf E))$.

We finally note that the operations of permutation transport and input weakening only affect the context variables. They leave the assignment and redirection sets unaffected. The gate operations do not depend on the position of the variables in the context and thus commute with both operations. The renaming of the variables in the context not present in the gates acts independently of the gate operations; therefore, it commutes as well.

\hfill $\square$

\begin{lemma}[The Metatheory Extends to Gated Signatures]\label{lem:gated-metatheory-extension}
Let $\Lambda$ declare gates as in (\cref{def:gate-coherence-laws}). Then \cref{lem:cartesian-weakening-for-sequents,lem:cartesian-contraction-for-sequents} hold for $\Lambda$, and \cref{prop:left-invertibility-of-f} and \cref{prop:right-invertibility-of-u} hold for $\Lambda$ whenever no gated rule of the derivation acts at the inverted variable itself. Every appeal to these results in the present subsection is of this form. The operations act at cartesian variables and the gates at linear ones.
\end{lemma}

\textit{Proof.}
We prove the extensions in three stages: first cartesian weakening, then the two invertibility results, and finally cartesian contraction.
The cases in which the last rule is one of the original rules are exactly the cases already treated in the proofs of \cref{lem:cartesian-weakening-for-sequents,lem:cartesian-contraction-for-sequents,prop:left-invertibility-of-f,prop:right-invertibility-of-u}. It remains in each induction to consider a derivation ending with one of the four gated rules of \cref{def:gated-structural-rules}. We write such a rule uniformly as
$$
\inferrule*
{
  \Theta\ ;\ \Gamma_G\vdash^{\Lambda}_{\mf E}\Delta_G
  \\
  r::\mf D\Rightarrow^{\Lambda}G(\mf E)
}
{
  \Theta\ ;\ \Gamma\vdash^{\Lambda}_{\mf D}\Delta
},
$$
where $G$ is one of the four gate operations and $(\Gamma_G,\Delta_G)$ and $(\Gamma,\Delta)$ are, respectively, the premise and conclusion boundaries of that operation. Thus, for example, Gate-weak removes its fresh input in the premise, Gate-contr replaces its conclusion input by the two inputs which it contracts, and the two output gates behave dually. The argument is uniform in $G$.

We start with cartesian weakening. Let $a:A$ be fresh at a cartesian mode $p$. Applying the induction hypothesis to the sequent premise of the displayed rule gives
$$
\Theta,a:A\ ;\ \Gamma_G
\vdash^{\Lambda}_{\mathrm{wk}^{a:p}_{\mathrm{in}}(\mf E)}
\Delta_G
$$
by a derivation of the same height as that premise. By the ambient weakening of gated $2$-cells, \cref{lem:gated-2-cell-metatheory}, the transformation premise becomes
$$
\mathrm{wk}^{a:p}_{\mathrm{in}}(r)::
\mathrm{wk}^{a:p}_{\mathrm{in}}(\mf D)
\Rightarrow^{\Lambda}
\mathrm{wk}^{a:p}_{\mathrm{in}}(G(\mf E)).
$$
The linear gate variables are distinct from the cartesian variable $a$. Hence \cref{lem:gate-operations-commute}(iv) identifies the target with
$$
\mathrm{wk}^{a:p}_{\mathrm{in}}(G(\mf E))
=G\bigl(\mathrm{wk}^{a:p}_{\mathrm{in}}(\mf E)\bigr).
$$
We can now reapply the same gated rule, with the same linear gate variables, and obtain
$$
\Theta,a:A\ ;\ \Gamma
\vdash^{\Lambda}_{\mathrm{wk}^{a:p}_{\mathrm{in}}(\mf D)}
\Delta.
$$
One rule has been reapplied to a premise of the same height, so the resulting derivation has the same height as the original one. This proves the gated extension of cartesian weakening.

We consider next left-invertibility of $F$. Suppose that the formula being inverted is $z:\mathsf F$, with $\mathsf F=F^y_{\mf D_F}(\Theta_F;\Gamma_F)$. By hypothesis the last gated rule does not act at $z$, so this occurrence of $z:\mathsf F$ is present in its sequent premise. The induction hypothesis applied to that premise gives its $F$-left premise, indexed by
$$
\mf D_F\circ_{y,z}\mf E.
$$
Using $2$-Subst on $r$ with the identity $2$-cell of $\mf D_F$ gives
$$
\mathbf 1_{\mf D_F}\circ_{y,z}r::
\mf D_F\circ_{y,z}\mf D
\Rightarrow^{\Lambda}
\mf D_F\circ_{y,z}G(\mf E).
$$
The cut variable $z$ is disjoint from the variables acted on by the gate. If $z$ is cartesian, this is automatic because all gate variables are linear; if $z$ is linear, it is exactly the hypothesis that the gated rule does not act at the inverted variable. Therefore \cref{lem:gate-operations-commute}(i) gives
$$
\mf D_F\circ_{y,z}G(\mf E)
=G\bigl(\mf D_F\circ_{y,z}\mf E\bigr).
$$
The transformed sequent premise and transformed $2$-cell are now the premises of the same gated rule over the zone in which $z:\mathsf F$ has been replaced by $\Theta_F;\Gamma_F$. Reapplying it yields the required $F$-left premise of the original conclusion, indexed by $\mf D_F\circ_{y,z}\mf D$.

Right-invertibility of $U$ is dual. Let the inverted output be $z:\mathsf U$, where $\mathsf U=U^{\mf D_U}_{x_U}(\Theta_U;\Gamma_U\mid\Delta_U)$. Since the gated rule does not act at $z$, the output occurs in its sequent premise, to which the induction hypothesis applies and gives the index
$$
\mf E\circ_{z,x_U}\mf D_U.
$$
Composing the transformation premise on the right with the identity of $\mf D_U$ gives
$$
r\circ_{z,x_U}\mathbf 1_{\mf D_U}::
\mf D\circ_{z,x_U}\mf D_U
\Rightarrow^{\Lambda}
G(\mf E)\circ_{z,x_U}\mf D_U.
$$
By the other instance of \cref{lem:gate-operations-commute}(i), its target is
$$
G(\mf E)\circ_{z,x_U}\mf D_U
=G\bigl(\mf E\circ_{z,x_U}\mf D_U\bigr).
$$
The same gated rule therefore reapplies and gives the desired $U$-right premise of the original conclusion. In both invertibility arguments the induction hypothesis produces a derivation of no greater height than the sequent premise and exactly one gated rule is reapplied. The resulting derivation is consequently of no greater height than the original one. The unchanged premises of any branching rule occurring in the original invertibility proofs can be moved to the enlarged cartesian zone using the gated weakening established in the first stage.

It remains to extend cartesian contraction. We use the same induction on the complexity of $A$ and the height of the derivation as in \cref{lem:cartesian-contraction-for-sequents}. Suppose the displayed gated conclusion carries the cartesian declarations $x:A,x':A$. Since every gate acts at a linear mode, both declarations occur unchanged in its sequent premise. The induction hypothesis contracts that premise to
$$
\Theta,x:A\ ;\ \Gamma_G
\vdash^{\Lambda}_{\mf E[x':=x]}
\Delta_G.
$$
For a fresh output $b:p$, let
$$
\mf I_x:=[\varnothing\mid\{x\mapsto b\}]
$$
be the identity c-descriptor used to realize the contraction. Applying $2$-Subst to $r$ with $\mathbf 1_{\mf I_x}$ gives
$$
\mathbf 1_{\mf I_x}\circ_{b,x'}r::
\mf D[x':=x]
\Rightarrow^{\Lambda}
G(\mf E)[x':=x],
$$
where the source and target cuts compute to the displayed substitutions by \cref{lem:identity-cut-realizes-input-contraction}. The c-substitution clause of \cref{lem:gate-operations-commute}(i) gives
$$
G(\mf E)[x':=x]=G\bigl(\mf E[x':=x]\bigr).
$$
Reapplying the same gated rule therefore produces
$$
\Theta,x:A\ ;\ \Gamma
\vdash^{\Lambda}_{\mf D[x':=x]}
\Delta,
$$
of no greater height. The only exceptional non-gated cases in the contraction proof are $F_{\mathsf c}$L and $F_{\mathsf c}$L-in-l principal at one of the two copies being contracted. By exchanging the names of the two copies we may take the principal one to be $x'$. The case then uses the gated left-invertibility established in the second stage and contracts the entries of $\Theta_F$ at strictly smaller formula complexity. The case in which $x$ is principal is the same after the opposite exchange. Thus the lexicographic measure still decreases.

These arguments apply to Gate-weak, Gate-coweak, Gate-contr and Gate-cocontr alike, since they use only their common one-premise form and the corresponding operation $G$. All the inductions retain the height bounds of the original results, which completes the proof. \hfill $\square$

\textit{Remark:} At the gate variable itself the situation is that of the principal gate-variable cases. A weakened-in hypothesis is inverted by discarding its producer along $\mathrm{disc}$, and a contracted one by duplicating it along $\mathrm{dup}$, at the price of the interface weakenings and recombinations, outside the height bounds of the propositions. No result of this chapter uses such an inversion, and at the level of identity of proofs the $\eta$-laws supply it for every signature (\cref{def:equational-theory-of-derivations}).

\begin{proposition}[Harmony, affine case]\label{prop:harmony-affine}
Let the gates declared by $\Lambda$ be exactly $\mathsf{weak}(l)$ and $\mathsf{coweak}(l)$ at every linear mode $l$, with the discard axioms of \cref{def:gate-coherence-laws}. Then \cref{thm:cut-admissibility} holds for $\Lambda$.
\end{proposition}

\textit{Proof.}
The original proof is extended by the cases where a derivation ends with a gated rule, only $\mathsf{weak}(l)$ and $\mathsf{coweak}(l)$ being declared here. The substructural operations and the inversions it invokes extend to the gated calculus by \cref{lem:gated-metatheory-extension}.

If the gate variable is not the cut variable, the cut commutes. Take the right derivation ending with $\mathsf{weak}(l)$ at a variable $x'\neq x$, the cut input $x:A$ staying in the premise, $\Gamma''\!,x'{:}A'=\Gamma'_L,x{:}A,\Gamma'_R$:
$$
\inferrule*[right=$\mathsf{weak}(l)$]
{
  \Theta\ ;\ \Gamma''\vdash_{\mf E}^{\Lambda}\Delta'
  \\
  r_R::\mf D_2\Rightarrow^{\Lambda}\mf E\cup w_l(x')
}
{
  \Theta\ ;\ \Gamma''\!,\,x':A'\vdash_{\mf D_2}^{\Lambda}\Delta'
}
$$
We can therefore apply the induction hypothesis to cut the premise against the left derivation and, using 2-Subst on $r_R$, get the composition on the transformation $\mathbf 1_{\mf D_1}\circ_{y,x}r_R::\mf D_1\circ_{y,x}\mf D_2\Rightarrow\mf D_1\circ_{y,x}(\mf E\cup w_l(x'))$, the adjoined node $w_l(x')$ disjoint from the cut ports $y,x$. Applying \cref{lem:gate-operations-commute} rearranges the descriptors:
$$\mf D_1\circ_{y,x}(\mf E\cup w_l(x'))=(\mf D_1\circ_{y,x}\mf E)\cup w_l(x'),$$
and $\mathsf{weak}(l)$ reapplies at $x'$. The rule $\mathsf{coweak}$ at an output $\neq y$ is the same on the output side, and for a cartesian cut formula the same commutation applies, the gate operations passing c-substitution by \cref{lem:gate-operations-commute}(i).

The new principal cases are those where the cut formula is the gated variable itself. Suppose the right derivation ends with $\mathsf{weak}(l)$ at the cut input $x:A$, its premise dropping $x$:
$$
\inferrule*[right=$\mathsf{weak}(l)$]
{
  \Theta\ ;\ \Gamma'_L,\Gamma'_R\vdash_{\mf E}^{\Lambda}\Delta'
  \\
  r_R::\mf D_2\Rightarrow^{\Lambda}\mf E\cup w_l(x)
}
{
  \Theta\ ;\ \Gamma'_L,x:A,\Gamma'_R\vdash_{\mf D_2}^{\Lambda}\Delta'
}
$$
Using 2-Subst on $r_R$ gives the composition on the transformation
$$\mathbf 1_{\mf D_1}\circ_{y,x}r_R::\mf D_1\circ_{y,x}\mf D_2\Rightarrow\mf D_1\circ_{y,x}(\mf E\cup w_l(x)),$$
whose target is exactly the source of the discard axiom of \cref{def:gate-coherence-laws},
$$\mathrm{disc}_{\mf D_1,\mf E}::\mf D_1\circ_{y,x}\bigl(\mf E\cup w_l(x)\bigr)\Rightarrow\mathrm{wk}^{w}_{\gamma}\bigl(\mathrm{cowk}^{w^\perp}_{\delta_L,\delta_R}(\mf E)\bigr),$$
which effectively weakens the descriptor $\mf E$ by the relevant context formulas contained in $\mf D_1$ while forgetting the weakened variable $x$ by way of cutting it out. The sequent carrying the codomain descriptor of this transformation is derivable from the premise. Indeed, starting from $\Theta;\Gamma'_L,\Gamma'_R\vdash_{\mf E}\Delta'$, we adjoin each variable of $\Gamma$ by $\mathsf{weak}$ at its mode and each output of $\Delta_L,\Delta_R$ by $\mathsf{coweak}$ at its mode, all gates being declared, with identity $2$-cells:
$$\Theta\ ;\ \Gamma'_L,\Gamma,\Gamma'_R\vdash_{\mathrm{wk}^{w}_{\gamma}(\mathrm{cowk}^{w^\perp}_{\delta_L,\delta_R}(\mf E))}^{\Lambda}\Delta_L,\Delta',\Delta_R,$$
its descriptor exactly the codomain of $\mathrm{disc}_{\mf D_1,\mf E}$.

Composing the two transformations together gets us $\mathrm{disc}_{\mf D_1,\mf E}\circ(\mathbf 1_{\mf D_1}\circ_{y,x}r_R)::\mf D_1\circ_{y,x}\mf D_2\Rightarrow\mathrm{wk}^{w}_{\gamma}(\mathrm{cowk}^{w^\perp}_{\delta_L,\delta_R}(\mf E))$, and respect for transformation (\cref{prop:respect-for-structural-transformation}) carries this sequent back to the descriptor $\mf D_1\circ_{y,x}\mf D_2$, which is the desired conclusion.
The case where the \emph{left} derivation ends with $\mathsf{coweak}$ at the cut output $y$ is dual, via $\mathrm{disc}^{\mathrm{op}}_{\mf D_2,\mf E}$. In all these cases the cut is eliminated outright, and the measure is respected trivially. \hfill $\square$

We extend finally the cut theorem to the full gate package. The main obstruction has been known since Gentzen \cite{Gen35}. The contraction operation is notoriously difficult to handle as the standard induction hypothesis is too weak for it. Common solutions include building contraction into the rules so that it is admissible, as in the G3-style calculi \cite{NvP01,TS00}, or replacing cut by Gentzen's mix, cutting several occurrences at once \cite{Gen35}. In our case, contraction is an addition to the signature in the form of a gate and axioms.  We solve the issue by showing that cut is eliminable in its \emph{multicut} form, i.e. one producer with several designated outputs against one consumer with several designated inputs, and induct on the rank of the gates. This rank counts redirection ancestry.

\begin{definition}[Ancestors]\label{def:ancestors}
  Let $\pi$ be a linear derivation and $w:A$ a variable occurring in its conclusion, on either the input (antecedent) or the output (succedent) side. The \emph{ancestor tree} of $w$ in $\pi$ is the finite tree with root $w$, defined by recursion on $\pi$. Writing $\mathcal R$ for the last rule of $\pi$, the subtrees of the root are:
  \begin{itemize}
    \item \textbf{(introduced)} none, i.e. the tree is the single node $w$, whenever $w$ is the principal formula introduced by $\mathcal R$, or the fresh variable of $\mathsf{weak}(l)$ or $\mathsf{coweak}(l)$;
    \item \textbf{((co)contraction)} the ancestor trees of the two premise occurrences $x',x''$ (resp.\ $b',b''$), when $w$ is the contracted input $x_0$ of $\mathsf{contr}(l)$ (resp.\ the merged output $b_0$ of $\mathsf{cocontr}(l)$);
    \item \textbf{(idempotence)} the ancestor tree of the unique premise occurrence carrying the same name, in every remaining case.
  \end{itemize}
\end{definition}

\begin{example}[Linear--Nonlinear Logic]\label{ex:linear-logic-ancestor-tree}
Let $P$ be a proposition at a linear mode and let $\pi$ be the derivation obtained by contracting the tensor introduction with $\mathsf{contr}(l)$ (\cref{def:gated-structural-rules}). Two instances of $\mathsf v_{\mathsf l}$ feed the two components of $F_{\mathsf l}$R, which introduces
$$\cdot\ ;\ x' : P,\, y' : P \vdash^{\Lambda}_{\mf D'_\otimes} z : P \otimes P,
\qquad
\mf D'_\otimes := [\{x', y' \xmapsto{\otimes_l} w\} \mid \{w \mapsto z\}],$$
and the gate merges $x', y'$ into $x_0$, concluding $\cdot\ ;\ x_0 : P \vdash^{\Lambda}_{c_l(x_0; x', y') \cdot \mf D'_\otimes} z : P \otimes P$. The ancestor tree of the input $x_0$ branches at its root, the last rule being the contraction gate at $x_0$, and its two subtrees are the trees of $x'$ and of $y'$ in the premise. Each of those passes the branching $F_{\mathsf l}$R by the idempotence clause, the name occurring in exactly one component premise, and stops at the variable rule introducing it. Writing every node with its subtrees in parentheses, the tree is
$$x_0\bigl(\,x'(x'),\ y'(y')\,\bigr).$$
The output $z$ has instead the unbranched tree $z(z)$, the contraction gate acting on the input side only, its single leaf the occurrence that $F_{\mathsf l}$R introduces as its principal formula. Had the last rule been $\mathsf{weak}(l)$ adjoining a fresh input, that input would be a leaf on the spot, no occurrence of it standing in the premise.
\end{example}

\begin{definition}[Gate rank]\label{def:gate-rank}
  The \emph{rank} of a variable $w$ in a derivation $\pi$ is the height of its ancestor tree (\cref{def:ancestors}), the single-node tree having height $1$. For a set of designated variables, the rank is the maximum of the ranks of its members.
\end{definition}

\textit{Remark:} Two facts carry the multicut case analysis. If a variable has one ancestor in a premise, the ancestor tree of that occurrence is the unique subtree under the root, so its rank is exactly one less. If the last rule is a contraction or cocontraction at the variable, there are two ancestor subtrees: the maximum of their ranks is exactly one less than the rank of the root, although one of the two branches may be shorter. Consequently, if a nonempty set of designated variables has rank at least $2$, replacing it by the full family of its premise ancestors lowers the maximum rank by exactly one. A variable has rank $1$ exactly when the last rule introduces it, since every other clause places at least one subtree under the root; such a variable has no premise ancestor and is treated in the principal cases.

\begin{example}[Linear--Nonlinear Logic]
In the derivation $\pi$ of the \hyperref[ex:linear-logic-ancestor-tree]{ancestor-tree example}, the ancestor tree of $x_0$ has height $3$, so $x_0$ has rank $3$, while the output $z$ has rank $2$. Passing to the premise of $\mathsf{contr}(l)$ replaces the tree of $x_0$ by its two subtrees, and $x', y'$ have rank $2$ there, one less, as the rank calculation predicts; the designated set $\{x', y'\}$ has then rank $2$ as well, the maximum of the two. Passing once more to the components brings both to rank $1$, the variable rule introducing them, and the two derivations are ready to meet in a principal case. This is the measure the multicut argument spends. The premise of the contraction gate carries the cut formula twice, so neither the complexity of $P \otimes P$ nor the height of the derivation controls the recursion there, while the rank of the designated inputs drops by one.
\end{example}

\begin{lemma}[Rank Stability]\label{lem:rank-stability}
Let $\pi$ be a linear derivation and let $\mathcal O(\pi)$ denote the derivation produced by one of the recursive constructions used for cartesian input weakening or contraction (\cref{lem:cartesian-weakening-for-sequents,lem:cartesian-contraction-for-sequents}), inversion at a variable not acted on by a gate (\cref{prop:left-invertibility-of-f,prop:right-invertibility-of-u,lem:gated-metatheory-extension}), boundary renaming, or permutation transport. For a boundary renaming $\sigma$, put $w^{\mathcal O}:=\sigma(w)$, and put $w^{\mathcal O}:=w$ for the other operations. If $w$ is a conclusion variable at a linear mode and its corresponding variable $w^{\mathcal O}$ survives in the conclusion of $\mathcal O(\pi)$, then
$$
\operatorname{rk}_{\mathcal O(\pi)}(w^{\mathcal O})
\leq
\operatorname{rk}_{\pi}(w).
$$
\end{lemma}

\textit{Proof.}
Write $T_{\pi}(w)$ for the ancestor tree of $w$ in $\pi$. Unfolding \cref{def:ancestors,def:gate-rank} gives this recursive computation:
$$
\operatorname{rk}_{\pi}(w)=
\begin{cases}
1,
& \text{if the last rule introduces }w,\\
1+\max\bigl\{\operatorname{rk}_{\pi_0}(w_1),
\operatorname{rk}_{\pi_0}(w_2)\bigr\},
& \text{if the last rule contracts or cocontracts }w,\\
1+\operatorname{rk}_{\pi_i}(w_i),
& \text{otherwise.}
\end{cases}
$$
Here $w_1,w_2$ are the two premise ancestors in the second case, and in the third case $w_i$ is the unique premise occurrence from which $w$ descends. Even when the last rule has several sequent premises, a linear variable lies in exactly one component unless it is the principal variable of a contraction or cocontraction. This formula allows us to compare the ranks one operation at a time.

We start with boundary renaming and permutation transport. Both are defined on a sequent derivation by re-instantiating every rule with the renamed, respectively permuted, boundary, using the corresponding operations on its descriptor indexes. No rule is added or removed. Under a renaming $\sigma$, every node labelled by a variable $u$ is relabelled by $\sigma(u)$, while a permutation merely moves declarations in the context and leaves their names unchanged. Thus
$$
T_{\mathrm{ren}_{\sigma}(\pi)}(\sigma(w))\cong T_{\pi}(w),
\qquad
T_{(\rho;\tau)_*\pi}(w)\cong T_{\pi}(w).
$$
The trees are isomorphic as rooted trees, and their heights are equal. Boundary renaming and permutation transport therefore preserve rank exactly.

We consider now cartesian input weakening. Its proof is by induction on the derivation and re-applies the same last rule after weakening its sequent premises. We compare the ancestor trees simultaneously along that induction. If the last rule introduces $w$, the two trees are both the single-node tree. If it contracts or cocontracts $w$, the induction hypothesis identifies the two ancestor subtrees before and after weakening, and the same binary root is placed above them. In every other case, the unique premise carrying $w$ is weakened by the induction hypothesis and the same unary root is put back. The added hypothesis is cartesian, so no new linear gate can occur on the thread of $w$. Hence
$$
T_{\mathrm{wk}(\pi)}(w)\cong T_{\pi}(w),
$$
and cartesian input weakening preserves the rank exactly.

We turn to inversion, and argue along the induction used in \cref{prop:left-invertibility-of-f,prop:right-invertibility-of-u}. Suppose first that the last rule is the matching principal $F$-left or $U$-right rule at the inverted variable $z$. Inversion returns its sequent premise. A surviving variable $w$ is different from $z$, so the principal rule was a unary node on the thread of $w$. If $\pi_0$ is its premise, then
$$
\operatorname{rk}_{\pi}(w)
=1+\operatorname{rk}_{\pi_0}(w),
\qquad
\operatorname{rk}_{\operatorname{Inv}_z(\pi)}(w)
=\operatorname{rk}_{\pi_0}(w).
$$
The rank drops by one in this case.

If the last rule is not principal at $z$, the inversion construction proceeds through its sequent premises and then re-applies that same rule. When $z$ is linear, it lies in exactly one component, and the induction hypothesis is applied to that component. When $z$ is cartesian, it belongs to the shared cartesian zone, and the induction hypothesis is applied to every premise in which it occurs. For a surviving variable $w$ in one of those premises, the induction hypothesis says that its premise rank does not increase; adding back the same unary or binary root preserves the inequality in this recursive formula. In the linear case, if $w$ lies in another component, that component is unchanged except possibly for cartesian weakening, which preserves rank exactly. If the last rule is a gated rule away from $z$, \cref{lem:gated-metatheory-extension} re-applies the same gate. In particular, when this gate contracts or cocontracts $w$, the induction hypothesis applies to both ancestors and taking their maximum preserves the inequality. Thus inversion may remove unary nodes but never introduces a new node on the thread of a surviving variable, and
$$
\operatorname{rk}_{\operatorname{Inv}_z(\pi)}(w)
\leq
\operatorname{rk}_{\pi}(w).
$$

Finally, consider cartesian contraction, using the lexicographic induction of \cref{lem:cartesian-contraction-for-sequents}. In every nonexceptional case the construction contracts the relevant sequent premises and re-applies the same last rule. This includes the retained-principal rules $U_{\mathsf c}$L, $U_{\mathsf c}$L-in-l and $U_{\mathsf{shift}}$L: their principal cartesian variable is renamed to the retained copy in the contracted premises, and the same rule is put back. The recursive rank formula and the induction hypothesis give the desired inequality exactly as in the inversion case: an introduced variable remains a leaf, a (co)contracted wire retains the same binary root, and every other surviving variable retains the same unary root over a subtree whose height has not increased. This also includes a last gated rule, since the gate acts at a linear mode and is reapplied unchanged by \cref{lem:gated-metatheory-extension}.

There remains the exceptional case in which the last rule is $F_{\mathsf c}$L or $F_{\mathsf c}$L-in-l at one of the two cartesian copies being contracted. After exchanging the copies we may write the original derivation schematically as
$$
\pi=F_{\mathsf c}\mathrm L_{x'}(\pi_0).
$$
The contraction construction first inverts $\pi_0$ at the retained copy $x$, then contracts the two exposed copies of $\Theta_F$ componentwise, and finally re-applies $F_{\mathsf c}$L at $x$. Writing $\operatorname{Inv}_x$ for the inversion and $\operatorname{Ctr}_{\Theta_F}$ for those componentwise contractions, the resulting derivation has the form
$$
\operatorname{Ctr}_{x'\to x}(\pi)
=
F_{\mathsf c}\mathrm L_x
\bigl(\operatorname{Ctr}_{\Theta_F}(\operatorname{Inv}_x(\pi_0))\bigr).
$$
For every surviving linear variable $w$, the outer $F$-left rule is a unary node in both derivations. The induction hypothesis at the strict subformulas in $\Theta_F$, followed by the inversion inequality already proved, gives
\begin{align*}
\operatorname{rk}_{\operatorname{Ctr}_{x'\to x}(\pi)}(w)
&=1+
\operatorname{rk}_{\operatorname{Ctr}_{\Theta_F}(\operatorname{Inv}_x(\pi_0))}(w)\\
&\leq 1+
\operatorname{rk}_{\operatorname{Inv}_x(\pi_0)}(w)\\
&\leq 1+\operatorname{rk}_{\pi_0}(w)\\
&=\operatorname{rk}_{\pi}(w).
\end{align*}
Either inequality can be strict. For example, if $\pi_0$ itself ends by eliminating the retained copy $x$, the inversion returns its premise and removes that unary $F$-left node. This explains why cartesian contraction is rank non-increasing but not rank preserving in general. All the stated cases are now covered. \hfill $\square$

\textit{(Co)Contraction Trees.} If $\mathsf{contr}(l)$ is declared and $\mf T$ is a well-formed ambient l-descriptor with inputs including $x_1,\ldots,x_k:q$, a \emph{contraction tree} $W$ is a rooted binary tree carrying those leaves, with root $x_0:q$. It acts on $\mf T$ by recursion, one instance of the contraction operation $c_l$ of \cref{def:gate-descriptor-generation} per internal node: a single leaf $x_i$ with root $r$ acts as the renaming $\mf T[x_i:=r]$, and a node with root $r$ joining two subtrees $W_a,W_b$ with roots $a,b$ acts as
$$
W\cdot\mf T\ :=\ c_l(r;a,b)\cdot\bigl(W_a\cdot(W_b\cdot\mf T)\bigr).
$$
The two subtrees consume disjoint inputs, so their order is immaterial: $W_a\cdot(W_b\cdot\mf T)=W_b\cdot(W_a\cdot\mf T)$. The root instance $r=x_0$ gives $W\cdot\mf T$, and for $k=1$ this degenerates to $\mf T[x_1:=x_0]$. Dually, a \emph{cocontraction tree} $M$ with leaves $y_1,\ldots,y_m$ and root $y_0$ merges the outputs by the mirror recursion, one instance of $c^\perp_l$ per node,
$$
\mf T\cdot M\ :=\ \bigl((\mf T\cdot M_a)\cdot M_b\bigr)\cdot c^\perp_l(a,b;s)
$$
at a node with root $s$ joining $M_a,M_b$ with roots $a,b$, with $\mf T\cdot M:=\mf T[y_1:=y_0]$ for $m=1$. For $m=k=1$ both trees are trivial and the mix descriptor is the plain cut, up to the renaming of the cut variable.

\begin{lemma}[Tree Transformation]\label{lem:tree-reshaping}
Under the (co)commutative (co)monoid laws in \cref{def:gate-coherence-laws}, any two contraction trees over the same contexts with the same root and the same set of leaves are related by an inverse pair of derivable $2$-cells. Similarly and dually for cocontraction trees.
\end{lemma}
\textit{Proof.}
Recall that a contraction tree over $\mf T$ is an iterated application of $c_l$-nodes, such that the root of the tree is the input $x_0$ and the leaves are the contracted inputs $x_1,\ldots,x_k$ of $\mf T$. Two such trees with the same root and the same leaves differ therefore only in the order of application of the contraction operator and in the order of the leaves. The gate $c_l$ is associative and commutative by \cref{def:gate-coherence-laws}, so this is the familiar situation of an associative and commutative operation: neither the order in which the contractions are applied nor the order of the leaves can change the result. We produce the pair of $2$-cells by rewriting one tree into the other with the laws, each rewrite an instance of an axiom and hence a derivable $2$-cell. The axioms being invertible pairs, the reverse rewriting gives the inverse.

The associativity law of \cref{def:gate-coherence-laws} settles the order of operations. The law
$$c_l(x_0;a,b)\cdot\bigl(c_l(a;c,d)\cdot\mf T\bigr)\Leftrightarrow c_l(x_0;c,e)\cdot\bigl(c_l(e;d,b)\cdot\mf T\bigr)$$
turns a contraction of $c,d$, followed by contraction with $b$, into a contraction of $d,b$, followed by contraction with $c$, the leaves keeping their order $c,d,b$. Reading it always in the same direction reassociates any tree to the one that brings in a single leaf at a time. Each step puts one more node into that form, and we induct on the number still to move. Two trees over the same leaves in the same order reduce therefore to that one tree, hence to one another.

The commutativity law of \cref{def:gate-coherence-laws} settles the order of the variables. The law $c_l(x_0;x,x')\cdot\mf T\Leftrightarrow c_l(x_0;x',x)\cdot\mf T$ swaps the two leaves meeting at a node. To exchange two neighbouring leaves elsewhere, reassociate until they meet at a common node, swap them, reassociate back. Every reordering of the leaves is a succession of such neighbour swaps, so commutativity brings the orders of any two trees into agreement, and we induct on the number of swaps. The bracketings reconciled by associativity and the orders by commutativity, any two trees over $\mf T$ are related.

It remains only to note that each rewrite is carried out inside the whole tree, not at its top alone. The law is applied at one node, everything below that node serving as the ambient $\mf T$ of the axiom. The contractions standing above it are put back one at a time by the gated former (\cref{def:gated-descriptor-2-cells}), which sends a $2$-cell $r$ to the $2$-cell $c_l(x_0;x,x')\cdot r$ between the two operated descriptors. Composing all the rewrites gives the $2$-cell of the statement, and reversing them gives its inverse. The cocontraction case is the same with $c^\perp_l$ in place of $c_l$, merging outputs where this one merges inputs. \hfill $\square$

\begin{theorem}[Multicut]\label{thm:multicut}
Let $\Lambda$ declare all four gates at every linear mode, with their axiom packages (\cref{def:gate-coherence-laws}). Let $A$ be a type at a linear mode $q$. If
$$
\Theta\ ;\ \Gamma\vdash^{\Lambda}_{\mf D_1}\Delta_L,\,y_1{:}A,\ldots,y_m{:}A,\,\Delta_R
\qquad\text{and}\qquad
\Theta\ ;\ \Gamma'_L,\,x_1{:}A,\ldots,x_k{:}A,\,\Gamma'_R\vdash^{\Lambda}_{\mf D_2}\Delta'
$$
are derivable ($m,k\geq 1$), then for every cocontraction tree $M$ merging $\bar y$ into $y_0$ and every contraction tree $W$ merging the inputs $\bar x$ into $x_0$, this sequent is derivable:
$$
\Theta\ ;\ \Gamma'_L,\Gamma,\Gamma'_R\vdash^{\Lambda}_{(\mf D_1\cdot M)\,\circ_{y_0,x_0}\,(W\cdot\mf D_2)}\Delta_L,\Delta',\Delta_R .
$$
For $m=k=1$ the trees are empty and the statement is the linear case of \cref{thm:cut-admissibility}.
\end{theorem}

\textit{Proof.}
By \cref{lem:tree-reshaping} and \cref{prop:respect-for-structural-transformation}, derivability for one choice of trees gives it for every choice. Each case fixes whichever shapes are convenient. The theorem is proven simultaneously with the extension of the cartesian case of \cref{thm:cut-admissibility} to $\Lambda$. The cartesian case gains only the gated rules as new cases, and those commute. The gate operations pass c-substitution by \cref{lem:gate-operations-commute}(i), the $2$-cells transform by $2$-Subst and the gated formers, and the substructural operations and inversions extend by \cref{lem:gated-metatheory-extension}. Every cross-appeal between the two cases strictly lowers the cut-formula complexity, so the outer induction on $\mathrm{cx}(A)$ keeps their inner measures independent: heights for the cartesian case, and for the multicut the lexicographic pair
$$
\bigl(\mathrm{cx}(A),\ \rho_L+\rho_R\bigr),
$$
where $\rho_L$ is the rank of the designated outputs $\bar y$ in the left derivation and $\rho_R$ the rank of the designated inputs $\bar x$ in the right one (\cref{def:gate-rank}). Throughout, a cartesian weakening or a boundary relabelling of the opposite derivation preserves the rank of its designated variables, while an inversion can only lower it, by \cref{lem:rank-stability}. Whenever the active rank is at least $2$, the rank recursion shows that passing to the full family of premise ancestors lowers it by exactly one. The rank sum used by the induction hypothesis is therefore strictly smaller than $\rho_L+\rho_R$; rank-$1$ variables are handled separately in the principal cases.

\emph{Case $\rho_R\geq 2$:} some input has an ancestor in a premise of the right derivation's last rule $\mathcal R$.

\emph{(R1) No designated input is principal for $\mathcal R$ and none is its gate variable.} Then every variable is present in the premises.
If $\mathcal R$ has a single sequent premise ($F_{\mathsf l}$L, $U_{\mathsf l}$R, the cartesian left rules, or a gated rule at other variables), the inputs $\bar x$ are all present in that premise; we start with $F_{\mathsf l}$L:
$$
\inferrule*[right=$F_{\mathsf l}$L]
{
  \Theta,\Theta_{F'}\ ;\ \Gamma'_L,\Gamma_{F'},\Gamma'_R\vdash_{\mf D_{F'}\circ_{y_{F'},z}\mf D_2}^{\Lambda}\Delta'
}
{
  \Theta\ ;\ \Gamma'_L,z:\mathsf F',\Gamma'_R\vdash_{\mf D_2}^{\Lambda}\Delta'
}
$$
We apply the induction hypothesis to the premises at the same inputs $\bar x$, of rank $\rho_R{-}1$, and re-apply $\mathcal R$. The connective descriptor passes the mix by \cref{lem:gate-operations-commute}, the trees $W,M$ acting on variables distinct from the rule's cut variable $z$, and \cref{lem:left-interchange-law}, the two inputs $z,x_0$ distinct, pulls it outside:
$$(\mf D_1\cdot M)\circ_{y_0,x_0}\bigl(W\cdot(\mf D_{F'}\circ_{y_{F'},z}\mf D_2)\bigr)=\mf D_{F'}\circ_{y_{F'},z}\bigl((\mf D_1\cdot M)\circ_{y_0,x_0}(W\cdot\mf D_2)\bigr),$$
so that $F_{\mathsf l}$L reapplies with the mix descriptor of the conclusion. For $U_{\mathsf l}$R the same holds with \cref{lem:associativity} in place of \cref{lem:left-interchange-law}, and the cartesian left rules pass as their counterparts in \cref{thm:cut-admissibility} through \cref{lem:gated-metatheory-extension}. The $2$-cell premises are transformed by $2$-Subst and the gated formers.
If $\mathcal R$ branches ($F_{\mathsf l}$R, $U_{\mathsf l}$L, $U_{\mathsf{shift}}$L), the designated inputs distribute over the component premises, $\bar x=\biguplus_c\bar x_c$. Let $J$ be the set of components with $\bar x_c\neq\varnothing$, and for $c\in J$ write the component premise as $\Theta\ ;\ \Gamma^{(c)}\vdash_{\mf D_c}\Delta^{(c)}$, its designated inputs among $\Gamma^{(c)}$. Reshape $W$ by \cref{lem:tree-reshaping} so that each class $\bar x_c$ of the partition is collected by its own subtree $W_c$ with root $x_0^{c}$, the classes joined by $|J|-1$ separating contraction nodes. For each $c\in J$, the induction hypothesis mixes the left derivation against component $c$ at the variables $\bar x_c$, of rank at most $\rho_R{-}1$, with the trees $W_c$ and $M$, the left derivation taken in a fresh boundary-renamed copy whose linear interface $\Gamma,\Delta_L,\Delta_R$ is replaced by disjoint fresh copies $\Gamma^{\langle c\rangle},\Delta_L^{\langle c\rangle},\Delta_R^{\langle c\rangle}$ and whose cartesian zone $\Theta$ is shared (\cref{thm:renaming-admissibility-for-descriptor-judgments}; these are the boundary-renamed copies of $(\mf D_1\cdot M)$ required by the duplication axiom, and the rank is unchanged by \cref{lem:rank-stability}). The modified component is
$$\Theta\ ;\ \Gamma^{(c)},\Gamma^{\langle c\rangle}\vdash_{(\mf D_1\cdot M)^{\langle c\rangle}\,\circ_{y_0^{\langle c\rangle},x_0^{c}}\,(W_c\cdot\mf D_c)}\Delta_L^{\langle c\rangle},\Delta^{(c)},\Delta_R^{\langle c\rangle},$$
carrying the renamed copy of the left context and side outputs. The component contexts are now genuinely pairwise disjoint, the copies for distinct classes sharing no variable, so $\mathcal R$ re-applies with the modified components, the untouched components and the family unchanged. Its conclusion carries the $|J|$ disjoint copies of every variable of $\Gamma$ and of $\Delta_L,\Delta_R$. We merge them back by the gated rules, $\mathsf{contr}(l)$ contracting the $|J|$ copies of each input of $\Gamma$ to the single variable and $\mathsf{cocontr}(l)$ cocontracting the copies of each side output, with identity $2$-cells, all modes being linear and all gates declared.

Regarding the descriptor, the $2$-cell of $\mathcal R$ is transformed by the gated formers and $2$-Subst as in the single-premise cases, reaching the mix of $(\mf D_1\cdot M)$ against $W$ applied to the composite target of $\mathcal R$. At each of the $|J|-1$ separating nodes the duplication axiom applies,
$$
\mathrm{dup}_{(\mf D_1\cdot M),\,\mf E}\ ::\ (\mf D_1\cdot M)\circ_{y_0,x_0}\bigl(c_l(x_0;x',x'')\cdot\mf E\bigr)\ \Rightarrow\ \mathrm{ctr}^{c}\Bigl(\mathrm{cctr}^{c^\perp}\bigl((\mf D_1\cdot M)\circ_{y_0,x'}((\mf D_1\cdot M)^{c}\circ_{y_0^{c},x''}\mf E)\bigr)\Bigr),
$$
with $\mf E$ the descriptor standing under the node. By induction on the number of separating nodes, the iterated axiom splits the single mix into one mix $(\mf D_1\cdot M)\circ_{y_0,x_0^{c}}(W_c\cdot-)$ per class, each localized to its component by \cref{lem:localization}, the whole recombined by the contraction trees of the copies of $\gamma_1$ and the cocontraction trees of the copies of $\delta_L,\delta_R$ --- exactly the index of the re-applied rule followed by the merging gated rules with identity $2$-cells. After composing the transformed $2$-cell of $\mathcal R$ with the iterated duplication $2$-cell, we conclude by applying respect for transformation (\cref{prop:respect-for-structural-transformation}).

\emph{(R2) $\mathcal R$ is a gated rule at a designated input $x_i$.}
If $\mathcal R=\mathsf{contr}(l)$ at $x_i$, its premise carries the ancestors $x',x''$ in place of $x_i$:
$$
\inferrule*[right=$\mathsf{contr}(l)$]
{
  \Theta\ ;\ \Gamma',x':A,x'':A\vdash_{\mf E}^{\Lambda}\Delta'
  \\
  r_R::\mf D_2\Rightarrow^{\Lambda}c_l(x_i;x',x'')\cdot\mf E
}
{
  \Theta\ ;\ \Gamma',x_i:A\vdash_{\mf D_2}^{\Lambda}\Delta'
}
$$
the induction hypothesis applies at the inputs $(\bar x\smallsetminus x_i)\cup\{x',x''\}$, with the tree $W'$ obtained from $W$ by splitting the leaf $x_i$ once, of rank $\rho_R{-}1$. Contracting the rule's $2$-cell gives
$$W\cdot\mf D_2\Rightarrow W\cdot\bigl(c_l(x_i;x',x'')\cdot\mf E\bigr)=W'\cdot\mf E,$$
an equality of classes, grafting the contraction node $x_i\mapsto x',x''$ under $W$ being the split tree $W'$, and Respect concludes.
If $\mathcal R=\mathsf{weak}(l)$ at $x_i$ and $k\geq 2$, its premise drops $x_i$ and carries the remaining variables:
$$
\inferrule*[right=$\mathsf{weak}(l)$]
{
  \Theta\ ;\ \Gamma'\vdash_{\mf E}^{\Lambda}\Delta'
  \\
  r_R::\mf D_2\Rightarrow^{\Lambda}\mf E\cup w_l(x_i)
}
{
  \Theta\ ;\ \Gamma',x_i:A\vdash_{\mf D_2}^{\Lambda}\Delta'
}
$$
the induction hypothesis applies at $\bar x\smallsetminus x_i$, of rank $\rho_R{-}1$. Contracting gives $W\cdot\mf D_2\Rightarrow W\cdot(\mf E\cup w_l(x_i))$. Reshaping brings the leaf $x_i$ to an outermost contraction node by \cref{lem:tree-reshaping}, where the ambient counit law $c_l(x_0;x_i,x)\cdot(\mf E\cup w_l(x_i))\Leftrightarrow\mathrm{ren}_{x\mapsto x_0}(\mf E)$ deletes the killed branch, giving $W_{\smallsetminus i}\cdot\mf E$, and Respect concludes. For $k=1$ the rank is $1$, which is handled by the rank-$1$ case.

\emph{(R3) $\mathcal R$ is principal at a variable $x_j$ ($F_{\mathsf l}$L or $U_{\mathsf l}$L at $x_j$; a $\mathsf v_{\mathsf l}$ conclusion declares no other variable, so its rank is $1$) while other variables are left untouched.} Take $F_{\mathsf l}$L at $x_j$, so $A=F^{y'}_{\mf D_F}(\Theta_F;\Gamma_F)$ and the rule premise contains $\Theta_F;\Gamma_F$, and the remaining untouched variables $\bar x\smallsetminus x_j$ stay among $\Gamma'_L,\Gamma'_R$:
$$
\inferrule*[right=$F_{\mathsf l}$L]
{
  \Theta,\Theta_F\ ;\ \Gamma'_L,\Gamma_F,\Gamma'_R\vdash_{\mf D_F\circ_{y_F,x_j}\mf D_2}^{\Lambda}\Delta'
}
{
  \Theta\ ;\ \Gamma'_L,x_j:A,\Gamma'_R\vdash_{\mf D_2}^{\Lambda}\Delta'
}
$$
We cut the premise at the variables $\bar x\smallsetminus x_j$ as in (R1), the single premise carrying also the context $\Gamma_F$, and re-apply $F_{\mathsf l}$L, reintroducing the variable $x_j$. Now mix the left derivation once more, against the re-applied conclusion, at the single variable $x_j$. In that derivation $x_j$ is ancestor-free, so the new cut has rank sum
$$\rho_L+1<\rho_L+\rho_R,$$
and the induction hypothesis applies. The second cut contributes one further copy of the left context, recombined by the gates as in (R1), and the descriptor result is the same duplication-and-reshaping computation, the tree $W$ split into the subtree at $x_j$ and the rest. The rule $U_{\mathsf l}$L at $x_j$ is the same, mixed per component.

\emph{Case $\rho_R=1<\rho_L$:} the mirror analysis, on the \emph{left} derivation's last rule. (L1) If no designated output is principal and none is a gate variable, the rule commutes. A single-premise rule passes the mix by \cref{lem:gate-operations-commute,lem:associativity,lem:left-interchange-law,lem:right-interchange-law}, and a branching rule distributes the designated outputs over its components, the tree $M$ reshaped so that each class is collected by its own subtree (\cref{lem:tree-reshaping}), the dual duplication axiom $\mathrm{dup}^{\mathrm{op}}$ iterated at each separating cocontraction node, and the copies of the right context and side outputs merged by the gated rules, dually to (R1). (L2) If the last rule is $\mathsf{cocontr}(l)$ at a designated output, the induction hypothesis applies at its two ancestors with the split tree $M'$, of rank $\rho_L-1$. If it is $\mathsf{coweak}(l)$ at a designated output and $m\geq 2$, the induction hypothesis applies at the remaining outputs and the ambient counit law deletes the killed branch of $M$, dually to (R2). $\mathrm{disc}^{\mathrm{op}}$ enters only at $m=1$, in the rank-$1$ case. (L3) If the rule is principal at a designated output, the mix is completed as in (R3), with a further cut at the freshly reintroduced output variable, of rank sum $1+\rho_R'<\rho_L+\rho_R$.

\emph{Case $\rho_L=\rho_R=1$:} every designated variable is introduced by the last rule of its derivation, and a rule introduces at most one designated input and at most one designated output --- its principal formula, or the fresh variable of a weakening gate --- so $m=k=1$, the trees are trivial, and the descriptor is the plain cut $\mf D_1\circ_{y,x}\mf D_2$, up to the renaming of the wire. Both variables are introduced by their respective last rules, which leaves these pairs (all others excluded by the shape of $A$):
\begin{itemize}
\item \emph{Right $\mathsf{weak}(l)$ at $x$:} the cut input $x$ is the fresh variable of the rule, and is not present in the premise:
$$
\inferrule*[right=$\mathsf{weak}(l)$]
{
  \Theta\ ;\ \Gamma'_L,\Gamma'_R\vdash_{\mf E}^{\Lambda}\Delta'
  \\
  r_R::\mf D_2\Rightarrow^{\Lambda}\mf E\cup w_l(x)
}
{
  \Theta\ ;\ \Gamma'_L,x:A,\Gamma'_R\vdash_{\mf D_2}^{\Lambda}\Delta'
}
$$
We use 2-Subst on the $2$-cell against $\mathbf 1_{\mf D_1}$, giving $\mathbf 1_{\mf D_1}\circ_{y,x}r_R::\mf D_1\circ_{y,x}\mf D_2\Rightarrow\mf D_1\circ_{y,x}(\mf E\cup w_l(x))$, and follow up with the discard axiom of \cref{def:gate-coherence-laws},
$$\mathrm{disc}_{\mf D_1,\mf E}::\mf D_1\circ_{y,x}\bigl(\mf E\cup w_l(x)\bigr)\Rightarrow\mathrm{wk}^{w}_{\gamma}\bigl(\mathrm{cowk}^{w^\perp}_{\delta_L,\delta_R}(\mf E)\bigr).$$
The codomain is derivable from the premise by rebuilding the interface. Starting from $\Theta;\Gamma'_L,\Gamma'_R\vdash_{\mf E}\Delta'$, we adjoin each variable of $\Gamma$ by $\mathsf{weak}$ at its mode and each output of $\Delta_L,\Delta_R$ by $\mathsf{coweak}$, with identity $2$-cells, exactly as in the principal case of \cref{prop:harmony-affine}. After composing the two $2$-cells, we apply Respect (\cref{prop:respect-for-structural-transformation}) to carry the sequent back to $\mf D_1\circ_{y,x}\mf D_2$. \emph{Left $\mathsf{coweak}$ at $y$:} dual, the left producer ending with $\mathsf{coweak}(l)$ at the cut output $y$ and its premise dropping it, reduced by $\mathrm{disc}^{\mathrm{op}}_{\mf D_2,\mf E}$.
\item \emph{$\mathsf v_{\mathsf l}$/$\mathsf v_{\mathsf l}$:} the type $A=P$ is atomic, the exact zones of the variable rule forcing a single input against a single output on either side:
$$
\inferrule*[right=$\mathsf v_{\mathsf l}$]
{
  P\in\mathrm{Prop}_{\Lambda,q} \\ q\model \\ r_L::\mf D_1\Rightarrow^{\Lambda}[\varnothing\mid\{a\mapsto y\}]
}
{
  \Theta\ ;\ a:P\vdash_{\mf D_1}^{\Lambda}y:P
}
$$
$$
\inferrule*[right=$\mathsf v_{\mathsf l}$]
{
  P\in\mathrm{Prop}_{\Lambda,q} \\ q\model \\ r_R::\mf D_2\Rightarrow^{\Lambda}[\varnothing\mid\{x\mapsto d\}]
}
{
  \Theta\ ;\ x:P\vdash_{\mf D_2}^{\Lambda}d:P
}
$$
By 2-Subst, $r_L\circ_{y,x}r_R::\mf D_1\circ_{y,x}\mf D_2\Rightarrow[\varnothing\mid\{a\mapsto y\}]\circ_{y,x}[\varnothing\mid\{x\mapsto d\}]$. Cutting the left output $y$ into the right input $x$ is a unit composition, the left redirection relabelling that input to $a$ by \cref{lem:units-for-composition}(i):
$$[\varnothing\mid\{a\mapsto y\}]\circ_{y,x}[\varnothing\mid\{x\mapsto d\}]=[\varnothing\mid\{a\mapsto d\}].$$
The target is again an identity class, so $\mathsf v_{\mathsf l}$ concludes $\Theta;a{:}P\vdash_{\mf D_1\circ_{y,x}\mf D_2}d{:}P$, exactly as in \cref{thm:cut-admissibility}.
\item \emph{Left $F_{\mathsf l}$R at $y$, right $F_{\mathsf l}$L at $x$} ($A=F^{y'}_{\mf D_F}(\Theta_F;\Gamma_F)$): this is the reduction of \cref{thm:cut-admissibility}, both sides principal so that the rule's own premise stands in place of the inversion. The left introduces $y:\mathsf F$, and the right consumes $x:\mathsf F$:
$$
\inferrule*[right=$F_{\mathsf l}$R]
{
  r_L::\mf D_1\Rightarrow^{\Lambda}\bigl(\mf D_{\bar x_i}\bigr)_i\circ_{(\bar x_i,x_i)}\bigl((\sigma_a)_{a\in\theta_F}\circ_{(b_a,a)}\mf D_F\bigr)
  \\
  \bigl(\Theta\vdash^{\Lambda}_{\sigma_a}A_a\bigr)_{a\in\theta_F}
  \\
  \bigl(\Theta\ ;\ \Gamma^{(i)}\vdash_{\mf D_{\bar x_i}}^{\Lambda}\Delta_L^{(i)},\bar x_i:A_i,\Delta_R^{(i)}\bigr)_{1\le i\le n}
}
{
  \Theta\ ;\ \Gamma^{(1)},\ldots,\Gamma^{(n)}\vdash_{\mf D_1}^{\Lambda}\Delta_L,y:\mathsf F,\Delta_R
}
$$
$$
\inferrule*[right=$F_{\mathsf l}$L]
{
  \Theta,\Theta_F\ ;\ \Gamma'_L,\Gamma_F,\Gamma'_R\vdash_{\mf D_F\circ_{y',x}\mf D_2}^{\Lambda}\Delta'
}
{
  \Theta\ ;\ \Gamma'_L,x:\mathsf F,\Gamma'_R\vdash_{\mf D_2}^{\Lambda}\Delta'
}
$$
We substitute the left rule's $\sigma$ into the $F_{\mathsf l}$L-premise by the cartesian case at the strictly smaller types of $\Theta_F$, turning its descriptor into $\bigl((\sigma_a)_a\circ\mf D_F\bigr)\circ_{y',x}\mf D_2$. We cut the left rule's $n$ components into the exposed inputs of $\Gamma_F$ by the multicut at the strictly smaller complexities $\mathrm{cx}(A_i)$, legal at any rank since the complexity leads the measure, and regroup by \cref{lem:localization,lem:associativity} into
$$\bigl((\mf D_{\bar x_i})_i\circ((\sigma_a)_a\circ\mf D_F)\bigr)\circ_{y,x}\mf D_2,$$
the bound label $y'$ $\alpha$-converted to $y$. The $2$-cell $r_L\circ_{y,x}\mathbf 1_{\mf D_2}$ from 2-Subst and \cref{prop:respect-for-structural-transformation} conclude. No context duplication occurs, the left derivation being spent exactly once through its components.
\item \emph{Left $U_{\mathsf l}$R at $y$, right $U_{\mathsf l}$L at $x$} ($A=U^{\mf D_U}_{x_U}(\Theta_U;\Gamma_U\mid\Delta_U)$): dual to the $F_{\mathsf l}$R/$F_{\mathsf l}$L shape, reducing through the left premise as in \cref{thm:cut-admissibility}. The left introduces $y:\mathsf U$, and the right consumes $x:\mathsf U$:
$$
\inferrule*[right=$U_{\mathsf l}$R]
{
  \Theta,\Theta_U\ ;\ \Gamma_U^L,\Gamma,\Gamma_U^R\vdash_{\mf D_1\circ_{y,x_U}\mf D_U}^{\Lambda}\Delta_L,\Delta_U,\Delta_R
}
{
  \Theta\ ;\ \Gamma\vdash_{\mf D_1}^{\Lambda}\Delta_L,y:\mathsf U,\Delta_R
}
$$
$$
\inferrule*[right=$U_{\mathsf l}$L]
{
  r_R::\mf D_2\Rightarrow^{\Lambda}\mf C
  \\
  \bigl(\Theta\vdash^{\Lambda}_{\sigma_a}A_a\bigr)_{a\in\theta_U}
  \\
  \bigl(\Theta\ ;\ \Gamma^{(i)}\vdash_{\mf D_{\bar x_i}}^{\Lambda}\Delta_L^{(i)},\bar x_i:A_i,\Delta_R^{(i)}\bigr)_{1\le i\le n}
  \\
  \bigl(\Theta\ ;\ \Gamma^{(n+j)},\bar y_j:B_j\vdash_{\mf D_{\bar y_j}}^{\Lambda}\Delta^{(n+j)}\bigr)_{1\le j\le m}
}
{
  \Theta\ ;\ \Gamma^{(1)},\ldots,\Gamma^{(n+m)},x:\mathsf U\vdash_{\mf D_2}^{\Lambda}\Delta'
}
$$
We substitute the right rule's family $\sigma$ into the $U_{\mathsf l}$R-premise by the cartesian case at the strictly smaller types of $\Theta_U$, and cut its $n$ input components into the exposed inputs $x_i:A_i$ of $\Gamma_U$ and the exposed outputs $y_j:B_j$ of $\Delta_U$ into its $m$ output components, by the multicut at the strictly smaller complexities $\mathrm{cx}(A_i),\mathrm{cx}(B_j)$, legal at any rank since the complexity leads the measure. We regroup by \cref{lem:localization,lem:associativity} into
$$\mf D_1\circ_{y,x}\mf C,$$
the instantiation $[x_U:=x]$ absorbed into the cut wire and the bound label $\alpha$-converted to $y$. The $2$-cell $\mathbf 1_{\mf D_1}\circ_{y,x}r_R$ from 2-Subst and \cref{prop:respect-for-structural-transformation} conclude. No context duplication occurs, the right derivation being spent exactly once through its components.
\end{itemize}
This exhausts all cases. Every step either lowers $\mathrm{cx}(A)$ outright or lowers $\rho_L+\rho_R$ at fixed complexity; the induction is therefore well founded. \hfill $\square$

\begin{proposition}[Harmony, classical case]\label{prop:harmony-classical}
Let $\Lambda$ declare all four gates at every linear mode, with their full axiom packages (\cref{def:gate-coherence-laws}). Then \cref{thm:cut-admissibility} holds for $\Lambda$.
\end{proposition}

\textit{Proof.}
The cartesian case extends within the simultaneous induction of \cref{thm:multicut}, its new cases (the gated rules) commuting by \cref{lem:gate-operations-commute}, and its principal cases unchanged. Its appeals to the linear cut at strictly smaller complexity are instances of \cref{thm:multicut}. The linear case is the instance $m=k=1$ of \cref{thm:multicut}, with empty trees. \hfill $\square$

\endgroup{}
\begingroup

\subsection{Equational Theories}

The descriptor $2$-cells of \cref{def:descriptor-2-cells} and the sequent rules of \cref{def:rule-schemas} are not yet proof-relevant. We introduce terms for both in this section. The categorical semantics of \cref{sec:categorical-semantics} requires identities between both kinds of terms. We impose those identities here.

\subsubsection{The Equational Theory of Descriptor \texorpdfstring{$2$}{2}-Cells}

The four rules of \cref{def:descriptor-2-cells} generate the $2$-cells as raw terms. Two given terms may denote the same structural transformation, and the equational theory identifies them.

\begin{definition}[Equational Theory of $2$-Cells]\label{def:equational-theory-of-2-cells}
Let $\Lambda\ \mathrm{Sig}$. We write $r\equiv r'$ for the least congruence with respect to $2$-Comp, $2$-Subst and the permutation transports of \cref{thm:exchange-admissibility-for-descriptor-2-cells} on parallel $2$-cell terms.
\begin{itemize}
  \item \textbf{Vertical laws.} For $r::\mf D\Rightarrow^{\Lambda}\mf E$,
  $$r\circ\mathbf 1_{\mf D}\equiv r\equiv\mathbf 1_{\mf E}\circ r,\qquad t\circ(s\circ r)\equiv(t\circ s)\circ r.$$
  \item \textbf{Functoriality of the cut.} The cut of identities is an identity and the cut of vertical composites is the vertical composite of the cuts,
  $$\mathbf 1_{\mf D}\circ_{y,x}\mathbf 1_{\mf E}\equiv\mathbf 1_{\mf D\circ_{y,x}\mf E},\qquad (s'\circ r')\circ_{y,x}(s\circ r)\equiv(s'\circ_{y,x}s)\circ(r'\circ_{y,x}r).$$
  \item \textbf{Units.} Cutting against the identity $2$-cell of an identity descriptor renames the boundary along the variable and nothing else,
  $$(\pi_a;\mathbf 1)_*\bigl(\mathbf 1_{[\varnothing\mid\{a\mapsto b\}]}\circ_{b,x}r\bigr)\equiv \mathrm{ren}_{x\mapsto a}(r),\qquad (\mathbf 1;\tau_b)_*\bigl(r\circ_{y,x_0}\mathbf 1_{[\varnothing\mid\{x_0\mapsto b\}]}\bigr)\equiv \mathrm{ren}_{y\mapsto b}(r),$$
  \item \textbf{Associativity and interchange.} $(r\circ_{y,x}s)\circ_{w,v}t\equiv r\circ_{y,x}(s\circ_{w,v}t)$ together with the left and right interchange equations for cuts at independent variables, in the configurations of \cref{lem:associativity}, \cref{lem:left-interchange-law} and \cref{lem:right-interchange-law}. 
  \item \textbf{Permutation equivariance.} For $2$-cells $t,s$ cut at the variables $y,x$, permutation transport distributes over the $2$-Subst cut,
  $$(\pi_1\cup\pi_2;\tau_1,\tau_2)_*(t\circ_{y,x}s)\equiv\bigl((\pi_1;\tau_y)_*t\bigr)\circ_{y,x}\bigl((\pi_x;\tau_2)_*s\bigr),$$
  in the configuration of \cref{lem:permutation-transport-commutes-with-cut}, the two sides parallel over $(\pi_1\cup\pi_2;\tau_1,\tau_2)_*(\mf D_1\circ_{y,x}\mf D_2)$.
  \item \textbf{Cartesian distribution and projection.} For a $2$-cell $u$ between c-descriptors producing $b$ and a cartesian input $a$ shared by the factors,
  $$u\circ_{b,a}(r\circ_{y,x}s)\equiv(u\circ_{b,a}r)\circ_{y,x}(u\circ_{b,a}s),$$
  the endpoints agreeing by \cref{lem:c-substitution-commutes-with-composition}
\end{itemize}
\end{definition}

\begin{example}[Linear--Nonlinear Logic]
In the signature of the \hyperref[ex:weakening-generating-2-cell]{weakening example}, the vertical laws identify the composite from the \hyperref[ex:linear-logic-descriptor-2-cell-terms]{descriptor $2$-cell example} with its generator,
$$\mathbf{1}_{\mf D_a} \circ \mathrm{gen}(\rho_{\otimes,a}) \equiv \mathrm{gen}(\rho_{\otimes,a}),$$
the two terms being parallel over $\gamma \Vdash \delta$. Functoriality of the cut identifies the cut of two identity terms with the identity of the composite class,
$$\mathbf{1}_{\mf D_c} \circ_{z',a'} \mathbf{1}_{\mf D_{\otimes}} \equiv \mathbf{1}_{\mf D_c \circ_{z',a'} \mf D_{\otimes}}.$$
The left-hand terms carry more syntax but no more structural content.
\end{example}

\subsubsection{Derivation Terms}

We turn to the derivation of terms for sequents in the calculus. This apparatus allows us to identify equality between derivations. Along the same lines as \cite{LSR17}, we write explicit terms for admissible operations such as transport and cut. Then we develop the equational theory on the terms attached to derivations. We treat the $\beta, \eta$ laws used in \cref{thm:cut-admissibility} and enforce the universal property of the connectives. We feel free to omit notation from the descriptor and sequent whenever they are not directly relevant for the term construction. For instance, we will usually omit the subscripts $x,y$ from compositions $(\--)\circ_{x,y}(\--)$, as they are determined by \cref{def:composition-of-descriptors}.
\begin{definition}[Derivation Terms]\label{def:derivation-terms}
Let $\Lambda\ \mathrm{Sig}$. The judgment $d::\Theta\ ;\ \Gamma\vdash^{\Lambda}_{\mf D}\Delta$ of \emph{derivation terms} is generated by these rules.
\begin{itemize}
  \item The connectives $F$-left, $F$-right, $U$-right and $U$-left generate the terms:
 $$\mathrm{FL}(d)  \qquad \mathrm{UR}(d)  \qquad \mathrm{FR}\bigl(r;(d_a)_a;(d_i)_i\bigr)  \qquad \mathrm{UL}\bigl(r;(d_a)_a;(d_i)_i;(d_j)_j\bigr) $$
  with these typings.

  $$
  \inferrule*
  {
    d :: \Theta, \Theta_F \; ; \; \Gamma_L, \Gamma_F, \Gamma_R \vdash_{\mf D_F \circ \mf D} \Delta
  }
  {
    \mathrm{FL}(d) :: \Theta \; ; \; \Gamma_L, z : \mathsf F, \Gamma_R \vdash_{\mf D} \Delta
  }
  \quad
  \inferrule*
  {
    d :: \Theta, \Theta_U \; ; \; \Gamma^L_U, \Gamma, \Gamma^R_U \vdash_{\mf D \circ \mf D_U} \Delta_L, \Delta_U, \Delta_R
  }
  {
    \mathrm{UR}(d) :: \Theta \; ; \; \Gamma \vdash_{\mf D} \Delta_L, z : \mathsf U, \Delta_R
  }
  $$

  $$
  \inferrule*
  {
    r :: \mf D \Rightarrow \left(\mf D_{\bar x_i}\right) \circ (\sigma_a \circ \mf D_F)\\
		d_a :: \Theta \vdash_{\sigma_a} A_a \\
		d_i :: \Theta \; ; \; \Gamma^i \vdash_{\mf D_{\bar x_i}} \Delta^i_L, \bar x_i : A_i, \Delta^i_R  
  }
  {
    \mathrm{FR}\bigl(r;(d_a)_a;(d_i)_i\bigr) :: \Theta \; ; \; \Gamma \vdash_{\mf D} \Delta_L, z : \mathsf F, \Delta_R 
  }
  $$

  $$
  \inferrule*
  {
    r :: \mf D \Rightarrow \left(\mf D_{\bar x_i}\right) \circ (\sigma_a \circ \mf D_U) \circ \left(\mf D_{\bar y_j}\right)\\
		d_a :: \Theta \vdash_{\sigma_a} A_a \\
		d_i :: \Theta \; ; \; \Gamma^i \vdash_{\mf D_{\bar x_i}} \Delta^i_L, \bar x_i : A_i, \Delta^i_R \\ 
		d_j :: \Theta \; ; \; \Gamma^{n + j}_L, \bar y_j : B_j,\Gamma^{n + j}_R \ \vdash_{\mf D_{\bar y_j}} \Delta^{n + j}  
  }
  {
    \mathrm{UL}\bigl(r;(d_a)_a;(d_i)_i;(d_j)_j\bigr) :: \Theta \; ; \; \Gamma,\, z : \mathsf U \vdash_{\mf D} \Delta
  }
  $$

  \item The gated structural rules of \cref{def:gated-structural-rules} when $\Lambda$ declares gates, written $\mathsf{weak}(r;d)$, $\mathsf{coweak}(r;d)$, $\mathsf{contr}(r;d)$ and $\mathsf{cocontr}(r;d)$.
  \item Three explicit rules internalizing the admissible operations of the metatheory,
  $$
  \inferrule*[Right=Id]
  {A\ \mathrm{type}_{\Lambda,q}}
  {\mathrm{id}_A::\Theta\ ;\ x:A\vdash^{\Lambda}_{[\varnothing\mid\{x\mapsto y\}]}y:A}
  \qquad\qquad
  \inferrule*[Right=Tr]
  {r::\mf D\Rightarrow^{\Lambda}\mf E \\ d::\Theta\ ;\ \Gamma\vdash^{\Lambda}_{\mf E}\Delta}
  {r_*(d)::\Theta\ ;\ \Gamma\vdash^{\Lambda}_{\mf D}\Delta}
  $$
  $$
  \inferrule*[Right=Cut]
  {d_1::\Theta\ ;\ \Gamma\vdash^{\Lambda}_{\mf D_1}\Delta_L,y:A,\Delta_R \\ d_2::\Theta\ ;\ \Gamma'_L,x:A,\Gamma'_R\vdash^{\Lambda}_{\mf D_2}\Delta' \\ \pi:\Theta\ ;\ \Gamma'_L,\Gamma,\Gamma'_R\cong\Theta_1\ ;\ \Gamma_1 \\ \tau:\Delta_L,\Delta',\Delta_R\cong\Delta_1}
  {(\pi;\tau)_*(d_1\circ_{y,x}d_2)::\Theta_1\ ;\ \Gamma_1\vdash^{\Lambda}_{(\pi;\tau)_*(\mf D_1\circ_{y,x}\mf D_2)}\Delta_1}
  $$
  with the boundaries of \cref{thm:identity}, \cref{prop:respect-for-structural-transformation} and \cref{thm:cut-admissibility}; the ambient context $\Theta$ ranges in each rule over arbitrary well-formed cartesian contexts.
\end{itemize}
\end{definition}

\textit{Remark (the variable rule):} The careful reader might have noticed that the variable rule is not among the generators. This is because it is recovered from the transport and the identity rule considering $r_*(\mathrm{id}_P)$.

\begin{lemma}[Boundary Renaming of Derivation Terms]\label{lem:boundary-renaming-of-derivation-terms}
Let $d::\Theta\ ;\ \Gamma\vdash^{\Lambda}_{\mf D}\Delta$ be a derivation term and let $\sigma$ be a boundary renaming of the underlying mode boundary (\cref{def:boundary-renaming}), lifted to the typed contexts. There is a derivation term
$$\mathrm{ren}_{\sigma}(d)::\Theta^{\sigma}\ ;\ \Gamma^{\sigma}\vdash^{\Lambda}_{\mathrm{ren}_{\sigma}(\mf D)}\Delta^{\sigma},$$
obtained by recursion on $d$ after choosing the descriptor $2$-cell witnesses
needed in the $2$-cell parameters. Every such recursion commutes with the term
constructors, sends derivations of the sequent calculus to derivations, and
preserves the equational theory of \cref{def:equational-theory-of-derivations}.
\end{lemma}

\textit{Proof.}
By recursion on $d$, over the formers of \cref{def:derivation-terms}, exactly as for \cref{lem:permutation-transport-of-derivation-terms}.
\begin{itemize}
  \item \textbf{Id.} The instance at the renamed boundary is again an instance, and its index is the identity class at the renamed variable, i.e. $\mathrm{ren}_{\sigma}$ of the identity class.
  \item \textbf{Transport.} If $d=r_*(d')$, we set $\mathrm{ren}_{\sigma}(r_*(d')):=(\mathrm{ren}_{\sigma}(r))_*(\mathrm{ren}_{\sigma}(d'))$, the $2$-cell renamed by \cref{lem:boundary-renaming-of-descriptor-2-cells} and the premise by the recursion, the endpoints matching by the same lemma.
  \item \textbf{Cut.} If $d=(\pi;\tau)_*(d_1\circ_{y,x}d_2)$, we rename the premises along the restrictions of $\sigma$ extended by the identity.
  \item \textbf{F-left and U-right.} The premise carries the connective blocks $\Theta_F$ and $\Gamma_F$ next to the ambient contexts. The recursion renames it along $\sigma$ extended by the identity on the blocks, the principal renamed with its zone, and the premise indexes agree since the renaming touches neither the connective descriptor nor the variable (\cref{lem:boundary-renaming-is-well-defined}). $U$-right is the mirror computation on the output side.
  \item \textbf{F-right and U-left.} The family premises rename at the renamed $\Theta$ by the recursion, each component renames within its slice, the fresh output labels kept fixed, and the $2$-cell premise renames by \cref{lem:boundary-renaming-of-descriptor-2-cells}, its target stage by stage through \cref{lem:boundary-renaming-is-well-defined}. $U$-left adds the output components, renamed in the same fashion.
  \item \textbf{Gated rules.} The $2$-cell renames by \cref{lem:gated-2-cell-metatheory}, the premise by the recursion, and the conclusion is the instance at the renamed boundary.
\end{itemize}
The recursion sends each rule to an instance of the same rule and therefore
sends derivations of the calculus to derivations. No raw equality between
different choices of recursively transported $2$-cell parameters is claimed.
Their identity and composition laws are imposed in
\cref{def:coherent-transport-completion}. \hfill $\square$

\begin{lemma}[Permutation Transport of Derivation Terms]\label{lem:permutation-transport-of-derivation-terms}
Let $d::\Theta\ ;\ \Gamma\vdash^{\Lambda}_{\mf D}\Delta$ be a derivation term and let $\pi:\Theta\ ;\ \Gamma\cong\Theta_1\ ;\ \Gamma_1$ and $\tau:\Delta\cong\Delta_1$ be context permutations of its boundary respecting the zones. There is a derivation term
$$(\pi;\tau)_*(d)::\Theta_1\ ;\ \Gamma_1\vdash^{\Lambda}_{(\pi;\tau)_*\mf D}\Delta_1,$$
obtained by recursion on $d$ after choosing the transported generating
$2$-cells used by its $2$-cell parameters. Each such recursion sends every
former to an instance of the same former and preserves the equational theory
of \cref{def:equational-theory-of-derivations}. Its identity and composition laws are imposed by
\cref{def:coherent-transport-completion}.
\end{lemma}

\textit{Proof.}
By recursion on $d$, over the formers of \cref{def:derivation-terms}.
\begin{itemize}
  \item \textbf{Cut.} The former carries closing permutations. If $d=(\pi_0;\tau_0)_*(d_1\circ_{y,x}d_2)$, we apply the same rule to the same two premises with the composite permutations $\pi\circ\pi_0$ and $\tau\circ\tau_0$. Its index is $((\pi\circ\pi_0);(\tau\circ\tau_0))_*(\mf D_1\circ_{y,x}\mf D_2)=(\pi;\tau)_*\mf D$ by \cref{prop:functoriality-of-permutation-transport}, exactly as in the Cut case of \cref{thm:exchange-admissibility-for-descriptor-judgments} and the $2$-Subst case of \cref{thm:exchange-admissibility-for-descriptor-2-cells}.
  \item \textbf{Id.} The ambient context $\Theta$ is arbitrary data of the rule, the instance at the permuted context is again an instance, and the principal boundary is a single declaration on each side, i.e. the identity descriptor transports to itself.
  \item \textbf{Transport.} If $d=r_*(d')$ with $r::\mf D\Rightarrow^{\Lambda}\mf E$, we set $(\pi;\tau)_*(r_*(d')):=((\pi;\tau)_*r)_*((\pi;\tau)_*(d'))$, the $2$-cell transported by \cref{thm:exchange-admissibility-for-descriptor-2-cells} and the premise by the recursion, the endpoints matching by the same theorem.
  \item \textbf{F-left and U-right.} For $d=\mathrm{FL}(d')$ the conclusion declares the principal $z:\mathsf F$ in the zone of its class, and the permuted boundary is again of the schema shape, the principal moved along $\pi$. The premise transports by the recursion along the permutation extending $\pi$ by the identity on the blocks $\Theta_F$ and $\Gamma_F$, kept at the position of the principal, and the premise indexes agree by \cref{lem:permutation-transport-commutes-with-cut}, the transport touching neither the connective descriptor nor the variable. We set $(\pi;\tau)_*(\mathrm{FL}(d')):=\mathrm{FL}((\pi';\tau)_*(d'))$. The $U$-right former is the mirror computation on the output side, the premise transported along the extension of $\tau$ by the identity on $\Delta_U$.
  \item \textbf{F-right and U-left.} For $d=\mathrm{FR}\bigl(r;(d_a)_a;(d_i)_i\bigr)$ the slice-interleaving convention (\cref{def:rule-schemas}) makes the arrangement of the component slices in the conclusion boundary instance data closed by the permutations; the schema constrains only how the side contexts split among the components. The permutation $(\pi;\tau)$ therefore redistributes the slices into the slicing data of the transported instance, each component $d_i$ transported within its slice by the recursion, each family premise $d_a$ transported at the permuted $\Theta$, and the $2$-cell premise transported by \cref{thm:exchange-admissibility-for-descriptor-2-cells}, its target rearranged stage by stage through \cref{lem:permutation-transport-commutes-with-cut}. The $U$-left former adds the output components $(d_j)_j$, transported within their slices in the same fashion.
  \item \textbf{Gated rules.} The four formers carry a $2$-cell and a premise term. The $2$-cell transports by \cref{lem:gated-2-cell-metatheory}, the premise by the recursion, and the conclusion is the instance at the moved principal, as for $F$-left.
\end{itemize}
The recursion sends each rule to an instance of the same rule. At descriptors,
the endpoints of successive and composite transports agree by
\cref{prop:functoriality-of-permutation-transport}; at generating $2$-cell
parameters the signature supplies existence but no equality between the
chosen witnesses. \hfill $\square$

\begin{lemma}[Cartesian Weakening of Derivation Terms]\label{lem:cartesian-weakening-of-derivation-terms}
Let $d::\Theta\ ;\ \Gamma\vdash^{\Lambda}_{\mf D}\Delta$ be a derivation term and let $x':A$ be a declaration at a cartesian mode $p$, its variable fresh for the boundary. There is a derivation term
$$\mathrm{wk}_{x':A}(d)::\Theta,x':A\ ;\ \Gamma\vdash^{\Lambda}_{\mathrm{wk}^{x':p}_{\mathrm{in}}(\mf D)}\Delta,$$
obtained by recursion on $d$ after choosing the weakened generating
$2$-cells used by its $2$-cell parameters. Every such recursion sends each
former to an instance of the same former and preserves the equational theory
of \cref{def:equational-theory-of-derivations}. We write $\mathrm{wk}_{\Theta'}(d)$ for iterated weakening in the
displayed order. Independence of the chosen witnesses and of alternative
orders is imposed in \cref{def:coherent-transport-completion}.
\end{lemma}

\textit{Proof.}
By recursion on $d$.
\begin{itemize}
  \item \textbf{Id.} The ambient context is arbitrary data of the rule and the instance at $\Theta,x':A$ absorbs the enlarged zone, its index at the enlarged zone being the input weakening of the identity descriptor.
  \item \textbf{Transport.} If $d=r_*(d')$, we set $\mathrm{wk}_{x':A}(r_*(d')):=\bigl(\mathrm{wk}^{x':p}_{\mathrm{in}}(r)\bigr)_*\bigl(\mathrm{wk}_{x':A}(d')\bigr)$, the $2$-cell weakened by \cref{lem:ambient-weakening-of-descriptor-2-cells} and the premise by the recursion, the endpoints matching by the same lemma.
  \item \textbf{Cut.} The two premises share the cartesian zone $\Theta$ and both weaken by the recursion, the closing permutations extending by the identity on $x'$ exactly as in the $2$-Subst case of \cref{lem:ambient-weakening-of-descriptor-2-cells}. The indexes agree by \cref{lem:weakening-commutes-with-composition}; this is the same-variable configuration in which both factors read the united zone.
  \item \textbf{F-left and U-right.} The premise of $\mathrm{FL}(d')$ carries the ambient zone $\Theta,\Theta_F$. The recursion weakens it at the end of the zone and the transport of \cref{lem:permutation-transport-of-derivation-terms} moves the fresh declaration across $\Theta_F$, the premise standing then over $(\Theta,x':A),\Theta_F$ as the instance at the enlarged conclusion requires. The premise indexes agree by \cref{lem:weakening-commutes-with-composition}, the union of the ambient contexts reading the fresh declaration once. $U$-right is the same computation, the connective blocks on the output side untouched.
  \item \textbf{F-right and U-left.} Every premise carries the ambient zone $\Theta$. The family premises, the components and, for $U$-left, the output components weaken by the recursion, the $2$-cell premise by \cref{lem:ambient-weakening-of-descriptor-2-cells}, and its target rearranges stage by stage through \cref{lem:weakening-commutes-with-composition}.
  \item \textbf{Gated rules.} The $2$-cell weakens by \cref{lem:gated-2-cell-metatheory}, the premise by the recursion, and the conclusion is the instance at the enlarged zone.
\end{itemize}

The right rule applied with identities produces the \emph{generic term} of a connective, the term-level version of the sequents that the invertibility propositions extract from the identity, and the analogue of the axiomatic $\mathrm{FR}^{*}$ and $\mathrm{UL}^{*}$ of \cite{LSR17}. In the sequent calculus these sequents are derived inside the proof of \cref{thm:identity} by the induction on the type. In the term calculus the rule Id is among the formers, the construction is a single instance of the right or left rule, and its well-formedness is a consequence of the definitions already in place. We record it as a proposition.

\begin{proposition}[Generic Terms]\label{prop:generic-terms}
Let $\mathsf F=F^{y}_{\mf D_F}(\Theta_F;\Gamma_F)$ with $\Theta_F=(a:A_a)_{a\in\theta_F}$ and $\Gamma_F=x_1:A_1,\ldots,x_n:A_n$, and let $\bar z$ be a fresh output variable. Choose a fresh boundary-renamed copy $\widetilde{\mf D}_F$ of $\mf D_F$, with inputs $\widetilde\Theta_F=(\tilde a:A_a)_{a\in\theta_F}$ and $\tilde x_1:A_1,\ldots,\tilde x_n:A_n$, and output $\tilde y$. By \hyperref[rem:context-variable-naming-invariance]{context variable naming invariance}, this copy presents the same type $\mathsf F$. Let
$$
\sigma_{\tilde a}:=[\varnothing\mid\{a\mapsto b_{\tilde a}\}],
\qquad
\mf D_{\bar x_i}:=[\varnothing\mid\{x_i\mapsto\bar x_i\}],
$$
where all $b_{\tilde a}$ and $\bar x_i$ are fresh. Each $\sigma_{\tilde a}$ is regarded over the common ambient context $\Theta_F$, the other declarations unused. Define
$$
\mf C_F:=
\left(
  \bigl(\mf D_{\bar x_i}\bigr)_i
  \circ_{(\bar x_i,\tilde x_i)_{1\le i\le n}}
  \left(
    \bigl(\sigma_{\tilde a}\bigr)_{a\in\theta_F}
    \circ_{(b_{\tilde a},\tilde a)_{a\in\theta_F}}
    \widetilde{\mf D}_F
  \right)
\right)[\tilde y:=\bar z].
$$
Then there is a derivation of the \emph{generic term} of $\mathsf F$:
$$
\eta_{\mathsf F}\ :=\ \mathrm{FR}\bigl(\mathbf 1_{\mf D_F[y:=\bar z]};(\mathrm{id}_{A_a})_a;(\mathrm{id}_{A_i})_i\bigr)\ ::\ \Theta_F\ ;\ \Gamma_F\vdash^{\Lambda}_{\mf D_F[y:=\bar z]}\bar z:\mathsf F,
$$

Dually, let $\mathsf U=U^{\mf D_U}_{x_U}(\Theta_U;\Gamma_U\mid\Delta_U)$ with $\Theta_U=(a:A_a)_{a\in\theta_U}$ and $\Gamma_U=\Gamma_U^L,\Gamma_U^R$, enumerate the latter context as $x_1:A_1,\ldots,x_n:A_n$, and let $\Delta_U=y_1:B_1,\ldots,y_m:B_m$. Let $x_0$ be fresh and write
$$
\widehat\Theta_U:=
\begin{cases}
\Theta_U,x_0:\mathsf U,& \mathsf U\text{ cartesian},\\
\Theta_U,& \mathsf U\text{ linear},
\end{cases}
\qquad
\widehat\Gamma_U:=
\begin{cases}
\Gamma_U^L,\Gamma_U^R,& \mathsf U\text{ cartesian},\\
\Gamma_U^L,x_0:\mathsf U,\Gamma_U^R,& \mathsf U\text{ linear}.
\end{cases}
$$
Choose a fresh boundary-renamed copy $\widetilde{\mf D}_U$ of $\mf D_U$, with inputs $\widetilde\Theta_U=(\tilde a:A_a)_{a\in\theta_U}$, inputs $\tilde x_i:A_i$ and $\tilde x_U$, and outputs $\tilde y_j:B_j$. By \hyperref[rem:context-variable-naming-invariance]{context variable naming invariance}, this copy presents the same type $\mathsf U$. Let
$$
\sigma_{\tilde a}:=[\varnothing\mid\{a\mapsto b_{\tilde a}\}],
\qquad
\mf D_{\bar x_i}:=[\varnothing\mid\{x_i\mapsto\bar x_i\}],
\qquad
\mf D_{\bar y_j}:=[\varnothing\mid\{\bar y_j\mapsto y_j\}],
$$
Each $\sigma_{\tilde a}$ is regarded over the common ambient context $\widehat\Theta_U$, the other declarations unused. When $\theta_U$ is empty, the displayed family composite is the empty family c-substitution of \cref{def:c-substitution}, and therefore retains $\widehat\Theta_U$ by ambient weakening. Define
$$
\mf G_U:=
\left(
  \bigl(\sigma_{\tilde a}\bigr)_{a\in\theta_U}
  \circ_{(b_{\tilde a},\tilde a)_{a\in\theta_U}}
  \widetilde{\mf D}_U
\right)[\tilde x_U:=x_0],
$$
$$
\mf C_U:=
\left(
  \bigl(\mf D_{\bar x_i}\bigr)_i
  \circ_{(\bar x_i,\tilde x_i)_{1\le i\le n}}
  \mf G_U
\right)
\circ_{(\tilde y_j,\bar y_j)_{1\le j\le m}}
\bigl(\mf D_{\bar y_j}\bigr)_j.
$$
Then there is a derivation of the \emph{generic term } of $\mathsf U$ :
$$
\varepsilon_{\mathsf U}\ :=\ \mathrm{UL}\bigl(\mathbf 1_{\mf D_U[x_U:=x_0]};(\mathrm{id}_{A_a})_a;(\mathrm{id}_{A_i})_i;(\mathrm{id}_{B_j})_j\bigr)\ ::\ \widehat\Theta_U\ ;\ \widehat\Gamma_U\vdash^{\Lambda}_{\mf D_U[x_U:=x_0]}\Delta_U,
$$
\end{proposition}

\textit{Proof.}
For $F$, use the Id instances
$$
\bigl(\mathrm{id}_{A_a}::\Theta_F\vdash^{\Lambda}_{\sigma_{\tilde a}}b_{\tilde a}:A_a\bigr)_{a\in\theta_F},
\qquad
\bigl(\mathrm{id}_{A_i}::\Theta_F\ ;\ x_i:A_i\vdash^{\Lambda}_{\mf D_{\bar x_i}}\bar x_i:A_i\bigr)_{1\le i\le n}.
$$
in the right introduction rule.

For $U$, take the Id instances
$$
\bigl(\mathrm{id}_{A_a}::\widehat\Theta_U\vdash^{\Lambda}_{\sigma_{\tilde a}}b_{\tilde a}:A_a\bigr)_{a\in\theta_U},
$$
$$
\bigl(\mathrm{id}_{A_i}::\widehat\Theta_U\ ;\ x_i:A_i\vdash^{\Lambda}_{\mf D_{\bar x_i}}\bar x_i:A_i\bigr)_{1\le i\le n},
\qquad
\bigl(\mathrm{id}_{B_j}::\widehat\Theta_U\ ;\ \bar y_j:B_j\vdash^{\Lambda}_{\mf D_{\bar y_j}}y_j:B_j\bigr)_{1\le j\le m},
$$
in the left introduction rules.  \hfill $\square$

\begin{example}[Linear--Nonlinear Logic, generic terms]
Recall from \cref{ex:linear-logic-type-formers,ex:linear-logic-multiplicative-units} the descriptors
$$
\mf D_\otimes=[\,x,y\xmapsto{\otimes_l}w\mid w\mapsto z\,],
\qquad
\mf D_{\parr}=[\,z\xmapsto{\parr}w_1,w_2\mid w_1\mapsto x,w_2\mapsto y\,]
$$
and the induced connectives $A\otimes B=F^z_{\mf D_\otimes}(\cdot\ ;x:A,y:B)$ and $A\parr B=U_z^{\mf D_{\parr}}(\cdot\ ;\cdot\mid x:A,y:B)$ at a linear mode. The generic terms of \cref{prop:generic-terms} specialize to these displays, in which the fresh bound copies of the connective descriptors are suppressed:
$$
\eta_{A\otimes B}
:=
\mathrm{FR}\bigl(\mathbf 1_{\mf D_\otimes[z:=\bar t]};();(\mathrm{id}_A,\mathrm{id}_B)\bigr)
::
\cdot\ ;\ x:A,y:B
\vdash^\Lambda_{\mf D_\otimes[z:=\bar t]}
\bar t:A\otimes B
$$
and
$$
\varepsilon_{A\parr B}
:=
\mathrm{UL}\bigl(\mathbf 1_{\mf D_{\parr}[z:=p_0]};();();(\mathrm{id}_A,\mathrm{id}_B)\bigr)
::
\cdot\ ;\ p_0:A\parr B
\vdash^\Lambda_{\mf D_{\parr}[z:=p_0]}
x:A,y:B.
$$
Here $()$ denotes an empty premise family. Thus the first term has two identity input components and names the familiar tensor introduction $A,B\vdash A\otimes B$, whereas the second has two identity output components and names the canonical exposure $A\parr B\vdash A,B$.

The same construction for $A\multimap B=U_y^{\mf D_\otimes}(\cdot\ ;x:A\mid z:B)$ gives the linear application term
$$
\operatorname{app}_{A,B}
:=
\mathrm{UL}\bigl(\mathbf 1_{\mf D_\otimes[y:=f]};();(\mathrm{id}_A);(\mathrm{id}_B)\bigr)
::
\cdot\ ;\ x:A,f:A\multimap B
\vdash^\Lambda_{\mf D_\otimes[y:=f]}
z:B.
$$
Applying $F$-left packages its two inputs and recovers the evaluation map from \cref{ex:linear-logic-type-formers},
$$
\operatorname{ev}_{A,B}:=\mathrm{FL}(\operatorname{app}_{A,B})
::
\cdot\ ;\ t:A\otimes(A\multimap B)
\vdash^\Lambda_{[\varnothing\mid\{t\mapsto z\}]}
z:B,
$$
where renaming of the variables is used by \hyperref[rem:context-variable-naming-invariance]{context variable naming invariance}.
\end{example}

\subsubsection{The Equational Theory of Derivations}

\begin{definition}[Equational Theory of Derivations]\label{def:equational-theory-of-derivations}
Let $\Lambda\ \mathrm{Sig}$. We write $d\equiv d'$ for the least congruence with respect to all the term formers of \cref{def:derivation-terms} and the operations of \cref{lem:boundary-renaming-of-derivation-terms}, \cref{lem:permutation-transport-of-derivation-terms} and \cref{lem:cartesian-weakening-of-derivation-terms}, on terms of a common boundary and descriptor, containing the equations listed here. Under each equation we display the two derivations it relates for visualization purposes.

\textbf{Cut and identity.}
\begin{itemize}
  \item \emph{Units.} $\mathrm{id}_A\circ_{y,x}d\equiv \mathrm{ren}_{x\mapsto a}(d)$ and $d\circ_{y,x_0}\mathrm{id}_A\equiv \mathrm{ren}_{y\mapsto b}(d)$, for the identity instances $\mathrm{id}_A::\Theta\ ;\ a:A\vdash^{\Lambda} y:A$ and $\mathrm{id}_A::\Theta\ ;\ x_0:A\vdash^{\Lambda} b:A$. The composition renames the boundary and nothing else, the two sides parallel by \cref{lem:boundary-renaming-of-derivation-terms}, the descriptors agreeing by \cref{lem:units-for-composition}. For $d::\Theta\ ;\ \Gamma'_L,x:A,\Gamma'_R\vdash^{\Lambda}_{\mf D}\Delta'$ the first law equates
  $$
  \inferrule*[Right=Cut]
  {
    \inferrule*[Right=Id]
    {A\ \mathrm{type}_{\Lambda,q}}
    {\mathrm{id}_A::\Theta\ ;\ a:A\vdash^{\Lambda}_{[\varnothing\mid\{a\mapsto y\}]}y:A}
    \\
    d::\Theta\ ;\ \Gamma'_L,x:A,\Gamma'_R\vdash^{\Lambda}_{\mf D}\Delta'
  }
  {
    \mathrm{id}_A\circ_{y,x}d::\Theta\ ;\ \Gamma'_L,a:A,\Gamma'_R\vdash^{\Lambda}_{[\varnothing\mid\{a\mapsto y\}]\circ_{y,x}\mf D}\Delta'
  }
  \quad\equiv\quad
  \mathrm{ren}_{x\mapsto a}(d),
  $$
  and for $d::\Theta\ ;\ \Gamma\vdash^{\Lambda}_{\mf D}\Delta_L,y:A,\Delta_R$ the second equates
  $$
  \inferrule*[Right=Cut]
  {
    d::\Theta\ ;\ \Gamma\vdash^{\Lambda}_{\mf D}\Delta_L,y:A,\Delta_R
    \\
    \inferrule*[Right=Id]
    {A\ \mathrm{type}_{\Lambda,q}}
    {\mathrm{id}_A::\Theta\ ;\ x_0:A\vdash^{\Lambda}_{[\varnothing\mid\{x_0\mapsto b\}]}b:A}
  }
  {
    d\circ_{y,x_0}\mathrm{id}_A::\Theta\ ;\ \Gamma\vdash^{\Lambda}_{\mf D\circ_{y,x_0}[\varnothing\mid\{x_0\mapsto b\}]}\Delta_L,b:A,\Delta_R
  }
  \quad\equiv\quad
  \mathrm{ren}_{y\mapsto b}(d),
  $$
  \item \emph{Associativity.} $(d_1\circ_{y,x}d_2)\circ_{w,v}d_3\equiv d_1\circ_{y,x}(d_2\circ_{w,v}d_3)$. Associativity in \cref{lem:associativity} supplies the equality of the indexes. These derivations display the two different ways to associate the descriptor:
  $$
  \inferrule*[Right=Cut]
  {
    \inferrule*[Right=Cut]
    {d_1 \\ d_2}
    {d_1\circ_{y,x}d_2::\Theta\ ;\ \Gamma'_L,\Gamma,\Gamma'_R\vdash^{\Lambda}_{\mf D_1\circ_{y,x}\mf D_2}\Delta_L,\Delta'_L,w:B,\Delta'_R,\Delta_R}
    \\
    d_3
  }
  {
    (d_1\circ_{y,x}d_2)\circ_{w,v}d_3::\Theta\ ;\ \Gamma''_L,\Gamma'_L,\Gamma,\Gamma'_R,\Gamma''_R\vdash^{\Lambda}_{(\mf D_1\circ_{y,x}\mf D_2)\circ_{w,v}\mf D_3}\Delta_L,\Delta'_L,\Delta'',\Delta'_R,\Delta_R
  }
  $$
  $$
  \inferrule*[Right=Cut]
  {
    d_1
    \\
    \inferrule*[Right=Cut]
    {d_2 \\ d_3}
    {d_2\circ_{w,v}d_3::\Theta\ ;\ \Gamma''_L,\Gamma'_L,x:A,\Gamma'_R,\Gamma''_R\vdash^{\Lambda}_{\mf D_2\circ_{w,v}\mf D_3}\Delta'_L,\Delta'',\Delta'_R}
  }
  {
    d_1\circ_{y,x}(d_2\circ_{w,v}d_3)::\Theta\ ;\ \Gamma''_L,\Gamma'_L,\Gamma,\Gamma'_R,\Gamma''_R\vdash^{\Lambda}_{\mf D_1\circ_{y,x}(\mf D_2\circ_{w,v}\mf D_3)}\Delta_L,\Delta'_L,\Delta'',\Delta'_R,\Delta_R
  }
  $$
  \item \emph{Interchange.} The left and right interchange equations for cuts at independent variables, in the laws \cref{lem:left-interchange-law} and \cref{lem:right-interchange-law}. This display equates the two different orders of cuts:
  $$
  \inferrule*[Right=Cut]
  {
    d_1
    \\
    \inferrule*[Right=Cut]
    {d_2 \\ e}
    {d_2\circ_{y_2,x_2}e::\Theta\ ;\ \Gamma_L,x_1:A_1,\Gamma_M,\Gamma_2,\Gamma_R\vdash^{\Lambda}_{\mf D_2\circ_{y_2,x_2}\mf E}\Delta_2,\Delta}
  }
  {
    d_1\circ_{y_1,x_1}(d_2\circ_{y_2,x_2}e)::\Theta\ ;\ \Gamma_L,\Gamma_1,\Gamma_M,\Gamma_2,\Gamma_R\vdash^{\Lambda}_{\mf D_1\circ_{y_1,x_1}(\mf D_2\circ_{y_2,x_2}\mf E)}\Delta_1,\Delta_2,\Delta
  }
  $$
  to the transported term of the same two cuts in the other order, $(\mathbf 1;\tau)_*\bigl(d_2\circ_{y_2,x_2}(d_1\circ_{y_1,x_1}e)\bigr)$, the closing permutation $\tau:\Delta_2,\Delta_1,\Delta\cong\Delta_1,\Delta_2,\Delta$ matching the two displayed conclusions so that the two sides are parallel. For consumers $d_i::\Theta\ ;\ x_i:A_i,\Gamma_i\vdash^{\Lambda}_{\mf D_i}\Delta_i$ of a producer $e::\Theta\ ;\ \Gamma\vdash^{\Lambda}_{\mf E}\Delta,y_1:A_1,y_2:A_2$, the right equation relates
  $$
  \inferrule*[Right=Cut]
  {
    \inferrule*[Right=Cut]
    {e \\ d_1}
    {e\circ_{y_1,x_1}d_1::\Theta\ ;\ \Gamma,\Gamma_1\vdash^{\Lambda}_{\mf E\circ_{y_1,x_1}\mf D_1}\Delta,\Delta_1,y_2:A_2}
    \\
    d_2
  }
  {
    (e\circ_{y_1,x_1}d_1)\circ_{y_2,x_2}d_2::\Theta\ ;\ \Gamma,\Gamma_1,\Gamma_2\vdash^{\Lambda}_{(\mf E\circ_{y_1,x_1}\mf D_1)\circ_{y_2,x_2}\mf D_2}\Delta,\Delta_1,\Delta_2
  }
  $$
  to the transported term $(\pi;\mathbf 1)_*\bigl((e\circ_{y_2,x_2}d_2)\circ_{y_1,x_1}d_1\bigr)$, the closing permutation $\pi:\Gamma,\Gamma_2,\Gamma_1\cong\Gamma,\Gamma_1,\Gamma_2$ matching the two displayed conclusions so that the two sides are parallel.
  \item \emph{Cartesian projection.} $u\circ_{b,x'}\mathrm{wk}_{x':A}(d)\equiv d$ for a cartesian sequent $u::\Theta\vdash^{\Lambda}_{\sigma}b:A$ and a term $d::\Theta\ ;\ \Gamma\vdash^{\Lambda}_{\mf D}\Delta$, the weakening being the operation of \cref{lem:cartesian-weakening-of-derivation-terms}. Cutting a variable that has been weakened discards the producer and retains the original term. The two sides are parallel over the shared zone $\Theta$, the indexes agreeing since a composition at an input without occurrences retains the consumer alone (\cref{def:raw-composition}), the ambient context of $\sigma$ already present in the union.
  $$
  \inferrule*[Right=Cut]
  {
    u::\Theta\vdash^{\Lambda}_{\sigma}b:A
    \\
    \mathrm{wk}_{x':A}(d)::\Theta,x':A\ ;\ \Gamma\vdash^{\Lambda}_{\mathrm{wk}^{x':p}_{\mathrm{in}}(\mf D)}\Delta
  }
  {
    u\circ_{b,x'}\mathrm{wk}_{x':A}(d)::\Theta\ ;\ \Gamma\vdash^{\Lambda}_{\sigma\circ_{b,x'}\mathrm{wk}^{x':p}_{\mathrm{in}}(\mf D)}\Delta
  }
  \quad\equiv\quad d,
  $$
  \item \emph{Cartesian distribution.} $u\circ_{b,a}(d_1\circ_{y,x}d_2)\equiv(u\circ_{b,a}d_1)\circ_{y,x}(u\circ_{b,a}d_2)$, in the case that the cartesian variable $a$ is shared by the two factors. The descriptors agree by \cref{lem:c-substitution-commutes-with-composition}. For $d_1::\Theta,a:A\ ;\ \Gamma_1\vdash^{\Lambda}_{\mf D_1}\Delta_1,y:B$ and $d_2::\Theta,a:A\ ;\ x:B,\Gamma_2\vdash^{\Lambda}_{\mf D_2}\Delta_2$ the two sides are the derivations
  $$
  \inferrule*[Right=Cut]
  {
    u::\Theta\vdash^{\Lambda}_{\sigma}b:A
    \\
    \inferrule*[Right=Cut]
    {d_1 \\ d_2}
    {d_1\circ_{y,x}d_2::\Theta,a:A\ ;\ \Gamma_1,\Gamma_2\vdash^{\Lambda}_{\mf D_1\circ_{y,x}\mf D_2}\Delta_1,\Delta_2}
  }
  {
    u\circ_{b,a}(d_1\circ_{y,x}d_2)::\Theta\ ;\ \Gamma_1,\Gamma_2\vdash^{\Lambda}_{(\mf D_1\circ_{y,x}\mf D_2)[a:=\sigma]}\Delta_1,\Delta_2
  }
  $$
  $$
  \inferrule*[Right=Cut]
  {
    \inferrule*[Right=Cut]
    {u \\ d_1}
    {u\circ_{b,a}d_1::\Theta\ ;\ \Gamma_1\vdash^{\Lambda}_{\mf D_1[a:=\sigma]}\Delta_1,y:B}
    \\
    \inferrule*[Right=Cut]
    {u \\ d_2}
    {u\circ_{b,a}d_2::\Theta\ ;\ x:B,\Gamma_2\vdash^{\Lambda}_{\mf D_2[a:=\sigma]}\Delta_2}
  }
  {
    (u\circ_{b,a}d_1)\circ_{y,x}(u\circ_{b,a}d_2)::\Theta\ ;\ \Gamma_1,\Gamma_2\vdash^{\Lambda}_{\mf D_1[a:=\sigma]\circ_{y,x}\mf D_2[a:=\sigma]}\Delta_1,\Delta_2
  }
  $$
\end{itemize}

\textbf{Transport.}
\begin{itemize}
  \item $(\mathbf 1_{\mf D})_*(d)\equiv d$, and $(s\circ r)_*(d)\equiv r_*(s_*(d))$ for composable $2$-cells. For $d::\Theta\ ;\ \Gamma\vdash^{\Lambda}_{\mf D}\Delta$ the first law equates
  $$
  \inferrule*[Right=Tr]
  {\mathbf 1_{\mf D}::\mf D\Rightarrow^{\Lambda}\mf D \\ d}
  {(\mathbf 1_{\mf D})_*(d)::\Theta\ ;\ \Gamma\vdash^{\Lambda}_{\mf D}\Delta}
  \quad\equiv\quad d,
  $$
  For $r::\mf D\Rightarrow^{\Lambda}\mf D'$, $s::\mf D'\Rightarrow^{\Lambda}\mf D''$ and $d::\Theta\ ;\ \Gamma\vdash^{\Lambda}_{\mf D''}\Delta$ the second equates
  $$
  \inferrule*[Right=Tr]
  {s\circ r::\mf D\Rightarrow^{\Lambda}\mf D'' \\ d}
  {(s\circ r)_*(d)::\Theta\ ;\ \Gamma\vdash^{\Lambda}_{\mf D}\Delta}
  \quad\equiv\quad
  \inferrule*[Right=Tr]
  {
    r::\mf D\Rightarrow^{\Lambda}\mf D'
    \\
    \inferrule*[Right=Tr]
    {s::\mf D'\Rightarrow^{\Lambda}\mf D'' \\ d}
    {s_*(d)::\Theta\ ;\ \Gamma\vdash^{\Lambda}_{\mf D'}\Delta}
  }
  {r_*(s_*(d))::\Theta\ ;\ \Gamma\vdash^{\Lambda}_{\mf D}\Delta}
  $$
  i.e. transporting by the identity gives the same sequent, and transporting by a composition is the composition of the transports.
  \item $(r\circ_{y,x}s)_*(d_1\circ_{y,x}d_2)\equiv r_*(d_1)\circ_{y,x}s_*(d_2)$. The transport of a cut along a $2$-Subst is the cut of the transports. For $r::\mf D_1\Rightarrow^{\Lambda}\mf E_1$, $s::\mf D_2\Rightarrow^{\Lambda}\mf E_2$, $d_1::\Theta\ ;\ \Gamma\vdash^{\Lambda}_{\mf E_1}\Delta_L,y:A,\Delta_R$ and $d_2::\Theta\ ;\ \Gamma'_L,x:A,\Gamma'_R\vdash^{\Lambda}_{\mf E_2}\Delta'$ the two sides are the derivations
  $$
  \inferrule*[Right=Tr]
  {
    r\circ_{y,x}s::\mf D_1\circ_{y,x}\mf D_2\Rightarrow^{\Lambda}\mf E_1\circ_{y,x}\mf E_2
    \\
    \inferrule*[Right=Cut]
    {d_1 \\ d_2}
    {d_1\circ_{y,x}d_2::\Theta\ ;\ \Gamma'_L,\Gamma,\Gamma'_R\vdash^{\Lambda}_{\mf E_1\circ_{y,x}\mf E_2}\Delta_L,\Delta',\Delta_R}
  }
  {
    (r\circ_{y,x}s)_*(d_1\circ_{y,x}d_2)::\Theta\ ;\ \Gamma'_L,\Gamma,\Gamma'_R\vdash^{\Lambda}_{\mf D_1\circ_{y,x}\mf D_2}\Delta_L,\Delta',\Delta_R
  }
  $$
  $$
  \inferrule*[Right=Cut]
  {
    \inferrule*[Right=Tr]
    {r \\ d_1}
    {r_*(d_1)::\Theta\ ;\ \Gamma\vdash^{\Lambda}_{\mf D_1}\Delta_L,y:A,\Delta_R}
    \\
    \inferrule*[Right=Tr]
    {s \\ d_2}
    {s_*(d_2)::\Theta\ ;\ \Gamma'_L,x:A,\Gamma'_R\vdash^{\Lambda}_{\mf D_2}\Delta'}
  }
  {
    r_*(d_1)\circ_{y,x}s_*(d_2)::\Theta\ ;\ \Gamma'_L,\Gamma,\Gamma'_R\vdash^{\Lambda}_{\mf D_1\circ_{y,x}\mf D_2}\Delta_L,\Delta',\Delta_R
  }
  $$
  \item \textbf{Permutation equivariance.} $(\pi_1\cup\pi_2;\tau_1,\tau_2)_*(d_1\circ_{y,x}d_2)\equiv\bigl((\pi_1;\tau_y)_*d_1\bigr)\circ_{y,x}\bigl((\pi_x;\tau_2)_*d_2\bigr)$. Permutation transport of a cut is the cut of the transported factors, in the configuration of \cref{lem:permutation-transport-commutes-with-cut}: $\pi_1,\tau_1$ permute the free input and output of the producer $d_1$, $\pi_2,\tau_2$ those of the consumer $d_2$, $\pi_x,\tau_y$ extend $\pi_2,\tau_1$ fixing the cut variables $x,y$, and $\pi_1\cup\pi_2$ is the common extension on the united input context. 
\end{itemize}

\textbf{The $\beta$-laws.} Let $\mathsf F=F^{y}_{\mf D_F}(\Theta_F;\Gamma_F)$, let $f=\mathrm{FR}\bigl(r;(d_a)_a;(d_i)_i\bigr)::\Theta\ ;\ \Gamma\vdash^{\Lambda}_{\mf D}\Delta_L,z:\mathsf F,\Delta_R$, and let $\mathrm{FL}(d)::\Theta\ ;\ \Gamma'_L,z':\mathsf F,\Gamma'_R\vdash^{\Lambda}_{\mf E}\Delta'$ with premise $d::\Theta,\Theta_F\ ;\ \Gamma'_L,\Gamma_F,\Gamma'_R\vdash^{\Lambda}_{\mf D_F\circ\mf E}\Delta'$. Recall that the beta laws effectively inform the logic how to compute from the connectives. Left introducing and right introducing a sequent leading to a cut results in the same conclusion sequent as cutting the premises. This corresponds to the existence aspect of the universal property of the connective in the categorical semantics. In the equational theory, it is captured by:
$$
f\circ\mathrm{FL}(d)\ \equiv\ (\pi_\beta;\tau_\beta)_*\bigl(r\circ\mathbf 1_{\mf E}\bigr)_*\Bigl(\bigl(d_i\bigr)_i\circ\bigl(\bigl(d_a\bigr)_a\circ d\bigr)\Bigr),
$$
the closing permutation $(\pi_\beta;\tau_\beta)$ returning the side outputs sandwiched by the nesting of the iterated cut to their displayed order $\Delta_L,\Delta',\Delta_R$, so that the two sides are parallel. These derivations witness the equality graphically over that displayed boundary. First we cut the connective with itself:
$$
\inferrule*[Right=Cut]
{
  \inferrule*[Right=$F$R]
  {
    r::\mf D\Rightarrow^{\Lambda}\bigl(\mf D_{\bar x_i}\bigr)_i\circ\Bigl(\bigl(\sigma_a\bigr)_{a\in\theta_F}\circ\mf D_F\Bigr)
    \\
    \bigl(d_a::\Theta\vdash^{\Lambda}_{\sigma_a}A_a\bigr)_{a\in\theta_F}
    \\
    \bigl(d_i::\Theta\ ;\ \Gamma^{(i)}\vdash^{\Lambda}_{\mf D_{\bar x_i}}\Delta_L^{(i)},\bar x_i:A_i,\Delta_R^{(i)}\bigr)_{1\le i\le n}
  }
  {f::\Theta\ ;\ \Gamma\vdash^{\Lambda}_{\mf D}\Delta_L,z:\mathsf F,\Delta_R}
  \\
  \inferrule*[Right=$F$L]
  {d::\Theta,\Theta_F\ ;\ \Gamma'_L,\Gamma_F,\Gamma'_R\vdash^{\Lambda}_{\mf D_F\circ\mf E}\Delta'}
  {\mathrm{FL}(d)::\Theta\ ;\ \Gamma'_L,z':\mathsf F,\Gamma'_R\vdash^{\Lambda}_{\mf E}\Delta'}
}
{
  f\circ\mathrm{FL}(d)::\Theta\ ;\ \Gamma'_L,\Gamma,\Gamma'_R\vdash^{\Lambda}_{\mf D\circ\mf E}\Delta_L,\Delta',\Delta_R
}
$$
and we equate the result to cutting the premises:
$$
\inferrule*[Right=Tr]
{
  r\circ\mathbf 1_{\mf E}::\mf D\circ\mf E\Rightarrow^{\Lambda}\bigl(\mf D_{\bar x_i}\bigr)_i\circ\Bigl(\bigl(\sigma_a\bigr)_a\circ\bigl(\mf D_F\circ\mf E\bigr)\Bigr)
  \\
  \inferrule*[Right=Cut]
  {
    \bigl(d_i\bigr)_i
    \\
    \inferrule*[Right=Cut]
    {\bigl(d_a\bigr)_a \\ d}
    {\bigl(d_a\bigr)_a\circ d::\Theta\ ;\ \Gamma'_L,\Gamma_F,\Gamma'_R\vdash^{\Lambda}_{(\sigma_a)_a\circ(\mf D_F\circ\mf E)}\Delta'}
  }
  {
    \bigl(d_i\bigr)_i\circ\bigl(\bigl(d_a\bigr)_a\circ d\bigr)::\Theta\ ;\ \Gamma'_L,\Gamma,\Gamma'_R\vdash^{\Lambda}_{(\mf D_{\bar x_i})_i\circ((\sigma_a)_a\circ(\mf D_F\circ\mf E))}\Delta_L,\Delta',\Delta_R
  }
}
{
  \bigl(r\circ\mathbf 1_{\mf E}\bigr)_*\Bigl(\bigl(d_i\bigr)_i\circ\bigl(\bigl(d_a\bigr)_a\circ d\bigr)\Bigr)::\Theta\ ;\ \Gamma'_L,\Gamma,\Gamma'_R\vdash^{\Lambda}_{\mf D\circ\mf E}\Delta_L,\Delta',\Delta_R
}
$$
Dually, let $\mathsf U=U^{\mf D_U}_{x_U}(\Theta_U;\Gamma_U\mid\Delta_U)$, let $\mathrm{UR}(d)::\Theta\ ;\ \Gamma\vdash^{\Lambda}_{\mf D}\Delta_L,z:\mathsf U,\Delta_R$ with premise $d::\Theta,\Theta_U\ ;\ \Gamma_U^L,\Gamma,\Gamma_U^R\vdash^{\Lambda}_{\mf D\circ\mf D_U}\Delta_L,\Delta_U,\Delta_R$, and let $h=\mathrm{UL}\bigl(r;(d_a)_a;(d_i)_i;(d_j)_j\bigr)::\Theta\ ;\ \Gamma'',z':\mathsf U\vdash^{\Lambda}_{\mf E}\Delta''$. Then
$$
\mathrm{UR}(d)\circ h\ \equiv\ (\pi_\beta;\tau_\beta)_*\bigl(\mathbf 1_{\mf D}\circ r\bigr)_*\Bigl(\Bigl(\bigl(d_i\bigr)_i\circ\bigl(\bigl(d_a\bigr)_a\circ d\bigr)\Bigr)\circ\bigl(d_j\bigr)_j\Bigr),
$$
the closing permutation $(\pi_\beta;\tau_\beta)$ returning the boundary sandwiched by the nesting of the iterated cuts to the displayed order $\Gamma'',\Gamma\vdash\Delta_L,\Delta'',\Delta_R$, so that the two sides are parallel, displayed by these two derivations over that boundary:
$$
\inferrule*[Right=Cut]
{
  \inferrule*[Right=$U$R]
  {d::\Theta,\Theta_U\ ;\ \Gamma_U^L,\Gamma,\Gamma_U^R\vdash^{\Lambda}_{\mf D\circ\mf D_U}\Delta_L,\Delta_U,\Delta_R}
  {\mathrm{UR}(d)::\Theta\ ;\ \Gamma\vdash^{\Lambda}_{\mf D}\Delta_L,z:\mathsf U,\Delta_R}
  \\
  \inferrule*[Right=$U$L]
  {
    r::\mf E\Rightarrow^{\Lambda}\mf C
    \\
    \bigl(d_a::\Theta\vdash^{\Lambda}_{\sigma_a}A_a\bigr)_{a\in\theta_U}
    \\
    \bigl(d_i::\Theta\ ;\ \Gamma^{(i)}\vdash^{\Lambda}_{\mf D_{\bar x_i}}\Delta_L^{(i)},\bar x_i:A_i,\Delta_R^{(i)}\bigr)_{1\le i\le n}
    \\
    \bigl(d_j::\Theta\ ;\ \Gamma^{(n+j)},\bar y_j:B_j\vdash^{\Lambda}_{\mf D_{\bar y_j}}\Delta^{(n+j)}\bigr)_{1\le j\le m}
  }
  {h::\Theta\ ;\ \Gamma'',z':\mathsf U\vdash^{\Lambda}_{\mf E}\Delta''}
}
{
  \mathrm{UR}(d)\circ h::\Theta\ ;\ \Gamma'',\Gamma\vdash^{\Lambda}_{\mf D\circ\mf E}\Delta_L,\Delta'',\Delta_R
}
$$
and
$$
\inferrule*[Right=Tr]
{
  \mathbf 1_{\mf D}\circ r::\mf D\circ\mf E\Rightarrow^{\Lambda}\mf D\circ\mf C
  \\
  \inferrule*[Right=Cut]
  {\bigl(d_a\bigr)_a \\ \bigl(d_i\bigr)_i \\ \bigl(d_j\bigr)_j \\ d}
  {k::\Theta\ ;\ \Gamma'',\Gamma\vdash^{\Lambda}_{\mf D\circ\mf C}\Delta_L,\Delta'',\Delta_R}
}
{
  \bigl(\mathbf 1_{\mf D}\circ r\bigr)_*(k)::\Theta\ ;\ \Gamma'',\Gamma\vdash^{\Lambda}_{\mf D\circ\mf E}\Delta_L,\Delta'',\Delta_R
}
$$
\textbf{The $\eta$-laws.} For every term $h::\Theta\ ;\ \Gamma_L,z:\mathsf F,\Gamma_R\vdash^{\Lambda}_{\mf D}\Delta$ with the principal in the zone of its class,
$$
h\ \equiv\ \mathrm{FL}\bigl(\eta_{\mathsf F}\circ_{\bar z,z}h\bigr),
$$
the right-hand side being the derivation
$$
\inferrule*[Right=$F$L]
{
  \inferrule*[Right=Cut]
  {
    \eta_{\mathsf F}::\Theta_F\ ;\ \Gamma_F\vdash^{\Lambda}_{\mf D_F[y:=\bar z]}\bar z:\mathsf F
    \\
    h::\Theta\ ;\ \Gamma_L,z:\mathsf F,\Gamma_R\vdash^{\Lambda}_{\mf D}\Delta
  }
  {\eta_{\mathsf F}\circ_{\bar z,z}h::\Theta,\Theta_F\ ;\ \Gamma_L,\Gamma_F,\Gamma_R\vdash^{\Lambda}_{\mf D_F\circ_{y,z}\mf D}\Delta}
}
{
  \mathrm{FL}\bigl(\eta_{\mathsf F}\circ_{\bar z,z}h\bigr)::\Theta\ ;\ \Gamma_L,z:\mathsf F,\Gamma_R\vdash^{\Lambda}_{\mf D}\Delta
}
$$
the cut landing on the boundary and index of the $F$-left premise, the factors read in the united zone $\Theta,\Theta_F$ (\cref{lem:cartesian-weakening-of-derivation-terms}), the variable absorbing the renaming of the generic term, and for every term $h::\Theta\ ;\ \Gamma\vdash^{\Lambda}_{\mf D}\Delta_L,z:\mathsf U,\Delta_R$,
$$
h\ \equiv\ \mathrm{UR}\bigl(h\circ_{z,x_0}\varepsilon_{\mathsf U}\bigr),
$$
the right-hand side being the derivation
$$
\inferrule*[Right=$U$R]
{
  \inferrule*[Right=Cut]
  {
    h::\Theta\ ;\ \Gamma\vdash^{\Lambda}_{\mf D}\Delta_L,z:\mathsf U,\Delta_R
    \\
    \varepsilon_{\mathsf U}::\Theta_U\ ;\ \Gamma_U^L,x_0:\mathsf U,\Gamma_U^R\vdash^{\Lambda}_{\mf D_U[x_U:=x_0]}\Delta_U
  }
  {h\circ_{z,x_0}\varepsilon_{\mathsf U}::\Theta,\Theta_U\ ;\ \Gamma_U^L,\Gamma,\Gamma_U^R\vdash^{\Lambda}_{\mf D\circ_{z,x_U}\mf D_U}\Delta_L,\Delta_U,\Delta_R}
}
{
  \mathrm{UR}\bigl(h\circ_{z,x_0}\varepsilon_{\mathsf U}\bigr)::\Theta\ ;\ \Gamma\vdash^{\Lambda}_{\mf D}\Delta_L,z:\mathsf U,\Delta_R
}
$$
the factors read in the united zone and the cartesian blocks of the second composite exchanged by the transport of \cref{lem:permutation-transport-of-derivation-terms} to display the ambient zone first, the one transport the ambient-first format of the right rule requires.
\end{definition}

\phantomsection\label{rem:generic-beta-special-instances}
\textit{Remark (special instances):} At the generic terms the $\beta$-laws collapse to
$$
\eta_{\mathsf F}\circ\mathrm{FL}(d)\equiv d,
\qquad\qquad
\mathrm{UR}(d)\circ\varepsilon_{\mathsf U}\equiv d,
$$
whose left-hand sides are the derivations
$$
\inferrule*[Right=Cut]
{
  \eta_{\mathsf F}
  \\
  \inferrule*[Right=$F$L]
  {d::\Theta,\Theta_F\ ;\ \Gamma_L,\Gamma_F,\Gamma_R\vdash^{\Lambda}_{\mf D_F\circ\mf D}\Delta}
  {\mathrm{FL}(d)::\Theta\ ;\ \Gamma_L,z:\mathsf F,\Gamma_R\vdash^{\Lambda}_{\mf D}\Delta}
}
{
  \eta_{\mathsf F}\circ\mathrm{FL}(d)::\Theta,\Theta_F\ ;\ \Gamma_L,\Gamma_F,\Gamma_R\vdash^{\Lambda}_{\mf D_F\circ\mf D}\Delta
}
$$
$$
\inferrule*[Right=Cut]
{
  \inferrule*[Right=$U$R]
  {d::\Theta,\Theta_U\ ;\ \Gamma_U^L,\Gamma,\Gamma_U^R\vdash^{\Lambda}_{\mf D\circ\mf D_U}\Delta_L,\Delta_U,\Delta_R}
  {\mathrm{UR}(d)::\Theta\ ;\ \Gamma\vdash^{\Lambda}_{\mf D}\Delta_L,z:\mathsf U,\Delta_R}
  \\
  \varepsilon_{\mathsf U}
}
{
  \mathrm{UR}(d)\circ\varepsilon_{\mathsf U}::\Theta,\Theta_U\ ;\ \Gamma_U^L,\Gamma,\Gamma_U^R\vdash^{\Lambda}_{\mf D\circ\mf D_U}\Delta_L,\Delta_U,\Delta_R
}
.$$

\begin{example}[Linear--Nonlinear Logic, $\beta\eta$-laws]
Let $A,B$ be types at the linear mode and suppose
$$
d::\Theta\ ;\ \Gamma_L,x:A,y:B,\Gamma_R
\vdash^\Lambda_{\mf D_\otimes\circ_{z,t}\mf E}\Delta,
\qquad
e::\Theta\ ;\ \Gamma
\vdash^\Lambda_{\mf C\circ_{p,z}\mf D_{\parr}}
\Delta_L,x:A,y:B,\Delta_R.
$$
The special $\beta$-laws say precisely that introducing the generic connective and immediately eliminating it changes no proof:
$$
\eta_{A\otimes B}\circ_{\bar t,t}\mathrm{FL}(d)\equiv d,
\qquad
\mathrm{UR}(e)\circ_{p,p_0}\varepsilon_{A\parr B}\equiv e.
$$
At the identity terms, the $\eta$-laws and the cut-unit equations give the familiar identity expansions
$$
\mathrm{id}_{A\otimes B}\equiv\mathrm{FL}(\eta_{A\otimes B}),
\qquad
\mathrm{id}_{A\parr B}\equiv\mathrm{UR}(\varepsilon_{A\parr B}),
$$
with the identity instances chosen at the displayed boundary names. In particular, the tensor $\beta$-law applied to the application term gives
$$
\eta_{A\otimes(A\multimap B)}
\circ_{\bar t,t}
\operatorname{ev}_{A,B}
\equiv
\operatorname{app}_{A,B}.
$$

The $U$-laws for implication are the usual linear function $\beta\eta$-laws. If
$$
q::\Theta\ ;\ x:A,\Gamma
\vdash^\Lambda_{\mf D\circ_{u,y}\mf D_\otimes}z:B,
\qquad
h::\Theta\ ;\ \Gamma\vdash^\Lambda_{\mf E}u:A\multimap B,
$$
then $\mathrm{UR}(q)$ is the abstraction of $q$, and
$$
\mathrm{UR}(q)\circ_{u,f}\operatorname{app}_{A,B}\equiv q,
\qquad
h\equiv\mathrm{UR}\bigl(h\circ_{u,f}\operatorname{app}_{A,B}\bigr).
$$
The first equation substitutes the generic linear argument into an abstraction, while the second states that abstracting the application of a term recovers that term. 
\end{example}

\subsubsection{Coherent Completion of Structural Transport}

The four closure clauses of \cref{def:signature} assert the mere existence of the transported generating $2$-cells. We introduce a choice of $2$-cell that we call a \emph{cleavage} in \cref{def:transport-cleavage}.

\begin{definition}[Transport Cleavage]\label{def:transport-cleavage}
Let $\Lambda\ \mathrm{Sig}$. For every generating $2$-cell $\rho$ and every admissible instance of the four closure operations of \cref{def:signature}, we define $\chi$, a \emph{transport cleavage}, to be a choice of witness denoted 
$$
(\pi;\tau)_*^\chi\rho,
\qquad
\mathrm{wk}^{a:p,\chi}_{\mathrm{in}}(\rho),
\qquad
\mathrm{ren}_{\sigma}^{\chi}(\rho)
$$
Contraction of an arbitrary descriptor $2$-cell remains derived. If
$r::\mf D\Rightarrow^\Lambda_{\gamma,x:p,x':p\Vdash\delta}\mf E$
with $p$ cartesian, choose a fresh $a:p$, put $\mf I_x=[\varnothing\mid\{x\mapsto a\}]$, and define
$$
\mathrm{ctr}^{\chi}_{x'\to x}(r)
:=
\mathbf 1_{\mf I_x}\circ_{a,x'}r,
$$
\end{definition}

\begin{definition}[Transport Presentation]\label{def:transport-presentation}
A \emph{transport presentation} is a finite composite of the specified permutation, weakening and boundary-renaming transports and the contractions derived in \cref{def:transport-cleavage}. It acts on a term. Two presentations have the same structural action when they induce the same map of boundary declarations and the same operation on descriptor classes.
\end{definition}

\begin{example}[two presentations of one weakening]\label{ex:two-presentations-one-weakening}
Let $r::\mf D\Rightarrow^{\Lambda}_{\gamma\Vdash\delta}\mf E$, let $p$ be
cartesian, and let $a:p$ and $b:p$ be fresh input declarations. Consider the
transport presentations
$$
P:=\mathrm{ren}_{a\mapsto b}\circ\mathrm{wk}^{a:p}_{\mathrm{in}},
\qquad
Q:=\mathrm{wk}^{b:p}_{\mathrm{in}}.
$$
The presentation $P$ first adds an unused input named $a$ and then renames it
to $b$, whereas $Q$ adds the input $b$ directly. Both carry
$\gamma\Vdash\delta$ to $\gamma,b:p\Vdash\delta$ and send every descriptor
class $\mf C$ to $\mathrm{wk}^{b:p}_{\mathrm{in}}(\mf C)$, so they have the
same structural action.
\end{example}

\begin{definition}[Coherent Transport Completion]\label{def:coherent-transport-completion}
Fix a transport cleavage $\chi$ for $\Lambda$.
\begin{itemize}
  \item The \emph{completed mode congruence}, written $\equiv^{\chi}_{\mathrm M}$, is the least congruence extending the congruence of \cref{def:equational-theory-of-2-cells}, closed under the specified transports, and satisfying
  $$
  P_*^{\chi}(r)\equiv^{\chi}_{\mathrm M}Q_*^{\chi}(r)
  $$
  whenever the transport presentations $P$ and $Q$ have the same structural action. In particular, identity presentations act as identities and successive transports agree with the presentation of their composite. For every generating $2$-cell $\rho$ at a cartesian pair $x,x'$, it also contains the comparison
  $$
  \mathrm{gen}\bigl(\mathrm{ctr}_{x'\to x}^{\chi}(\rho)\bigr)
  \equiv^{\chi}_{\mathrm M}
  \mathrm{ctr}_{x'\to x}^{\chi}\bigl(\mathrm{gen}(\rho)\bigr)
  =
  \mathbf 1_{\mf I_x}\circ_{a,x'}\mathrm{gen}(\rho),
  $$
  with the same closing convention as in \cref{def:transport-cleavage}.

  \item The \emph{completed derivation congruence}, written $\equiv^{\chi}_{\mathrm D}$, is the least congruence extending the congruence of \cref{def:equational-theory-of-derivations}, closed under the specified term transports, and satisfying
  $$
  P_*^{\chi}(d)\equiv^{\chi}_{\mathrm D}Q_*^{\chi}(d)
  $$
  whenever $P$ and $Q$ have the same structural action. It is moreover cross-sorted: if $r\equiv^{\chi}_{\mathrm M}s$ and $d\equiv^{\chi}_{\mathrm D}e$, then
  $$r_*(d)\equiv^{\chi}_{\mathrm D}s_*(e),$$
  and replacing a descriptor $2$-cell parameter by an $\equiv^{\chi}_{\mathrm M}$-equivalent one in $\mathrm{FR}$, $\mathrm{UL}$ or any gated former produces $\equiv^{\chi}_{\mathrm D}$-equivalent derivation terms.
\end{itemize}
\end{definition}

\textit{Remark.} The completed mode congruence for the \hyperref[ex:two-presentations-one-weakening]{two weakening presentations} therefore contains
$$
\mathrm{ren}^{\chi}_{a\mapsto b}
\bigl(\mathrm{wk}^{a:p,\chi}_{\mathrm{in}}(r)\bigr)
=P_*^{\chi}(r)
\equiv^{\chi}_{\mathrm M}
Q_*^{\chi}(r)
=\mathrm{wk}^{b:p,\chi}_{\mathrm{in}}(r).
$$

\textit{Remark:} The completion changes no derivability statement and adds no generator to the signature. It turns the existential closure witnesses into coherent operations only after terms have been formed. The resulting presentation is relative to $\chi$.
\endgroup{}

\begingroup

\section{Categorical Semantics}
\label{sec:categorical-semantics}

In this section, we present the categorical semantics associated with our syntax. The syntax consists of two types of objects: linear and cartesian. The categorical structure that presents this data turns out to be that of the \emph{Linear-Non-Linear-Polycategories} of \cite{Shu23}, abbreviated LNL-Polycategories. They generalize the judgmental structure of $\star$-autonomous categories of \cite{Bar79} and \cite{Bla23} as well as the judgmental structure of intuitionistic type theories of \cite{LSR17}. The key idea of LNL-Polycategories is the separation between cartesian and linear resources as a way to express both judgmental structures in one unified categorical concept. Moreover, the cartesian and linear objects and morphisms interact with each other, offering a primitive presentation of monadic and comonadic modalities. The definition of an LNL polycategory is recalled in \cref{def:lnl-polycategory} of \cref{subsec:lnl-polycategories}.

For the rest of this section, fix a signature $\Lambda$ and a transport cleavage $\chi$ (\cref{def:transport-cleavage}). We use the completed congruences $\equiv^{\chi}_{\mathrm M}$ and $\equiv^{\chi}_{\mathrm D}$ of \cref{def:coherent-transport-completion}; when the sort is clear, we write simply $\equiv$, and we suppress $\chi$ from the notation $\mc M_\Lambda$ and $\mc D_\Lambda$.

\begin{proposition}\label{prop:mode-theory-presents-lnl-polycategory}
  Relative to the fixed transport cleavage, the mode theory syntax presents an LNL $2$-polycategory $\mc M_\Lambda$.
\end{proposition}

\begin{proof}
  We start by defining the components of the LNL $2$-polycategory and finish by showing it is indeed one.

  A hom of an LNL polycategory is indexed by a list of modes, while a sequent of the mode theory declares named variables. We therefore fix for each list of modes a canonical context of pairwise distinct variables, and we take as morphisms the well-formed descriptor classes over those variables. Two sequents differing by a renaming of their variables present the same morphism, the identification being well-defined by \cref{thm:renaming-admissibility-for-descriptor-judgments} and \cref{lem:boundary-renaming-of-descriptor-2-cells}. This convention makes all three compositions well-defined. We rename the representatives apart before cutting; the union $\gamma_1 \cup \gamma_2$ taken by the Cut rule is then a disjoint union.

  \begin{itemize}
    \item \textbf{Nonlinear Objects.} Those are the cartesian modes.
    \item \textbf{Nonlinear Morphisms.} A nonlinear morphism between the list of cartesian objects $\theta$ and the cartesian object $p$ is a well-formed descriptor class $\mf D$ and a sequent $\theta \Vdash_{\mf D} x : p$ (\cref{prop:sorts-of-well-formed-descriptors}).
    \item \textbf{Cartesian Composition.} The composition between $\theta \Vdash_{\mf D} x : p$ and $\theta_1, y : p, \theta_2 \Vdash_{\mf D'} z : q$ is the cut operation $\mf D \circ_{x,y} \mf D'$ (\cref{def:well-formed-composition-cut}), giving rise to the sequent $$\theta_1, \theta, \theta_2 \Vdash_{\mf D \circ_{x,y} \mf D'} z : q.$$
    \item \textbf{Cartesian identity.} The identity is the sequent $x : p \Vdash_{[\varnothing \mid \{x \mapsto y\}]}  y : p$ (\cref{def:well-formed-descriptors-identity}).
    \item \textbf{Linear Objects.} Those are the linear modes.
    \item \textbf{Linear Morphisms.} For a list of cartesian objects $\theta$ and lists of linear objects $\gamma, \delta$, a linear morphism is a well-formed descriptor class $\mf D$ and a sequent $\theta \; ; \; \gamma \Vdash_{\mf D} \delta$.
    \item \textbf{Linear Composition.} The composition between $\theta \; ; \; \gamma \Vdash_{\mf D} \delta_1, x : p, \delta_2$ and $\theta' \; ; \; \gamma_1, y : p, \gamma_2 \Vdash_{\mf D'} \delta$ is the cut operation $\mf D \circ_{x,y} \mf D'$, giving rise to the sequent $$\theta', \theta \; ; \; \gamma_1, \gamma, \gamma_2 \Vdash_{\mf D \circ_{x,y} \mf D'} \delta_1, \delta, \delta_2.$$ The zone of the sequent consuming the cut variable stands first, as in \cref{def:lnl-polycategory}. The closing permutation of the Cut rule arranges the boundary so.
    \item \textbf{Linear identity.} The identity is the sequent $x : l \Vdash_{[\varnothing \mid \{x \mapsto y\}]}  y : l$ at a linear mode $l$ with an empty cartesian zone, i.e. the instance of \cref{def:well-formed-descriptors-identity} at $\theta = \varnothing$.
    \item \textbf{Cartesian-Linear Composition.} For a cartesian morphism $\theta \Vdash_{\mf D} x : p$ and a linear morphism $\theta_1, y : p, \theta_2 \; ; \; \gamma \Vdash_{\mf D'} \delta$, the composition is $$\theta_1, \theta, \theta_2 \; ; \; \gamma \Vdash_{\mf D \circ_{x,y} \mf D'} \delta.$$
    \item \textbf{Structural Actions.} Both sorts of morphisms carry an action of context permutations on their domain and codomain, given by the permutation transport of \cref{thm:exchange-admissibility-for-descriptor-judgments} and functorial by \cref{prop:functoriality-of-permutation-transport}. The cartesian zone carries in addition the reindexing along an arbitrary structural map. Let $\sigma : \{1, \ldots, n\} \to \{1, \ldots, m\}$ be a function, let $b_1 : p_1, \ldots, b_m : p_m$ be the canonical context of a list of cartesian modes, and let $a_1 : p_{\sigma 1}, \ldots, a_n : p_{\sigma n}$ be the canonical context of the reindexed list. For a morphism $\mf D$ of either sort whose cartesian zone is $a_1, \ldots, a_n$, we define $\mf D^{\sigma}$ over $b_1, \ldots, b_m$ by the simultaneous input substitution $[a_i := b_{\sigma i}]$, i.e. by relabelling every occurrence of the input $a_i$ as $b_{\sigma i}$. The relabelled class is a morphism since the relabelling factors as the contractions identifying the inputs sent to a common variable, the boundary renamings carrying the surviving inputs to the corresponding $b_j$, the weakenings by the variables $b_j$ with $j$ outside the image of $\sigma$, and a final permutation. Each step is admissible by \cref{thm:input-contraction-admissibility-for-well-formed-descriptors}, \cref{thm:renaming-admissibility-for-descriptor-judgments}, \cref{thm:input-weakening-admissibility-for-well-formed-descriptors} and \cref{thm:exchange-admissibility-for-descriptor-judgments}. Notice that the factorization is used for derivability only; the class $\mf D^{\sigma}$ itself is defined by the relabelling. The boundary usage property forbids weakening and contraction at linear modes; the linear zones therefore keep their permutation actions only.
    \item \textbf{2-Cells.} The hom-category over a boundary has the well-formed classes as objects and the descriptor $2$-cell terms $r :: \mf D \Rightarrow^{\Lambda} \mf D'$ over the canonical contexts, modulo $\equiv^{\chi}_{\mathrm M}$, as morphisms. $2$-Identity and $2$-Comp provide the identities and the composition.
  \end{itemize}
  We show that this data defines an LNL $2$-polycategory, considering the one-dimensional laws first. The unit laws are the computations of \cref{lem:units-for-composition}. Cutting an identity descriptor on either side of a morphism renames a boundary variable. The resulting sequent presents the same morphism. Since the operation is sort-agnostic, the two unit laws hold for the cartesian, the linear and the cartesian-linear compositions.

  For associativity, we refer to \cref{lem:associativity}, and for the interchange laws to \cref{lem:left-interchange-law,lem:right-interchange-law}. It remains to check that composition is equivariant for the structural actions. It commutes with context permutations by \cref{lem:permutation-transport-commutes-with-cut}, with boundary renaming by \cref{lem:boundary-renaming-is-well-defined}(iii), and with weakening by \cref{lem:weakening-commutes-with-composition}. Contraction is realized as a cut against an identity descriptor by \cref{lem:identity-cut-realizes-input-contraction}; when the contracted declaration is shared by both factors, its compatibility with composition is the cartesian distribution equality of \cref{lem:c-substitution-commutes-with-composition}, while associativity and interchange handle the remaining placements.

  Functoriality of the reindexing remains. On a representative, the relabelling along $\sigma$ touches only the labels of the input occurrences. Relabelling along $\sigma'$ and then along $\sigma$ therefore relabels every occurrence directly along the composite function, i.e. $(\mf D^{\sigma'})^{\sigma}=\mf D^{\sigma\circ\sigma'}$, and the identity function relabels nothing. Equivariance for an arbitrary structural map then reduces to the three generator cases. When the relabelling is written as contractions, weakenings and a permutation, each step commutes with the compositions by \cref{lem:permutation-transport-commutes-with-cut,lem:weakening-commutes-with-composition,lem:identity-cut-realizes-input-contraction,lem:c-substitution-commutes-with-composition}, and functoriality reassembles the steps.

  We turn to the two-dimensional structure. The hom-categories are categories by the vertical laws of \cref{def:equational-theory-of-2-cells}, inherited by $\equiv^{\chi}_{\mathrm M}$. The three compositions extend to functors. On $2$-cells, the composition is $2$-Subst, $r \circ_{y,x} s$, with one-sided composites whiskered by identity $2$-cells, and the functoriality clause of \cref{def:equational-theory-of-2-cells} states exactly that identities go to identities and vertical composites to vertical composites. The polycategory laws on objects follow from \cref{lem:units-for-composition,lem:associativity,lem:left-interchange-law,lem:right-interchange-law}, and on $2$-cells from the units, associativity, interchange, distribution and projection clauses of the theory, each imposed over the corresponding strict law of the classes. The reindexing extends to functors as well. Its permutation, weakening and boundary-renaming stages are the recursions selected by $\chi$, and its contraction stage is the $2$-Subst against the identity $2$-cell of an identity descriptor (\cref{def:transport-cleavage}). The path equations of \cref{def:coherent-transport-completion} identify any two presentations of the same structural action and give the identity and composition laws; the contraction comparison identifies the selected contraction generator with the derived action. Compatibility with the three compositions follows from the units, associativity, interchange, functoriality, distribution, projection and permutation-equivariance clauses of the mode theory. This is the LNL $2$-polycategory presented by the mode theory, denoted $\mc M_\Lambda$.
\end{proof}

We turn now to the type theory. Its morphisms are the derivation terms of \cref{def:derivation-terms} modulo $\equiv^{\chi}_{\mathrm D}$. Recall that the terms are the derivations of the calculus extended by explicit identity, transport and cut, that their indexes are well-formed classes combined only by the operations of the mode theory, and that the strict coherence laws in \cref{lem:associativity,lem:left-interchange-law,lem:right-interchange-law} therefore apply throughout. The conventions of the \hyperref[anchor:metatheory-conventions]{metatheory preliminaries} remain in force. The rules and the axioms respect context permutations; the judgments, the terms and the operations on classes are stable under boundary renamings, and we reorder and rename by the $\chi$-selected transports of \cref{lem:permutation-transport-of-derivation-terms,lem:boundary-renaming-of-derivation-terms}.

With the cleavage fixed, write $\mc M_\Lambda$ for the LNL $2$-polycategory the mode theory presents. We show that the type theory syntax presents a functor of LNL $2$-polycategories.

\begin{proposition}[Completeness: the Syntactic Fibration]\label{prop:type-theory-presents-lnl-polycategory}
  Relative to the fixed transport cleavage, the syntax presents a functor of LNL $2$-polycategories $\pi : \mc D_\Lambda \to \mc M_\Lambda$ that is a local discrete fibration and a fibration, and admits an opcartesian lift of every morphism of the base with a single output.
\end{proposition}

\begin{proof}
  We start by defining the objects and morphisms of $\mc D_\Lambda$. As in the mode theory, we fix for each list of types a context of pairwise distinct variables, lying over the context of the underlying list of modes, we take the morphisms over those contexts, and we identify along boundary renamings (\cref{lem:boundary-renaming-of-derivation-terms}). Every derivable sequent carries a well-formed descriptor of the sort of its boundary (\cref{prop:sequent-indexes-are-well-formed}).

  \begin{itemize}
    \item \textbf{Cartesian Objects.} Those are the cartesian types over a mode $(A,p)$ where $A$ is a cartesian type and $p$ is a cartesian mode.
    \item \textbf{Cartesian Morphisms.} A nonlinear morphism between the list of cartesian objects
\[
((\Theta_1, p_1), \dots, (\Theta_n, p_n))
\]
and the cartesian object $(B,q)$ is a well-formed c-descriptor class $\mf D$ such that $x_1 : p_1, \dots, x_n : p_n \vdash_{\mf D} y : q$ is derivable in the mode theory, together with an $\equiv$-class of derivation terms $d :: \Theta \vdash^{\Lambda}_{\mf D} y : A$.
    \item \textbf{Cartesian Composition.} The composition between $d :: \Theta \vdash^{\Lambda}_{\mf D} y : A$ and $d' :: \Theta_1, x : A, \Theta_2 \vdash^{\Lambda}_{\mf D'} C$ at $y,x$ is the term $$d \circ_{y,x} d' :: \Theta_1, \Theta, \Theta_2 \vdash^{\Lambda}_{\mf D \circ_{y,x} \mf D'} C.$$
    \item \textbf{Cartesian identity.} The identity is the term $\mathrm{id}_A :: x : A \vdash^{\Lambda}_{[\varnothing \mid \{x \mapsto y\}]}  y : A$.
    \item \textbf{Linear Objects.} Those are the linear types over a mode $(A,p)$ where $A$ is a linear type and $p$ is a linear mode.
    \item \textbf{Linear Morphisms.} For a list of cartesian objects $(\Theta_1,\theta_1), \dots, (\Theta_n, \theta_n)$ and lists of linear objects $(\Gamma_1,\gamma_1), \dots, (\Gamma_k, \gamma_k)$ and $(\Delta_1,\delta_1), \dots, (\Delta_k, \delta_m)$, a linear morphism is a well-formed l-descriptor class $\mf D$ such that $\theta \; ; \; \gamma \Vdash_{\mc D} \delta$, together with an $\equiv$-class of derivation terms $d :: \Theta \; ; \; \Gamma \vdash^{\Lambda}_{\mf D} \Delta$.
    \item \textbf{Linear Composition.} The composition between $d :: \Theta \; ; \; \Gamma \vdash^{\Lambda}_{\mf D} \Delta_1, y : A, \Delta_2$ and $d' :: \Theta' \; ; \; \Gamma_1, x : A, \Gamma_2 \vdash^{\Lambda}_{\mf D'} \Delta$ is the term $$d \circ_{y,x} d' :: \Theta', \Theta \; ; \; \Gamma_1, \Gamma, \Gamma_2 \vdash^{\Lambda}_{\mf D \circ_{y,x} \mf D'} \Delta_1, \Delta, \Delta_2,$$ the two terms read in the united cartesian zone (\cref{lem:cartesian-weakening-of-derivation-terms}) and the closing permutations of the Cut former arranging the zone of the consuming term first, exactly as in the mode theory.
    \item \textbf{Linear identity.} The identity is the term $\mathrm{id}_A :: \varnothing \; ; \; x : A \vdash_{[\varnothing \mid \{x \mapsto y\}]} y : A$ at a linear type, with an empty cartesian zone.
    \item \textbf{Cartesian-Linear Composition.} For a cartesian morphism $d :: \Theta \vdash^{\Lambda}_{\mf D} y : A$ and a linear morphism $d' :: \Theta_1, x : A, \Theta_2 \; ; \; \Gamma \vdash^{\Lambda}_{\mf D'} \Delta$, the composition is the term $$d \circ_{y,x} d' :: \Theta_1, \Theta, \Theta_2 \; ; \; \Gamma \vdash^{\Lambda}_{\mf D \circ_{y,x} \mf D'} \Delta.$$
    \item \textbf{Structural Actions.} Both sorts of morphisms carry the $\chi$-selected action of context permutations, given by \cref{lem:permutation-transport-of-derivation-terms} on the terms and by \cref{thm:exchange-admissibility-for-descriptor-judgments} on their descriptors, functorial modulo the completed congruences. The cartesian zone carries in addition the reindexing along an arbitrary structural map between lists of cartesian types, realized on terms exactly as on their descriptors. The weakenings are the $\chi$-selected operations of \cref{lem:cartesian-weakening-of-derivation-terms}; the contraction of two declarations $x', x''$ into a shared declaration $x_0$ is the double cut against identity terms $$\mathrm{id}_A \circ_{b',x'}\bigl(\mathrm{id}_A \circ_{b'',x''} d\bigr),$$ with both principals named $x_0$ in the united cartesian zone. By \cref{lem:identity-cut-realizes-input-contraction}, this computes the descriptor; a final permutation closes the factorization. The contractions identify variables carrying the same type since the reindexed list repeats the types of the original one. The linear zones keep their permutation actions only.
    \item \textbf{2-Cells.} A morphism of the hom-category from $(\mf D, d)$ to $(\mf D', d')$ is an $\equiv^{\chi}_{\mathrm M}$-class of $2$-cells $r :: \mf D \Rightarrow^{\Lambda} \mf D'$ satisfying $$r_*(d') \equiv^{\chi}_{\mathrm D} d.$$
  \end{itemize}
  We show that this data defines $\mc D_\Lambda$ as an LNL $2$-polycategory. First, we note that the morphisms form a category, the identity morphisms are the identity $2$-cells, and composition and laws are verified by the transport clauses of \cref{def:equational-theory-of-derivations}. The condition of the identity $2$-cell at $(\mf D,d)$ is the transport law for the identity, $(\mathbf 1_{\mf D})_*(d)\equiv d$. For $r$ from $(\mf D,d)$ to $(\mf D',d')$ and $s$ from $(\mf D',d')$ to $(\mf D'',d'')$, the vertical composite of the base satisfies
  $$(s\circ r)_*(d'')\ \equiv\ r_*\bigl(s_*(d'')\bigr)\ \equiv\ r_*(d')\ \equiv\ d,$$
  by the transport law for a composition first, the condition of $s$ under the congruence next and the condition of $r$ last; $s\circ r$ is a morphism from $(\mf D,d)$ to $(\mf D'',d'')$. The cross-sorted clause of \cref{def:coherent-transport-completion} makes the condition stable under both completed congruences, so the homs are well-defined on classes. The unit and associativity laws hold as they hold in the base. A morphism of the hom-category is a $2$-cell of $\mc M_\Lambda$ carrying a property; the identities and the composition are those of the base, and the vertical laws of \cref{def:equational-theory-of-2-cells} apply.

  We verify next the one-dimensional laws. The operations are defined on representatives and are respected by the $\equiv$-classes. The unit laws are the units clause. Cutting an identity term on either side of a term renames its boundary variables and nothing else, up to $\equiv$; under the fixed boundary-renaming convention, the renamed term presents the same morphism (\cref{lem:boundary-renaming-of-derivation-terms}), and the descriptors agree by \cref{lem:units-for-composition}. The clause as defined is agnostic to whether the cut variables are cartesian or linear; therefore, it holds for the cartesian, the linear and the cartesian-linear compositions. Associativity is the associativity clause over \cref{lem:associativity}, and the two interchange laws are the interchange clause over \cref{lem:left-interchange-law,lem:right-interchange-law}, the closing transports of the interchange clause matching the conclusions of the two orders.

  It remains to check the structural clauses. Composition commutes with the context permutations, the terms agreeing by the permutation-equivariance clause of \cref{def:equational-theory-of-derivations} and the indexes by \cref{lem:permutation-transport-commutes-with-cut}, and it commutes with the weakening by the Cut clause of \cref{lem:cartesian-weakening-of-derivation-terms}, the Cut former reading its premises in the united cartesian zone, the indexes agreeing by \cref{lem:weakening-commutes-with-composition}. For the contraction, we compute on its realization. The identity term at a cartesian type is a cartesian sequent through the uniform display, and its cut at a declaration $x_0 : A$ shared by the two factors of a composition distributes over the composition by the cartesian distribution clause,
  $$
  \mathrm{id}_A\circ_{b,x_0}(d_1\circ_{y,x}d_2)\ \equiv\ (\mathrm{id}_A\circ_{b,x_0}d_1)\circ_{y,x}(\mathrm{id}_A\circ_{b,x_0}d_2),
  $$
  the indexes agreeing by \cref{lem:c-substitution-commutes-with-composition}. At a declaration that a factor carries by weakening alone, the cut is discarded on that factor by the cartesian projection clause, with the identity input re-weakening it under the shared name; the double cuts reorder among themselves and across a composition by \cref{lem:associativity,lem:left-interchange-law,lem:right-interchange-law}. The contraction of two declarations of a composite is therefore the composite of the contracted or re-weakened factors, exactly as \cref{lem:identity-cut-realizes-input-contraction} computes it on the indexes.

  Functoriality of the reindexing remains. A factorization into contractions, boundary renamings, weakenings and a permutation determines a transport presentation, and \cref{def:coherent-transport-completion} identifies the terms obtained from any two presentations of the same structural map. In particular, the empty presentation acts as the identity and $(d^{\sigma'})^{\sigma}\equiv d^{\sigma\circ\sigma'}$. Compatibility with composition reduces to the same generator cases as in the mode theory: weakening uses the recursive Cut clause, contraction shared by both factors uses distribution, contraction of an introduced weakening uses projection, and the remaining rearrangements use the associativity, interchange, and permutation-equivariance clauses of \cref{def:equational-theory-of-derivations}.

  We turn to the two-dimensional structure. The three compositions extend to functors. The composite of $r$ from $(\mf D_1,d_1)$ to $(\mf E_1,e_1)$ with $s$ from $(\mf D_2,d_2)$ to $(\mf E_2,e_2)$ is the $2$-Subst of the underlying transformations. The transport of a cut is the cut of the transports,
  $$(r\circ_{y,x}s)_*\bigl(e_1\circ_{y,x}e_2\bigr)\ \equiv\ r_*(e_1)\circ_{y,x}s_*(e_2)\ \equiv\ d_1\circ_{y,x}d_2.$$
  Every law that remains is an equation between $2$-cells of $\mc D_\Lambda$. A morphism of a hom-category is its underlying transformation together with its endpoints; two parallel $2$-cells are therefore equal exactly when their transformations are equal in the theory of the base, and each remaining law descends from \cref{prop:mode-theory-presents-lnl-polycategory} as soon as its two sides connect the same pairs, which the one-dimensional laws allow. The compositions preserve in this way the identities and the vertical composites by the functoriality clause of \cref{def:equational-theory-of-2-cells}, with the one-sided composites whiskered by identity $2$-cells whose condition is the transport law for the identity; the units, associativity and interchange laws hold on $2$-cells as they hold in $\mc M_\Lambda$.

  The reindexing extends to functors in the same fashion. The permutation and ambient-weakening transports of a $2$-cell of $\mc D_\Lambda$ are the $\chi$-selected transports of \cref{thm:exchange-admissibility-for-descriptor-2-cells,lem:ambient-weakening-of-descriptor-2-cells}, and the contraction of two cartesian declarations is the derived $2$-Subst against the identity $2$-cells of identity descriptors. The condition of each acted-on pair is a computation by the transport laws and the cross-sorted clause of \cref{def:coherent-transport-completion}. For the contraction of $r$ from $(\mf D, d)$ to $(\mf D', d')$, the chain
  $$
  \bigl(\mathbf 1\circ_{b',x'}(\mathbf 1\circ_{b'',x''}r)\bigr)_*\Bigl(\mathrm{id}_A\circ_{b',x'}\bigl(\mathrm{id}_A\circ_{b'',x''}d'\bigr)\Bigr)
  \ \equiv\ \mathrm{id}_A\circ_{b',x'}\bigl(\mathrm{id}_A\circ_{b'',x''}r_*(d')\bigr)
  \ \equiv\ \mathrm{id}_A\circ_{b',x'}\bigl(\mathrm{id}_A\circ_{b'',x''}d\bigr)
  $$
  applies the transport of a cut twice, the transport law for the identity on the identity factors, and the condition of $r$ under the congruence last. The path equations give the functor laws and the equivariance of the assembled action, while the units, associativity, interchange, functoriality, distribution and projection clauses give compatibility with composition. This is the LNL $2$-polycategory presented by the type theory, denoted $\mc D_\Lambda$.

  The two structures assemble now into the fibrational statement. Write $\pi$ for the assignment sending a type over a mode to the mode, a canonical typed context to the canonical mode context over the same variables, a morphism $(\mf D, d)$ to its descriptor $\mf D$, and a $2$-cell to its underlying structural transformation. The type formation rules of \cref{def:type-formation-rules} attach every type to a mode of its class; a cartesian type therefore projects to a cartesian mode and a linear type to a linear mode, and \cref{prop:sequent-indexes-are-well-formed} makes the descriptors well-formed classes of the appropriate sort. Functoriality holds by construction. The compositions of $\mc D_\Lambda$ are computed on the indexes by the Cut rule of the mode theory, the identity terms are indexed by the identity descriptors, the structural actions act on the indexes by the permutation transports, $\mathrm{wk}^{x' : p}_{\mathrm{in}}$ and $[x' := x_0]$ of the mode theory, and the $2$-cells of $\mc D_\Lambda$ are $2$-cells of $\mc M_\Lambda$ carrying a property, composed and acted on as such. The projection therefore preserves every component on the nose, and its actions on the hom-categories are functors.

  For local discreteness, let $r :: \mf D \Rightarrow^{\Lambda} \mf D'$ be a $2$-cell of the base and $(\mf D', d')$ a morphism of $\mc D_\Lambda$ above its target. The rule Tr produces a term above $\mf D$,
  $$
  \inferrule*[Right=Tr]
  {r :: \mf D \Rightarrow^{\Lambda} \mf D' \\ d' :: \Theta\ ;\ \Gamma \vdash^{\Lambda}_{\mf D'} \Delta}
  {r_*(d') :: \Theta\ ;\ \Gamma \vdash^{\Lambda}_{\mf D} \Delta},
  $$
  and the pair of $r$ with domain $(\mf D, r_*(d'))$ and codomain $(\mf D', d')$ is a $2$-cell of $\mc D_\Lambda$ above $r$, its condition holding on the nose, and every $2$-cell above $r$ with this codomain has domain $\equiv r_*(d')$ by its own condition. Every $2$-cell of the base with a chosen lift of its codomain therefore admits a unique lift, i.e. every hom-functor of $\pi$ is a discrete fibration. In particular, a $2$-cell of $\mc D_\Lambda$ is determined by its underlying transformation and its codomain, a fact used in both lift constructions. Throughout those constructions, we write the domain and codomain in the general way, the declarations standing in the zones of their classes, and we verify the \hyperref[anchor:lnl-local-discrete-elementwise]{two elementwise clauses from the prerequisites}, the sort instances of the schemas covering the nonlinear and the linear configurations of the definitions uniformly.

  \textbf{Opcartesian lifts.} We show that the data presents opcartesian lifts of mode multimorphisms (i.e. mode morphisms with only one output). Let $\mathsf F = F^{y}_{\mf D_F}(\Theta_F ; \Gamma_F)$ be an $F$-type at a mode $q$, with $\theta_F \; ; \; \gamma_F \Vdash_{\mf D_F} y : q$ of either sort, and let
  $$\eta_{\mathsf F} :: \Theta_F \; ; \; \Gamma_F \vdash^{\Lambda}_{\mf D_F[y := \bar z]} \bar z : \mathsf F$$
  be its generic term (\cref{prop:generic-terms}). Base morphisms are identified under boundary renamings, so $\pi\eta_{\mathsf F}$ is $\mf D_F$ and $\eta_{\mathsf F}$ lies above it. We show that $\eta_{\mathsf F}$ is opcartesian at $\bar z$.

  Let $\mf D$ be a well-formed class over $(\gamma_L, z : q, \gamma_R \; ; \; \delta)$ and let $g$ be a term above the composite, $g :: \Theta, \Theta_F \; ; \; \Gamma_L, \Gamma_F, \Gamma_R \vdash^{\Lambda}_{\mf D_F \circ_{y,z} \mf D} \Delta$. The $F$-left rule, in the sort instance of the connective and of the consumer (\cref{prop:rule-specialization}), applies to $g$ and produces a term above $\mf D$,
  $$
  \inferrule*[Right=$F$L]
  {g :: \Theta, \Theta_F \; ; \; \Gamma_L, \Gamma_F, \Gamma_R \vdash^{\Lambda}_{\mf D_F \circ_{y,z} \mf D} \Delta}
  {\mathrm{FL}(g) :: \Theta \; ; \; \Gamma_L, z : \mathsf F, \Gamma_R \vdash^{\Lambda}_{\mf D} \Delta},
  $$
  and cutting the generic term back into it, the two terms weakened into the united cartesian zone (\cref{lem:cartesian-weakening-of-derivation-terms}), lands on the boundary and index of $g$,
  $$
  \inferrule*[Right=Cut]
  {
    \eta_{\mathsf F} :: \Theta_F \; ; \; \Gamma_F \vdash^{\Lambda}_{\mf D_F[y := \bar z]} \bar z : \mathsf F
    \\
    \mathrm{FL}(g) :: \Theta \; ; \; \Gamma_L, z : \mathsf F, \Gamma_R \vdash^{\Lambda}_{\mf D} \Delta
  }
  {
    \eta_{\mathsf F} \circ_{\bar z, z} \mathrm{FL}(g) :: \Theta, \Theta_F \; ; \; \Gamma_L, \Gamma_F, \Gamma_R \vdash^{\Lambda}_{\mf D_F \circ_{y,z} \mf D} \Delta
  }.
  $$
  The $\beta$-law at the generic term, in the \hyperref[rem:generic-beta-special-instances]{recorded special instance}, gives the factorization
  $$\eta_{\mathsf F} \circ_{\bar z, z} \mathrm{FL}(g)\ \equiv\ g,$$ which establishes the existence part. For the uniqueness, let $\varphi$ be a term above $\mf D$ with $\eta_{\mathsf F} \circ_{\bar z, z} \varphi \equiv g$. The $\eta$-law and the congruence under $\mathrm{FL}$ give
  $$\varphi\ \equiv\ \mathrm{FL}\bigl(\eta_{\mathsf F} \circ_{\bar z, z} \varphi\bigr)\ \equiv\ \mathrm{FL}(g),$$
  and the factor is unique. Where the invertibility propositions of the metatheory produced some inversion of a consumer of $\mathsf F$, the $\eta$-law states that every consumer is the left rule applied to its own composite with the generic term, up to $\equiv$. Existence of the factorization is the $\beta$-law and its uniqueness the $\eta$-law, in exact parallel with \cite{LSR17}.

  Now we show that the lift respects $2$-cells. Let $b :: \mf D \Rightarrow^{\Lambda} \mf D'$ be a $2$-cell of the base between two factor descriptors and let $t$ be a $2$-cell of $\mc D_\Lambda$ from $(\mf D_F \circ_{y,z} \mf D, g)$ to $(\mf D_F \circ_{y,z} \mf D', g')$ above $\mathbf 1_{\mf D_F} \circ_{y,z} b$, i.e. with the condition $(\mathbf 1_{\mf D_F} \circ_{y,z} b)_*(g') \equiv g$. The factorizations of its endpoints are $\mathrm{FL}(g)$ and $\mathrm{FL}(g')$, and the candidate lift of $b$ is $b$ itself. To verify that this candidate is a $2$-cell between the factorizations, it remains to prove $b_*(\mathrm{FL}(g'))\equiv\mathrm{FL}(g)$. Consider this chain:
  $$
  \begin{aligned}
  b_*\bigl(\mathrm{FL}(g')\bigr)
  &\ \equiv\ \mathrm{FL}\Bigl(
      \eta_{\mathsf F}\circ_{\bar z,z}b_*\bigl(\mathrm{FL}(g')\bigr)
    \Bigr)
  && \text{by the $F$-$\eta$-law}
  \\
  &\ \equiv\ \mathrm{FL}\Bigl(
      \bigl(\mathbf 1_{\mf D_F[y:=\bar z]}\bigr)_*(\eta_{\mathsf F})
      \circ_{\bar z,z}b_*\bigl(\mathrm{FL}(g')\bigr)
    \Bigr)
  && \text{by identity transport}
  \\
  &\ \equiv\ \mathrm{FL}\Bigl(
      \bigl(\mathbf 1_{\mf D_F[y:=\bar z]}\circ_{\bar z,z}b\bigr)_*
      \bigl(\eta_{\mathsf F}\circ_{\bar z,z}\mathrm{FL}(g')\bigr)
    \Bigr)
  && \text{by transport of a cut}
  \\
  &\ \equiv\ \mathrm{FL}\Bigl(
      \bigl(\mathbf 1_{\mf D_F[y:=\bar z]}\circ_{\bar z,z}b\bigr)_*(g')
    \Bigr)
  && \text{by the $F$-$\beta$-law under transport}
  \\
  &\ \equiv\ \mathrm{FL}\Bigl(
      \bigl(\mathbf 1_{\mf D_F}\circ_{y,z}b\bigr)_*(g')
    \Bigr)
  && \text{by cut-variable renaming}
  \\
  &\ \equiv\ \mathrm{FL}(g)
  && \text{by the condition on $t$.}
  \end{aligned}
  $$
  Hence $\eta_{\mathsf F}$ is opcartesian at $\bar z$ above $\mf D_F$.

  \textbf{Cartesian lifts.} We show that the syntax presents every cartesian lift. Let $\mathsf U = U^{\mf D_U}_{x_U}(\Theta_U ; \Gamma_U \mid \Delta_U)$ be a $U$-type at a mode $q$, with connective descriptor $\mf D_U$ of either sort declaring the input $x_U : q$ in the zone of its class, and let
  $$\varepsilon_{\mathsf U} :: \Theta_U \; ; \; \Gamma^L_U, x_0 : \mathsf U, \Gamma^R_U \vdash^{\Lambda}_{\mf D_U[x_U := x_0]} \Delta_U$$
  be its generic term (\cref{prop:generic-terms}), lying above $\mf D_U$ because base morphisms are identified under boundary renamings. We show that $\varepsilon_{\mathsf U}$ is cartesian at $x_0$.

  Let $\mf D$ be a well-formed class over $(\gamma \; ; \; \delta_L, z : q, \delta_R)$ and let $g$ be a term above the composite, $g :: \Theta_U, \Theta \; ; \; \Gamma^L_U, \Gamma, \Gamma^R_U \vdash^{\Lambda}_{\mf D \circ_{z, x_U} \mf D_U} \Delta_L, \Delta_U, \Delta_R$. Applying the $U$-right rule, we get
  $$
  \inferrule*[Right=$U$R]
  {g :: \Theta, \Theta_U \; ; \; \Gamma^L_U, \Gamma, \Gamma^R_U \vdash^{\Lambda}_{\mf D \circ_{z, x_U} \mf D_U} \Delta_L, \Delta_U, \Delta_R}
  {\mathrm{UR}(g) :: \Theta \; ; \; \Gamma \vdash^{\Lambda}_{\mf D} \Delta_L, z : \mathsf U, \Delta_R},
  $$
  and cutting it back into the generic term gets us:
  $$
  \inferrule*[Right=Cut]
  {
    \mathrm{UR}(g) :: \Theta \; ; \; \Gamma \vdash^{\Lambda}_{\mf D} \Delta_L, z : \mathsf U, \Delta_R
    \\
    \varepsilon_{\mathsf U} :: \Theta_U \; ; \; \Gamma^L_U, x_0 : \mathsf U, \Gamma^R_U \vdash^{\Lambda}_{\mf D_U[x_U := x_0]} \Delta_U
  }
  {
    \mathrm{UR}(g) \circ_{z, x_0} \varepsilon_{\mathsf U} :: \Theta_U, \Theta \; ; \; \Gamma^L_U, \Gamma, \Gamma^R_U \vdash^{\Lambda}_{\mf D \circ_{z, x_U} \mf D_U} \Delta_L, \Delta_U, \Delta_R
  }.
  $$
  The $\beta$-law at the generic term gives
  $$\mathrm{UR}(g) \circ_{z, x_0} \varepsilon_{\mathsf U}\ \equiv\ g,$$ which establishes the existence part. For the uniqueness, a term $\varphi$ above $\mf D$ with $\varphi \circ_{z, x_0} \varepsilon_{\mathsf U} \equiv g$ satisfies, by the $\eta$-law and the congruence under $\mathrm{UR}$,
  $$\varphi\ \equiv\ \mathrm{UR}\bigl(\varphi \circ_{z, x_0} \varepsilon_{\mathsf U}\bigr)\ \equiv\ \mathrm{UR}(g).$$
  Now let $b :: \mf D \Rightarrow^{\Lambda} \mf D'$ be a $2$-cell of the base and let $t$ be a $2$-cell from $(\mf D \circ_{z, x_U} \mf D_U, g)$ to $(\mf D' \circ_{z, x_U} \mf D_U, g')$ above $b \circ_{z, x_U} \mathbf 1_{\mf D_U}$, i.e. with the condition $(b \circ_{z, x_U} \mathbf 1_{\mf D_U})_*(g')\equiv g$. The candidate lift between the factors is $b$. To verify its defining condition, it remains to prove $b_*(\mathrm{UR}(g'))\equiv\mathrm{UR}(g)$. The equality is established by this chain:
  $$
  \begin{aligned}
  b_*\bigl(\mathrm{UR}(g')\bigr)
  &\ \equiv\ \mathrm{UR}\Bigl(
      b_*\bigl(\mathrm{UR}(g')\bigr)\circ_{z,x_0}\varepsilon_{\mathsf U}
    \Bigr)
  && \text{by the $U$-$\eta$-law}
  \\
  &\ \equiv\ \mathrm{UR}\Bigl(
      b_*\bigl(\mathrm{UR}(g')\bigr)\circ_{z,x_0}
      \bigl(\mathbf 1_{\mf D_U[x_U:=x_0]}\bigr)_*(\varepsilon_{\mathsf U})
    \Bigr)
  && \text{by identity transport}
  \\
  &\ \equiv\ \mathrm{UR}\Bigl(
      \bigl(b\circ_{z,x_0}\mathbf 1_{\mf D_U[x_U:=x_0]}\bigr)_*
      \bigl(\mathrm{UR}(g')\circ_{z,x_0}\varepsilon_{\mathsf U}\bigr)
    \Bigr)
  && \text{by transport of a cut}
  \\
  &\ \equiv\ \mathrm{UR}\Bigl(
      \bigl(b\circ_{z,x_0}\mathbf 1_{\mf D_U[x_U:=x_0]}\bigr)_*(g')
    \Bigr)
  && \text{by the $U$-$\beta$-law under transport}
  \\
  &\ \equiv\ \mathrm{UR}\Bigl(
      \bigl(b\circ_{z,x_U}\mathbf 1_{\mf D_U}\bigr)_*(g')
    \Bigr)
  && \text{by cut-variable renaming}
  \\
  &\ \equiv\ \mathrm{UR}(g)
  && \text{by the condition on $t$.}
  \end{aligned}
  $$

\end{proof}

\endgroup{}

\setcounter{secnumdepth}{2}%
\subfilebibliography

\endgroup{}
\begingroup

\chapter{Examples}\label{chap:examples}
We now treat three representative examples for the framework. The first is a direct generalization of the LSR framework that was concerned with intuitionistic type systems \cite{LSR17}. The precise correspondence is \cref{prop:logical-adequacy-lsr}. The cartesian data in our syntax is necessary and sufficient to interpret the judgmental structure of LSR. As a bonus, this means that our framework is able to interpret the judgmental structure of all the logics that the LSR framework can interpret, giving them all effectively for free.
The second is an interpretation of the judgmental structure of classical linear logic of Blanco \cite{Bla23}, established in \cref{prop:logical-adequacy-cmll}. This is achieved by only considering linear modes in the framework. The third combines cartesian and linear modes to describe linear logic with exponentials. The relationship between the two modes gives the modalities their structural behavior, as we expose in \cref{sec:examples-linear-logic-exponentials}.
The reader should focus on this section while being familiar with the bifibrational framework of \cref{chap:fibrational-framework}. \hyperref[anchor:examples-overview]{The table below} summarizes the three examples' properties.

\phantomsection\label{anchor:examples-overview}
\begin{center}
\small
\renewcommand{\arraystretch}{1.3}
\setlength{\tabcolsep}{4pt}
\begin{tabular}{|l|l|p{0.31\textwidth}|>{\raggedright\arraybackslash}p{0.30\textwidth}|}
\hline
\textbf{Example} & \textbf{Modes} & \textbf{Descriptors and sequents} & \textbf{Categorical semantics} \\
\hline
LSR \cite{LSR17} & all $\modec$ & cartesian descriptors; one output & Bifibration of cartesian  $2$-multicategories \\
\hline
Blanco CMLL \cite{Bla23} & one $l\model$ & linear descriptors; multiple inputs and outputs & Pure linear presentation over the terminal base (\cref{prop:logical-adequacy-cmll}) \\
\hline
Linear logic & \shortstack[l]{$c\modec$\\$l\model$} & mixed descriptors; reusable inputs and multiple linear outputs & Linear--nonlinear adjunctions; exponential modalities $!$ and $?$ \\
\hline
\end{tabular}
\end{center}

\section{The LSR Framework}\label{sec:examples-lsr}

Let $\Sigma_{\mathrm{LSR}}$ be an LSR mode theory and construct its \emph{exact cartesian presentation} $\Lambda_{\mathrm{LSR}}$ in the present framework as follows. Declare every LSR mode cartesian and represent every LSR context-descriptor constant by a c-generator. Take as generating descriptor $2$-cells those exposed in LSR together with the closure instances required by \cref{def:signature}. For each LSR descriptor-equation class $[\alpha]_{\equiv}$, choose a raw representative $\bar\alpha$ for use in translated types and judgments. Finally, no linear modes, $\ell$-generators or gates are required.

The LSR descriptor judgments have the form
$$\theta\vdash\alpha:q$$
for some mode context $\theta$, mode $q$, and context-descriptor $\alpha$. The equivalent judgment in our framework has the form
$$\theta \Vdash_{\mf D_\alpha} b:q,$$
where the descriptor $\mf D_\alpha$ has one redirection $\{r_\alpha \mapsto b\}$ and an assignment set $\mc S_\alpha$ mirroring the tree structure of $\alpha$. Formally, $\mf D_\alpha := [\mc S_\alpha \mid \{r_\alpha \mapsto b\}]$. We show that there is a one-to-one correspondence between LSR descriptors and our own descriptors.
Write $\operatorname{RawExpr}_{\Sigma_{\mathrm{LSR}}}(\theta;q)$ for the raw terms generated by the variable and generator rules, before quotienting by the LSR descriptor equations.
\begin{proposition}\label{prop:lsr-raw-descriptor-bijection}
  Given a mode context $\theta$, a mode $q$ and a variable $b$, there is a bijection between the set of raw LSR descriptor expressions and the set of c-descriptors.
  $$
  \operatorname{RawExpr}_{\Sigma_{\mathrm{LSR}}}(\theta;q)
  \cong
  \left\{\mf D\mid\theta\Vdash_{\mf D}b:q\right\}.
  $$
\end{proposition}
\textit{Proof.}
We construct maps in both directions and prove that they are mutually inverse.

We first recall the two rules generating LSR context descriptors: the variable rule and the generator rule
$$
\inferrule*[Right=LSR-Var]
{
  a:q\in\theta
}
{
  \theta\vdash a:q
},
\qquad\qquad\qquad
\inferrule*[Right=LSR-Gen]
{
  f:p_1,\ldots,p_n\to q\in\Sigma_{\mathrm{LSR}} \\
  \theta\vdash\alpha_1:p_1 \\
  \cdots \\
  \theta\vdash\alpha_n:p_n
}
{
  \theta\vdash f(\alpha_1,\ldots,\alpha_n):q
}.
$$

\begin{itemize}
\item \textbf{LSR descriptors} $\Rightarrow$ \textbf{c-descriptors.} For every output variable $b$, define
$$
\Phi_b:\operatorname{RawExpr}_{\Sigma_{\mathrm{LSR}}}(\theta;q)
\longrightarrow
\left\{\mf D\mid\theta\Vdash_{\mf D}b:q\right\}
$$
by induction on the derivation of the LSR context descriptor $\alpha$ associated to the context $\theta$ and the mode $q$.

\begin{itemize}
  \item If $\alpha=a$ for some declaration $a:q\in\theta$, then the LSR sequent is $\theta', a : q \vdash a : q$; set
  $$
  \Phi_b(a):=[\varnothing\mid\{a\mapsto b\}].
  $$
  An application of the identity rule gives the desired sequent $\theta\Vdash_{\Phi_b(a)}b:q$. 

\item Suppose now that the LSR descriptor $\alpha$ has been constructed using a morphism $f : p_1, \ldots, p_n \to q$ and a list of LSR descriptors $\alpha_1, \ldots, \alpha_n$ where
$$
\alpha=f(\alpha_1,\ldots,\alpha_n),
$$
with $\theta\vdash\alpha_i:p_i$ for every $i$. Let $b_1, \ldots, b_n$ be pairwise distinct output variables. By the induction hypothesis, there are c-descriptors associated to LSR descriptors $\alpha_1, \ldots, \alpha_n$ given by
$$
\Phi_{b_i}(\alpha_i)
=
[\mc S_i\mid\{r_i\mapsto b_i\}].
$$
Now, let $w$ be a fresh internal variable and set 
$$
\Phi_b\big(f(\alpha_1,\ldots,\alpha_n)\big)
:=
\left[
\bigcup_{i=1}^{n}\mc S_i
\cup
\{r_1,\ldots,r_n\xmapsto{f}w\}
\ \middle|\
\{w\mapsto b\}
\right].
$$
An application of the rule c-Comp gives the sequent $\theta\Vdash_{\Phi_b(f(\alpha_1, \ldots, \alpha_n))} b:q$. This finishes the definition of the map from LSR descriptors to c-descriptors.
\end{itemize}

\item \textbf{c-descriptors} $\Rightarrow$ \textbf{LSR descriptors.} Let $\mf D$ be a c-descriptor so that $\theta\Vdash_{\mf D}b:q$. Let $(\mc S\mid\mc T)\in\mf D$ be a representative of $\mf D$. A well-formed descriptor is regular by \cref{thm:well-formed-descriptors-are-regular}. The boundary usage property gives a unique redirection $r\mapsto b$, and the dependent-tree recursion of \cref{def:dependent-tree} therefore defines
$$
\Psi_b(\mf D):=\llbracket b\rrbracket_{(\mc S\mid\mc T)}
=\llbracket r\rrbracket_{(\mc S\mid\mc T)}.
$$
Let $\mu$ be the mode assignment witnessing mode-validity of $(\mc S\mid\mc T)$. Since $\mf D$ is a c-descriptor, every variable of $(\mc S\mid\mc T)$ is at a cartesian mode and every assignment in $\mc S$ is a c-generator assignment. Thus, by the unique internal target occurrence property, every internal variable $w$ occurring in the recursion has a unique producing assignment of the form
$$
s_1,\ldots,s_n\xmapsto{f}w
$$
for a c-generator $f:p_1,\ldots,p_n\to p$, and we have
$
\llbracket w\rrbracket
=
f(\llbracket s_1\rrbracket,\ldots,\llbracket s_n\rrbracket).
$
By the acyclicity of the dependent tree, every recursive application of a c-generator terminates either at an input variable or at a nullary c-generator.

We now show a stronger claim: for every input or internal variable $v$ in the dependent tree of $b$, the following LSR sequent is derivable
$$
\theta\vdash\llbracket v\rrbracket_{(\mc S\mid\mc T)}:
\mathrm{mode}_{\mu}(v).
$$
\begin{itemize}
  \item Suppose first that $v=a$ is an input variable. This means that $a:p\in\theta$, where $p=\mathrm{mode}_{\mu}(a)$. Using the definition of the input case of the formation of the dependent tree, we have that $\llbracket a \rrbracket_{(\mc S \mid \mc T)} = a$. Using the variable rule for the LSR context descriptor, we can form $\theta \vdash a : p$ and so
  $$
    \theta \vdash \llbracket a \rrbracket_{(\mc S \mid \mc T)} : \mathrm{mode}_mu(a)
  $$
  \item Suppose now that $v=w$ is an internal variable so that
  $
  s_1,\ldots,s_n\xmapsto{f}w,$ for a morphism $f:p_1,\ldots,p_n\to p.$
  Mode-validity gives that 
  $\mathrm{mode}_{\mu}(s_i)=p_i$ for every $1\leq i\leq n$ as well as $\mathrm{mode}_{\mu}(w)=p$.
  By the induction hypothesis on the variables $s_1, \ldots, s_n$ (which are either input or internal but never output), we have the following collection of LSR sequents
  $$
  \theta\vdash\llbracket s_i\rrbracket_{(\mc S\mid\mc T)}:p_i
  \qquad\text{for every }1\leq i\leq n.
  $$
  We now apply the second generator rule for LSR descriptors to those previous sequents taken as premises to get
  $$
  \theta\vdash
  f(\llbracket s_1\rrbracket_{(\mc S\mid\mc T)},\ldots,
    \llbracket s_n\rrbracket_{(\mc S\mid\mc T)}):p.
  $$
  This descriptor is exactly the one generated by the generator clause of the dependent tree at $w$, and so
  $
  \theta\vdash\llbracket w\rrbracket_{(\mc S\mid\mc T)}:
  \mathrm{mode}_{\mu}(w),
  $
  as desired. 
\end{itemize}

It remains to verify that this expression does not depend on the chosen representative. Let $(\mc S'\mid\mc T')\in\mf D$ be another representative. By the definition of $\alpha$-equivalence, there is an internal renaming
$$
\phi:\mathrm{Int}(\mc S\mid\mc T)
\longrightarrow
\mathrm{Int}(\mc S'\mid\mc T')
$$
such that $(\mc S\mid\mc T)^\phi=(\mc S'\mid\mc T')$. Its extension fixes every input and output variable. In particular, the root redirection $r\mapsto b$ is carried to $r^\phi\mapsto b$. We prove, again by induction on the dependent tree, that
$$
\llbracket v^\phi\rrbracket_{(\mc S'\mid\mc T')}
=
\llbracket v\rrbracket_{(\mc S\mid\mc T)}
$$
for every input or internal variable $v$ in the tree. If $v=a$ is an input, then $a^\phi=a$ and both sides are equal to $a$. Suppose now that $v=w$ is internal and is produced in $\mc S$ by
$
s_1,\ldots,s_n\xmapsto{f}w.
$
The corresponding assignment in $\mc S'$ is
$
s_1^\phi,\ldots,s_n^\phi\xmapsto{f}w^\phi.
$
The induction hypothesis applies to every source subtree; we therefore get
\begin{align*}
\llbracket w^\phi\rrbracket_{(\mc S'\mid\mc T')}
&=f(\llbracket s_1^\phi\rrbracket_{(\mc S'\mid\mc T')},\ldots,
       \llbracket s_n^\phi\rrbracket_{(\mc S'\mid\mc T')})\\
&=f(\llbracket s_1\rrbracket_{(\mc S\mid\mc T)},\ldots,
       \llbracket s_n\rrbracket_{(\mc S\mid\mc T)})\\
&=\llbracket w\rrbracket_{(\mc S\mid\mc T)}.
\end{align*}

\item \textbf{Correspondence.} We now show that the two constructed maps are inverses of each other; we first prove
$$
\Psi_b(\Phi_b(\alpha))=\alpha
$$
by induction on $\alpha$. For a variable $a$, the unique redirection is $a\mapsto b$, and hence $\llbracket b\rrbracket=a$. For $\alpha=f(\alpha_1,\ldots,\alpha_n)$, the defining assignment and redirection of $\Phi_b(\alpha)$ give
\begin{align*}
\Psi_b(\Phi_b(\alpha))
&=\llbracket b\rrbracket
=\llbracket w\rrbracket\\
&=f(\llbracket r_1\rrbracket,\ldots,\llbracket r_n\rrbracket)\\
&=f\big(\Psi_{b_1}(\Phi_{b_1}(\alpha_1)),\ldots,
        \Psi_{b_n}(\Phi_{b_n}(\alpha_n))\big)\\
&=f(\alpha_1,\ldots,\alpha_n),
\end{align*}
where we used the induction hypothesis for the last equality.

Conversely, let $\mf D=[\mc S\mid\mc T]$ and put $\alpha:=\Psi_b(\mf D)$. We prove
$
\Phi_b(\Psi_b(\mf D))=\mf D
$
by induction on the dependent tree of $b$. By the boundary usage property, there exists a unique redirection. Set $\mc T=\{r\mapsto b\}$ for a unique input or internal variable $r$. For every input or internal variable $v$ in the dependent tree, define
\begin{align*}
\mc S_{\downarrow v}
&:=
\left\{
E\in\mc S
\ \middle|\ 
\mathrm{tgt}(E)=\{z\}
\text{ and }
(z=v\text{ or }z\rightsquigarrow v)
\right\},\\
\mf D_v^c
&:=
\left[
\mc S_{\downarrow v}
\ \middle|\ 
\{v\mapsto c\}
\right],
\end{align*}
where $c$ is an output variable at the mode of $v$. The set $\mc S_{\downarrow v}$ consists of all the assignments in the subtree rooted at $v$.

Note that if $a$ is an input variable, no assignment targets $a$ and no variable reaches $a$; therefore, $\mc S_{\downarrow a}=\varnothing.$
Furthermore, note that if $w$ is internal and its unique assignment is $ E_w:=\big(s_1,\ldots,s_n\xmapsto{f}w\big), $
then every nonempty path ending at $w$ has a final edge $s_i\leadsto w$ for some $i$. Consequently,
\begin{equation}
\mc S_{\downarrow w}
=
\{E_w\}
\cup
\bigcup_{i=1}^{n}\mc S_{\downarrow s_i}.
\label{eq:rooted-assignment-decomposition}
\end{equation}
Moreover, using regularity and, more particularly, the unique internal source occurrence and acyclicity of the dependent tree, we get
$\mathrm{Int}(\mc S_{\downarrow s_i}) \cap
\mathrm{Int}(\mc S_{\downarrow s_j})
=\varnothing$ for $i\neq j$ and $w \notin \bigcup_{i=1}^{n}\mathrm{Var}(\mc S_{\downarrow s_i})$.

We now prove the stronger equality
\begin{equation}
\Phi_c\big(\llbracket v\rrbracket_{(\mc S\mid\mc T)}\big)
=
\mf D_v^c
\label{eq:phi-reconstructs-rooted-subtree}
\end{equation}
for every node $v$ of the dependent tree and every output variable $c$ at its mode. If $v=a$ is an input, LSR-Var and the definition of $\Phi_c$ give
\begin{align*}
\Phi_c\big(\llbracket a\rrbracket_{(\mc S\mid\mc T)}\big)
&=\Phi_c(a)\\
&=[\varnothing\mid\{a\mapsto c\}]\\
&=[\mc S_{\downarrow a}\mid\{a\mapsto c\}]\\
&=\mf D_a^c.
\end{align*}
Suppose now that $v=w$ is internal with assignment
$$
E_w=\big(s_1,\ldots,s_n\xmapsto{f}w\big),
\qquad
f:p_1,\ldots,p_n\to p.
$$
Let $c_1,\ldots,c_n$ be pairwise distinct output variables at modes $p_1,\ldots,p_n$. By the induction hypothesis,
$$
\Phi_{c_i}\big(\llbracket s_i\rrbracket_{(\mc S\mid\mc T)}\big)
=
\left[
\mc S_{\downarrow s_i}
\ \middle|\ 
\{s_i\mapsto c_i\}
\right]
\qquad (1\leq i\leq n).
$$
Using these representatives in the defining clause of $\Phi_c$, and choosing $w$ as its fresh internal variable, we calculate
\begin{align*}
\Phi_c\big(\llbracket w\rrbracket_{(\mc S\mid\mc T)}\big)
&=
\Phi_c\big(
f(\llbracket s_1\rrbracket_{(\mc S\mid\mc T)},\ldots,
  \llbracket s_n\rrbracket_{(\mc S\mid\mc T)})
\big)\\
&=
\left[
\bigcup_{i=1}^{n}\mc S_{\downarrow s_i}
\cup
\{s_1,\ldots,s_n\xmapsto{f}w\}
\ \middle|\ 
\{w\mapsto c\}
\right]\\
&=
\left[
\mc S_{\downarrow w}
\ \middle|\ 
\{w\mapsto c\}
\right]\\
&=\mf D_w^c,
\end{align*}
where the third equality is \eqref{eq:rooted-assignment-decomposition}.  This proves \eqref{eq:phi-reconstructs-rooted-subtree}.
\end{itemize}
This finishes the proof.
\hfill $\square$

Under this bijection, descriptor cut is expression substitution by \cref{cor:cartesian-cut-is-term-substitution}.

The same correspondence extends to typed sequents. After choosing representatives, an LSR sequent
$$
\Theta\vdash_\alpha A
$$
is represented in the present framework by
$$
\Theta\vdash^{\Lambda_{\mathrm{LSR}}}_{\mf D^y_{\bar\alpha}}y:A.
$$

Since we are only considering c-descriptors, the type formers' premises for the connectives $F,U$ specialize to $\theta_\alpha\Vdash_{\mf D^y_{\bar\alpha}}y:q$ and
$\theta_\beta,x:q\Vdash_{\mf D^b_{\bar\beta}}b:r,$ and the rules give rise to the connectives
$$
\mathsf F_{\mathsf c}:=F^y_{\mf D^y_{\bar\alpha}}(\Theta_\alpha;\cdot),
\qquad
\mathsf U_{\mathsf c}:=U^{\mf D^b_{\bar\beta}}_{x}(\Theta_\beta;\cdot\mid b:B),
$$
which correspond to the LSR connectives $F_\alpha(\Theta_\alpha)$ and $U_{x.\beta}(\Theta_\beta \mid B)$. The variables $b$ and $y$ are omitted in the LSR framework because they only name the unique output.
The cartesian instances of the five rule schemas in \cref{prop:rule-specialization} are exactly LSR's variable, $\mathsf{FL}$, $\mathsf{FR}$, $\mathsf{UL}$, and $\mathsf{UR}$ rules \cite{LSR17}. The generic terms of \cref{prop:generic-terms} are the universal $\mathsf{FR}^{*}$ and $\mathsf{UL}^{*}$ terms, and the $\beta\eta$ equations give their universal properties.

We now make the correspondence precise at the level of derivability. For every raw LSR descriptor expression $\alpha$ and fresh output variable $y$ of its codomain mode, write
$$
\mf D^y_\alpha:=\Phi_y(\alpha).
$$
Then, the descriptor specification extends to the transformations as follows
\begin{equation}
\theta\vdash \alpha\Rightarrow_q\beta
\quad\Longleftrightarrow\quad
\mf D^y_\alpha\Rightarrow^{\Lambda_{\mathrm{LSR}}}_{\theta\Vdash y:q}\mf D^y_\beta.
\label{eq:lsr-structural-correspondence}
\end{equation}

Identify the atomic types of the two calculi and define the translation $(-)^\dagger$ of LSR types by
\begin{align*}
P^\dagger
&:=P,\\
\bigl(F_\alpha(\Theta_\alpha)\bigr)^\dagger
&:=F^y_{\mf D^y_{\bar\alpha}}(\Theta_\alpha^\dagger;\cdot),\\
\bigl(U_{x.\beta}(\Theta_\beta\mid B)\bigr)^\dagger
&:=U_x^{\mf D^b_{\bar\beta}}(\Theta_\beta^\dagger;\cdot\mid b:B^\dagger),
\end{align*}
where $y$ and $b$ are chosen fresh. Their names are immaterial by \hyperref[rem:context-variable-naming-invariance]{context variable naming invariance}. Extend the translation pointwise to contexts and set
$$
\bigl(\Theta\vdash_\alpha A\bigr)^\dagger
:=
\Theta^\dagger\vdash^{\Lambda_{\mathrm{LSR}}}_{\mf D^y_{\bar\alpha}}y:A^\dagger.
$$
An LSR substitution judgment $\Theta\vdash_\gamma\Theta_0$, where $\Theta_0=(a:A_a)_{a\in\theta_0}$ and $\gamma=(\gamma_a/a)_{a\in\theta_0}$, is translated pointwise to the family
$$
\left(
\Theta^\dagger
\vdash^{\Lambda_{\mathrm{LSR}}}_{\mf D^{b_a}_{\bar\gamma_a}}
b_a:A_a^\dagger
\right)_{a\in\theta_0},
$$
with the $b_a$ pairwise fresh.

\begin{proposition}[Logical adequacy for LSR]\label{prop:logical-adequacy-lsr}
For every well-formed sequent $\Theta\vdash_\alpha A$ of the LSR calculus,
$$
\Theta\vdash_\alpha A
\text{ is derivable in LSR}
\quad\Longleftrightarrow\quad
\Theta^\dagger
\vdash^{\Lambda_{\mathrm{LSR}}}_{\mf D^y_{\bar\alpha}}
y:A^\dagger
\text{ is derivable},
$$
where the derivation on the right is in the cartesian fragment of the framework.
\end{proposition}

\textit{Proof.}
We first define the translation of an LSR context descriptor to a c-descriptor. There are three cases to consider: the variable descriptor $x$ for any variable, the substitution descriptor $\beta[\alpha / z]$ for context descriptors $\beta, \alpha$ and a variable $z$, and the context substitution $\alpha[\gamma]$ for two context descriptors $\alpha$ and $\gamma$. The new variables introduced for c-descriptors are chosen fresh.

\begin{align}
\mf D_x^y
&=[\varnothing\mid\{x\mapsto y\}],
&\text{(variable translation)}\label{eq:lsr-variable-translation}\\
\mf D_{\beta[\alpha/z]}^u
&=\mf D_\alpha^y\circ_{y,z}\mf D_\beta^u,
&\text{(substitution)}\label{eq:lsr-single-substitution}\\
\mf D_{\alpha[\gamma]}^z
&=
\left(
  \bigl(\mf D_{\gamma_a}^{b_a}\bigr)_{a\in\theta_0}
  \circ_{(b_a,a)_{a\in\theta_0}}
  \mf D_\alpha^y
\right)[y:=z],
&\text{(context substitution)}\label{eq:lsr-simultaneous-substitution}
\end{align}
Retaining the distinguished variable $x$ in a $U$-descriptor gives the further identity
\begin{equation}
\left(
  \bigl(\mf D_{\gamma_a}^{b_a}\bigr)_{a\in\theta_0}
  \circ_{(b_a,a)_{a\in\theta_0}}
  \mf D_\alpha^b
\right)[x:=z]
=\mf D_{\alpha[\gamma,z/x]}^b.
\label{eq:lsr-u-substitution}
\end{equation}
For the forward implication, proceed by induction on the LSR derivation, mutually with the pointwise induction on substitution judgments.
\begin{itemize}
\item \textbf{Variable.} An LSR variable rule has $x:P\in\Theta$ and a structural transformation $\beta\Rightarrow x$. By \eqref{eq:lsr-structural-correspondence} and \eqref{eq:lsr-variable-translation}, it translates to
$$
\mf D_\beta^y\Rightarrow^{\Lambda_{\mathrm{LSR}}}
[\varnothing\mid\{x\mapsto y\}],
$$
so the cartesian variable rule gives the translated conclusion.

\item \textbf{F-left.} The induction hypothesis translates the premise indexed by $\beta[\alpha/z]$ to one indexed by $\mf D_\alpha^y\circ_{y,z}\mf D_\beta^u$, by \eqref{eq:lsr-single-substitution}. The rule $F_{\mathsf c}$L gives the translated conclusion.

\item \textbf{F-right.} Write $\sigma_a:=\mf D_{\gamma_a}^{b_a}$ for the translations of the components of the LSR substitution premise. The structural premise $\beta\Rightarrow\alpha[\gamma]$ translates, by \eqref{eq:lsr-structural-correspondence} and \eqref{eq:lsr-simultaneous-substitution}, to
$$
\mf D_\beta^z\Rightarrow^{\Lambda_{\mathrm{LSR}}}
\left(
  (\sigma_a)_{a\in\theta_\alpha}
  \circ_{(b_a,a)_{a\in\theta_\alpha}}
  \mf D_\alpha^y
\right)[y:=z].
$$
Together with the translated family of substitution premises, these are exactly the premises of $F_{\mathsf c}$R.

\item \textbf{U-right.} The induction hypothesis translates the premise indexed by $\alpha[\beta/x]$. Equation \eqref{eq:lsr-single-substitution}, with the cut variables renamed, identifies this index with $\mf D_\beta^z\circ_{z,x}\mf D_\alpha^b$. The rule $U_{\mathsf c}$R gives the translated conclusion.

\item \textbf{U-left.} Let $z:U_{x.\alpha}(\Theta_\alpha\mid B)\in\Theta$ and write $x':B$ for the output variable. The LSR transformation premise has the form
$$
\beta\Rightarrow
\beta'\bigl[\alpha[\gamma,z/x]/x'\bigr].
$$
Translate the substitution premise pointwise and put $\sigma_a:=\mf D_{\gamma_a}^{b_a}$. Equations \eqref{eq:lsr-single-substitution} and \eqref{eq:lsr-u-substitution} identify the translated target with
$$
\mf D_{\beta'}^u
\left[
x':=
\left(
  (\sigma_a)_{a\in\theta_\alpha}
  \circ_{(b_a,a)_{a\in\theta_\alpha}}
  \mf D_\alpha^b
\right)[x:=z]
\right].
$$
All the translated premises are therefore exactly the ones needed for the $U_{\mathsf c}$L rule.
\end{itemize}

For the converse statement, if $\mf E$ is a c-descriptor with output $w$, we define its associated LSR context descriptor by
$$
\langle\mf E\rangle_w:=\Psi_w(\mf E).
$$
According to the correspondence of descriptors established in \cref{prop:lsr-raw-descriptor-bijection}, we have that
$\mf E=\mf D^w_{\langle\mf E\rangle_w}.$
We prove the stronger statement
\begin{equation}
\Pi:\Theta^\dagger
\vdash^{\Lambda_{\mathrm{LSR}}}_{\mf E}
w:A^\dagger
\quad\Longrightarrow\quad
\Theta\vdash_{\langle\mf E\rangle_w}A
\text{ is derivable in LSR}
\label{eq:lsr-readback-induction}
\end{equation}
by induction on the height of the derivation $\Pi$.

\begin{itemize}
\item \textbf{Variable.} Suppose the last rule of the derivation is an instance of the variable rule
$$
\inferrule*[Right=$\mathsf v$]
{x:P\in\Theta^\dagger \\
 r::\mf E\Rightarrow^{\Lambda_{\mathrm{LSR}}}
 [\varnothing\mid\{x\mapsto w\}]}
{\Theta^\dagger\vdash^{\Lambda_{\mathrm{LSR}}}_{\mf E}w:P}.
$$
Let $\beta:=\langle\mf E\rangle_w$; the premise becomes
$\mf D^w_\beta\Rightarrow\mf D^w_x$. This corresponds to the LSR transformation 
$\beta\Rightarrow x$ by \eqref{eq:lsr-structural-correspondence}. The LSR variable rule then gives the result
$$
\inferrule*[Right=$\mathsf v$]
{x:P\in\Theta \\ \beta\Rightarrow x}
{\Theta\vdash_\beta P}.
$$

\item \textbf{F-left.}
Suppose the final rule is an instance of the left introduction rule for the F-connective
$$
\inferrule*[Right=$F_{\mathsf c}$L]
{\Pi_0:\Theta^\dagger,\Theta_\alpha^\dagger
 \vdash^{\Lambda_{\mathrm{LSR}}}_{
   \mf D^y_{\bar\alpha}\circ_{y,z}\mf E}
 u:C^\dagger}
{\Theta^\dagger,z:\bigl(F_\alpha(\Theta_\alpha)\bigr)^\dagger
 \vdash^{\Lambda_{\mathrm{LSR}}}_{\mf E}u:C^\dagger}.
$$
Write $\beta:=\langle\mf E\rangle_u$; by
\eqref{eq:lsr-single-substitution}, the premise descriptor reads as
$\beta[\alpha/z]$. The induction hypothesis applied to $\Pi_0$, followed by LSR's $\mathsf{FL}$ rule, gives
$$
\inferrule*[Right=$\mathsf{FL}$]
{\Theta,\Theta_\alpha\vdash_{\beta[\alpha/z]}C}
{\Theta,z:F_\alpha(\Theta_\alpha)\vdash_\beta C}.
$$

\item \textbf{$F$-right.}
Suppose the final rule is
$$
\inferrule*[Right=$F_{\mathsf c}$R]
{r::\mf E\Rightarrow^{\Lambda_{\mathrm{LSR}}}
 \left(
   (\sigma_a)_{a\in\theta_\alpha}
   \circ_{(b_a,a)_{a\in\theta_\alpha}}
   \mf D^y_{\bar\alpha}
 \right)[y:=w]
 \\
 \left(
   \Pi_a:\Theta^\dagger
   \vdash^{\Lambda_{\mathrm{LSR}}}_{\sigma_a}
   b_a:A_a^\dagger
 \right)_{a\in\theta_\alpha}}
{\Theta^\dagger\vdash^{\Lambda_{\mathrm{LSR}}}_{\mf E}
 w:\bigl(F_\alpha(\Theta_\alpha)\bigr)^\dagger}.
$$
Set
$$
\beta:=\langle\mf E\rangle_w,
\qquad
\gamma_a:=\langle\sigma_a\rangle_{b_a}.
$$
By \eqref{eq:lsr-simultaneous-substitution}, the target of $r$ reads as $\alpha[\gamma]$. Therefore
$\beta\Rightarrow\alpha[\gamma]$ by
\eqref{eq:lsr-structural-correspondence}. The induction hypotheses give
$\Theta\vdash_{\gamma_a}A_a$ for every $a$. The LSR substitution rules give
$\Theta\vdash_\gamma\Theta_\alpha$. Hence LSR's $\mathsf{FR}$ rule gives
$$
\inferrule*[Right=$\mathsf{FR}$]
{\beta\Rightarrow\alpha[\gamma]
 \\ \Theta\vdash_\gamma\Theta_\alpha}
{\Theta\vdash_\beta F_\alpha(\Theta_\alpha)}.
$$

\item \textbf{U-right.}
Suppose the final rule is an instance of the right introduction rule for the U connective
$$
\inferrule*[Right=$U_{\mathsf c}$R]
{\Pi_0:\Theta^\dagger,\Theta_\alpha^\dagger
 \vdash^{\Lambda_{\mathrm{LSR}}}_{
   \mf E\circ_{w,x}\mf D^b_{\bar\alpha}}
 b:B^\dagger}
{\Theta^\dagger\vdash^{\Lambda_{\mathrm{LSR}}}_{\mf E}
 w:\bigl(U_{x.\alpha}(\Theta_\alpha\mid B)\bigr)^\dagger}.
$$
Let $\beta:=\langle\mf E\rangle_w$; by
\eqref{eq:lsr-single-substitution}, the premise descriptor is
$\alpha[\beta/x]$. The induction hypothesis and LSR's $\mathsf{UR}$ rule give
$$
\inferrule*[Right=$\mathsf{UR}$]
{\Theta,\Theta_\alpha\vdash_{\alpha[\beta/x]}B}
{\Theta\vdash_\beta U_{x.\alpha}(\Theta_\alpha\mid B)}.
$$

\item \textbf{$U$-left.}
Suppose the final rule is
$$
\inferrule*[Right=$U_{\mathsf c}$L]
{z:\bigl(U_{x.\alpha}(\Theta_\alpha\mid B)\bigr)^\dagger
 \in\Theta^\dagger
 \\
 r::\mf E\Rightarrow^{\Lambda_{\mathrm{LSR}}}
 \mf E'\left[
   x':=
   \left(
     (\sigma_a)_{a\in\theta_\alpha}
     \circ_{(b_a,a)_{a\in\theta_\alpha}}
     \mf D^b_{\bar\alpha}
   \right)[x:=z]
 \right]
 \\
 \left(
   \Pi_a:\Theta^\dagger
   \vdash^{\Lambda_{\mathrm{LSR}}}_{\sigma_a}
   b_a:A_a^\dagger
 \right)_{a\in\theta_\alpha}
 \\
 \Pi_0:\Theta^\dagger,x':B^\dagger
 \vdash^{\Lambda_{\mathrm{LSR}}}_{\mf E'}u:C^\dagger}
{\Theta^\dagger\vdash^{\Lambda_{\mathrm{LSR}}}_{\mf E}u:C^\dagger}.
$$
Set
$$
\beta:=\langle\mf E\rangle_u,
\qquad
\beta':=\langle\mf E'\rangle_u,
\qquad
\gamma_a:=\langle\sigma_a\rangle_{b_a}.
$$
Equations \eqref{eq:lsr-u-substitution} and
\eqref{eq:lsr-single-substitution} show that the target of $r$ is $ \beta'\bigl[\alpha[\gamma,z/x]/x'\bigr]. $
Thus \eqref{eq:lsr-structural-correspondence} translates $r$ to $ \beta\Rightarrow \beta'\bigl[\alpha[\gamma,z/x]/x'\bigr]. $ The induction hypotheses give
$\Theta\vdash_\gamma\Theta_\alpha$ and
$\Theta,x':B\vdash_{\beta'}C$. Applying LSR's $\mathsf{UL}$ rule yields
$$
\inferrule*[Right=$\mathsf{UL}$]
{z:U_{x.\alpha}(\Theta_\alpha\mid B)\in\Theta
 \\
 \beta\Rightarrow
 \beta'\bigl[\alpha[\gamma,z/x]/x'\bigr]
 \\
 \Theta\vdash_\gamma\Theta_\alpha
 \\
 \Theta,x':B\vdash_{\beta'}C}
{\Theta\vdash_\beta C}.
$$
\end{itemize}
Finally, $\langle\mf D^y_{\bar\alpha}\rangle_y=\bar\alpha\equiv\alpha$, so applying
\eqref{eq:lsr-readback-induction} to the derivation in \cref{prop:logical-adequacy-lsr} gives
$\Theta\vdash_\alpha A$.
\hfill $\square$

\begin{corollary}[Transfer of LSR adequacy]\label{cor:lsr-adequacy-transfer}
Let $\mc L$ be an object logic with a logically adequate translation $(-)^\star$ into the LSR calculus, so that every judgment $J$ of $\mc L$ satisfies
$$
J\text{ is derivable in }\mc L
\quad\Longleftrightarrow\quad
J^\star\text{ is derivable in LSR}.
$$
Then the composite translation $J\mapsto(J^\star)^\dagger$ is logically adequate for the cartesian fragment of the present framework.
\end{corollary}

\textit{Proof.} Immediate. \hfill $\square$

\Cref{prop:logical-adequacy-lsr,cor:lsr-adequacy-transfer} establish the adequacy of the current framework for every logic that is adequate in \cite{LSR17}. This is an equivalence of derivability.

\section{Blanco's Classical Multiplicative Linear Logic}\label{sec:examples-blanco-cmll}

We now treat how the framework encapsulates the judgmental structure of classical linear logic \cite{Bla23}. Let $\Lambda_{\mathrm{CMLL}}$ be pure linear, with one mode $l\model$, no cartesian modes, and no gates.

A presentation of the calculus uses the generators
$$
\otimes_l:l,l\to l,
\qquad
\mathbf 1_l:\cdot\to l,
\qquad
\parr_l:l\to l,l,
\qquad
\bot_l:l\to\cdot,
$$
which encode the information needed for the multiplicative conjunction and the multiplicative disjunction alongside their units.

Every sequent in this setup is linear since there are no declared cartesian modes. Every sequent then has the form
$$
\Gamma\vdash^{\Lambda_{\mathrm{CMLL}}}_{\mf D}\Delta
$$
where the cartesian formulas in $\Theta$ are not needed. All formulas in the context $\Gamma$ are linear. Moreover, every formula is used exactly once. Cut composes one output formula with one input formula and copies the side formulas. This is the definition of the polycategorical cuts of \cite[Definition~2.2.1]{Bla23} in their symmetric variant.

The mode theory generates the following descriptors from the generators
\begin{align*}
\mf D_{\otimes}
&:=[\{x,y\xmapsto{\otimes_l}w\}\mid\{w\mapsto z\}], &
\mf D_{\parr}
&:=[\{z\xmapsto{\parr_l}w_1,w_2\}\mid\{w_1\mapsto x,w_2\mapsto y\}],\\
\mf D_{\mathbf 1}
&:=[\{\cdot\xmapsto{\mathbf 1_l}w\}\mid\{w\mapsto z\}], &
\mf D_{\bot}
&:=[\{x\xmapsto{\bot_l}\cdot\}\mid\cdot].
\end{align*}
Now, we proceed to construct and choose a canonical representative for well-formed descriptors. The idea is that for a mode sequent $\gamma\Vdash\delta$, many non-equivalent well-formed descriptors can exist. We relate them using $2$-cells to a canonical construction. This is necessary to prove an adequacy result at the level of derivability. We construct a canonical
well-formed descriptor as follows. The context $\gamma$ gives rise to $\otimes_l \gamma$ by way of applying the generator to the variables. We set a fresh output $z$ to be this expression and use $\mf D_{\mathbf 1}$ when $\gamma$ is empty and the identity descriptor
when it is a singleton. Let the result be denoted $\mf L^z_\gamma$. Dually, for the fresh input $x$, we can apply the generator $\parr_l$ on the output variables, using $\mf D_{\bot}$ when $\delta$ is empty and
the identity descriptor when it is a singleton.  Let the result be denoted $\mf R_x^\delta$.
Set
$$
\mf S_{\gamma;\delta}
:=\mf L^z_\gamma\circ_{z,x}\mf R_x^\delta.
$$
Induction on the two list lengths, followed by the Cut rule, proves that this
descriptor is well formed over $\gamma\Vdash\delta$. \phantomsection\label{anchor:cmll-singleton-canonical-descriptor}For a singleton input $a:l$ and singleton output $b:l$, the two
trees are identities, and \cref{lem:units-for-composition}(ii) gives
$$
\mf S_{a:l;b:l}
=\mf I_{a,z}\circ_{z,x}\mf I_{x,b}
=\mf I_{a,b}.
$$
Now, for every well-formed descriptor $\mf T$ over the same boundary,
we include in the signature generating $2$-cells
\begin{equation}
\kappa_{\mf T}::\mf T\Rightarrow^{\Lambda_{\mathrm{CMLL}}}
\mf S_{\gamma;\delta},
\qquad
\bar\kappa_{\mf T}::\mf S_{\gamma;\delta}
\Rightarrow^{\Lambda_{\mathrm{CMLL}}}\mf T
\label{eq:cmll-terminal-comparisons}
\end{equation}
such that
$\bar\kappa_{\mf T}\circ\kappa_{\mf T}=\mathbf 1_{\mf T}$ and
$\kappa_{\mf T}\circ\bar\kappa_{\mf T}=\mathbf 1_{\mf S_{\gamma;\delta}}$. This effectively makes every well-formed descriptor isomorphic to the canonical one we just constructed.

The type theory uses those descriptors to generate the following connectives
\phantomsection\label{anchor:cmll-bottom-definition}
\begin{align*}
A\otimes B
&:=F^z_{\mf D_{\otimes}}(\cdot;x:A,y:B), &
\mathbf 1
&:=F^z_{\mf D_{\mathbf 1}}(\cdot;\cdot),\\
A\parr B
&:=U^{\mf D_{\parr}}_z(\cdot;\cdot\mid x:A,y:B), &
\bot
&:=U^{\mf D_{\bot}}_x(\cdot;\cdot\mid\cdot).
\end{align*}
\phantomsection\label{anchor:cmll-negation-definition}
Multiplicative negation is another instance of $U$. Compose the positive and negative generators to obtain
$$
\mf D_{\mathrm{cap}}
:=
[\{x,x'\xmapsto{\otimes_l}w,\ w\xmapsto{\bot_l}\cdot\}\mid\cdot]
$$
over $x:l,x':l\Vdash\cdot$, and define
$$
A^\perp:=U^{\mf D_{\mathrm{cap}}}_{x}(\cdot;x':A\mid\cdot).
$$
As an example of the behavior of the framework, let us show how to derive $A \otimes A^\perp \vdash \bot$ in \cref{prop:cmll-negation-evaluation}.

\begin{proposition}[Negation evaluation]\label{prop:cmll-negation-evaluation}
For every linear type $A$ and fresh labels $t,b$, there is a derivation
$$
t:A\otimes A^\perp
\vdash^{\Lambda_{\mathrm{CMLL}}}_{\mf I_{t,b}}
b:\bot,
\qquad
\mf I_{t,b}:=[\varnothing\mid\{t\mapsto b\}].
$$
\end{proposition}

\textit{Proof.}
Choose fresh input labels $a,f,r$, an output label $b$, and internal labels
as needed. We build the following descriptors. 
\begin{align*}
\mf D_{a \otimes f}^{b}
&:=[\{a,f\xmapsto{\otimes_l}w\}\mid\{w\mapsto b\}], &
\mf D_{f \otimes a}^{b}
&:=[\{f,a\xmapsto{\otimes_l}w\}\mid\{w\mapsto b\}],\\
\mf D_{\bot}^{r}
&:=[\{r\xmapsto{\bot_l}\cdot\}\mid\cdot], &
\mf C_{a \otimes f}
&:=\mf D_{a \otimes f}^{b}\circ_{b,r}\mf D_{\bot}^{r},\\
\mf C_{f \otimes a}
&:=\mf D_{f \otimes a}^{b}\circ_{b,r}\mf D_{\bot}^{r}.
\end{align*}
Thus
\begin{align*}
\mf C_{a \otimes f}
&=[\{a,f\xmapsto{\otimes_l}w,
       w\xmapsto{\bot_l}\cdot\}\mid\cdot],\\
\mf C_{f \otimes a}
&=[\{f,a\xmapsto{\otimes_l}w,
       w\xmapsto{\bot_l}\cdot\}\mid\cdot].
\end{align*}
By the \hyperref[anchor:cmll-negation-definition]{negation definition above}, $ A^\perp=U_f^{\mf C_{f \otimes a}}(\cdot;a:A\mid\cdot). $
Its generic $U$-term from \cref{prop:generic-terms} is
\begin{equation}
\operatorname{cap}_A
:=
\mathrm{UL}\bigl(
  \mathbf 1_{\mf C_{f \otimes a}};();(\mathrm{id}_A);()
\bigr)
::
f:A^\perp,a:A
\vdash^{\Lambda_{\mathrm{CMLL}}}_{\mf C_{f \otimes a}}
\cdot.
\label{eq:cmll-generic-cap-term}
\end{equation}
By the terminal comparison cells of \eqref{eq:cmll-terminal-comparisons}, the descriptors $\mf D_{f \otimes a}^{b}$ and $\mf D_{a\otimes f }^{b}$ over the same boundary contexts admit a transformation; thus, we write
$$
\rho_{\otimes}^{a,f;b}
:=
\bar\kappa_{\mf D_{f \otimes a}^{b}}
\circ
\kappa_{\mf D_{a \otimes f}^{b}}
::
\mf D_{a \otimes f}^{b}
\Rightarrow^{\Lambda_{\mathrm{CMLL}}}
\mf D_{f \otimes a}^{b}.
$$
Applying $2$-Subst to this comparison and the identity of
$\mf D_{\bot}^{r}$ gives
\begin{equation}
\chi_{a,f}
:=
\rho_{\otimes}^{a,f;b}
\circ_{b,r}
\mathbf 1_{\mf D_{\bot}^{r}}
::
\mf C_{a \otimes f}
\Rightarrow^{\Lambda_{\mathrm{CMLL}}}
\mf C_{f \otimes a}.
\label{eq:cmll-cap-exchange-comparison}
\end{equation}
Respect for structural transformation along \eqref{eq:cmll-cap-exchange-comparison} then
gives
$$
d_0
:=
(\chi_{a,f})_*
\bigl(\operatorname{Ex}(\operatorname{cap}_A)\bigr)
::
a:A,f:A^\perp
\vdash^{\Lambda_{\mathrm{CMLL}}}_{\mf C_{a \otimes f}}
\cdot.
$$
Recalling the \hyperref[anchor:cmll-bottom-definition]{definition of $\bot$ above}, namely
$ \bot_r:=U_r^{\mf D_{\bot}^{r}}(\cdot;\cdot\mid\cdot),$ we compute the needed composition for the final derivation
\begin{align*}
\mf D_{a \otimes f}^{b}\circ_{b,r}\mf D_{\bot}^{r}
&=\mf C_{a \otimes f},\\
\mf D_{a \otimes f}^{z}\circ_{z,t}\mf I_{t,b}
&=\mf D_{a \otimes f}^{b},
\end{align*}
and finally observe that applying the introduction rules for the connective and closing the leaf of the following derivation tree using respect for transformation give us
$$
\inferrule*[right=$F_{\mathsf l}$L]
{
  \inferrule*[right=$U_{\mathsf l}$R]
  {
    d_0::
    a:A,f:A^\perp
    \vdash^{\Lambda_{\mathrm{CMLL}}}_{
      \mf D_{a \otimes f}^{b}\circ_{b,r}\mf D_{\bot}^{r}}
    \cdot
  }
  {
    \mathrm{UR}(d_0)::
    a:A,f:A^\perp
    \vdash^{\Lambda_{\mathrm{CMLL}}}_{\mf D_{a \otimes f}^{b}}
    b:\bot_r
  }
}
{
  \mathrm{FL}(\mathrm{UR}(d_0))::
  t:F^z_{\mf D_{a \otimes f}^{z}}(\cdot;a:A,f:A^\perp)
  \vdash^{\Lambda_{\mathrm{CMLL}}}_{\mf I_{t,b}}
  b:\bot_r
},
$$
which is exactly what we desired.\hfill $\square$

We compare these rules with a symmetric two-sided presentation of CMLL \cite{Bla23}. Its formulas are
$$
A,B::=P\mid A\otimes B\mid\mathbf 1\mid A\parr B\mid\bot\mid A^\perp,
$$
and its sequents are $\Gamma\vdash\Delta$, for finite lists $\Gamma$ and $\Delta$. Independent exchange on both lists is allowed. Besides identity and cut, its multiplicative rules are
$$
\inferrule*[right=$\otimes$L]
{\Gamma_L,A,B,\Gamma_R\vdash\Delta}
{\Gamma_L,A\otimes B,\Gamma_R\vdash\Delta}
\qquad
\inferrule*[right=$\otimes$R]
{\Gamma_1\vdash\Delta_L^1,A,\Delta_R^1 \\
 \Gamma_2\vdash\Delta_L^2,B,\Delta_R^2}
{\Gamma_1,\Gamma_2\vdash
 \Delta_L^1,\Delta_L^2,A\otimes B,\Delta_R^1,\Delta_R^2},
$$
$$
\inferrule*[right=$\parr$L]
{\Gamma_1,A\vdash\Delta_1 \\
 \Gamma_2,B\vdash\Delta_2}
{\Gamma_1,\Gamma_2,A\parr B\vdash\Delta_1,\Delta_2}
\qquad
\inferrule*[right=$\parr$R]
{\Gamma\vdash\Delta_L,A,B,\Delta_R}
{\Gamma\vdash\Delta_L,A\parr B,\Delta_R}.
$$
The unit and negation rules are
$$
\inferrule*[right=$\mathbf 1$L]{\Gamma\vdash\Delta}{\Gamma,\mathbf 1\vdash\Delta}
\qquad
\inferrule*[right=$\mathbf 1$R]{ }{\cdot\vdash\mathbf 1}
\qquad
\inferrule*[right=$\bot$L]{ }{\bot\vdash\cdot}
\qquad
\inferrule*[right=$\bot$R]{\Gamma\vdash\Delta}{\Gamma\vdash\Delta,\bot},
$$
$$
\inferrule*[right=$(-)^\perp$L]
{\Gamma\vdash\Delta,A}
{\Gamma,A^\perp\vdash\Delta}
\qquad
\inferrule*[right=$(-)^\perp$R]
{\Gamma,A\vdash\Delta}
{\Gamma\vdash\Delta,A^\perp}.
$$
Note that the order of the formulas is not of importance since exchange is admissible in the calculus. The identity rule is $A\vdash A$, and cut has the same two-sided, one-formula shape as \cref{thm:cut-admissibility}. Equivalently, one may take atomic identity and cut-free derivations, since identity expansion and cut elimination do not change derivability.

Identify the atomic formulas of the two calculi. Define the formula translation $(-)^\ddagger$ recursively by
\begin{align*}
P^\ddagger&:=P,\\
(A\otimes B)^\ddagger
&:=F^z_{\mf D_{\otimes}}(\cdot;x:A^\ddagger,y:B^\ddagger), &
\mathbf 1^\ddagger
&:=F^z_{\mf D_{\mathbf 1}}(\cdot;\cdot),\\
(A\parr B)^\ddagger
&:=U^{\mf D_{\parr}}_z(\cdot;\cdot\mid x:A^\ddagger,y:B^\ddagger), &
\bot^\ddagger
&:=U^{\mf D_{\bot}}_x(\cdot;\cdot\mid\cdot),\\
(A^\perp)^\ddagger
&:=U^{\mf D_{\mathrm{cap}}}_x(\cdot;x':A^\ddagger\mid\cdot).
\end{align*}
Extend $(-)^\ddagger$ pointwise to contexts. For a sequent $\Gamma\vdash\Delta$, choose pairwise distinct boundary labels, write $\gamma_\Gamma\Vdash\delta_\Delta$ for the resulting mode boundary, and set
\begin{equation}
(\Gamma\vdash\Delta)^\ddagger
:=
\Gamma^\ddagger
\vdash^{\Lambda_{\mathrm{CMLL}}}_{\mf S_{\gamma_\Gamma;\delta_\Delta}}
\Delta^\ddagger.
\label{eq:cmll-sequent-translation}
\end{equation}
The comparison cells act on derivations in both directions. If
$d::\Gamma^\ddagger\vdash_{\mf D}\Delta^\ddagger$, then Respect applied to
$\bar\kappa_{\mf D}::\mf S_{\gamma_\Gamma;\delta_\Delta}\Rightarrow\mf D$
gives $(\bar\kappa_{\mf D})_*(d)$ at the canonical index. Conversely, if $e$
is at the canonical index, then Respect applied to
$\kappa_{\mf D}::\mf D\Rightarrow\mf S_{\gamma_\Gamma;\delta_\Delta}$ gives
$(\kappa_{\mf D})_*(e)$ at $\mf D$. Hence
\begin{equation}
\Gamma^\ddagger\vdash^{\Lambda_{\mathrm{CMLL}}}_{\mf D}\Delta^\ddagger
\quad\Longleftrightarrow\quad
\Gamma^\ddagger
\vdash^{\Lambda_{\mathrm{CMLL}}}_{\mf S_{\gamma_\Gamma;\delta_\Delta}}
\Delta^\ddagger.
\label{eq:cmll-index-invariance}
\end{equation}

\begin{proposition}[Logical adequacy for CMLL]\label{prop:logical-adequacy-cmll}
For every well-formed two-sided CMLL sequent $\Gamma\vdash\Delta$,
$$
\begin{gathered}
\Gamma\vdash\Delta
\text{ is derivable in CMLL}
\\
\Longleftrightarrow
\\
(\Gamma\vdash\Delta)^\ddagger
\text{ is derivable in the linear fragment of the framework}.
\end{gathered}
$$
\end{proposition}

\textit{Proof.}
All boundary labels below are chosen pairwise fresh. For a displayed labelled
CMLL sequent $Q$, write $\mf S_Q$ for its chosen canonical descriptor. In an
individual rule instance, we abbreviate the canonical descriptor of the
conclusion by $\mf S$ and those of its premises by
$\mf S_0,\mf S_1,\mf S_2$. Write
$\mf T^u_{a,b}$, $\mf J^u$, $\mf P_u^{a,b}$, $\mf B_u$, and
$\mf N_{u,a}$ for boundary-renamed copies of
$\mf D_{\otimes}$, $\mf D_{\mathbf 1}$, $\mf D_{\parr}$,
$\mf D_{\bot}$, and $\mf D_{\mathrm{cap}}$, respectively.
Thus $\mf T^u_{a,b}$ has boundary $a:l,b:l\Vdash u:l$,
$\mf P_u^{a,b}$ has boundary $u:l\Vdash a:l,b:l$, and in
$\mf N_{u,a}$ the input $u$ is distinguished, while $a$ is the
non-distinguished input. Every composite below includes the closing boundary
permutation determined by the displayed order. If $\mf E$ has the boundary
of a premise with canonical index $\mf S_i$, Respect has the instance
\begin{equation}
\inferrule*[right=Respect]
{\kappa_{\mf E}::\mf E\Rightarrow\mf S_i
 \\
 d::\Gamma^\ddagger\vdash_{\mf S_i}\Delta^\ddagger}
{(\kappa_{\mf E})_*(d)::
 \Gamma^\ddagger\vdash_{\mf E}\Delta^\ddagger}.
\label{eq:cmll-premise-reindexing}
\end{equation}
This fixes the direction of every reindexing used below.

For preservation, proceed by induction on the height of a CMLL derivation
$\pi$. For every proper premise $\pi_i$, the induction hypothesis produces a
framework derivation $\Pi_i:=\operatorname{IH}(\pi_i)$ at its canonical
index.

\begin{itemize}
\item \textbf{Identity.}
If $\pi$ is $A\vdash A$, the \hyperref[anchor:cmll-singleton-canonical-descriptor]{singleton computation above} identifies the
canonical conclusion index with $\mf I_{a,b}$. Identity expansion therefore
gives the translated conclusion directly:
$$
\mathrm{id}_{A^\ddagger}::
a:A^\ddagger
\vdash^{\Lambda_{\mathrm{CMLL}}}_{\mf I_{a,b}}
b:A^\ddagger,
\qquad
\mf S_{a:l;b:l}=\mf I_{a,b},
\qquad
\mf I_{a,b}:=[\varnothing\mid\{a\mapsto b\}].
$$

\item \textbf{Conjunction left introduction.}
Suppose the final rule is
$$
\inferrule*[right=$\otimes$L]
{\pi_0:\Gamma_L,A,B,\Gamma_R\vdash\Delta}
{\Gamma_L,A\otimes B,\Gamma_R\vdash\Delta}.
$$
We apply the induction hypothesis to the premise to get the sequent
 $  \Pi_0::(\Gamma_L)^\ddagger,a:A^\ddagger,b:B^\ddagger,
 (\Gamma_R)^\ddagger\vdash_{\mf S_0}\Delta^\ddagger
 $;
then an instance of left introduction gives the result
$$
\inferrule*[right=$F_{\mathsf l}$L]
{
  \Pi_0::(\Gamma_L)^\ddagger,a:A^\ddagger,b:B^\ddagger,
  (\Gamma_R)^\ddagger\vdash_{\mf S_0}\Delta^\ddagger
}
{(\Gamma_L)^\ddagger,t:(A\otimes B)^\ddagger,
 (\Gamma_R)^\ddagger\vdash_{\mf S}\Delta^\ddagger}.
$$

\item \textbf{Conjunction right introduction.}
Suppose the final rule is
$$
\inferrule*[right=$\otimes$R]
{\pi_1:\Gamma_1\vdash\Delta_L^1,A,\Delta_R^1
 \\
 \pi_2:\Gamma_2\vdash\Delta_L^2,B,\Delta_R^2}
{\Gamma_1,\Gamma_2\vdash
 \Delta_L^1,\Delta_L^2,A\otimes B,\Delta_R^1,\Delta_R^2}.
$$
Set 
$
\mf E:=
(\mf S_1,\mf S_2)
\circ_{(\bar a,a),(\bar b,b)}
\mf T^t_{a,b}
$
and apply the induction hypotheses to the two premises. An application of the right introduction rule
$F_{\mathsf l}$R gives the result, while
$\bar\kappa_{\mf E}::\mf S\Rightarrow\mf E$ is its structural premise:
$$
\inferrule*[Right=$F_{\mathsf l}$R, rightskip=0.2em]
{
 \bar\kappa_{\mf E}::\mf S\Rightarrow\mf E
 \\
 \Pi_1::\Gamma_1^\ddagger\vdash_{\mf S_1}
   (\Delta_L^1)^\ddagger,\bar a:A^\ddagger,
   (\Delta_R^1)^\ddagger
 \\
 \Pi_2::\Gamma_2^\ddagger\vdash_{\mf S_2}
   (\Delta_L^2)^\ddagger,\bar b:B^\ddagger,
   (\Delta_R^2)^\ddagger
}
{\Gamma_1^\ddagger,\Gamma_2^\ddagger\vdash_{\mf S}
 (\Delta_L^1)^\ddagger,(\Delta_L^2)^\ddagger,
 t:(A\otimes B)^\ddagger,
 (\Delta_R^1)^\ddagger,(\Delta_R^2)^\ddagger}.
$$

\item \textbf{Conjunction unit.}
Let
$\mf E:=\mf J^u\circ_{u,t}\mf S$. The induction hypothesis on the
premise and the nullary-input instance of $F_{\mathsf l}$L give
$$
\inferrule*[right=$F_{\mathsf l}$L]
{
 \inferrule*[right=Respect]
 {\kappa_{\mf E}::\mf E\Rightarrow\mf S_0
  \\
  \Pi_0::\Gamma^\ddagger\vdash_{\mf S_0}\Delta^\ddagger}
 {\Gamma^\ddagger\vdash_{\mf E}\Delta^\ddagger}
}
{(\Gamma_L)^\ddagger,t:\mathbf 1^\ddagger,
 (\Gamma_R)^\ddagger\vdash_{\mf S}\Delta^\ddagger},
$$
where $\Gamma=\Gamma_L,\Gamma_R$. For $\mathbf 1$R, there are no induction
hypotheses. The nullary instance of $F_{\mathsf l}$R is
$$
\inferrule*[right=$F_{\mathsf l}$R]
{\bar\kappa_{\mf J^t}::\mf S\Rightarrow\mf J^t}
{\cdot\vdash_{\mf S}t:\mathbf 1^\ddagger}.
$$

\item \textbf{Disjunction right introduction.}
Suppose the final rule is
$$
\inferrule*[right=$\parr$R]
{\pi_0:\Gamma\vdash\Delta_L,A,B,\Delta_R}
{\Gamma\vdash\Delta_L,A\parr B,\Delta_R}.
$$
As in the case of the conjunction, we apply the induction hypothesis to the premise and then apply an instance of the right introduction rule for $U$.
$$
\inferrule*[right=$U_{\mathsf l}$R]
{
 \inferrule*[right=Respect]
 {\kappa_{\mf E}::\mf E\Rightarrow\mf S_0
  \\
  \Pi_0::\Gamma^\ddagger\vdash_{\mf S_0}
  \Delta_L^\ddagger,a:A^\ddagger,b:B^\ddagger,
  \Delta_R^\ddagger}
 {\Gamma^\ddagger\vdash_{\mf E}
  \Delta_L^\ddagger,a:A^\ddagger,b:B^\ddagger,
  \Delta_R^\ddagger}
}
{\Gamma^\ddagger\vdash_{\mf S}
 \Delta_L^\ddagger,t:(A\parr B)^\ddagger,\Delta_R^\ddagger}.
$$

\item \textbf{Disjunction left introduction.}
Suppose the final rule is
$$
\inferrule*[right=$\parr$L]
{\pi_1:\Gamma_1,A\vdash\Delta_1
 \\
 \pi_2:\Gamma_2,B\vdash\Delta_2}
{\Gamma_1,\Gamma_2,A\parr B\vdash\Delta_1,\Delta_2}.
$$
Set
$
\mf E:=
\bigl(\mf P_t^{a,b}\circ_{a,\bar a}\mf S_1\bigr)
\circ_{b,\bar b}\mf S_2
$
and apply the induction hypothesis to the two premises. An application of the left introduction rule terminates the proof.
$$
\inferrule*[right=$U_{\mathsf l}$L]
{
 \bar\kappa_{\mf E}::\mf S\Rightarrow\mf E
 \\
 \Pi_1::\Gamma_1^\ddagger,\bar a:A^\ddagger
   \vdash_{\mf S_1}\Delta_1^\ddagger
 \\
 \Pi_2::\Gamma_2^\ddagger,\bar b:B^\ddagger
   \vdash_{\mf S_2}\Delta_2^\ddagger
}
{\Gamma_1^\ddagger,\Gamma_2^\ddagger,
 t:(A\parr B)^\ddagger
 \vdash_{\mf S}\Delta_1^\ddagger,\Delta_2^\ddagger}.
$$

\item \textbf{Disjunction unit.}
Let $\mf E:=\mf S\circ_{t,u}\mf B_u$. Applying the induction hypothesis to the
 premise gives
$$
\inferrule*[right=$U_{\mathsf l}$R]
{
 \inferrule*[right=Respect]
 {\kappa_{\mf E}::\mf E\Rightarrow\mf S_0
  \\
  \Pi_0::\Gamma^\ddagger\vdash_{\mf S_0}\Delta^\ddagger}
 {\Gamma^\ddagger\vdash_{\mf E}\Delta^\ddagger}
}
{\Gamma^\ddagger\vdash_{\mf S}
 \Delta^\ddagger,t:\bot^\ddagger}.
$$
For $\bot$L, $U_{\mathsf l}$L has no component premises, and so we have

$$
\inferrule*[right=$U_{\mathsf l}$L]
{\bar\kappa_{\mf B_t}::\mf S\Rightarrow\mf B_t}
{t:\bot^\ddagger\vdash_{\mf S}\cdot}.
$$

\item \textbf{Negation-right.}
Suppose the last rule has premise
$\pi_0:\Gamma,A\vdash\Delta$. Set
$\mf E:=\mf S\circ_{t,u}\mf N_{u,a}$. The non-distinguished input of
$\mf N_{u,a}$ carries $A^\ddagger$, so the required $U_{\mathsf l}$R
instance is
$$
\inferrule*[right=$U_{\mathsf l}$R]
{
 \inferrule*[right=Respect]
 {\kappa_{\mf E}::\mf E\Rightarrow\mf S_0
  \\
  \Pi_0::\Gamma^\ddagger,a:A^\ddagger
  \vdash_{\mf S_0}\Delta^\ddagger}
 {\Gamma^\ddagger,a:A^\ddagger
  \vdash_{\mf E}\Delta^\ddagger}
}
{\Gamma^\ddagger\vdash_{\mf S}\Delta^\ddagger,
 t:(A^\perp)^\ddagger}.
$$

\item \textbf{Negation-left.}
Suppose the last rule has premise
$\pi_0:\Gamma\vdash\Delta,A$. Set
$\mf E:=\mf S_0\circ_{\bar a,a}\mf N_{t,a}$. The induction hypothesis is
the unique input-component premise of $U_{\mathsf l}$L:
$$
\inferrule*[right=$U_{\mathsf l}$L]
{
 \bar\kappa_{\mf E}::\mf S\Rightarrow\mf E
 \\
 \Pi_0::\Gamma^\ddagger\vdash_{\mf S_0}
 \Delta^\ddagger,\bar a:A^\ddagger
}
{\Gamma^\ddagger,t:(A^\perp)^\ddagger
 \vdash_{\mf S}\Delta^\ddagger}.
$$
\end{itemize}
This finishes the proof.

For the other direction, we prove the stronger statement
\begin{equation}
\Pi::\Gamma^\ddagger
\vdash^{\Lambda_{\mathrm{CMLL}}}_{\mf D}
\Delta^\ddagger
\quad\Longrightarrow\quad
\Gamma\vdash\Delta
\text{ is derivable in CMLL}.
\label{eq:cmll-readback-induction}
\end{equation}
We proceed again by induction on the derivation.

\begin{itemize}
\item \textbf{Variable.}
If the last rule is
$$
\inferrule*[right=$\mathsf v_{\mathsf l}$]
{P\in\mathrm{Prop}_{\Lambda_{\mathrm{CMLL}},l}
 \\
 r::\mf D\Rightarrow[\varnothing\mid\{a\mapsto b\}]}
{a:P\vdash_{\mf D}b:P},
$$
then its formula boundary reads back as $P\vdash P$. The CMLL identity rule
therefore gives the required conclusion. 

\item \textbf{$F_{\mathsf l}$-left.}
A principal translated $F$-formula is either a tensor or its unit. In the
tensor case, the last rule has the form
$$
\inferrule*[right=$F_{\mathsf l}$L]
{\Pi_0::(\Gamma_L)^\ddagger,a:A^\ddagger,b:B^\ddagger,
 (\Gamma_R)^\ddagger
 \vdash_{\mf T^u_{a,b}\circ_{u,t}\mf D}\Delta^\ddagger}
{(\Gamma_L)^\ddagger,t:(A\otimes B)^\ddagger,
 (\Gamma_R)^\ddagger\vdash_{\mf D}\Delta^\ddagger}.
$$
The induction hypothesis applies to the proper premise $\Pi_0$. Reapplying
the corresponding CMLL rule gives
$$
\inferrule*[right=$\otimes$L]
{\operatorname{IH}(\Pi_0)::
 \Gamma_L,A,B,\Gamma_R\vdash\Delta}
{\Gamma_L,A\otimes B,\Gamma_R\vdash\Delta}.
$$
For a principal $\mathbf 1^\ddagger$, the premise contains no exposed
connective input:
$$
\inferrule*[right=$F_{\mathsf l}$L]
{\Pi_0::\Gamma^\ddagger
 \vdash_{\mf J^u\circ_{u,t}\mf D}\Delta^\ddagger}
{(\Gamma_L)^\ddagger,t:\mathbf 1^\ddagger,
 (\Gamma_R)^\ddagger\vdash_{\mf D}\Delta^\ddagger}.
$$
The induction hypothesis on $\Pi_0$, followed by $\mathbf 1$L, gives the
source conclusion.

\item \textbf{$F_{\mathsf l}$-right.}
For a principal tensor, the last rule necessarily has two component
premises. With
$$
\mf E:=(\mf D_1,\mf D_2)
\circ_{(\bar a,a),(\bar b,b)}\mf T^t_{a,b},
$$
it is
$$
\inferrule*[right=$F_{\mathsf l}$R]
{r::\mf D\Rightarrow\mf E
 \\
 \Pi_1::\Gamma_1^\ddagger\vdash_{\mf D_1}
   (\Delta_L^1)^\ddagger,\bar a:A^\ddagger,
   (\Delta_R^1)^\ddagger
 \\
 \Pi_2::\Gamma_2^\ddagger\vdash_{\mf D_2}
   (\Delta_L^2)^\ddagger,\bar b:B^\ddagger,
   (\Delta_R^2)^\ddagger}
{\Gamma_1^\ddagger,\Gamma_2^\ddagger\vdash_{\mf D}
 (\Delta_L^1)^\ddagger,(\Delta_L^2)^\ddagger,
 t:(A\otimes B)^\ddagger,
 (\Delta_R^1)^\ddagger,(\Delta_R^2)^\ddagger}.
$$
Applying the induction hypothesis separately to $\Pi_1$ and $\Pi_2$ gives
the two premises of
$$
\inferrule*[right=$\otimes$R]
{\operatorname{IH}(\Pi_1)::
  \Gamma_1\vdash\Delta_L^1,A,\Delta_R^1
 \\
 \operatorname{IH}(\Pi_2)::
  \Gamma_2\vdash\Delta_L^2,B,\Delta_R^2}
{\Gamma_1,\Gamma_2\vdash
 \Delta_L^1,\Delta_L^2,A\otimes B,\Delta_R^1,\Delta_R^2}.
$$
For a principal $\mathbf 1^\ddagger$, the connective has no input
components. The zero-component slicing conditions force all side contexts to
be empty, so the last rule is
$$
\inferrule*[right=$F_{\mathsf l}$R]
{r::\mf D\Rightarrow\mf J^t}
{\cdot\vdash_{\mf D}t:\mathbf 1^\ddagger}.
$$
It reads back directly as the nullary rule $\cdot\vdash\mathbf 1$.

\item \textbf{$U_{\mathsf l}$-right.}
There are three possible principal translated formulas. For par, the last
rule and its readback are
$$
\inferrule*[right=$U_{\mathsf l}$R]
{\Pi_0::\Gamma^\ddagger
 \vdash_{\mf D\circ_{t,u}\mf P_u^{a,b}}
 \Delta_L^\ddagger,a:A^\ddagger,b:B^\ddagger,
 \Delta_R^\ddagger}
{\Gamma^\ddagger\vdash_{\mf D}
 \Delta_L^\ddagger,t:(A\parr B)^\ddagger,\Delta_R^\ddagger},
$$
$$
\inferrule*[right=$\parr$R]
{\operatorname{IH}(\Pi_0)::
 \Gamma\vdash\Delta_L,A,B,\Delta_R}
{\Gamma\vdash\Delta_L,A\parr B,\Delta_R}.
$$
For bottom, the connective exposes no formula in its sole premise:
$$
\inferrule*[right=$U_{\mathsf l}$R]
{\Pi_0::\Gamma^\ddagger
 \vdash_{\mf D\circ_{t,u}\mf B_u}\Delta^\ddagger}
{\Gamma^\ddagger\vdash_{\mf D}\Delta^\ddagger,t:\bot^\ddagger}.
$$
The induction hypothesis gives $\Gamma\vdash\Delta$, and $\bot$R gives
$\Gamma\vdash\Delta,\bot$. Finally, for negation, the non-distinguished input
is exposed on the left:
$$
\inferrule*[right=$U_{\mathsf l}$R]
{\Pi_0::\Gamma^\ddagger,a:A^\ddagger
 \vdash_{\mf D\circ_{t,u}\mf N_{u,a}}\Delta^\ddagger}
{\Gamma^\ddagger\vdash_{\mf D}\Delta^\ddagger,
 t:(A^\perp)^\ddagger}.
$$
The induction hypothesis followed by $(-)^\perp$R gives
$$
\inferrule*[right=$(-)^\perp$R]
{\operatorname{IH}(\Pi_0)::\Gamma,A\vdash\Delta}
{\Gamma\vdash\Delta,A^\perp}.
$$

\item \textbf{$U_{\mathsf l}$-left.}
For a principal par, there are no input components and exactly two output
components. With
$$
\mf E:=
\bigl(\mf P_t^{a,b}\circ_{a,\bar a}\mf D_1\bigr)
\circ_{b,\bar b}\mf D_2,
$$
the last rule is
$$
\inferrule*[right=$U_{\mathsf l}$L]
{r::\mf D\Rightarrow\mf E
 \\
 \Pi_1::\Gamma_1^\ddagger,\bar a:A^\ddagger
  \vdash_{\mf D_1}\Delta_1^\ddagger
 \\
 \Pi_2::\Gamma_2^\ddagger,\bar b:B^\ddagger
  \vdash_{\mf D_2}\Delta_2^\ddagger}
{\Gamma_1^\ddagger,\Gamma_2^\ddagger,
 t:(A\parr B)^\ddagger
 \vdash_{\mf D}\Delta_1^\ddagger,\Delta_2^\ddagger}.
$$
The induction hypothesis applies independently to both component premises,
and the source rule is
$$
\inferrule*[right=$\parr$L]
{\operatorname{IH}(\Pi_1)::\Gamma_1,A\vdash\Delta_1
 \\
 \operatorname{IH}(\Pi_2)::\Gamma_2,B\vdash\Delta_2}
{\Gamma_1,\Gamma_2,A\parr B\vdash\Delta_1,\Delta_2}.
$$
For bottom, there are no input or output components, so the framework rule is
$$
\inferrule*[right=$U_{\mathsf l}$L]
{r::\mf D\Rightarrow\mf B_t}
{t:\bot^\ddagger\vdash_{\mf D}\cdot}.
$$
It reads back directly as $\bot\vdash\cdot$. For negation, there is one input
component and no output component. With
$\mf E:=\mf D_0\circ_{\bar a,a}\mf N_{t,a}$, the last rule is
$$
\inferrule*[right=$U_{\mathsf l}$L]
{r::\mf D\Rightarrow\mf E
 \\
 \Pi_0::\Gamma^\ddagger\vdash_{\mf D_0}
 \Delta^\ddagger,\bar a:A^\ddagger}
{\Gamma^\ddagger,t:(A^\perp)^\ddagger
 \vdash_{\mf D}\Delta^\ddagger}.
$$
The induction hypothesis and the source negation rule give
$$
\inferrule*[right=$(-)^\perp$L]
{\operatorname{IH}(\Pi_0)::\Gamma\vdash\Delta,A}
{\Gamma,A^\perp\vdash\Delta}.
$$
\end{itemize}

\section{Linear Logic with Exponentials}\label{sec:examples-linear-logic-exponentials}

We now treat the multiplicative-exponential fragment of classical linear logic \cite{Gir87}. The preceding example describes its multiplicative connectives using a single linear mode. To obtain the exponential modalities, we add a cartesian mode and two generators relating the two modes. This is the linear--nonlinear presentation of the logic, following the adjunction of Benton \cite{Ben95} and its multiple-output setting in \cite{Shu23}. We develop here the mixed instance introduced in \cref{ex:linear-logic-exponential-modality}.

\subsection{The Mode Theory}

Let $\Lambda_{\mathrm{LL}}$ have two modes
$$
c\modec,
\qquad
l\model.
$$
At the cartesian mode, declare the c-generators
$$
\otimes_c:c,c\to c,
\qquad
\mathbf 1_c:\cdot\to c.
$$
At the linear mode, retain the four generators of \cref{sec:examples-blanco-cmll}:
$$
\otimes_l:l,l\to l,
\qquad
\mathbf 1_l:\cdot\to l,
\qquad
\parr_l:l\to l,l,
\qquad
\bot_l:l\to\cdot.
$$
We add the two mixed generators
$$
g:(c\mid\cdot)\to l,
\qquad
h:(c\mid l)\to\cdot.
$$
The generator $g$ takes one cartesian input and produces one linear output. Its descriptor is
\begin{equation}
\mf D_g
:=
\left[\{(a\mid\cdot)\xmapsto{g}w\}\mid\{w\mapsto y\}\right],
\qquad
a:c\ ;\ \cdot\Vdash_{\mf D_g}y:l.
\label{eq:ll-mixed-descriptor}
\end{equation}
The generator $h$ takes one cartesian input and one linear input and has no outputs. Its descriptor is
\begin{equation}
\mf D_h
:=
\left[\{(a\mid x)\xmapsto{h}\cdot\}\mid\varnothing\right],
\qquad
a:c\ ;\ x:l\Vdash_{\mf D_h}\cdot.
\label{eq:ll-negative-mixed-descriptor}
\end{equation}
This is an instance of the co-nullary generation rule of \cref{def:well-formed-descriptor-negative}. The cartesian input can be reused, while the linear input is consumed once. We will use the two input slots of this descriptor to construct the why not modality.

For the structural part of $\Lambda_{\mathrm{LL}}$, include invertible comparison $2$-cells between every pair of well-formed descriptors over the same boundary, both for cartesian and for mixed judgments, together with the closure instances of \cref{def:signature}. This extends the comparison data of \cref{sec:examples-blanco-cmll}. In particular, $\mf D_h$ is isomorphic to the descriptor
$$
\left[
\{(a\mid\cdot)\xmapsto{g}w,\ w,x\xmapsto{\otimes_l}v,\ v\xmapsto{\bot_l}\cdot\}
\ \middle|\ \varnothing
\right]
$$
over the same boundary. This relates the two mixed generators in the classical presentation. We suppress descriptor indices and boundary labels in the logical rules below, using the comparison cells for the structural premises.

We declare no gates at $l$. Thus the use of a formula at the linear mode continues to be governed by the linear rules, while a cartesian input can be used any number of times.

\subsection{The Two Judgments and the Multiplicative Connectives}

Write $X,Y$ for types at the cartesian mode and $A,B$ for types at the linear mode. The judgments have the two forms
$$
\Theta\vdash X,
\qquad
\Theta\ ;\ \Gamma\vdash\Delta.
$$
All the formulas in $\Theta$ are cartesian, and all the formulas in $\Gamma,\Delta$ are linear. The first judgment has a single cartesian conclusion. The second has any finite number of linear conclusions, including none. Exchange is allowed in each context. Weakening and contraction apply to the cartesian context, giving in particular
$$
\inferrule*[right=W$_c$]
{\Theta\ ;\ \Gamma\vdash\Delta}
{\Theta,X\ ;\ \Gamma\vdash\Delta}
\qquad
\inferrule*[right=C$_c$]
{\Theta,X,X\ ;\ \Gamma\vdash\Delta}
{\Theta,X\ ;\ \Gamma\vdash\Delta}.
$$
The two occurrences of $X$ in the contraction premise carry distinct labels before those labels are suppressed. The same structural rules apply to cartesian sequents.

The c-generators give cartesian conjunction and its unit. Explicitly, set
$$
\mf D_{\otimes_c}:=[\{a,b\xmapsto{\otimes_c}w\}\mid\{w\mapsto d\}],
\qquad
\mf D_{\mathbf 1_c}:=[\{\cdot\xmapsto{\mathbf 1_c}w\}\mid\{w\mapsto d\}].
$$
The associated types are
$$
X\times Y:=F^d_{\mf D_{\otimes_c}}(a:X,b:Y;\cdot),
\qquad
\top_c:=F^d_{\mf D_{\mathbf 1_c}}(\cdot;\cdot).
$$
The rules for $X\times Y$ share the entire context when introducing it on the right and expose $X,Y$ when using it on the left. In particular,
$$
\inferrule*[right=$\times$R]
{\Theta\vdash X \\ \Theta\vdash Y}
{\Theta\vdash X\times Y}
\qquad
\inferrule*[right=$\times$L]
{\Theta,X,Y\ ;\ \Gamma\vdash\Delta}
{\Theta,X\times Y\ ;\ \Gamma\vdash\Delta}.
$$
The left rule also applies when its conclusion is cartesian. The unit has the rule $\Theta\vdash\top_c$ and can be inserted into the cartesian input context.

The linear connectives use the descriptors already defined in \cref{sec:examples-blanco-cmll}:
\begin{align*}
A\otimes B
&:=F^z_{\mf D_{\otimes}}(\cdot;x:A,y:B), &
\mathbf 1
&:=F^z_{\mf D_{\mathbf 1}}(\cdot;\cdot),\\
A\parr B
&:=U^{\mf D_{\parr}}_z(\cdot;\cdot\mid x:A,y:B), &
\bot
&:=U^{\mf D_{\bot}}_x(\cdot;\cdot\mid\cdot),\\
A^\perp
&:=U^{\mf D_{\mathrm{cap}}}_x(\cdot;x':A\mid\cdot).
\end{align*}
Their rules now carry the shared cartesian context $\Theta$:
\begin{mathpar}
\inferrule*[right=$\otimes$L]
{\Theta\ ;\ \Gamma,A,B\vdash\Delta}
{\Theta\ ;\ \Gamma,A\otimes B\vdash\Delta}
\and
\inferrule*[right=$\otimes$R]
{\Theta\ ;\ \Gamma_1\vdash\Delta_1,A \\
 \Theta\ ;\ \Gamma_2\vdash\Delta_2,B}
{\Theta\ ;\ \Gamma_1,\Gamma_2\vdash\Delta_1,\Delta_2,A\otimes B}
\and
\inferrule*[right=$\parr$L]
{\Theta\ ;\ \Gamma_1,A\vdash\Delta_1 \\
 \Theta\ ;\ \Gamma_2,B\vdash\Delta_2}
{\Theta\ ;\ \Gamma_1,\Gamma_2,A\parr B\vdash\Delta_1,\Delta_2}
\and
\inferrule*[right=$\parr$R]
{\Theta\ ;\ \Gamma\vdash\Delta,A,B}
{\Theta\ ;\ \Gamma\vdash\Delta,A\parr B}.
\end{mathpar}
Note that the two-premise rules partition both linear contexts. Each formula in those contexts belongs to exactly one premise. The cartesian context, on the other hand, is available in both premises. The unit and negation rules are
\begin{mathpar}
\inferrule*[right=$\mathbf 1$L]
{\Theta\ ;\ \Gamma\vdash\Delta}
{\Theta\ ;\ \Gamma,\mathbf 1\vdash\Delta}
\and
\inferrule*[right=$\mathbf 1$R]
{ }
{\Theta\ ;\ \cdot\vdash\mathbf 1}
\and
\inferrule*[right=$\bot$L]
{ }
{\Theta\ ;\ \bot\vdash\cdot}
\and
\inferrule*[right=$\bot$R]
{\Theta\ ;\ \Gamma\vdash\Delta}
{\Theta\ ;\ \Gamma\vdash\Delta,\bot}
\and
\inferrule*[right=$(-)^\perp$L]
{\Theta\ ;\ \Gamma\vdash\Delta,A}
{\Theta\ ;\ \Gamma,A^\perp\vdash\Delta}
\and
\inferrule*[right=$(-)^\perp$R]
{\Theta\ ;\ \Gamma,A\vdash\Delta}
{\Theta\ ;\ \Gamma\vdash\Delta,A^\perp}.
\end{mathpar}
Linear implication is another instance of $U$ over the tensor descriptor:
$$
A\multimap B:=U^{\mf D_{\otimes}}_y(\cdot;x:A\mid z:B).
$$
Its rules expose its argument on the left when introducing the implication, and use that argument together with a continuation for its result when eliminating it:
\begin{mathpar}
\inferrule*[right=$\multimap$R]
{\Theta\ ;\ \Gamma,A\vdash\Delta,B}
{\Theta\ ;\ \Gamma\vdash\Delta,A\multimap B}
\and
\inferrule*[right=$\multimap$L]
{\Theta\ ;\ \Gamma_1\vdash\Delta_1,A \\
 \Theta\ ;\ \Gamma_2,B\vdash\Delta_2}
{\Theta\ ;\ \Gamma_1,\Gamma_2,A\multimap B\vdash\Delta_1,\Delta_2}.
\end{mathpar}

Identity has the forms $\Theta,X\vdash X$ and $\Theta\ ;\ A\vdash A$. The two ways of composing the judgments are cartesian substitution and linear cut:
\begin{mathpar}
\inferrule*[right=Subst]
{\Theta\vdash X \\ \Theta,X\ ;\ \Gamma\vdash\Delta}
{\Theta\ ;\ \Gamma\vdash\Delta}
\and
\inferrule*[right=Cut]
{\Theta\ ;\ \Gamma_1\vdash\Delta_1,A \\
 \Theta\ ;\ \Gamma_2,A\vdash\Delta_2}
{\Theta\ ;\ \Gamma_1,\Gamma_2\vdash\Delta_1,\Delta_2}.
\end{mathpar}
Cartesian substitution also has a cartesian conclusion when its second premise is cartesian. These are the instances of the cut operation of the framework at $c$ and $l$, respectively.

\subsection{The Connectives Relating the Modes}

The same descriptor $\mf D_g$ gives two connectives, depending on whether we use $F$ or $U$. For a cartesian type $X$ and a linear type $A$, define
\begin{equation}
\mathsf F(X):=F^y_{\mf D_g}(a:X;\cdot),
\qquad
G(A):=U^{\mf D_g}_a(\cdot;\cdot\mid y:A).
\label{eq:ll-shift-types}
\end{equation}
The type $\mathsf F(X)$ is at mode $l$ because $y:l$ is the output of $\mf D_g$. The type $G(A)$ is at mode $c$ because $a:c$ is its distinguished input. Thus $G(A)$ can occur in the cartesian context, while $\mathsf F(X)$ occurs in a linear context.

The introduction rules for $\mathsf F$ specialize to
$$
\inferrule*[right=$\mathsf F$L]
{\Theta,X\ ;\ \Gamma\vdash\Delta}
{\Theta\ ;\ \Gamma,\mathsf F(X)\vdash\Delta}
\qquad
\inferrule*[right=$\mathsf F$R]
{\Theta\vdash X}
{\Theta\ ;\ \cdot\vdash\mathsf F(X)}.
$$
The left rule places the exposed input $X$ in the cartesian zone. The right rule has a cartesian premise and an empty linear input context. Its sole output is the displayed formula $\mathsf F(X)$, since its descriptor has no linear input components to carry side contexts.

For $G$, the two rules are the shift instances of \cref{prop:rule-specialization}:
$$
\inferrule*[right=$G$R]
{\Theta\ ;\ \cdot\vdash A}
{\Theta\vdash G(A)}
\qquad
\inferrule*[right=$G$L]
{\Theta,G(A)\ ;\ \Gamma,A\vdash\Delta}
{\Theta,G(A)\ ;\ \Gamma\vdash\Delta}.
$$
Notice that $G(A)$ remains in the cartesian context of the left rule. An application of this rule makes one linear occurrence of $A$ available while retaining the possibility of using $G(A)$ again. The right rule has exactly one linear output and no linear inputs. It therefore places in the cartesian layer a proof whose assumptions are all reusable.

In the indexed version, the latter restriction is explicit. If $\theta\Vdash_\sigma b:c$, the right rule is
$$
\inferrule*[right=$U_{\mathsf{shift}}$R]
{\Theta\ ;\ \cdot
 \vdash^{\Lambda_{\mathrm{LL}}}_{\sigma\circ_{b,a}\mf D_g}y:A}
{\Theta\vdash^{\Lambda_{\mathrm{LL}}}_{\sigma}b:G(A)}.
$$
The index $\sigma$ is a c-descriptor and hence has no linear inputs or additional outputs. The comparison cells in the signature allow a premise at any descriptor over $\theta\ ;\ \cdot\Vdash y:l$ to be reindexed to the displayed composite.

These connectives express the linear--nonlinear adjunction $\mathsf F\dashv G$. In its ordinary categorical notation, the adjunction has the form
$$
\mc L(\mathsf F(X),A)\cong\mc C(X,G(A)),
$$
where $\mc C$ is the cartesian category and $\mc L$ is the linear category of the intended model. In the multiple-output setting, the left rule for $\mathsf F$ also allows arbitrary linear side contexts, as the displayed sequent rule records.

\subsection{The Of Course Modality}

The exponential modality $!$ is the composite of these two connectives:
\begin{equation}
!A:=\mathsf F(G(A))
=F^y_{\mf D_g}\bigl(a:G(A);\cdot\bigr).
\label{eq:ll-of-course}
\end{equation}
The type $!A$ is linear. Using it on the left exposes the cartesian formula $G(A)$, giving
$$
\inferrule*[right=$!$L]
{\Theta,G(A)\ ;\ \Gamma\vdash\Delta}
{\Theta\ ;\ \Gamma,!A\vdash\Delta}.
$$
Thus the ability to reuse the contents of $!A$ comes from the mode of the exposed formula. The linear occurrence of $!A$ itself is consumed by this rule.

The right rule is promotion. It consists of $G$R followed by $\mathsf F$R:
$$
\inferrule*[right=$\mathsf F$R]
{
  \inferrule*[right=$G$R]
  {\Theta\ ;\ \cdot\vdash A}
  {\Theta\vdash G(A)}
}
{\Theta\ ;\ \cdot\vdash !A}.
$$
Both steps require a proof depending only on the cartesian context. In particular, promotion has no arbitrary linear side formulas. This condition is part of the sorts of the two judgments.

The familiar structural behavior of $!$ is expressed by dereliction, weakening, and contraction:
\begin{mathpar}
\inferrule*[right=Der$_!$]
{\Theta\ ;\ \Gamma,A\vdash\Delta}
{\Theta\ ;\ \Gamma,!A\vdash\Delta}
\and
\inferrule*[right=W$_!$]
{\Theta\ ;\ \Gamma\vdash\Delta}
{\Theta\ ;\ \Gamma,!A\vdash\Delta}
\and
\inferrule*[right=C$_!$]
{\Theta\ ;\ \Gamma,!A,!A\vdash\Delta}
{\Theta\ ;\ \Gamma,!A\vdash\Delta}.
\end{mathpar}
For dereliction, expose $G(A)$ and use $G$L to make one occurrence of $A$ available. For weakening, the exposed cartesian formula can be ignored. For contraction, one cartesian hypothesis $G(A)$ can supply two occurrences of $\mathsf F(G(A))$ by $\mathsf F$R, each with its own linear output. Cutting these into the two displayed occurrences of $!A$ and applying $\mathsf F$L gives the conclusion. The sharing takes place entirely in the cartesian zone.

For example, using the same cartesian hypothesis in both tensor premises gives
$$
\inferrule*[right=$!$L]
{
  \inferrule*[right=$\otimes$R]
  {G(A)\ ;\ \cdot\vdash !A \\ G(A)\ ;\ \cdot\vdash !A}
  {G(A)\ ;\ \cdot\vdash !A\otimes !A}
}
{\cdot\ ;\ !A\vdash !A\otimes !A}.
$$
Each leaf is $\mathsf F$R applied to the cartesian identity $G(A)\vdash G(A)$. Similarly, ignoring $G(A)$ in the unit rule gives $\cdot\ ;\ !A\vdash\mathbf 1$. Dereliction and promotion also give the sequents
$$
\cdot\ ;\ !A\vdash A,
\qquad
\cdot\ ;\ !A\vdash !!A.
$$
For the second, first obtain $G(A)\ ;\ \cdot\vdash !A$, promote it, and then apply $!$L. In the adjunction semantics, these two maps are the counit and comultiplication of the comonad $\mathsf F G$. If $\eta$ is the unit of the adjunction and $\varepsilon$ its counit, they are written
$$
\varepsilon_A:!A\to A,
\qquad
\delta_A:=\mathsf F(\eta_{G(A)}):!A\to !!A.
$$
The maps $!A\to\mathbf 1$ and $!A\to !A\otimes !A$ record its weakening and contraction structure.

\subsection{The Why Not Modality and the Two-Sided Rules}

Besides the of course modality, the framework also expresses the why not modality directly through its judgmental structure. The reuse of resources on the target side is obtained by internalizing cartesian data as a linear conclusion. We now construct this modality using the descriptor $\mf D_h$ of \eqref{eq:ll-negative-mixed-descriptor}, following the negative modalities of \cite[Definition~2.14]{Shu23}.

The two input slots of $\mf D_h$ give two instances of $U$. For a linear type $A$ and a cartesian type $X$, define
\begin{equation}
H(A):=U^{\mf D_h}_a(\cdot;x:A\mid\cdot),
\qquad
K(X):=U^{\mf D_h}_x(a:X;\cdot\mid\cdot).
\label{eq:ll-negative-shift-types}
\end{equation}
The type $H(A)$ is at mode $c$, since its distinguished input is $a:c$. The type $K(X)$ is at mode $l$, since its distinguished input is $x:l$. Both types have an empty output context in their formation data. Thus the same descriptor relates a cartesian formula and a linear formula by consuming them together.

The rules for $H$ are the shift instances of \cref{prop:rule-specialization}:
$$
\inferrule*[right=$H$R]
{\Theta\ ;\ A\vdash\cdot}
{\Theta\vdash H(A)}
\qquad
\inferrule*[right=$H$L]
{\Theta,H(A)\ ;\ \Gamma\vdash\Delta,A}
{\Theta,H(A)\ ;\ \Gamma\vdash\Delta}.
$$
The right rule has one linear input and no outputs in its premise. The left rule consumes one conclusion $A$ using the cartesian hypothesis $H(A)$. Notice that this hypothesis remains in the cartesian zone, so that it can be used again.

The rules for $K$ are linear instances of the same $U$-schemas:
$$
\inferrule*[right=$K$L]
{\Theta\vdash X}
{\Theta\ ;\ K(X)\vdash\cdot}
\qquad
\inferrule*[right=$K$R]
{\Theta,X\ ;\ \Gamma\vdash\Delta}
{\Theta\ ;\ \Gamma\vdash\Delta,K(X)}.
$$
The left rule has only a cartesian premise. Its conclusion has no linear side formulas, since there are no linear components to carry them. The right rule internalizes the cartesian assumption $X$ into the output $K(X)$ while retaining the linear side contexts.

We can make the role of the empty output explicit in the indexed rules. If $\theta\Vdash_\sigma b:c$, then $H$R is
$$
\inferrule*[right=$U_{\mathsf{shift}}$R]
{\Theta\ ;\ x:A
 \vdash^{\Lambda_{\mathrm{LL}}}_{\sigma\circ_{b,a}\mf D_h}\cdot}
{\Theta\vdash^{\Lambda_{\mathrm{LL}}}_{\sigma}b:H(A)}.
$$
For $K$R, if $\theta\ ;\ \gamma\Vdash_{\mf D}\delta,z:l$, the instance is
$$
\inferrule*[right=$U_{\mathsf l}$R]
{\Theta,a:X\ ;\ \Gamma
 \vdash^{\Lambda_{\mathrm{LL}}}_{\mf D\circ_{z,x}\mf D_h}\Delta}
{\Theta\ ;\ \Gamma
 \vdash^{\Lambda_{\mathrm{LL}}}_{\mf D}\Delta,z:K(X)}.
$$
The composition consumes the output $z$ and introduces the cartesian input $a$. The comparison cells of the signature allow the premise to be reindexed to this composite.

The why not modality is now the composite
\begin{equation}
?A:=K(H(A))
=U^{\mf D_h}_x\bigl(a:H(A);\cdot\mid\cdot\bigr).
\label{eq:ll-why-not}
\end{equation}
It is a linear type. Its right rule exposes the cartesian formula $H(A)$:
$$
\inferrule*[right=$?$R]
{\Theta,H(A)\ ;\ \Gamma\vdash\Delta}
{\Theta\ ;\ \Gamma\vdash\Delta,?A}.
$$
Its left promotion rule consists of $H$R followed by $K$L:
$$
\inferrule*[right=$K$L]
{
  \inferrule*[right=$H$R]
  {\Theta\ ;\ A\vdash\cdot}
  {\Theta\vdash H(A)}
}
{\Theta\ ;\ ?A\vdash\cdot}.
$$
The premise has exactly one linear input and no linear outputs. All the other assumptions are in the cartesian zone. This is the counterpart of the restriction on the right promotion rule for $!$.

Its structural rules act on the output side:
\begin{mathpar}
\inferrule*[right=Der$_?$]
{\Theta\ ;\ \Gamma\vdash\Delta,A}
{\Theta\ ;\ \Gamma\vdash\Delta,?A}
\and
\inferrule*[right=W$_?$]
{\Theta\ ;\ \Gamma\vdash\Delta}
{\Theta\ ;\ \Gamma\vdash\Delta,?A}
\and
\inferrule*[right=C$_?$]
{\Theta\ ;\ \Gamma\vdash\Delta,?A,?A}
{\Theta\ ;\ \Gamma\vdash\Delta,?A}.
\end{mathpar}
For dereliction, add $H(A)$ to the cartesian context by weakening, apply $H$L to consume the displayed conclusion $A$, and then apply $K$R. For weakening, add the cartesian hypothesis $H(A)$ and apply $K$R without using it. For contraction, the right-invertibility of $U$ exposes the two conclusions $?A$ as two cartesian hypotheses $H(A)$. Cartesian contraction merges these hypotheses, and $K$R returns the single conclusion $?A$. Thus the structural behavior on the output side comes from the cartesian input zone.

For example, the identity sequent gives the unit of the modality:
$$
\inferrule*[right=$K$R]
{
  \inferrule*[right=$H$L]
  {H(A)\ ;\ A\vdash A}
  {H(A)\ ;\ A\vdash\cdot}
}
{\cdot\ ;\ A\vdash ?A}.
$$
Weakening and contraction also give the sequents
$$
\cdot\ ;\ \bot\vdash ?A,
\qquad
\cdot\ ;\ ?A\parr ?A\vdash ?A.
$$
For the first, add $?A$ to the empty output of $\bot$L. For the second, apply $\parr$L to two identity premises and then contract the two outputs. Left promotion applied to $H(A)\ ;\ ?A\vdash\cdot$, followed by $K$R, gives
$$
\cdot\ ;\ ??A\vdash ?A,
$$
where the starting sequent is $K$L applied to $H(A)\vdash H(A)$.

We can now write the familiar two-sided exponential rules using only linear formulas. For finite lists $\Gamma=A_1,\ldots,A_n$ and $\Delta=B_1,\ldots,B_m$, put
$$
!\Gamma:=!A_1,\ldots,!A_n,
\qquad
?\Delta:=?B_1,\ldots,?B_m.
$$
Suppressing the empty cartesian zone, the two promotion rules are
$$
\inferrule*[right=Prom$_!$]
{!\Gamma\vdash ?\Delta,A}
{!\Gamma\vdash ?\Delta,!A}
\qquad
\inferrule*[right=Prom$_?$]
{!\Gamma,A\vdash ?\Delta}
{!\Gamma,?A\vdash ?\Delta}.
$$
Every side formula is exponential. To read the first rule in the mixed presentation, use the left-invertibility of $\mathsf F$ to expose the assumptions $!A_i$ as cartesian hypotheses $G(A_i)$, and the right-invertibility of $K$ to expose the conclusions $?B_i$ as cartesian hypotheses $H(B_i)$. The premise then has the form
$$
G(\Gamma),H(\Delta)\ ;\ \cdot\vdash A,
$$
where $G$ and $H$ are applied pointwise to the lists. Right promotion for $!$ applies to this judgment. The rules $\mathsf F$L and $K$R then restore the displayed two-sided conclusion.

For the second rule, exposing the same side formulas gives
$$
G(\Gamma),H(\Delta)\ ;\ A\vdash\cdot.
$$
Left promotion for $?$ applies, and the same two rules restore the side formulas. Both conventional promotion rules therefore arise by placing all exponential side data in the shared cartesian zone. Their derivation uses the four shifts $\mathsf F,G,H,K$ directly.

The formulas of the multiplicative-exponential fragment are therefore
$$
A,B::=P\mid A\otimes B\mid\mathbf 1\mid A\parr B\mid\bot
\mid A^\perp\mid !A\mid ?A,
$$
with linear implication available as above. The multiplicative rules split the linear resources, and the exponential rules expose cartesian data through $g$ and $h$. This gives a mixed example in which the two classes of modes each retain their own structural behavior and the modalities express the relationship between them.

\subfilebibliography

\endgroup{}
\begingroup

\chapter{Representability}\label{chap:representability}

\begingroup

The goal of this chapter is to derive, step by step, representability results for the multicategories and polycategories involved in this thesis. We start with the notion of a monoidal category, which will roughly encode multicategories; we then move on to polycategories and their analogues, weakly distributive categories; and we finish with explicit data for a one-way construction of planar polycategories over a base. The resulting projection has canonical liftings and is a planar bifibration (\cref{thm:planar-cartesianness}). The one-way construction is novel, as the previous results can be found in \cite{Lam69,Her00} and \cite{CS97,Bla23}. Our exposition is slightly different from these sources; thus we take special care in providing all the required diagrams and details wherever needed.

\section{Monoidal Categories and Multicategorical Representability}\label{sec:multicategorical-representability}

\phantomsection\label{anchor:multicategorical-representability-roadmap}
We start our journey by trying to find a category with enough structure to contain the same amount of information as a multicategory. The key aspect of a multicategory is that a multimorphism possesses multiple inputs. In a category, we internalize this as a monoidal operation $\otimes$ with a unit $\top$, so as to read a multimap $\Gamma \to A$ as a map $\otimes \Gamma \to A$. The picture of the current situation can be extracted from the following diagram:
\begin{center}
  \begin{tikzcd}[row sep=large, column sep=huge]
    \mathbf{MonCat} \arrow[r, hook] & \mathbf{MultiCat} \\
    \mathbf{MonCat} \arrow[r, "\simeq"] \arrow[u, equals]  & \mathbf{MultiCat_{t}} \arrow[u, hook ] \\
    \mathbf{ClosedMonCat} \arrow[r, "\simeq"] \arrow[u, hook ]  & \mathbf{MultiCat_{bi}} \arrow[u, hook ] \\
  \end{tikzcd}
\end{center}
where the horizontal hook arrow represents a fully faithful functor that is not essentially surjective, and the vertical hook arrows represent forgetful functors. Let us fix the objects and morphisms of these categories once and for all. We regard a closed monoidal category as equipped with chosen left and right internal-hom adjunctions. The morphisms of $\mathbf{MonCat}$ are the \emph{lax} monoidal functors, the morphisms of $\mathbf{ClosedMonCat}$ are the lax monoidal functors between closed monoidal categories, with no requirement that they preserve the chosen adjunctions, and the morphisms of $\mathbf{MultiCat}$, $\mathbf{MultiCat_{t}}$ and $\mathbf{MultiCat_{bi}}$ are the functors of multicategories, with no preservation condition on the universal objects. This choice is what makes the horizontal functors fully faithful. Under the top functor and the middle equivalence, strong monoidal functors correspond to multifunctors preserving tensor-universal maps. The same statement holds for the tensor universals in the bottom row, but preservation of the cartesian universals, equivalently of the internal-hom structure, is an additional condition not implied by strong monoidality.

\subsection{The Underlying Multicategory}\label{subsec:underlying-multicategory}

Let us start with the notion of a monoidal category. The next investigation is concerned with the top row of the \hyperref[anchor:multicategorical-representability-roadmap]{multicategorical roadmap}:
\begin{center}
  \begin{tikzcd}[row sep=large, column sep=huge]
    \mathbf{MonCat} \arrow[r, hook] & \mathbf{MultiCat}
  \end{tikzcd}
\end{center}
i.e.\ we will build a functor between the category of monoidal categories and the category of multicategories.

A monoidal category seems to have enough structure to interpret a list of objects in the domain and/or codomain of a polycategory. It will turn out to be true for the domain, but not for the codomain, and the reason is rather straightforward to see. Given two polymaps $f : A \to B,C$ and $g : B,C \to D$ in a standard polycategory, two compositions are possible: one along the object $B$ and one along $C$. Respectively, these give the two maps $g \circ_B f : A,C \to D, C$ and $g \circ_C f : B,A \to B, D$. There is no way a priori to cut both $B,C$ at the same time. In a monoidal category, those maps would be represented as $A \to B \otimes C$ and $B \otimes C \to D$, and only one composition is possible. Therefore, we must find a way to differentiate objects in the domain and objects in the codomain in order to preserve this partial composition (cut) operation. One way to do this is to introduce yet another monoidal operator $\oplus$ and interpret the maps as $A \to B \oplus C$ and $B \otimes C \to D$, with an appropriate structure for the cut operation.\\
Before diving into this idea, we note that the best we can do with the present data is to interpret multicategories.

At this point of the investigation, an equivalence between monoidal categories and multicategories is not yet reachable, as a general multicategory need not have objects representing its lists of inputs. To solve this issue, we will introduce the tensor multicategories of \cref{subsec:tensor-multicategories}, which internalize each list of objects into an object itself, allowing the monoidal structure to be in equivalence. Without this representability structure, we would only get a fully faithful functor between the two, without essential surjectivity. In other words, some multicategories are not isomorphic to the underlying multicategory of any monoidal category. The multicategory with a single object $x$ and the identity as its only multimap is such an example; a monoidal category with a single object forces $x \otimes x = x$, and its underlying multicategory then carries an element of $\mc M(x, x; x)$, which is empty here. Adding a tensor operation might seem a superficial addition, one that merely forces the same monoidal structure onto multicategories. Two key differences separate the notions:
\begin{itemize}
  \item In the case of monoidal categories, $\otimes$ is a functor from the cartesian product of the category $\mc C \times \mc C$ to the category $\mc C$ and, in some sense, is external to the category. In a multicategory, the tensor $\otimes \Gamma$ is an internal object of the multicategory.
  \item This tensor will be a special case of the universal multimaps of \cref{def:universal-multimap}, which unify several constructions in multi- and polycategories.
\end{itemize}

\phantomsection\label{anchor:multicategorical-associator-convention}
\textit{Remark (convention):}
  For a list of objects $\Gamma = \Gamma_1, \ldots, \Gamma_n$ of a monoidal category we write $\otimes \Gamma$ for the right-associated tensor $\Gamma_1 \otimes (\Gamma_2 \otimes (\ldots (\Gamma_{n-1} \otimes \Gamma_n)\ldots))$, with the conventions $\otimes() := \top$ and $\otimes(A) := A$ for the empty and singleton lists. By the coherence theorem for monoidal categories \cite{Mac98}, between any two ways to associate a tensor of the same list there is a unique isomorphism built from associators and unitors; we denote any such canonical isomorphism by $a$ (decorated with its endpoints when needed) and we allow ourselves to re-associate silently along these isomorphisms. In particular, when a list $\Gamma$ appears inside a tensor expression, as in $\Gamma_1 \otimes (\Gamma \otimes \Gamma_2)$, it stands for the right-associated tensor of the corresponding sublist, and when one of the outer lists $\Gamma_1, \Gamma_2$ is empty the corresponding tensor factor is removed by the unitors. The same conventions apply later to $\oplus \Delta$ with unit $\bot$.

Before that, let us build the promised functor between monoidal categories and multicategories.
\begin{proposition}\label{prop:moncat-to-multicat}
  There is a fully faithful functor
\[\mathbf{MonCat} \xto{\mc F} \mathbf{MultiCat}\]
\end{proposition}
\textit{Proof:} For a monoidal category $\mc M$, its underlying multicategory $\mc F \mc M := \mc M_M$ is defined as follows:
	\begin{itemize}
  	\item The objects of $\mc M_M$ are the same as the ones in $\mc M$.
  	\item $\Gamma \to A$ is a morphism in $\mc M_M$ if and only if $\otimes \Gamma \to A$ is a morphism in $\mc M$. More descriptively, the collection of multimaps is $\mc M_M(\Gamma; A) := \mc M(\otimes \Gamma, A)$, where $\otimes \Gamma$ is the right-associated tensor.
	\end{itemize}
\phantomsection\label{anchor:underlying-multicategory-composition}
We need to define the composition operation in $\mc M_M$. Given multimaps $\Gamma \xto{f_M} A$ and $\Gamma_1, A, \Gamma_2 \xto{g_M} B$, there are morphisms $\otimes \Gamma \xto{f} A$ and $\Gamma_1 \otimes  (A \otimes \Gamma_2) \xto{g}B$ in $\mc M$. We construct the map $\Gamma_1 \otimes (\Gamma \otimes \Gamma_2) \to B$ as
\[  \Gamma_1 \otimes (\Gamma \otimes \Gamma_2) \xto{a} \Gamma_1 \otimes (\otimes\Gamma \otimes \Gamma_2) \xto{\Gamma_1 \otimes (f \otimes \Gamma_2)} \Gamma_1 \otimes (A \otimes \Gamma_2) \xto{g} B\]
where the first map is the canonical coherence isomorphism, and so the composition $g_M \circ_A f_M$ is defined as $g_M \circ_A f_M := (g \circ (\Gamma_1 \otimes (f \otimes \Gamma_2)) \circ a)_M$. Here and below, we use $\Gamma_1 \otimes (f \otimes \Gamma_2)$ to denote the map $\1_{\Gamma_1} \otimes (f \otimes \1_{\Gamma_2})$. For the right unit law, let $g_M : \Gamma_1,A,\Gamma_2 \to B$; then $g_M \circ_A (\1_A)_M$ is represented by $g$ composed with the canonical rebracketing of $\otimes(\Gamma_1,A,\Gamma_2)$ and the corresponding componentwise identity map. This composite is $g$ by monoidal coherence. For the left unit law, if $f_M : \Gamma \to A$, then $(\1_A)_M \circ_A f_M$ is represented by $f$ followed by $\1_A$, with the canonical unitors removing the empty outer contexts, and is therefore $f$. Thus both unit laws hold, including for empty input contexts. For associativity of composition we look at the following diagram and argue its commutativity. Given multimaps $\Gamma \xto{f_M} A$, $\Gamma_1, A, \Gamma_2 \xto{g_M}  B$ and $\Delta_1, B, \Delta_2 \xto{h_M} C$, the two composites $h_M \circ_B (g_M \circ_A f_M)$ and $(h_M \circ_B g_M) \circ_A f_M$ are the two outer paths of:

\begin{center}
  \begin{tikzcd}
		\Delta_1 \otimes (\Gamma_1 \otimes (\otimes\Gamma \otimes (\Gamma_2 \otimes \Delta_2))) \arrow[rr, "\Delta_1 \otimes (\Gamma_1 \otimes (f \otimes (\Gamma_2 \otimes \Delta_2)))"]  \arrow[d, "\Delta_1 \otimes (\Gamma_1 \otimes a^{-1})", swap] & & \Delta_1 \otimes (\Gamma_1 \otimes (A \otimes (\Gamma_2 \otimes \Delta_2))) \arrow[d, "\Delta_1 \otimes (\Gamma_1 \otimes a^{-1})"]   \\
		\Delta_1 \otimes (\Gamma_1 \otimes ((\otimes\Gamma \otimes \Gamma_2) \otimes \Delta_2)) \arrow[rr, "\Delta_1 \otimes (\Gamma_1 \otimes ((f \otimes \Gamma_2) \otimes \Delta_2))"] \arrow[d, "\Delta_1 \otimes a^{-1}", swap]   & & \Delta_1 \otimes (\Gamma_1 \otimes ((A \otimes \Gamma_2) \otimes \Delta_2)) \arrow[d, "\Delta_1 \otimes a^{-1}"]  \\
		\Delta_1 \otimes ((\Gamma_1 \otimes (\otimes\Gamma \otimes \Gamma_2)) \otimes \Delta_2) \arrow[rr, "\Delta_1 \otimes ((\Gamma_1 \otimes (f \otimes \Gamma_2)) \otimes \Delta_2)"]  \arrow[d, "\Delta_1 \otimes ((\Gamma_1 \otimes (f \otimes \Gamma_2)) \otimes \Delta_2)", swap ]& & \Delta_1 \otimes ((\Gamma_1 \otimes (A \otimes \Gamma_2)) \otimes \Delta_2) \arrow[d, equals] \\
		\Delta_1 \otimes ((\Gamma_1 \otimes (A \otimes \Gamma_2)) \otimes \Delta_2) \arrow[rr, equals] \arrow[dr, "\Delta_1 \otimes (g \otimes \Delta_2)", swap ] & & \Delta_1 \otimes ((\Gamma_1 \otimes (A \otimes \Gamma_2)) \otimes \Delta_2) \arrow[dl, "\Delta_1 \otimes (g \otimes \Delta_2)" ] \\
		& \Delta_1 \otimes (B \otimes \Delta_2) \arrow[d, "h"]  & \\
		& C & \\
  \end{tikzcd}
\end{center}
Both composites start with the canonical isomorphism $\otimes(\Delta_1,\Gamma_1,\Gamma,\Gamma_2,\Delta_2) \to \Delta_1 \otimes (\Gamma_1 \otimes (\otimes\Gamma \otimes (\Gamma_2 \otimes \Delta_2)))$, which is unique by coherence. The two upper rectangles commute by naturality of the canonical isomorphisms in the $\otimes\Gamma$-component, the third by definition, and the bottom square trivially; thus we have associativity in the multicategory as well.

It remains to check the exchange law: for $f_M : \Gamma \to A$, $g_M : \Delta \to B$ and $h_M : \Sigma_1, A, \Sigma_2, B, \Sigma_3 \to C$, one must have $(h_M \circ_A f_M) \circ_B g_M = (h_M \circ_B g_M) \circ_A f_M$. Unfolding the definitions, both sides are composites of $h$ with the two componentwise tensor maps $\otimes(\Sigma_1, f, \Sigma_2, B, \Sigma_3)$ and $\otimes(\Sigma_1, A, \Sigma_2, g, \Sigma_3)$ and canonical isomorphisms. These maps act on disjoint tensor factors, so by functoriality of $\otimes$ in each variable,
\[
\begin{aligned}
&\otimes(\Sigma_1, A, \Sigma_2, g, \Sigma_3) \circ \otimes(\Sigma_1, f, \Sigma_2, \otimes\Delta, \Sigma_3)
\\
&\quad= \otimes(\Sigma_1, f, \Sigma_2, g, \Sigma_3)
\\
&\quad= \otimes(\Sigma_1, f, \Sigma_2, B, \Sigma_3) \circ \otimes(\Sigma_1, \otimes\Gamma, \Sigma_2, g, \Sigma_3),
\end{aligned}
\]
and together with naturality of the canonical isomorphisms this gives the exchange law.

Now we would like to show that $\mc F$ can be promoted to a functor. Its action on objects is $\mc F(\mc M) := \mc M_M$ where $\mc M$ is a monoidal category. To define its action on morphisms (i.e.\ functors), let $\mc M \xto{F} \mc M'$ be a lax monoidal functor between the two monoidal categories $\mc M, \mc M'$. We want to define a functor $\mc M_M \xto{F_M} \mc M'_M$ between the underlying multicategories. The action on objects is $F_M(A_M) := FA$. For the action on morphisms, let $\Gamma_M \xto{f_M} A_M$ be a multimap in $\mc M_M$, i.e.\ a map $f : \otimes \Gamma \to A$ in $\mc M$. Since $F$ is lax monoidal, it comes with a lax monoidal map $\eta^{lax}_{A,B} : FA \otimes FB \to F(A \otimes B)$ and a unit $\eta^{lax}_{0} : \top \to F\top$, which assemble into an $n$-ary laxator
\[
  \eta^{lax}_{\Gamma} : \otimes F\Gamma \to F(\otimes \Gamma)
\]
defined recursively along the right-associated tensor by $\eta^{lax}_{[]} := \eta^{lax}_{0}$, $\eta^{lax}_{[A]} := \1_{FA}$, and, for every nonempty list $\Sigma$,
\[
  \eta^{lax}_{(A,\Sigma)} := \eta^{lax}_{A,\,\otimes\Sigma} \circ (\1_{FA} \otimes \eta^{lax}_{\Sigma}).
\]
Here the subscript $(A,\Sigma)$ on the left denotes concatenation of lists, while the first subscript on the binary laxator on the right is the object $A$. We define the image of $f_M$ by the composite
\[
  \otimes F\Gamma \xto{\eta^{lax}_{\Gamma}} F(\otimes \Gamma) \xto{Ff} FA
\]
and set $F_M(f_M) := (Ff \circ \eta^{lax}_{\Gamma})_M$. We check that $F_M$ is a functor of multicategories. Identities are preserved since $F_M((\1_A)_M) = (F\1_A \circ \eta^{lax}_{[A]})_M = (\1_{FA} \circ \1_{FA})_M = (\1_{FA})_M$. For preservation of composition we use naturality and coherence of $\eta^{lax}$.
Let $f_M : \Gamma \to A$ and $g_M : \Gamma_1, A, \Gamma_2 \to B$, so that $f : \otimes\Gamma \to A$ and $g : \Gamma_1 \otimes (A \otimes \Gamma_2) \to B$ in $\mc M$. By the \hyperref[anchor:underlying-multicategory-composition]{composition construction above}, $g_M \circ_A f_M = (g \circ (\Gamma_1 \otimes (f \otimes \Gamma_2)) \circ a)_M$. Unfolding the definition of $F_M$ on morphisms,
\begin{align*}
  F_M(g_M \circ_A f_M) &= \big(F(g \circ (\Gamma_1 \otimes (f \otimes \Gamma_2)) \circ a) \circ \eta^{lax}_{\Gamma_1, \Gamma, \Gamma_2}\big)_M \\
  &= \big(Fg \circ F(\Gamma_1 \otimes (f \otimes \Gamma_2)) \circ Fa \circ \eta^{lax}_{\Gamma_1, \Gamma, \Gamma_2}\big)_M,
\end{align*}
where $\eta^{lax}_{\Gamma_1, \Gamma, \Gamma_2}$ denotes the $n$-ary laxator on the list $\Gamma_1, \Gamma, \Gamma_2$. Unfolding instead the right-hand side and using the composition rule of $\mc M'_M$,
\begin{align*}
  F_M(g_M) \circ_{FA} F_M(f_M) &= \big(Fg \circ \eta^{lax}_{\Gamma_1, A, \Gamma_2}\big)_M \circ_{FA} \big(Ff \circ \eta^{lax}_{\Gamma}\big)_M \\
  &= \big(Fg \circ \eta^{lax}_{\Gamma_1, A, \Gamma_2} \circ (F\Gamma_1 \otimes ((Ff \circ \eta^{lax}_\Gamma) \otimes F\Gamma_2)) \circ a'\big)_M,
\end{align*}
where $a'$ is the canonical isomorphism of $\mc M'$. The two expressions coincide because the $n$-ary laxator is natural in each component and coherent with the monoidal associativity, which together yield the identity
\[
  F(\Gamma_1 \otimes (f \otimes \Gamma_2)) \circ Fa \circ \eta^{lax}_{\Gamma_1, \Gamma, \Gamma_2}
  \;=\;
  \eta^{lax}_{\Gamma_1, A, \Gamma_2} \circ (F\Gamma_1 \otimes ((Ff \circ \eta^{lax}_{\Gamma}) \otimes F\Gamma_2)) \circ a'.
\]
Hence $F_M(g_M \circ_A f_M) = F_M(g_M) \circ_{FA} F_M(f_M)$, so $F_M$ preserves composition. Together with preservation of identities this shows that $F_M$ is a functor of multicategories, which finishes the specification of $\mc F$ on $1$-cells. Functoriality of $\mc F$ ($(G \circ F)_M = G_M \circ F_M$) follows since the laxator of a composite lax monoidal functor is $G\eta^{F} \circ \eta^{G}$, matching the composite of the two actions.

It remains to verify that $\mc F$ is fully faithful, i.e.\ that for monoidal categories $\mc M, \mc M'$ the map $\mathbf{MonCat}(\mc M, \mc M') \to \mathbf{MultiCat}(\mc M_M, \mc M'_M)$, $F \mapsto F_M$, is a bijection. Suppose $H : \mc M_M \to \mc M'_M$ is any functor of multicategories. We recover a lax monoidal functor $F$ with $F_M = H$ as follows. On objects and on unary multimaps, $F$ is $H$ itself; since $\mc M_M(A; B) = \mc M(A, B)$ by the singleton convention, this defines a functor $F : \mc M \to \mc M'$. For the laxator, consider the multimap $u_{A,B} : A, B \to A \otimes B$ of $\mc M_M$ corresponding to the identity of $A \otimes B$; its image $H(u_{A,B}) : HA, HB \to H(A \otimes B)$ is a multimap of $\mc M'_M$, i.e.\ a map
\[ \eta^{lax}_{A,B} := H(u_{A,B}) : HA \otimes HB \to H(A \otimes B), \]
and similarly the nullary multimap $() \to \top$ corresponding to $\1_\top$ yields $\eta^{lax}_0 : \top' \to H\top$. Naturality of $\eta^{lax}$ follows by applying $H$ to the equality $u_{A',B'} \circ (f, g) = (f \otimes g)_M \circ_{A \otimes B} u_{A,B}$ of multimaps $A, B \to A' \otimes B'$, valid for unary $f : A \to A'$, $g : B \to B'$. The associativity coherence law follows by applying $H$ to
\[
  u_{A\otimes B, C} \circ_{A \otimes B} u_{A,B}
  = (a^{-1})_M \circ_{A \otimes (B \otimes C)}
    \big(u_{A, B\otimes C} \circ_{B \otimes C} u_{B,C}\big),
\]
where again $a$ is the associator. Applying $H$ to the analogous equations involving $u_{()}$ and the identity multimaps gives the two unit axioms; explicitly,
\[
\begin{aligned}
  F\lambda_A \circ \eta^{lax}_{\top,A}
    \circ (\eta^{lax}_0 \otimes \1_{FA}) &= \lambda'_{FA}, \\
  F\rho_A \circ \eta^{lax}_{A,\top}
    \circ (\1_{FA} \otimes \eta^{lax}_0) &= \rho'_{FA}.
\end{aligned}
\]
Finally, to conclude that $F_M = H$, we note that every multimap $f_M : \Gamma \to A$ of $\mc M_M$ factors as $f_M = (f)_M \circ_{\otimes\Gamma} u_\Gamma$ where $u_\Gamma : \Gamma \to \otimes\Gamma$ corresponds to $\1_{\otimes\Gamma}$ and $(f)_M$ is unary, and $H$ preserves this factorization while $H(u_\Gamma) = \eta^{lax}_\Gamma$ by construction and coherence. Uniqueness of $F$ is forced by the same computation: any lax monoidal $F$ with $F_M = H$ has $F = H$ on objects and unary maps, and $\eta^{lax}_{A,B} = F_M(u_{A,B}) = H(u_{A,B})$. $\square$

\subsection{Tensor Multicategories}\label{subsec:tensor-multicategories}

We move on to the middle row of the \hyperref[anchor:multicategorical-representability-roadmap]{multicategorical roadmap}. The multicategories that arise from monoidal categories are exactly the ones equipped with universal objects. Those are objects that internalize the domain of a multimap:
\begin{center}
  \begin{tikzcd}[row sep=large, column sep=huge]
    \mathbf{MonCat} \arrow[r, "\simeq"] & \mathbf{MultiCat_{t}}
  \end{tikzcd}
\end{center}
We recall the definition of a multicategory with universal objects. The subscript in $\mathbf{MultiCat_{t}}$ records that its objects are tensor multicategories; the universal property below is the one that the opcartesian morphisms of our final fibrational structure will generalize.

\begin{definition}\label{def:universal-multimap}
  Let $\Gamma$ be a list of objects in a multicategory $\mc M$. A multimap $u_\Gamma : \Gamma \to \otimes\Gamma$ is \emph{universal} if for every pair of lists $\Gamma_1, \Gamma_2$ and every multimap $f : \Gamma_1, \Gamma, \Gamma_2 \to A$ there exists a unique multimap $u_\Gamma^f : \Gamma_1, \otimes\Gamma, \Gamma_2 \to A$ such that $u_\Gamma^f \circ_{\otimes\Gamma} u_\Gamma = f$. In other words, precomposition with $u_\Gamma$ induces bijections
  \[ - \circ_{\otimes\Gamma} u_\Gamma \; : \; \mc M(\Gamma_1, \otimes\Gamma, \Gamma_2; A) \xto{\;\simeq\;} \mc M(\Gamma_1, \Gamma, \Gamma_2; A). \]
  A \emph{tensor multicategory} is a multicategory equipped with a choice of universal multimap $u_\Gamma : \Gamma \to \otimes\Gamma$ for every finite list $\Gamma$.
\end{definition}
\textit{Remark:}
  The universal object $\otimes\Gamma$ is unique up to a unique invertible unary multimap. This means that given two universal maps $u : \Gamma \to T$ and $u' : \Gamma \to T'$, the factorizations $u^{u'}$ and $u'^{u}$ are mutually inverse by the uniqueness clauses.

\begin{lemma}\label{lem:universal-composition}
  In any multicategory, identities are universal, and universal multimaps are closed under composition: if $u_i : \Gamma_i \to A_i$ are universal for $1 \leq i \leq n$ and $u : A_1, \ldots, A_n \to B$ is universal, then the composite $u \circ (u_1, \ldots, u_n) : \Gamma_1, \ldots, \Gamma_n \to B$ is universal.
\end{lemma}
\textit{Proof:} For the identity, every multimap factors uniquely through it as itself. For composition, factoring a multimap $f : \Delta_1, \Gamma_1, \ldots, \Gamma_n, \Delta_2 \to C$ successively through $u_1$, then $u_2$, \ldots, then $u_n$, and finally through $u$ produces the required unique factorization; uniqueness at each stage follows from the uniqueness clause of each universal map, cf.\ \cite[Proposition~1.3.4]{Bla23}. $\square$

\begin{theorem}\label{thm:moncat-multicat-equivalence}
  There is an equivalence of categories between the category of monoidal categories with lax monoidal functors and the category of tensor multicategories with functors of multicategories:
\[\mathbf{MonCat} \simeq \mathbf{MultiCat_{t}}\]
\end{theorem}
\textit{Proof:} We will construct functors in both directions and prove that they are inverse to each other up to natural isomorphism.

First we check that the functor $\mc F$ of \cref{prop:moncat-to-multicat} lands in tensor multicategories, i.e.\ that for a monoidal category $\mc M$ the multicategory $\mc M_M$ has all universal objects. For a list $\Gamma$ in $\mc M_M$ we take $\otimes\Gamma$ in the multicategory to be the object $\otimes\Gamma$ in the monoidal category represented as $\Gamma_1 \otimes (\dots \otimes (\Gamma_{n - 1} \otimes \Gamma_n)\dots)$, with universal multimap $u_\Gamma : \Gamma \to \otimes\Gamma$ the multimap corresponding to the identity $\1_{\otimes\Gamma}$. Universality is checked as follows: a multimap $f : \Gamma_1, \Gamma, \Gamma_2 \to A$ is a map $f : \otimes(\Gamma_1,\Gamma,\Gamma_2) \to A$ of $\mc M$, and we set
\[ u_\Gamma^f := f \circ a^{-1} : \otimes(\Gamma_1, \otimes\Gamma, \Gamma_2) \to A \]
where $a : \otimes(\Gamma_1,\Gamma,\Gamma_2) \to \otimes(\Gamma_1,\otimes\Gamma,\Gamma_2)$ is the canonical isomorphism. Unfolding the composition of $\mc M_M$, $u_\Gamma^f \circ_{\otimes\Gamma} u_\Gamma = f \circ a^{-1} \circ (\Gamma_1 \otimes (\1_{\otimes\Gamma} \otimes \Gamma_2)) \circ a = f$, and since $- \circ a$ is a bijection this factorization is unique.

\begin{itemize}

\item \phantomsection\label{anchor:tensor-multicategory-underlying-monoidal-category}$\mathbf{MultiCat_{t}} \xto{\mc G} \mathbf{MonCat}$: We would like now to define the underlying monoidal category of a tensor multicategory. For this we will extract the ``category'' part of the multicategory, which means that we will only consider multimaps that are just maps in the sense of having only one input. Given a tensor multicategory $\mc M$ we define the underlying monoidal category $\mc M_{MO}$ as follows:
\begin{itemize}
  \item The objects of $\mc M_{MO}$ are the objects of $\mc M$.
  \item $A \xto{f_{MO}} B$ is a morphism in $\mc M_{MO}$ if and only if $A \xto{f} B$ is a morphism in $\mc M$. More descriptively, the collection of morphisms is $\mc M_{MO}(A,B) = \mc M(A; B)$ where $A, B$ are objects in $\mc M$. Composition and identities are those of $\mc M$ restricted to unary multimaps.
  \item The tensor of objects is given by the chosen universal objects: $A \otimes_{MO} B := \otimes(A,B)$, with unit $\top_{MO} := \otimes()$.
  \item The tensor of morphisms $f : A \to B$, $g : C \to D$ is given by the universal property: $f \otimes_{MO} g := u_{A,C}^{(u_{B,D} \circ_D g) \circ_B f}$ is the unique unary map satisfying
  \[
    (f \otimes_{MO} g) \circ_{A \otimes C} u_{A,C}
    = (u_{B,D} \circ_D g) \circ_B f : A,C \to B \otimes D.
  \]
  \item The associator $a_{A,B,C} : (A \otimes B) \otimes C \to A \otimes (B \otimes C)$ is the unique factorization of the composite universal map $u_{A, B\otimes C} \circ_{B \otimes C} u_{B,C} : A, B, C \to A \otimes (B \otimes C)$ through the composite universal map $u_{A \otimes B, C} \circ_{A \otimes B} u_{A,B} : A, B, C \to (A \otimes B) \otimes C$; it is invertible because both composites are universal for the list $A,B,C$ by \cref{lem:universal-composition}, and universal objects are unique up to unique isomorphism. For the unitors, define the universal unary maps
  \[
    v_A^L := u_{\top,A}\circ_{\top}u_{()} : A\to\top\otimes A,
    \qquad
    v_A^R := u_{A,\top}\circ_{\top}u_{()} : A\to A\otimes\top.
  \]
  Then $\lambda_A:\top\otimes A\to A$ and $\rho_A:A\otimes\top\to A$ are the unique maps satisfying $\lambda_A\circ v_A^L=\1_A$ and $\rho_A\circ v_A^R=\1_A$. They are invertible because $v_A^L,v_A^R$ and $\1_A$ are universal for the singleton list.
\end{itemize}
Functoriality of $\otimes_{MO}$, naturality of the associator and unitors, and the pentagon and triangle laws all follow from the uniqueness clause of the universal property: in each diagram, both composites are factorizations of one and the same multimap out of a universal map, hence they coincide. For instance $\1_A \otimes_{MO} \1_B = \big(u_{A,B}^{(u_{A,B} \circ_B \1_B) \circ_A \1_A}\big) = u_{A,B}^{u_{A,B}} = \1_{A \otimes B}$, since the identity also factors $u_{A,B}$ through itself, and for composable $f, f', g, g'$ both $(f' \otimes_{MO} g') \circ (f \otimes_{MO} g)$ and $(f' \circ f) \otimes_{MO} (g' \circ g)$ factor the multimap $(u \circ g' g) \circ f' f$ through $u_{A,C}$.

We now define the functor $\mc G$. Its action on objects is $\mc G(\mc M) := \mc M_{MO}$. To define its action on morphisms, suppose we are given a functor $\mc M \xto{F} \mc M'$ of multicategories. We define $F_{MO} : \mc M_{MO} \to \mc M'_{MO}$ as follows:
\begin{itemize}
  \item For any object $A \in \mc M_{MO}$ we set $F_{MO}(A) = FA$.
  \item For any morphism $A \xto{f} B$ we set $F_{MO}(f) = Ff$.
\end{itemize}
Note that $F_{MO}(\1_{A}) = F\1_A = \1_{FA}$, and composition is preserved because morphisms in $\mc M_{MO}$ are the unary multimaps of $\mc M$, on which multicategorical composition restricts to ordinary composition. Therefore $F_{MO}$ is a functor.
\end{itemize}
It remains to equip $F_{MO}$ with a lax monoidal structure. We cannot expect $F$ to preserve the universal object strictly: a functor of multicategories sends the universal object $\otimes(A,B)$ of $\mc M$ to the object $F(\otimes(A,B))$ of $\mc M'$, which has no reason to coincide with the universal object $\otimes(FA,FB)$ of $\mc M'$. The comparison morphism is instead produced by the universal property.

In the \hyperref[anchor:tensor-multicategory-underlying-monoidal-category]{construction above}, the tensor $A \otimes_{MO} B$ in $\mc M_{MO}$ is the universal object $\otimes(A,B)$, equipped with its universal multimap $u_{A,B} : A, B \to \otimes(A,B)$. Applying the functor $F$ gives a multimap
\[ Fu_{A,B} : FA, FB \to F(\otimes(A,B)) \]
in $\mc M'$. By the universal property of $\otimes(FA,FB)$, with universal multimap $u'_{FA,FB} : FA, FB \to \otimes(FA,FB)$, this multimap factors uniquely through a unary morphism
\[ \phi_{A,B} : F_{MO}A \otimes F_{MO}B = \otimes(FA,FB) \;\longrightarrow\; F(\otimes(A,B)) = F_{MO}(A \otimes B) \]
satisfying $\phi_{A,B} \circ_{\otimes(FA,FB)} u'_{FA,FB} = Fu_{A,B}$. Likewise, the unit $\top$ of $\mc M_{MO}$ is the universal object of the empty list, with universal (nullary) multimap $u_{()}$, and applying $F$ then factoring through the universal map of the unit $\top'$ of $\mc M'_{MO}$ gives the unit comparison $\phi_0 : \top' \to F\top$.

The comparison $\phi$ is natural. Indeed, for $f:A\to A'$ and $g:B\to B'$, we must prove the equality of unary maps
\[
  F(f\otimes_{MO}g)\circ\phi_{A,B}
  = \phi_{A',B'}\circ(Ff\otimes'_{MO}Fg)
  : \otimes'(FA,FB)\longrightarrow F(\otimes(A',B')).
\]
The universal property of $u'_{FA,FB}:FA,FB\to\otimes'(FA,FB)$ says that precomposition induces a bijection
\[
  -\circ_{\otimes'(FA,FB)}u'_{FA,FB}:
  \mc M'\big(\otimes'(FA,FB);F(\otimes(A',B'))\big)
  \xto{\;\simeq\;}
  \mc M'\big(FA,FB;F(\otimes(A',B'))\big).
\]
In particular, this function is injective, so it suffices to show that the two unary maps have the same composite with $u'_{FA,FB}$. For the left-hand side, associativity of cut, the defining equation $\phi_{A,B}\circ u'_{FA,FB}=Fu_{A,B}$, preservation of cut by $F$, and the definition of $f\otimes_{MO}g$ give
\begin{align*}
  &\big(F(f\otimes_{MO}g)\circ\phi_{A,B}\big)
    \circ_{\otimes'(FA,FB)}u'_{FA,FB} \\
  &\quad=F(f\otimes_{MO}g)\circ_{F(\otimes(A,B))}
    \big(\phi_{A,B}\circ_{\otimes'(FA,FB)}u'_{FA,FB}\big) \\
  &\quad=F(f\otimes_{MO}g)\circ_{F(\otimes(A,B))}Fu_{A,B} \\
  &\quad=F\big((f\otimes_{MO}g)\circ_{\otimes(A,B)}u_{A,B}\big) \\
  &\quad=F\big((u_{A',B'}\circ_{B'}g)\circ_{A'}f\big).
\end{align*}
For the right-hand side, first use the defining equation for the tensor of the morphisms $Ff,Fg$ in $\mc M'_{MO}$, then use associativity of cut and the defining equation for $\phi_{A',B'}$, and finally use preservation of the two cuts by $F$:
\begin{align*}
  &\big(\phi_{A',B'}\circ(Ff\otimes'_{MO}Fg)\big)
    \circ_{\otimes'(FA,FB)}u'_{FA,FB} \\
  &\quad=\phi_{A',B'}\circ_{\otimes'(FA',FB')}
    \big((Ff\otimes'_{MO}Fg)\circ_{\otimes'(FA,FB)}u'_{FA,FB}\big) \\
  &\quad=\phi_{A',B'}\circ_{\otimes'(FA',FB')}
    \big((u'_{FA',FB'}\circ_{FB'}Fg)\circ_{FA'}Ff\big) \\
  &\quad=\big((\phi_{A',B'}\circ_{\otimes'(FA',FB')}u'_{FA',FB'})
    \circ_{FB'}Fg\big)\circ_{FA'}Ff \\
  &\quad=\big(Fu_{A',B'}\circ_{FB'}Fg\big)\circ_{FA'}Ff \\
  &\quad=F\big((u_{A',B'}\circ_{B'}g)\circ_{A'}f\big).
\end{align*}
Thus both precomposites are the same multimap $FA,FB\to F(\otimes(A',B'))$. Injectivity of precomposition with $u'_{FA,FB}$ gives the required naturality equation. The associativity and unit coherence axioms for $(\phi_{A,B}, \phi_0)$ follow by the same method: in each coherence diagram, precompose both paths with the relevant universal map and use associativity of cut to see that they give the same multimap. Therefore $F_{MO}$ is a lax monoidal functor.

This assignment is functorial as an assignment to lax monoidal functors. Its underlying functors preserve identities and composition, so it remains to check that their lax monoidal comparisons do as well. For the identity multifunctor $\1_{\mc M}$, the reconstructed binary comparison $\phi^{\1}_{A,B}:\otimes(A,B)\to\otimes(A,B)$ is characterized by
\[
  \phi^{\1}_{A,B}\circ_{\otimes(A,B)}u_{A,B}=u_{A,B}.
\]
The identity of $\otimes(A,B)$ has this property, so uniqueness in the universal property of $u_{A,B}$ gives $\phi^{\1}_{A,B}=\1_{\otimes(A,B)}$. The same argument with $u_{()}$ shows that the nullary comparison of the identity multifunctor is $\1_\top$.

Now let $\mc M \xto{F} \mc M' \xto{K} \mc M''$ be functors of tensor multicategories. The binary comparison reconstructed directly from the composite $KF$ is the unique map
\[
  \phi^{KF}_{A,B}:\otimes''(KFA,KFB)\longrightarrow KF(\otimes(A,B))
\]
satisfying
\[
  \phi^{KF}_{A,B}\circ_{\otimes''(KFA,KFB)}u''_{KFA,KFB}=KF(u_{A,B}).
\]
On the other hand, the composite of the comparisons of $K_{MO}$ and $F_{MO}$ is
\[
  \otimes''(KFA,KFB) \xto{\phi^K_{FA,FB}} K(\otimes'(FA,FB))
  \xto{K\phi^F_{A,B}} KF(\otimes(A,B)).
\]
Precomposing this composite with $u''_{KFA,KFB}$ and using successively the defining equations for $\phi^K$ and $\phi^F$, together with preservation of cut by $K$, gives
\begin{align*}
  &\big(K\phi^F_{A,B}\circ\phi^K_{FA,FB}\big)
    \circ_{\otimes''(KFA,KFB)}u''_{KFA,KFB} \\
  &\quad=K\phi^F_{A,B}\circ_{K(\otimes'(FA,FB))}
    \big(\phi^K_{FA,FB}\circ_{\otimes''(KFA,KFB)}u''_{KFA,KFB}\big) \\
  &\quad=K\phi^F_{A,B}\circ_{K(\otimes'(FA,FB))}K(u'_{FA,FB}) \\
  &\quad=K\big(\phi^F_{A,B}\circ_{\otimes'(FA,FB)}u'_{FA,FB}\big) \\
  &\quad=K(Fu_{A,B})=KF(u_{A,B}).
\end{align*}
Thus $K\phi^F_{A,B}\circ\phi^K_{FA,FB}$ satisfies the same defining equation as $\phi^{KF}_{A,B}$. Since precomposition with the universal map $u''_{KFA,KFB}$ is injective, we obtain
\[
  \phi^{KF}_{A,B}=K\phi^F_{A,B}\circ\phi^K_{FA,FB}.
\]
For the nullary comparisons, the same calculation is
\begin{align*}
  \big(K\phi^F_0\circ\phi^K_0\big)\circ_{\top''}u''_{()}
  &=K\phi^F_0\circ_{K\top'}K(u'_{()}) \\
  &=K\big(\phi^F_0\circ_{\top'}u'_{()}\big) \\
  &=K(Fu_{()})=KF(u_{()}).
\end{align*}
Universality of $u''_{()}$ therefore gives $\phi^{KF}_0=K\phi^F_0\circ\phi^K_0$. Hence $\mc G$ preserves identities and composition as a functor into $\mathbf{MonCat}$. Finally, $F_{MO}$ is strong monoidal precisely when $F$ carries the chosen nullary and binary tensor-universal maps to universal maps. By closure of universal maps under composition and uniqueness of universal objects, this is equivalent to preservation of all finite tensor-universal maps.

\phantomsection\label{anchor:iterated-chosen-tensor-comparison}
Before comparing the two composites, we distinguish the tensor obtained by iterating the chosen binary tensor from the independently chosen universal object for an arbitrary list. Define
\[
  \widehat{\otimes}() := \otimes(), \qquad
  \widehat{\otimes}(A) := A, \qquad
  \widehat{\otimes}(A,\Sigma) := \otimes(A,\widehat{\otimes}\Sigma)
\]
for every nonempty list $\Sigma$. Thus $\widehat{\otimes}\Gamma$ is exactly the right-associated tensor of $\Gamma$ in $\mc M_{MO}$. Define a multimap $\widehat u_\Gamma : \Gamma \to \widehat{\otimes}\Gamma$ recursively by
\[
  \widehat u_{()} := u_{()}, \qquad
  \widehat u_{(A)} := \1_A, \qquad
  \widehat u_{(A,\Sigma)} := u_{A,\widehat{\otimes}\Sigma}
  \circ_{\widehat{\otimes}\Sigma} \widehat u_\Sigma.
\]
Every $\widehat u_\Gamma$ is universal by \cref{lem:universal-composition}. Consequently there is a unique isomorphism
\[
  \theta_\Gamma : \widehat{\otimes}\Gamma \xto{\cong} \otimes\Gamma
  \qquad\text{such that}\qquad
  \theta_\Gamma\circ_{\widehat{\otimes}\Gamma}\widehat u_\Gamma=u_\Gamma.
\]
In general $\theta_\Gamma$ need not be an identity.

Now we show that the two constructed functors are inverse to each other up to natural isomorphism.

$\mc F \circ \mc G \simeq \1_{\mathbf{MultiCat_{t}}}$: let $\mc M$ be a tensor multicategory and observe that
\[(\mc F \circ \mc G)(\mc M) = \mc F(\mc M_{MO}) = (\mc M_{MO})_M.\]
We exhibit an isomorphism of multicategories $\eta_{\mc M} : (\mc M_{MO})_M \to \mc M$. On objects, $\eta_{\mc M}$ is the identity. A multimap $\Gamma \to A$ of $(\mc M_{MO})_M$ is by definition a map $\widehat{\otimes}\Gamma \to A$ of $\mc M_{MO}$, i.e.\ a unary multimap $g : \widehat{\otimes}\Gamma \to A$ of $\mc M$; we set
\[
  \eta_{\mc M}(g) := g \circ_{\widehat{\otimes}\Gamma} \widehat u_\Gamma : \Gamma \to A.
\]
Universality of $\widehat u_\Gamma$ says precisely that this is a bijection
\[
  \mc M(\widehat{\otimes}\Gamma; A) \xto{\;\simeq\;} \mc M(\Gamma; A).
\]
For $\Gamma=(A)$ the map $\widehat u_\Gamma$ is $\1_A$, so identities are preserved. Composition is preserved because the definition of the tensor on morphisms and of the coherence maps of $\mc M_{MO}$ makes both sides, after precomposition with the relevant $\widehat u$, equal to the same iterated cut in $\mc M$; associativity and exchange permit the cuts at distinct inputs to be reordered. Hence $\eta_{\mc M}$ is an isomorphism of multicategories.

For naturality, let $F : \mc M \to \mc M'$ be a functor of tensor multicategories. Write
\[
  \widehat\phi^F_\Gamma : \widehat{\otimes}' F\Gamma \to F(\widehat{\otimes}\Gamma)
\]
for the $n$-ary laxator recursively assembled from the binary and nullary comparisons of $F_{MO}$. Induction on $\Gamma$, using the defining factorization of each binary comparison, gives
\[
  F\widehat u_\Gamma
  = \widehat\phi^F_\Gamma \circ_{\widehat{\otimes}'F\Gamma}
    \widehat u'_{F\Gamma}.
\]
Therefore, for $g : \widehat{\otimes}\Gamma \to A$,
\[
\begin{aligned}
  F(\eta_{\mc M}(g))
  &= Fg \circ_{F(\widehat{\otimes}\Gamma)} F\widehat u_\Gamma \\
  &= Fg \circ \widehat\phi^F_\Gamma
     \circ_{\widehat{\otimes}'F\Gamma} \widehat u'_{F\Gamma} \\
  &= \eta_{\mc M'}\big((\mc F\mc G F)(g)\big),
\end{aligned}
\]
since $(\mc F\mc G F)(g)=Fg\circ\widehat\phi^F_\Gamma$. Hence the square
\begin{center}
  \begin{tikzcd}[row sep=large, column sep=huge]
    (\mc F \circ \mc G)\mc M \arrow[r, "\eta_{\mc M}"] \arrow[d, "(\mc F \circ \mc G)F", swap]  & \mc M \arrow[d, "F"]  \\
    (\mc F \circ \mc G)\mc M' \arrow[r, "\eta_{\mc M'}"] &  \mc M'
  \end{tikzcd}
\end{center}
commutes.

$\mc G \circ \mc F \simeq \1_{\mathbf{MonCat}}$: for a monoidal category $\mc M$, the category $(\mc M_M)_{MO}$ has the same objects as $\mc M$ and morphisms $(\mc M_M)_{MO}(A,B) = \mc M_M(A;B) = \mc M(\otimes(A), B) = \mc M(A,B)$ by the singleton convention, so it is $\mc M$ itself as a category. Its monoidal structure is given by the chosen universal objects of $\mc M_M$, namely $A \otimes_{MO} B = \otimes(A,B) = A \otimes B$ and $\top_{MO} = \otimes() = \top$; the tensor of morphisms and the coherence isomorphisms agree with those of $\mc M$ by the uniqueness clause of the universal property, since $f \otimes g$ also factors the relevant multimap. Hence the identity-on-objects strong monoidal functor $\varepsilon_{\mc M} : (\mc M_M)_{MO} \to \mc M$ is an isomorphism, and its naturality in lax monoidal functors is checked exactly as above (the laxator of $(F_M)_{MO}$ is recovered as $F_M(u_{A,B}) = \eta^{lax}_{A,B}$). This establishes the equivalence. $\square$

\subsection{Bitensor Multicategories}\label{subsec:bitensor-multicategories}

It is possible to add an extra notion to both monoidal categories and multicategories allowing us to recover the right adjoint to $A \otimes (\--)$, representing the left implication of linear logic. This structure is mirrored in multicategories by allowing a universal property for elements in the domain of multimaps, i.e.\ a cartesian counterpart of the universal objects in \cref{def:universal-multimap}. We will do so here for the sake of exposition and to make the correspondence tighter in the universe of multicategories, as it will map very conveniently when we move to polycategories. We are thus interested in the last row of the \hyperref[anchor:multicategorical-representability-roadmap]{multicategorical roadmap}:
\begin{center}
  \begin{tikzcd}[row sep=large, column sep=huge]
    \mathbf{ClosedMonCat} \arrow[r, "\simeq"] & \mathbf{MultiCat_{bi}}
  \end{tikzcd}
\end{center}

\begin{definition}\label{def:cartesian-universal-object}
  Let $\Gamma_1, \Gamma_2$ be lists of objects and $B$ an object of a multicategory $\mc M$. A \emph{cartesian universal object at position $(\Gamma_1 \mid \-- \mid \Gamma_2)$ for $B$} is an object $U(\Gamma_1 \mid B \mid \Gamma_2)$ equipped with an evaluation multimap
  \[ u^{\Gamma_1, B, \Gamma_2} : \Gamma_1,\, U(\Gamma_1 \mid B \mid \Gamma_2),\, \Gamma_2 \to B \]
  such that for every list $\Gamma$ and multimap $f : \Gamma_1, \Gamma, \Gamma_2 \to B$ there exists a unique multimap $u^{\Gamma_1, B, \Gamma_2}_f : \Gamma \to U(\Gamma_1 \mid B \mid \Gamma_2)$ with $u^{\Gamma_1, B, \Gamma_2} \circ_{U} u^{\Gamma_1, B, \Gamma_2}_f = f$.
  A \emph{bitensor multicategory} is a tensor multicategory equipped with a choice of cartesian universal object $U(\Gamma_1 \mid B \mid \Gamma_2)$ for all $\Gamma_1, B, \Gamma_2$.
\end{definition}
We denote the category of bitensor multicategories and their functors by $\mathbf{MultiCat_{bi}}$. Note the direction of the universality: maps \emph{into} $U(\Gamma_1 \mid B \mid \Gamma_2)$ correspond to maps $\Gamma_1, \--, \Gamma_2 \to B$; this is the ``cartesian'' behaviour that the cartesian morphisms of a fibration will generalize, in the same way that the tensor $\otimes\Gamma$ displays the ``opcartesian'' behaviour.

\begin{theorem}\label{thm:closedmoncat-multicat-equivalence}
  There is an equivalence of categories between the category of closed monoidal categories and the category of bitensor multicategories:
  \[\mathbf{ClosedMonCat} \simeq \mathbf{MultiCat_{bi}}\]
\end{theorem}
\textit{Proof:} To exhibit the functor $\mathbf{ClosedMonCat} \xto{\mc F}  \mathbf{MultiCat_{bi}}$, we take the functor of \cref{thm:moncat-multicat-equivalence} and show that its image can be equipped with the additional structure. So let $\mc M$ be a closed monoidal category, with its chosen adjunctions $A \otimes (\--) \dashv A \multimap (\--)$ and $(\--) \otimes A \dashv (\--) \multimapinv A$. The underlying multicategory and its tensors are exactly as defined in \cref{subsec:tensor-multicategories}. We now need to exhibit the cartesian universal objects. For lists $\Gamma_1, \Gamma_2$ and an object $B$ we take
\[ U(\Gamma_1 \mid B \mid \Gamma_2) := (\otimes \Gamma_1 \multimap B) \multimapinv \otimes \Gamma_2. \]
We first exhibit the evaluation multimap $\Gamma_1, (\otimes\Gamma_1 \multimap B) \multimapinv \otimes\Gamma_2, \Gamma_2 \to B$, i.e.\ a map $\otimes\Gamma_1 \otimes \big((( \otimes\Gamma_1 \multimap B) \multimapinv \otimes\Gamma_2) \otimes \otimes\Gamma_2\big) \to B$ in $\mc M$. We start with the identity map and transpose twice:
\begin{align*}
  (\otimes\Gamma_1 \multimap B) \multimapinv \otimes\Gamma_2 \to (\otimes\Gamma_1 \multimap B) \multimapinv \otimes\Gamma_2  &\Longrightarrow ((\otimes\Gamma_1 \multimap B) \multimapinv \otimes\Gamma_2) \otimes \otimes\Gamma_2 \to \otimes\Gamma_1 \multimap B \\
  &\Longrightarrow \otimes\Gamma_1 \otimes \big(((\otimes\Gamma_1 \multimap B) \multimapinv \otimes\Gamma_2) \otimes \otimes\Gamma_2\big) \to  B
\end{align*}
where the first step uses the counit of $(\--) \otimes \otimes\Gamma_2 \dashv (\--) \multimapinv \otimes\Gamma_2$ and the second the counit of $\otimes\Gamma_1 \otimes (\--) \dashv \otimes\Gamma_1 \multimap (\--)$.
It remains to show universality. For any list $\Gamma$ we have the following sequence of natural bijections
\begin{align*}
  \mc M_M\big(\Gamma;\, U(\Gamma_1 \mid B \mid \Gamma_2)\big)
  &= \mc M\big(\otimes\Gamma,\, (\otimes\Gamma_1 \multimap B) \multimapinv \otimes\Gamma_2\big) \\
  &\cong \mc M\big(\otimes\Gamma \otimes \otimes\Gamma_2,\, \otimes\Gamma_1 \multimap B\big) \\
  &\cong \mc M\big(\otimes\Gamma_1 \otimes (\otimes\Gamma \otimes \otimes\Gamma_2),\, B\big) \\
  &\cong \mc M\big(\otimes(\Gamma_1, \Gamma, \Gamma_2),\, B\big) = \mc M_M(\Gamma_1, \Gamma, \Gamma_2;\, B),
\end{align*}
where the last step is the canonical coherence isomorphism. By naturality of the adjunction bijections, the composite bijection is implemented by cutting with the evaluation multimap constructed above (transposing $g : \Gamma \to U$ twice and composing is the same as composing with the double transpose of the identity); hence the evaluation multimap satisfies the required universal property, with $u^{\Gamma_1,B,\Gamma_2}_f$ the double transpose of $f$.

In the other direction, let $\mc M$ be a bitensor multicategory. Its underlying monoidal category $\mc M_{MO}$ is the one from the \hyperref[anchor:tensor-multicategory-underlying-monoidal-category]{tensor-multicategory construction}; we must show it is closed. For objects $A, B$ set
\[ A \multimap B := U(A \mid B \mid ()), \qquad B \multimapinv A := U(() \mid B \mid A), \]
each equipped with its evaluation multimap, e.g.\ $u^{A,B} : A, A \multimap B \to B$. For fixed $A$, the assignment $B \mapsto A \multimap B$ extends to a functor $\mc M_{MO} \to \mc M_{MO}$: for $h : B \to B'$, the map $A \multimap h$ is the unique factorization $u^{A,B'}_{h \circ_B u^{A,B}}$ of the composite $A, A\multimap B \xto{u^{A,B}} B \xto{h} B'$ through the evaluation of $A \multimap B'$; functoriality follows from uniqueness. The adjunction bijection
\[ \mc M_{MO}(A \otimes X, B) \cong \mc M(A, X; B) \cong \mc M_{MO}(X, A \multimap B) \]
is the composite of the universal property of the tensor $u_{A,X}$ and the universal property of the evaluation $u^{A,B}$; naturality in $X$ and $B$ again follows from the uniqueness clauses. The right closure $(\--) \multimapinv A$ is obtained symmetrically from $U(() \mid B \mid A)$. Hence $\mc M_{MO}$ is a closed monoidal category, and a functor of bitensor multicategories induces a lax monoidal functor exactly as before (a lax monoidal functor between closed monoidal categories needs no additional data to be a morphism of $\mathbf{ClosedMonCat}$, as the comparison $F(A \multimap B) \to FA \multimap FB$ is definable from the laxator and evaluation).

Finally, the two natural isomorphisms $\mc F \circ \mc G \simeq \1$ and $\mc G \circ \mc F \simeq \1$ are those of \cref{thm:moncat-multicat-equivalence}: the isomorphisms $\eta_{\mc M}$ and $\varepsilon_{\mc M}$ constructed there are compatible with the additional structure, because on the reconstructed side the object $(\widehat{\otimes}\Gamma_1 \multimap B) \multimapinv \widehat{\otimes}\Gamma_2$ satisfies the universal property of $U(\Gamma_1 \mid B \mid \Gamma_2)$, and cartesian universal objects are unique up to unique isomorphism. The canonical isomorphisms $\theta_{\Gamma_i}$ from the \hyperref[anchor:iterated-chosen-tensor-comparison]{iterated-tensor comparison} compare these recursively built context tensors with the independently chosen $\otimes\Gamma_i$. $\square$

\section{Weakly Distributive Categories and Polycategorical Representability}\label{sec:polycategorical-representability}

\phantomsection\label{anchor:polycategorical-representability-roadmap}
Multicategories are a special case of polycategories, namely those whose polymaps have exactly one output. If monoidal categories represent tensor multicategories, then we hope to see a similar result for polycategories. The right objects to investigate are categories with two monoidal structures $\otimes, \oplus$, so that we can interpret the domain and codomain of polymaps correctly. We will use a similar realization: to get an equivalence, we will need a way to internalize lists in the domain and lists in the codomain using universal objects $\otimes \Gamma$ and $\oplus \Delta$. Finally, a bit of extra structure will be needed in the doubly monoidal category to interpret the cut operation of polycategories. These will be associativity and distributivity properties of the monoidal operations. The path forward for the next part is encapsulated in the following diagram:
\begin{center}
  \begin{tikzcd}[row sep=large, column sep=huge]
    \mathbf{WDC} \arrow[r, hook] & \mathbf{SymPolyCat} \\
    \mathbf{WDC} \arrow[r, "\simeq"] \arrow[u, equals]  & \mathbf{SymPolyCat_t} \arrow[u, hook ] \\
    \mathbf{\star\text{-}Aut} \arrow[r, "\simeq"] \arrow[u, hook ]  & \mathbf{SymPolyCat_{bi}} \arrow[u, hook ] \\
  \end{tikzcd}
\end{center}
\textit{Convention.}
We use symmetric polycategories in the sense of \cref{def:symmetric-polycategory}. Thus cuts are unrestricted, input and output contexts carry permutation transports, and cut is equivariant under those transports. Throughout this chapter, \emph{polycategory} means symmetric polycategory and \emph{polyfunctor} means a functor preserving the permutation transports. We write $\mathbf{SymPolyCat}$ for the category of polycategories with polyfunctors.

\subsection{The Underlying Polycategory}\label{subsec:underlying-symmetric-polycategory}

We start by defining the category that will allow us to speak of lists of objects in the domain and codomain. In \cref{subsec:underlying-multicategory}, a monoidal operation $\otimes$ with unit $\top$ was enough to make a list $\Gamma$ behave like the object $\otimes \Gamma$. We also saw there why the same monoidal operation is not enough for both sides of a polymap. We now add a monoidal operation $\oplus$ with unit $\bot$ to treat codomains. Thus we arrive at a notion of doubly monoidal category, usually called a \emph{weakly distributive category} \cite{CS97} (nowadays also called a \emph{linearly distributive category}, e.g.\ in \cite{Bla23}).

\begin{definition}\label{def:weakly-distributive-category}
  A \emph{weakly distributive category} (WDC) is a category $\C$ equipped with two monoidal structures $(\otimes, \top)$ and $(\oplus, \bot)$ together with two natural transformations, the \emph{linear distributors}
  \[ a^{\otimes\oplus}_{X,Y,Z} : X \otimes (Y \oplus Z) \to (X \otimes Y) \oplus Z, \qquad a^{\oplus\otimes}_{X,Y,Z} : (X \oplus Y) \otimes Z \to X \oplus (Y \otimes Z), \]
  subject to coherence laws (six pentagons and four triangles) making the two monoidal structures and the distributors compatible; we refer to \cite{CS97} for the complete list of diagrams. A WDC is \emph{symmetric} when both monoidal structures are symmetric, with symmetries $\sigma^\otimes, \sigma^\oplus$ compatible with the distributors.
\end{definition}

Throughout we assume both monoidal structures $(\otimes, \top)$ and $(\oplus, \bot)$ to be \emph{symmetric}. This guarantees that, besides the two basic linear distributors above, their symmetric variants
\[ d^{\otimes}_{X,Y,Z} : (X\oplus Y)\otimes Z \to (X\otimes Z)\oplus Y, \qquad d_{\otimes\,X,Y,Z} : X\otimes(Y\oplus Z) \to Y\oplus(X\otimes Z) \]
are available, each obtained by conjugating the corresponding basic distributor with the symmetry $\sigma^\oplus$:
\[ d^{\otimes}_{X,Y,Z} = \sigma^\oplus \circ a^{\oplus\otimes}_{Y,X,Z} \circ (\sigma^\oplus \otimes Z), \qquad d_{\otimes\,X,Y,Z} = \sigma^\oplus \circ a^{\otimes\oplus}_{X,Z,Y} \circ (X \otimes \sigma^\oplus). \]
The \hyperref[anchor:wdc-polycategory-cut-construction]{cut construction below} and its \hyperref[anchor:wdc-polycategory-cut-associativity-diagram]{associativity diagram} make essential use of the symmetric variants $d^{\otimes}$ and $d_{\otimes}$.

The \hyperref[anchor:multicategorical-associator-convention]{associator convention of the multicategory part} is retained here, now for both $\otimes$ and $\oplus$: lists appearing in tensor or par expressions denote their right-associated products, $\otimes() = \top$, $\oplus() = \bot$, singleton lists denote their object, and canonical coherence isomorphisms of either monoidal structure are denoted $a$ and used implicitly. When an outer list is empty, the corresponding factor is removed by the unitors; this convention is what makes the cut composite below specialize correctly to cuts with empty contexts, such as the identity laws.

The part of the \hyperref[anchor:polycategorical-representability-roadmap]{polycategorical roadmap} we are now focusing on is
\begin{center}
  \begin{tikzcd}[row sep=large, column sep=huge]
    \mathbf{WDC} \arrow[r, hook] & \mathbf{SymPolyCat}
  \end{tikzcd}
\end{center}

\begin{proposition}\label{prop:wdc-to-polycat}
  There is a fully faithful functor from the category of weakly distributive categories with Frobenius linear functors to the category of symmetric polycategories:
  \[\mathbf{WDC} \xto{\mc F} \mathbf{SymPolyCat}\]
\end{proposition}
\textit{Proof:} We need to specify an underlying polycategory for a given weakly distributive category $\C$. Let $\C_{poly}$ be the polycategory defined by:
\begin{itemize}
  \item Objects are the objects of $\C$.
  \item Polymaps $A_1, \ldots, A_n \to B_1, \ldots, B_m$ are maps $A_1 \otimes (\ldots (A_{n-1} \otimes A_n)\ldots) \to   B_1 \oplus (\ldots (B_{m-1} \oplus B_m)\ldots)$, i.e.\ $\C_{poly}(\Gamma; \Delta) := \C(\otimes\Gamma, \oplus\Delta)$.
  \item For a positional bijection $\sigma:[n]\to[n]$, let
  \[
    s^\otimes_{\Gamma,\sigma}:\otimes\Gamma_\sigma
    \xrightarrow{\cong}\otimes\Gamma
  \]
  be the canonical isomorphism assembled from the associators, unitors, and symmetries of $\otimes$, and define $s^\oplus_{\Delta,\tau}:\oplus\Delta_\tau\xrightarrow{\cong}\oplus\Delta$ similarly. For $f\in\C_{poly}(\Gamma_\sigma;\Delta_\tau)=\C(\otimes\Gamma_\sigma,\oplus\Delta_\tau),$
  define the permutation reindexing by
  \[
    f^{\sigma\mid\tau}
    :=s^\oplus_{\Delta,\tau}\circ f\circ
      \bigl(s^\otimes_{\Gamma,\sigma}\bigr)^{-1}
    \quad\in\quad\C_{poly}(\Gamma;\Delta).
  \]
  Symmetric monoidal coherence gives
  $
    s^\otimes_{\Gamma,\sigma\circ\sigma'}
      =s^\otimes_{\Gamma,\sigma}\circ
        s^\otimes_{\Gamma_\sigma,\sigma'},$ and
    $s^\oplus_{\Delta,\tau\circ\tau'}
      =s^\oplus_{\Delta,\tau}\circ
        s^\oplus_{\Delta_\tau,\tau'},
  $
  and the identity instances are identities. Consequently
  \[
    f^{\1\mid\1}=f,
    \qquad
    \bigl(f^{\sigma'\mid\tau'}\bigr)^{\sigma\mid\tau}
      =f^{\sigma\circ\sigma'\mid\tau\circ\tau'}.
  \]
  \item \phantomsection\label{anchor:wdc-polycategory-cut-construction}Given two polymaps $f : \Gamma \to \Delta_1, A, \Delta_2$ and $g : \Gamma_1, A, \Gamma_2 \to \Delta$, i.e.\ maps $f : \otimes\Gamma \to \Delta_1 \oplus (A \oplus \Delta_2)$ and $g : \Gamma_1 \otimes (A \otimes \Gamma_2) \to \oplus\Delta$, we define the polymap $g \circ_A f : \Gamma_1, \Gamma, \Gamma_2 \to \Delta_1, \Delta, \Delta_2$ by the composite:
\begin{align*}
  \otimes(\Gamma_1, \Gamma, \Gamma_2) &\xto{a} \Gamma_1 \otimes (\otimes\Gamma \otimes \Gamma_2)\\
	&\xto{\Gamma_1 \otimes (f \otimes \Gamma_2)} \Gamma_1 \otimes ((\Delta_1 \oplus (A \oplus \Delta_2)) \otimes \Gamma_2)\\
	&\xto{\Gamma_1 \otimes a^{\oplus\otimes}_{\Delta_1,\, A \oplus \Delta_2,\, \Gamma_2}} \Gamma_1 \otimes (\Delta_1 \oplus ((A \oplus \Delta_2) \otimes \Gamma_2)) \\
	&\xto{\Gamma_1 \otimes (\Delta_1 \oplus d^{\otimes}_{A,\Delta_2,\Gamma_2})} \Gamma_1 \otimes (\Delta_1 \oplus ((A \otimes \Gamma_2) \oplus \Delta_2)) \\
	&\xto{d_{\otimes\,\Gamma_1,\, \Delta_1,\, (A \otimes \Gamma_2) \oplus \Delta_2}} \Delta_1 \oplus (\Gamma_1 \otimes ((A \otimes \Gamma_2) \oplus \Delta_2)) \\
	&\xto{\Delta_1 \oplus a^{\otimes\oplus}_{\Gamma_1,\, A \otimes \Gamma_2,\, \Delta_2}} \Delta_1 \oplus ((\Gamma_1 \otimes (A \otimes \Gamma_2)) \oplus \Delta_2) \\
	&\xto{\Delta_1 \oplus (g \oplus \Delta_2)} \Delta_1 \oplus (\oplus\Delta \oplus \Delta_2) \xto{a} \oplus(\Delta_1, \Delta, \Delta_2)
\end{align*}
The key to this big composite, and what allows us to interpret the cut rule in our category, is the distributivity and mixed associativity properties of the monoidal structures.

The cut is equivariant under permutation reindexing. More precisely, for
\[
  f\in\C_{poly}\bigl(\Gamma_\sigma;(\Delta_1,A,\Delta_2)_\tau\bigr),
  \qquad
  g\in\C_{poly}\bigl((\Gamma_1,A,\Gamma_2)_\rho;\Delta_\upsilon\bigr),
\]
where the displayed cut occurrences have their unique preimages under the bijections $\tau$ and $\rho$, one has
\[
  \bigl(g^{\rho\mid\upsilon}\bigr)\circ_A
  \bigl(f^{\sigma\mid\tau}\bigr)
  =
  (g\circ_A f)^{\rho\star_A\sigma\mid\tau\star_A\upsilon}.
\]

	\item The identity polymap for an object $A$ is the map $\1_A : A \to A$ of the category.
  	Suppose $f : \Gamma \to \Delta_1, A, \Delta_2$. The identity respects the identity laws as shown in the diagram below:
	\begin{center}
\begin{tikzcd}[row sep=large, column sep=huge]
   \otimes\Gamma \arrow[rr, bend right=20, swap,  "f" ]   \arrow[r, "f"] & \Delta_1 \oplus (A \oplus \Delta_2)  \arrow[r, "\Delta_1 \oplus (\1_A \oplus \Delta_2)"] & \Delta_1 \oplus (A \oplus \Delta_2)
  \end{tikzcd}
  \end{center}
  The top composite is the cut $\1_A \circ_A f$. With $\Gamma_1 = \Gamma_2 = ()$, the distributor-unit triangles, after the canonical unitor identifications, reduce its structural part to the identity, leaving $(\Delta_1 \oplus (\1_A \oplus \Delta_2)) \circ f$. Remembering that $\Delta_1$ and $\Delta_2$ in arrow labels are shorthands for $\1_{\Delta_1},\1_{\Delta_2}$, we see that
\[\Delta_1 \oplus (\1_A \oplus \Delta_2) = \1_{\Delta_1} \oplus (\1_A \oplus \1_{\Delta_2}) = \1_{\Delta_1 \oplus (A \oplus \Delta_2)} \]
using functoriality of $\oplus$. The composite finalizes as $(\Delta_1 \oplus (\1_A \oplus \Delta_2)) \circ f = \1_{\Delta_1 \oplus (A \oplus \Delta_2)} \circ f = f$.\\
	For the second identity law, let $f : \Gamma_1, A, \Gamma_2 \to \Delta$. The cut $f \circ_A \1_A$ has $\Delta_1 = \Delta_2 = ()$; the dual distributor-unit triangles and unitors reduce its structural part to the identity, so it reduces to
\[f \circ (\Gamma_1 \otimes (\1_A \otimes \Gamma_2)) = f \circ \1_{\Gamma_1 \otimes (A \otimes \Gamma_2)} = f\]
using functoriality of $\otimes$.

	\item \phantomsection\label{anchor:wdc-polycategory-cut-associativity-diagram}\textbf{Associativity of the cut}: suppose we are given three polymaps:
	\begin{itemize}
  	\item $h: \Gamma \to \Lambda_1, B, \Lambda_2$, i.e.\ $h : \otimes\Gamma \to \Lambda_1 \oplus (B \oplus \Lambda_2)$
  	\item $f: \Gamma_1, B, \Gamma_2 \to \Delta_1, A, \Delta_2$, i.e.\ $f: \Gamma_1 \otimes (B \otimes \Gamma_2) \to \Delta_1 \oplus (A \oplus \Delta_2)$
  	\item $g: \Theta_1, A, \Theta_2 \to \Lambda$, i.e.\ $g: \Theta_1 \otimes (A \otimes \Theta_2) \to \oplus\Lambda$
	\end{itemize}
	We must show $(g \circ_A f) \circ_B h = g \circ_A (f \circ_B h)$, both polymaps $\Theta_1, \Gamma_1, \Gamma, \Gamma_2, \Theta_2 \to \Lambda_1, \Delta_1, \Lambda, \Delta_2, \Lambda_2$. Both composites begin with the same moves: apply $h$, then extract the blocks $\Lambda_1, \Lambda_2$ past $\Gamma_1, \Gamma_2$. They then diverge: in $g \circ_A (f \circ_B h)$ the map $f$ is applied next (left branch below), while in $(g \circ_A f) \circ_B h$ the blocks $\Lambda_1, \Lambda_2$ are first extracted past $\Theta_1, \Theta_2$ as well (right branch); the two branches reunite through four naturality parallelograms, and the composites share their remaining tail. We use the label $a$ for the associativity steps.
	\begin{center}
	\adjustbox{scale=0.62}{
  \begin{tikzcd}[row sep=large, column sep=tiny]
    & \Theta_1 \otimes ((\Gamma_1 \otimes (\otimes\Gamma \otimes \Gamma_2)) \otimes \Theta_2) \arrow[d, "\Theta_1 \otimes ((\Gamma_1 \otimes (h \otimes \Gamma_2)) \otimes \Theta_2)"]   & \\
    & \Theta_1 \otimes ((\Gamma_1 \otimes ((\Lambda_1 \oplus (B \oplus \Lambda_2)) \otimes \Gamma_2)) \otimes \Theta_2) \arrow[d, "\Theta_1 \otimes ((\Gamma_1 \otimes a^{\oplus\otimes}_{\Lambda_1,\, B \oplus \Lambda_2,\, \Gamma_2}) \otimes \Theta_2)"] & \\
		& \Theta_1 \otimes ((\Gamma_1 \otimes (\Lambda_1 \oplus ((B \oplus \Lambda_2) \otimes \Gamma_2))) \otimes \Theta_2) \arrow[d, "\Theta_1 \otimes ((\Gamma_1 \otimes (\Lambda_1 \oplus d^{\otimes}_{B, \Lambda_2, \Gamma_2})) \otimes \Theta_2)"] & \\
		& \Theta_1 \otimes ((\Gamma_1 \otimes (\Lambda_1 \oplus ((B \otimes \Gamma_2) \oplus \Lambda_2))) \otimes \Theta_2) \arrow[d, "\Theta_1 \otimes (d_{\otimes\,\Gamma_1,\, \Lambda_1,\, (B \otimes \Gamma_2) \oplus \Lambda_2} \otimes \Theta_2)"]  & \\
		& \Theta_1 \otimes ((\Lambda_1 \oplus (\Gamma_1 \otimes ((B \otimes \Gamma_2) \oplus \Lambda_2))) \otimes \Theta_2) \arrow[d, "\Theta_1 \otimes ((\Lambda_1 \oplus a^{\otimes\oplus}_{\Gamma_1,\, B \otimes \Gamma_2,\, \Lambda_2}) \otimes \Theta_2)"] & \\
		& \Theta_1 \otimes ((\Lambda_1 \oplus ((\Gamma_1 \otimes (B \otimes \Gamma_2)) \oplus \Lambda_2)) \otimes \Theta_2)  \arrow[dl, "\Theta_1 \otimes ((\Lambda_1 \oplus (f \oplus \Lambda_2)) \otimes \Theta_2)", swap]  \arrow[dr, "\Theta_1 \otimes a^{\oplus\otimes}_{\Lambda_1,\, (\Gamma_1 \otimes (B \otimes \Gamma_2)) \oplus \Lambda_2,\, \Theta_2}"]  & \\
		\Theta_1 \otimes ((\Lambda_1 \oplus ((\Delta_1 \oplus (A \oplus \Delta_2)) \oplus \Lambda_2)) \otimes \Theta_2) \arrow[d, "\Theta_1 \otimes a^{\oplus\otimes}_{\Lambda_1,\, (\Delta_1 \oplus (A \oplus \Delta_2)) \oplus \Lambda_2,\, \Theta_2}", swap] & & \Theta_1 \otimes (\Lambda_1 \oplus (((\Gamma_1 \otimes (B \otimes \Gamma_2)) \oplus \Lambda_2) \otimes \Theta_2)) \arrow[dll, "\Theta_1 \otimes (\Lambda_1 \oplus ((f \oplus \Lambda_2) \otimes \Theta_2))" description] \arrow[d, "\Theta_1 \otimes (\Lambda_1 \oplus d^{\otimes}_{\Gamma_1 \otimes (B \otimes \Gamma_2),\, \Lambda_2,\, \Theta_2})"]  \\
		\Theta_1 \otimes (\Lambda_1 \oplus (((\Delta_1 \oplus (A \oplus \Delta_2)) \oplus \Lambda_2) \otimes \Theta_2)) \arrow[d, "\Theta_1 \otimes (\Lambda_1 \oplus d^{\otimes}_{\Delta_1 \oplus (A \oplus \Delta_2),\, \Lambda_2,\, \Theta_2})", swap] & & \Theta_1 \otimes (\Lambda_1 \oplus (((\Gamma_1 \otimes (B \otimes \Gamma_2)) \otimes \Theta_2) \oplus \Lambda_2)) \arrow[dll, "\Theta_1 \otimes (\Lambda_1 \oplus ((f \otimes \Theta_2) \oplus \Lambda_2))" description] \arrow[d, "d_{\otimes\,\Theta_1,\, \Lambda_1,\, ((\Gamma_1 \otimes (B \otimes \Gamma_2)) \otimes \Theta_2) \oplus \Lambda_2}"]  \\
		\Theta_1 \otimes (\Lambda_1 \oplus (((\Delta_1 \oplus (A \oplus \Delta_2)) \otimes \Theta_2) \oplus \Lambda_2)) \arrow[d, "d_{\otimes\,\Theta_1,\, \Lambda_1,\, ((\Delta_1 \oplus (A \oplus \Delta_2)) \otimes \Theta_2) \oplus \Lambda_2}", swap] & & \Lambda_1 \oplus (\Theta_1 \otimes (((\Gamma_1 \otimes (B \otimes \Gamma_2)) \otimes \Theta_2) \oplus \Lambda_2)) \arrow[dll, "\Lambda_1 \oplus (\Theta_1 \otimes ((f \otimes \Theta_2) \oplus \Lambda_2))" description] \arrow[d, "\Lambda_1 \oplus a^{\otimes\oplus}_{\Theta_1,\, (\Gamma_1 \otimes (B \otimes \Gamma_2)) \otimes \Theta_2,\, \Lambda_2}"] \\
		\Lambda_1 \oplus (\Theta_1 \otimes (((\Delta_1 \oplus (A \oplus \Delta_2)) \otimes \Theta_2) \oplus \Lambda_2)) \arrow[d, "\Lambda_1 \oplus a^{\otimes\oplus}_{\Theta_1,\, (\Delta_1 \oplus (A \oplus \Delta_2)) \otimes \Theta_2,\, \Lambda_2}", swap] & & \Lambda_1 \oplus ((\Theta_1 \otimes ((\Gamma_1 \otimes (B \otimes \Gamma_2)) \otimes \Theta_2)) \oplus \Lambda_2) \arrow[dll, "\Lambda_1 \oplus ((\Theta_1 \otimes (f \otimes \Theta_2)) \oplus \Lambda_2)" description] \\
		\Lambda_1 \oplus ((\Theta_1 \otimes ((\Delta_1 \oplus (A \oplus \Delta_2)) \otimes \Theta_2)) \oplus \Lambda_2) \arrow[d, "\Lambda_1 \oplus ((\Theta_1 \otimes a^{\oplus\otimes}_{\Delta_1,\, A \oplus \Delta_2,\, \Theta_2}) \oplus \Lambda_2)", swap] & & \\
		\Lambda_1 \oplus ((\Theta_1 \otimes (\Delta_1 \oplus ((A \oplus \Delta_2) \otimes \Theta_2))) \oplus \Lambda_2) \arrow[d, "\Lambda_1 \oplus ((\Theta_1 \otimes (\Delta_1 \oplus d^{\otimes}_{A, \Delta_2, \Theta_2})) \oplus \Lambda_2)", swap] & & \\
		\Lambda_1 \oplus ((\Theta_1 \otimes (\Delta_1 \oplus ((A \otimes \Theta_2) \oplus \Delta_2))) \oplus \Lambda_2) \arrow[d, "\Lambda_1 \oplus (d_{\otimes\,\Theta_1,\, \Delta_1,\, (A \otimes \Theta_2) \oplus \Delta_2} \oplus \Lambda_2)", swap] & & \\
		\Lambda_1 \oplus ((\Delta_1 \oplus (\Theta_1 \otimes ((A \otimes \Theta_2) \oplus \Delta_2))) \oplus \Lambda_2) \arrow[d, "\Lambda_1 \oplus ((\Delta_1 \oplus a^{\otimes\oplus}_{\Theta_1,\, A \otimes \Theta_2,\, \Delta_2}) \oplus \Lambda_2)", swap] & & \\
		\Lambda_1 \oplus ((\Delta_1 \oplus ((\Theta_1 \otimes (A \otimes \Theta_2)) \oplus \Delta_2)) \oplus \Lambda_2) \arrow[d, "\Lambda_1 \oplus ((\Delta_1 \oplus (g \oplus \Delta_2)) \oplus \Lambda_2)", swap]   & & \\
		\Lambda_1 \oplus ((\Delta_1 \oplus (\oplus\Lambda \oplus \Delta_2)) \oplus \Lambda_2) \arrow[d, "a", swap]   & & \\
		\oplus(\Lambda_1, \Delta_1, \Lambda, \Delta_2, \Lambda_2)  & & \\
		\end{tikzcd}
	}
	\end{center}
	The middle of the diagram (parallelograms) commutes using naturality of $a^{\oplus\otimes}$, $d^{\otimes}$, $d_{\otimes}$ and $a^{\otimes\oplus}$.

  \item \phantomsection\label{anchor:wdc-polycategory-input-interchange-proof}\textbf{Interchange laws}: Let
  \[
    f:\Gamma\to\Delta_1,A,\Delta_2,
    \qquad
    g:\Gamma'\to\Delta'_1,B,\Delta'_2,
    \qquad
    h:\Sigma_1,A,\Sigma_2,B,\Sigma_3\to\Delta.
  \]
  Put
  \[
  \begin{aligned}
    S&:=(\Sigma_1,\Gamma,\Sigma_2,\Gamma',\Sigma_3),\\
    L_{\mathrm{out}}&:=(\Delta'_1,\Delta_1,\Delta,\Delta_2,\Delta'_2),\\
    R_{\mathrm{out}}&:=(\Delta_1,\Delta'_1,\Delta,\Delta'_2,\Delta_2).
  \end{aligned}
  \]
  Let $\tau$ be the permutation such that $(L_{\mathrm{out}})_\tau=R_{\mathrm{out}}$. The two cut orders form the diamond
  \begin{center}
  \adjustbox{max width=\textwidth}{
  \begin{tikzcd}[row sep=large, column sep=7em]
    & h
      \arrow[dl, bend right=8]
      \arrow[dr, bend left=8] & \\
    h\circ_A f
      \arrow[d ]
    &&
    h\circ_B g
      \arrow[d, ] \\
    (h\circ_A f)\circ_B g
    &&
    (h\circ_B g)\circ_A f
      \arrow[ll, dashed, "{(\--)^{\1\mid\tau}}"']
  \end{tikzcd}}
  \end{center}
  Before the dashed transport, the two bottom polymaps have types
  \[
  \begin{aligned}
    (h\circ_A f)\circ_B g &: S\longrightarrow L_{\mathrm{out}},\\
    (h\circ_B g)\circ_A f &: S\longrightarrow R_{\mathrm{out}},\\
    \bigl((h\circ_B g)\circ_A f\bigr)^{\1\mid\tau}
      &: S\longrightarrow L_{\mathrm{out}}.
  \end{aligned}
  \]
  Thus the input-interchange law says precisely that the diamond commutes:
  \[
    (h\circ_A f)\circ_B g
    =
    \bigl((h\circ_B g)\circ_A f\bigr)^{\1\mid\tau}.
  \]
  We now verify this commutativity. Abbreviate
  \[
  \begin{aligned}
    P&:=\oplus(\Delta_1,A,\Delta_2),
    &Q&:=\oplus(\Delta'_1,B,\Delta'_2),\\
    X_0&:=\otimes S,
    &X_f&:=\otimes(\Sigma_1,P,\Sigma_2,\Gamma',\Sigma_3),\\
    X_g&:=\otimes(\Sigma_1,\Gamma,\Sigma_2,Q,\Sigma_3),
    &X&:=\otimes(\Sigma_1,P,\Sigma_2,Q,\Sigma_3).
  \end{aligned}
  \]
  The two possible orders of applying $f$ and $g$ form the square
  \begin{center}
  \adjustbox{max width=\textwidth}{
  \begin{tikzcd}[row sep=large, column sep=huge]
    X_0
      \arrow[r, "{\otimes(\1,f,\1,\1,\1)}"]
      \arrow[d, "{\otimes(\1,\1,\1,g,\1)}"']
    & X_f
      \arrow[d, "{\otimes(\1,\1,\1,g,\1)}"] \\
    X_g
      \arrow[r, "{\otimes(\1,f,\1,\1,\1)}"']
    & X .
  \end{tikzcd}}
  \end{center}
  It commutes by functoriality of the tensor product, since both boundary composites are
  \[
  \begin{aligned}
    &\otimes(\1,\1,\1,g,\1)
       \circ\otimes(\1,f,\1,\1,\1)\\
    &\qquad=\otimes(\1,f,\1,g,\1)\\
    &\qquad=\otimes(\1,f,\1,\1,\1)
       \circ\otimes(\1,\1,\1,g,\1).
  \end{aligned}
  \]

  We make the structural arrows explicit. For lists $\Theta_1,\Theta_2,\Pi_1,\Pi_2$ and an object $D$, let
  \[
    \mathsf e^D_{\Theta_1;\Pi_1,\Pi_2;\Theta_2}:
    \otimes\bigl(\Theta_1,\oplus(\Pi_1,D,\Pi_2),\Theta_2\bigr)
    \longrightarrow
    \oplus\bigl(\Pi_1,\otimes(\Theta_1,D,\Theta_2),\Pi_2\bigr)
  \]
  denote the following composite, with the canonical reassociations at its two ends displayed as $a$:
  \[
  \begin{aligned}
  &\otimes\bigl(\Theta_1,\oplus(\Pi_1,D,\Pi_2),\Theta_2\bigr)\\
  &\xto{a}
    \Theta_1\otimes
      \bigl((\Pi_1\oplus(D\oplus\Pi_2))\otimes\Theta_2\bigr)\\
  &\xto{\Theta_1\otimes
      a^{\oplus\otimes}_{\Pi_1,\,D\oplus\Pi_2,\,\Theta_2}}
    \Theta_1\otimes
      \bigl(\Pi_1\oplus((D\oplus\Pi_2)\otimes\Theta_2)\bigr)\\
  &\xto{\Theta_1\otimes
      (\Pi_1\oplus d^{\otimes}_{D,\Pi_2,\Theta_2})}
    \Theta_1\otimes
      \bigl(\Pi_1\oplus((D\otimes\Theta_2)\oplus\Pi_2)\bigr)\\
  &\xto{d_{\otimes\,\Theta_1,\,\Pi_1,\,(D\otimes\Theta_2)\oplus\Pi_2}}
    \Pi_1\oplus
      \bigl(\Theta_1\otimes((D\otimes\Theta_2)\oplus\Pi_2)\bigr)\\
  &\xto{\Pi_1\oplus
      a^{\otimes\oplus}_{\Theta_1,\,D\otimes\Theta_2,\,\Pi_2}}
    \Pi_1\oplus
      \bigl((\Theta_1\otimes(D\otimes\Theta_2))\oplus\Pi_2\bigr)\\
  &\xto{a}
    \oplus\bigl(\Pi_1,\otimes(\Theta_1,D,\Theta_2),\Pi_2\bigr).
  \end{aligned}
  \]
  The list notation means the corresponding right-associated tensor or par. 

  Now set
  \[
  \begin{aligned}
    K&:=\otimes(\Sigma_1,A,\Sigma_2,B,\Sigma_3),\\
    X_A&:=\oplus\bigl(\Delta_1,
      \otimes(\Sigma_1,A,\Sigma_2,Q,\Sigma_3),\Delta_2\bigr),\\
    X_B&:=\oplus\bigl(\Delta'_1,
      \otimes(\Sigma_1,P,\Sigma_2,B,\Sigma_3),\Delta'_2\bigr),\\
    \widetilde L&:=(\Delta'_1,\Delta_1,K,\Delta_2,\Delta'_2),\\
    \widetilde R&:=(\Delta_1,\Delta'_1,K,\Delta'_2,\Delta_2),
  \end{aligned}
  \]
  and define
  \[
  \begin{aligned}
    E_A&:=\mathsf e^A_{\Sigma_1;\Delta_1,\Delta_2;
      (\Sigma_2,Q,\Sigma_3)}:X\to X_A,\\
    E_B&:=\mathsf e^B_{(\Sigma_1,P,\Sigma_2);
      \Delta'_1,\Delta'_2;\Sigma_3}:X\to X_B,\\
    e_{B\mid A}&:=\mathsf e^B_{(\Sigma_1,A,\Sigma_2);
      \Delta'_1,\Delta'_2;\Sigma_3},\\
    e_{A\mid B}&:=\mathsf e^A_{\Sigma_1;\Delta_1,\Delta_2;
      (\Sigma_2,B,\Sigma_3)}.
  \end{aligned}
  \]
  Applying the last two maps in the middle output positions gives
  \[
  \begin{aligned}
    F_B^A&:=\oplus(\1_{\Delta_1},e_{B\mid A},\1_{\Delta_2}):
      X_A\to\oplus\widetilde R,\\
    F_A^B&:=\oplus(\1_{\Delta'_1},e_{A\mid B},\1_{\Delta'_2}):
      X_B\to\oplus\widetilde L,
  \end{aligned}
  \]
  with canonical par reassociations understood. Let $\widetilde\tau$ be the permutation induced by $\tau$, so that $(\widetilde L)_{\widetilde\tau}=\widetilde R$. The structural part of the two cuts is
  \begin{center}
  \begin{tikzcd}[row sep=large, column sep=huge]
    & X
      \arrow[dl, "E_A"']
      \arrow[dr, "E_B"] & \\
    X_A
      \arrow[d, "F_B^A"']
    &&
    X_B
      \arrow[d, "F_A^B"] \\
    \oplus\widetilde R
      \arrow[rr, "s^\oplus_{\widetilde L,\widetilde\tau}"']
    &&
    \oplus\widetilde L .
  \end{tikzcd}
  \end{center}
  This diagram commutes by the WDC coherence equation
  \[
    F_A^B\circ E_B
    =s^\oplus_{\widetilde L,\widetilde\tau}
       \circ F_B^A\circ E_A:
       X\longrightarrow\oplus\widetilde L.
  \]
  Finally, define
  \[
  \begin{aligned}
    H_L&:=\oplus(\1_{\Delta'_1},\1_{\Delta_1},h,
      \1_{\Delta_2},\1_{\Delta'_2}):
      \oplus\widetilde L\to\oplus L_{\mathrm{out}},\\
    H_R&:=\oplus(\1_{\Delta_1},\1_{\Delta'_1},h,
      \1_{\Delta'_2},\1_{\Delta_2}):
      \oplus\widetilde R\to\oplus R_{\mathrm{out}}.
  \end{aligned}
  \]
  Naturality of the canonical output symmetry is exactly the commutativity of
  \begin{center}
  \begin{tikzcd}[row sep=large, column sep=huge]
    \oplus\widetilde R
      \arrow[r, "H_R"]
      \arrow[d, "s^\oplus_{\widetilde L,\widetilde\tau}"']
    & \oplus R_{\mathrm{out}}
      \arrow[d, "s^\oplus_{L_{\mathrm{out}},\tau}"] \\
    \oplus\widetilde L
      \arrow[r, "H_L"']
    & \oplus L_{\mathrm{out}} .
  \end{tikzcd}
  \end{center}
  Write $C:=\otimes(\1,f,\1,g,\1):X_0\to X$ for the common composite in the first square. The preceding two diagrams now give the required equality directly:
  \[
  \begin{aligned}
    (h\circ_A f)\circ_B g
      &=H_L\circ F_A^B\circ E_B\circ C\\
      &=H_L\circ s^\oplus_{\widetilde L,\widetilde\tau}
          \circ F_B^A\circ E_A\circ C\\
      &=s^\oplus_{L_{\mathrm{out}},\tau}\circ H_R
          \circ F_B^A\circ E_A\circ C\\
      &=\bigl((h\circ_B g)\circ_A f\bigr)^{\1\mid\tau}.
  \end{aligned}
  \]

  Dually, let
  \[
    f:\Gamma\to\Delta_1,A,\Delta_2,B,\Delta_3,
    \qquad
    g:\Gamma'_1,A,\Gamma'_2\to\Delta',
    \qquad
    h:\Gamma''_1,B,\Gamma''_2\to\Delta''.
  \]
  Put
  \[
  \begin{aligned}
    L_{\mathrm{in}}&:=(\Gamma''_1,\Gamma'_1,\Gamma,\Gamma'_2,\Gamma''_2),\\
    R_{\mathrm{in}}&:=(\Gamma'_1,\Gamma''_1,\Gamma,\Gamma''_2,\Gamma'_2),\\
    T&:=(\Delta_1,\Delta',\Delta_2,\Delta'',\Delta_3).
  \end{aligned}
  \]
  Let $\bar\sigma_{g,h}$ be the positional permutation of the tagged occurrences induced by the displayed block shuffle, order-preserving inside every block, such that $(L_{\mathrm{in}})_{\bar\sigma_{g,h}}=R_{\mathrm{in}}$. The dual cut-order diamond is
  \begin{center}
  \adjustbox{max width=\textwidth}{
  \begin{tikzcd}[row sep=large, column sep=7em]
    & f
      \arrow[dl, bend right=8]
      \arrow[dr, bend left=8] & \\
    g\circ_A f
      \arrow[d]
    &&
    h\circ_B f
      \arrow[d] \\
    h\circ_B(g\circ_A f)
    &&
    g\circ_A(h\circ_B f)
      \arrow[ll, dashed, "{(\--)^{\bar\sigma_{g,h}\mid\1}}"']
  \end{tikzcd}}
  \end{center}
  Here the bottom polymaps have types
  \[
  \begin{aligned}
    h\circ_B(g\circ_A f) &: L_{\mathrm{in}}\longrightarrow T,\\
    g\circ_A(h\circ_B f) &: R_{\mathrm{in}}\longrightarrow T,\\
    \bigl(g\circ_A(h\circ_B f)\bigr)^{\bar\sigma_{g,h}\mid\1}
      &: L_{\mathrm{in}}\longrightarrow T.
  \end{aligned}
  \]
  Hence output-interchange is the commuting-diamond equation
  \[
    h\circ_B(g\circ_A f)
    =
    \bigl(g\circ_A(h\circ_B f)\bigr)^{\bar\sigma_{g,h}\mid\1}.
  \]
  This time the two raw composites already have the same output list $T$, but their input blocks occur in the orders $L_{\mathrm{in}}$ and $R_{\mathrm{in}}$. Precomposition by
  \[
    \bigl(s^\otimes_{L_{\mathrm{in}},\bar\sigma_{g,h}}\bigr)^{-1}
  \]
  supplies the dashed input transport. Applying the \hyperref[anchor:wdc-polycategory-input-interchange-proof]{input-interchange argument} in the opposite category proves the equation.
\end{itemize}
This establishes the symmetric polycategory $\C_{poly}$, i.e.\ the action of $\mc F$ on objects. For the action on morphisms, suppose we are given two weakly distributive categories $\C, \mc D$ and a functor $F : \C \to \mc D$; we would like to construct a functor $F_{poly} : \C_{poly} \to \mc D_{poly}$ between the respective polycategories.
\begin{itemize}
  \item \textbf{Action on objects:} for $A \in \C_{poly}$ we define $F_{poly}(A) := FA$.
  \item \textbf{Action on polymaps:} a polymap $\Gamma \xto{f_{poly}} \Delta$ in $\C_{poly}$ is by definition a map $\bar f : \otimes \Gamma \to \oplus \Delta$ in $\C$, and its image must be a map $\otimes (F\Gamma) \to \oplus (F\Delta)$ in $\mc D$. We unfortunately cannot use $F\bar f : F(\otimes \Gamma) \to F(\oplus \Delta)$ directly; instead we precompose with the $n$-ary lax comparison $m_\Gamma : \otimes (F\Gamma) \to F(\otimes \Gamma)$ and postcompose with the $n$-ary oplax comparison $n_\Delta : F(\oplus \Delta) \to \oplus (F\Delta)$, and set \[F_{poly}(f_{poly}) := \big(n_\Delta \circ F\bar f \circ m_\Gamma\big)_{poly} : \otimes (F\Gamma) \to \oplus (F\Delta).\]
\end{itemize}
Compatibility of the comparisons with the symmetries means, for every pair of positional bijections $\sigma,\tau$,
\[
  F(s^\otimes_{\Gamma,\sigma})\circ m_{\Gamma_\sigma}
    =m_\Gamma\circ s^\otimes_{F\Gamma,\sigma},
  \qquad
  s^\oplus_{F\Delta,\tau}\circ n_{\Delta_\tau}
    =n_\Delta\circ F(s^\oplus_{\Delta,\tau}).
\]
Using these two equations in the definition above gives
\[
  F_{poly}\bigl(f^{\sigma\mid\tau}\bigr)
  =F_{poly}(f)^{\sigma\mid\tau},
\]
so $F_{poly}$ preserves the permutation reindexings.
For $F_{poly}$ to preserve identities we need $m, n$ to be compatible with the unitors, and for $F_{poly}$ to preserve the cut the comparisons $m_\Gamma, n_\Delta$ must moreover cohere with the linear distributors, namely make the following hexagons commute for all objects $X, Y, Z$:
\begin{center}
  \begin{tikzcd}[row sep=large, column sep=large]
    FX \otimes F(Y \oplus Z) \arrow[r, "FX \otimes n_{Y,Z}"] \arrow[d, "m_{X,\, Y\oplus Z}", swap] & FX \otimes (FY \oplus FZ) \arrow[r, "a^{\otimes\oplus}"] & (FX \otimes FY) \oplus FZ \arrow[d, "m_{X,Y} \oplus FZ"] \\
    F(X \otimes (Y \oplus Z)) \arrow[r, "F(a^{\otimes\oplus})", swap] & F((X \otimes Y) \oplus Z) \arrow[r, "n_{X \otimes Y,\, Z}", swap] & F(X \otimes Y) \oplus FZ \\
  \end{tikzcd}
\end{center}
together with the mirror hexagon for $a^{\oplus\otimes}$ (and similarly for the symmetric variants, which follows if $m, n$ are compatible with the symmetries). A single functor $F$ equipped with such a pair $(m, n)$ is called a \emph{Frobenius linear functor} \cite{CS99,Bla23}. After a cut is expanded, naturality moves the comparisons through the component maps, the two Frobenius hexagons move them through the distributor stages, and lax/oplax monoidal coherence combines the comparisons at the two boundaries. Thus
\[
  F_{poly}(g\circ_A f)=F_{poly}(g)\circ_{FA}F_{poly}(f),
\]
while singleton normalization $m_{(A)}=n_{(A)}=\1_{FA}$ gives preservation of identities; this is the biased form of \cite[Proposition~2.3.2]{Bla23}. The composite of Frobenius linear functors has comparisons
\[
  m^{KF}_\Gamma=K(m^F_\Gamma)\circ m^K_{F\Gamma},
  \qquad
  n^{KF}_\Delta=n^K_{F\Delta}\circ K(n^F_\Delta),
\]
so substitution in the definition of $F_{poly}$ proves $(KF)_{poly}=K_{poly}F_{poly}$; identities are immediate. Hence this assignment is a functor.

For the proof of full faithfulness, see \cite[Proposition~2.3.12]{Bla23}.

\subsection{Two-Tensor Polycategories}\label{subsec:two-tensor-polycategories}

At this point the functor being built is not essentially surjective: there are polycategories that are not isomorphic to the image of any weakly distributive category, for exactly the same reason as before: the data contained in a polycategory is strictly larger than that of a weakly distributive category. The polycategories that arise from a WDC need a little more structure. This is where the notion of a two-tensor polycategory comes to shine: in those, for any lists of objects $\Gamma$ and $\Delta$ there are universal objects $\otimes \Gamma$ and $\oplus \Delta$ which internalize the inputs and outputs respectively. This allows us to only consider polymaps with one input and one output, collapsing the cut to standard composition and providing essential surjectivity of the functor.

\begin{definition}\label{def:tensor-par-universal}
  Let $\Gamma, \Delta$ be lists of objects in a polycategory $\mc P$.
  \begin{itemize}
    \item A \emph{tensor product} of $\Gamma$ is an object $\otimes\Gamma$ equipped with a polymap $u_\Gamma : \Gamma \to \otimes\Gamma$ such that for all lists $\Gamma_1, \Gamma_2, \Delta$ and every polymap $f : \Gamma_1, \Gamma, \Gamma_2 \to \Delta$ there is a unique polymap $u_\Gamma^f : \Gamma_1, \otimes\Gamma, \Gamma_2 \to \Delta$ with $u_\Gamma^f \circ_{\otimes\Gamma} u_\Gamma = f$.
    \item A \emph{par product} of $\Delta$ is an object $\oplus\Delta$ equipped with a polymap $u^\Delta : \oplus\Delta \to \Delta$ such that for all lists $\Gamma, \Delta_1, \Delta_2$ and every polymap $f : \Gamma \to \Delta_1, \Delta, \Delta_2$ there is a unique polymap $u^\Delta_f : \Gamma \to \Delta_1, \oplus\Delta, \Delta_2$ with $u^\Delta \circ_{\oplus\Delta} u^\Delta_f = f$.
  \end{itemize}
  A \emph{two-tensor polycategory} is a symmetric polycategory equipped with a choice of tensor product $u_\Gamma$ and par product $u^\Delta$ for all finite lists $\Gamma, \Delta$. We write $\top := \otimes()$ and $\bot := \oplus()$ for the tensor and par of the empty list. The category of two-tensor polycategories and all functors of symmetric polycategories is denoted $\mathbf{SymPolyCat_t}$.
\end{definition}

\begin{lemma}\label{lem:identities-universal}
  In any polycategory, identities are both tensor- and par-universal for their singleton list, tensor-universal polymaps are closed under composition, and par-universal polymaps are closed under composition.
\end{lemma}
\textit{Proof:} The same successive-factorization argument as in \cref{lem:universal-composition} applies; cf.\ also \cite[Proposition~2.4.4]{Bla23}. $\square$

\begin{lemma}[flattening test]\label{lem:flattening-test}
  Let $v:\Gamma\to X$ be tensor-universal and let $w:Y\to\Delta$ be par-universal. For unary co-unary polymaps $r,s:X\to Y$, one has
  \[
    r=s
    \quad\Longleftrightarrow\quad
    w\circ_Y(r\circ_Xv)=w\circ_Y(s\circ_Xv).
  \]
  The same statement applies when $v$ or $w$ is an iterated compatible composite of universal polymaps.
\end{lemma}
\textit{Proof:} Precomposition with $v$ and postcomposition with $w$ are bijections by the two universal properties, hence their composite is injective. Iterated compatible composites are universal by \cref{lem:identities-universal}. $\square$

We now describe the underlying WDC of a two-tensor polycategory. Suppose $\mc P$ is a two-tensor polycategory; we define $\mc P_w$ to be the WDC specified as follows:

\begin{itemize}
	\item \textbf{Objects:} the objects of $\mc P_w$ are those of $\mc P$.
	\item \textbf{Morphisms:} the morphisms of $\mc P_w$ are the polymaps of $\mc P$ with exactly one input and one output: $\mc P_w(A, B) := \mc P(A;B)$.
	\item \textbf{Composition:} given morphisms $f : A \to B$ and $g : B \to C$ we define $g \circ f := g \circ_{B} f$. There is no ambiguity about which cut to perform, as domain and codomain consist of single objects; unit and associativity laws are inherited from the polycategory axioms.
	\item \textbf{Monoidal operations:} on objects, $A \otimes_w B := \otimes(A,B)$ and $A \oplus_w B := \oplus(A,B)$, with units $\top = \otimes()$ and $\bot = \oplus()$, using the chosen universal objects.
\end{itemize}
It remains to show that this specification defines a weakly distributive category: we must produce the functorial actions of $\otimes_w, \oplus_w$, the associators, unitors and symmetries of both structures, and the distributors, and verify the coherence laws.

\begin{itemize}
  \item \textbf{Functoriality of $\otimes_w, \oplus_w$:}
  	we already defined the action on objects; for morphisms $f : A \to B$ and $g : C \to D$ define
  	\[ f \otimes_w g := u_{A,C}^{(u_{B,D} \circ_D g) \circ_B f} : A \otimes C \to B \otimes D, \]
		the unique factorization through $u_{A,C}$ of the composite polymap displayed on the left below, and dually $f \oplus_w g := \big(u^{B,D}\big)_{g \circ_C (f \circ_A u^{A,C})}$ as on the right:
  	\begin{center}
    \begin{tikzcd}[row sep=large, column sep=huge]
      	A, C  \arrow[d, "u_{A,C}", swap] \arrow[rr, "(u_{B,D} \circ_D g) \circ_B f"]     &  &  B, D \arrow[d, "u_{B,D}"]  \\
      	A \otimes C \arrow[rr, dashed, "u_{A,C}^{(u_{B,D} \circ_D g) \circ_B f} ", swap]  & & B \otimes D
    	\end{tikzcd}
    	\quad
    \begin{tikzcd}[row sep=large, column sep=huge]
      A \oplus C \arrow[rr, dashed, "u^{B,D}_{g \circ_C (f \circ_A u^{A,C})}"] \arrow[d, "u^{A,C}", swap]  & & B \oplus D \arrow[d, "u^{B,D}"] \\
      A, C \arrow[rr, "g \circ_C (f \circ_A u^{A,C})", swap]  & & B, D
     	\end{tikzcd}
  	\end{center}
  	Note that $\1_A \otimes_w \1_B = u_{A,B}^{(u_{A,B} \circ_B \1_B) \circ_A \1_A} = u_{A,B}^{u_{A,B}} = \1_{A \otimes B}$, by the uniqueness of the factorization of $u_{A,B}$ through itself.
    For composable $f, f'$ and $g, g'$, both $(f' \otimes_w g') \circ (f \otimes_w g)$ and $(f' \circ f) \otimes_w (g' \circ g)$ factor the polymap $(u_{B',D'} \circ_{D'} (g' \circ g)) \circ_{B'} (f' \circ f) : A, C \to B' \otimes D'$ through $u_{A,C}$ — the first by pasting the two defining squares, using the interchange law of $\mc P$ to commute the cuts of $f$-side and $g$-side maps — hence they are equal by uniqueness. The argument for $\oplus_w$ is dual.

  \item \textbf{Associators and unitors:} the composites $u_{A \otimes B, C} \circ_{A \otimes B} u_{A,B}$ and $u_{A, B \otimes C} \circ_{B \otimes C} u_{B,C}$ are both tensor-universal polymaps $A, B, C \to \cdot$ by \cref{lem:identities-universal}, with codomains $(A \otimes B) \otimes C$ and $A \otimes (B \otimes C)$ respectively. We define $a_{\otimes(A,B,C)} : (A \otimes B) \otimes C \to A \otimes (B \otimes C)$ by factoring the second through the first, in two steps:
\begin{center}
    \begin{tikzcd}[row sep=large, column sep=2in]
       A, B, C \arrow[d, "u_{A,B}", swap] \arrow[ddr, bend left=60, "u_{A, B \otimes C} \circ_{B \otimes C} u_{B, C} "] &   \\
      A \otimes B, C \arrow[d, "u_{A \otimes B, C}", swap] \arrow[dr, dashed, "u_{A,B}^{\,u_{A, B \otimes C} \circ_{B \otimes C} u_{B, C}} "] \\
      (A \otimes B) \otimes C \arrow[r, dashed, swap, "a_{\otimes(A,B,C)}"] & A \otimes (B \otimes C)
  	\end{tikzcd}
	\end{center}
	In human terms: to produce a map $(A \otimes B) \otimes C \to A \otimes (B \otimes C)$ it suffices, by the universal property of $u_{A \otimes B, C}$, to provide a map $A \otimes B, C \to A \otimes (B \otimes C)$ (the lower triangle), and for that it suffices, by the universal property of $u_{A,B}$, to provide a map $A, B, C \to A \otimes (B \otimes C)$ (the upper triangle), for which we take the composite universal map. The inverse map is obtained by the mirrored construction, factoring $u_{A \otimes B, C} \circ_{A \otimes B} u_{A,B}$ through $u_{B,C}$ and then $u_{A, B \otimes C}$. Pasting the two constructions,
\begin{center}
    \begin{tikzcd}[row sep=large, column sep=4em]
       & A, B, C \arrow[dr, "u_{B,C}"] \arrow[dl, swap, "u_{B,C}"] \arrow[d, "u_{A,B}"] & \\
			A, B \otimes C \arrow[d, "u_{A, B \otimes C}", swap] & A\otimes B, C \arrow[d, "u_{A \otimes B, C}"] & A, B \otimes C \arrow[d, "u_{A, B \otimes C}"] \\
			A \otimes (B \otimes C) \arrow[r, "g", swap] & (A \otimes B) \otimes C \arrow[r, "a_{\otimes(A,B,C)}", swap]  & A \otimes (B \otimes C)
  	\end{tikzcd}
	\end{center}
	where $g$ is the inverse candidate: the left and right regions commute by construction, so the outer composite satisfies
\[a_{\otimes(A,B,C)} \circ g \circ (u_{A, B \otimes C} \circ_{B \otimes C} u_{B,C}) = u_{A, B \otimes C} \circ_{B \otimes C} u_{B,C} = \1_{A \otimes (B \otimes C) } \circ (u_{A, B \otimes C} \circ_{B \otimes C} u_{B,C})\]
	and since precomposition with the composite universal map is injective, $a_{\otimes(A,B,C)} \circ g =\1_{A \otimes (B \otimes C)}$; symmetrically $g \circ a_{\otimes(A,B,C)} = \1_{(A \otimes B) \otimes C}$, which shows that $a_{\otimes(A,B,C)}$ is an isomorphism. Naturality in $A, B, C$ follows from uniqueness: both composites around a naturality square factor one and the same polymap $A,B,C \to A' \otimes (B' \otimes C')$.
	For the unitors, universality of $u_{()} : () \to \top$ gives, for any $f : \Gamma_1, \Gamma_2 \to \Delta$, a unique $u_{()}^f : \Gamma_1, \top, \Gamma_2 \to \Delta$ with $u_{()}^f \circ_\top u_{()} = f$. Taking $f = \1_A : A \to A$ (with $\Gamma_1 = (A)$, $\Gamma_2 = ()$) gives $u_{()}^{\1_A} : A, \top \to A$, and factoring through $u_{A, \top}$ yields $\rho_A : A \otimes \top \to A$; its inverse candidate is the composite $u_{A,\top} \circ_{\top} u_{()} : A \to A \otimes \top$, a one-input one-output polymap and hence a morphism of $\mc P_w$, and the two are mutually inverse by the uniqueness clauses, exactly as for the associator. The left unitor $\lambda_A : \top \otimes A \to A$ is constructed symmetrically. Dually, the par unitors are
  \[
    \rho_A^\oplus:=u^{()}\circ_\bot u^{A,\bot}:A\oplus\bot\to A,
    \qquad
    \lambda_A^\oplus:=u^{()}\circ_\bot u^{\bot,A}:\bot\oplus A\to A.
  \]
  Their inverse candidates and the monoidal triangle laws follow by the same uniqueness argument: both legs factor, through the same composite universal map, one and the same polymap built from the nullary and binary universals.

  The two unit coherences are worth displaying. All triangles below commute by the universal properties of the (op)cartesian maps, which is what it means for $\top$ to behave correctly with respect to $\otimes$, and dually for $\bot$ with respect to $\oplus$:

  \begin{center}
    \begin{tikzcd}[row sep=large, column sep=1.5em]
			A \otimes (\top \otimes B) \arrow[ddrr, ""]
			& A, (\top \otimes B) \arrow[ddr, bend right=10, ""] \arrow[l, ""]
			& \underset{\top}{A  , \top, B} \arrow[dd, bend right=30, ""] \arrow[dd, bend left=30, ""] \arrow[l, ""] \arrow[r, ""]  
			& A \otimes \top, B \arrow[ddl, bend left=10, ""] \arrow[r, ""] & (A \otimes \top)\otimes B \arrow[ddll, ""] \\
      & & \overset{()}{A ,B}  \arrow[d, ""] \arrow[u, ""]   & & \\
      & & A \otimes B & &
    \end{tikzcd}
  \end{center}

  \begin{center}
    \begin{tikzcd}[row sep=large, column sep=1.5em]
			A \oplus (\bot \oplus B)  \arrow[r, ""] 
			& A, (\bot \oplus B)  \arrow[r, ""]
			& \underset{\bot}{A  , \bot, B}    \arrow[d, ""]  
			& A \oplus \bot, B \arrow[l, ""]   & (A \oplus \bot)\oplus B \arrow[l, ""]  \\
      & & \overset{()}{A ,B}      & & \\
      & & A \oplus B \arrow[u, ""] \arrow[uu, bend right=30, ""] \arrow[uu, bend left=30, ""] \arrow[uur, ""]\arrow[uul, ""]\arrow[uurr, ""]\arrow[uull, ""] &  &
    \end{tikzcd}
  \end{center}
  \item \textbf{Symmetries:} since $\mc P$ is a symmetric polycategory, precomposing $u_{A,B}$ with the exchange gives a tensor-universal polymap $B, A \to A \otimes B$, so $A \otimes B$ satisfies the universal property of $B \otimes A$ and we obtain the canonical isomorphism $\sigma^{\otimes}_{A,B} : A \otimes B \to B \otimes A$; dually for $\sigma^\oplus$. Coherence with associators and unitors is again by uniqueness.
  \item \textbf{Distributors:} we seek a map
		\[ a^{\otimes\oplus}_{A,B,C} : A \otimes_w (B \oplus_w C) \longrightarrow (A \otimes_w B) \oplus_w C, \]
		that is, a one-input one-output polymap $\otimes(A, \oplus(B,C)) \to \oplus(\otimes(A,B), C)$ in $\mc P$. Cutting the universal maps $u^{B,C} : B \oplus C \to B, C$ and $u_{A,B} : A, B \to A \otimes B$ along $B$ gives the polymap
		\[ h := u_{A,B} \circ_B u^{B,C} : A,\, B \oplus C \longrightarrow A \otimes B,\, C. \]
		Factoring the input list of $h$ through the universal map $u_{A, B\oplus C} : A, B\oplus C \to A\otimes(B\oplus C)$ gives $h' := u_{A, B\oplus C}^{\,h} : A\otimes(B\oplus C) \to A \otimes B,\, C$, and factoring the output list of $h'$ through the universal map $u^{A\otimes B, C} : (A\otimes B)\oplus C \to A\otimes B, C$ gives the desired distributor
		\[ a^{\otimes\oplus}_{A,B,C} := \big(u^{A\otimes B,\, C}\big)_{h'} : A\otimes(B\oplus C) \longrightarrow (A\otimes B)\oplus C. \]
		The remaining basic distributor $a^{\oplus\otimes}$ is produced identically. We define the symmetric variants by the conjugation formulas at the start of this section; exchange equivariance shows equivalently that directly factoring the exchanged cut cores produces those same conjugates.

		That direct factorization is the content of the following two diagrams, which produce $d_{\otimes} : A\otimes(B\oplus C)\to B\oplus(A\otimes C)$ and $d^{\otimes} : (B\oplus C)\otimes A\to (B\otimes A)\oplus C$ from the exchanged cut core $u_{A,C}\circ_C u^{B,C}$, by factoring its input list through the tensor universal and its output list through the par universal:

  \begin{center}
    \begin{tikzcd}[row sep=large, column sep=huge]
      A \otimes (B \oplus C) \arrow[rr, "u_{A, B \oplus C}^{u^{B, A \otimes C}_{u_{A,C} \circ_C u^{B,C} } } "] & & B \oplus (A \otimes C) \arrow[d, "u^{B, A \otimes C}"]  \\
      \overset{A, B \oplus C}{B \oplus C} \arrow[rr, bend left=10, yshift=5, "u_{A,C} \circ_C u^{B,C} ", swap]   \arrow[u, "u_{A, B \oplus C}"] \arrow[urr, yshift=5, "u^{B, A \otimes C}_{u_{A,C} \circ_C u^{B,C} } "] \arrow[r, "u^{B,C}", swap] & B, C \quad A, C \arrow[r, "u_{A,C}", swap] & \overset{B, A \otimes C}{A \otimes C} 
    \end{tikzcd}
  \end{center}

  \begin{center}
    \begin{tikzcd}[row sep=large, column sep=huge]
       (B \oplus C) \otimes A \arrow[rr, "u^{u_{u_{A,C} \circ_C u^{B,C}}^{B \otimes A, C}}_{B \oplus C, A}"] & & (B \otimes A) \oplus C \arrow[d, "u^{B \otimes A, C}"]  \\
      \overset{ B \oplus C, A}{B \oplus C} \arrow[rr, bend left=10, yshift=5, "u_{A,C} \circ_C u^{B,C} ", swap]   \arrow[u, "u_{B \oplus C, A}"] \arrow[urr, yshift=5, "u_{u_{A,C} \circ_C u^{B,C}}^{B \otimes A, C}  "] \arrow[r, "u^{B,C}", swap] & B, C \quad B, A \arrow[r, "u_{B,A}", swap] & \overset{B \otimes A, C}{B \otimes A} 
    \end{tikzcd}
  \end{center}

    It remains to check the mixed axioms. Expanding the definition of each distributor, and then applying \cref{lem:flattening-test} at the outer tensor source and par target, reduces the two paths in the six mixed pentagons to the following six polymaps, listed in the standard order of the connectives:
    \begin{align*}
      \oplus\otimes\otimes:\quad&
        (u_{B,C\otimes D}\circ_{C\otimes D}u_{C,D})
          \circ_Bu^{A,B},\\
      \otimes\otimes\oplus:\quad&
        (u_{A\otimes B,C}\circ_{A\otimes B}u_{A,B})
          \circ_Cu^{C,D},\\
      \otimes\oplus\oplus:\quad&
        u^{C,D}\circ_{C\oplus D}
          (u_{A,B}\circ_Bu^{B,C\oplus D}),\\
      \oplus\oplus\otimes:\quad&
        u^{A,B}\circ_{A\oplus B}
          (u_{C,D}\circ_Cu^{A\oplus B,C}),\\
      \otimes\oplus\otimes:\quad&
        u_{C,D}\circ_C(u_{A,B}\circ_Bu^{B,C}),\\
      \oplus\otimes\oplus:\quad&
        (u_{B,C}\circ_Cu^{C,D})\circ_Bu^{A,B}.
    \end{align*}
    Their respective boundaries are
    \begin{gather*}
      A\oplus B,C,D\to A,B\otimes(C\otimes D),\qquad
      A,B,C\oplus D\to (A\otimes B)\otimes C,D,\\
      A,B\oplus(C\oplus D)\to A\otimes B,C,D,\qquad
      (A\oplus B)\oplus C,D\to A,B,C\otimes D,\\
      A,B\oplus C,D\to A\otimes B,C\otimes D,\qquad
      A\oplus B,C\oplus D\to A,B\otimes C,D.
    \end{gather*}
    For example, the $\oplus\otimes\otimes$ pentagon starts at $((A\oplus B)\otimes C)\otimes D$ and ends at $A\oplus(B\otimes(C\otimes D))$; precomposing with its iterated tensor universal and postcomposing with its par universal turns either path into the first displayed polymap. The other five expansions are the same cancellation, with associativity and interchange of cuts supplying the displayed parenthesizations. Thus the two paths in every mixed pentagon agree by injectivity, rather than merely by having common endpoints.

    Finally, insert the nullary universals $u_{()}:()\to\top$ and $u^{()}:\bot\to()$ in the four mixed unit triangles. In the standard order of the four triangles, their two paths flatten respectively to
    \[
      u^{A,B},\qquad
      u^{A,B},\qquad
      u_{A,B}\circ_B\rho_B^\oplus,\qquad
      u_{A,B}\circ_A\lambda_A^\oplus.
    \]
    This completes the construction of a symmetric WDC. Its planar unbiased antecedent is \cite[Proposition~2.3.8]{Bla23}, while the strong-universal restriction is treated in \cite[Definitions~2.3.13--2.3.14 and Proposition~2.3.15]{Bla23}; the biased and symmetric checks are the explicit calculations above.
\end{itemize}
This completely defines the underlying weakly distributive category of a two-tensor polycategory. Again this has been possible because of the existence of universal objects $\otimes \Gamma$ and $\oplus \Delta$, so that the cut becomes ``deterministic'' in the sense that it collapses to a cut on only one object, which is what is needed to get a category with a standard composition operation.

\textit{Remark:}
  In a two-tensor polycategory there is a bijective correspondence between polymaps $\Gamma \to \Delta$ and maps $\otimes \Gamma \to \oplus \Delta$, obtained by factoring through $u_\Gamma$ and $u^\Delta$. This correspondence identifies the cut along the distinguished objects with composition in $\mc P_w$, but a cut along an object \emph{inside} $\Gamma$ or $\Delta$ corresponds on the categorical side to the composite built from distributors; for this reason the comparison of \cref{thm:wdc-sympolycat-equivalence} is an equivalence of categories and not an isomorphism, while its sharpest form is $2$-categorical \cite[Proposition~2.3.15]{Bla23}.

The underlying polycategory construction of \cref{prop:wdc-to-polycat} extends to land in two-tensor polycategories. The objects and polymaps are unchanged; the only remaining specification is the choice of $\otimes \Gamma$, $\oplus \Delta$ and of the universal maps $\Gamma \to \otimes \Gamma$ and $\oplus \Delta \to \Delta$. Fortunately, this is a straightforward task, as the symbols we used are very suggestive: we take
\begin{align*}
	\otimes \Gamma &:= \Gamma_1 \otimes (\Gamma_2 \otimes (\ldots (\Gamma_{n-1} \otimes \Gamma_n)\ldots )) \\
	\oplus \Delta &:= \Delta_1 \oplus (\Delta_2 \oplus (\ldots (\Delta_{m-1} \oplus \Delta_m)\ldots ))
\end{align*}
and we claim the identity maps $u_\Gamma := \1_{\otimes \Gamma}$ and $u^\Delta := \1_{\oplus \Delta}$, viewed as polymaps $\Gamma\to \otimes \Gamma $ and $\oplus \Delta\to \Delta$, are universal. We show the universal property for $\otimes$ only, since the proof for $\oplus$ is dual. Suppose we are given a polymap $f : \Gamma_1, \Gamma, \Gamma_2 \to \Delta$, i.e.\ a map $f : \otimes(\Gamma_1,\Gamma,\Gamma_2) \to \oplus\Delta$ of $\C$; we take
\[ u_\Gamma^f := f \circ a^{-1} : \otimes(\Gamma_1, \otimes\Gamma, \Gamma_2) \to \oplus\Delta \]
where $a$ is the canonical isomorphism, exactly as in the multicategory case: unfolding the cut composite, $u_\Gamma^f \circ_{\otimes\Gamma} u_\Gamma = f \circ a^{-1} \circ (\Gamma_1 \otimes (\1_{\otimes\Gamma} \otimes \Gamma_2)) \circ a = f$, and uniqueness holds because $- \circ a$ is a bijection.
This was the only additional structure we needed to exhibit. Now we are able to do the following:
 \begin{itemize}
   \item From a weakly distributive category $\C$, construct the underlying two-tensor polycategory $\C_{poly}$ (with the above choices).
   \item From a two-tensor polycategory $\mc P$, construct the underlying weakly distributive category $\mc P_w$.
 \end{itemize}
We therefore have two functors $\mc F : \mathbf{WDC} \to \mathbf{SymPolyCat_t}$ and $\mc G : \mathbf{SymPolyCat_t} \to \mathbf{WDC}$, where $\mc F$ is the functor of \cref{prop:wdc-to-polycat} (with the tensor choices above) and the action of $\mc G$ on morphisms is defined as follows. Recall that a morphism in $\mathbf{WDC}$ is a Frobenius linear functor, while a morphism of $\mathbf{SymPolyCat_t}$ is an arbitrary functor of symmetric polycategories.
Suppose we are given a morphism $G : \mc P \to \mc Q$ of $\mathbf{SymPolyCat_t}$, i.e.\ an arbitrary symmetry-preserving polyfunctor between the underlying symmetric polycategories; we define the Frobenius linear functor $G_w : \mc P_w \to \mc Q_w$ in the following way:
\begin{itemize}
  \item \textbf{Action on objects:} given $A \in \mc P_w$ set $G_w(A) := GA$.
  \item \textbf{Action on morphisms:} given $g : A \to B$ in $\mc P_w$ set $G_w(g) := Gg$.
  \item \textbf{Comparisons:} the lax comparison $m_{A,B} : G_wA \otimes G_wB \to G_w(A \otimes B)$ is the factorization of the polymap $G(u_{A,B}) : GA, GB \to G(A \otimes B)$ through the universal map $u_{GA, GB}$, and the oplax comparison $n_{A,B} : G_w(A \oplus B) \to G_wA \oplus G_wB$ is the factorization of $G(u^{A,B}) : G(A \oplus B) \to GA, GB$ through $u^{GA,GB}$. The same argument applies to the units.
\end{itemize}
Note that $G_w(1_{A}) = G\1_A = \1_{GA}$ and $G_w(f \circ g) = G(f \circ_B g) = Gf \circ_{GB} Gg = G_w(f) \circ G_w(g)$, so $G_w$ is a functor. 

For naturality, let $f:A\to A'$ and $g:B\to B'$. Precomposing the two sides of the naturality square for $m^G$ with $u^{\mc Q}_{GA,GB}$ reduces both to
\[
  G\bigl((u_{A',B'}\circ_{B'}g)\circ_{A'}f\bigr).
\]
Dually, postcomposing the naturality square for $n^G$ with $u_{\mc Q}^{GA',GB'}$ reduces both sides to
\[
  G\bigl(g\circ_B(f\circ_Au^{A,B})\bigr).
\]
Thus $m^G$ and $n^G$ are natural.

For lax associativity, precompose the two paths with the left-associated composite tensor universal of $\mc Q$. Preservation of cuts by $G$ and the two defining equations reduce both paths to
\[
  G\bigl(u_{A,B\otimes C}\circ_{B\otimes C}u_{B,C}\bigr).
\]
The nullary calculation reduces both lax unit paths to $G(u_{()})$. The oplax associativity and unit diagrams are dual. Since $G$ preserves exchanges, the symmetry diagrams flatten to the images under $G$ of the correspondingly exchanged universal maps. Hence $(G_w,m^G)$ is lax symmetric monoidal and $(G_w,n^G)$ is oplax symmetric monoidal.

For the first Frobenius hexagon, precompose with the source tensor universal of $\mc Q$ and postcompose with the target par universal. Both routes then reduce to the image of the defining distributor core
\[
  G\bigl(u_{A,B}\circ_Bu^{B,C}\bigr).
\]
The mirror hexagon reduces in the same way to
\[
  G\bigl(u_{B,C}\circ_Bu^{A,B}\bigr).
\]
The hexagons for the symmetric variants follow by exchange preservation. Injectivity in \cref{lem:flattening-test} therefore proves all the Frobenius equations, so $G_w$ is a Frobenius linear functor.

It remains to check functoriality of this structured assignment. For the identity polyfunctor, each comparison factors a universal map through itself and is therefore an identity. If $\mc P\xto{G}\mc Q\xto{K}\mc R$, precomposition with $u^{\mc R}_{KGA,KGB}$ shows that both sides of
\[
  m^{KG}_{A,B}=K(m^G_{A,B})\circ m^K_{GA,GB}
\]
factor $(KG)(u^{\mc P}_{A,B})$; dually, postcomposition with $u_{\mc R}^{KGA,KGB}$ gives
\[
  n^{KG}_{A,B}=n^K_{GA,GB}\circ K(n^G_{A,B}).
\]
The same calculations with the nullary universals prove the unit-comparison equations. Thus $\mc G$ is a functor into $\mathbf{WDC}$. The comparison-level reconstruction is \cite[Proposition~2.3.9]{Bla23}, its functoriality is \cite[Proposition~2.3.11]{Bla23}, and the object-level restriction from weak to strong universals is \cite[Proposition~2.3.15]{Bla23}. We now show that $\mc F$ and $\mc G$ define an equivalence.

Before comparing the two composites, we must again distinguish the right-associated products obtained by iterating the chosen binary products from the independently chosen universal objects for arbitrary lists. For a two-tensor polycategory $\mc P$, define
\[
  \widehat\otimes() := \otimes(), \qquad
  \widehat\otimes(A) := A, \qquad
  \widehat\otimes(A,\Sigma) := \otimes(A,\widehat\otimes\Sigma),
\]
and dually
\[
  \widehat\oplus() := \oplus(), \qquad
  \widehat\oplus(A) := A, \qquad
  \widehat\oplus(A,\Sigma) := \oplus(A,\widehat\oplus\Sigma),
\]
for every nonempty list $\Sigma$. These are exactly the right-associated products of the reconstructed WDC $\mc P_w$. Define polymaps recursively by
\[
\begin{aligned}
  \widehat u_{()}&:=u_{()}, &
  \widehat u_{(A)}&:=\1_A, &
  \widehat u_{(A,\Sigma)}&:=
    u_{A,\widehat\otimes\Sigma}
    \circ_{\widehat\otimes\Sigma}\widehat u_\Sigma,\\
  \widehat u^{()}&:=u^{()}, &
  \widehat u^{(A)}&:=\1_A, &
  \widehat u^{(A,\Sigma)}&:=
    \widehat u^\Sigma
    \circ_{\widehat\oplus\Sigma}u^{A,\widehat\oplus\Sigma}.
\end{aligned}
\]
Thus $\widehat u_\Gamma:\Gamma\to\widehat\otimes\Gamma$ and $\widehat u^\Delta:\widehat\oplus\Delta\to\Delta$. They are universal by \cref{lem:identities-universal}. Comparison with the independently chosen finite universals gives unique isomorphisms
\[
\begin{aligned}
  \theta^\otimes_\Gamma &: \widehat\otimes\Gamma
    \xto{\cong}\otimes\Gamma,
  &\theta^\otimes_\Gamma
    \circ_{\widehat\otimes\Gamma}\widehat u_\Gamma&=u_\Gamma,\\
  \theta^\oplus_\Delta &: \widehat\oplus\Delta
    \xto{\cong}\oplus\Delta,
  &u^\Delta\circ_{\oplus\Delta}\theta^\oplus_\Delta&=\widehat u^\Delta.
\end{aligned}
\]
In general neither comparison is an identity; this is already visible for a singleton list.

We first exhibit a natural isomorphism $\eta : \mc F \circ \mc G \simeq \1_{\mathbf{SymPolyCat_t}}$. Let $\mc P \in \mathbf{SymPolyCat_t}$. A polymap $\Gamma \to \Delta$ of $(\mc F \circ \mc G)(\mc P) = (\mc P_w)_{poly}$ is by definition a map $\widehat\otimes\Gamma \to \widehat\oplus\Delta$ of $\mc P_w$, hence a one-input one-output polymap $g : \widehat\otimes\Gamma \to \widehat\oplus\Delta$ of $\mc P$. Define $\eta_{\mc P}$ to be the identity on objects and
\[
  \eta_{\mc P}(g)
  :=\widehat u^\Delta\circ_{\widehat\oplus\Delta}
    \bigl(g\circ_{\widehat\otimes\Gamma}\widehat u_\Gamma\bigr)
  :\Gamma\to\Delta.
\]
The two universal properties make this assignment a bijection on every polymap collection: its inverse first factors a polymap through $\widehat u_\Gamma$ and then through $\widehat u^\Delta$. Since the singleton components of both recursive universal maps are identities, $\eta_{\mc P}$ preserves identities. For a cut, expanding the cut of $(\mc P_w)_{poly}$ and then composing with the relevant recursive universal maps gives on both sides the same iterated cut in $\mc P$; the definitions of the tensor and par actions and of the distributors in $\mc P_w$ are precisely their unique factorizations through these universal maps. Injectivity of the universal factorizations, together with associativity and interchange in $\mc P$, therefore gives preservation of cut. The canonical symmetries of $\mc P_w$ are likewise the unique factorizations of the permuted recursive universal maps, so $\eta_{\mc P}$ preserves the input and output permutation actions. Hence it is an isomorphism of symmetric polycategories.

For naturality, let $F:\mc P\to\mc Q$ be a morphism of $\mathbf{SymPolyCat_t}$. Write
\[
  \widehat m^F_\Gamma:
    \widehat\otimes_{\mc Q}F\Gamma\to
    F(\widehat\otimes_{\mc P}\Gamma),
  \qquad
  \widehat n^F_\Delta:
    F(\widehat\oplus_{\mc P}\Delta)\to
    \widehat\oplus_{\mc Q}F\Delta
\]
for the comparisons recursively assembled from the binary and nullary comparisons of $F_w$. Induction on the lists, using the defining universal factorizations of those comparisons, gives
\[
\begin{aligned}
  F\widehat u^{\mc P}_\Gamma
    &=\widehat m^F_\Gamma
      \circ_{\widehat\otimes_{\mc Q}F\Gamma}
      \widehat u^{\mc Q}_{F\Gamma},\\
  F\widehat u_{\mc P}^\Delta
    &=\widehat u_{\mc Q}^{F\Delta}
      \circ_{\widehat\oplus_{\mc Q}F\Delta}
      \widehat n^F_\Delta.
\end{aligned}
\]
Consequently, for $g:\widehat\otimes_{\mc P}\Gamma\to\widehat\oplus_{\mc P}\Delta$,
\[
\begin{aligned}
  F\bigl(\eta_{\mc P}(g)\bigr)
  &=F\widehat u_{\mc P}^\Delta
    \circ\bigl(Fg\circ F\widehat u^{\mc P}_\Gamma\bigr)\\
  &=\widehat u_{\mc Q}^{F\Delta}
    \circ\Bigl(
      \bigl(\widehat n^F_\Delta\circ Fg\circ
      \widehat m^F_\Gamma\bigr)
      \circ\widehat u^{\mc Q}_{F\Gamma}
    \Bigr)\\
  &=\eta_{\mc Q}\bigl((\mc F\mc G F)(g)\bigr),
\end{aligned}
\]
where the uniquely determined cuts are suppressed in the middle line. Hence the naturality square commutes.

For the other direction, $\varepsilon : \mc G \circ \mc F \simeq \1_{\mathbf{WDC}}$: for a WDC $\C$, the category $(\C_{poly})_w$ has the same objects, and morphisms $(\C_{poly})_w(A,B) = \C_{poly}(A;B) = \C(A, B)$ by the singleton conventions, so it is $\C$ itself; its monoidal operations are the chosen universal objects of $\C_{poly}$, which for pairs are $A \otimes B$ and $A \oplus B$ on the nose, and the induced associators, unitors, symmetries and distributors coincide with those of $\C$ by the uniqueness clauses. Hence the identity functor is an isomorphism of WDCs $(\C_{poly})_w \cong \C$, natural in Frobenius linear functors by the same computation as above. Therefore we have the following theorem:

\begin{theorem}\label{thm:wdc-sympolycat-equivalence}
  There is an equivalence between the category of weakly distributive categories with Frobenius linear functors and the category of two-tensor symmetric polycategories with arbitrary symmetry-preserving polyfunctors:
  \[\mathbf{WDC} \simeq \mathbf{SymPolyCat_t}\]
\end{theorem}

\subsection{Bicartesian Polycategories}\label{subsec:bicartesian-polycategories}

To integrate fibrational semantics, we strengthen a two-tensor polycategory by requiring universal maps at arbitrary positions in a context. These contextual universal maps are precisely the liftings for the terminal-base bifibration of \cref{thm:bicartesian-terminal-bifibration}; liftings against a nonterminal base are treated later.\\
Let us start by defining the slightly more general notion of universal object.

\begin{definition}\label{def:contextual-universal}
  Let $\Gamma, \Delta_1, \Delta_2$ (resp.\ $\Gamma_1, \Gamma_2, \Delta$) be lists of objects in a polycategory $\mc P$.
  \begin{itemize}
    \item An \emph{opcartesian universal object for $\Gamma$ in output context $(\Delta_1 \mid \-- \mid \Delta_2)$} is an object $F(\Delta_1 \mid \Gamma \mid \Delta_2)$ equipped with a polymap $u : \Gamma \to \Delta_1, F(\Delta_1 \mid \Gamma \mid \Delta_2), \Delta_2$ such that for all lists $\Gamma_1, \Gamma_2, \Delta$ and every polymap $f : \Gamma_1, \Gamma, \Gamma_2 \to \Delta_1, \Delta, \Delta_2$ there is a unique polymap $u^f : \Gamma_1, F(\Delta_1 \mid \Gamma \mid \Delta_2), \Gamma_2 \to \Delta$ with $u^f \circ_{F} u = f$.
    \item A \emph{cartesian universal object for $\Delta$ in input context $(\Gamma_1 \mid \-- \mid \Gamma_2)$} is an object $U(\Gamma_1 \mid \Delta \mid \Gamma_2)$ equipped with a polymap $u : \Gamma_1, U(\Gamma_1 \mid \Delta \mid \Gamma_2), \Gamma_2 \to \Delta$ such that for all lists $\Gamma, \Delta_1, \Delta_2$ and every polymap $f : \Gamma_1, \Gamma, \Gamma_2 \to \Delta_1, \Delta, \Delta_2$ there is a unique polymap $u_f : \Gamma \to \Delta_1, U(\Gamma_1 \mid \Delta \mid \Gamma_2), \Delta_2$ with $u \circ_{U} u_f = f$.
  \end{itemize}
  A \emph{bicartesian polycategory} is a symmetric polycategory equipped with a choice of all opcartesian and cartesian universal objects. The category of bicartesian polycategories and all functors of symmetric polycategories is denoted $\mathbf{SymPolyCat_{bi}}$.
\end{definition}

\textit{Remark:}
  The tensor and par products are the special cases with empty surrounding contexts: $\otimes\Gamma = F(\, \mid \Gamma \mid \,)$ and $\oplus\Delta = U(\, \mid \Delta \mid \,)$, so a bicartesian polycategory is in particular two-tensor. The universal objects with empty middle list will give the duals: $A^\star := F(A \mid\, \mid\,)$, with universal map $() \to A, A^\star$. As before, universal polymaps of either kind are closed under composition (same proof as \cref{lem:identities-universal}, cf.\ \cite[Proposition~2.4.4]{Bla23}), and each universal object is unique up to a unique isomorphism compatible with the universal maps. In \cite[Definition~2.4.3]{Bla23} bicartesian polycategories are called \emph{birepresentable} (a term due to Shulman), and in \cite{BZ20} $\star$-representable; \cref{thm:bicartesian-terminal-bifibration} identifies $F$ and $U$ with the opcartesian and cartesian liftings for the unique functor to the terminal polycategory.

The corresponding structure in a weakly distributive category turns out to be the notion of a \emph{dual}. Indeed, the universal object in a polycategory records the ``position'' of the surrounding context formulas. Weakly distributive categories do not have the same power: to recover it, it is very convenient to be able to ``move'' objects between the left and the right of the morphism arrow. This is akin, in logic, to the negation operator: proving $B$ from $A$ is the same as proving $\neg A$ or $B$. We thus introduce the notion of duals in a weakly distributive category.

\begin{definition}\label{def:dual-object}
  In a symmetric weakly distributive category, a \emph{dual} of an object $A$ is an object $A^\star$ together with maps
  \[ \eta_A : \top \to A \oplus A^\star \quadt{and} \varepsilon_A : A^\star \otimes A \to \bot \]
  such that the two ``snake'' composites are identities:
  \begin{align*}
    A \xto{\lambda^{-1}} \top \otimes A \xto{\eta_A \otimes A} (A \oplus A^\star) \otimes A \xto{a^{\oplus\otimes}} A \oplus (A^\star \otimes A) \xto{A \oplus \varepsilon_A} A \oplus \bot \xto{\rho^\oplus} A \;&=\; \1_A, \\
    A^\star \xto{(\rho^\otimes)^{-1}} A^\star \otimes \top \xto{A^\star \otimes \eta_A} A^\star \otimes (A \oplus A^\star) \xto{a^{\otimes\oplus}} (A^\star \otimes A) \oplus A^\star \xto{\varepsilon_A \oplus A^\star} \bot \oplus A^\star \xto{\lambda^\oplus} A^\star \;&=\; \1_{A^\star},
  \end{align*}
  where $\lambda, \rho$ denote the unitors of the relevant monoidal structure. A symmetric weakly distributive category in which every object has a dual is a $\star$-\emph{autonomous category} \cite{CS97}; the category of $\star$-autonomous categories and Frobenius linear functors is denoted $\mathbf{\star\text{-}Aut}$. We also write $\eta'_A := \sigma^\oplus \circ \eta_A : \top \to A^\star \oplus A$ and $\varepsilon'_A := \varepsilon_A \circ \sigma^\otimes : A \otimes A^\star \to \bot$ for the symmetric variants, and for a list $\Delta$ we abbreviate $\Delta^\star := (\oplus\Delta)^\star$.
\end{definition}

\textit{Remark:}
  Since the weakly distributive categories in this section are symmetric, both monoidal operations carry symmetries $\sigma^\otimes_{A,B} : A \otimes B \to B \otimes A$ and $\sigma^\oplus_{A,B} : A \oplus B \to B \oplus A$ as part of their assumed structure. In the $\star$-autonomous case of \cref{def:dual-object}, these symmetries identify the left and right duals of the non-symmetric theory, ${}^\star A \cong A^\star$, so the prefix/suffix distinction in the dual notation may be dropped. In a general non-symmetric $\star$-autonomous category this identification need not hold; nor is symmetry of $\otimes$ a consequence of duality alone.

\begin{theorem}\label{thm:star-autonomous-equivalence}
  There is an equivalence between the category of $\star$-autonomous categories with Frobenius linear functors and the category of bicartesian polycategories with functors of polycategories:
  \[\mathbf{\star\text{-}Aut} \simeq \mathbf{SymPolyCat_{bi}} \]
\end{theorem}

\textit{Proof:} The equivalence of \cref{thm:wdc-sympolycat-equivalence} between weakly distributive categories and two-tensor polycategories is our starting point. For a $\star$-autonomous category, the addition of a dual structure needs to be accounted for in the underlying polycategory. Informally, the dual structure gives a way to move objects between the left and right sides of the arrow. In classical systems, this corresponds to moving a formula between the left and right sides of the turnstile. This effectively adds a negation operator to the formula. To replicate this behavior in the polycategorical context, we define a special type of universal object. Universality will be the property available to us to mimic the behavior.

\begin{itemize}
  \item \textbf{Underlying bicartesian polycategory of a $\star$-autonomous category:} starting from a $\star$-autonomous category $\C$, we equip the two-tensor polycategory $\C_{poly}$ with the remaining universal objects. For lists $\Gamma, \Delta_1, \Delta_2$ we define the opcartesian universal object as
  \[ F(\Delta_1 \mid \Gamma \mid \Delta_2) := \Delta_1^\star \otimes (\otimes\Gamma \otimes \Delta_2^{\star}). \]
  We construct the universal map $u : \Gamma \to \Delta_1, F(\Delta_1 \mid \Gamma \mid \Delta_2), \Delta_2$, i.e.\ the map $u_b : \otimes\Gamma \to \Delta_1 \oplus \big[(\Delta_1^\star \otimes (\otimes\Gamma \otimes \Delta_2^{\star})) \oplus \Delta_2\big]$, as follows:
	\[
	\begin{array}{rll}
  	\otimes \Gamma &\xto{(\rho^{\otimes})^{-1} } & \otimes\Gamma \otimes \top \\
		& \xto{\otimes\Gamma \otimes \eta'_{\Delta_2}} & \otimes\Gamma \otimes (\Delta_2^{\star} \oplus \oplus\Delta_2) \\
		& \xto{a^{\otimes\oplus}_{\otimes\Gamma,\, \Delta_2^\star,\, \oplus\Delta_2}} & (\otimes\Gamma \otimes \Delta_2^{\star}) \oplus \Delta_2 \\
		& \xto{(\lambda^{\otimes})^{-1}} & \top \otimes \big[(\otimes\Gamma \otimes \Delta_2^{\star}) \oplus \Delta_2\big] \\
		& \xto{\eta_{\Delta_1}  \otimes \1} & (\oplus\Delta_1 \oplus \Delta_1^{\star}) \otimes \big[(\otimes\Gamma \otimes \Delta_2^{\star}) \oplus \Delta_2\big] \\
		& \xto{a^{\oplus\otimes}_{\oplus\Delta_1,\, \Delta_1^\star,\, (\otimes\Gamma \otimes \Delta_2^{\star}) \oplus \Delta_2}} & \Delta_1 \oplus \big[\Delta_1^{\star} \otimes ((\otimes\Gamma \otimes \Delta_2^{\star}) \oplus \Delta_2)\big] \\
		& \xto{\Delta_1 \oplus a^{\otimes\oplus}_{\Delta_1^\star,\, \otimes\Gamma \otimes \Delta_2^{\star},\, \oplus\Delta_2}} & \Delta_1 \oplus \big[(\Delta_1^{\star} \otimes (\otimes\Gamma \otimes \Delta_2^{\star})) \oplus \Delta_2\big] \\
	\end{array}
	\]
We now show that this map is universal. Suppose we are given a polymap $g : \Gamma_1, \Gamma, \Gamma_2 \to \Delta_1, \Delta, \Delta_2$, i.e.\ the map $g_b : \otimes(\Gamma_1, \Gamma, \Gamma_2) \to \oplus(\Delta_1, \Delta, \Delta_2)$; we must produce $u^{g} : \Gamma_1, F(\Delta_1 \mid \Gamma \mid \Delta_2), \Gamma_2 \to \Delta$, i.e.\ the map $(u^{g})_b : \otimes\big(\Gamma_1, \Delta_1^{\star} \otimes (\otimes\Gamma \otimes \Delta_2^{\star}), \Gamma_2\big) \to \oplus\Delta$. We construct it as follows, writing $s$ for the canonical symmetric-monoidal shuffle (unique for its underlying permutation, by symmetric coherence):
\[
\begin{array}{rll}
\otimes\big(\Gamma_1, \Delta_1^{\star} \otimes (\otimes\Gamma \otimes \Delta_2^{\star}), \Gamma_2\big) & \xto{s} & \big(\Delta_1^{\star}\otimes \otimes(\Gamma_1, \Gamma, \Gamma_2)\big)\otimes \Delta_2^{\star}  \\
&\xto{(\Delta_1^\star \otimes g_b) \otimes \Delta_2^\star} & \big(\Delta_1^{\star}\otimes (\Delta_1  \oplus (\oplus\Delta \oplus \Delta_2))\big)\otimes \Delta_2^{\star}  \\
&\xto{a^{\otimes\oplus}_{\Delta_1^\star,\, \oplus\Delta_1,\, \oplus\Delta \oplus \oplus\Delta_2} \otimes \Delta_2^\star} & \big((\Delta_1^{\star}\otimes \oplus\Delta_1)  \oplus (\oplus\Delta \oplus \Delta_2)\big)\otimes \Delta_2^{\star}  \\
&\xto{(\varepsilon_{\Delta_1} \oplus \1) \otimes \Delta_2^\star} & \big(\bot  \oplus (\oplus\Delta \oplus \Delta_2)\big)\otimes \Delta_2^{\star}  \\
&\xto{\lambda^{\oplus} \otimes \Delta_2^\star} & (\oplus\Delta \oplus \Delta_2)\otimes \Delta_2^{\star}  \\
&\xto{a^{\oplus\otimes}_{\oplus\Delta,\, \oplus\Delta_2,\, \Delta_2^\star}} & \oplus\Delta \oplus (\Delta_2\otimes \Delta_2^{\star})  \\
&\xto{\oplus\Delta \oplus \varepsilon'_{\Delta_2}} & \oplus\Delta \oplus \bot  \\
&\xto{\rho^{\oplus}} & \oplus\Delta  \\
\end{array}
\]
It remains to show that $g \mapsto u^g$ is inverse to cutting with $u$, i.e.\ that $u^{g} \circ_{F} u = g$ and that $u^g$ is the unique such polymap. Both operations are built from the cups and caps of the duals $\Delta_1^\star, \Delta_2^\star$: cutting with $u$ inserts the cups $\eta_{\Delta_1}, \eta'_{\Delta_2}$, while $g \mapsto u^g$ inserts the caps $\varepsilon_{\Delta_1}, \varepsilon'_{\Delta_2}$, in each case transported to the correct position by distributors and canonical shuffles. Unfolding the two composites and cancelling each adjacent cup/cap pair with the corresponding snake identity (using naturality of the distributors to bring the pairs together, and the coherence theorem \cite{BCST96} to identify the rearrangements) gives $u^{g} \circ_F u = g$. Conversely, for any $h$ with $h \circ_F u = g$, applying the cap-insertion construction to both sides and cancelling with the other snake identity gives $h = u^g$, which proves uniqueness. In other words, the two constructions are the polycategorical form of the adjunction bijections $\C(X, A \oplus Y) \cong \C(A^\star \otimes X, Y)$ furnished by the duals, whose unit and counit are the snake identities.

The cartesian universal objects are constructed dually: for lists $\Gamma_1, \Gamma_2, \Delta$ we set
\[ U(\Gamma_1 \mid \Delta \mid \Gamma_2) := \Gamma_1^{\star\otimes} \oplus (\oplus\Delta \oplus \Gamma_2^{\star\otimes}), \qquad \Gamma_i^{\star\otimes} := (\otimes\Gamma_i)^\star, \]
with universal map $\Gamma_1, U, \Gamma_2 \to \Delta$ obtained by inserting the caps $\varepsilon$, and factorization $u_g$ obtained by inserting the cups $\eta$; universality follows by the same snake-cancellation argument.
This shows that the underlying polycategory of a $\star$-autonomous category is indeed bicartesian.

  \item \textbf{Underlying $\star$-autonomous category of a bicartesian polycategory:} given a bicartesian polycategory $\mc P$, we define its underlying $\star$-autonomous category $\mc P_\star$ as follows.
  \begin{itemize}
    \item The objects of $\mc P_\star$ are those of $\mc P$, and the morphisms are the polymaps with exactly one input and one output, $\mc P_\star(A,B) := \mc P(A;B)$, with composition the unary cut.
    \item The weakly distributive structure ($\otimes, \top, \oplus, \bot$, associators, unitors, symmetries, distributors) is that of the two-tensor polycategory underlying $\mc P$, i.e.\ the construction $\mc P_w$ in the proof of \cref{thm:wdc-sympolycat-equivalence}, applied to the universal objects $\otimes\Gamma = F(\,\mid \Gamma \mid\,)$ and $\oplus\Delta = U(\,\mid \Delta \mid\,)$.
    \item For any object $A$ we define the dual $A^\star := F(A \mid\, \mid\,)$, the opcartesian universal object of the empty list in output context $(A \mid \-- \mid\,)$, with universal map $cup_A : () \to A, A^\star$.
  \end{itemize}
  It remains to produce the cups and caps in $\mc P_\star$ and verify the snake identities. First, at the level of the polycategory, the universal property of $cup_A$ applied to the identity $\1_A : A \to A$ (taking $\Gamma_1 = ()$, $\Gamma_2 = (A)$, $\Delta = ()$ in the definition) yields a unique polymap
  \[ cap_A := u^{\1_A} : A^\star, A \to () \qquad\text{with}\qquad cap_A \circ_{A^\star} cup_A = \1_A. \]
  For the other snake equation, note that both $cap_A \circ_A cup_A$ and $\1_{A^\star}$ factor the polymap $cup_A$ through itself:
  \[
    (cap_A \circ_A cup_A) \circ_{A^\star} cup_A
    = (cap_A \circ_{A^\star} cup_A) \circ_A cup_A
    = \1_A \circ_A cup_A
    = cup_A.
  \]
  The first equality is the independent-cut law; its prescribed block transport is the identity in this nullary configuration. Thus uniqueness for $cup_A$ gives $cap_A \circ_A cup_A = \1_{A^\star}$. In a polycategory the snake identities are therefore literally the two well-typed cut equations
  \[
    cap_A \circ_{A^\star} cup_A=\1_A,
    \qquad
    cap_A \circ_A cup_A=\1_{A^\star},
  \]
  cf.\ \cite[Definition~2.3.16, Proposition~2.4.6]{Bla23}.
  Now we transport this data to $\mc P_\star$. To obtain $\eta_A : \top \to A \oplus A^\star$, recall that to map out of $\top$ it suffices to map out of the empty list: factoring $cup_A : () \to A, A^\star$ through $u_{()} : () \to \top$ gives $\top \to A, A^\star$, and factoring the output through $u^{A, A^\star} : A \oplus A^\star \to A, A^\star$ gives $\eta_A : \top \to A \oplus A^\star$. Dually, factoring $cap_A : A^\star, A \to ()$ through $u^{()} : \bot \to ()$ and $u_{A^\star\!, A}$ gives $\varepsilon_A : A^\star \otimes A \to \bot$. Expand the first WDC snake composite using these definitions and the universally induced unitors and distributors, then cancel each binary and nullary universal map against its defining factorization: the resulting polymap is $cap_A \circ_{A^\star} cup_A:A\to A$. The same expansion of the second WDC snake composite gives $cap_A \circ_A cup_A:A^\star\to A^\star$. The two equations above therefore make the WDC composites identities; injectivity of the universal factorizations reflects those identities back to the unary maps of $\mc P_\star$. Hence $\mc P_\star$ is a $\star$-autonomous category.
\end{itemize}

Now that we have presented the respective underlying structures, we exhibit the functors between the category of $\star$-autonomous categories and bicartesian polycategories and show that they are inverse to each other up to natural isomorphism.

\begin{itemize}
  \item  $\mathbf{\star\text{-}Aut}\xto{\mc F} \mathbf{SymPolyCat_{bi}}$: on objects we set $\mc F(\C) = \C_{poly}$ with the universal objects chosen above, and on a Frobenius linear functor $\C \xto{F} \C'$ we define the functor $\C_{poly} \xto{F_{poly}} \C'_{poly}$ as in the WDC case:
  \begin{itemize}
    \item on objects, $F_{poly}(A) = FA$;
    \item on polymaps, $F_{poly}(f) = (n_\Delta \circ F\bar f \circ m_\Gamma)_{poly}$, where $m, n$ are the lax and oplax comparisons of $F$.
  \end{itemize}
  Identities are preserved since $F_{poly}(\1_{A}) = (n_{[A]} \circ F\1_A \circ m_{[A]})_{poly} = (\1_{FA})_{poly}$ by the singleton conventions, and preservation of the cut $g \circ_A f$ is precisely the statement that $(m, n)$ cohere with the distributors and symmetries, i.e.\ the Frobenius linear functor axioms: unfolding both sides, the comparisons $m, n$ must slide past the distributor stages of the cut composite, which is exactly what the coherence hexagons permit.

  \item  $\mathbf{SymPolyCat_{bi}}\xto{\mc G} \mathbf{\star\text{-}Aut}$: for a bicartesian polycategory $\mc P$, set $\mc G(\mc P) = \mc P_\star$. For a morphism $\mc P \xto{G} \mc P'$ of $\mathbf{SymPolyCat_{bi}}$, define the Frobenius linear functor $\mc P_\star \xto{G_\star} \mc P'_\star$ as follows:
	\begin{itemize}
  	\item on objects, $G_\star(A) = GA$;
  	\item on morphisms, $G_\star(f) = Gf$; identity and composition preservation follow as $G$ preserves identities and unary cuts.
  	\item The lax comparison $m_\Gamma : \otimes G\Gamma \to G(\otimes \Gamma)$: the universal map $u_\Gamma : \Gamma \to \otimes \Gamma$ of $\mc P$ is sent by $G$ to a polymap $G\Gamma \to G(\otimes \Gamma)$, which factors through the universal map $u_{G\Gamma}$ of $\mc P'$ to give $m_\Gamma$. Oplaxity follows the same blueprint: applying $G$ to $u^\Delta : \oplus\Delta \to \Delta$ and factoring through $u^{G\Delta}$ produces $n_\Delta : G(\oplus \Delta) \to \oplus G\Delta$. The coherence of $(m,n)$ with the unitors, symmetries and distributors holds by the uniqueness clauses, exactly as in the two-tensor case; no comparison for the duals is needed, since the cups and caps of $\mc P'_\star$ already mediate between $G(A^\star)$ and $(GA)^\star$.
	\end{itemize}
\end{itemize}
The natural isomorphisms $\mc F \circ \mc G \simeq \1$ and $\mc G \circ \mc F \simeq \1$ are those of \cref{thm:wdc-sympolycat-equivalence}: the isomorphisms $\eta_{\mc P}$ and $\varepsilon_{\C}$ constructed in its proof are compatible with the additional universal objects, because on each side the chosen $F(\Delta_1 \mid \Gamma \mid \Delta_2)$ and $U(\Gamma_1 \mid \Delta \mid \Gamma_2)$ satisfy the same universal property and universal objects are unique up to unique isomorphism. $\square$

\subsection{Universal Objects}\label{subsec:terminal-base-liftings}

The next major goal of this thesis is to derive a representability result for bifibrations of polycategories. Recall that, for ordinary categories, the Grothendieck construction identifies contravariant pseudofunctors into $\mathbf{Cat}$ with fibrations \cite{Gro71}; \cref{sec:multi-indexed-bifibrations,sec:poly-indexed-planar-bifibrations} develop the polycategorical counterpart of this correspondence, starting from \cref{thm:bicartesian-terminal-bifibration}, which expresses everything done so far in fibrational terms. A \emph{bifibration of polycategories} $\fib[T]{M}$ is a functor of polycategories along which all cartesian and opcartesian liftings exist; the precise polycategorical definitions are given in \cref{sec:poly-indexed-planar-bifibrations}, while the multicategorical analogues are \cref{def:cartesian-multimorphism,def:opcartesian-multimorphism,def:multicategory-fibration,def:multicategory-opfibration}.

\begin{theorem}\label{thm:bicartesian-terminal-bifibration}
  A polycategory $\mc P$ is bicartesian if and only if the unique functor $\mc P \to \1$ to the terminal polycategory is a bifibration. Consequently there is an equivalence between the category of bicartesian polycategories and the category of bifibrations of polycategories over $\1$:
  \[\mathbf{SymPolyCat_{bi}} \simeq \mathbf{BiFib}(\1) \]
\end{theorem}
\textit{Proof sketch:} over the terminal polycategory there is exactly one polymap of each arity, so the base data of an (op)cartesian lifting problem carries no information: the lifting property against $\mc P \to \1$ collapses to the plain factorization property of the universal polymaps, opcartesian liftings corresponding to the objects $F(\Delta_1 \mid \Gamma \mid \Delta_2)$ and cartesian liftings to the objects $U(\Gamma_1 \mid \Delta \mid \Gamma_2)$. The full proof follows from the formal definitions developed in \cref{sec:multi-indexed-bifibrations,sec:poly-indexed-planar-bifibrations}; cf.\ \cite[Propositions~4.2.1--4.2.2 and Theorem~4.2.3]{Bla23}, where combined with \cref{thm:star-autonomous-equivalence} this exhibits $\star$-autonomous categories as bifibrations over $\1$.

We next pass from the terminal-base case to indexed structures over nonterminal bases. \Cref{sec:multi-indexed-bifibrations} proves a representability equivalence for multicategories over the underlying multicategory of a fixed unbiased monoidal category. \Cref{sec:poly-indexed-planar-bifibrations} instead starts from explicit poly-indexed coordinate data and constructs a planar polycategory over its weakly distributive base; it does not prove a converse for arbitrary polycategorical bifibrations.

\endgroup{}
\begingroup

\section{Multi-Indexed Categories and Multicategorical Bifibrations}\label{sec:multi-indexed-bifibrations}

This section proves a representation theorem for multicategories opfibered over a common base. We assume that this base is the underlying multicategory of an unbiased monoidal category. Given an unbiased monoidal category $\mc M$, we write $\mathbf{Multi_{\mc M_m}}$ for the category of cloven opfibrations of multicategories over its underlying multicategory $\mc M_m$, with morphisms as in \cref{def:category-cloven-multicategory-opfibrations}. A multi-indexed category is a category over $\mc M$ equipped with tensor functors indexed by the multimorphisms of $\mc M_m$. \Cref{thm:multi-indexed-cloven-opfibration-equivalence} proves that these form a category $\mathbf{MIdx_{\mc M}}$ equivalent to $\mathbf{Multi_{\mc M_m}}$. We first fix our conventions and then define multi-indexed categories.

\subsection{Cloven Opfibrations}\label{subsec:unbiased-monoidal-base}

We fix an unbiased monoidal category $(\mc M, (\otimes_n)_{n \in \mathbb{N}}, (\eta_{n,m})_{n,m \in \mathbb{N}}, \iota)$ in the sense of \cref{def:unbiased-monoidal-category}. We use the \hyperref[anchor:unbiased-monoidal-associator-notation]{notation $\eta_{n,m}$ introduced after that definition}. We omit the insertion position when it is determined by the displayed lists. We use the following names for the coherence laws of \cref{def:unbiased-monoidal-category}:
\begin{itemize}
	\item \textbf{Nested associativity:} associating an inner block first and then the surrounding block agrees with associating the surrounding block first and then the inner one, i.e. $\eta_{n, m_1 + k + m_2} $ followed by $\otimes_n(\1, \eta_{m,k}, \1)$ agrees with the corresponding path through $\eta_{n + m - 1, k}$ and $\eta_{n,m}$;
	\item \textbf{Disjoint zones:} associating two non-overlapping blocks can be done in either order, i.e. the square built from the four evident $\eta$'s commutes;
	\item \textbf{Unit laws:} $\iota_{\otimes_n(a_1, \ldots, a_n)} \circ \eta_{1,n} = \1_{\otimes_n(a_1, \ldots, a_n)}$ and $\otimes_n(\1_{a_1}, \ldots, \iota_{a_i}, \ldots, \1_{a_n}) \circ \eta_{n,1} = \1_{\otimes_n(a_1, \ldots, a_n)}$.
\end{itemize}

\phantomsection\label{anchor:unbiased-monoidal-circledcirc}
\textbf{The operation $\circledcirc$.} Given morphisms $\otimes_n(\gamma, a, \gamma') \xto{\alpha} c$ and $\otimes_m \delta \xto{\beta} a$ in $\mc M$, where $\gamma, \gamma', \delta$ are lists of objects and the displayed occurrence of $a$ is marked, we write
$$\alpha \circledcirc \beta := \alpha \circ \otimes_n(\1_\gamma, \beta, \1_{\gamma'}) \circ \eta_{n,m} \ : \ \otimes_{n + m - 1}(\gamma, \delta, \gamma') \to c.$$
The notation omits the composition position. When confusion can arise, we explicitly mark it using a subscript such as $\alpha \circledcirc_a \beta$.

\begin{lemma}[Units]\label{lem:circledcirc-units}
	For every morphism $\otimes_n(\gamma, a, \gamma') \xto{\alpha} c$ in $\mc M$ we have
	$$\iota_c \circledcirc \alpha = \alpha \quadt{and} \alpha \circledcirc \iota_a = \alpha .$$
\end{lemma}
\textit{Proof:} For the first equation, unfolding the definition with outer morphism $\iota_c : \otimes_1(c) \to c$ gives $\iota_c \circledcirc \alpha = \iota_c \circ \otimes_1(\alpha) \circ \eta_{1,n}$. By naturality of $\iota$ at $\alpha$ we have $\iota_c \circ \otimes_1(\alpha) = \alpha \circ \iota_{\otimes_n(\gamma, a, \gamma')}$, and by the first unit law $\iota_{\otimes_n(\gamma,a,\gamma')} \circ \eta_{1,n} = \1$. Hence $\iota_c \circledcirc \alpha = \alpha$. For the second equation, unfolding the definition with inner morphism $\iota_a : \otimes_1(a) \to a$ gives $\alpha \circledcirc \iota_a = \alpha \circ \otimes_n(\1_\gamma, \iota_a, \1_{\gamma'}) \circ \eta_{n,1}$. The second unit law gives $\otimes_n(\1_\gamma, \iota_a, \1_{\gamma'}) \circ \eta_{n,1} = \1$. \hfill $\square$

\begin{lemma}[Associativity]\label{lem:circledcirc-associativity}
	Let $\otimes_n(\gamma, a, \gamma') \xto{\alpha} c$ be a morphism in $\mc M$.
	\begin{itemize}
		\item \textbf{Nested configuration:} for $\otimes_m(\delta, b, \delta') \xto{\beta} a$ and $\otimes_k \lambda \xto{\omega} b$,
		$$(\alpha \circledcirc \beta) \circledcirc \omega = \alpha \circledcirc (\beta \circledcirc \omega).$$
        \item \textbf{Disjoint configuration:} if the domain list of $\alpha$ carries two marked objects,
\[
\otimes_n(\gamma, a, \gamma', b, \gamma'') \xto{\alpha} c,
\]
then for $\otimes_m\delta \xto{\beta} a$ and $\otimes_k\lambda \xto{\omega} b$,
		$$(\alpha \circledcirc_a \beta) \circledcirc_b \omega = (\alpha \circledcirc_b \omega) \circledcirc_a \beta,$$
	\end{itemize}
\end{lemma}
\textit{Proof:} For the nested configuration, unfold both sides:
\begin{align*}
	(\alpha \circledcirc \beta) \circledcirc \omega &= \alpha \circ \otimes_n(\1_\gamma, \beta, \1_{\gamma'}) \circ \eta_{n,m} \circ \otimes_{n+m-1}(\1_\gamma, \1_\delta, \omega, \1_{\delta'}, \1_{\gamma'}) \circ \eta_{n+m-1, k} \\
	\alpha \circledcirc (\beta \circledcirc \omega) &= \alpha \circ \otimes_n(\1_\gamma, \beta, \1_{\gamma'}) \circ \otimes_n\big(\1_\gamma, \otimes_m(\1_\delta, \omega, \1_{\delta'}), \1_{\gamma'}\big) \circ \otimes_n(\1_\gamma, \eta_{m,k}, \1_{\gamma'}) \circ \eta_{n, m+k-1}
\end{align*}
where the second line uses functoriality of $\otimes_n$ to split $\otimes_n(\1_\gamma, \beta \circ \otimes_m(\1_\delta, \omega, \1_{\delta'}) \circ \eta_{m,k}, \1_{\gamma'})$ into three factors. By naturality of $\eta_{n,m}$ at the tuple $(\1_\gamma, (\1_\delta, \omega, \1_{\delta'}), \1_{\gamma'})$ we have
$$\eta_{n,m} \circ \otimes_{n+m-1}(\1_\gamma, \1_\delta, \omega, \1_{\delta'}, \1_{\gamma'}) = \otimes_n\big(\1_\gamma, \otimes_m(\1_\delta, \omega, \1_{\delta'}), \1_{\gamma'}\big) \circ \eta_{n,m},$$
where on the right the component of $\eta_{n,m}$ is taken at the list $(\gamma, (\delta, \otimes_k\lambda, \delta'), \gamma')$. Substituting, the two sides agree if and only if
$$\eta_{n,m} \circ \eta_{n + m - 1, k} = \otimes_n(\1_\gamma, \eta_{m,k}, \1_{\gamma'}) \circ \eta_{n, m + k - 1}.$$
This is the nested associativity coherence of the unbiased monoidal category. For the disjoint configuration, the same calculation reduces the claim to the disjoint zones coherence. Here we use naturality of $\eta$ at the tuples inserting $\beta$ and $\omega$, and functoriality of $\otimes_n$ to commute $\otimes_n(\1, \beta, \1, \1, \1)$ past $\otimes_n(\1, \1, \1, \omega, \1)$:
$$\eta_{n,m} \circ \eta_{n + m - 1, k} = \eta_{n, k} \circ \eta_{n + k - 1, m}$$
 which holds by assumption. \hfill $\square$

\textbf{The underlying multicategory $\mc M_m$.} We write $\mc M_m$ for the underlying multicategory of $\mc M$ and we recall:
\begin{itemize}
	\item the objects of $\mc M_m$ are the objects of $\mc M$, an object $a$ being written $a_m$ when regarded in $\mc M_m$;
	\item a multimorphism $\gamma_m \xto{\alpha_m} a_m$, for a list $\gamma := \gamma_1, \ldots, \gamma_n$, is a morphism $\otimes_n \gamma \xto{\alpha} a$ of $\mc M$;
	\item the identity of $a_m$ is $(\iota_a)_m$, for $\iota_a : \otimes_1(a) \to a$;
	\item the composition of $\gamma_m \xto{\alpha_m} a_m$ into $\delta_m, a_m, \delta'_m \xto{\beta_m} c_m$ is $\beta_m \circ \alpha_m := (\beta \circledcirc \alpha)_m$.
\end{itemize}

\begin{proposition}\label{prop:underlying-multicategory-unbiased-monoidal}
	$\mc M_m$ is a multicategory.
\end{proposition}
\textit{Proof:} The unit laws $\beta_m \circ (\iota_a)_m = \beta_m$ and $(\iota_c)_m \circ \beta_m = \beta_m$ follow from \cref{lem:circledcirc-units}, and the nested and disjoint associativity laws follow from \cref{lem:circledcirc-associativity}. $\square$

\begin{definition}
For a functor $\pi : \fib[T]{M}$ of categories and an object $\gamma \in \mc M$, the \textit{fibre} $\mc T_\gamma$ is the subcategory of $\mc T$ whose objects are the objects $A$ with $\pi A = \gamma$ and whose morphisms are the morphisms $f$ with $\pi f = \1_\gamma$. Morphisms lying in a fibre are called \textit{vertical}. Similarly, for a functor of multicategories $\pi : \fib[T]{M_m}$, the fibre over $a_m$ has the objects over $a_m$ and the unary multimorphisms over the identity $(\iota_a)_m$ as morphisms.
\end{definition}

\begin{definition}\label{def:cloven-multicategory-opfibration}
	We use \emph{opcartesian multimorphism} and \emph{opfibration of multicategories} in the senses of \cref{def:opcartesian-multimorphism,def:multicategory-opfibration}. A \textit{cloven opfibration of multicategories} is an opfibration $\pi : \fib[T]{M_m}$ equipped, for every list $\Gamma$ in $\mc T$ and every multimorphism $\pi\Gamma \xto{\alpha} a$ in $\mc M_m$, with a specified object $\alpha_*(\Gamma)$ over $a$ and a specified opcartesian lift
	$$\Gamma \xto{\overline{\alpha}(\Gamma)} \alpha_*(\Gamma)$$
	over $\alpha$. This chosen family of lifts is the \textit{cleavage} and is part of the data.
\end{definition}

\begin{definition}\label{def:category-cloven-multicategory-opfibrations}
	The category $\mathbf{Multi_{\mc M_m}}$ has as objects the cloven opfibrations of multicategories $\pi : \fib[T]{M_m}$ and as morphisms $\pi \to \pi'$ the functors of multicategories $F : \mc T \to \mc T'$ such that $\pi' \circ F = \pi$ and such that $F$ sends opcartesian multimorphisms to opcartesian multimorphisms. 
\end{definition}
\textit{Remark.} Morphisms are not required to preserve the chosen cleavages. Identities and composites satisfy both conditions, so this is a category.

\begin{lemma}[Opcartesian properties]\label{lem:opcartesian-properties}
	Let $\pi : \fib[T]{M_m}$ be a functor of multicategories.
	\begin{enumerate}
		\item\label{item:opcartesian-vertical-isomorphism} \textbf{Vertical isomorphisms preserve opcartesianness:} if $\Gamma \xto{f} A$ is opcartesian and $A \xto{v} B$ is a vertical isomorphism, then $v \circ f$ is opcartesian.
		\item\label{item:opcartesian-essential-uniqueness} \textbf{Essential uniqueness:} if $\Gamma \xto{f} A$ and $\Gamma \xto{f'} A'$ are opcartesian with $\pi f = \pi f'$, then there is a unique vertical isomorphism $A \xto{v} A'$ with $v \circ f = f'$.
		\item\label{item:opcartesian-identity-lift} \textbf{Lifts of identities are invertible:} if $A \xto{f} B$ is an opcartesian unary multimorphism with $\pi f$ an identity, then $f$ is invertible, with vertical inverse.
		\item\label{item:opcartesian-composition} \textbf{Opcartesian morphisms compose:} if $\Gamma \xto{f} A$ is opcartesian and $\Gamma_1, A, \Gamma_2 \xto{g} B$ is opcartesian, then $\Gamma_1, \Gamma, \Gamma_2 \xto{g \circ f} B$ is opcartesian.
	\end{enumerate}
\end{lemma}
\textit{Proof:}
\begin{enumerate}
	\item Note $\pi(v \circ f) = \1 \circ \pi f = \pi f$. Given $\Gamma_1, \Gamma, \Gamma_2 \xto{g} C$ and $\beta$ with $\beta \circ \pi(v \circ f) = \pi g$, the factorizations $\phi$ of $g$ through $v \circ f$ over $\beta$ and the factorizations $\psi$ of $g$ through $f$ over $\beta \circ \pi(v)= \beta$ are in bijection via $\phi \mapsto \phi \circ v$ and $\psi \mapsto \psi \circ v^{-1}$. Since $\psi$ is unique we conclude that $\phi$ is unique.
	\item Factor $f'$ through $f$ over the identity to get a vertical $v$ with $v \circ f = f'$, and symmetrically $v'$ with $v' \circ f' = f$. Then $(v' \circ v) \circ f = f = \1 \circ f$, and both $v' \circ v$ and $\1$ are factorizations of $f$ through $f$ over the identity, so $v' \circ v = \1$ by uniqueness. Symmetrically $v \circ v' = \1$.
	\item Factor $\1_A$ through $f$ over the identity. There is a unique vertical $g$ with $g \circ f = \1_A$. Then $(f \circ g) \circ f = f = \1_B \circ f$, and both $f \circ g$ and $\1_B$ are factorizations of $f$ through $f$ over the identity. Therefore, they are equal.
	\item Note $\pi(g \circ f) = \pi g \circ \pi f$. Given $\Delta_1, \Gamma_1, \Gamma, \Gamma_2, \Delta_2 \xto{h} C$ and $\beta$ with $\beta \circ \pi(g \circ f) = \pi h$, first factor $h$ through $f$ over $\beta \circ \pi g$: there is a unique $\psi$ with $\psi \circ f = h$ and $\pi \psi = \beta \circ \pi g$. Then factor $\psi$ through $g$ over $\beta$: there is a unique $\phi$ with $\phi \circ g = \psi$ and $\pi \phi = \beta$. Hence $\phi \circ (g \circ f) = h$. For uniqueness, if $\phi'$ also satisfies $\phi' \circ (g \circ f) = h$ over $\beta$, then $\phi' \circ g$ is a factorization of $h$ through $f$ over $\beta \circ \pi g$, so $\phi' \circ g = \psi$ by uniqueness, and then $\phi' = \phi$ by uniqueness of the factorization through $g$. $\square$
	\end{enumerate}

By \hyperref[item:opcartesian-vertical-isomorphism]{vertical-isomorphism invariance} and \hyperref[item:opcartesian-essential-uniqueness]{essential uniqueness}, every opcartesian multimorphism $\Gamma \to A$ over $\alpha$ can be factored through $\overline{\alpha}(\Gamma)$ and a vertical morphism. This factorization is used in the \hyperref[anchor:vertical-isomorphism-opcartesian-criterion]{proof that the constructed functors preserve opcartesian multimorphisms}.

\phantomsection\label{anchor:one-categorical-opcartesian-morphism}
We also use the corresponding $1$-categorical notion. For a functor of categories $\pi : \fib[T]{M}$, a morphism $A \xto{f} B$ in $\mc T$ with $\pi f = \alpha$ is \textit{opcartesian over $\alpha$} if for every morphism $A \xto{g} C$ in $\mc T$ and every $\pi B \xto{\beta} \pi C$ in $\mc M$ with $\beta \circ \alpha = \pi g$ there is a unique $B \xto{h} C$ with $\pi h = \beta$ and $h \circ f = g$. 

\subsection{Multi-Indexed Categories}\label{subsec:multi-indexed-categories}

\begin{definition}\label{def:multi-indexed-category}
  A \textit{multi-indexed category} $\mc T$ is a category with the following data:
  \begin{itemize}
    \item An unbiased monoidal category $(\mc M, (\otimes_n)_{n \in \mathbb{N}}, (\eta_{n,m})_{n,m \in \mathbb{N} }, \iota)$;
    \item  A functor $\pi : \fib[T]{M}$ of categories;
    \item For every list of objects $\gamma := \gamma_1, \ldots, \gamma_n$ and every morphism $\otimes_n\gamma \xto{\alpha} c \in \mc M$ , a functor $\otimes_\alpha : \mc T_{\gamma_1} \times \ldots \times \mc T_{\gamma_n}  \to \mc T_c$ between the fibres. The index of $\otimes_\alpha$ is really the pair $(\gamma, \alpha)$, i.e.\ the multimorphism $\gamma_m \xto{\alpha_m} c_m$ of $\mc M_m$. 
    \item For every morphism $a \xto{\alpha} c \in \mc M$ and every object $A \in \mc T_{a}$, an opcartesian lift of $\alpha$ : $$A \xto{\alpha_A^\star} \otimes_{\alpha \circ \iota_{a} }(A)$$ living over $\alpha$, i.e.\ a morphism with $\pi(\alpha^\star_A) = \alpha$ that is opcartesian over $\alpha$ in the sense of the \hyperref[anchor:one-categorical-opcartesian-morphism]{one-categorical definition above}. Here $\alpha \circ \iota_a : \otimes_1(a) \to c$.
    \item For every choice of:
            \begin{itemize}
             \item lists of objects $\gamma $ in $\mc M$ and $\Gamma $ in $\mc T$ such that $\gamma_i := \pi \Gamma_i $,
             \item lists of objects $\gamma' $ in $\mc M$  and $\Gamma' $ in $\mc T$  such that $\gamma'_i := \pi \Gamma'_i$,
             \item lists of objects $\delta $ in $\mc M$  and $\Delta $ in $\mc T$ such that $\delta_i := \pi \Delta_i$,

                 \item morphisms $\otimes_{|\gamma| + 1 + |\gamma'|}(\gamma, a, \gamma') \xto{\alpha} c$ and $\otimes_{|\delta|}\delta \xto{\beta} a$ in $\mc M$
         \end{itemize}
                a natural isomorphism :
                \[\zeta_{\alpha,\beta} : \otimes_{\alpha \circledcirc \beta }(\Gamma,\Delta, \Gamma')  \to \otimes_\alpha(\Gamma, \otimes_\beta\Delta, \Gamma') ;\]
            where $\alpha \circledcirc \beta = \alpha \circ \otimes_{|\gamma| + 1 + |\gamma'|}(\1_{\gamma}, \beta, \1_{\gamma'}) \circ \eta_{|\gamma| + 1 + |\gamma'|,|\delta|}$. 
  \end{itemize}
  These data must satisfy the following coherence relations.
  \begin{itemize}
    \item \textbf{Coherence between $\zeta$ and the opcartesian lifts of identities.} For any object $A \in \mc T$ such that $\pi A = a$, let $(1_a)^\star_A : A \to \otimes_{\iota_a}(A)$ be the opcartesian lift of the identity morphism $\1_a : a \to a$.
    The following diagrams commute for every list of objects $\Gamma$ over $\gamma$ and every morphism $\otimes_{|\gamma|}\gamma \xto{\alpha} c$ and for all lists $\Gamma, \Gamma'$ and every object $A$ over $a$ and every morphism $\otimes_{|\gamma| + 1 + |\gamma'| }(\gamma, a, \gamma') \xto{\alpha} c$ :
  \begin{center}
    \begin{tikzcd}[row sep=huge, column sep=huge]
      \otimes_\alpha(\Gamma) \arrow[r, "\zeta_{\iota_c, \alpha}"] \arrow[dr, "{(1_c)^\star_{\otimes_\alpha(\Gamma)}}", swap] & \otimes_{\iota_c}(\otimes_\alpha(\Gamma)) \\
      & \otimes_{\iota_c}(\otimes_\alpha(\Gamma)) \arrow[u, equals]
    \end{tikzcd}
    \quad
    \begin{tikzcd}[row sep=huge, column sep=huge]
      \otimes_\alpha(\Gamma, A, \Gamma') \arrow[r, "\zeta_{\alpha, \iota_a}"] \arrow[dr, "{\otimes_{\alpha}(\1_{\Gamma}, (1_a)^\star_A, \1_{\Gamma'})}", swap] & \otimes_{\alpha}(\Gamma, \otimes_{\iota_a}(A), \Gamma') \\
      & \otimes_{\alpha}(\Gamma, \otimes_{\iota_a}(A), \Gamma') \arrow[u, equals]
    \end{tikzcd}
  \end{center}
  Equationally, these diagrams translate to:
  $$\zeta_{\iota_c, \alpha} = (1_c)^\star_{\otimes_\alpha(\Gamma)} \quad \text{and} \quad \zeta_{\alpha, \iota_a} = \otimes_{\alpha}(\1_\Gamma, (1_a)^\star_A, \1_{\Gamma'}).$$
  Note that these diagrams type-check because $\iota_c \circledcirc \alpha = \alpha$ and $\alpha \circledcirc \iota_a = \alpha$ hold strictly, by \cref{lem:circledcirc-units}; also, $(1_c)^\star_A$ has codomain $\otimes_{\1_c \circ \iota_c}(A) = \otimes_{\iota_c}(A)$ as shown.

  \item \textbf{Coherence relation between $\zeta$ and $\otimes$, nested configuration.}  For any
            \begin{itemize}
             \item lists of objects $\gamma $ in $\mc M$ and $\Gamma $ in $\mc T$ such that $\gamma_i := \pi \Gamma_i $,
             \item lists of objects $\gamma' $ in $\mc M$ and $\Gamma' $ in $\mc T$ such that $\gamma'_i := \pi \Gamma'_i $,
               \item lists of objects $\delta $ and $\Delta $ such that $\delta_i := \pi \Delta_i $,
               \item lists of objects $\delta' $ and $\Delta' $ such that $\delta'_i := \pi \Delta'_i $,
               \item lists of objects $\lambda $ and $\Lambda $ such that $\lambda_i := \pi \Lambda_i $,
                 \item objects $a,b$ in $\mc M$ and morphisms  $\otimes_{|\gamma| + 1 + |\gamma'|}(\gamma, a, \gamma')  \xto{\alpha} c$, $\otimes_{|\delta| + 1 + |\delta'|} (\delta, b, \delta') \xto{\beta} a$ and $\otimes_{|\lambda|} \lambda \xto{\omega} b$  in $\mc M$,
         \end{itemize}
         the following diagram commutes:
      \begin{center}
      \adjustbox{scale=0.9}{
        \begin{tikzcd}[row sep=huge, column sep=huge]
          \otimes_{\alpha \circledcirc \beta\circledcirc \omega}(\Gamma, \Delta, \Lambda, \Delta', \Gamma') \arrow[r, "{\zeta_{\alpha, \beta \circledcirc \omega}}"] \arrow[d, "{\zeta_{\alpha \circledcirc \beta, \omega }}"] & \otimes_\alpha(\Gamma, \otimes_{\beta \circledcirc \omega}(\Delta, \Lambda, \Delta' ), \Gamma') \arrow[d, "{\otimes_\alpha(\1_{\Gamma}, \zeta_{\beta, \omega}, \1_{\Gamma'})}"]  \\
          \otimes_{\alpha \circledcirc \beta} (\Gamma, \Delta, \otimes_\omega\Lambda, \Delta' , \Gamma') \arrow[r, "{\zeta_{\alpha, \beta}}"] & \otimes_\alpha(\Gamma, \otimes_\beta(\Delta, \otimes_\omega\Lambda, \Delta' ), \Gamma')
        \end{tikzcd}
      }
      \end{center}
      Here the top left corner is unambiguous since $(\alpha \circledcirc \beta) \circledcirc \omega = \alpha \circledcirc (\beta \circledcirc \omega)$ by \cref{lem:circledcirc-associativity}.

      \item \textbf{Coherence relation between $\zeta$ and $\otimes$, disjoint configuration.} For any lists of objects $\Gamma, \Gamma', \Gamma''$ over $\gamma, \gamma', \gamma''$, lists $\Delta$ over $\delta$ and $\Lambda$ over $\lambda$, and morphisms
\[
\otimes_{|\gamma| + 1 + |\gamma'| + 1 + |\gamma''|}(\gamma, a, \gamma', b, \gamma'') \xto{\alpha} c,
\]
$\otimes_{|\delta|}\delta \xto{\beta} a$ and $\otimes_{|\lambda|}\lambda \xto{\omega} b$ in $\mc M$, the following diagram commutes :
      \begin{center}
      \adjustbox{scale=0.9}{
        \begin{tikzcd}[row sep=huge, column sep=huge]
          \otimes_{(\alpha \circledcirc \beta)\circledcirc \omega}(\Gamma, \Delta, \Gamma', \Lambda, \Gamma'') \arrow[r, "{\zeta_{\alpha \circledcirc \beta, \omega}}"] \arrow[d, "{\zeta_{\alpha \circledcirc \omega, \beta }}"] & \otimes_{\alpha \circledcirc \beta}(\Gamma, \Delta, \Gamma', \otimes_\omega\Lambda, \Gamma'') \arrow[d, "{\zeta_{\alpha, \beta}}"]  \\
          \otimes_{\alpha \circledcirc \omega} (\Gamma, \otimes_\beta\Delta, \Gamma', \Lambda , \Gamma'') \arrow[r, "{\zeta_{\alpha, \omega}}"] & \otimes_\alpha(\Gamma, \otimes_\beta\Delta, \Gamma', \otimes_\omega\Lambda, \Gamma'')
        \end{tikzcd}
      }
      \end{center}
      where $\beta$ is composed at the position of $a$ and $\omega$ at the position of $b$ throughout; the top left corner is unambiguous since $(\alpha \circledcirc_a \beta) \circledcirc_b \omega = (\alpha \circledcirc_b \omega) \circledcirc_a \beta$ by \cref{lem:circledcirc-associativity}.

      \item \textbf{Naturality of the opcartesian lifts.} For every morphism $a \xto{\alpha} c$ in $\mc M$ and every vertical morphism $f : A \to B$ in $\mc T_a$, the following diagram commutes :
      \begin{center}
        \begin{tikzcd}[row sep=large, column sep=huge]
          A \arrow[r, "\alpha^\star_A"] \arrow[d, "f", swap] & \otimes_{\alpha \circ \iota_a}(A) \arrow[d, "{\otimes_{\alpha \circ \iota_a}(f)}"] \\
          B \arrow[r, "\alpha^\star_B"] & \otimes_{\alpha \circ \iota_a}(B)
        \end{tikzcd}
      \end{center}
      Writing $j_a : \mc T_a \hookrightarrow \mc T$ and $j_c : \mc T_c \hookrightarrow \mc T$ for the fibre inclusions, the family $(\alpha^\star_A)_{A \in \mc T_a}$ is a natural transformation
      $$\alpha^\star : j_a \Longrightarrow j_c \circ \otimes_{\alpha \circ \iota_a}$$
      between functors $\mc T_a \to \mc T$, whose components live over $\alpha$.

      \item \textbf{The opcartesian lifts compose along $\zeta$.} For all morphisms $a \xto{\alpha} b \xto{\beta} c$ in $\mc M$ and every object $A \in \mc T_a$, the following diagram commutes :
      \begin{center}
        \begin{tikzcd}[row sep=large, column sep=huge]
          A \arrow[r, "{(\beta \circ \alpha)^\star_A}"] \arrow[d, "\alpha^\star_A", swap] & \otimes_{\beta \circ \alpha \circ \iota_a}(A) \arrow[d, "{\zeta_{\beta \circ \iota_b, \alpha \circ \iota_a}}"] \\
          \otimes_{\alpha \circ \iota_a}(A) \arrow[r, "{\beta^\star_{\otimes_{\alpha \circ \iota_a}(A)}}"] & \otimes_{\beta \circ \iota_b}(\otimes_{\alpha \circ \iota_a}(A))
        \end{tikzcd}
      \end{center}
      \phantomsection\label{anchor:unary-lift-composition-base}This type-checks because $(\beta \circ \iota_b) \circledcirc (\alpha \circ \iota_a) = \beta \circ \alpha \circ \iota_a$: indeed, unfolding gives $\beta \circ \iota_b \circ \otimes_1(\alpha \circ \iota_a) \circ \eta_{1,1} = \beta \circ \alpha \circ \iota_a \circ \iota_{\otimes_1(a)} \circ \eta_{1,1} = \beta \circ \alpha \circ \iota_a$, using naturality of $\iota$ at $\alpha \circ \iota_a$ and the first unit law.
      \end{itemize}
\end{definition}

We will often use the following natural isomorphism $\tau$. For $a \in \mc M$ and $A \in \mc T_a$, the lift $(1_a)^\star_A : A \to \otimes_{\iota_a}(A)$ is vertical and opcartesian over an identity. It is therefore invertible by \hyperref[item:opcartesian-identity-lift]{the identity-lift property} (in its $1$-categorical form). We define
$$\tau_a(A) := \left((1_a)^\star_A\right)^{-1} : \otimes_{\iota_a}(A) \to A.$$
By the naturality coherence applied with $\alpha = \1_a$, the family $((1_a)^\star_A)_A$ is a natural transformation $\1_{\mc T_a} \to \otimes_{\iota_a}$, and therefore
$$\tau_a : \otimes_{\iota_a}(\--) \to (\--)$$
is a natural isomorphism on the fibre $\mc T_a$. In terms of $\tau$, the coherence between $\zeta$ and the lifts of identities reads
$$\tau_c(\otimes_\alpha(\Gamma)) \circ \zeta_{\iota_c, \alpha} = \1_{\otimes_\alpha(\Gamma)} \quad \text{and} \quad \otimes_\alpha(\1_\Gamma, \tau_a(A), \1_{\Gamma'}) \circ \zeta_{\alpha, \iota_a} = \1_{\otimes_\alpha(\Gamma, A, \Gamma')}.$$

All multi-indexed categories and all morphisms considered in this section lie over the fixed unbiased monoidal category $\mc M$. We therefore define only the fixed-base notion of morphism needed below:

\begin{definition}\label{def:strict-functor-multi-indexed-categories}
    A \textit{strict functor}
\[
(\mc T,(\mc M, (\otimes_n)_{n \in \mathbb{N}}, (\eta_{n,m})_{n,m \in \mathbb{N} }, \iota), \pi) \xto{F} (\mc T',(\mc M, (\otimes_n)_{n \in \mathbb{N}}, (\eta_{n,m})_{n,m \in \mathbb{N} }, \iota), \pi')
\]
 between two multi-indexed categories over the same unbiased monoidal category consists of:

\begin{itemize}
  \item A functor $F  : \mc T \to \mc T'$ of categories such that $\pi' \circ F = \pi$. Note that $F$ then sends the fibre $\mc T_c$ into the fibre $\mc T'_c$.
	\item For every list $\gamma := (\gamma_1, \ldots, \gamma_n)$ and every morphism $\otimes_n\gamma \xto{\alpha} c$ in $\mc M$, a natural isomorphism whose component at objects $\Gamma_i \in \mc T_{\gamma_i}$ is vertical and has the form
	\[\kappa_\alpha : \otimes'_{\alpha}(F\Gamma_1, \ldots, F\Gamma_n) \to F(\otimes_\alpha(\Gamma_1, \ldots, \Gamma_n))\]

	\item Coherence equalities for $\tau$ and $\tau'$. For every $c \in \mc M$ and $A \in \mc T_c$, the following diagram commutes :
		\begin{center}
    \begin{tikzcd}[row sep=huge, column sep=huge]
    		F(\otimes_{\iota_c} A) \arrow[r, "F\tau_c"]& FA \\
				\otimes'_{\iota_{c}}FA \arrow[u, "\kappa_{\iota_c}"] \arrow[ur, "\tau'_{c}", swap]&
  		\end{tikzcd}
  		\quad i.e. $\tau'_{c} = F\tau_c \circ \kappa_{\iota_c}$
		\end{center}
	\item Coherence between the opcartesian lifts and $\kappa$. For every $a \xto{\alpha} c$ in $\mc M$ and $A \in \mc T_a$ :
	\begin{center}
    \begin{tikzcd}[row sep=huge, column sep=huge]
    	F A \arrow[r, "F(\alpha^\star_A)"] \arrow[d, "{\alpha^\star_{F A}}", swap] & F(\otimes_{\alpha \circ \iota_a}(A)) \\
    	\otimes'_{\alpha \circ \iota_a}(F A) \arrow[ur, "\kappa_{\alpha \circ \iota_a}", swap] &
  	\end{tikzcd}
  	\quad i.e. $F(\alpha^\star_A) = \kappa_{\alpha \circ \iota_a} \circ \alpha^\star_{FA}$
	\end{center}
	Taking $\alpha = \1_c$ and inverting recovers the $\tau$-coherence of the previous bullet. We state it separately because it is used below.
	\item Coherence equalities between $\zeta$ and $\zeta'$. Let $\alpha : \otimes_n(\gamma_1, \ldots, \gamma_n) \to c$ have a marked input $\gamma_i=a$, let $\beta : \otimes_m(\delta_1, \ldots, \delta_m) \to a$, and take objects $\Gamma_j \in \mc T_{\gamma_j}$ for $j \neq i$ and $\Delta_k \in \mc T_{\delta_k}$. The following diagram commutes, with the list $\Delta_1, \ldots, \Delta_m$ inserted at the marked position :

	\begin{center}
	\adjustbox{scale=0.8}{
    \begin{tikzcd}[row sep=huge, column sep=huge]
    	\otimes'_{\alpha \circledcirc \beta }(F\Gamma_1, \ldots, F\Delta_1, \ldots, F\Delta_m, \ldots, F\Gamma_n) \arrow[r, "\kappa_{\alpha \circledcirc \beta}"] \arrow[d, "\zeta'_{\alpha, \beta }"]   & F(\otimes_{\alpha \circledcirc \beta}(\Gamma_1, \ldots, \Delta_1, \ldots, \Delta_m, \ldots, \Gamma_n)) \arrow[d, "F\zeta_{\alpha,\beta}"]  \\
    	 \otimes'_{\alpha }(F\Gamma_1, \ldots, \otimes'_{\beta }(F\Delta_1, \ldots, F\Delta_m), \ldots, F\Gamma_n) \arrow[d, "{\otimes'_{\alpha }(\1_{F\Gamma_1}, \ldots, \kappa_\beta, \ldots, \1_{F\Gamma_n})}" ] & F(\otimes_\alpha(\Gamma_1, \ldots, \otimes_\beta(\Delta_1, \ldots, \Delta_m), \ldots, \Gamma_n)) \\
    	  \otimes'_{\alpha }(F\Gamma_1, \ldots, F(\otimes_\beta(\Delta_1, \ldots, \Delta_m)), \ldots, F\Gamma_n) \arrow[ur, "\kappa_\alpha", bend right=10, swap]&
  	\end{tikzcd}
	}
	\end{center}
\end{itemize}
\end{definition}

\begin{proposition}\label{prop:category-multi-indexed-categories}
  For a fixed unbiased monoidal category $(\mc M, (\otimes_n)_{n \in \mathbb{N}}, (\eta_{n,m})_{n,m \in \mathbb{N} }, \iota)$, the collection of multi-indexed categories over $\mc M$ and strict functors between them forms a category denoted $\mathbf{MIdx_{\mc M}}$.
\end{proposition}

\textit{Proof:} We describe the category structure.
\begin{itemize}
  \item \textbf{Objects : } The objects are multi-indexed categories $(\mc T,(\mc M, (\otimes_n)_{n \in \mathbb{N} }), \pi, \otimes, (\--)^\star, \zeta)$. We write simply $\mc T$ when no confusion can arise.
	\item \textbf{Morphisms : } The morphisms are the strict functors $(\mc T,(\mc M, (\otimes_n)_{n \in \mathbb{N} }), \pi, \otimes, (\--)^\star, \zeta) \xto{F} (\mc T',(\mc M, (\otimes_n)_{n \in \mathbb{N} }), \pi', \otimes', (\--)^\star, \zeta')$.
	\item \textbf{Identity : } The identity morphism $\mc T \xto{\1_{\mc T}} \mc T$ is the identity on the underlying category. Explicitly:
		\begin{itemize}
  		\item \textbf{Action on objects :} We set $\1_{\mc T}(A) = A$ for any object $A \in \mc T$
  		\item \textbf{Action on morphisms :} We set $\1_{\mc T}(A \xto{f} B) = f$ for any morphism $f \in \mc T$.
  		\item \textbf{Natural isomorphism $\kappa_\alpha$ } We set $\kappa_\alpha = \1_{\otimes_\alpha(\--)}$
		\item \textbf{Functoriality : } The functor $\1_{\mc T}$ is strict. Its coherence diagrams reduce to the following identities:
		\begin{center}
    \begin{tikzcd}[row sep=huge, column sep=huge]
    		\otimes_{\iota_c} A \arrow[r, "\tau_c"]& A \\
				\otimes_{\iota_{c}}A \arrow[u, "\1_{\otimes_{\iota_c}A}"] \arrow[ur, "\tau_{c}", swap]&
  		\end{tikzcd}
  		\quad $\tau_{c} = \tau_c \circ \1_{\otimes_{\iota_c}A}$

    	\adjustbox{scale=0.8}{
        \begin{tikzcd}[row sep=huge, column sep=huge]
        	\otimes_{\alpha \circledcirc \beta }(\Gamma_1, \ldots, \Delta_1, \ldots, \Delta_m, \ldots, \Gamma_n) \arrow[r, "\1_{\otimes_{\alpha \circledcirc \beta}(\--)}"] \arrow[d, "\zeta_{\alpha, \beta }"]   & \otimes_{\alpha \circledcirc \beta}(\Gamma_1, \ldots, \Delta_1, \ldots, \Delta_m, \ldots, \Gamma_n) \arrow[d, "\zeta_{\alpha,\beta}"]  \\
        	 \otimes_{\alpha }(\Gamma_1, \ldots, \otimes_{\beta }(\Delta_1, \ldots, \Delta_m), \ldots, \Gamma_n) \arrow[d, "{\otimes_{\alpha }(\1_{\Gamma_1}, \ldots, \1_{\otimes_\beta(\--)} , \ldots, \1_{\Gamma_n}) = \1_{\otimes_\alpha(\Gamma_1, \ldots, \otimes_\beta(\--), \ldots, \Gamma_n)}}" ] & \otimes_\alpha(\Gamma_1, \ldots, \otimes_\beta(\Delta_1, \ldots, \Delta_m), \ldots, \Gamma_n) \\
        	  \otimes_{\alpha }(\Gamma_1, \ldots, \otimes_\beta(\Delta_1, \ldots, \Delta_m), \ldots, \Gamma_n) \arrow[ur, "\1_{\otimes_\alpha(\--)}", bend right=10, swap]&
      	\end{tikzcd}
    	}
		\end{center}
		The coherence with the opcartesian lifts is also immediate: $\1_{\mc T}(\alpha^\star_A) = \alpha^\star_A = \1_{\otimes_{\alpha \circ \iota_a}(A)} \circ \alpha^\star_{A} = \kappa_{\alpha \circ \iota_a} \circ \alpha^\star_{\1_{\mc T}A}$.

	\end{itemize}
	\item \textbf{Composition: } Given strict functors $\mc T \xto{F} \mc T'$ and $\mc T' \xto{G} \mc T''$, define $G \circ F$ as follows:
	\begin{itemize}
		\item \textbf{Action on objects and morphisms:} This is ordinary functor composition. For any object $A$ and morphism $f$ in $\mc T$:
		\[ (G \circ F)(A) = G(FA) \quad \text{and} \quad (G \circ F)(f) = G(Ff) \]
		Note that $\pi'' \circ (G \circ F) = \pi' \circ F = \pi$.
		\item \textbf{Natural isomorphism $\kappa^{G \circ F}_\alpha$:} Compose the structural isomorphisms $\kappa^G_\alpha : \otimes_\alpha''(GF\Gamma) \to G(\otimes_\alpha'(F\Gamma))$ and $\kappa^F_\alpha : \otimes_\alpha'(F\Gamma) \to F(\otimes_\alpha(\Gamma))$. Thus:
		\[ \kappa^{G \circ F}_\alpha := G(\kappa^F_\alpha) \circ \kappa^G_\alpha : \otimes_\alpha''(GF\Gamma_1, \ldots, GF\Gamma_n) \to GF(\otimes_\alpha(\Gamma_1, \ldots, \Gamma_n)). \]
		This is a natural isomorphism with vertical components.
		\item \textbf{The composite satisfies the coherences:} For the coherence with the opcartesian lifts, compute:
		\begin{align*}
			(G \circ F)(\alpha^\star_A) &= G(F(\alpha^\star_A)) \\
			&= G(\kappa^F_{\alpha \circ \iota_a} \circ \alpha^\star_{FA}) &\text{Coherence for } F\\
			&= G(\kappa^F_{\alpha \circ \iota_a}) \circ G(\alpha^\star_{FA}) &\text{Functoriality of } G\\
			&= G(\kappa^F_{\alpha \circ \iota_a}) \circ \kappa^G_{\alpha \circ \iota_a} \circ \alpha^\star_{GFA} &\text{Coherence for } G\\
			&= \kappa^{G \circ F}_{\alpha \circ \iota_a} \circ \alpha^\star_{(G\circ F)A} &\text{By definition}
		\end{align*}
		The case $\alpha = \1_c$ gives the $\tau$-coherence. For the coherence between $\zeta$ and $\zeta''$, compute:
		\begin{align*}
&GF(\zeta_{\alpha, \beta}) \circ \kappa^{G\circ F}_{\alpha \circledcirc \beta} \\
&\quad = GF(\zeta_{\alpha,\beta}) \circ G(\kappa^F_{\alpha \circledcirc \beta}) \circ \kappa^G_{\alpha \circledcirc \beta} \\
&\quad \qquad \text{By definition} \\
&\quad = G\big(F(\zeta_{\alpha,\beta}) \circ \kappa^F_{\alpha \circledcirc \beta}\big) \circ \kappa^G_{\alpha \circledcirc \beta} \\
&\quad \qquad \text{Functoriality of } G \\
&\quad = G\big(\kappa^F_\alpha \circ \otimes'_\alpha(\1, \kappa^F_\beta, \1) \circ \zeta'_{\alpha, \beta}\big) \circ \kappa^G_{\alpha \circledcirc \beta} \\
&\quad \qquad \text{Coherence for } F \\
&\quad = G(\kappa^F_\alpha) \circ G(\otimes'_\alpha(\1, \kappa^F_\beta, \1)) \circ G(\zeta'_{\alpha,\beta}) \circ \kappa^G_{\alpha \circledcirc \beta} \\
&\quad \qquad \text{Functoriality of } G \\
&\quad = G(\kappa^F_\alpha) \circ G(\otimes'_\alpha(\1, \kappa^F_\beta, \1)) \circ \kappa^G_\alpha \circ \otimes''_\alpha(\1, \kappa^G_\beta, \1) \circ \zeta''_{\alpha, \beta} \\
&\quad \qquad \text{Coherence for } G \text{ at } \zeta' \\
&\quad = G(\kappa^F_\alpha) \circ \kappa^G_\alpha \circ \otimes''_\alpha(\1, G(\kappa^F_\beta), \1) \circ \otimes''_\alpha(\1, \kappa^G_\beta, \1) \circ \zeta''_{\alpha, \beta} \\
&\quad \qquad \text{Naturality of } \kappa^G_\alpha \\
&\quad = \kappa^{G \circ F}_\alpha \circ \otimes''_\alpha(\1, G(\kappa^F_\beta) \circ \kappa^G_\beta, \1) \circ \zeta''_{\alpha, \beta} \\
&\quad \qquad \text{Functoriality of } \otimes''_\alpha \\
&\quad = \kappa^{G \circ F}_\alpha \circ \otimes''_\alpha(\1, \kappa^{G \circ F}_\beta, \1) \circ \zeta''_{\alpha, \beta} \\
&\quad \qquad \text{By definition}
\end{align*}
		This proves that $G \circ F$ is a strict functor of multi-indexed categories.
	\end{itemize}
	\item \textbf{Identity laws:} Let $\mc T \xto{F} \mc T'$ be a strict functor. The identity laws hold for the underlying functors. It remains to check that $\kappa_\alpha^{F \circ \1_{\mc T}} = \kappa_\alpha^F$ and $\kappa^{\1_{\mc T'} \circ F}_\alpha = \kappa_\alpha^F$.
		\begin{itemize}
			\item \textbf{Right identity:} For $F \circ \1_{\mc T}$, the isomorphism is computed as :
			\begin{align*}
				\kappa^{F \circ \1_{\mc T}}_\alpha  &= F(\kappa^{\1_{\mc T}}_\alpha) \circ \kappa^F_\alpha \\
				&= F(\1_{\otimes_\alpha(\--) }) \circ \kappa^F_\alpha \\
				&= \1_{F(\otimes_\alpha(\--) )} \circ \kappa^F_\alpha \\
				&= \kappa^F_\alpha
			\end{align*}
			\item \textbf{Left identity:} For $\1_{\mc T'} \circ F$, the isomorphism is computed as :
			\begin{align*}
				\kappa^{\1_{\mc T'} \circ F}_\alpha  &= \1_{\mc T'}(\kappa^F_\alpha) \circ \kappa^{\1_{\mc T'}}_\alpha \\
				&= \kappa^F_\alpha \circ \1_{\otimes'_\alpha(F(\--)) } \\
				&= \kappa^F_\alpha
			\end{align*}
		\end{itemize}
		The identity laws follow.

	\item \textbf{Associativity law:} Let $\mc T \xto{F} \mc T' \xto{G} \mc T'' \xto{H} \mc T'''$ be strict functors. Associativity holds for the underlying functors. For the structural isomorphisms, we must show $\kappa^{(H \circ G) \circ F}_\alpha = \kappa^{H \circ (G \circ F)}_\alpha$. We compute:

		\begin{align*}
&\kappa^{(H \circ G) \circ F}_\alpha \\
&\quad = (H \circ G)(\kappa^F_\alpha) \circ \kappa^{H \circ G}_\alpha \\
&\quad \qquad \text{By definition} \\
&\quad = H(G(\kappa^F_\alpha)) \circ \Big( H(\kappa^G_\alpha) \circ \kappa^H_\alpha \Big) \\
&\quad \qquad \text{By definition} \\
&\quad = \Big(H(G(\kappa^F_\alpha)) \circ  H(\kappa^G_\alpha) \Big ) \circ \kappa^H_\alpha \\
&\quad \qquad \text{Composition of morphisms is associative} \\
&\quad = H\Big(G(\kappa^F_\alpha) \circ  \kappa^G_\alpha \Big ) \circ \kappa^H_\alpha \\
&\quad \qquad \text{By functoriality} \\
&\quad = H(\kappa^{G \circ F}_\alpha) \circ \kappa^H_\alpha \\
&\quad \qquad \text{} \\
&\quad = \kappa^{H \circ (G \circ F)}_\alpha \\
&\quad \qquad \text{By definition}
\end{align*}
	$\square$
\end{itemize}

\subsection{From Multi-Indexed Categories to Cloven Opfibrations}\label{subsec:multi-indexed-to-opfibrations}

\begin{proposition}\label{prop:multi-indexed-to-cloven-opfibrations}
	There is a functor between the category of multi-indexed categories over an unbiased monoidal category $\mc M$ and the category of cloven opfibrations of multicategories over the underlying multicategory $\mc M_m$.
	\[\mathbf{MIdx_{\mc M}} \to \mathbf{Multi_{\mc M_m}}\]
\end{proposition}

 Let $(\mc T,(\mc M, (\otimes_n)_{n \in \mathbb{N} }), \pi)$ be a multi-indexed category. We define its underlying cloven opfibration $\pi_m: \fib[T_m]{M_m}$ by specifying $\mc T_m$ and its cleavage.

\begin{itemize}
  \item \textbf{Objects : } The objects of $\mc T_m$ are the objects of $\mc T$. We denote by $A_m$ the object in $\mc T_m$ whenever $A$ is an object in $\mc T$.
  \item \textbf{Morphisms :} For a list of objects $\Gamma_m := (\Gamma_1)_m, \ldots, (\Gamma_n)_m$, a multimorphism $\Gamma_m \xto{f_m} A_m$ in $\mc T_m$  consists of a tuple $f_m = (f, \alpha)$ where $\otimes_n( \pi \Gamma_1, \ldots, \pi \Gamma_n)  \xto{\alpha} \pi A$ is a morphism in $\mc M$ and $\otimes_\alpha(\Gamma_1, \ldots, \Gamma_n) \xto{f} A$ is a \textit{vertical} morphism of $\mc T$, i.e.\ a morphism of the fibre $\mc T_{\pi A}$. We will use the shorthand notation $\Gamma$ for the list $\Gamma_1, \ldots, \Gamma_n$ and $\pi \Gamma$ for $\pi \Gamma_1, \ldots, \pi \Gamma_n$ when no confusion arises.
  \item \textbf{Functor $\mc T_m \xto{\pi_m} \mc M_m$ } :
		\begin{itemize}
  		\item The action of $\pi_m$ on objects is $\pi_m(A_m) = (\pi A)_m$
  		\item The action of $\pi_m$ on morphisms is $\pi_m(\Gamma_m \xto{(f, \alpha)} A_m) = \alpha_m$ where $\alpha_m$ is the underlying multimorphism $\pi_m \Gamma_m \xto{\alpha_m} \pi_m A_m $ in $\mc M_m$ of $\otimes_n(\pi \Gamma) \xto{\alpha} \pi A$ in $\mc M$.
		\item Once identities and composition are defined below, $\pi_m$ is a functor of multicategories: $\pi_m(\1_{A_m}) = (\iota_{\pi A})_m = \1_{(\pi A)_m}$, and $\pi_m\big((f, \beta) \circ (g, \alpha)\big) = (\beta \circledcirc \alpha)_m = \beta_m \circ \alpha_m = \pi_m(f, \beta) \circ \pi_m(g, \alpha)$, by the definition of composition in $\mc M_m$.
		\end{itemize}

    \item \textbf{Identities :} Given an object $A_m \in \mc T_m$ we define the identity morphism as
\[
\1_{A_m} = (\tau_{\pi A}(A), \iota_{\pi A}).
\]
Note that
\[
\tau_{\pi A}(A) : \otimes_{\iota_{\pi A}}(A) \to A
\]
 is vertical, as required.
  \item \textbf{Composition :}

		\begin{itemize}
        \item \textbf{Definition : } Given multimorphisms $\Gamma_m \xto{(g, \alpha)} A_m$ with $|\Gamma_m| = n$ and
\[
(\Gamma_1)_m, A_m, (\Gamma_2)_m \xto{(f, \beta)} B_m
\]
 with $|(\Gamma_1)_m, A_m, (\Gamma_2)_m| = k$, define $(f, \beta) \circ (g , \alpha)$ by
        \[
\begin{aligned}
&(f, \beta) \circ (g , \alpha)
\\
&\quad := (f \circ \otimes_{\beta}(\1_{\Gamma_1}, g, \1_{\Gamma_2}) \circ \zeta_{\beta, \alpha}, \beta \circ \otimes_{k}(\1_{\pi \Gamma_1}, \alpha, \1_{\pi \Gamma_2}) \circ \eta_{k,n})
\\
&\quad = (f \circ \otimes_{\beta}(\1_{\Gamma_1}, g, \1_{\Gamma_2}) \circ \zeta_{\beta, \alpha}, \beta \circledcirc \alpha).
\end{aligned}
\]
			The first component is constructed as follows. The maps $f_m$ and $g_m$ give vertical maps $$\otimes_\alpha\Gamma \xto{g} A \quadt{and} \otimes_\beta(\Gamma_1, A, \Gamma_2) \xto{f} B \quadt{in} \mc T$$ as well as $\otimes_n(\pi\Gamma) \xto{\alpha} \pi A$ and $
 \otimes_{k}(\pi\Gamma_1, \pi A, \pi\Gamma_2) \xto{\beta} \pi B$ in $\mc M$. The associator $\eta_{k,n}$ gives $ \otimes_{k + n - 1}(\pi\Gamma_1, \pi\Gamma, \pi\Gamma_2) \xto{\eta_{k,n}}  \otimes_{k}(\pi\Gamma_1, \otimes_n(\pi\Gamma), \pi\Gamma_2)$. The corresponding base morphism is \[
\begin{aligned}
&\otimes_{k + n - 1}(\pi\Gamma_1, \pi\Gamma, \pi\Gamma_2)\\
&\quad \xto{\eta_{k,n}}  \otimes_{k }(\pi\Gamma_1, \otimes_n(\pi\Gamma), \pi\Gamma_2)\\
&\quad \xto{\otimes_{k}(\1_{\pi\Gamma_1}, \alpha, \1_{\pi\Gamma_2})} \otimes_{k}(\pi\Gamma_1, \pi A, \pi\Gamma_2)\\
&\quad \xto{\beta} \pi B.
\end{aligned}
\]
 The structure on $\mc T$ provides a functor
 $$\otimes_{\beta \circledcirc \alpha } : \left(\prod_{X \in \Gamma_1} \mc T_{\pi X}\right) \times \left(\prod_{X \in \Gamma} \mc T_{\pi X}\right) \times \left(\prod_{X \in \Gamma_2} \mc T_{\pi X}\right) \to \mc T_{\pi B},$$
 where each product is over the entries of the indicated list and an empty product is $\mathbf 1$. The first component is:
$$\otimes_{\beta \circledcirc \alpha }(\Gamma_1, \Gamma, \Gamma_2) \xto{\zeta_{\beta, \alpha}} \otimes_{\beta}(\Gamma_1, \otimes_{\alpha}\Gamma, \Gamma_2) \xto{\otimes_{\beta}(\1_{\Gamma_1}, g, \1_{\Gamma_2})} \otimes_{\beta}(\Gamma_1, A, \Gamma_2) \xto{f} B.$$
All three morphisms in this composite are vertical, so the first component of $(f, \beta) \circ (g, \alpha)$ is indeed a morphism of the fibre $\mc T_{\pi B}$.
		\item \textbf{Identity laws :} Given morphisms $\Gamma_m \xto{g_m} A_m$ and $(\Gamma_1)_m, A_m, (\Gamma_2)_m \xto{f_m} B_m$, we show that $f_m \circ 1_{A_m} = f_m$ and $1_{A_m} \circ g_m = g_m$. By definition,
		$$\1_{A_m} \circ g_m = (\tau_{\pi A}(A), \iota_{\pi A}) \circ (g, \alpha) = (\tau_{\pi A}(A) \circ \otimes_{\iota_{\pi A}}(g)  \circ \zeta_{\iota_{\pi A}, \alpha}, \iota_{\pi A} \circledcirc \alpha).$$
		The isomorphism $\tau_{\pi A} : \otimes_{\iota_{\pi A}}(\--) \to (\--)$ is natural, so the following diagram commutes:
		\begin{center}
      \begin{tikzcd}[row sep=large, column sep=huge]
        \otimes_{\iota_{\pi A}}(\otimes_\alpha\Gamma) \arrow[r, "\tau_{\pi A}(\otimes_\alpha\Gamma)"] \arrow[d, "\otimes_{\iota_{\pi A}}(g)"] & \otimes_\alpha\Gamma \arrow[d, "g"] \\
        \otimes_{\iota_{\pi A}}(A) \arrow[r, "\tau_{\pi A}(A)"]  & A \\
      \end{tikzcd}
		\end{center}
		Thus, $\tau_{\pi A}(A) \circ \otimes_{\iota_{\pi A}}(g) = g \circ \tau_{\pi A}(\otimes_\alpha\Gamma)$. Substituting this gives:
		$$\1_{A_m} \circ g_m = (g \circ \tau_{\pi A}(\otimes_\alpha\Gamma) \circ \zeta_{\iota_{\pi A}, \alpha}, \iota_{\pi A} \circledcirc \alpha)$$
		The coherence between $\zeta$ and the lifts of identities, in its $\tau$ form, gives $\tau_{\pi A}(\otimes_\alpha\Gamma) \circ \zeta_{\iota_{\pi A}, \alpha} = \1_{\otimes_\alpha\Gamma}$. Hence:
		$$\1_{A_m} \circ g_m = (g \circ \1_{\otimes_\alpha\Gamma}, \iota_{\pi A} \circledcirc \alpha) = (g, \iota_{\pi A} \circledcirc \alpha)$$
		\Cref{lem:circledcirc-units} gives $\iota_{\pi A} \circledcirc \alpha = \alpha$, so $\1_{A_m} \circ g_m = (g, \alpha) = g_m$.
		For the right unit law, we compute by definition of the composition :
		$$f_m \circ \1_{A_m} = (f, \beta) \circ (\tau_{\pi A}(A), \iota_{\pi A}) = (f \circ \otimes_\beta(\1_{\Gamma_1}, \tau_{\pi A}(A), \1_{\Gamma_2}) \circ \zeta_{\beta, \iota_{\pi A}}, \beta \circledcirc \iota_{\pi A}).$$
		The second coherence equation between $\zeta$ and the lifts of identities, in its $\tau$ form, reads $\otimes_\beta(\1_{\Gamma_1}, \tau_{\pi A}(A), \1_{\Gamma_2}) \circ \zeta_{\beta, \iota_{\pi A}} = \1_{\otimes_\beta(\Gamma_1, A, \Gamma_2)}$, and \cref{lem:circledcirc-units} gives $\beta \circledcirc \iota_{\pi A} = \beta$. Hence $f_m \circ \1_{A_m} = (f, \beta) = f_m$.

    \item \textbf{Associativity, nested configuration:} For composable $g_m, f_m, h_m$ in $\mc T_m$, we must show:
\[h_m \circ (f_m \circ g_m) = (h_m \circ f_m) \circ g_m\]

Write their domains, codomains, and base morphisms as follows:
\begin{itemize}
    \item $g_m = (g, \alpha) : \Gamma_m \to A_m$ where $\otimes_\alpha(\Gamma) \xto{g} A$ and $\otimes_{n}(\gamma) \xto{\alpha} a$.
    \item $f_m = (f, \beta) : (\Gamma_1)_m, A_m, (\Gamma_2)_m \to B_m$ where $\otimes_\beta(\Gamma_1, A, \Gamma_2) \xto{f} B$ and
\[
\otimes_{|\gamma_1| + 1 + |\gamma_2|}(\gamma_1, a, \gamma_2) \xto{\beta} b.
\]

    \item $h_m = (h, \omega) : (\Delta_1)_m, B_m, (\Delta_2)_m \to C_m$ where $\otimes_\omega(\Delta_1, B, \Delta_2) \xto{h} C$ and
\[
\otimes_{|\delta_1| + 1 + |\delta_2|}(\delta_1, b, \delta_2) \xto{\omega} c.
\]

\end{itemize}

Recall the \hyperref[anchor:unbiased-monoidal-circledcirc]{shorthand for the underlying composition} in the unbiased monoidal category:
\begin{align*}
    \beta \circledcirc \alpha &:= \beta \circ \otimes_{|\gamma_1| + 1 + |\gamma_2|}(\1_{\gamma_1}, \alpha, \1_{\gamma_2}) \circ \eta_{|\gamma_1| + 1 + |\gamma_2|, n} \\
    \omega \circledcirc \beta &:= \omega \circ \otimes_{|\delta_1| + 1 + |\delta_2|}(\1_{\delta_1}, \beta, \1_{\delta_2}) \circ \eta_{|\delta_1| + 1 + |\delta_2|, |\gamma_1| + 1 + |\gamma_2|}
\end{align*}

By definition, $f_m \circ g_m$ is $(f \circ \otimes_\beta(\1_{\Gamma_1}, g, \1_{\Gamma_2}) \circ \zeta_{\beta, \alpha}, \ \beta \circledcirc \alpha)$. Composing this with $h_m$ gives:
\begin{align*}
    h_m \circ &(f_m \circ g_m) \\
    &= \left(h \circ \otimes_\omega\left(\1_{\Delta_1}, f \circ \otimes_\beta(\1_{\Gamma_1}, g, \1_{\Gamma_2}) \circ \zeta_{\beta, \alpha}, \1_{\Delta_2}\right) \circ \zeta_{\omega, \beta \circledcirc \alpha}, \ \omega \circledcirc (\beta \circledcirc \alpha)\right)
\end{align*}

\Cref{lem:circledcirc-associativity} gives $\omega \circledcirc (\beta \circledcirc \alpha) = (\omega \circledcirc \beta) \circledcirc \alpha$. The second components therefore agree.

For the first component, functoriality of $\otimes_\omega$ gives:
\begin{align*}
    = \Bigg(h \circ \otimes_\omega(\1_{\Delta_1}, f, \1_{\Delta_2}) \circ \otimes_\omega\left(\1_{\Delta_1}, \otimes_\beta(\1_{\Gamma_1}, g, \1_{\Gamma_2}), \1_{\Delta_2}\right) \circ \otimes_\omega(\1_{\Delta_1}, \zeta_{\beta, \alpha}, \1_{\Delta_2}) \circ \zeta_{\omega, \beta \circledcirc \alpha}, \\
    \quad (\omega \circledcirc \beta) \circledcirc \alpha\Bigg)
\end{align*}

The nested coherence law between $\zeta$ and $\otimes$ is the following commuting diagram:
\begin{center}
\begin{adjustbox}{scale=0.85}
    \begin{tikzcd}[row sep=huge, column sep=huge]
        \otimes_{\omega \circledcirc (\beta \circledcirc \alpha)}(\Delta_1, \Gamma_1, \Gamma, \Gamma_2, \Delta_2) \arrow[r, "{\zeta_{\omega, \beta \circledcirc \alpha}}"] \arrow[d, "{\zeta_{\omega \circledcirc \beta, \alpha}}", swap] & \otimes_\omega\left(\Delta_1, \otimes_{\beta \circledcirc \alpha}(\Gamma_1, \Gamma, \Gamma_2), \Delta_2\right) \arrow[d, "{\otimes_\omega(\1_{\Delta_1}, \zeta_{\beta, \alpha}, \1_{\Delta_2})}"] \\
        \otimes_{\omega \circledcirc \beta}\left(\Delta_1, \Gamma_1, \otimes_\alpha(\Gamma), \Gamma_2, \Delta_2\right) \arrow[r, "{\zeta_{\omega, \beta}}", swap] & \otimes_\omega\left(\Delta_1, \otimes_\beta(\Gamma_1, \otimes_\alpha(\Gamma), \Gamma_2), \Delta_2\right)
    \end{tikzcd}
\end{adjustbox}
\end{center}

The diagram gives $\otimes_\omega(\1_{\Delta_1}, \zeta_{\beta, \alpha}, \1_{\Delta_2}) \circ \zeta_{\omega, \beta \circledcirc \alpha} = \zeta_{\omega, \beta} \circ \zeta_{\omega \circledcirc \beta, \alpha}$. The composition becomes:
\begin{align*}
    = \Bigg(h \circ \otimes_\omega(\1_{\Delta_1}, f, \1_{\Delta_2}) \circ \otimes_\omega\left(\1_{\Delta_1}, \otimes_\beta(\1_{\Gamma_1}, g, \1_{\Gamma_2}), \1_{\Delta_2}\right) \circ \zeta_{\omega, \beta} \circ \zeta_{\omega \circledcirc \beta, \alpha}, \\
    \quad (\omega \circledcirc \beta) \circledcirc \alpha\Bigg)
\end{align*}

Naturality of $\zeta_{\omega, \beta}$ at $g$ gives the commuting square:
\begin{center}
\begin{adjustbox}{scale=0.85}
    \begin{tikzcd}[row sep=huge, column sep=huge]
        \otimes_{\omega \circledcirc \beta}(\Delta_1, \Gamma_1, \otimes_\alpha(\Gamma), \Gamma_2, \Delta_2) \arrow[r, "{\zeta_{\omega, \beta}}"] \arrow[d, "{\otimes_{\omega \circledcirc \beta}(\1_{\Delta_1}, \1_{\Gamma_1}, g, \1_{\Gamma_2}, \1_{\Delta_2})}", swap] & \otimes_\omega\left(\Delta_1, \otimes_\beta(\Gamma_1, \otimes_\alpha(\Gamma), \Gamma_2), \Delta_2\right) \arrow[d, "{\otimes_\omega\left(\1_{\Delta_1}, \otimes_\beta(\1_{\Gamma_1}, g, \1_{\Gamma_2}), \1_{\Delta_2}\right)}"] \\
        \otimes_{\omega \circledcirc \beta}(\Delta_1, \Gamma_1, A, \Gamma_2, \Delta_2) \arrow[r, "{\zeta_{\omega, \beta}}", swap] & \otimes_\omega\left(\Delta_1, \otimes_\beta(\Gamma_1, A, \Gamma_2), \Delta_2\right)
    \end{tikzcd}
\end{adjustbox}
\end{center}

This gives $\otimes_\omega\left(\1_{\Delta_1}, \otimes_\beta(\1_{\Gamma_1}, g, \1_{\Gamma_2}), \1_{\Delta_2}\right) \circ \zeta_{\omega, \beta} = \zeta_{\omega, \beta} \circ \otimes_{\omega \circledcirc \beta}(\1_{\Delta_1}, \1_{\Gamma_1}, g, \1_{\Gamma_2}, \1_{\Delta_2})$, so the expression becomes:
\begin{align*}
    = \Bigg(h \circ \otimes_\omega(\1_{\Delta_1}, f, \1_{\Delta_2}) \circ \zeta_{\omega, \beta} \circ \otimes_{\omega \circledcirc \beta}(\1_{\Delta_1}, \1_{\Gamma_1}, g, \1_{\Gamma_2}, \1_{\Delta_2}) \circ \zeta_{\omega \circledcirc \beta, \alpha}, \\
    \quad (\omega \circledcirc \beta) \circledcirc \alpha\Bigg)
\end{align*}

The first three factors form the first component of $h_m \circ f_m$. The last two are those used to compose with $g_m$. The resulting composite is $(h_m \circ f_m) \circ g_m$, so
$$h_m \circ (f_m \circ g_m) = (h_m \circ f_m) \circ g_m.$$

		\item \textbf{Associativity, disjoint configuration:} Suppose now that we are given multimorphisms $f_m = (f, \alpha) : \Gamma_m \to A_m$ and $g_m = (g, \beta) : \Delta_m \to B_m$, and a multimorphism $h_m = (h, \chi) : (\Lambda_1)_m, A_m, (\Lambda_2)_m, B_m, (\Lambda_3)_m \to C_m$ having both $A_m$ and $B_m$ in its domain. We must show
		$$(h_m \circ_{A_m} f_m) \circ_{B_m} g_m = (h_m \circ_{B_m} g_m) \circ_{A_m} f_m.$$
		We compute the left-hand side. First,
		$$h_m \circ_{A_m} f_m = \big(h \circ \otimes_\chi(\1_{\Lambda_1}, f, \1_{\Lambda_2}, \1_B, \1_{\Lambda_3}) \circ \zeta_{\chi, \alpha}, \ \chi \circledcirc_a \alpha\big)$$
		where the subscript on $\circledcirc$ records the position at which the composition happens. Composing with $g_m$ at the position of $B_m$ gives the first component
		$$h \circ \otimes_\chi(\1_{\Lambda_1}, f, \1_{\Lambda_2}, \1_B, \1_{\Lambda_3}) \circ \zeta_{\chi, \alpha} \circ \otimes_{\chi \circledcirc \alpha}(\1_{\Lambda_1}, \1_{\Gamma}, \1_{\Lambda_2}, g, \1_{\Lambda_3}) \circ \zeta_{\chi \circledcirc \alpha, \beta}$$
		over the base $(\chi \circledcirc_a \alpha) \circledcirc_b \beta$. By naturality of $\zeta_{\chi, \alpha}$ in the argument sitting at the position of $b$ we have
		$$\zeta_{\chi, \alpha} \circ \otimes_{\chi \circledcirc \alpha}(\1_{\Lambda_1}, \1_{\Gamma}, \1_{\Lambda_2}, g, \1_{\Lambda_3}) = \otimes_\chi(\1_{\Lambda_1}, \1_{\otimes_\alpha\Gamma}, \1_{\Lambda_2}, g, \1_{\Lambda_3}) \circ \zeta_{\chi, \alpha}$$
		where on the left the component of $\zeta_{\chi,\alpha}$ is taken at $(\Lambda_1, \Gamma, \Lambda_2, B, \Lambda_3)$ and on the right at $(\Lambda_1, \Gamma, \Lambda_2, \otimes_\beta\Delta, \Lambda_3)$. Using this and the functoriality of $\otimes_\chi$ to merge the two tuples of morphisms, the first component becomes
		$$h \circ \otimes_\chi(\1_{\Lambda_1}, f, \1_{\Lambda_2}, g, \1_{\Lambda_3}) \circ \zeta_{\chi, \alpha} \circ \zeta_{\chi \circledcirc_a \alpha, \beta}.$$
		The computation of the right-hand side is symmetric and gives
		$$h \circ \otimes_\chi(\1_{\Lambda_1}, f, \1_{\Lambda_2}, g, \1_{\Lambda_3}) \circ \zeta_{\chi, \beta} \circ \zeta_{\chi \circledcirc_b \beta, \alpha}$$
		over the base $(\chi \circledcirc_b \beta) \circledcirc_a \alpha$. The two bases agree by the disjoint configuration of \cref{lem:circledcirc-associativity}. The first components agree by the disjoint configuration coherence of the multi-indexed category:
		$$\zeta_{\chi, \alpha} \circ \zeta_{\chi \circledcirc_a \alpha, \beta} = \zeta_{\chi, \beta} \circ \zeta_{\chi \circledcirc_b \beta, \alpha}.$$
		The two composites are equal.

	\end{itemize}
\end{itemize}
We next show that this specification defines a cloven opfibration of multicategories. We construct the chosen opcartesian lifts and prove their universal property.

\begin{itemize}
  \item \textbf{Existence of opcartesian morphisms :} Let $\Gamma_m$ be a list of objects in $\mc T_m$ and let $\pi_m \Gamma_m \xto{\alpha_m} q_m$ be a multimorphism in $\mc M_m$, as in the following diagram:
  \begin{center}
    \begin{tikzcd}[row sep=large, column sep=huge]
      \Gamma_m & \\
      \pi_m \Gamma_m \arrow[r, "\alpha_m"] & q_m
    \end{tikzcd}
  \end{center}
 We seek an object $B_m$ and an opcartesian map $\Gamma_m \xto{g_m} B_m$ over $\alpha_m$:
  \begin{center}
    \begin{tikzcd}[row sep=large, column sep=huge]
      \Gamma_m \arrow[r, "g_m"] & B_m \\
      \pi_m \Gamma_m \arrow[r, "\alpha_m"] & q_m
    \end{tikzcd}
  \end{center}
   Take $B_m = (\otimes_\alpha \Gamma)_m$. We claim that there is a map $g_m$ as follows:
  \begin{center}
    \begin{tikzcd}[row sep=large, column sep=huge]
      \Gamma_m \arrow[r, "g_m"] & (\otimes_\alpha \Gamma)_m \\
      \pi_m \Gamma_m \arrow[r, "\alpha_m"] & q_m
    \end{tikzcd}
  \end{center}
  To give such a map $g_m$ is to give a vertical map $\otimes_\alpha \Gamma \xto{g} \otimes_\alpha \Gamma$ in $\mc T$ together with $\otimes_n(\pi \Gamma) \xto{\alpha} q$ . We take $g = \1_{\otimes_\alpha\Gamma}$. Thus $g_m = (\1_{\otimes_\alpha\Gamma}, \alpha)$, and $\pi_m(g_m) = \alpha_m$ as required.

	For the universal property, let $C_m$ be an object and let $(\Gamma_1)_m, (\Gamma_2)_m$ be lists of objects. Let $(\Gamma_1)_m, \Gamma_m,(\Gamma_2)_m \xto{(f, \sigma)} C_m$ be a morphism in $\mc T_m$. Also let $\pi_m(\Gamma_1)_m, q_m, \pi_m(\Gamma_2)_m \xto{\beta_m} \pi_m(C_m)$ satisfy $\beta_m \circ \alpha_m = \sigma_m$, as in:
  \begin{center}
    \begin{tikzcd}[row sep=large, column sep=huge]
      \underset{(\Gamma_1)_m, \Gamma_m, (\Gamma_2)_m}{\Gamma_m} \arrow[rr, "{(\1_{\otimes_\alpha\Gamma}, \alpha)}"] \arrow[dr, swap,  "{(f, \sigma)}"] & & (\otimes_\alpha \Gamma)_m \\
      & C_m   & \\
      \underset{\pi_m (\Gamma_1)_m, \pi_m \Gamma_m, \pi_m (\Gamma_2)_m}{\pi_m \Gamma_m} \arrow[rr, "\alpha_m"] \arrow[dr, swap,  "\sigma_m"]  &  & \underset{\pi_m(\Gamma_1)_m, q_m, \pi_m(\Gamma_2)_m}{q_m} \arrow[dl, "\beta_m"]  \\
      & \pi_m C_m &
    \end{tikzcd}
  \end{center}

	We want to find a map $(\Gamma_1)_m, (\otimes_\alpha \Gamma)_m, (\Gamma_2)_m \to C_m$ over $\beta_m$  such that the top diagram commutes.

  \begin{center}
    \begin{tikzcd}[row sep=large, column sep=huge]
      \underset{(\Gamma_1)_m, \Gamma_m, (\Gamma_2)_m}{\Gamma_m} \arrow[rr, "{(\1_{\otimes_\alpha\Gamma}, \alpha)}"] \arrow[dr, swap,  "{(f, \sigma)}"] & & \underset{(\Gamma_1)_m, (\otimes_\alpha \Gamma)_m, (\Gamma_2)_m}{(\otimes_\alpha \Gamma)_m} \arrow[dl, ""]  \\
      & C_m   & \\
      \underset{\pi_m (\Gamma_1)_m, \pi_m \Gamma_m, \pi_m (\Gamma_2)_m}{\pi_m \Gamma_m} \arrow[rr, "\alpha_m"] \arrow[dr, swap,  "\sigma_m"]  &  & \underset{\pi_m(\Gamma_1)_m, q_m, \pi_m(\Gamma_2)_m}{q_m} \arrow[dl, "\beta_m"]  \\
      & \pi_m C_m &
    \end{tikzcd}
  \end{center}

  It suffices to give a vertical map $\otimes_\beta(\Gamma_1, \otimes_\alpha \Gamma, \Gamma_2) \to C$. We have:
  $$
  \otimes_\beta(\Gamma_1, \otimes_\alpha\Gamma, \Gamma_2) \xto{\zeta_{\beta, \alpha}^{-1}} \otimes_{\beta \circledcirc \alpha}(\Gamma_1, \Gamma, \Gamma_2)
  $$
  The equation $\beta_m \circ \alpha_m = \sigma_m$ in $\mc M_m$ means that
  $$\sigma = \beta \circ \otimes_k(\1_{\pi \Gamma_1}, \alpha, \1_{\pi \Gamma_2}) \circ \eta_{k, n} = \beta \circledcirc \alpha,$$
  where $|\pi \Gamma_1, q, \pi \Gamma_2| = k$ and $|\pi \Gamma| = n$. We can therefore form the composite
  $$
  \otimes_\beta(\Gamma_1, \otimes_\alpha\Gamma, \Gamma_2) \xto{\zeta_{\beta, \alpha}^{-1}} \otimes_{\beta \circledcirc \alpha}(\Gamma_1, \Gamma, \Gamma_2) = \otimes_\sigma(\Gamma_1, \Gamma, \Gamma_2) \xto{f} C
  $$
  We take this as the map $\otimes_\beta(\Gamma_1, \otimes_\alpha\Gamma, \Gamma_2) \to C$. It is vertical because both factors are vertical. The completed diagram is:
  \begin{center}
    \begin{tikzcd}[row sep=large, column sep=huge]
      \underset{(\Gamma_1)_m, \Gamma_m, (\Gamma_2)_m}{\Gamma_m} \arrow[rr, "{(\1_{\otimes_\alpha\Gamma}, \alpha)}"] \arrow[dr, swap,  "{(f, \sigma)}"] & & \underset{(\Gamma_1)_m, (\otimes_\alpha \Gamma)_m, (\Gamma_2)_m}{(\otimes_\alpha \Gamma)_m} \arrow[dl, "{(f \circ \zeta_{\beta, \alpha}^{-1}, \beta)}"]  \\
      & C_m   &
    \end{tikzcd}
  \end{center}

  It remains to show that $(f \circ \zeta_{\beta,\alpha}^{-1}, \beta) \circ (\1_{\otimes_\alpha \Gamma}, \alpha) = (f, \sigma)$. By definition of composition we have :
		$$ (f \circ \zeta_{\beta,\alpha}^{-1}, \beta) \circ (\1_{\otimes_\alpha \Gamma}, \alpha) = (f \circ \zeta_{\beta, \alpha}^{-1}\ \circ \otimes_\beta(\1_{\Gamma_1}, \1_{\otimes_\alpha\Gamma}, \1_{\Gamma_2}) \circ \zeta_{\beta, \alpha}, \beta \circledcirc \alpha).$$
		The first component reduces to $f$, since $\zeta_{\beta, \alpha}^{-1}\ \circ \otimes_\beta(\1_{\Gamma_1}, \1_{\otimes_\alpha\Gamma}, \1_{\Gamma_2}) \circ \zeta_{\beta, \alpha} = \1_{\otimes_{\beta \circledcirc \alpha}(\Gamma_1, \Gamma, \Gamma_2)}$. The second component is $\sigma = \beta \circledcirc \alpha$. Hence
		$$(f \circ \zeta_{\beta,\alpha}^{-1}, \beta) \circ (\1_{\otimes_\alpha \Gamma}, \alpha) = (f, \sigma)$$
		\item \textbf{Uniqueness of opcartesian morphisms :} Suppose that $(h, \beta) : (\Gamma_1)_m, (\otimes_\alpha\Gamma)_m, (\Gamma_2)_m \to C_m$ is a multimorphism over $\beta_m$ satisfying $(h, \beta) \circ (\1_{\otimes_\alpha\Gamma}, \alpha) = (f, \sigma)$. Unfolding the composition and using functoriality of $\otimes_\beta$ on the tuple of identities gives:
		$$(h, \beta) \circ (\1_{\otimes_\alpha\Gamma}, \alpha) = (h \circ \otimes_\beta(\1_{\Gamma_1}, \1_{\otimes_\alpha\Gamma}, \1_{\Gamma_2}) \circ \zeta_{\beta, \alpha}, \beta \circledcirc \alpha) = (h \circ \zeta_{\beta, \alpha}, \beta \circledcirc \alpha).$$
		The equality of first components $h \circ \zeta_{\beta, \alpha} = f$ forces $h = f \circ \zeta_{\beta, \alpha}^{-1}$ since $\zeta_{\beta, \alpha}$ is invertible. Hence the factorization is unique and $(\1_{\otimes_\alpha\Gamma}, \alpha)$ is opcartesian.
  \end{itemize}
For every list $\Gamma_m$ and every $\alpha_m$ we choose $\overline{\alpha_m}(\Gamma_m) := (\1_{\otimes_\alpha\Gamma}, \alpha)$ as the opcartesian lift, with $(\alpha_m)_*(\Gamma_m) = (\otimes_\alpha\Gamma)_m$. This exhibits $\pi_m : \fib[T_m]{M_m}$ as a cloven opfibration of multicategories.

This defines $\mc F$ on objects by sending a multi-indexed category $\mc T$ over $\mc M$ to its underlying cloven opfibration $\pi_m : \fib[T_m]{M_m}$. It remains to define $\mc F$ on strict functors.

\textit{Construction:} We construct a functor $\mathbf{MIdx_{\mc M}} \xto{\mc F} \mathbf{Multi_{\mc M_m}}$.
	\begin{itemize}
    \item Given a multi-indexed category $\mc T$ over the unbiased monoidal category
\[
(\mc M, (\otimes_n)_{n \in \mathbb{N} }, (\eta_{n,m})_{n,m \in \mathbb{N}}, \iota),
\]
define $\mc F(\mc T) = \pi_m : \fib[T_m]{M_m}$ as above.
	\item A strict functor $(\mc T,(\mc M, (\otimes_n)_{n \in \mathbb{N} }), \pi) \xto{F} (\mc T',(\mc M, (\otimes_n)_{n \in \mathbb{N} }), \pi')$ induces a morphism $\mc T_m \xto{F_m} \mc T_m'$ between the underlying cloven opfibrations:
	\begin{center}
  	\begin{tikzcd}
    	\mc T \arrow[rr, "F"] \arrow[dr,swap, "\pi" ] & & \mc T' \arrow[dl, "\pi'" ] \\
    	& \mc M & \\
  	\end{tikzcd}
  	$\xto{\mc F}$
  	\begin{tikzcd}
    	\mc T_m \arrow[rr, "F_m"] \arrow[dr,swap, "\pi_m" ] & & \mc T_m' \arrow[dl, "\pi_m'" ] \\
    	& \mc M_m & \\
  	\end{tikzcd}
	\end{center}
	We specify its action on objects and morphisms.
	\begin{itemize}
  	\item \textbf{Action on objects:} For any object $A_m \in \mc T_m$ we define $F_m(A_m) := (FA)_m$.
	\item \textbf{Action on morphisms:} A morphism $(\Gamma_1)_m, \ldots, (\Gamma_n)_m \xto{(f, \alpha)} A_m$ in $\mc T_m$ gives
	$$\otimes'_{\alpha}(F\Gamma_1, \ldots, F\Gamma_n) \xto{Ff \circ \kappa_{\alpha}}FA.$$
	Define $$F_m(f, \alpha) := (Ff \circ \kappa_{\alpha}, \alpha).$$
  	Since $\pi' \circ F = \pi$, the functor $F$ maps the fibre $\mc T_{\pi A}$ into the fibre $\mc T'_{\pi A}$; as $\kappa_\alpha$ is vertical, the first component $Ff \circ \kappa_\alpha$ is a morphism of the fibre $\mc T'_{\pi A}$, as required.

	\end{itemize}
	We must show that $\mc T_m \xto{F_m} \mc T_m'$ is a morphism of $\mathbf{Multi_{\mc M_m}}$. Thus $F_m$ must be a functor of multicategories over $\mc M_m$ and must preserve opcartesian morphisms.

	\begin{itemize}
	\item \textbf{The projections and the functor commute:} For an object $A_m \in \mc T_m$,
	$$(\pi_m' \circ F_m)(A_m) = \pi_m'((FA)_m) = (\pi'(FA))_m = (\pi A)_m = \pi_m(A_m).$$

		For a morphism $\Gamma_m \xto{(f, \alpha)} A_m$ in $\mc T_m$, the definitions of $\pi_m$ and $\pi'_m$ give $$\pi'_m(Ff \circ \kappa_\alpha, \alpha) = \alpha_m = \pi_m(f, \alpha).$$

	\item \textbf{Respect for composition:} Let $\Gamma_m \xto{(g, \alpha)} A_m$ and $(\Gamma_1)_m, A_m, (\Gamma_2)_m \xto{(f, \beta)} B_m$ be composable, with $|\Gamma_m| = n$ and $|(\Gamma_1)_m, A_m, (\Gamma_2)_m | = k$. We show that $F_m((f, \beta) \circ (g , \alpha)) = F_m(f, \beta) \circ F_m(g , \alpha)$. By definition,
	$$F_m((f, \beta) \circ (g, \alpha)) =  \left(F\big(f \circ \otimes_\beta(\1_{\Gamma_1}, g, \1_{\Gamma_2}) \circ \zeta_{\beta, \alpha}\big) \circ \kappa_{\beta \circledcirc \alpha}, \beta \circledcirc \alpha\right).$$
	This map is pictured below:

	\begin{center}
    \begin{tikzcd}[row sep=large, column sep=huge]
    	\otimes'_{\beta \circledcirc \alpha}(F\Gamma_1, F\Gamma, F\Gamma_2) \arrow[r, "{\kappa_{\beta \circledcirc \alpha}}"]&  F(\otimes_{\beta \circledcirc \alpha}(\Gamma_1, \Gamma, \Gamma_2)) \arrow[d, "F\zeta_{\beta, \alpha}"]  \\
    	& F(\otimes_\beta(\Gamma_1, \otimes_\alpha\Gamma, \Gamma_2)) \arrow[d, swap, "{F(\otimes_\beta(\1_{\Gamma_1},g,\1_{\Gamma_2}))}"] \\
    	&  F(\otimes_\beta(\Gamma_1, A, \Gamma_2)) \arrow[d, "Ff"]  \\
			&  FB
  	\end{tikzcd}
	\end{center}
	The individual maps $F_m(f, \beta) = (Ff \circ \kappa_\beta, \beta)$ and $F_m(g, \alpha) = (Fg \circ \kappa_\alpha, \alpha)$ are shown below:

	\begin{center}
  	\begin{tikzcd}
    	F(\otimes_\alpha \Gamma) \arrow[r, "Fg"] & FA \\
			\otimes'_\alpha F\Gamma \arrow[u, "\kappa_\alpha"] \arrow[ur, "Fg \circ \kappa_\alpha", swap]
  	\end{tikzcd}
  	\quad
  	\begin{tikzcd}
    	F(\otimes_\beta (\Gamma_1, A, \Gamma_2)) \arrow[r, "Ff"] & FB \\
			\otimes'_\beta (F\Gamma_1, FA, F\Gamma_2) \arrow[u, "\kappa_\beta"] \arrow[ur, "Ff \circ \kappa_\beta", swap]
  	\end{tikzcd}
	\end{center}
	Their composite is
	\[
\begin{aligned}
&(Ff \circ \kappa_\beta, \beta) \circ (Fg \circ \kappa_\alpha, \alpha)
\\
&\quad = (Ff \circ \kappa_\beta \circ \otimes'_\beta(\1_{F\Gamma_1}, Fg \circ \kappa_\alpha, \1_{F\Gamma_2}) \circ \zeta'_{\beta, \alpha}, \beta \circledcirc \alpha).
\end{aligned}
\]
    Functoriality of $\otimes'_\beta$ gives
\[
\otimes'_\beta(\1_{F\Gamma_1}, Fg \circ \kappa_\alpha, \1_{F\Gamma_2}) = \otimes'_\beta(\1_{F\Gamma_1}, Fg, \1_{F\Gamma_2}) \circ \otimes'_\beta(\1_{F\Gamma_1},\kappa_\alpha, \1_{F\Gamma_2}).
\]
Hence the composite is represented by the following diagram:

	\begin{center}
    \begin{tikzcd}[row sep=large, column sep=huge]
    	\otimes'_{\beta \circledcirc \alpha}(F\Gamma_1, F\Gamma, F\Gamma_2) \arrow[d, swap,  "\zeta'_{\beta, \alpha}"] & &    \\
			\otimes'_\beta(F\Gamma_1, \otimes'_\alpha F\Gamma, F\Gamma_2) \arrow[d, swap,  "{\otimes'_\beta(\1_{F\Gamma_1}, \kappa_\alpha, \1_{F\Gamma_2})}"] & &   \\
			\otimes'_\beta(F\Gamma_1, F(\otimes_\alpha \Gamma), F\Gamma_2) \arrow[d, swap,  "{\otimes'_\beta(\1_{F\Gamma_1}, Fg, \1_{F\Gamma_2})}"] & &  \\
			\otimes'_\beta(F\Gamma_1, FA, F\Gamma_2) \arrow[d, swap,  "\kappa_\beta"] & & \\
			F(\otimes_\beta(\Gamma_1, A, \Gamma_2)) \arrow[rr, swap,  "Ff"]  & & FB
  	\end{tikzcd}
	\end{center}
	Combining the two diagrams gives:

	\begin{center}
    \begin{tikzcd}[row sep=large, column sep=huge]
    	\otimes'_{\beta \circledcirc \alpha}(F\Gamma_1, F\Gamma, F\Gamma_2) \arrow[d, swap,  "\zeta'_{\beta, \alpha}"] \arrow[r, "{\kappa_{\beta \circledcirc \alpha}}"]&  F(\otimes_{\beta \circledcirc \alpha}(\Gamma_1, \Gamma, \Gamma_2)) \arrow[d, "F\zeta_{\beta, \alpha}"]    \\
			\otimes'_\beta(F\Gamma_1, \otimes'_\alpha F\Gamma, F\Gamma_2) \arrow[d, swap,  "{\otimes'_\beta(\1_{F\Gamma_1}, \kappa_\alpha, \1_{F\Gamma_2})}"] &  F(\otimes_\beta(\Gamma_1, \otimes_\alpha\Gamma, \Gamma_2)) \arrow[d,  swap, "{F(\otimes_\beta(\1_{\Gamma_1},g,\1_{\Gamma_2}))}"] \\
			\otimes'_\beta(F\Gamma_1, F(\otimes_\alpha \Gamma), F\Gamma_2) \arrow[d, swap,  "{\otimes'_\beta(\1_{F\Gamma_1}, Fg, \1_{F\Gamma_2})}"] \arrow[ur, "\kappa_\beta"]  & F(\otimes_\beta(\Gamma_1, A, \Gamma_2)) \arrow[dd, "Ff"]  \\
			\otimes'_\beta(F\Gamma_1, FA, F\Gamma_2) \arrow[d, swap,  "\kappa_\beta"] \arrow[ur, "\kappa_\beta"] &  \\
			F(\otimes_\beta(\Gamma_1, A, \Gamma_2)) \arrow[r, swap,  "Ff"]  &  FB
  	\end{tikzcd}
	\end{center}
	The top region commutes by the coherence law between $\zeta, \zeta'$ and $\kappa$. The middle square commutes by naturality of $\kappa_\beta$ at $(\1_{\Gamma_1}, g, \1_{\Gamma_2})$. The bottom square is immediate. Hence the outer paths agree: $F_m((f,\beta) \circ (g, \alpha)) = F_m(f, \beta) \circ F_m(g, \alpha)$.

	\item \textbf{Identity laws:} We show that $F_m(\1_{A_m}) = \1_{F_mA_m}$. The identity in $\mc T_m$ is $\1_{A_m} = (\tau_{\pi A}(A), \iota_{\pi A})$, so
	$$F_m(\1_{A_m}) = (F(\tau_{\pi A}(A)) \circ \kappa_{\iota_{\pi A}}, \iota_{\pi A}).$$
	On the other hand, $\1_{F_mA_m} = (\tau'_{\pi A}(FA), \iota_{\pi A})$. These agree by the coherence equation $F(\tau_{\pi A}(A)) \circ \kappa_{\iota_{\pi A}} = \tau'_{\pi A}(FA)$.

	\item \phantomsection\label{anchor:vertical-isomorphism-opcartesian-criterion}\textbf{Opcartesianness is preserved:} In $\mc T_m$, every multimorphism $(w, \alpha) : \Gamma_m \to X_m$ with $w : \otimes_\alpha\Gamma \to X$ a vertical isomorphism is opcartesian. Given $(f, \sigma) : (\Gamma_1)_m, \Gamma_m, (\Gamma_2)_m \to C_m$ and $\beta_m$ with $\beta_m \circ \alpha_m = \sigma_m$, a factorization $(h, \beta)$ must satisfy
	$$(h, \beta) \circ (w, \alpha) = (h \circ \otimes_\beta(\1_{\Gamma_1}, w, \1_{\Gamma_2}) \circ \zeta_{\beta, \alpha}, \beta \circledcirc \alpha) = (f, \sigma),$$
	and since $\otimes_\beta(\1_{\Gamma_1}, w, \1_{\Gamma_2}) \circ \zeta_{\beta, \alpha}$ is invertible there is exactly one solution, $h = f \circ \zeta_{\beta, \alpha}^{-1} \circ \otimes_\beta(\1_{\Gamma_1}, w^{-1}, \1_{\Gamma_2})$. The chosen lifts are the case $w = \1$.

	Now let $\psi$ be an opcartesian multimorphism in $\mc T_m$ over $\alpha_m$, with domain $\Gamma_m$. By \hyperref[item:opcartesian-essential-uniqueness]{essential uniqueness of opcartesian lifts}, $\psi = \phi \circ (\1_{\otimes_\alpha\Gamma}, \alpha)$ for a vertical isomorphism $\phi$ of $\mc T_m$. Applying the functor $F_m$ gives:
	$$F_m(\psi) = F_m(\phi) \circ F_m(\1_{\otimes_\alpha\Gamma}, \alpha) = F_m(\phi) \circ (\kappa_\alpha, \alpha),$$
	using $F(\1_{\otimes_\alpha\Gamma}) = \1$. The map $(\kappa_\alpha, \alpha)$ is opcartesian in $\mc T'_m$ by the \hyperref[anchor:vertical-isomorphism-opcartesian-criterion]{vertical-isomorphism criterion}, since $\kappa_\alpha : \otimes'_\alpha(F\Gamma) \to F(\otimes_\alpha\Gamma)$ is a vertical isomorphism. The map $F_m(\phi)$ is also a vertical isomorphism of $\mc T'_m$. By \hyperref[item:opcartesian-vertical-isomorphism]{invariance under vertical isomorphisms}, $F_m(\psi)$ is opcartesian.
	\end{itemize}
	\item \textbf{Functoriality of $\mc F$:} For the identity strict functor $\1_{\mc T}$ (whose $\kappa$'s are identities) we get $(\1_{\mc T})_m(f, \alpha) = (f \circ \1_{\otimes_\alpha(\--)}, \alpha) = (f, \alpha)$, so $\mc F(\1_{\mc T}) = \1_{\mc T_m}$. For a composite $\mc T \xto{F} \mc T' \xto{G} \mc T''$ :
	\[
\begin{aligned}
&(G \circ F)_m(f, \alpha)
\\
&\quad = \big((G F)(f) \circ \kappa^{G \circ F}_\alpha, \alpha\big)
\\
&\quad = \big(G(Ff) \circ G(\kappa^F_\alpha) \circ \kappa^G_\alpha, \alpha\big)
\\
&\quad = \big(G(Ff \circ \kappa^F_\alpha) \circ \kappa^G_\alpha, \alpha\big)
\\
&\quad = G_m(F_m(f, \alpha)),
\end{aligned}
\]
	and the actions on objects agree as well, so $\mc F(G \circ F) = \mc F(G) \circ \mc F(F)$.
\end{itemize}

\subsection{From Cloven Opfibrations to Multi-Indexed Categories}\label{subsec:opfibrations-to-multi-indexed}
\phantomsection\label{anchor:underlying-multi-indexed-category-construction}

\begin{proposition}\label{prop:cloven-opfibrations-to-multi-indexed}
	There is a functor between the category of cloven opfibrations of multicategories over $\mc M_m$ and the category of multi-indexed categories over $\mc M$.
	\[\mathbf{Multi_{\mc M_m}}\to \mathbf{MIdx_{\mc M}} \]
\end{proposition}
We first construct the underlying multi-indexed category of a cloven opfibration $\pi : \fib[T]{M_m}$. Define $(\mc T_o,(\mc M, (\otimes_n)_{n \in \mathbb{N} }), \pi_o)$ as follows:

\begin{itemize}
  \item The objects of $\mc T_o$ are the objects of $\mc T$. We write $A_o$ for the object $A$ regarded in $\mc T_o$.
	\item $A_o \to B_o$ is a morphism in $\mc T_o$ if and only if it is a unary multimorphism $A \to B$ of $\mc T$. Thus $\mc T_o$ is the unary part of $\mc T$.
  \item We define the functor $\pi_o : \fib[T_o]{M}$ by $\pi_o(A_o) := \pi A$ on objects. For a morphism $f : A_o \to B_o$, its image $\pi(f)$ is a unary multimorphism $\pi A \to \pi B$ of $\mc M_m$, i.e.\ a morphism $\otimes_1(\pi A) \to \pi B$ of $\mc M$, and we set $$\pi_o(f) := \pi(f) \circ \iota_{\pi A}^{-1} \ : \ \pi A \to \pi B.$$
  This defines a functor $\pi_o$. The identity $\1_{A_o}$ is the identity unary multimorphism of $\mc T$. Its image under $\pi$ is $\iota_{\pi A}$, the identity multimorphism of $\pi A$ in $\mc M_m$. Hence $\pi_o(\1_{A_o}) = \iota_{\pi A} \circ \iota_{\pi A}^{-1} = \1_{\pi A}$. For a composite $A_o \xto{f} B_o \xto{g} C_o$, we have $\pi(g \circ f) = \pi g \circ \pi f = \pi g \circledcirc \pi f$ in $\mc M_m$, so
  \begin{align*}
  	\pi_o(g \circ f) &= \pi g \circ \otimes_1(\pi f) \circ \eta_{1,1} \circ \iota_{\pi A}^{-1} & \text{Definition of } \circledcirc \\
  	&= \pi g \circ \iota_{\pi B}^{-1} \circ \pi f \circ \iota_{\otimes_1 (\pi A)} \circ \eta_{1,1} \circ \iota_{\pi A}^{-1} & \text{Naturality of } \iota \text{ at } \pi f \\
  	&= \pi g \circ \iota_{\pi B}^{-1} \circ \pi f \circ \iota_{\pi A}^{-1} & \text{First unit law} \\
  	&= \pi_o(g) \circ \pi_o(f). &
  \end{align*}
  Note that a morphism $f$ of $\mc T_o$ is vertical for $\pi_o$ if and only if $\pi f = \iota_{\pi A}$, i.e.\ if and only if $f$ is a morphism of the fibre of the multicategory $\mc T$ over $\pi A$. Hence the fibres of $\pi_o$ are exactly the fibres of $\pi$.
	\item For a list of objects $\gamma \in \mc M$ and an object $a \in \mc M$, a morphism $\otimes_n \gamma \xto{\alpha} a$ in $\mc M$ is the same thing as a multimorphism $\gamma_m \xto{\alpha_m} a_m$ in $\mc M_m$. Let $\Gamma$ be a list of objects in $\mc T$ with $\pi \Gamma_i = \gamma_i$. Since $\pi$ is cloven, we have the chosen opcartesian map $\Gamma \xto{\overline{\alpha_m}(\Gamma)} (\alpha_m)_*(\Gamma) $ over $\alpha_m$. Define the functor $\otimes_{\alpha} : \mc T_{\gamma_1}  \times \ldots \times \mc T_{\gamma_n} \to \mc T_a$ on objects by
  \[\otimes_{\alpha}((\Gamma_1)_o, \ldots, (\Gamma_n)_o) := ((\alpha_m)_*(\Gamma))_o.\]
	We use the convention $\prod_{i=1}^0 \mc T_{\gamma_i} = \mathbf 1$. Thus, if $n=0$, this formula defines $\otimes_\alpha : \mathbf 1 \to \mc T_a$ on its unique object by $\otimes_\alpha(*) := ((\alpha_m)_*())_o$.
  The objects $\otimes_{\alpha}(\Gamma)$ are the codomains of the chosen opcartesian lifts along $\alpha_m$. Also $\pi_o\big(((\alpha_m)_*\Gamma)_o\big) = \pi((\alpha_m)_*\Gamma) = a$, as required. We now define the action on morphisms.
	\item If $n=0$, the domain is $\mathbf 1$, and we define $\otimes_\alpha$ on its unique morphism by
	$$\otimes_\alpha(\1_*) := (\1_{(\alpha_m)_*()})_o = \1_{((\alpha_m)_*())_o}.$$
	Equivalently, this identity is the unique vertical map $u$ satisfying $u \circ \overline{\alpha_m}() = \overline{\alpha_m}()$, by the opcartesianness of $\overline{\alpha_m}()$. Now suppose $n>0$ and let $(\Gamma_i)_o \xto{(f_i)_o} (\Delta_i)_o$, $i = 1, \ldots, n$, be vertical morphisms in the fibres $\mc T_{\gamma_i}$. We need a fibre map $(\alpha_m)_*(\Gamma_1, \ldots, \Gamma_n) \to (\alpha_m)_*(\Delta_1, \ldots, \Delta_n)$ in $\mc T$. We have the opcartesian map
	\[\Delta_1, \ldots, \Delta_n \xto{\overline{\alpha_m}(\Delta_1, \ldots, \Delta_n)} (\alpha_m)_*(\Delta_1, \ldots, \Delta_n).\]
	Substituting the maps $\Gamma_i \xto{f_i} \Delta_i$ in sequence gives a map
	\[\Gamma_1, \ldots, \Gamma_n \xto{(\ldots ((\overline{\alpha_m}(\Delta_1, \ldots, \Delta_n) \circ f_1) \circ f_2) \circ \ldots ) \circ f_n} (\alpha_m)_*(\Delta_1, \ldots, \Delta_n)\]
	(the order in which the $f_i$ are plugged in is irrelevant, by the associativity axiom of $\mc T$ for disjoint positions), a map lying over $\alpha_m$ since each $\pi f_i$ is an identity. By the universal property of the opcartesian map $\overline{\alpha_m}(\Gamma_1, \ldots, \Gamma_n)$ we get a unique vertical morphism
	\[(\alpha_m)_*(\Gamma_1, \ldots, \Gamma_n) \xto{u} (\alpha_m)_*(\Delta_1, \ldots, \Delta_n)\] as shown in the universal diagram below
	\begin{center}
    \begin{tikzcd}[row sep=large, column sep=huge]
			\Gamma_1, \ldots, \Gamma_n \arrow[r , "{\overline{\alpha_m}(\Gamma_1, \ldots, \Gamma_n)}" ] \arrow[dr, swap, "{(\ldots ((\overline{\alpha_m}(\Delta_1, \ldots, \Delta_n) \circ f_1) \circ f_2) \circ \ldots ) \circ f_n}"]  & (\alpha_m)_*(\Gamma_1, \ldots, \Gamma_n) \arrow[d, "u", dotted]  \\
			& (\alpha_m)_*(\Delta_1, \ldots, \Delta_n)
  	\end{tikzcd}
	\end{center}

 Define the action of $\otimes_{\alpha}$ on morphisms by
\[\otimes_{\alpha}\left((f_1)_o:\overset{\Gamma_1}{\underset{\Delta_1}{\downarrow}}, \ldots, (f_n)_o:\overset{\Gamma_n}{\underset{\Delta_n}{\downarrow}}\right) := u_o, \] i.e.\ the unique vertical map with
\[
 \otimes_\alpha(f_1, \ldots, f_n) \circ \overline{\alpha_m}(\Gamma) = \overline{\alpha_m}(\Delta) \circ (f_1, \ldots, f_n)\]
where $\overline{\alpha_m}(\Delta) \circ (f_1, \ldots, f_n)$ abbreviates the sequential plugging displayed above. For an empty tuple, simultaneous substitution performs no substitution, so $\overline{\alpha_m}() \circ () := \overline{\alpha_m}()$.

The functor laws follow from uniqueness. For $n=0$ they hold by the definition on the terminal category. For $n>0$, $\overline{\alpha_m}(\Gamma) \circ (\1_{\Gamma_1}, \ldots, \1_{\Gamma_n}) = \overline{\alpha_m}(\Gamma)$. The identity $\1_{(\alpha_m)_*(\Gamma)}$ therefore satisfies the defining equation, so $\otimes_\alpha(\1, \ldots, \1) = \1$. For vertical tuples $(f_i) : \Gamma \to \Delta$ and $(g_i) : \Delta \to \Lambda$, both $\otimes_\alpha(g_1 \circ f_1, \ldots, g_n \circ f_n)$ and $\otimes_\alpha(g_1, \ldots, g_n) \circ \otimes_\alpha(f_1, \ldots, f_n)$ satisfy $u \circ \overline{\alpha_m}(\Gamma) = \overline{\alpha_m}(\Lambda) \circ (g_1 \circ f_1, \ldots, g_n \circ f_n)$. They coincide by uniqueness of factorizations through $\overline{\alpha_m}(\Gamma)$.

	\item For a morphism $a \xto{\alpha} c$ in $\mc M$ and an object $A \in \mc T_a$, the morphism $\alpha \circ \iota_a : \otimes_1(a) \to c$ of $\mc M$ is a unary multimorphism $(\alpha \circ \iota_a)_m : a_m \to c_m$ of $\mc M_m$, and we define the lift
	$$\alpha^\star_A := \overline{(\alpha \circ \iota_a)_m}(A) \ : \ A \to \otimes_{\alpha \circ \iota_a}(A),$$
	the chosen opcartesian lift in the multicategory $\mc T$. Since it is unary, it is a morphism of $\mc T_o$. Its projection is $\pi_o(\alpha^\star_A) = (\alpha \circ \iota_a) \circ \iota_a^{-1} = \alpha$. It is also opcartesian for $\pi_o$ in the $1$-categorical sense. Given $g : A \to C$ in $\mc T_o$ and $c \xto{\beta} \pi C$ in $\mc M$ with $\beta \circ \alpha = \pi_o(g)$, the factorizations of $g$ through $\alpha^\star_A$ over $\beta$ (for $\pi_o$) are exactly the factorizations of $g$ through $\overline{(\alpha \circ \iota_a)_m}(A)$ over $(\beta \circ \iota_c)_m$ (for $\pi$). Indeed, $\pi_o(h) = \beta$ if and only if $\pi(h) = (\beta \circ \iota_c)_m$. The bases are compatible because $(\beta \circ \iota_c) \circledcirc (\alpha \circ \iota_a) = \beta \circ \alpha \circ \iota_a = \pi_o(g) \circ \iota_a = \pi(g)$. Opcartesianness of $\overline{(\alpha \circ \iota_a)_m}(A)$ gives a unique factorization.

	\item In particular, for $\alpha = \1_c$ the lift $(1_c)^\star_A = \overline{(\iota_c)_m}(A) : A \to \otimes_{\iota_c}(A)$ is an opcartesian lift of the identity multimorphism of $c_m$, hence invertible by \hyperref[item:opcartesian-identity-lift]{the identity-lift property}. Its inverse, the derived isomorphism $\tau_c(A) : \otimes_{\iota_c}(A) \to A$, is explicitly the unique vertical factorization $u$ of $\1_A$ through it :
	\begin{center}
    \begin{tikzcd}[row sep=large, column sep=huge]
      A \arrow[r, "\overline{(\iota_c)_m}(A)"] \arrow[dr,swap, "\1_{A}"]   & ((\iota_c)_m)_*(A) \arrow[d, "u", dotted]  \\
      & A
  	\end{tikzcd}
	\end{center}

	\item We define the natural isomorphism \[
\begin{aligned}
&\zeta_{\alpha, \beta} : \otimes_{\alpha \circledcirc \beta}((\Gamma_1)_o, \ldots, (\Delta_1)_o, \ldots, (\Delta_k)_o, \ldots, (\Gamma_n)_o)\\
&\quad\to \otimes_{\alpha}((\Gamma_1)_o, \ldots, \otimes_{\beta}((\Delta_1)_o, \ldots, (\Delta_k)_o), \ldots, (\Gamma_n)_o).
\end{aligned}
\]
	By definition, the domain and codomain of this map are (the images in $\mc T_o$ of) codomains of opcartesian lifts in the multicategory $\mc T$. We need to construct a fibre map $$((\alpha \circledcirc \beta)_m)_*(\Gamma_1, \ldots, \Delta_1, \ldots, \Delta_k, \ldots, \Gamma_n) \to (\alpha_m)_*(\Gamma_1, \ldots, (\beta_m)_*(\Delta_1, \ldots, \Delta_k) , \ldots, \Gamma_n),$$
	where $(\alpha \circledcirc \beta)_m = \alpha_m \circ \beta_m$ is the composition in $\mc M_m$. We have the opcartesian map \[
\begin{aligned}
&\Gamma_1, \ldots, \Delta_1, \ldots, \Delta_k, \ldots, \Gamma_n \\
&\quad \xto{\overline{\alpha_m \circ \beta_m}(\Gamma_1, \ldots, \Delta_1, \ldots, \Delta_k, \ldots, \Gamma_n)} (\alpha_m \circ \beta_m)_*(\Gamma_1, \ldots, \Delta_1, \ldots, \Delta_k, \ldots, \Gamma_n)
\end{aligned}
\]
	as well as \[
\begin{aligned}
&\Gamma_1, \ldots, (\beta_m)_*(\Delta_1, \ldots, \Delta_k), \ldots, \Gamma_n \\
&\quad \xto{\overline{\alpha_m}(\Gamma_1, \ldots, (\beta_m)_*(\Delta_1, \ldots, \Delta_k), \ldots, \Gamma_n)}  (\alpha_m)_*(\Gamma_1, \ldots, (\beta_m)_*(\Delta_1, \ldots, \Delta_k), \ldots, \Gamma_n).
\end{aligned}
\]
	Consider the composite
	\[
\begin{aligned}
&\Gamma_1, \ldots, \Delta_1, \ldots, \Delta_k, \ldots, \Gamma_n \\
&\quad \xto{\overline{\alpha_m}(\Gamma_1, \ldots, (\beta_m)_*(\Delta), \ldots, \Gamma_n) \circ \overline{\beta_m}(\Delta)} (\alpha_m)_*(\Gamma_1, \ldots, (\beta_m)_*(\Delta_1, \ldots, \Delta_k), \ldots, \Gamma_n)
\end{aligned}
\]
	obtained by plugging $\overline{\beta_m}(\Delta)$ into the marked input of $\overline{\alpha_m}(\Gamma_1, \ldots, (\beta_m)_*(\Delta), \ldots, \Gamma_n)$. By \hyperref[item:opcartesian-composition]{closure of opcartesian morphisms under composition} this composite is opcartesian, and it lives over $\alpha_m \circ \beta_m$. The chosen lift $\overline{\alpha_m \circ \beta_m}(\Gamma_1, \ldots, \Delta, \ldots, \Gamma_n)$ is another opcartesian map over $\alpha_m \circ \beta_m$ with the same domain; by \hyperref[item:opcartesian-essential-uniqueness]{essential uniqueness} there is a unique vertical isomorphism between their codomains commuting with them, and we take it as our definition :
	Let $\zeta_{\alpha,\beta}$ be the unique vertical isomorphism such that
\[
\begin{aligned}
&\zeta_{\alpha, \beta} \circ \overline{\alpha_m \circ \beta_m}(\Gamma_1, \ldots, \Delta, \ldots, \Gamma_n) \\
&\quad = \overline{\alpha_m}(\Gamma_1, \ldots, (\beta_m)_*(\Delta), \ldots, \Gamma_n) \circ \overline{\beta_m}(\Delta).
\end{aligned}
\]
	For the following diagram, write
\begin{align*}
N_{1}&:=\Gamma_1, \ldots, \Delta_1, \ldots, \Delta_k, \ldots, \Gamma_n \\
N_{2}&:=(\alpha_m \circ \beta_m)_*(\Gamma_1, \ldots, \Delta_1, \ldots, \Delta_k, \ldots, \Gamma_n) \\
N_{3}&:=(\alpha_m)_*(\Gamma_1, \ldots, (\beta_m)_*(\Delta_1, \ldots, \Delta_k), \ldots, \Gamma_n) \\
a_{1}&:=\overline{\alpha_m \circ \beta_m}(\Gamma_1, \ldots, \Delta_1, \ldots, \Delta_k, \ldots, \Gamma_n) \\
a_{2}&:=\overline{\alpha_m}(\Gamma_1, \ldots,(\beta_m)_*(\Delta), \ldots, \Gamma_n ) \circ  \overline{\beta_m}(\Delta)
\end{align*}
\begin{center}
\begin{tikzcd}[row sep=large, column sep=4em]

        N_{1} \arrow[rr, "a_{1}"] \arrow[dr, "a_{2}", swap]  & &  N_{2}  \arrow[dl, "{\zeta_{\alpha, \beta}}", dotted] \\
        &  N_{3} &

\end{tikzcd}
\end{center}
	\textbf{Naturality of $\zeta_{\alpha, \beta}$ :} Let $(f_i) : \Gamma \to \Gamma'$ and $(g_j) : \Delta \to \Delta'$ be tuples of vertical morphisms. Here $\Gamma$ denotes the outer list with its marked position removed. We must show
	$$\otimes_\alpha\big(f_1, \ldots, \otimes_\beta(g_1, \ldots, g_k), \ldots, f_n\big) \circ \zeta_{\alpha, \beta} = \zeta_{\alpha, \beta} \circ \otimes_{\alpha \circledcirc \beta}(f_1, \ldots, g_1, \ldots, g_k, \ldots, f_n).$$
	Both sides are vertical; we precompose with the opcartesian map $\overline{\alpha_m \circ \beta_m}(\Gamma_1, \ldots, \Delta, \ldots, \Gamma_n)$ and check that the results agree, which suffices by uniqueness of factorizations. On the right-hand side, the defining equation of $\otimes_{\alpha \circledcirc \beta}$ on morphisms followed by the defining equation of $\zeta$ gives
	$$\overline{\alpha_m}(\Gamma'_1, \ldots, (\beta_m)_*(\Delta'), \ldots, \Gamma'_n) \circ \overline{\beta_m}(\Delta') \circ (f_1, \ldots, g_1, \ldots, g_k, \ldots, f_n)$$
	where the tuple is plugged in sequentially. On the left-hand side, the defining equation of $\zeta$ followed by the defining equation of $\otimes_\alpha$ on the tuple $(f_1, \ldots, \otimes_\beta(g), \ldots, f_n)$ gives
	$$\overline{\alpha_m}(\Gamma'_1, \ldots, (\beta_m)_*(\Delta'), \ldots, \Gamma'_n) \circ \big(f_1, \ldots, \otimes_\beta(g_1, \ldots, g_k), \ldots, f_n\big) \circ \overline{\beta_m}(\Delta),$$
	and plugging the defining equation of $\otimes_\beta(g_1, \ldots, g_k)$, namely $\otimes_\beta(g) \circ \overline{\beta_m}(\Delta) = \overline{\beta_m}(\Delta') \circ (g_1, \ldots, g_k)$, into the marked position turns this into the same composite as on the right-hand side (the pluggings at disjoint positions commute by the associativity axioms of $\mc T$). Hence the two sides coincide.

	\item \textbf{Coherence axioms :} In each case, we compare the two vertical morphisms by precomposing with a chosen opcartesian lift.
	\begin{itemize}
		\item $\zeta$ \textbf{against the lifts of identities :} for the first equation, note that $\iota_c \circledcirc \alpha = \alpha$, so $\zeta_{\iota_c, \alpha}$ is by definition the unique vertical map with $\zeta_{\iota_c, \alpha} \circ \overline{\alpha_m}(\Gamma) = \overline{(\iota_c)_m}\big((\alpha_m)_*(\Gamma)\big) \circ \overline{\alpha_m}(\Gamma)$, and $(1_c)^\star_{\otimes_\alpha(\Gamma)} = \overline{(\iota_c)_m}((\alpha_m)_*(\Gamma))$ satisfies this equation on the nose; by uniqueness they are equal. For the second equation, $\alpha \circledcirc \iota_a = \alpha$ and both $\zeta_{\alpha, \iota_a}$ and $\otimes_\alpha(\1_\Gamma, (1_a)^\star_A, \1_{\Gamma'})$ satisfy $(\--) \circ \overline{\alpha_m}(\Gamma, A, \Gamma') = \overline{\alpha_m}(\Gamma, \otimes_{\iota_a}(A), \Gamma') \circ \overline{(\iota_a)_m}(A)$ --- the first by its defining triangle, the second by the defining equation of $\otimes_\alpha$ on the tuple $(\1_\Gamma, (1_a)^\star_A, \1_{\Gamma'})$; by uniqueness they are equal.
        \item \textbf{Nested configuration :} precomposing either path of the square with
\[
\overline{(\alpha \circledcirc \beta \circledcirc \omega)_m}(\Gamma, \Delta, \Lambda, \Delta', \Gamma')
\]
 and unfolding the defining triangles of the three $\zeta$'s involved together with the defining equation of $\otimes_\alpha(\1_\Gamma, \zeta_{\beta,\omega}, \1_{\Gamma'})$, we reduce both paths to the same composite
		$$\overline{\alpha_m}\big(\Gamma, (\beta_m)_*(\Delta, \otimes_\omega\Lambda, \Delta'), \Gamma'\big) \circ \overline{\beta_m}(\Delta, \otimes_\omega\Lambda, \Delta') \circ \overline{\omega_m}(\Lambda)$$
		of the three chosen opcartesian lifts; by uniqueness of factorizations the two paths agree.
        \item \textbf{Disjoint configuration :} Likewise, both paths of the square, precomposed with
\[
\overline{((\alpha \circledcirc \beta) \circledcirc \omega)_m}(\Gamma, \Delta, \Gamma', \Lambda, \Gamma''),
\]
reduce to the composite
\[
\overline{\alpha_m}(\Gamma, \otimes_\beta\Delta, \Gamma', \otimes_\omega\Lambda, \Gamma'') \circ \overline{\beta_m}(\Delta) \circ \overline{\omega_m}(\Lambda)
\]
 of the chosen lifts plugged at the two disjoint positions (in either order, these being equal by the associativity axiom of $\mc T$); by uniqueness the two paths agree.
		\item \textbf{Naturality of the lifts :} For a vertical $f : A \to B$ in $\mc T_a$ and $a \xto{\alpha} c$, the defining equation of $\otimes_{\alpha \circ \iota_a}$ on $f$ is $\otimes_{\alpha \circ \iota_a}(f) \circ \overline{(\alpha \circ \iota_a)_m}(A) = \overline{(\alpha \circ \iota_a)_m}(B) \circ f$. Equivalently, $\otimes_{\alpha \circ \iota_a}(f) \circ \alpha^\star_A = \alpha^\star_B \circ f$. This is the naturality axiom.
		\item \textbf{The lifts compose along} $\zeta$: for $a \xto{\alpha} b \xto{\beta} c$, the \hyperref[anchor:unary-lift-composition-base]{base calculation above} gives $(\beta \circ \iota_b) \circledcirc (\alpha \circ \iota_a) = \beta \circ \alpha \circ \iota_a$, so the defining triangle of $\zeta_{\beta \circ \iota_b, \alpha \circ \iota_a}$ reads
		$$\zeta_{\beta \circ \iota_b, \alpha \circ \iota_a} \circ \overline{(\beta \circ \alpha \circ \iota_a)_m}(A) = \overline{(\beta \circ \iota_b)_m}\big(\otimes_{\alpha \circ \iota_a}(A)\big) \circ \overline{(\alpha \circ \iota_a)_m}(A),$$
		which is the required axiom $\zeta_{\beta \circ \iota_b, \alpha \circ \iota_a} \circ (\beta \circ \alpha)^\star_A = \beta^\star_{\otimes_{\alpha \circ \iota_a}(A)} \circ \alpha^\star_A$.
	\end{itemize}
\end{itemize}
This defines the underlying multi-indexed category of a cloven opfibration. We now specify the functor on morphisms.

\textit{Construction:} We define a functor \[\mathbf{Multi_{\mc M_m}} \xto{\mc D} \mathbf{MIdx_{\mc M}} \] as follows :

\begin{itemize}
  \item On objects, set $\mc D\left(\pi : \fib[T]{M_m}\right) = \pi_o : \fib[T_o]{M}$, with $\mc T_o$ from the \hyperref[anchor:underlying-multi-indexed-category-construction]{construction above}.
  \item A morphism between cloven opfibrations induces a strict functor of multi-indexed categories as follows:
  \begin{center}
    \begin{tikzcd}[row sep=large, column sep=huge]
      \mc T \arrow[rr, "F"] \arrow[dr, swap, "\pi"]  & & \mc T' \arrow[dl, "\pi'"]  \\
      & \mc M_m &
    \end{tikzcd}
    \quad
    \begin{tikzcd}[row sep=large, column sep=huge]
      \mc T_o \arrow[rr, "F_o"] \arrow[dr, swap, "\pi_o"]  & & \mc T_o' \arrow[dl, "\pi_o'"]  \\
      & \mc M &
    \end{tikzcd}
  \end{center}

  \begin{itemize}
    \item \textbf{Action on objects:} We define $F_o(A_o) := (FA)_o$
    \item \textbf{Action on morphisms:} We define $F_o(A_o \xto{f_o} B_o) := (Ff)_o$
  \end{itemize}
  On objects, $(\pi_o' \circ F_o)(A_o) = \pi_o'((FA)_o) = \pi'(FA) = \pi A = \pi_o(A_o)$. On a morphism $f_o$, the equality $\pi'(Ff) = \pi(f)$ gives $$\pi'_o(F_o f_o) = \pi'(Ff) \circ \iota_{\pi' F A}^{-1} = \pi(f) \circ \iota_{\pi A}^{-1} = \pi_o(f_o).$$ Thus the triangle commutes. Also $F_o(\1_{A_o}) = (F\1_A)_o = (\1_{FA})_o = \1_{F_oA_o}$. For composable maps $A_o \xto{f_o} B_o \xto{g_o} C_o$:
  \[F_o(g_o \circ f_o) = (F(g \circ f))_o = (Fg \circ Ff)_o = (Fg)_o \circ (Ff)_o = F_o(g_o) \circ F_o(f_o).\]
  We now construct the isomorphisms $\kappa_{\alpha} : \otimes'_\alpha(F_o\Gamma_o) \to F_o(\otimes_{\alpha}\Gamma_o)$. For a list $\Gamma$ over $\gamma$ and a morphism $\otimes_n\gamma \xto{\alpha} a$, the object $\otimes_\alpha(\Gamma_o)$ is $((\alpha_m)_*(\Gamma))_o$. It is the codomain of the chosen lift $\overline{\alpha_m}(\Gamma)$ in $\mc T$. Similarly, $\otimes'_{\alpha}(F_o\Gamma_o) = ((\alpha_m)_*(F\Gamma))_o$ is the codomain of $\overline{\alpha_m}(F\Gamma)$ in $\mc T'$. Since $F$ preserves opcartesian morphisms, $F(\overline{\alpha_m}(\Gamma))$ is opcartesian over $\alpha_m$ and has the same domain as $\overline{\alpha_m}(F\Gamma)$. \hyperref[item:opcartesian-essential-uniqueness]{Essential uniqueness} gives a unique vertical isomorphism between their codomains. We define $\kappa_\alpha$ to be this isomorphism:
  \begin{center}
    \begin{tikzcd}[row sep=large, column sep=huge]
      F\Gamma \arrow[rr, "\overline{\alpha_m}(F\Gamma)"] \arrow[dr, swap, "F(\overline{\alpha_m}(\Gamma))"]& & (\alpha_m)_*(F\Gamma) \arrow[dl, "{\exists! \kappa_{\alpha}}", dotted]  \\
			& F((\alpha_m)_*(\Gamma))
    \end{tikzcd}
  \end{center}
  Thus $\kappa_\alpha$ is the unique vertical isomorphism with $\kappa_\alpha \circ \overline{\alpha_m}(F\Gamma) = F(\overline{\alpha_m}(\Gamma))$. It is natural in $\Gamma_o$. Indeed, for a vertical tuple $(f_i) : \Gamma \to \Delta$, precomposing either naturality composite with $\overline{\alpha_m}(F\Gamma)$ gives $F(\overline{\alpha_m}(\Delta)) \circ (Ff_1, \ldots, Ff_n)$. The two composites agree by uniqueness of factorizations through $\overline{\alpha_m}(F\Gamma)$.

  It remains to verify that $(F_o, \kappa)$ satisfies the coherence diagrams of a strict functor of multi-indexed categories.
  \begin{itemize}
	\item \textbf{Coherence with the opcartesian lifts :} For $a \xto{\alpha} c$ and $A \in \mc T_a$, the lifts are $\alpha^\star_A = \overline{(\alpha \circ \iota_a)_m}(A)$ and $\alpha^\star_{F_oA} = \overline{(\alpha \circ \iota_a)_m}(FA)$. The required equation $F_o(\alpha^\star_A) = \kappa_{\alpha \circ \iota_a} \circ \alpha^\star_{F_o A}$ is the defining triangle of $\kappa_{\alpha \circ \iota_a}$ at $A$. The $\tau$-coherence is the case $\alpha = \1_c$, inverted.
	\item \textbf{Coherence between $\zeta$ and $\zeta'$ :} we must show $F_o(\zeta_{\alpha, \beta}) \circ \kappa_{\alpha \circledcirc \beta} = \kappa_\alpha \circ \otimes'_\alpha(\1, \kappa_\beta, \1) \circ \zeta'_{\alpha, \beta}$. Precompose both sides with the opcartesian $\overline{(\alpha_m \circ \beta_m)}(F\Gamma_1, \ldots, F\Delta, \ldots, F\Gamma_n)$. The left-hand side gives
  	$$F(\zeta_{\alpha,\beta}) \circ F(\overline{\alpha_m \circ \beta_m}(\Gamma_1, \ldots, \Delta, \ldots, \Gamma_n)) = F\big(\overline{\alpha_m}(\Gamma_1, \ldots, (\beta_m)_*(\Delta), \ldots, \Gamma_n) \circ \overline{\beta_m}(\Delta)\big)$$
	using the defining triangle of $\kappa_{\alpha \circledcirc \beta}$ and then the defining triangle of $\zeta_{\alpha, \beta}$, transported by $F$. The right-hand side gives, using the defining triangle of $\zeta'_{\alpha, \beta}$, then the defining equation of $\otimes'_\alpha$ on the tuple $(\1, \kappa_\beta, \1)$, then the defining triangles of $\kappa_\beta$ and $\kappa_\alpha$ :
  	\begin{align*}
&\kappa_\alpha \circ \otimes'_\alpha(\1, \kappa_\beta, \1) \circ \overline{\alpha_m}(F\Gamma_1, \ldots, (\beta_m)_*(F\Delta), \ldots, F\Gamma_n) \circ \overline{\beta_m}(F\Delta) \\
&\quad = \kappa_\alpha \circ \overline{\alpha_m}(F\Gamma_1, \ldots, F((\beta_m)_*\Delta), \ldots, F\Gamma_n) \circ \kappa_\beta \circ \overline{\beta_m}(F\Delta) \\
&\quad = F\big(\overline{\alpha_m}(\Gamma_1, \ldots, (\beta_m)_*(\Delta), \ldots, \Gamma_n)\big) \circ F(\overline{\beta_m}(\Delta)),
\end{align*}
  	which is the same composite. By uniqueness of factorizations through the opcartesian map, the two sides agree.
  \end{itemize}
  \item \textbf{Functoriality of $\mc D$ :} for the identity functor of $\mc T$, the defining triangle of $\kappa_\alpha$ is satisfied by the identity, so $\mc D(\1) = \1$. For a composite $\mc T \xto{F} \mc T' \xto{G} \mc T''$, both $\kappa^{G \circ F}_\alpha$ and $G(\kappa^F_\alpha) \circ \kappa^G_\alpha$ are vertical morphisms $u$ satisfying $u \circ \overline{\alpha_m}(GF\Gamma) = GF(\overline{\alpha_m}(\Gamma))$ --- the first by definition, the second since $G(\kappa^F_\alpha) \circ \kappa^G_\alpha \circ \overline{\alpha_m}(GF\Gamma) = G(\kappa^F_\alpha) \circ G(\overline{\alpha_m}(F\Gamma)) = G\big(\kappa^F_\alpha \circ \overline{\alpha_m}(F\Gamma)\big) = G(F(\overline{\alpha_m}(\Gamma)))$. By uniqueness of factorizations through $\overline{\alpha_m}(GF\Gamma)$ they coincide, so $\mc D(G \circ F) = \mc D(G) \circ \mc D(F)$.
\end{itemize}

\subsection{The Equivalence}\label{subsec:multi-indexed-equivalence}

We now have two functors
\begin{center}
  \begin{tikzcd}
		\mathbf{MIdx_{\mc M}} \arrow[r, yshift=3, "\mc F"]  & \mathbf{Multi_{\mc M_m}} \arrow[l, yshift=-3, "\mc D"]
  \end{tikzcd}
\end{center}
\begin{theorem}\label{thm:multi-indexed-cloven-opfibration-equivalence}
  There is an equivalence of categories:
	$$\mathbf{MIdx_{\mc M}} \simeq \mathbf{Multi_{\mc M_m} }.$$
\end{theorem}
\textit{Remark:} A sharper result would be to establish an equivalence over arbitrary unbiased monoidal categories with morphisms allowed to change the base, or equivalently over arbitrary multicategories; this might require the base multicategory to be representable. We leave this to future work.

\textit{Proof:} We show that there exist natural isomorphisms $\mc F \circ \mc D \simeq \1_{\mathbf{Multi_{\mc M_m}}}$ and $\1_{\mathbf{MIdx_{\mc M}}} \simeq \mc D \circ \mc F $.

\begin{itemize}
  \item We construct the natural isomorphism $\mc F \circ \mc D \xto{\eta} \1_{\mathbf{Multi_{\mc M_m}}}$ by defining the functor $\eta_{\mc T} : \mc F(\mc D(\mc T)) \to \mc T$, for a cloven opfibration $\pi : \fib[T]{M_m}$, as follows:
	\begin{itemize}
  	\item The action on objects is defined as $\eta_{\mc T}((A_o)_m) = A$.
  	\item The action on morphisms is defined as follows. A multimorphism $(\Gamma_o)_m \xto{(f, \alpha)} (A_o)_m$ of $(\mc T_o)_m$ consists of a morphism $\otimes_n(\pi\Gamma) \xto{\alpha} \pi A$ of $\mc M$ together with a vertical morphism $f : \otimes_\alpha(\Gamma_o) \to A_o$ of $\mc T_o$, i.e.\ a fibre map $f : (\alpha_m)_*(\Gamma) \to A$ of $\mc T$. We set $$\eta_{\mc T}(f, \alpha) := f \circ \overline{\alpha_m}(\Gamma),$$ where $\overline{\alpha_m}(\Gamma)$ is the chosen opcartesian lift of $\pi \Gamma \xto{\alpha_m} \pi A$ in $\mc T$.
	\end{itemize}
	We check that $\eta_{\mc T}$ is a morphism of $\mathbf{Multi_{\mc M_m}}$.
	\begin{itemize}
		\item \textbf{Over $\mc M_m$ :} $\pi(\eta_{\mc T}(f, \alpha)) = \pi(f) \circ \pi(\overline{\alpha_m}(\Gamma)) = \1_{(\pi A)_m} \circ \alpha_m = \alpha_m = \pi_m(f, \alpha)$, since $f$ is a fibre map.
		\item \textbf{Identities :} the identity of $(A_o)_m$ is $(\tau_{\pi A}(A), \iota_{\pi A})$ where, by construction of the underlying multi-indexed category, $\tau_{\pi A}(A)$ is the unique vertical factorization of $\1_A$ through $\overline{(\iota_{\pi A})_m}(A)$. Hence $\eta_{\mc T}(\tau_{\pi A}(A), \iota_{\pi A}) = \tau_{\pi A}(A) \circ \overline{(\iota_{\pi A})_m}(A) = \1_A$.
		\item \textbf{Composition :} given $(g, \alpha) : \Gamma_m \to (A_o)_m$ and $(f, \beta) : (\Gamma_1)_m, (A_o)_m, (\Gamma_2)_m \to (B_o)_m$ in $(\mc T_o)_m$, we compute, using the defining triangle of $\zeta_{\beta, \alpha}$ and the defining equation of $\otimes_\beta$ on the tuple $(\1_{\Gamma_1}, g, \1_{\Gamma_2})$ :
		\begin{align*}
&\eta_{\mc T}\big((f, \beta) \circ (g, \alpha)\big) \\
&\quad = f \circ \otimes_\beta(\1_{\Gamma_1}, g, \1_{\Gamma_2}) \circ \zeta_{\beta, \alpha} \circ \overline{(\beta \circledcirc \alpha)_m}(\Gamma_1, \Gamma, \Gamma_2) \\
&\quad \qquad \text{Definition} \\
&\quad = f \circ \otimes_\beta(\1_{\Gamma_1}, g, \1_{\Gamma_2}) \circ \overline{\beta_m}(\Gamma_1, (\alpha_m)_*(\Gamma), \Gamma_2) \circ \overline{\alpha_m}(\Gamma) \\
&\quad \qquad \text{Triangle of } \zeta_{\beta, \alpha} \\
&\quad = f \circ \overline{\beta_m}(\Gamma_1, A, \Gamma_2) \circ g \circ \overline{\alpha_m}(\Gamma) \\
&\quad \qquad \text{Equation of } \otimes_\beta(\1, g, \1) \\
&\quad = \eta_{\mc T}(f, \beta) \circ \eta_{\mc T}(g, \alpha),
\end{align*}
		where the last step is the associativity of composition in $\mc T$ (the map $g \circ \overline{\alpha_m}(\Gamma)$ is plugged at the position of $A$).
		\item \textbf{Opcartesianness is preserved :} the chosen opcartesian lift of $(\mc T_o)_m$ over $\alpha_m$ at $\Gamma_m$ is $(\1_{(\alpha_m)_*(\Gamma)}, \alpha)$, whose image is $\eta_{\mc T}(\1, \alpha) = \overline{\alpha_m}(\Gamma)$, opcartesian by definition. A general opcartesian multimorphism over $\alpha_m$ factors as $\phi \circ (\1, \alpha)$ with $\phi$ a vertical isomorphism by \hyperref[item:opcartesian-essential-uniqueness]{essential uniqueness}, and its image $\eta_{\mc T}(\phi) \circ \overline{\alpha_m}(\Gamma)$ is opcartesian by \hyperref[item:opcartesian-vertical-isomorphism]{invariance under vertical isomorphisms}, $\eta_{\mc T}(\phi)$ being a vertical isomorphism of $\mc T$.
	\end{itemize}
	$\eta_{\mc T}$ \textbf{is an isomorphism :} It is the identity on objects. On multimorphisms, the map $(f, \alpha) \mapsto f \circ \overline{\alpha_m}(\Gamma)$ is a bijection from the multimorphisms $(\Gamma_o)_m \to (A_o)_m$ over $\alpha_m$ to the multimorphisms $\Gamma \to A$ over $\alpha_m$. Indeed, opcartesianness of $\overline{\alpha_m}(\Gamma)$ gives a unique fibre map $f$ with $h = f \circ \overline{\alpha_m}(\Gamma)$ for every $h : \Gamma \to A$ over $\alpha_m$. Hence $\eta_{\mc T}$ has an inverse functor, which also commutes with the projections.

	It remains to check that this inverse preserves opcartesian multimorphisms. Let $h : \Gamma \to A$ be opcartesian over $\alpha_m$. Essential uniqueness, applied to $h$ and the chosen lift $\overline{\alpha_m}(\Gamma)$, gives a vertical isomorphism $v : (\alpha_m)_*(\Gamma) \to A$ such that $h = v \circ \overline{\alpha_m}(\Gamma)$. Hence $\eta_{\mc T}^{-1}(h) = (v,\alpha)$. This is opcartesian in $(\mc T_o)_m$ by the \hyperref[anchor:vertical-isomorphism-opcartesian-criterion]{vertically-invertible-first-component criterion}. Therefore $\eta_{\mc T}^{-1}$ is a morphism of $\mathbf{Multi_{\mc M_m}}$, and $\eta_{\mc T}$ is invertible in that category.

	We next prove naturality. Let $\mc T \xto{F} \mc T'$ be a morphism between cloven multicategory opfibrations over $\mc M_m$.
	\begin{center}
    \begin{tikzcd}[row sep=large, column sep=huge]
    	\mc F(\mc D(\mc T)) \arrow[r, "\eta_{\mc T}"] \arrow[d, "\mc F(\mc D(F))", swap ]   & \mc T \arrow[d, "F"]  \\
    	\mc F(\mc D(\mc T')) \arrow[r, "\eta_{\mc T'}"]  & \mc T' \\
  	\end{tikzcd}
	\end{center}
	Using the shorthands $\mc F(\mc D(\mc T')) = (\mc T'_o)_m$ and $\mc F(\mc D(\mc T)) = (\mc T_o)_m$, the diagram is :
	\begin{center}
    \begin{tikzcd}[row sep=large, column sep=huge]
    	 (\mc T_o)_m \arrow[r, "\eta_{\mc T}"] \arrow[d, "(F_o)_m", swap ]   & \mc T \arrow[d, "F"]\\
    	(\mc T'_o)_m \arrow[r, "\eta_{\mc T'}"]  & \mc T' \\
  	\end{tikzcd}
	\end{center}
	We compare $F \circ \eta_{\mc T}$ and $\eta_{\mc T'} \circ (F_o)_m$ on objects and morphisms.

	\begin{itemize}
		\item \textbf{Action on objects : } For an object $(A_o)_m$, both composites give $FA$:
		$$F(\eta_{\mc T}((A_o)_m)) = FA = \eta_{\mc T'}((F_o)_m((A_o)_m)).$$
		\item \textbf{Action on morphisms : } Let $(\Gamma_o)_m \xto{(f, \alpha)} (A_o)_m$ be a multimorphism. By definition,
		$$F(\eta_{\mc T}(f, \alpha)) = Ff \circ F(\overline{\alpha_m}(\Gamma))$$
		and
		$$(\eta_{\mc T'} \circ (F_o)_m)(f, \alpha) = Ff \circ \kappa_{\alpha} \circ \overline{\alpha_m}(F\Gamma).$$
		The defining triangle for $\kappa_\alpha$ is $\kappa_\alpha \circ \overline{\alpha_m}(F\Gamma) = F(\overline{\alpha_m}(\Gamma))$:
		\begin{center}
      \begin{tikzcd}[row sep=large, column sep=huge]
    		F\Gamma \arrow[rr, "F(\overline{\alpha_m}(\Gamma) )"] \arrow[dr, swap, "\overline{\alpha_m}(F\Gamma)"]   & &  F((\alpha_m)_*(\Gamma)) \\
    		& (\alpha_m)_*(F\Gamma) \arrow[ur, "\kappa_\alpha", swap ]
  		\end{tikzcd}
		\end{center}
		The triangle gives $$F(\eta_{\mc T}(f, \alpha)) = Ff \circ \kappa_\alpha \circ \overline{\alpha_m}(F\Gamma) = (\eta_{\mc T'} \circ (F_o)_m)(f, \alpha),$$ so the actions on morphisms agree.
	\end{itemize}
	Thus $\mc F \circ \mc D \xto{\eta} \1_{\mathbf{Multi_{\mc M_m}} }$ is a natural transformation whose components are isomorphisms.

	\item We construct the natural isomorphism $\1_{\mathbf{MIdx_{\mc M}}} \xto{\sigma} \mc D \circ \mc F$. For a multi-indexed category $\mc T$, define the strict functor $\sigma_{\mc T} : \mc T \to (\mc T_m)_o$ as follows.
	\begin{itemize}
		\item \textbf{Action on objects : } Set $\sigma_{\mc T}A = (A_m)_o$.
		\item \textbf{Action on morphisms : } Let $A \xto{f} B$ be a morphism over $\varphi := \pi f : a \to b$. The opcartesian lift $\varphi^\star_A : A \to \otimes_{\varphi \circ \iota_a}(A)$ gives a unique \textit{vertical} morphism $u^f : \otimes_{\varphi \circ \iota_a}(A) \to B$ such that $u^f \circ \varphi^\star_A = f$. Set
  	$$\sigma_{\mc T}(f) := (u^f, \varphi \circ \iota_a),$$
  	a well-formed unary multimorphism of $\mc T_m$, i.e.\ a morphism $(A_m)_o \to (B_m)_o$ of $(\mc T_m)_o$. When $f$ is vertical ($\varphi = \1_a$), uniqueness gives $u^f = f \circ \tau_a(A)$, recovering the formula $\sigma_{\mc T}(f) = (f \circ \tau_{\pi A}(A), \iota_{\pi A})$ on the fibres.
	\end{itemize}
	$\sigma_{\mc T}$ \textbf{ is a functor :} For an identity, $\1_A = \tau_a(A) \circ (1_a)^\star_A$ by definition of $\tau$, so $u^{\1_A} = \tau_a(A)$ and $\sigma_{\mc T}(\1_A) = (\tau_a(A), \iota_a) = \1_{(A_m)_o}$. For a composite $A \xto{f} B \xto{g} C$ over $a \xto{\varphi} b \xto{\psi} c$, we compute the composition in $(\mc T_m)_o$ :
	$$\sigma_{\mc T}(g) \circ \sigma_{\mc T}(f) = (u^g, \psi \circ \iota_b) \circ (u^f, \varphi \circ \iota_a) = \big(u^g \circ \otimes_{\psi \circ \iota_b}(u^f) \circ \zeta_{\psi \circ \iota_b, \varphi \circ \iota_a},\ (\psi \circ \iota_b) \circledcirc (\varphi \circ \iota_a)\big).$$
	The base component is $(\psi \circ \iota_b) \circledcirc (\varphi \circ \iota_a) = \psi \circ \varphi \circ \iota_a$ by the \hyperref[anchor:unary-lift-composition-base]{unary-lift base calculation}. For the first component, we check that the vertical morphism $v := u^g \circ \otimes_{\psi \circ \iota_b}(u^f) \circ \zeta_{\psi \circ \iota_b, \varphi \circ \iota_a}$ satisfies the defining property of $u^{g \circ f}$ :
	\begin{align*}
		v \circ (\psi \circ \varphi)^\star_A &= u^g \circ \otimes_{\psi \circ \iota_b}(u^f) \circ \zeta_{\psi \circ \iota_b, \varphi \circ \iota_a} \circ (\psi \circ \varphi)^\star_A & \\
		&= u^g \circ \otimes_{\psi \circ \iota_b}(u^f) \circ \psi^\star_{\otimes_{\varphi \circ \iota_a}(A)} \circ \varphi^\star_A &\text{The lifts compose along } \zeta \\
		&= u^g \circ \psi^\star_B \circ u^f \circ \varphi^\star_A &\text{Naturality of the lifts at } u^f \\
		&= g \circ f. &\text{Definition of } u^f, u^g
	\end{align*}
	Uniqueness of the vertical factorization through $(\psi \circ \varphi)^\star_A$ gives $v = u^{g \circ f}$. Hence $\sigma_{\mc T}(g \circ f) = \sigma_{\mc T}(g) \circ \sigma_{\mc T}(f)$. Also $\pi_o(\sigma_{\mc T}(f)) = (\varphi \circ \iota_a) \circ \iota_a^{-1} = \varphi = \pi(f)$, so $\sigma_{\mc T}$ commutes with the projections.

	$\sigma_{\mc T}$ \textbf{ is a strict functor of multi-indexed categories, with $\kappa_\alpha = \1$ :} The multi-indexed category structure of $(\mc T_m)_o = \mc D(\mc F(\mc T))$ is constructed from the chosen lifts $\overline{\alpha_m}(\Gamma_m) = (\1_{\otimes_\alpha\Gamma}, \alpha)$ of $\mc T_m$. On objects,
	$$\otimes^{\mc{DF}}_\alpha(\sigma_{\mc T}\Gamma_1, \ldots, \sigma_{\mc T}\Gamma_n) = ((\alpha_m)_*(\Gamma_m))_o = ((\otimes_\alpha\Gamma)_m)_o = \sigma_{\mc T}(\otimes_\alpha(\Gamma_1, \ldots, \Gamma_n)),$$
	so we may take $\kappa_\alpha := \1$. We verify the required compatibilities. 
	\begin{itemize}
		\item \textbf{Naturality}, i.e.\ $\otimes^{\mc{DF}}_\alpha(\sigma_{\mc T}f_1, \ldots, \sigma_{\mc T}f_n) = \sigma_{\mc T}(\otimes_\alpha(f_1, \ldots, f_n))$ for vertical $f_i : \Gamma_i \to \Delta_i$. The left-hand side is the unique vertical morphism $u$ of $(\mc T_m)_o$ with $u \circ (\1_{\otimes_\alpha\Gamma}, \alpha) = (\1_{\otimes_\alpha\Delta}, \alpha) \circ (\sigma_{\mc T}f_1, \ldots, \sigma_{\mc T}f_n)$. We check the candidate $\sigma_{\mc T}(\otimes_\alpha(f)) = (\otimes_\alpha(f_1, \ldots, f_n) \circ \tau_c(\otimes_\alpha\Gamma), \iota_c)$. On one hand :
		$$\sigma_{\mc T}(\otimes_\alpha(f)) \circ (\1_{\otimes_\alpha\Gamma}, \alpha) = \big(\otimes_\alpha(f) \circ \tau_c(\otimes_\alpha\Gamma) \circ \zeta_{\iota_c, \alpha}, \iota_c \circledcirc \alpha\big) = (\otimes_\alpha(f_1, \ldots, f_n), \alpha)$$
		by the first coherence between $\zeta$ and the lifts of identities (in $\tau$ form) and \cref{lem:circledcirc-units}. On the other hand, plugging $\sigma_{\mc T}f_i = (f_i \circ \tau_{\gamma_i}(\Gamma_i), \iota_{\gamma_i})$ into the $i$-th input of $(\1_{\otimes_\alpha\Delta}, \alpha)$ gives
		\[
\begin{aligned}
&\big(\otimes_\alpha(\1, \ldots, f_i \circ \tau_{\gamma_i}(\Gamma_i), \ldots, \1) \circ \zeta_{\alpha, \iota_{\gamma_i}}, \alpha \circledcirc \iota_{\gamma_i}\big)
\\
&\quad = \big(\otimes_\alpha(\1, \ldots, f_i, \ldots, \1), \alpha\big),
\end{aligned}
\]
        using the second coherence between $\zeta$ and the lifts of identities,
\[
\zeta_{\alpha, \iota_{\gamma_i}} = \otimes_\alpha(\1, \ldots, (1_{\gamma_i})^\star_{\Gamma_i}, \ldots, \1),
\]
the functoriality of $\otimes_\alpha$, $\tau_{\gamma_i}(\Gamma_i) \circ (1_{\gamma_i})^\star_{\Gamma_i} = \1$, and \cref{lem:circledcirc-units}. Iterating over $i = 1, \ldots, n$ gives
\[
(\1_{\otimes_\alpha\Delta}, \alpha) \circ (\sigma_{\mc T}f_1, \ldots, \sigma_{\mc T}f_n) = (\otimes_\alpha(f_1, \ldots, f_n), \alpha).
\]
Both computations agree, so by uniqueness $\otimes^{\mc{DF}}_\alpha(\sigma f) = \sigma(\otimes_\alpha(f))$.
		\item \textbf{Coherence with the opcartesian lifts :} the lift of $(\mc T_m)_o$ over $a \xto{\alpha} c$ at $(A_m)_o$ is $\overline{(\alpha \circ \iota_a)_m}(A_m) = (\1_{\otimes_{\alpha \circ \iota_a}(A)}, \alpha \circ \iota_a)$. On the other side, $\sigma_{\mc T}(\alpha^\star_A) = (u^{\alpha^\star_A}, \alpha \circ \iota_a)$ where $u^{\alpha^\star_A}$ is the unique vertical map with $u \circ \alpha^\star_A = \alpha^\star_A$; hence $u^{\alpha^\star_A} = \1$ and $\sigma_{\mc T}(\alpha^\star_A) = (\1, \alpha \circ \iota_a) = \kappa \circ \alpha^\star_{\sigma_{\mc T}A}$ with $\kappa = \1$, as required. The $\tau$-coherence is the case $\alpha = \1_c$.
		\item \textbf{Coherence between $\zeta$ and $\zeta^{\mc{DF}}$ :} with $\kappa = \1$ the coherence diagram reduces to $\sigma_{\mc T}(\zeta_{\alpha, \beta}) = \zeta^{\mc{DF}}_{\alpha, \beta}$. The right-hand side is the unique vertical morphism $u$ of $(\mc T_m)_o$ with $u \circ (\1_{\otimes_{\alpha \circledcirc \beta}(\Gamma, \Delta, \Gamma')}, \alpha \circledcirc \beta) = (\1_{\otimes_\alpha(\Gamma, \otimes_\beta\Delta, \Gamma')}, \alpha) \circ (\1_{\otimes_\beta\Delta}, \beta)$. The composite on the right is $(\otimes_\alpha(\1, \1, \1) \circ \zeta_{\alpha, \beta}, \alpha \circledcirc \beta) = (\zeta_{\alpha, \beta}, \alpha \circledcirc \beta)$, and the candidate satisfies
		$$\sigma_{\mc T}(\zeta_{\alpha, \beta}) \circ (\1, \alpha \circledcirc \beta) = \big(\zeta_{\alpha, \beta} \circ \tau_c(\otimes_{\alpha \circledcirc \beta}(\Gamma, \Delta, \Gamma')) \circ \zeta_{\iota_c, \alpha \circledcirc \beta}, \alpha \circledcirc \beta \big) = (\zeta_{\alpha, \beta}, \alpha \circledcirc \beta)$$
		by the first coherence in $\tau$ form. By uniqueness the two agree.
	\end{itemize}

	We now prove naturality. For every strict functor $F : \mc T \to \mc T'$, the following diagram must commute:
	\begin{center}
    \begin{tikzcd}[row sep=large, column sep=huge]
    	 \mc T \arrow[r, "\sigma_{\mc T}"] \arrow[d, "F", swap ]   & (\mc T_m)_o \arrow[d, "(F_m)_o"]\\
    	\mc T' \arrow[r, "\sigma_{\mc T'}"]  & (\mc T'_m)_o \\
  	\end{tikzcd}
	\end{center}
	\begin{itemize}
		\item \textbf{Agreement on objects : } For $A \in \mc T$, both composites give $((FA)_m)_o$.
  	\item \textbf{Agreement on morphisms : } Let $A \xto{f} B$ be a morphism of $\mc T$ over $\varphi := \pi f$. On one hand
  	$$(F_m)_o(\sigma_{\mc T} f) = (F_m)_o(u^f, \varphi \circ \iota_a) = (F(u^f) \circ \kappa_{\varphi \circ \iota_a}, \varphi \circ \iota_a),$$
		while $\sigma_{\mc T'}(Ff) = (u^{Ff}, \varphi \circ \iota_a)$, since $\pi'(Ff) = \pi(f) = \varphi$. It remains to show that $u^{Ff} = F(u^f) \circ \kappa_{\varphi \circ \iota_a}$. We check the defining property of $u^{Ff}$:
  	\begin{align*}
  		F(u^f) \circ \kappa_{\varphi \circ \iota_a} \circ \varphi^\star_{FA} &= F(u^f) \circ F(\varphi^\star_A) &\text{Coherence of } F \text{ with the lifts}\\
  		&= F(u^f \circ \varphi^\star_A) &\text{Functoriality of } F\\
  		&= F(f). &\text{Definition of } u^f
  	\end{align*}
  	Since $F(u^f) \circ \kappa_{\varphi \circ \iota_a}$ is vertical, uniqueness of the factorization through the opcartesian $\varphi^\star_{FA}$ gives $u^{Ff} = F(u^f) \circ \kappa_{\varphi \circ \iota_a}$, so the two composites agree as functors.
  	\item \textbf{Agreement of the structural isomorphisms : } since the $\kappa$'s of $\sigma_{\mc T}$ and $\sigma_{\mc T'}$ are identities, the composite structural isomorphisms of the two paths of the square are $\kappa^{(F_m)_o}_\alpha$ and $\sigma_{\mc T'}(\kappa^F_\alpha)$ respectively, and we must show that they are equal. Both are vertical morphisms $u$ of $(\mc T'_m)_o$ satisfying $u \circ (\1_{\otimes'_\alpha(F\Gamma)}, \alpha) = (\kappa^F_\alpha, \alpha)$ : the first by its defining triangle in the $\mc D$-construction, since the chosen opcartesian lift of $\mc T'_m$ is $(\1_{\otimes'_\alpha F\Gamma}, \alpha)$ and $F_m(\1_{\otimes_\alpha\Gamma}, \alpha) = (F\1 \circ \kappa^F_\alpha, \alpha) = (\kappa^F_\alpha, \alpha)$; the second by the vertical-case formula $\sigma_{\mc T'}(\kappa^F_\alpha) = (\kappa^F_\alpha \circ \tau'_c(\otimes'_\alpha F\Gamma), \iota_c)$ together with the computation
  	$$(\kappa^F_\alpha \circ \tau'_c(\otimes'_\alpha F\Gamma), \iota_c) \circ (\1_{\otimes'_\alpha F\Gamma}, \alpha) = \big(\kappa^F_\alpha \circ \tau'_c(\otimes'_\alpha F\Gamma) \circ \zeta'_{\iota_c, \alpha}, \iota_c \circledcirc \alpha\big) = (\kappa^F_\alpha, \alpha),$$
		using the first coherence between $\zeta'$ and the lifts of identities in $\tau$ form, and \cref{lem:circledcirc-units}. By uniqueness of factorizations through the opcartesian $(\1_{\otimes'_\alpha F\Gamma}, \alpha)$ they coincide. Hence $(F_m)_o \circ \sigma_{\mc T} = \sigma_{\mc T'} \circ F$ as strict functors of multi-indexed categories.
	\end{itemize}
	$\sigma_{\mc T}$ \textbf{ is an isomorphism :} It is bijective on objects. On morphisms, send $(v, \alpha) : (A_m)_o \to (B_m)_o$, where $\alpha : \otimes_1(a) \to b$ in $\mc M$ and $v : \otimes_\alpha(A) \to B$ is vertical, to $v \circ (\alpha \circ \iota_a^{-1})^\star_A : A \to B$. Write $\varphi := \alpha \circ \iota_a^{-1}$, so that $\alpha = \varphi \circ \iota_a$. Then $\sigma_{\mc T}(v \circ \varphi^\star_A) = (u^{v \circ \varphi^\star_A}, \varphi \circ \iota_a) = (v, \alpha)$. Conversely, $u^f \circ \varphi^\star_A = f$ by definition. Thus the two assignments are mutually inverse, and $\sigma_{\mc T}$ is an isomorphism of categories over $\mc M$.

	Let $\rho_{\mc T} : (\mc T_m)_o \to \mc T$ denote this inverse functor, so that $\rho_{\mc T}((A_m)_o) = A$ and
	$$\rho_{\mc T}(v,\alpha) = v \circ (\alpha \circ \iota_a^{-1})^\star_A.$$
	We now verify that it is a strict functor of multi-indexed categories. On objects we have
	$$\rho_{\mc T}\big(\otimes^{\mc{DF}}_\alpha(((\Gamma_1)_m)_o, \ldots, ((\Gamma_n)_m)_o)\big) = \otimes_\alpha(\Gamma_1, \ldots, \Gamma_n),$$
	so we take $\kappa^{\rho}_\alpha = \1$. To check its naturality, let $q_i$ be vertical morphisms of $(\mc T_m)_o$. Since $\sigma_{\mc T}$ and $\rho_{\mc T}$ are mutually inverse on morphisms, there are unique vertical $f_i := \rho_{\mc T}(q_i)$ with $q_i = \sigma_{\mc T}(f_i)$. The naturality of the identity comparisons of $\sigma_{\mc T}$, proved above, then gives
	\begin{align*}
		\rho_{\mc T}\big(\otimes^{\mc{DF}}_\alpha(q_1, \ldots, q_n)\big)
		&= \rho_{\mc T}\big(\otimes^{\mc{DF}}_\alpha(\sigma_{\mc T}f_1, \ldots, \sigma_{\mc T}f_n)\big) \\
		&= \rho_{\mc T}\big(\sigma_{\mc T}(\otimes_\alpha(f_1, \ldots, f_n))\big) \\
		&= \otimes_\alpha(\rho_{\mc T}q_1, \ldots, \rho_{\mc T}q_n).
	\end{align*}
	For a nullary $\alpha$ the same assertion is the identity equation for the unique morphism of the terminal domain.

	For the coherence with the opcartesian lifts, the lift in $(\mc T_m)_o$ over $a \xto{\varphi} c$ at $(A_m)_o$ is $(\1_{\otimes_{\varphi \circ \iota_a}(A)}, \varphi \circ \iota_a)$, and hence
	$$\rho_{\mc T}(\1_{\otimes_{\varphi \circ \iota_a}(A)}, \varphi \circ \iota_a) = \varphi^\star_A.$$
	This is the required lift coherence for $\kappa^\rho = \1$. Taking $\varphi = \1_a$ and inverting gives the $\tau$-coherence. The preceding verification for $\sigma_{\mc T}$ gives $\sigma_{\mc T}(\zeta_{\alpha,\beta}) = \zeta^{\mc{DF}}_{\alpha,\beta}$. Applying $\rho_{\mc T}$ and using $\rho_{\mc T}\sigma_{\mc T}=\1$ gives
	$$\rho_{\mc T}(\zeta^{\mc{DF}}_{\alpha,\beta}) = \zeta_{\alpha,\beta},$$
	which is the $\zeta$-coherence for $\rho_{\mc T}$. Thus $\rho_{\mc T}$ is a strict functor of multi-indexed categories. Both composites $\rho_{\mc T}\sigma_{\mc T}$ and $\sigma_{\mc T}\rho_{\mc T}$ are identity functors. Their structural isomorphisms are also identities. Therefore $\sigma_{\mc T}$ is invertible in $\mathbf{MIdx_{\mc M}}$.
\end{itemize}
This proves the theorem. $\square$

\subsection{Cloven Bifibrations}\label{subsec:biclosed-multi-indexed}

We now turn to cloven bifibrations of multicategories. On the multi-indexed category side, the cartesian universal property says that each one-variable functor $\otimes_{\alpha}(\Gamma_1, (\--), \Gamma_2)$ has a right adjoint, denoted $(\Gamma_1 \Rightarrow_{\alpha} (\--) \Leftarrow \Gamma_2)$.

\begin{definition}\label{def:biclosed-multi-indexed-category}
  A multi-indexed category $(\mc T,(\mc M, (\otimes_n)_{n \in \mathbb{N}}, (\eta_{n,m})_{n,m \in \mathbb{N}},\iota), \pi)$ is \textit{biclosed} if every functor $\otimes_\alpha(\Gamma_1, (\--), \Gamma_2) : \mc T_a \to \mc T_c$, for $\otimes_{|\gamma_1| + 1 + |\gamma_2|}(\gamma_1, a, \gamma_2) \xto{\alpha} c$ in $\mc M$ and lists of objects $\Gamma_1, \Gamma_2$ of $\mc T$ over $\gamma_1, \gamma_2$, has a right adjoint  \[(\Gamma_1 \Rightarrow_{\alpha}(\--) \Leftarrow  \Gamma_2) : \mc T_c \to \mc T_a.\]
  For such a right adjoint, we denote the unit and the counit by $$\eta_\otimes : \1_{\mc T_a} \to (\Gamma_1 \Rightarrow_{\alpha} \otimes_\alpha(\Gamma_1, (\--), \Gamma_2)  \Leftarrow \Gamma_2) \quadt{and} \epsilon_\otimes : \otimes_\alpha(\Gamma_1, (\Gamma_1 \Rightarrow_{\alpha}(\--) \Leftarrow \Gamma_2 ), \Gamma_2) \to \1_{\mc T_c}.$$
\end{definition}
We write $\mathbf{MIdx^{bi}_{\mc M}}$ for the full subcategory of $\mathbf{MIdx_{\mc M}}$ whose objects are the biclosed multi-indexed categories. Its morphisms are the ambient strict functors; no compatibility with the chosen right adjoints is imposed.
\begin{definition}\label{def:cloven-bifibration-multicategories}
	We use \emph{cartesian multimorphism} and \emph{fibration of multicategories} in the senses of \cref{def:cartesian-multimorphism,def:multicategory-fibration}. A \textit{cloven bifibration of multicategories} is a cloven opfibration whose underlying functor is also a fibration of multicategories. 
\end{definition}

We now restrict $\mc F$ and $\mc D$ to biclosed multi-indexed categories and cloven bifibrations. We begin by showing that $\mc F$ maps biclosed multi-indexed categories to cloven bifibrations.

\begin{proposition}\label{prop:biclosed-to-cloven-bifibration}
  Given an unbiased monoidal category $(\mc M, (\otimes_n)_{n \in \mathbb{N}}, (\eta_{n,m})_{n,m \in \mathbb{N}},\iota)$ and its underlying multicategory $\mc M_m$, the functor $\mc F$ restricts to a functor between the category of biclosed multi-indexed categories over $\mc M$ and the category of cloven bifibrations of multicategories over $\mc M_m$:
	$$\mathbf{MIdx^{bi}_{\mc M}} \xto{\mc F}\mathbf{Multi^{bi}_{\mc M_m}} $$
\end{proposition}

\textit{Proof: }
Since the morphisms of both subcategories are those of the ambient categories, it suffices to show that the underlying cloven opfibration $\pi_m : \fib[T_m]{M_m}$ of a biclosed multi-indexed category $\mc T$ is also a fibration. We construct the required cartesian morphisms.

Let $(\Gamma_1)_m, (\Gamma_2)_m$ be lists of objects, let $B_m$ be an object, and let
\[
\pi_m(\Gamma_1)_m, q_m,\pi_m(\Gamma_2)_m \xto{\alpha_m} \pi_m B_m
\]
 be a multimorphism in $\mc M_m$. We seek an object $D$ and a cartesian multimorphism
\[
(\Gamma_1)_m, D,(\Gamma_2)_m \to B_m
\]
over $\alpha_m$. The base multimorphism corresponds to $\otimes_{k}(\pi \Gamma_1, q, \pi \Gamma_2) \xto{\alpha} \pi B$ in $\mc M$, where $k := |\Gamma_1| + 1 + |\Gamma_2|$. It therefore gives a functor $\otimes_\alpha$. Since $\mc T$ is biclosed, this functor has the right adjoint $(\Gamma_1 \Rightarrow_{\alpha} (\--) \Leftarrow \Gamma_2)$. Define
\[D := (\Gamma_1 \Rightarrow_{\alpha} B \Leftarrow \Gamma_2)_m.\]
We first construct a map $(\Gamma_1)_m, (\Gamma_1 \Rightarrow_{\alpha} B \Leftarrow \Gamma_2)_m, (\Gamma_2)_m \to B_m$. It is enough to give a vertical map
\[\otimes_\alpha\big(\Gamma_1, (\Gamma_1 \Rightarrow_{\alpha} B \Leftarrow \Gamma_2), \Gamma_2\big) \to B.\]
Under the adjunction, this is equivalent to a vertical map
\[(\Gamma_1 \Rightarrow_{\alpha} B \Leftarrow \Gamma_2) \to (\Gamma_1 \Rightarrow_{\alpha} B \Leftarrow \Gamma_2).\]
Take the identity. Its transpose is the counit component $\epsilon_\otimes(B) : \otimes_\alpha(\Gamma_1, (\Gamma_1 \Rightarrow_{\alpha} B \Leftarrow \Gamma_2), \Gamma_2) \to B$. Thus our map is
$$(\Gamma_1)_m, (\Gamma_1 \Rightarrow_\alpha B \Leftarrow \Gamma_2)_m, (\Gamma_2)_m \xto{(\epsilon_\otimes(B), \alpha)}B_m.$$
We show that it is cartesian at $(\Gamma_1 \Rightarrow_{\alpha} B \Leftarrow \Gamma_2)_m$. Let
$(\Gamma_1)_m, \Gamma_m, (\Gamma_2)_m \xto{(g, \rho)} B_m$ be a multimorphism in $\mc T_m$, as in
\begin{center}
  \begin{tikzcd}
    (\Gamma_1)_m, (\Gamma_1 \Rightarrow_\alpha B \Leftarrow \Gamma_2)_m, (\Gamma_2)_m \arrow[rr, "{(\epsilon_\otimes(B), \alpha)}"]  & & B_m \\
		&  (\Gamma_1)_m, \Gamma_m, (\Gamma_2)_m \arrow[ur, "{(g, \rho)}", swap ] &
  \end{tikzcd}
\end{center}
together with a multimorphism $\beta_m : \pi_m\Gamma_m \to q_m$ in $\mc M_m$ such that the following diagram commutes:
\begin{center}
  \begin{tikzcd}[row sep=large, column sep=huge]
    \underset{\pi_m(\Gamma_1)_m,\, q_m,\, \pi_m(\Gamma_2)_m}{q_m} \arrow[rr, "\alpha_m"]  & & \pi_m B_m \\
		\underset{\pi_m(\Gamma_1)_m,\, \pi_m\Gamma_m,\, \pi_m(\Gamma_2)_m}{\pi_m \Gamma_m} \arrow[u, "\beta_m"] \arrow[urr, "\rho_m", swap] &  &
  \end{tikzcd}
\end{center}
Thus $\alpha_m \circ \beta_m = \rho_m$, with composition at the marked position $q_m$. We need a map $\Gamma_m \to (\Gamma_1 \Rightarrow_{\alpha} B \Leftarrow \Gamma_2)_m$ over $\beta_m$ that makes the triangle commute. It is enough to construct a vertical map $\otimes_\beta(\Gamma) \to (\Gamma_1 \Rightarrow_\alpha B \Leftarrow \Gamma_2)$. By adjunction, this is equivalent to a vertical map
\[\otimes_\alpha(\Gamma_1, \otimes_\beta(\Gamma), \Gamma_2) \to B.\]
The equation $\alpha_m \circ \beta_m = \rho_m$ gives $\alpha \circledcirc \beta = \alpha \circ \otimes_k(\1_{\pi \Gamma_1}, \beta, \1_{\pi \Gamma_2}) \circ \eta_{k,n} = \rho$, where $n := |\Gamma|$. Define the required map by
$$\otimes_\alpha(\Gamma_1, \otimes_\beta\Gamma, \Gamma_2) \xto{\zeta_{\alpha, \beta}^{-1}} \otimes_{\alpha \circledcirc \beta}(\Gamma_1, \Gamma, \Gamma_2) = \otimes_{\rho}(\Gamma_1, \Gamma, \Gamma_2) \xto{g} B.$$
The candidate is its adjoint transpose:
$$\otimes_\beta(\Gamma) \xto{(\Gamma_1 \Rightarrow_{\alpha} (g \circ \zeta_{\alpha, \beta}^{-1}) \Leftarrow \Gamma_2) \circ \eta_\otimes(\otimes_\beta\Gamma)}(\Gamma_1 \Rightarrow_{\alpha} B \Leftarrow \Gamma_2)$$
We show that the triangle commutes. By the definition of composition in $\mc T_m$:
\begin{align*}
&(\epsilon_\otimes(B), \alpha) \circ \big((\Gamma_1 \Rightarrow_{\alpha} (g \circ \zeta_{\alpha, \beta}^{-1}) \Leftarrow \Gamma_2) \circ \eta_\otimes(\otimes_\beta\Gamma), \beta\big) \\
&\quad = (\epsilon_\otimes(B) \circ \otimes_\alpha(\1_{\Gamma_1}, (\Gamma_1 \Rightarrow_{\alpha} (g \circ \zeta_{\alpha, \beta}^{-1}) \Leftarrow \Gamma_2) \circ \eta_\otimes(\otimes_\beta\Gamma)  , \1_{\Gamma_2}) \circ \zeta_{\alpha, \beta}, \alpha \circledcirc \beta )
\end{align*}
The second component is $\alpha \circledcirc \beta = \rho$. For the first component, use counit naturality at $g \circ \zeta_{\alpha, \beta}^{-1} : \otimes_\alpha(\Gamma_1, \otimes_\beta(\Gamma), \Gamma_2) \to B$:

\begin{center}
	\adjustbox{scale=0.8}{
    \begin{tikzcd}[row sep=large, column sep=huge]
  	\otimes_\alpha(\Gamma_1, (\Gamma_1 \Rightarrow_{\alpha} \otimes_\alpha(\Gamma_1, \otimes_\beta(\Gamma), \Gamma_2) \Leftarrow \Gamma_2), \Gamma_2) \arrow[rr, "{\epsilon_{\otimes}(\otimes_\alpha(\Gamma_1, \otimes_\beta(\Gamma), \Gamma_2))}"] \arrow[d, "{\otimes_\alpha(\1_{\Gamma_1}, (\Gamma_1 \Rightarrow_{\alpha} (g \circ \zeta_{\alpha, \beta}^{-1}) \Leftarrow \Gamma_2), \1_{\Gamma_2})}", swap]   & & \otimes_\alpha(\Gamma_1, \otimes_\beta(\Gamma), \Gamma_2) \arrow[d, "g \circ \zeta_{\alpha, \beta}^{-1}"] \\
  \otimes_\alpha(\Gamma_1, (\Gamma_1 \Rightarrow_{\alpha} B \Leftarrow \Gamma_2), \Gamma_2) \arrow[rr, "\epsilon_{\otimes}(B)"]  & & B
  \end{tikzcd}
  }
\end{center}
This gives:

\begin{align*}
	\epsilon_\otimes(B) &\circ \otimes_\alpha(\1_{\Gamma_1}, (\Gamma_1 \Rightarrow_{\alpha} (g \circ \zeta_{\alpha, \beta}^{-1}) \Leftarrow \Gamma_2) \circ \eta_\otimes(\otimes_\beta\Gamma)  , \1_{\Gamma_2}) \circ \zeta_{\alpha, \beta} \\
	&= \epsilon_\otimes(B) \circ \otimes_\alpha(\1_{\Gamma_1}, (\Gamma_1 \Rightarrow_{\alpha} (g \circ \zeta_{\alpha, \beta}^{-1}) \Leftarrow \Gamma_2) , \1_{\Gamma_2}) \circ \otimes_\alpha(\1_{\Gamma_1}, \eta_\otimes(\otimes_\beta\Gamma)  , \1_{\Gamma_2}) \circ \zeta_{\alpha, \beta} \\
	&= g \circ \zeta_{\alpha, \beta}^{-1} \circ \epsilon_\otimes(\otimes_\alpha(\Gamma_1, \otimes_\beta\Gamma, \Gamma_2)) \circ \otimes_\alpha(\1_{\Gamma_1}, \eta_\otimes(\otimes_\beta\Gamma)  , \1_{\Gamma_2}) \circ \zeta_{\alpha, \beta}
\end{align*}

The triangle identity of the adjunction, evaluated at $X = \otimes_\beta\Gamma$, now gives
\begin{align*}
	g \circ \zeta_{\alpha, \beta}^{-1} \circ \epsilon_\otimes(\otimes_\alpha(\Gamma_1, \otimes_\beta\Gamma, \Gamma_2)) \circ \otimes_\alpha(\1_{\Gamma_1}, \eta_\otimes(\otimes_\beta\Gamma)  , \1_{\Gamma_2}) \circ \zeta_{\alpha, \beta} &=  g \circ \zeta_{\alpha, \beta}^{-1} \circ \zeta_{\alpha, \beta} \\
	&= g
\end{align*}
as required.
It remains to prove the uniqueness of the factorization. Suppose $(\chi, \beta)$, with $\chi : \otimes_\beta(\Gamma) \to (\Gamma_1 \Rightarrow_\alpha B \Leftarrow \Gamma_2)$ vertical, satisfies $(\epsilon_\otimes(B), \alpha) \circ (\chi, \beta) = (g, \rho)$. The first components give
$$\epsilon_\otimes(B) \circ \otimes_\alpha(\1_{\Gamma_1}, \chi, \1_{\Gamma_2}) \circ \zeta_{\alpha, \beta} = g, \quad \text{i.e.} \quad \epsilon_\otimes(B) \circ \otimes_\alpha(\1_{\Gamma_1}, \chi, \1_{\Gamma_2}) = g \circ \zeta_{\alpha, \beta}^{-1}.$$
The assignment $\chi \mapsto \epsilon_\otimes(B) \circ \otimes_\alpha(\1_{\Gamma_1}, \chi, \1_{\Gamma_2})$ is the adjunction bijection between vertical maps $\otimes_\beta(\Gamma) \to (\Gamma_1 \Rightarrow_\alpha B \Leftarrow \Gamma_2)$ and vertical maps $\otimes_\alpha(\Gamma_1, \otimes_\beta(\Gamma), \Gamma_2) \to B$. It is injective, so $\chi$ is the transpose of $g \circ \zeta_{\alpha, \beta}^{-1}$. Thus $(\epsilon_\otimes(B), \alpha)$ is cartesian at $(\Gamma_1 \Rightarrow_\alpha B \Leftarrow \Gamma_2)_m$, and $\pi_m$ is a cloven bifibration. $\square$

\begin{proposition}\label{prop:cloven-bifibration-to-biclosed}
	Given a cloven bifibration of multicategories $\pi : \fib[T]{M_m}$, the underlying multi-indexed category $\pi_o : \fib[T_o]{M}$ is biclosed. Consequently the functor $\mc D$ restricts to a functor
	$$\mathbf{Multi^{bi}_{\mc M_m}} \xto{\mc D} \mathbf{MIdx^{bi}_{\mc M}}.$$
\end{proposition}
\textit{Proof:} Fix a morphism $\otimes_k(\gamma_1, a, \gamma_2) \xto{\alpha} c$ in $\mc M$ and lists $\Gamma_1, \Gamma_2$ of $\mc T$ over $\gamma_1, \gamma_2$, and write $L := \otimes_\alpha(\Gamma_1, (\--), \Gamma_2) : \mc T_a \to \mc T_c$ for the corresponding functor of the underlying multi-indexed category. We must exhibit a right adjoint of $L$.

\textbf{The functor $(\Gamma_1 \Rightarrow_{\alpha}(\--) \Leftarrow \Gamma_2)$ :} since $\pi$ is a fibration of multicategories, for every object $B \in \mc T_c$ there are an object $R(B) \in \mc T_a$ and a cartesian multimorphism
$$\Gamma_1, R(B), \Gamma_2 \xto{e_B} B$$
over $\alpha_m$, cartesian at $R(B)$. Choose one such pair $(R(B),e_B)$ for every $B \in \mc T_c$, and set $(\Gamma_1 \Rightarrow_{\alpha} B \Leftarrow \Gamma_2) := R(B)$. For a vertical morphism $v : B \to B'$ in $\mc T_c$, the composite $v \circ e_B : \Gamma_1, R(B), \Gamma_2 \to B'$ lives over $\alpha_m$, so by cartesianness of $e_{B'}$ there is a unique vertical morphism $R(v) : R(B) \to R(B')$ with
$$e_{B'} \circ R(v) = v \circ e_B.$$
Functoriality follows from the uniqueness of these factorizations: $\1_{R(B)}$ satisfies the equation defining $R(\1_B)$, and for composable $v, v'$ both $R(v' \circ v)$ and $R(v') \circ R(v)$ satisfy $e_{B''} \circ (\--) = v' \circ v \circ e_B$; therefore, they are equal.

\textbf{The adjunction $L \dashv R$ :} for $X \in \mc T_a$ and $B \in \mc T_c$ we construct a bijection between the vertical maps $L(X) = \otimes_\alpha(\Gamma_1, X, \Gamma_2) \to B$ and the vertical maps $X \to R(B)$.
\begin{itemize}
	\item Given a vertical $w : \otimes_\alpha(\Gamma_1, X, \Gamma_2) \to B$, the composite $w \circ \overline{\alpha_m}(\Gamma_1, X, \Gamma_2) : \Gamma_1, X, \Gamma_2 \to B$ lives over $\alpha_m$; by cartesianness of $e_B$ there is a unique vertical $\Phi(w) : X \to R(B)$ with $e_B \circ \Phi(w) = w \circ \overline{\alpha_m}(\Gamma_1, X, \Gamma_2)$.
	\item Given a vertical $u : X \to R(B)$, the composite $e_B \circ u : \Gamma_1, X, \Gamma_2 \to B$ lives over $\alpha_m$; by opcartesianness of $\overline{\alpha_m}(\Gamma_1, X, \Gamma_2)$ there is a unique vertical $\Psi(u) : \otimes_\alpha(\Gamma_1, X, \Gamma_2) \to B$ with $\Psi(u) \circ \overline{\alpha_m}(\Gamma_1, X, \Gamma_2) = e_B \circ u$.
\end{itemize}
The two assignments are mutually inverse. The equation $e_B \circ \Phi(\Psi(u)) = \Psi(u) \circ \overline{\alpha_m}(\Gamma_1, X, \Gamma_2) = e_B \circ u$ gives $\Phi(\Psi(u)) = u$ by cartesian uniqueness. Similarly, $\Psi(\Phi(w)) \circ \overline{\alpha_m}(\Gamma_1, X, \Gamma_2) = e_B \circ \Phi(w) = w \circ \overline{\alpha_m}(\Gamma_1, X, \Gamma_2)$ gives $\Psi(\Phi(w)) = w$ by opcartesian uniqueness. The bijection $\Phi$ is natural:
\begin{itemize}
	\item in $X$ : for a vertical $x : X' \to X$, both $\Phi(w) \circ x$ and $\Phi(w \circ \otimes_\alpha(\1_{\Gamma_1}, x, \1_{\Gamma_2}))$ are vertical maps $u$ satisfying $e_B \circ u = w \circ \otimes_\alpha(\1_{\Gamma_1}, x, \1_{\Gamma_2}) \circ \overline{\alpha_m}(\Gamma_1, X', \Gamma_2)$ since $e_B \circ \Phi(w) \circ x = w \circ \overline{\alpha_m}(\Gamma_1, X, \Gamma_2) \circ x = w \circ \otimes_\alpha(\1_{\Gamma_1}, x, \1_{\Gamma_2}) \circ \overline{\alpha_m}(\Gamma_1, X', \Gamma_2)$.
	\item in $B$ : for a vertical $v : B \to B'$, both $R(v) \circ \Phi(w)$ and $\Phi(v \circ w)$ are vertical maps $u$ satisfying $e_{B'} \circ u = v \circ w \circ \overline{\alpha_m}(\Gamma_1, X, \Gamma_2)$, using $e_{B'} \circ R(v) = v \circ e_B$. They agree by cartesian uniqueness.
\end{itemize}
Hence $L \dashv R$, with unit $\eta_\otimes(X) := \Phi(\1_{L(X)})$ and counit $\epsilon_\otimes(B) := \Psi(\1_{R(B)})$, i.e.\ $\epsilon_\otimes(B)$ is the unique vertical map with $\epsilon_\otimes(B) \circ \overline{\alpha_m}(\Gamma_1, R(B), \Gamma_2) = e_B$. This shows that $\mc T_o$ is biclosed. $\square$

\begin{theorem}\label{thm:biclosed-cloven-bifibration-equivalence}
  There is an equivalence of categories:
	$$\mathbf{MIdx^{bi}_{\mc M}} \simeq \mathbf{Multi^{bi}_{\mc M_m} }.$$
\end{theorem}
\textit{Proof:}
By \cref{prop:biclosed-to-cloven-bifibration,prop:cloven-bifibration-to-biclosed}, the functors $\mc F$ and $\mc D$ restrict to the two full subcategories.
\endgroup{}
\begingroup

\section{Poly-Indexed Categories and Planar Bifibrations}\label{sec:poly-indexed-planar-bifibrations}

The purpose of this section is to apply the analysis of multicategories over unbiased monoidal categories to polycategories over weakly distributive categories. For multicategories, we developed the notion of a \emph{multi-indexed category} (\cref{def:multi-indexed-category}). It describes cloven opfibrations with a family of tensor functors $\otimes_\alpha$ indexed by a morphism in the base. In the polycategorical case, morphisms in the base weakly distributive category have the form $\otimes_n \gamma \xto{\alpha} \oplus_m \delta$. This suggests two families of functors $\otimes_\alpha, \oplus_\alpha$. We generalize them to a collection of functors indexed by a single object. Each functor depends on the surrounding objects in the base morphism.

\subsection{The Planar Base}\label{subsec:planar-wdc-base}

\textbf{Setup and recalled definitions.} For this section, we fix an \textit{unbiased weakly distributive category} denoted
$$(\mc M,\ (\otimes_n)_{n \in \mathbb{N}},\ (\eta^\otimes_{p,m,q})_{p,m,q \in \mathbb{N}},\ \iota_{(\--)},\ (\oplus_n)_{n \in \mathbb{N}},\ (\eta^\oplus_{p,m,q})_{p,m,q \in \mathbb{N}},\ \iota^{(\--)},\ (a_i)_i,\ d^L,\ d^R)$$
in the sense of \cref{def:unbiased-weakly-distributive-category}. We do not assume symmetry and work in the planar case. We recall the conventions attached to the unbiased data, paying attention to the direction of the structural maps:
\begin{itemize}
	\item for every $n \in \mathbb{N}$, functors $\otimes_n, \oplus_n : \mc M \times \ldots \times \mc M \to \mc M$ where the domain is a cartesian product of $n$ factors;
	\item natural isomorphisms
\[
\begin{gathered}
\eta^\otimes_{p,m,q} : \otimes_{p + m + q}(\gamma, \delta, \gamma') \to \otimes_{p + 1 + q}(\gamma, \otimes_m\delta, \gamma')
	\\ \text{and}\quad
	\eta^\oplus_{p,m,q} : \oplus_{p + 1 + q}(\delta, \oplus_m\lambda, \delta') \to \oplus_{p + m + q}(\delta, \lambda, \delta').
\end{gathered}
\]
	Note the chosen directions. The $\otimes$-operator \textit{nests} itself while the $\oplus$-operator \textit{flattens} itself;
	\item natural isomorphisms $\iota_a : \otimes_1(a) \to a$ and $\iota^a : a \to \oplus_1(a)$;
	\item for every list $\lambda$,
	$$a_i : \otimes(\gamma, \oplus\lambda, \gamma') \to \oplus\big(\otimes(\gamma, \oplus\lambda_{\leq i}),\ \otimes(\oplus\lambda_{>i}, \gamma')\big);$$
	\item natural \emph{distributor} transformations
\[
\begin{gathered}
d^L : \otimes_{n+1}(\gamma, \oplus_{m+1}(\delta, a)) \to \oplus_{m+1}(\delta, \otimes_{n+1}(\gamma, a))
	\\ \text{and}\quad
	d^R : \otimes_{n+1}(\oplus_{m+1}(a, \delta), \gamma) \to \oplus_{m+1}(\otimes_{n+1}(a, \gamma), \delta).
\end{gathered}
\]
\end{itemize}

\textbf{Planarity condition.} A pair of composable morphisms at $a$
$\otimes_n \gamma \xto{\alpha} \oplus(\delta_1,a,\delta_2)$ and $\otimes(\gamma_1,a,\gamma_2) \xto{\beta} \oplus_{m'}\delta$
 is \textit{planar} if
$$\big(\delta_1=()\ \text{or}\ \gamma_1=()\big) \qquad\text{and}\qquad \big(\delta_2=()\ \text{or}\ \gamma_2=()\big).$$
An iterated configuration is admissible when every composition occurring in each factor is planar. This is exactly the planarity condition of \cite[Definition~2.2.1]{Bla23}.

\textbf{The two-sided distributor.} For every planar configuration consisting of lists $\gamma_1, \gamma_2, \delta_1, \delta_2$ of objects of $\mc M$ and an object $a$, there is a natural family
$$d_{\gamma_1;\delta_1,a,\delta_2;\gamma_2} \ : \ \otimes_{n'}(\gamma_1, \oplus_m(\delta_1, a, \delta_2), \gamma_2) \to \oplus_m(\delta_1, \otimes_{n'}(\gamma_1, a, \gamma_2), \delta_2),$$
where $n' := |\gamma_1| + 1 + |\gamma_2|$ and $m := |\delta_1| + 1 + |\delta_2|$. The \hyperref[anchor:two-sided-distributor-recovery]{recovery discussion following \cref*{def:unbiased-weakly-distributive-category}} explains how to obtain these maps. It uses a sequence of $\eta^\otimes$ and $\eta^\oplus$, as well as $a_i$ and the maps $d^L,d^R$.

\textbf{Coherences.} We require the two unbiased monoidal structures together with this family $d$ (referred to by Blanco as $\delta$) to satisfy the naturality, unit, nesting and distributivity coherence diagrams of a lax unbiased linearly distributive category \cite[Definition~2.1.18]{Bla23}.

\phantomsection\label{anchor:planar-circledcirc}
\textbf{The operation $\circledcirc$.} Given a planar pair as above, where $n':=|\gamma_1|+1+|\gamma_2|$ and $m:=|\delta_1|+1+|\delta_2|$, we write
\[
\begin{aligned}
&\beta \circledcirc \alpha \ := \ \eta^\oplus_{|\delta_1|, m', |\delta_2|} \circ \oplus_m(\1_{\delta_1}, \beta, \1_{\delta_2}) \circ d \circ \otimes_{n'}(\1_{\gamma_1}, \alpha, \1_{\gamma_2}) \circ \eta^\otimes_{|\gamma_1|, n, |\gamma_2|} \\
&\quad : \ \otimes_{n + n' - 1}(\gamma_1, \gamma, \gamma_2) \to \oplus_{m + m' - 1}(\delta_1, \delta, \delta_2).
\end{aligned}
\]
\Cref{subsec:unbiased-monoidal-base} explains in detail how to recover the multicategorical analogue of this map. The corresponding unit, associativity, and interchange laws are \cref{lem:poly-circledcirc-units,lem:poly-circledcirc-associativity,lem:poly-circledcirc-interchange}.

\begin{lemma}[units]\label{lem:poly-circledcirc-units}
    For every morphism $\otimes_n\gamma \xto{\alpha} \oplus_m(\delta_1, a, \delta_2)$ and every morphism
\[
\otimes_{n'}(\gamma_1, a, \gamma_2) \xto{\beta} \oplus_{m'}\delta
\]
 in $\mc M$, we have
	$$(\iota^a \circ \iota_a) \circledcirc \alpha = \alpha \quadt{and} \beta \circledcirc (\iota^a \circ \iota_a) = \beta .$$
\end{lemma}
\textit{Proof:} The displayed operation $\circledcirc$ is term-for-term the cut constructed from an unbiased linearly distributive category in \cite[Proposition~2.3.1]{Bla23}: $\eta^\otimes$ nests the inner input list, $d$ is Blanco's $\delta$, and $\eta^\oplus$ flattens the resulting output list. The identity in that construction is $\iota^a \circ \iota_a$. \hfill $\square$

\begin{lemma}[associativity]\label{lem:poly-circledcirc-associativity}
    Let $\otimes_n\gamma \xto{\alpha} \oplus_m(\delta_1, a, \delta_2)$,
\[
\otimes_{n'}(\gamma_1, a, \gamma_2) \xto{\beta} \oplus_{m'}(\lambda_1, b, \lambda_2)
\]
 and $\otimes_{n''}(\theta_1, b, \theta_2) \xto{\omega} \oplus_{m''}\xi$ be morphisms in $\mc M$. Then
	$$\omega \circledcirc (\beta \circledcirc \alpha) = (\omega \circledcirc \beta) \circledcirc \alpha.$$
	
\end{lemma}
\textit{Proof:} Under the identifications in the proof of \cref{lem:poly-circledcirc-units}, this is the associativity law for the cut constructed in \cite[Proposition~2.3.1]{Bla23}. The admissibility hypothesis ensures that every distributor component used in either parenthesisation is among the components allowed by \cite[Definition~2.1.18]{Bla23}. $\square$

\begin{lemma}[interchange]\label{lem:poly-circledcirc-interchange}
  	The interchange laws hold for the operation $\circledcirc$:
	\begin{itemize}
        \item \textbf{Output configuration:} let
\[
\otimes_n\gamma \xto{\alpha} \oplus(\delta_1, a, \delta_2, b, \delta_3)
\]
  and let $\otimes(\gamma', a) \xto{\beta} \oplus_{m'}\delta'$ and $\otimes(b, \gamma'') \xto{\omega} \oplus_{m''}\delta''$. Assume $\delta_1=()$ or $\gamma'=()$, and $\delta_3=()$ or $\gamma''=()$. Then
		$$\omega \circledcirc_b (\beta \circledcirc_a \alpha) = \beta \circledcirc_a (\omega \circledcirc_b \alpha).$$
		\item \textbf{Input configuration:} let $\otimes(\theta_1, a, \theta_2, b, \theta_3) \xto{\omega} \oplus_{m''}\xi$ and let $\otimes_n\gamma \xto{\beta} \oplus(\delta_1, a)$ and $\otimes_{n'}\gamma' \xto{\beta'} \oplus(b, \delta'_2)$. Assume $\delta_1=()$ or $\theta_1=()$, and $\delta'_2=()$ or $\theta_3=()$. Then
		$$(\omega \circledcirc_a \beta) \circledcirc_b \beta' = (\omega \circledcirc_b \beta') \circledcirc_a \beta.$$
	\end{itemize}
\end{lemma}
\textit{Proof:} Under the identification used in the proof of \cref{lem:poly-circledcirc-units}, these are the two interchange laws for the planar polycategory constructed in \cite[Proposition~2.3.1]{Bla23}. The empty-list conditions are exactly those needed for the configurations to be planar. \hfill $\square$

\textbf{The underlying planar polycategory $\mc M_p$.} We write $\mc M_p$ for the polycategory underlying the unbiased weakly distributive category $\mc M$:
\begin{itemize}
	\item the objects of $\mc M_p$ are the objects of $\mc M$, an object $a$ being written $a_p$ when regarded in $\mc M_p$;
	\item a polymap $\gamma_p \xto{\alpha_p} \delta_p$, for lists $\gamma := \gamma_1, \ldots, \gamma_n$ and $\delta := \delta_1, \ldots, \delta_m$, is a morphism $\otimes_n\gamma \xto{\alpha} \oplus_m\delta$ of $\mc M$;
	\item the identity of $a_p$ is $(\iota^a \circ \iota_a)_p$, for $\iota^a \circ \iota_a : \otimes_1(a) \to \oplus_1(a)$;
	\item for a planar configuration, the cut of $\gamma_p \xto{\alpha_p} (\delta_1, a, \delta_2)_p$ into $(\gamma_1, a, \gamma_2)_p \xto{\beta_p} \delta_p$ at $a$ is $\beta_p \circ_{a} \alpha_p := (\beta \circledcirc \alpha)_p$.
\end{itemize}

\begin{proposition}\label{prop:base-planar-polycategory}
	$\mc M_p$ is a planar polycategory.
\end{proposition}
\textit{Proof:} By the identification used in the proof of \cref{lem:poly-circledcirc-units}, this is precisely the planar polycategory construction of \cite[Proposition~2.3.1]{Bla23}. Equivalently, its unit, associativity and interchange axioms are \cref{lem:poly-circledcirc-units,lem:poly-circledcirc-associativity,lem:poly-circledcirc-interchange}. $\square$

\textbf{Fibres.} For a functor $\pi : \fib[T]{M}$ of categories and an object $a \in \mc M$, the \textit{fibre} $\mc T_a$ is the subcategory of $\mc T$ whose objects are the objects $A$ with $\pi A = a$. Its morphisms are the morphisms $f$ with $\pi f = \1_a$. Such morphisms are called \textit{vertical}. For a list $\gamma := (\gamma_1, \ldots, \gamma_n)$ we write
$$\mc T_\gamma:=\prod_{k=1}^n\mc T_{\gamma_k} \quadt{and} \mc T^{op}_\gamma:=\prod_{k=1}^n\mc T^{op}_{\gamma_k},$$
with both products equal to $\mathbf 1$ when $n=0$; thus $\mc T_{()}=\mc T^{op}_{()}=\mathbf 1$. For a list of objects $\Gamma := (\Gamma_1, \ldots, \Gamma_n)$ of $\mc T$ we write $\pi \Gamma := (\pi\Gamma_1, \ldots, \pi\Gamma_n)$.

\textbf{Cartesian polymorphisms.} Let $\pi : \fib[T]{M_p}$ be a functor of planar polycategories. A polymap $\Gamma \xto{\phi} \Delta_1, S, \Delta_2$ in $\mc T$ is \textit{out-cartesian at $S$} if the following condition holds. Let $\Theta_1, \Gamma, \Theta_2 \xto{\xi} \Delta_1, \Lambda, \Delta_2$ be any polymap in $\mc T$, and let $\pi\Theta_1, \pi S, \pi\Theta_2 \xto{\beta} \pi\Lambda$ be any polymap in $\mc M_p$. Assume that $\Delta_1=()$ or $\Theta_1=()$, and that $\Delta_2=()$ or $\Theta_2=()$. If $\beta \circ_{\pi S} \pi\phi = \pi\xi$, then there is a unique polymap $\Theta_1, S, \Theta_2 \xto{\psi} \Lambda$ with $\pi \psi = \beta$ and $\psi \circ_S \phi = \xi$. Dually, a polymap $\Gamma_1, R, \Gamma_2 \xto{\psi} \Delta$ is \textit{in-cartesian at $R$} if the following condition holds. Let $\Gamma_1, \Gamma, \Gamma_2 \xto{\xi} \Lambda_1, \Delta, \Lambda_2$ be any polymap, and let $\pi\Gamma \xto{\alpha} \pi\Lambda_1, \pi R, \pi\Lambda_2$ be any polymap. Assume that $\Lambda_1=()$ or $\Gamma_1=()$, and that $\Lambda_2=()$ or $\Gamma_2=()$. If $\pi\psi \circ_{\pi R} \alpha = \pi\xi$, then there is a unique $\Gamma \xto{\phi} \Lambda_1, R, \Lambda_2$ with $\pi\phi = \alpha$ and $\psi \circ_R \phi = \xi$.\\
We say that the functor $\pi$ is a \textit{bifibration of planar polycategories} if cartesian maps always exist:
\begin{itemize}
  \item  for every polymap $\pi\Gamma \xto{\alpha} \delta_1, b, \delta_2$ of $\mc M_p$ and all lists $\Gamma, \Delta_1, \Delta_2$ over $\pi\Gamma, \delta_1, \delta_2$, there exist an object over $b$ and an out-cartesian polymap $\Gamma \to \Delta_1, (\--), \Delta_2$ over $\alpha$;
  \item for every $\gamma_1, a, \gamma_2 \xto{\beta} \pi\Delta$ and all lists $\Gamma_1, \Gamma_2, \Delta$ over $\gamma_1, \gamma_2, \pi\Delta$, there exist an object over $a$ and an in-cartesian polymap $\Gamma_1, (\--), \Gamma_2 \to \Delta$ over $\beta$. 
\end{itemize}

\subsection{Planar Poly-Indexed Categories}\label{subsec:planar-poly-indexed-categories}

To simplify the exposition, we specify a poly-indexed category using some shorthand notation that will be easier to track and digest.
\begin{itemize}
	\item For any list of objects $\Gamma_1, \ldots, \Gamma_n$, we denote $\Gamma := \Gamma_1, \ldots, \Gamma_n$ and $|\Gamma| := n$.
	\item For any list $\Gamma$, we denote $\Gamma_{<i} := \Gamma_1, \ldots, \Gamma_{i-1}$ and $\Gamma_{>i} := \Gamma_{i+1}, \ldots, \Gamma_n$.
	\item For concatenated lists we will constantly use the projection notation. For example, when $\Gamma_i$ lies over $\gamma_i$, we use
	\begin{align*}
		(\Gamma_1, A, \Gamma_2)_{|\gamma_1| + 1} &= A,\\
		(\Gamma_1, A, \Gamma_2)_{< |\gamma_1| + 1} &= \Gamma_1,\\
		(\Gamma_1, A, \Gamma_2)_{> |\gamma_1| + 1} &= \Gamma_2.
	\end{align*}
\end{itemize}

Positions will be treated uniformly: given lists $\gamma, \delta$ with $|\gamma| = n$ and $|\delta| = m$, a \textit{position} of the pair $(\gamma; \delta)$ is either an \textit{output position} $i^+$ for $1 \leq i \leq m$ or an \textit{input position} $j^-$ for $1 \leq j \leq n$.

\begin{definition}\label{def:planar-poly-indexed-category}
	Given an unbiased weakly distributive category $\mc M$, a \textit{planar poly-indexed category} $\mc T$ \textit{over} $\mc M$, denoted $(\mc T, \mc M, \pi)$, consists of
	\begin{itemize}
		\item An $\mc M$-indexed family of categories $(\mc T_a)_{a\in\mc M}$. For notational convenience, let
		$$\mc T:=\coprod_{a\in\mc M}\mc T_a$$
		be the disjoint union category and let $\pi:\mc T\to\mc M$ send every object of $\mc T_a$ to $a$ and every morphism to $\1_a$. 
		\item For every pair of lists $(\gamma; \delta)$ with $|\gamma| = n$, $|\delta| = m$ and every morphism $\otimes_n\gamma \xto{\alpha} \oplus_m\delta$, a family of functors between the fibres, one for each position:
		\begin{align*}
			F_\alpha^i &: \mc T_{\delta_{<i}}^{op} \times \mc T_{\gamma} \times \mc T_{\delta_{>i}}^{op} \to \mc T_{\delta_i} & \text{for every output position } i^+, \\
			U_\alpha^j &: \mc T_{\gamma_{<j}}^{op} \times \mc T_{\delta} \times \mc T_{\gamma_{>j}}^{op} \to \mc T_{\gamma_j} & \text{for every input position } j^-.
		\end{align*}
		We will use the bar notation for these functors, as in $F^i_\alpha(\Delta_{<i} \mid \Gamma \mid \Delta_{>i})$ and $U^j_\alpha(\Gamma_{<j} \mid \Delta \mid \Gamma_{>j})$.
		\item \textbf{Representations.} For lists $\Gamma, \Delta$ over $\gamma, \delta$, write
\[
\begin{gathered}
\mathrm{Rep}^{i^+}_\alpha(\Gamma; \Delta) := \mc T_{\delta_i}\big(F^i_\alpha(\Delta_{<i} \mid \Gamma \mid \Delta_{>i}),\, \Delta_i\big)
		\\ \text{and}\quad
		\mathrm{Rep}^{j^-}_\alpha(\Gamma; \Delta) := \mc T_{\gamma_j}\big(\Gamma_j,\, U^j_\alpha(\Gamma_{<j} \mid \Delta \mid \Gamma_{>j})\big)
\end{gathered}
\]
		for the \textit{representation sets} of $\alpha$. An element of $\mathrm{Rep}^p_\alpha(\Gamma;\Delta)$ is a presentation \textit{at the position} $p$ of a map from $\Gamma$ to $\Delta$ over $\alpha$. No position is preferred. The data therefore includes a family of natural isomorphisms called \emph{change of representative}:
		$$\varphi^{p, q}_\alpha : \mathrm{Rep}^{p}_\alpha(\Gamma; \Delta) \to \mathrm{Rep}^{q}_\alpha(\Gamma; \Delta).$$
		There is one such isomorphism for every ordered pair of positions of $(\gamma;\delta)$. These isomorphisms are subject to the \textit{change of representative conditions}
		$$\varphi^{p,p}_\alpha = \1 \quadt{and} \varphi^{q,r}_\alpha \circ \varphi^{p,q}_\alpha = \varphi^{p,r}_\alpha \quad \text{for all positions } p, q, r.$$
		In particular, each $\varphi^{p,q}_\alpha$ is a bijection, with inverse $\varphi^{q,p}_\alpha$.

		A \textit{representation of $\alpha$ at $(\Gamma;\Delta)$} is a family
		$$x=(x_p)_p, \qquad x_p \in \mathrm{Rep}^p_\alpha(\Gamma;\Delta) \ \text{ for every position } p \text{ of } (\gamma;\delta).$$
		Its members all present the same map, in the sense that $\varphi^{p,q}_\alpha(x_p)=x_q$ for all positions $p,q$. By the change of representative conditions, if $\alpha$ has at least one position, then every element $h$ of every $\mathrm{Rep}^p_\alpha(\Gamma;\Delta)$ extends to exactly one representation. We write it as
		$$\langle h\rangle^p, \qquad \langle h\rangle^p_q := \varphi^{p,q}_\alpha(h).$$
		Every representation arises this way from any one of its members. The superscript is dropped when the position is clear. If $\alpha$ has no position, there is exactly one representation, the empty family. The whole section turns on representations. They are what \cref{def:underlying-planar-polycategory} takes as the polymorphisms of $\mc T_p$. The two correspondence axioms below say how they behave under identities and under cutting.
		\item For every $a \in \mc M$, natural isomorphisms
		$$\tau^F_a : F^1_{\iota^a \circ \iota_a} \to \1_{\mc T_a} \quadt{and} \tau^U_a : \1_{\mc T_a} \to U^1_{\iota^a \circ \iota_a}$$
        \item \textbf{Decomposition isomorphisms.} For every planar pair $\otimes_n\gamma \xto{\alpha} \oplus_m(\delta_1, a, \delta_2)$ and
\[
\otimes_{n'}(\gamma_1, a, \gamma_2) \xto{\beta} \oplus_{m'}\delta,
\]
set $c := |\delta_1| + 1$ and $c' := |\gamma_1| + 1$. Then $a$ is at position $c$ in $\alpha$ and at position $c'$ in $\beta$. For all lists $\Gamma, \Gamma_1, \Gamma_2, \Delta, \Delta_1, \Delta_2$ of objects of $\mc T$ over $\gamma, \gamma_1, \gamma_2, \delta, \delta_1, \delta_2$, respectively, the data includes the following natural isomorphisms:
		\begin{itemize}
			\item \textbf{Covariant families:}
			\begin{align*}
&\epsilon^F_i[\beta,\alpha] : F^{|\delta_1| + i}_{\beta \circledcirc \alpha}(\Delta_1, \Delta_{<i} \mid \Gamma_1, \Gamma, \Gamma_2 \mid \Delta_{>i}, \Delta_2)  \\
&\quad\xto{} F^i_\beta\big(\Delta_{<i} \mid \Gamma_1, F^{c}_\alpha(\Delta_1 \mid \Gamma \mid \Delta_2), \Gamma_2 \mid \Delta_{>i}\big)\qquad 1 \leq i \leq |\delta| \\[0.6em]
&\epsilon^U_j[\beta,\alpha] : U^j_\alpha\big(\Gamma_{<j} \mid \Delta_1, U^{c'}_\beta(\Gamma_1 \mid \Delta \mid \Gamma_2), \Delta_2 \mid \Gamma_{>j}\big)  \\
&\quad\xto{} U^{|\gamma_1| + j}_{\beta \circledcirc \alpha}(\Gamma_1, \Gamma_{<j} \mid \Delta_1, \Delta, \Delta_2 \mid \Gamma_{>j}, \Gamma_2)\qquad 1 \leq j \leq |\gamma|
			\end{align*}
			\item \textbf{Contravariant families:}
			{\small
			\begin{align*}
&\eta^F_{i \to}[\beta,\alpha] : F^i_{\beta \circledcirc \alpha}\big((\Delta_1)_{<i} \mid \Gamma_1, \Gamma, \Gamma_2 \mid (\Delta_1)_{>i}, \Delta, \Delta_2\big)  \\
&\quad\xto{} F^{i}_\alpha\big((\Delta_1)_{<i} \mid \Gamma \mid (\Delta_1)_{>i}, U^{c'}_\beta(\Gamma_1 \mid \Delta \mid \Gamma_2), \Delta_2\big)\qquad 1 \leq i \leq |\delta_1| \\[0.6em]
&\eta^F_{\to i}[\beta,\alpha] : F^{|\delta_1| + |\delta| + i}_{\beta \circledcirc \alpha}\big(\Delta_1, \Delta, (\Delta_2)_{<i} \mid \Gamma_1, \Gamma, \Gamma_2 \mid (\Delta_2)_{>i}\big)  \\
&\quad\xto{} F^{|\delta_1| + 1 + i}_\alpha\big(\Delta_1, U^{c'}_\beta(\Gamma_1 \mid \Delta \mid \Gamma_2), (\Delta_2)_{<i} \mid \Gamma \mid (\Delta_2)_{>i}\big)\qquad 1 \leq i \leq |\delta_2| \\[0.6em]
&\eta^U_{j \to}[\beta,\alpha] : U^{j}_\beta\big((\Gamma_1)_{<j} \mid \Delta \mid (\Gamma_1)_{>j}, F_\alpha^{c}(\Delta_1 \mid \Gamma \mid \Delta_2), \Gamma_2\big)  \\
&\quad\xto{} U^j_{\beta \circledcirc \alpha}\big((\Gamma_1)_{<j} \mid \Delta_1, \Delta, \Delta_2 \mid (\Gamma_1)_{>j}, \Gamma, \Gamma_2\big)\qquad 1 \leq j \leq |\gamma_1| \\[0.6em]
&\eta^U_{\to j}[\beta,\alpha] : U^{|\gamma_1| + 1 + j}_\beta\big(\Gamma_1, F_\alpha^{c}(\Delta_1 \mid \Gamma \mid \Delta_2), (\Gamma_2)_{<j} \mid \Delta \mid (\Gamma_2)_{>j}\big)  \\
&\quad\xto{} U^{|\gamma_1| + |\gamma| + j}_{\beta \circledcirc \alpha}\big(\Gamma_1, \Gamma, (\Gamma_2)_{<j} \mid \Delta_1, \Delta, \Delta_2 \mid (\Gamma_2)_{>j}\big)\qquad 1 \leq j \leq |\gamma_2|
			\end{align*}
			}
		\end{itemize}
	\end{itemize}
	Moreover, a poly-indexed category is required to satisfy the coherence axioms in \cref{subsec:poly-indexed-coherence}.
\end{definition}

\subsection{Coordinates of Cut}\label{subsec:coordinate-calculus-cut}

For the rest of this section, the central operation of cut and the objects representing morphisms occur constantly. We describe their interaction here and fix more notation. The cut in the planar base is the operation $\circledcirc$ defined in the \hyperref[anchor:planar-circledcirc]{preceding setup}. In a linearly distributive category, the only operation available is composition. The operation of cut is more general. Two polymorphisms can be composed (cut) at different objects, and the resulting cuts need not lead to the same composite. Moreover, composites at different objects land in different collections of morphisms and are therefore incomparable.

\Cref{def:planar-poly-indexed-category,def:cut-configuration} formalize this operation of cut in a category. The definition of $\circledcirc$ is the stepping stone of this project. The second ingredient is the presence of the functors $F,U$, which are designed to represent polymorphisms. The idea is the following. For any polymorphism $\Gamma \to \Delta$ in a two-tensor polycategory, there exist objects $\otimes \Gamma, \oplus \Delta$ that represent the list $\Gamma, \Delta$. The polymorphisms $\Gamma \to \otimes \Gamma$ and $\oplus \Delta \to \Delta$ are universal. Thus, there is a one-to-one correspondence between maps $\Gamma \to \Delta$ and maps $\otimes \Gamma \to \Delta$. In turn, the maps $\otimes \Gamma \to \Delta$ are in one-to-one correspondence with maps $\otimes \Gamma \to \oplus \Delta$. This last map is unary, with one object in the domain and one in the codomain. Hence a polymorphism can be represented by a unary map in the two-tensor polycategory.

We extend this realization to a bifibration, where representing objects exist at every slot of the domain and codomain. This makes the analysis rather tedious. The idea, however, is the same. A polymorphism $\Gamma \to \Delta$ has many representing objects, namely $|\Gamma| + |\Delta|$ objects. In the following, we consider these objects and the maps between them as representing the original map. Most of the definitions below ensure that this collection of maps is well behaved and that its representatives are unique.

\begin{definition}[cut configuration]\label{def:cut-configuration}
	A \textit{cut configuration} consists of a planar pair
	$$\otimes_n\gamma \xto{\alpha} \oplus(\delta_1,a,\delta_2), \qquad \otimes(\gamma_1,a,\gamma_2) \xto{\beta} \oplus_{m'}\delta$$
	of morphisms of $\mc M$, together with lists $\Gamma,\Gamma_1,\Gamma_2,\Delta,\Delta_1,\Delta_2$ of objects of $\mc T$ over $\gamma,\gamma_1,\gamma_2,\delta,\delta_1,\delta_2$, respectively. We write
	$$c:=|\delta_1|+1, \qquad c':=|\gamma_1|+1$$
	for the two \textit{cut positions}. Throughout, the composite is written $\beta\circledcirc\alpha$. The two \textit{representative objects} of the configuration are
	$$\widehat F:=F^c_\alpha(\Delta_1\mid\Gamma\mid\Delta_2) \quadt{and} \widehat U:=U^{c'}_\beta(\Gamma_1\mid\Delta\mid\Gamma_2).$$
\end{definition}

\begin{example}
    In the multicategorical case, we can take the two morphisms $\otimes \gamma \to a$ and
\[
\otimes(\gamma_1, a, \gamma_2) \to b.
\]
The two cut positions are $1$ in the first and $|\gamma_1| + 1$ in the second. Given $\Gamma$ over $\gamma$, the representative at the output position is $\widehat F := F^1(\mid \Gamma \mid)$, which represents the universal tensor in a tensor multicategory, $\otimes \Gamma$.
\end{example}

\begin{definition}[relative positions, cut set]\label{def:inherited-position}
	For any cut configuration, the positions of $\beta\circledcirc\alpha$ are exactly the positions in $\delta_1,\delta,\delta_2$ and $\gamma_1,\gamma,\gamma_2$. For a position $p$ of $\beta\circledcirc\alpha$, we write $\mathsf{fac}(p)\in\{\alpha,\beta\}$ for the morphism carrying the original position and $\bar p$ for the original position.

	The two positions not present in $\beta \circledcirc \alpha$ are the cut positions. They occur only in the original morphisms $\alpha, \beta$. Their representation sets agree, and their common value is called the \emph{cut set}:
	$$\mathrm{Rep}^{c^+}_\alpha(\Gamma;\Delta_1,\widehat U,\Delta_2)\ =\ \mc T_a(\widehat F,\widehat U)\ =\ \mathrm{Rep}^{c'^-}_\beta(\Gamma_1,\widehat F,\Gamma_2;\Delta).$$
\end{definition}

\textit{Remark.}
	An element $u:\widehat F\to\widehat U$ of the cut set therefore sits at a position of $\alpha$ and at a position of $\beta$, and so generates one representation of each factor. We write
	$$u_\alpha:=\langle u\rangle^{c^+} \quadt{and} u_\beta:=\langle u\rangle^{c'^-},$$
	with the change of representative map $(u_\alpha)_r=\varphi^{c^+,r}_\alpha(u)$ at a position $r$ of $\alpha$ and $(u_\beta)_s=\varphi^{c'^-,s}_\beta(u)$ at a position $s$ of $\beta$. These are two different families. 

\begin{example}\label{ex:minimal-cut}
	Consider a cut at one output and one input: $y \xto{\alpha} a$ and $a \xto{\beta} z$. The composition $y \xto{\beta\circledcirc\alpha} z$ has one input and one output. Write $Y$ for the object of $\mc T$ at the input of $\alpha$ and $Z$ for the object at the output of $\beta$, so that $\Gamma=(Y)$ and $\Delta=(Z)$. The representative objects are then
	$$\widehat F=F^1_\alpha(\,\mid Y\mid\,) \quadt{and} \widehat U=U^1_\beta(\,\mid Z\mid\,),$$
	both in the fibre $\mc T_a$.
	Each factor has two positions, and so does $\beta\circledcirc\alpha$:
	$$\begin{array}{lll}
		\text{position of } \beta\circledcirc\alpha & \text{zone} & (\mathsf{fac}(p),\bar p) \\[3pt]
		1^+ \text{ over } z  & z & (\beta,\,1^+) \\[3pt]
		1^- \text{ over } y & y & (\alpha,\,1^-)
	\end{array}$$
	The two remaining positions of the factors are the cut positions, the output $1^+$ of $\alpha$ and the input $1^-$ of $\beta$. They carry the same set:
	$$\mathrm{Rep}^{1^+}_\alpha(Y;\widehat U)=\mc T_a\big(F^1_\alpha(\,\mid Y\mid\,),\widehat U\big)=\mc T_a(\widehat F,\widehat U)=\mc T_a\big(\widehat F,U^1_\beta(\,\mid Z\mid\,)\big)=\mathrm{Rep}^{1^-}_\beta(\widehat F;Z).$$
	An element $u:\widehat F\to\widehat U$ of this cut set generates $u_\alpha=\langle u\rangle^{1^+}$ and $u_\beta=\langle u\rangle^{1^-}$.
\end{example}

\begin{definition}[coordinate change]\label{def:coordinate-change}
	Fix a cut configuration and a position $p$ of $\beta\circledcirc\alpha$. The representative functor of the composite at $p$ is related to that of $\mathsf{fac}(p)$ at $\bar p$. For each of the two factors, we set a bijection $\kappa_p$, called the \textit{coordinate change} at $p$:
	\begin{align*}
		\kappa_p &: \mathrm{Rep}^{\bar p}_\alpha\big(\Gamma;\ \Delta_1,\widehat U,\Delta_2\big) \xrightarrow{\ \cong\ } \mathrm{Rep}^{p}_{\beta\circledcirc\alpha}\big(\Gamma_1,\Gamma,\Gamma_2;\ \Delta_1,\Delta,\Delta_2\big) & \text{when } \mathsf{fac}(p)=\alpha, \\
		\kappa_p &: \mathrm{Rep}^{\bar p}_\beta\big(\Gamma_1,\widehat F,\Gamma_2;\ \Delta\big) \xrightarrow{\ \cong\ } \mathrm{Rep}^{p}_{\beta\circledcirc\alpha}\big(\Gamma_1,\Gamma,\Gamma_2;\ \Delta_1,\Delta,\Delta_2\big) & \text{when } \mathsf{fac}(p)=\beta.
	\end{align*}
	The following table gathers the construction of the bijection according to the zone in which $p$ lies.
	$$\begin{array}{llll}
		\text{zone} & p & (\mathsf{fac}(p),\bar p) & \kappa_p(h) \\[3pt]
		\delta_1,\ i & i^+ & (\alpha,i^+) & h\circ\eta^F_{i\to} \\[3pt]
		\delta,\ i & (|\delta_1|+i)^+ & (\beta,i^+) & h\circ\epsilon^F_i \\[3pt]
		\delta_2,\ i & (|\delta_1|+|\delta|+i)^+ & (\alpha,(|\delta_1|+1+i)^+) & h\circ\eta^F_{\to i} \\[3pt]
		\gamma_1,\ j & j^- & (\beta,j^-) & \eta^U_{j\to}\circ h \\[3pt]
		\gamma,\ j & (|\gamma_1|+j)^- & (\alpha,j^-) & \epsilon^U_j\circ h \\[3pt]
		\gamma_2,\ j & (|\gamma_1|+|\gamma|+j)^- & (\beta,(|\gamma_1|+1+j)^-) & \eta^U_{\to j}\circ h
	\end{array}$$
\end{definition}

\begin{example}
	In the configuration of \cref{ex:minimal-cut}, with $Y$ at the unique input and $Z$ at the unique output, there are two positions and hence two coordinate changes:
	$$\kappa_{1^+}:\mathrm{Rep}^{1^+}_\beta(\widehat F;Z)\to\mathrm{Rep}^{1^+}_{\beta\circledcirc\alpha}(Y;Z) \quadt{and} \kappa_{1^-}:\mathrm{Rep}^{1^-}_\alpha(Y;\widehat U)\to\mathrm{Rep}^{1^-}_{\beta\circledcirc\alpha}(Y;Z).$$
	Both are covariant:
	\begin{align*}
		\kappa_{1^+}(h) &= h\circ\epsilon^F_1, & \epsilon^F_1 &: F^1_{\beta\circledcirc\alpha}(\,\mid Y\mid\,)\to F^1_\beta(\,\mid\widehat F\mid\,), && \text{so } h:F^1_\beta(\,\mid\widehat F\mid\,)\to Z \ \mapsto \ F^1_{\beta\circledcirc\alpha}(\,\mid Y\mid\,)\to Z,\\
		\kappa_{1^-}(h) &= \epsilon^U_1\circ h, & \epsilon^U_1 &: U^1_\alpha(\,\mid\widehat U\mid\,)\to U^1_{\beta\circledcirc\alpha}(\,\mid Z\mid\,), && \text{so } h:Y\to U^1_\alpha(\,\mid\widehat U\mid\,) \ \mapsto \ Y\to U^1_{\beta\circledcirc\alpha}(\,\mid Z\mid\,).
	\end{align*}
\end{example}

\begin{definition}[cut reading]\label{def:cut-reading}
	Fix a cut configuration. For every position $p$ of $\beta\circledcirc\alpha$, put
	$$K_p \ : \ \mc T_a(\widehat F,\widehat U)\longrightarrow\mathrm{Rep}^p_{\beta\circledcirc\alpha}(\Gamma_1,\Gamma,\Gamma_2;\Delta_1,\Delta,\Delta_2), \qquad K_p(u):=\kappa_p\Big(\big(u_{\mathsf{fac}(p)}\big)_{\bar p}\Big).$$
	The map $K_p$ is a bijection. It is $\kappa_p$ composed with the change of representative $\varphi^{c^+,\bar p}_\alpha$ or $\varphi^{c'^-,\bar p}_\beta$, according to whether $\mathsf{fac}(p)$ is $\alpha$ or $\beta$. The \textit{cut reading} of an element $u:\widehat F\to \widehat U$ of the cut set is the family
	$$\kappa\langle u\rangle:=\big(K_p(u)\big)_p$$
	indexed by the positions $p$ of $\beta\circledcirc\alpha$.
\end{definition}

\textit{Remark.} The operation $\kappa\langle u \rangle$ generates a collection of representative maps of the composition $\beta \circledcirc \alpha$ from a cut map $u \in \mc T_a(\widehat F, \widehat U)$. This collection is generated from $\alpha$ or $\beta$, depending on the position $p$. Nothing guarantees that its members cohere as a representation of the composition in the sense of the maps $\varphi$. A further axiom is therefore needed to generate a representation of a polymorphism.

\begin{example}
	Still in \cref{ex:minimal-cut}, $Y$ lies in $\mc T_{y}$, $Z$ in $\mc T_{z}$, and $\widehat F,\widehat U$ in $\mc T_a$. The cut set is
	$$\mathrm{Rep}^{1^+}_\alpha(Y;\widehat U)=\mc T_a(\widehat F,\widehat U)=\mathrm{Rep}^{1^-}_\beta(\widehat F;Z),$$
	and the four other representation sets in play are
	\begin{align*}
		\mathrm{Rep}^{1^+}_\beta(\widehat F;Z) &= \mc T_{z}\big(F^1_\beta(\,\mid\widehat F\mid\,),\ Z\big), &
		\mathrm{Rep}^{1^+}_{\beta\circledcirc\alpha}(Y;Z) &= \mc T_{z}\big(F^1_{\beta\circledcirc\alpha}(\,\mid Y\mid\,),\ Z\big), \\
		\mathrm{Rep}^{1^-}_\alpha(Y;\widehat U) &= \mc T_{y}\big(Y,\ U^1_\alpha(\,\mid\widehat U\mid\,)\big), &
		\mathrm{Rep}^{1^-}_{\beta\circledcirc\alpha}(Y;Z) &= \mc T_{y}\big(Y,\ U^1_{\beta\circledcirc\alpha}(\,\mid Z\mid\,)\big).
	\end{align*}
	The cut reading of an element $u:\widehat F\to\widehat U$ of the cut set has exactly two components, one per position of $\beta\circledcirc\alpha$. Each component is obtained by applying change of representative to $u$ within its factor and then changing coordinates. As maps of sets, the two are the composites
	\begin{align*}
		K_{1^+} &: \mc T_a(\widehat F,\widehat U) \xto{\ \varphi^{1^-,1^+}_\beta\ } \mathrm{Rep}^{1^+}_\beta(\widehat F;Z) \xto{\ \kappa_{1^+}\ } \mathrm{Rep}^{1^+}_{\beta\circledcirc\alpha}(Y;Z), \\
		K_{1^-} &: \mc T_a(\widehat F,\widehat U) \xto{\ \varphi^{1^+,1^-}_\alpha\ } \mathrm{Rep}^{1^-}_\alpha(Y;\widehat U) \xto{\ \kappa_{1^-}\ } \mathrm{Rep}^{1^-}_{\beta\circledcirc\alpha}(Y;Z).
	\end{align*}
	The first reads $u$ inside $\beta$. The second reads $u$ inside $\alpha$. On elements, they are
	\begin{align*}
		K_{1^+}(u) &= \kappa_{1^+}\big((u_\beta)_{1^+}\big) = \varphi^{1^-,1^+}_\beta(u)\circ\epsilon^F_1 && : \ F^1_{\beta\circledcirc\alpha}(\,\mid Y\mid\,)\to Z, \\
		K_{1^-}(u) &= \kappa_{1^-}\big((u_\alpha)_{1^-}\big) = \epsilon^U_1\circ\varphi^{1^+,1^-}_\alpha(u) && : \ Y\to U^1_{\beta\circledcirc\alpha}(\,\mid Z\mid\,).
	\end{align*}
	The first composite is formed in $\mc T_{z}$, and the second is formed in $\mc T_{y}$. They are formed from
	\begin{align*}
		\epsilon^F_1 &: F^1_{\beta\circledcirc\alpha}(\,\mid Y\mid\,)\to F^1_\beta(\,\mid\widehat F\mid\,), & \varphi^{1^-,1^+}_\beta(u) &: F^1_\beta(\,\mid\widehat F\mid\,)\to Z, \\
		\varphi^{1^+,1^-}_\alpha(u) &: Y\to U^1_\alpha(\,\mid\widehat U\mid\,), & \epsilon^U_1 &: U^1_\alpha(\,\mid\widehat U\mid\,)\to U^1_{\beta\circledcirc\alpha}(\,\mid Z\mid\,).
	\end{align*}
	The first is formed through $\beta$, and the second is formed through $\alpha$. Nothing said so far forces them to present the same map. The correspondence--decomposition axiom requires that they do. Here, this requirement is the single equation $\varphi^{1^+,1^-}_{\beta\circledcirc\alpha}\big(K_{1^+}(u)\big)=K_{1^-}(u)$ in $\mathrm{Rep}^{1^-}_{\beta\circledcirc\alpha}(Y;Z)$. The comparison map is
	$$\varphi^{1^+,1^-}_{\beta\circledcirc\alpha}:\mathrm{Rep}^{1^+}_{\beta\circledcirc\alpha}(Y;Z)\longrightarrow\mathrm{Rep}^{1^-}_{\beta\circledcirc\alpha}(Y;Z).$$
	This axiom makes $\kappa\langle u\rangle$ a polymorphism $Y\to Z$ over $\beta\circledcirc\alpha$. When $u$ arises as $u=g'_1\circ f_1$ from polymorphisms $f$ over $\alpha$ at $(Y;A)$ and $g$ over $\beta$ at $(A;Z)$ meeting at an object $A$ of $\mc T_a$, we have
	$$f_1\in\mathrm{Rep}^{1^+}_\alpha(Y;A)=\mc T_a(\widehat F,A) \quadt{and} g'_1\in\mathrm{Rep}^{1^-}_\beta(A;Z)=\mc T_a(A,\widehat U).$$
	The resulting polymorphism is precisely their cut.
\end{example}

\textit{Remark.} A cut reading is a collection of potential representative maps for the composition, not yet necessarily a representation. The following coherence laws contain the axiom that makes it a representation.

\subsection{Coherence Laws}\label{subsec:poly-indexed-coherence}

\begin{definition}[coherence laws of a poly-indexed category]\label{def:poly-indexed-coherence}
	The following coherence laws relate the components of poly-indexed categories.
	\begin{itemize}
        \item \textbf{Contravariant Units.} For every
\[
\otimes_n\gamma \xto{\alpha} \oplus_m(\delta_1, a, \delta_2)
\]
 and objects $\Gamma, \Delta_1, \Delta_2, A$ over $\gamma, \delta_1, \delta_2, a$:
		\begin{center}
			\adjustbox{scale=0.8}{
			\begin{tikzcd}[row sep=large, column sep=huge]
				F^i_\alpha\big((\Delta_1)_{<i} \mid \Gamma \mid (\Delta_1)_{>i}, A, \Delta_2\big) \arrow[rr, bend left=15, "\eta^F_{i \to}"] & & F^i_\alpha\big((\Delta_1)_{<i} \mid \Gamma \mid (\Delta_1)_{>i}, U^1_{\iota^a \circ \iota_a}(A), \Delta_2\big) \arrow[d, "{F^i_\alpha(\1_{(\Delta_1)_{<i}} \mid \1_\Gamma \mid \1_{(\Delta_1)_{>i}}, \tau^U_a(A), \1_{\Delta_2})}"] \\
				& & F^i_\alpha\big((\Delta_1)_{<i} \mid \Gamma \mid (\Delta_1)_{>i}, A, \Delta_2\big) \arrow[ull, equals, bend left=10]
			\end{tikzcd}
			}
		\end{center}
\begin{samepage}
		For this diagram, write
\begin{align*}
N_{1}&:=F^{|\delta_1| + 1 + i}_\alpha\big(\Delta_1, A, (\Delta_2)_{<i} \mid \Gamma \mid (\Delta_2)_{>i}\big) \\
N_{2}&:=F^{|\delta_1| + 1 + i}_\alpha\big(\Delta_1, U^1_{\iota^a \circ \iota_a}(A), (\Delta_2)_{<i} \mid \Gamma \mid (\Delta_2)_{>i}\big) \\
a_{1}&:=F^{|\delta_1| + 1 + i}_\alpha(\1_{\Delta_1}, \tau^U_a(A), \1_{(\Delta_2)_{<i}} \mid \1_\Gamma \mid \1_{(\Delta_2)_{>i}})
\end{align*}
\begin{center}
\begin{tikzcd}[row sep=large, column sep=4em]

				N_{1} \arrow[rr, bend left=15, "\eta^F_{\to i}"] & & N_{2} \arrow[d, "a_{1}"] \\
				& & N_{1} \arrow[ull, equals, bend left=10]

\end{tikzcd}
\end{center}
\end{samepage}
		\begin{center}
			\adjustbox{scale=0.8}{
			\begin{tikzcd}[row sep=large, column sep=huge]
				U^j_\alpha\big(\Gamma_{<j} \mid \Delta_1, A, \Delta_2 \mid \Gamma_{>j}\big) \arrow[rr, bend left=15, "{U^j_\alpha(\1_{\Gamma_{<j}} \mid \1_{\Delta_1}, \tau^U_a(A), \1_{\Delta_2} \mid \1_{\Gamma_{>j}})}"] & & U^j_\alpha\big(\Gamma_{<j} \mid \Delta_1, U^1_{\iota^a \circ \iota_a}(A), \Delta_2 \mid \Gamma_{>j}\big) \arrow[d, "\epsilon^U_j"] \\
				& & U^j_\alpha\big(\Gamma_{<j} \mid \Delta_1, A, \Delta_2 \mid \Gamma_{>j}\big) \arrow[ull, equals, bend left=10]
			\end{tikzcd}
			}
			\qquad
			\adjustbox{scale=0.8}{
			\begin{tikzcd}[row sep=large, column sep=huge]
				\widehat F \arrow[rr, bend left=15, "\epsilon^F_1"] & & F^1_{\iota^a \circ \iota_a}\big(\widehat F\big) \arrow[d, "\tau^F_a(\widehat F)"] \\
				& & \widehat F \arrow[ull, equals, bend left=10]
			\end{tikzcd}
			}
		\end{center}
		In the last diagram $\epsilon^F_1$ is the covariant decomposition at the unique output of $\iota^a \circ \iota_a$.
		\item \textbf{Covariant Units.} Dually, for every $\otimes_{n'}(\gamma_1, a, \gamma_2) \xto{\beta} \oplus_{m'}\delta$ and objects $\Gamma_1, \Gamma_2, \Delta, A$ over $\gamma_1, \gamma_2, \delta, a$:
		\begin{center}
			\adjustbox{scale=0.8}{
			\begin{tikzcd}[row sep=large, column sep=huge]
				F^i_\beta\big(\Delta_{<i} \mid \Gamma_1, A, \Gamma_2 \mid \Delta_{>i}\big) \arrow[rr, bend left=15, "\epsilon^F_i"] & & F^i_\beta\big(\Delta_{<i} \mid \Gamma_1, F^1_{\iota^a \circ \iota_a}(A), \Gamma_2 \mid \Delta_{>i}\big) \arrow[d, "{F^i_\beta(\1_{\Delta_{<i}} \mid \1_{\Gamma_1}, \tau^F_a(A), \1_{\Gamma_2} \mid \1_{\Delta_{>i}})}"] \\
				& & F^i_\beta\big(\Delta_{<i} \mid \Gamma_1, A, \Gamma_2 \mid \Delta_{>i}\big) \arrow[ull, equals, bend left=10]
			\end{tikzcd}
			}
		\end{center}
		\begin{center}
			\adjustbox{scale=0.8}{
			\begin{tikzcd}[row sep=large, column sep=huge]
				U^j_\beta\big((\Gamma_1)_{<j} \mid \Delta \mid (\Gamma_1)_{>j}, A, \Gamma_2\big) \arrow[rr, bend left=15, "{U^j_\beta(\1_{(\Gamma_1)_{<j}} \mid \1_\Delta \mid \1_{(\Gamma_1)_{>j}}, \tau^F_a(A), \1_{\Gamma_2})}"] & & U^j_\beta\big((\Gamma_1)_{<j} \mid \Delta \mid (\Gamma_1)_{>j}, F^1_{\iota^a \circ \iota_a}(A), \Gamma_2\big) \arrow[d, "\eta^U_{j \to}"] \\
				& & U^j_\beta\big((\Gamma_1)_{<j} \mid \Delta \mid (\Gamma_1)_{>j}, A, \Gamma_2\big) \arrow[ull, equals, bend left=10]
			\end{tikzcd}
			}
		\end{center}
		\begin{center}
			\adjustbox{scale=0.8}{
			\begin{tikzcd}[row sep=large, column sep=huge]
				U^{|\gamma_1| + 1 + j}_\beta\big(\Gamma_1, A, (\Gamma_2)_{<j} \mid \Delta \mid (\Gamma_2)_{>j}\big) \arrow[rr, bend left=15, "{U^{|\gamma_1| + 1 + j}_\beta(\1_{\Gamma_1}, \tau^F_a(A), \1_{(\Gamma_2)_{<j}} \mid \1_\Delta \mid \1_{(\Gamma_2)_{>j}})}"] & & U^{|\gamma_1| + 1 + j}_\beta\big(\Gamma_1, F^1_{\iota^a \circ \iota_a}(A), (\Gamma_2)_{<j} \mid \Delta \mid (\Gamma_2)_{>j}\big) \arrow[d, "\eta^U_{\to j}"] \\
				& & U^{|\gamma_1| + 1 + j}_\beta\big(\Gamma_1, A, (\Gamma_2)_{<j} \mid \Delta \mid (\Gamma_2)_{>j}\big) \arrow[ull, equals, bend left=10]
			\end{tikzcd}
			}
			\qquad
			\adjustbox{scale=0.8}{
			\begin{tikzcd}[row sep=large, column sep=huge]
				\widehat U \arrow[rr, bend left=15, "\tau^U_a(\widehat U)"] & & U^1_{\iota^a \circ \iota_a}\big(\widehat U\big) \arrow[d, "\epsilon^U_1"] \\
				& & \widehat U \arrow[ull, equals, bend left=10]
			\end{tikzcd}
			}
		\end{center}
		\item \textbf{Associativity Invariance.} For every planar configuration
		$$\otimes_n\gamma \xto{\alpha} \oplus_m(\delta_1, a, \delta_2), \qquad \otimes_{n'}(\gamma_1, a, \gamma_2) \xto{\beta} \oplus_{m'}(\lambda_1, b, \lambda_2), \qquad \otimes_{n''}(\theta_1, b, \theta_2) \xto{\omega} \oplus_{m''}\xi$$
		and all lists of objects $\Gamma, \Gamma_1, \Gamma_2, \Theta_1, \Theta_2, \Delta_1, \Delta_2, \Lambda_1, \Lambda_2, \Xi$ over the corresponding lists, the following diagrams commute:
		\begin{itemize}
			\item \textbf{Covariant pentagons ($\xi$- and $\gamma$-zones):} for $1 \leq i \leq |\xi|$ and $1 \leq j \leq |\gamma|$,
			\begin{align*}
				\epsilon^F_i[\omega,\beta] \circ \epsilon^F_{|\lambda_1| + i}[\omega \circledcirc \beta,\alpha] &= F^i_\omega\big(\1_{\Xi_{<i}} \mid \1_{\Theta_1}, \epsilon^F_{|\lambda_1| + 1}[\beta,\alpha], \1_{\Theta_2} \mid \1_{\Xi_{>i}}\big) \circ \epsilon^F_i[\omega,\beta \circledcirc \alpha], \\
				\epsilon^U_{|\gamma_1| + j}[\omega,\beta \circledcirc \alpha] \circ \epsilon^U_j[\beta,\alpha] &= \epsilon^U_j[\omega \circledcirc \beta,\alpha] \circ U^j_\alpha\big(\1_{\Gamma_{<j}} \mid \1_{\Delta_1}, \epsilon^U_{|\gamma_1| + 1}[\omega,\beta], \1_{\Delta_2} \mid \1_{\Gamma_{>j}}\big),
			\end{align*}
			witnessed by the following commuting diagrams.
			For the first diagram, write
\begin{align*}
N_{1}&:=F^{|\lambda_1| + i}_{\omega \circledcirc \beta}(\Lambda_1, \Xi_{<i} \mid \Theta_1, \Gamma_1, \widehat F, \Gamma_2, \Theta_2 \mid \Xi_{>i}, \Lambda_2) \\
N_{2}&:=F^{|\delta_1| + |\lambda_1| + i}_{\omega \circledcirc \beta \circledcirc \alpha}(\Delta_1, \Lambda_1, \Xi_{<i} \mid \Theta_1, \Gamma_1, \Gamma, \Gamma_2, \Theta_2 \mid \Xi_{>i}, \Lambda_2, \Delta_2) \\
N_{3}&:=F^i_\omega\big(\Xi_{<i} \mid \Theta_1, F^{|\lambda_1| + 1}_\beta(\Lambda_1 \mid \Gamma_1, \widehat F, \Gamma_2 \mid \Lambda_2), \Theta_2 \mid \Xi_{>i}\big) \\
N_{4}&:=F^{i}_\omega\big(\Xi_{<i} \mid \Theta_1, F^{|\delta_1| + |\lambda_1| + 1}_{\beta \circledcirc \alpha}(\Delta_1, \Lambda_1 \mid \Gamma_1, \Gamma, \Gamma_2 \mid \Lambda_2, \Delta_2), \Theta_2 \mid \Xi_{>i}\big) \\
a_{1}&:=F^i_\omega(\1_{\Xi_{<i}} \mid \1_{\Theta_1}, \epsilon^F_{|\lambda_1| + 1}[\beta,\alpha], \1_{\Theta_2} \mid \1_{\Xi_{>i}})
\end{align*}
\begin{center}
\begin{tikzcd}[row sep=large, column sep=4em]

					N_{1} \arrow[dr, bend left = 12, "{\epsilon^F_i[\omega,\beta]}"] & \\
					N_{2} \arrow[u, "{\epsilon^F_{|\lambda_1| + i}[\omega \circledcirc \beta,\alpha]}"] \arrow[d, "{\epsilon^F_i[\omega,\beta \circledcirc \alpha]}", swap] & N_{3} \\
					N_{4} \arrow[ur, swap, bend right=12, "a_{1}"] &

\end{tikzcd}
\end{center}
			For the second diagram, write
\begin{align*}
N_{1}&:=U^j_\alpha\big(\Gamma_{<j} \mid \Delta_1, U^{|\gamma_1| + 1}_\beta(\Gamma_1 \mid \Lambda_1, \widehat U, \Lambda_2 \mid \Gamma_2), \Delta_2 \mid \Gamma_{>j}\big) \\
N_{2}&:=U^{|\gamma_1| + j}_{\beta \circledcirc \alpha}(\Gamma_1, \Gamma_{<j} \mid \Delta_1, \Lambda_1, \widehat U, \Lambda_2, \Delta_2 \mid \Gamma_{>j}, \Gamma_2) \\
N_{3}&:=U^j_\alpha\big(\Gamma_{<j} \mid \Delta_1, U^{|\theta_1| + |\gamma_1| + 1}_{\omega \circledcirc \beta}(\Theta_1, \Gamma_1 \mid \Lambda_1, \Xi, \Lambda_2 \mid \Gamma_2, \Theta_2), \Delta_2 \mid \Gamma_{>j}\big) \\
N_{4}&:=U^{|\theta_1| + |\gamma_1| + j}_{\omega \circledcirc \beta \circledcirc \alpha}(\Theta_1, \Gamma_1, \Gamma_{<j} \mid \Delta_1, \Lambda_1, \Xi, \Lambda_2, \Delta_2 \mid \Gamma_{>j}, \Gamma_2, \Theta_2) \\
a_{1}&:=U^j_\alpha(\1_{\Gamma_{<j}} \mid \1_{\Delta_1}, \epsilon^U_{|\gamma_1| + 1}[\omega,\beta], \1_{\Delta_2} \mid \1_{\Gamma_{>j}})
\end{align*}
\begin{center}
\begin{tikzcd}[row sep=large, column sep=4em]

					N_{1} \arrow[dr, bend left=12, "a_{1}"] \arrow[d, "{\epsilon^U_j[\beta,\alpha]}", swap] & \\
					N_{2} \arrow[d, "{\epsilon^U_{|\gamma_1| + j}[\omega,\beta \circledcirc \alpha]}", swap] & N_{3} \arrow[dl, bend left=12, "{\epsilon^U_j[\omega \circledcirc \beta,\alpha]}"] \\
					N_{4} &

\end{tikzcd}
\end{center}
			where $\widehat U := U^{|\theta_1|+1}_\omega(\Theta_1 \mid \Xi \mid \Theta_2)$.
			\item \textbf{Contravariant pentagons ($\delta_1$-, $\delta_2$-, $\theta_1$- and $\theta_2$-zones):} for $1 \leq i \leq |\delta_1|$, resp.\ $1 \leq i \leq |\delta_2|$, resp.\ $1 \leq j \leq |\theta_1|$, resp.\ $1 \leq j \leq |\theta_2|$,
			\begin{align*}
&\eta^F_{i \to}[\beta,\alpha] \circ \eta^F_{i \to}[\omega,\beta \circledcirc \alpha] \\
&\quad= F^i_\alpha\big(\1_{(\Delta_1)_{<i}} \mid \1_\Gamma \mid \1_{(\Delta_1)_{>i}}, \epsilon^U_{|\gamma_1| + 1}[\omega,\beta], \1_{\Delta_2}\big) \circ \eta^F_{i \to}[\omega \circledcirc \beta,\alpha], \\[0.6em]
&\eta^F_{\to i}[\beta,\alpha] \circ \eta^F_{\to (|\lambda_2| + i)}[\omega,\beta \circledcirc \alpha] \\
&\quad= F^{|\delta_1| + 1 + i}_\alpha\big(\1_{\Delta_1}, \epsilon^U_{|\gamma_1| + 1}[\omega,\beta], \1_{(\Delta_2)_{<i}} \mid \1_\Gamma \mid \1_{(\Delta_2)_{>i}}\big) \circ \eta^F_{\to i}[\omega \circledcirc \beta,\alpha], \\[0.6em]
&\eta^U_{j \to}[\omega \circledcirc \beta,\alpha] \circ \eta^U_{j \to}[\omega,\beta] \\
&\quad= \eta^U_{j \to}[\omega,\beta \circledcirc \alpha] \circ U^j_\omega\big(\1_{(\Theta_1)_{<j}} \mid \1_\Xi \mid \1_{(\Theta_1)_{>j}}, \epsilon^F_{|\lambda_1| + 1}[\beta,\alpha], \1_{\Theta_2}\big), \\[0.6em]
&\eta^U_{\to (|\gamma_2| + j)}[\omega \circledcirc \beta,\alpha] \circ \eta^U_{\to j}[\omega,\beta] \\
&\quad= \eta^U_{\to j}[\omega,\beta \circledcirc \alpha] \circ U^{|\theta_1| + 1 + j}_\omega\big(\1_{\Theta_1}, \epsilon^F_{|\lambda_1| + 1}[\beta,\alpha], \1_{(\Theta_2)_{<j}} \mid \1_\Xi \mid \1_{(\Theta_2)_{>j}}\big),
			\end{align*}
			the first two witnessed by the following diagrams.
			For the first diagram, write
\begin{align*}
N_{1}&:=F^i_\alpha\big((\Delta_1)_{<i} \mid \Gamma \mid (\Delta_1)_{>i}, U^{|\theta_1| + |\gamma_1| + 1}_{\omega \circledcirc \beta}(\Theta_1, \Gamma_1 \mid \Lambda_1, \Xi, \Lambda_2 \mid \Gamma_2, \Theta_2), \Delta_2\big) \\
N_{2}&:=F^i_{\omega \circledcirc \beta \circledcirc \alpha}\big((\Delta_1)_{<i} \mid \Theta_1, \Gamma_1, \Gamma, \Gamma_2, \Theta_2 \mid (\Delta_1)_{>i}, \Lambda_1, \Xi, \Lambda_2, \Delta_2\big) \\
N_{3}&:=F^i_\alpha\big((\Delta_1)_{<i} \mid \Gamma \mid (\Delta_1)_{>i}, U^{|\gamma_1| + 1}_\beta(\Gamma_1 \mid \Lambda_1, \widehat U, \Lambda_2 \mid \Gamma_2), \Delta_2\big) \\
N_{4}&:=F^i_{\beta \circledcirc \alpha}\big((\Delta_1)_{<i} \mid \Gamma_1, \Gamma, \Gamma_2 \mid (\Delta_1)_{>i}, \Lambda_1, \widehat U, \Lambda_2, \Delta_2\big) \\
a_{1}&:=F^i_\alpha(\1_{(\Delta_1)_{<i}} \mid \1_\Gamma \mid \1_{(\Delta_1)_{>i}}, \epsilon^U_{|\gamma_1| + 1}[\omega,\beta] , \1_{\Delta_2})
\end{align*}
\begin{center}
\begin{tikzcd}[row sep=large, column sep=4em]

					N_{1} \arrow[dr, bend left=12, "a_{1}"] & \\
					N_{2} \arrow[u, "{\eta^F_{i \to}[\omega \circledcirc \beta,\alpha]}"] \arrow[d, "{\eta^F_{i \to}[\omega,\beta \circledcirc \alpha]}", swap] & N_{3} \\
					N_{4} \arrow[ur, swap, bend right=12, "{\eta^F_{i \to}[\beta,\alpha]}"] &

\end{tikzcd}
\end{center}
			For the second diagram, write
\begin{align*}
N_{1}&:=F^{|\delta_1| + |\lambda_1| + 1 + |\lambda_2| + i}_{\beta \circledcirc \alpha}\big(\Delta_1, \Lambda_1, \widehat U, \Lambda_2, (\Delta_2)_{<i} \mid \Gamma_1, \Gamma, \Gamma_2 \mid (\Delta_2)_{>i}\big) \\
N_{2}&:=F^{|\delta_1| + |\lambda_1| + |\xi| + |\lambda_2| + i}_{\omega \circledcirc \beta \circledcirc \alpha}\big(\Delta_1, \Lambda_1, \Xi, \Lambda_2, (\Delta_2)_{<i} \mid \Theta_1, \Gamma_1, \Gamma, \Gamma_2, \Theta_2 \mid (\Delta_2)_{>i}\big) \\
N_{3}&:=F^{|\delta_1| + 1 + i}_\alpha\big(\Delta_1, U^{|\gamma_1| + 1}_\beta(\Gamma_1 \mid \Lambda_1, \widehat U, \Lambda_2 \mid \Gamma_2), (\Delta_2)_{<i} \mid \Gamma \mid (\Delta_2)_{>i}\big) \\
N_{4}&:=F^{|\delta_1| + 1 + i}_\alpha\big(\Delta_1, U^{|\theta_1| + |\gamma_1| + 1}_{\omega \circledcirc \beta}(\Theta_1, \Gamma_1 \mid \Lambda_1, \Xi, \Lambda_2 \mid \Gamma_2, \Theta_2), (\Delta_2)_{<i} \mid \Gamma \mid (\Delta_2)_{>i}\big) \\
a_{1}&:=F^{|\delta_1| + 1 + i}_\alpha(\1_{\Delta_1}, \epsilon^U_{|\gamma_1| + 1}[\omega,\beta], \1_{(\Delta_2)_{<i}} \mid \1_\Gamma \mid \1_{(\Delta_2)_{>i}})
\end{align*}
\begin{center}
\begin{tikzcd}[row sep=large, column sep=4em]

					N_{1} \arrow[dr, bend left=12, "{\eta^F_{\to i}[\beta,\alpha]}"] & \\
					N_{2} \arrow[u, "{\eta^F_{\to (|\lambda_2| + i)}[\omega,\beta \circledcirc \alpha]}"] \arrow[d, "{\eta^F_{\to i}[\omega \circledcirc \beta,\alpha]}", swap] & N_{3} \\
					N_{4} \arrow[ur, swap, bend right=12, "a_{1}"] &

\end{tikzcd}
\end{center}
			The last two are witnessed by the following mirror diagrams, where $Q := F^{|\lambda_1| + 1}_\beta(\Lambda_1 \mid \Gamma_1, \widehat F, \Gamma_2 \mid \Lambda_2)$.
			For the first diagram, write
\begin{align*}
N_{1}&:=U^j_{\omega \circledcirc \beta}\big((\Theta_1)_{<j} \mid \Lambda_1, \Xi, \Lambda_2 \mid (\Theta_1)_{>j}, \Gamma_1, \widehat F, \Gamma_2, \Theta_2\big) \\
N_{2}&:=U^j_\omega\big((\Theta_1)_{<j} \mid \Xi \mid (\Theta_1)_{>j}, Q, \Theta_2\big) \\
N_{3}&:=U^j_{\omega \circledcirc \beta \circledcirc \alpha}\big((\Theta_1)_{<j} \mid \Delta_1, \Lambda_1, \Xi, \Lambda_2, \Delta_2 \mid (\Theta_1)_{>j}, \Gamma_1, \Gamma, \Gamma_2, \Theta_2\big) \\
N_{4}&:=U^j_\omega\big((\Theta_1)_{<j} \mid \Xi \mid (\Theta_1)_{>j}, F^{|\delta_1| + |\lambda_1| + 1}_{\beta \circledcirc \alpha}(\Delta_1, \Lambda_1 \mid \Gamma_1, \Gamma, \Gamma_2 \mid \Lambda_2, \Delta_2), \Theta_2\big) \\
a_{1}&:=U^j_\omega(\1_{(\Theta_1)_{<j}} \mid \1_\Xi \mid \1_{(\Theta_1)_{>j}}, \epsilon^F_{|\lambda_1| + 1}[\beta,\alpha], \1_{\Theta_2})
\end{align*}
\begin{center}
\begin{tikzcd}[row sep=large, column sep=4em]

					N_{1} \arrow[dr, bend left=12, "{\eta^U_{j \to}[\omega \circledcirc \beta,\alpha]}"] & \\
					N_{2} \arrow[u, "{\eta^U_{j \to}[\omega,\beta]}"] \arrow[d, "a_{1}", swap] & N_{3} \\
					N_{4} \arrow[ur, swap, bend right=12, "{\eta^U_{j \to}[\omega,\beta \circledcirc \alpha]}"] &

\end{tikzcd}
\end{center}
			For the second diagram, write
\begin{align*}
N_{1}&:=U^{|\theta_1| + |\gamma_1| + 1 + |\gamma_2| + j}_{\omega \circledcirc \beta}\big(\Theta_1, \Gamma_1, \widehat F, \Gamma_2, (\Theta_2)_{<j} \mid \Lambda_1, \Xi, \Lambda_2 \mid (\Theta_2)_{>j}\big) \\
N_{2}&:=U^{|\theta_1| + 1 + j}_\omega\big(\Theta_1, Q, (\Theta_2)_{<j} \mid \Xi \mid (\Theta_2)_{>j}\big) \\
N_{3}&:=U^{|\theta_1| + |\gamma_1| + |\gamma| + |\gamma_2| + j}_{\omega \circledcirc \beta \circledcirc \alpha}\big(\Theta_1, \Gamma_1, \Gamma, \Gamma_2, (\Theta_2)_{<j} \mid \Delta_1, \Lambda_1, \Xi, \Lambda_2, \Delta_2 \mid (\Theta_2)_{>j}\big) \\
N_{4}&:=U^{|\theta_1| + 1 + j}_\omega\big(\Theta_1, F^{|\delta_1| + |\lambda_1| + 1}_{\beta \circledcirc \alpha}(\Delta_1, \Lambda_1 \mid \Gamma_1, \Gamma, \Gamma_2 \mid \Lambda_2, \Delta_2), (\Theta_2)_{<j} \mid \Xi \mid (\Theta_2)_{>j}\big) \\
a_{1}&:=U^{|\theta_1| + 1 + j}_\omega(\1_{\Theta_1}, \epsilon^F_{|\lambda_1| + 1}[\beta,\alpha], \1_{(\Theta_2)_{<j}} \mid \1_\Xi \mid \1_{(\Theta_2)_{>j}})
\end{align*}
\begin{center}
\begin{tikzcd}[row sep=large, column sep=4em]

					N_{1} \arrow[dr, bend left=12, "{\eta^U_{\to (|\gamma_2| + j)}[\omega \circledcirc \beta,\alpha]}"] & \\
					N_{2} \arrow[u, "{\eta^U_{\to j}[\omega,\beta]}"] \arrow[d, "a_{1}", swap] & N_{3} \\
					N_{4} \arrow[ur, swap, bend right=12, "{\eta^U_{\to j}[\omega,\beta \circledcirc \alpha]}"] &

\end{tikzcd}
\end{center}
			\item \textbf{Mixed squares ($\lambda_1$-, $\lambda_2$-, $\gamma_1$- and $\gamma_2$-zones).} For $1 \leq j \leq |\lambda_1|$, resp.\ $|\lambda_2|$, resp.\ $|\gamma_1|$, resp.\ $|\gamma_2|$, the four identities below hold.
			\begin{align*}
				\eta^F_{j \to}[\omega,\beta] \circ \epsilon^F_j[\omega \circledcirc \beta,\alpha] &= \epsilon^F_j[\beta,\alpha] \circ \eta^F_{(|\delta_1| + j) \to}[\omega,\beta \circledcirc \alpha], \\
				\epsilon^F_{|\lambda_1| + 1 + j}[\beta,\alpha] \circ \eta^F_{\to j}[\omega,\beta \circledcirc \alpha] &= \eta^F_{\to j}[\omega,\beta] \circ \epsilon^F_{|\lambda_1| + |\xi| + j}[\omega \circledcirc \beta,\alpha], \\
				\eta^U_{(|\theta_1| + j) \to}[\omega \circledcirc \beta,\alpha] \circ \epsilon^U_j[\omega,\beta] &= \epsilon^U_j[\omega,\beta \circledcirc \alpha] \circ \eta^U_{j \to}[\beta,\alpha], \\
				\eta^U_{\to j}[\omega \circledcirc \beta,\alpha] \circ \epsilon^U_{|\gamma_1| + 1 + j}[\omega,\beta] &= \epsilon^U_{|\gamma_1| + |\gamma| + j}[\omega,\beta \circledcirc \alpha] \circ \eta^U_{\to j}[\beta,\alpha],
			\end{align*}
			The first and third are witnessed by the following commuting squares.
			\begin{center}
				\adjustbox{scale=0.62}{
				\begin{tikzcd}[row sep=huge, column sep=large]
					F^{|\delta_1| + j}_{\omega \circledcirc \beta \circledcirc \alpha}\big(\Delta_1, (\Lambda_1)_{<j} \mid \Theta_1, \Gamma_1, \Gamma, \Gamma_2, \Theta_2 \mid (\Lambda_1)_{>j}, \Xi, \Lambda_2, \Delta_2\big) \arrow[r, "{\eta^F_{(|\delta_1| + j) \to}[\omega,\beta \circledcirc \alpha]}"] \arrow[d, "{\epsilon^F_j[\omega \circledcirc \beta,\alpha]}", swap] & F^{|\delta_1| + j}_{\beta \circledcirc \alpha}\big(\Delta_1, (\Lambda_1)_{<j} \mid \Gamma_1, \Gamma, \Gamma_2 \mid (\Lambda_1)_{>j}, \widehat U, \Lambda_2, \Delta_2\big) \arrow[d, "{\epsilon^F_j[\beta,\alpha]}"] \\
					F^j_{\omega \circledcirc \beta}\big((\Lambda_1)_{<j} \mid \Theta_1, \Gamma_1, \widehat F, \Gamma_2, \Theta_2 \mid (\Lambda_1)_{>j}, \Xi, \Lambda_2\big) \arrow[r, "{\eta^F_{j \to}[\omega,\beta]}", swap] & F^j_\beta\big((\Lambda_1)_{<j} \mid \Gamma_1, \widehat F, \Gamma_2 \mid (\Lambda_1)_{>j}, \widehat U, \Lambda_2\big)
				\end{tikzcd}
				}
			\end{center}
			\begin{center}
				\adjustbox{scale=0.62}{
				\begin{tikzcd}[row sep=huge, column sep=large]
					U^j_\beta\big((\Gamma_1)_{<j} \mid \Lambda_1, \widehat U, \Lambda_2 \mid (\Gamma_1)_{>j}, \widehat F, \Gamma_2\big) \arrow[r, "{\epsilon^U_j[\omega,\beta]}"] \arrow[d, "{\eta^U_{j \to}[\beta,\alpha]}", swap] & U^{|\theta_1| + j}_{\omega \circledcirc \beta}\big(\Theta_1, (\Gamma_1)_{<j} \mid \Lambda_1, \Xi, \Lambda_2 \mid (\Gamma_1)_{>j}, \widehat F, \Gamma_2, \Theta_2\big) \arrow[d, "{\eta^U_{(|\theta_1| + j) \to}[\omega \circledcirc \beta,\alpha]}"] \\
					U^j_{\beta \circledcirc \alpha}\big((\Gamma_1)_{<j} \mid \Delta_1, \Lambda_1, \widehat U, \Lambda_2, \Delta_2 \mid (\Gamma_1)_{>j}, \Gamma, \Gamma_2\big) \arrow[r, "{\epsilon^U_j[\omega,\beta \circledcirc \alpha]}", swap] & U^{|\theta_1| + j}_{\omega \circledcirc \beta \circledcirc \alpha}\big(\Theta_1, (\Gamma_1)_{<j} \mid \Delta_1, \Lambda_1, \Xi, \Lambda_2, \Delta_2 \mid (\Gamma_1)_{>j}, \Gamma, \Gamma_2, \Theta_2\big)
				\end{tikzcd}
				}
			\end{center}
			The second and fourth are witnessed by their mirror images.
			\begin{center}
				\adjustbox{scale=0.62}{
				\begin{tikzcd}[row sep=huge, column sep=large]
					F^{|\delta_1| + |\lambda_1| + |\xi| + j}_{\omega \circledcirc \beta \circledcirc \alpha}\big(\Delta_1, \Lambda_1, \Xi, (\Lambda_2)_{<j} \mid \Theta_1, \Gamma_1, \Gamma, \Gamma_2, \Theta_2 \mid (\Lambda_2)_{>j}, \Delta_2\big) \arrow[r, "{\eta^F_{\to j}[\omega,\beta \circledcirc \alpha]}"] \arrow[d, "{\epsilon^F_{|\lambda_1| + |\xi| + j}[\omega \circledcirc \beta,\alpha]}", swap] & F^{|\delta_1| + |\lambda_1| + 1 + j}_{\beta \circledcirc \alpha}\big(\Delta_1, \Lambda_1, \widehat U, (\Lambda_2)_{<j} \mid \Gamma_1, \Gamma, \Gamma_2 \mid (\Lambda_2)_{>j}, \Delta_2\big) \arrow[d, "{\epsilon^F_{|\lambda_1| + 1 + j}[\beta,\alpha]}"] \\
					F^{|\lambda_1| + |\xi| + j}_{\omega \circledcirc \beta}\big(\Lambda_1, \Xi, (\Lambda_2)_{<j} \mid \Theta_1, \Gamma_1, \widehat F, \Gamma_2, \Theta_2 \mid (\Lambda_2)_{>j}\big) \arrow[r, "{\eta^F_{\to j}[\omega,\beta]}", swap] & F^{|\lambda_1| + 1 + j}_\beta\big(\Lambda_1, \widehat U, (\Lambda_2)_{<j} \mid \Gamma_1, \widehat F, \Gamma_2 \mid (\Lambda_2)_{>j}\big)
				\end{tikzcd}
				}
			\end{center}
			\begin{center}
				\adjustbox{scale=0.62}{
				\begin{tikzcd}[row sep=huge, column sep=large]
					U^{|\gamma_1| + 1 + j}_\beta\big(\Gamma_1, \widehat F, (\Gamma_2)_{<j} \mid \Lambda_1, \widehat U, \Lambda_2 \mid (\Gamma_2)_{>j}\big) \arrow[r, "{\epsilon^U_{|\gamma_1| + 1 + j}[\omega,\beta]}"] \arrow[d, "{\eta^U_{\to j}[\beta,\alpha]}", swap] & U^{|\theta_1| + |\gamma_1| + 1 + j}_{\omega \circledcirc \beta}\big(\Theta_1, \Gamma_1, \widehat F, (\Gamma_2)_{<j} \mid \Lambda_1, \Xi, \Lambda_2 \mid (\Gamma_2)_{>j}, \Theta_2\big) \arrow[d, "{\eta^U_{\to j}[\omega \circledcirc \beta,\alpha]}"] \\
					U^{|\gamma_1| + |\gamma| + j}_{\beta \circledcirc \alpha}\big(\Gamma_1, \Gamma, (\Gamma_2)_{<j} \mid \Delta_1, \Lambda_1, \widehat U, \Lambda_2, \Delta_2 \mid (\Gamma_2)_{>j}\big) \arrow[r, "{\epsilon^U_{|\gamma_1| + |\gamma| + j}[\omega,\beta \circledcirc \alpha]}", swap] & U^{|\theta_1| + |\gamma_1| + |\gamma| + j}_{\omega \circledcirc \beta \circledcirc \alpha}\big(\Theta_1, \Gamma_1, \Gamma, (\Gamma_2)_{<j} \mid \Delta_1, \Lambda_1, \Xi, \Lambda_2, \Delta_2 \mid (\Gamma_2)_{>j}, \Theta_2\big)
				\end{tikzcd}
				}
			\end{center}
		\end{itemize}
		\item \textbf{Disjoint Invariance, output configuration.} For every
		$$\otimes_n\gamma \xto{\alpha} \oplus(\delta_1, a, \delta_2, b, \delta_3), \qquad \otimes(\gamma', a) \xto{\beta} \oplus_{m'}\delta', \qquad \otimes(b, \gamma'') \xto{\omega} \oplus_{m''}\delta''$$
		with $\delta_1=()$ or $\gamma'=()$ and $\delta_3=()$ or $\gamma''=()$, let $\Gamma, \Delta_1, \Delta_2, \Delta_3, \Gamma', \Delta', \Gamma'', \Delta''$ be lists over $\gamma, \delta_1, \delta_2, \delta_3, \gamma', \delta', \gamma'', \delta''$, respectively. Write
		$$\widehat U' := U^{|\gamma'|+1}_\beta(\Gamma' \mid \Delta' \mid \,), \qquad \widehat U'' := U^{1}_\omega(\, \mid \Delta'' \mid \Gamma''), \qquad c_a := |\delta_1| + 1, \qquad c_b := |\delta_1| + |\delta_2| + 2 .$$
		Write $\chi:=\omega\circledcirc_b(\beta\circledcirc_a\alpha)=\beta\circledcirc_a(\omega\circledcirc_b\alpha)$ for the common composite of the two cut orders, unambiguous by \cref{lem:poly-circledcirc-interchange}. The following eight squares commute, one for each zone of $\chi$. 

		($\delta_1$\textit{-zone, $1 \leq i \leq |\delta_1|$):}
		\begin{center}
			\adjustbox{scale=0.6}{
			\begin{tikzcd}[row sep=huge, column sep=large]
				F^i_\chi\big((\Delta_1)_{<i} \mid \Gamma', \Gamma, \Gamma'' \mid (\Delta_1)_{>i}, \Delta', \Delta_2, \Delta'', \Delta_3\big) \arrow[r, "{\eta^F_{i \to}\ (\omega,\, \beta \circledcirc_a \alpha)}"] \arrow[d, "{\eta^F_{i \to}\ (\beta,\, \omega \circledcirc_b \alpha)}", swap] & F^i_{\beta \circledcirc_a \alpha}\big((\Delta_1)_{<i} \mid \Gamma', \Gamma \mid (\Delta_1)_{>i}, \Delta', \Delta_2, \widehat U'', \Delta_3\big) \arrow[d, "{\eta^F_{i \to}\ (\beta,\, \alpha)}"] \\
				F^i_{\omega \circledcirc_b \alpha}\big((\Delta_1)_{<i} \mid \Gamma, \Gamma'' \mid (\Delta_1)_{>i}, \widehat U', \Delta_2, \Delta'', \Delta_3\big) \arrow[r, "{\eta^F_{i \to}\ (\omega,\, \alpha)}", swap] & F^i_\alpha\big((\Delta_1)_{<i} \mid \Gamma \mid (\Delta_1)_{>i}, \widehat U', \Delta_2, \widehat U'', \Delta_3\big)
			\end{tikzcd}
			}
		\end{center}
		($\delta'$\textit{-zone, $1 \leq i \leq |\delta'|$):}
		\begin{center}
			\adjustbox{scale=0.6}{
			\begin{tikzcd}[row sep=huge, column sep=large]
				F^{|\delta_1| + i}_\chi\big(\Delta_1, (\Delta')_{<i} \mid \Gamma', \Gamma, \Gamma'' \mid (\Delta')_{>i}, \Delta_2, \Delta'', \Delta_3\big) \arrow[r, "{\eta^F_{(|\delta_1| + i) \to}\ (\omega,\, \beta \circledcirc_a \alpha)}"] \arrow[d, "{\epsilon^F_i\ (\beta,\, \omega \circledcirc_b \alpha)}", swap] & F^{|\delta_1| + i}_{\beta \circledcirc_a \alpha}\big(\Delta_1, (\Delta')_{<i} \mid \Gamma', \Gamma \mid (\Delta')_{>i}, \Delta_2, \widehat U'', \Delta_3\big) \arrow[d, "{\epsilon^F_i\ (\beta,\, \alpha)}"] \\
				F^i_\beta\big((\Delta')_{<i} \mid \Gamma', F^{c_a}_{\omega \circledcirc_b \alpha}(\Delta_1 \mid \Gamma, \Gamma'' \mid \Delta_2, \Delta'', \Delta_3) \mid (\Delta')_{>i}\big) \arrow[r, "{F^i_\beta(\1 \mid \1,\, \eta^F_{c_a \to}\ (\omega, \alpha) \mid \1)}", swap] & F^i_\beta\big((\Delta')_{<i} \mid \Gamma', F^{c_a}_\alpha(\Delta_1 \mid \Gamma \mid \Delta_2, \widehat U'', \Delta_3) \mid (\Delta')_{>i}\big)
			\end{tikzcd}
			}
		\end{center}
		($\delta_2$\textit{-zone, $1 \leq i \leq |\delta_2|$):}
		\begin{center}
			\adjustbox{scale=0.6}{
			\begin{tikzcd}[row sep=huge, column sep=large]
				F^{|\delta_1| + |\delta'| + i}_\chi\big(\Delta_1, \Delta', (\Delta_2)_{<i} \mid \Gamma', \Gamma, \Gamma'' \mid (\Delta_2)_{>i}, \Delta'', \Delta_3\big) \arrow[r, "{\eta^F_{(|\delta_1| + |\delta'| + i) \to}\ (\omega,\, \beta \circledcirc_a \alpha)}"] \arrow[d, "{\eta^F_{\to i}\ (\beta,\, \omega \circledcirc_b \alpha)}", swap] & F^{|\delta_1| + |\delta'| + i}_{\beta \circledcirc_a \alpha}\big(\Delta_1, \Delta', (\Delta_2)_{<i} \mid \Gamma', \Gamma \mid (\Delta_2)_{>i}, \widehat U'', \Delta_3\big) \arrow[d, "{\eta^F_{\to i}\ (\beta,\, \alpha)}"] \\
				F^{|\delta_1| + 1 + i}_{\omega \circledcirc_b \alpha}\big(\Delta_1, \widehat U', (\Delta_2)_{<i} \mid \Gamma, \Gamma'' \mid (\Delta_2)_{>i}, \Delta'', \Delta_3\big) \arrow[r, "{\eta^F_{(|\delta_1| + 1 + i) \to}\ (\omega,\, \alpha)}", swap] & F^{|\delta_1| + 1 + i}_\alpha\big(\Delta_1, \widehat U', (\Delta_2)_{<i} \mid \Gamma \mid (\Delta_2)_{>i}, \widehat U'', \Delta_3\big)
			\end{tikzcd}
			}
		\end{center}
		($\delta''$\textit{-zone, $1 \leq i \leq |\delta''|$):}
		\begin{center}
			\adjustbox{scale=0.6}{
			\begin{tikzcd}[row sep=huge, column sep=large]
				F^{|\delta_1| + |\delta'| + |\delta_2| + i}_\chi\big(\Delta_1, \Delta', \Delta_2, (\Delta'')_{<i} \mid \Gamma', \Gamma, \Gamma'' \mid (\Delta'')_{>i}, \Delta_3\big) \arrow[r, "{\epsilon^F_i\ (\omega,\, \beta \circledcirc_a \alpha)}"] \arrow[d, "{\eta^F_{\to (|\delta_2| + i)}\ (\beta,\, \omega \circledcirc_b \alpha)}", swap] & F^i_\omega\big((\Delta'')_{<i} \mid F^{|\delta_1| + |\delta'| + |\delta_2| + 1}_{\beta \circledcirc_a \alpha}(\Delta_1, \Delta', \Delta_2 \mid \Gamma', \Gamma \mid \Delta_3), \Gamma'' \mid (\Delta'')_{>i}\big) \arrow[d, "{F^i_\omega(\1 \mid \eta^F_{\to (|\delta_2| + 1)}\ (\beta, \alpha),\, \1 \mid \1)}"] \\
				F^{|\delta_1| + 1 + |\delta_2| + i}_{\omega \circledcirc_b \alpha}\big(\Delta_1, \widehat U', \Delta_2, (\Delta'')_{<i} \mid \Gamma, \Gamma'' \mid (\Delta'')_{>i}, \Delta_3\big) \arrow[r, "{\epsilon^F_i\ (\omega,\, \alpha)}", swap] & F^i_\omega\big((\Delta'')_{<i} \mid F^{c_b}_\alpha(\Delta_1, \widehat U', \Delta_2 \mid \Gamma \mid \Delta_3), \Gamma'' \mid (\Delta'')_{>i}\big)
			\end{tikzcd}
			}
		\end{center}
		($\delta_3$\textit{-zone, $1 \leq i \leq |\delta_3|$):}
		\begin{center}
			\adjustbox{scale=0.6}{
			\begin{tikzcd}[row sep=huge, column sep=large]
				F^{|\delta_1| + |\delta'| + |\delta_2| + |\delta''| + i}_\chi\big(\Delta_1, \Delta', \Delta_2, \Delta'', (\Delta_3)_{<i} \mid \Gamma', \Gamma, \Gamma'' \mid (\Delta_3)_{>i}\big) \arrow[r, "{\eta^F_{\to i}\ (\omega,\, \beta \circledcirc_a \alpha)}"] \arrow[d, "{\eta^F_{\to (|\delta_2| + |\delta''| + i)}\ (\beta,\, \omega \circledcirc_b \alpha)}", swap] & F^{|\delta_1| + |\delta'| + |\delta_2| + 1 + i}_{\beta \circledcirc_a \alpha}\big(\Delta_1, \Delta', \Delta_2, \widehat U'', (\Delta_3)_{<i} \mid \Gamma', \Gamma \mid (\Delta_3)_{>i}\big) \arrow[d, "{\eta^F_{\to (|\delta_2| + 1 + i)}\ (\beta,\, \alpha)}"] \\
				F^{|\delta_1| + 1 + |\delta_2| + |\delta''| + i}_{\omega \circledcirc_b \alpha}\big(\Delta_1, \widehat U', \Delta_2, \Delta'', (\Delta_3)_{<i} \mid \Gamma, \Gamma'' \mid (\Delta_3)_{>i}\big) \arrow[r, "{\eta^F_{\to i}\ (\omega,\, \alpha)}", swap] & F^{c_b + i}_\alpha\big(\Delta_1, \widehat U', \Delta_2, \widehat U'', (\Delta_3)_{<i} \mid \Gamma \mid (\Delta_3)_{>i}\big)
			\end{tikzcd}
			}
		\end{center}
		($\gamma'$\textit{-zone, $1 \leq j \leq |\gamma'|$):}
		\begin{center}
			\adjustbox{scale=0.6}{
			\begin{tikzcd}[row sep=huge, column sep=large]
				U^j_\beta\big((\Gamma')_{<j} \mid \Delta' \mid (\Gamma')_{>j}, F^{c_a}_\alpha(\Delta_1 \mid \Gamma \mid \Delta_2, \widehat U'', \Delta_3)\big) \arrow[r, "{\eta^U_{j \to}\ (\beta,\, \alpha)}"] \arrow[d, "{U^j_\beta(\1 \mid \1 \mid \1,\, \eta^F_{c_a \to}\ (\omega, \alpha))}", swap] & U^j_{\beta \circledcirc_a \alpha}\big((\Gamma')_{<j} \mid \Delta_1, \Delta', \Delta_2, \widehat U'', \Delta_3 \mid (\Gamma')_{>j}, \Gamma\big) \arrow[d, "{\epsilon^U_j\ (\omega,\, \beta \circledcirc_a \alpha)}"] \\
				U^j_\beta\big((\Gamma')_{<j} \mid \Delta' \mid (\Gamma')_{>j}, F^{c_a}_{\omega \circledcirc_b \alpha}(\Delta_1 \mid \Gamma, \Gamma'' \mid \Delta_2, \Delta'', \Delta_3)\big) \arrow[r, "{\eta^U_{j \to}\ (\beta,\, \omega \circledcirc_b \alpha)}", swap] & U^j_\chi\big((\Gamma')_{<j} \mid \Delta_1, \Delta', \Delta_2, \Delta'', \Delta_3 \mid (\Gamma')_{>j}, \Gamma, \Gamma''\big)
			\end{tikzcd}
			}
		\end{center}
		($\gamma$\textit{-zone, $1 \leq j \leq |\gamma|$):}
		\begin{center}
			\adjustbox{scale=0.6}{
			\begin{tikzcd}[row sep=huge, column sep=large]
				U^j_\alpha\big(\Gamma_{<j} \mid \Delta_1, \widehat U', \Delta_2, \widehat U'', \Delta_3 \mid \Gamma_{>j}\big) \arrow[r, "{\epsilon^U_j\ (\beta,\, \alpha)}"] \arrow[d, "{\epsilon^U_j\ (\omega,\, \alpha)}", swap] & U^{|\gamma'| + j}_{\beta \circledcirc_a \alpha}\big(\Gamma', \Gamma_{<j} \mid \Delta_1, \Delta', \Delta_2, \widehat U'', \Delta_3 \mid \Gamma_{>j}\big) \arrow[d, "{\epsilon^U_{|\gamma'| + j}\ (\omega,\, \beta \circledcirc_a \alpha)}"] \\
				U^j_{\omega \circledcirc_b \alpha}\big(\Gamma_{<j} \mid \Delta_1, \widehat U', \Delta_2, \Delta'', \Delta_3 \mid \Gamma_{>j}, \Gamma''\big) \arrow[r, "{\epsilon^U_j\ (\beta,\, \omega \circledcirc_b \alpha)}", swap] & U^{|\gamma'| + j}_{\chi}\big(\Gamma', \Gamma_{<j} \mid \Delta_1, \Delta', \Delta_2, \Delta'', \Delta_3 \mid \Gamma_{>j}, \Gamma''\big)
			\end{tikzcd}
			}
		\end{center}
		($\gamma''$\textit{-zone, $1 \leq j \leq |\gamma''|$):}
		\begin{center}
			\adjustbox{scale=0.6}{
			\begin{tikzcd}[row sep=huge, column sep=large]
				U^{1 + j}_\omega\big(F^{c_b}_\alpha(\Delta_1, \widehat U', \Delta_2 \mid \Gamma \mid \Delta_3), (\Gamma'')_{<j} \mid \Delta'' \mid (\Gamma'')_{>j}\big) \arrow[r, "{\eta^U_{\to j}\ (\omega,\, \alpha)}"] \arrow[d, "{U^{1 + j}_\omega(\eta^F_{\to (|\delta_2| + 1)}\ (\beta, \alpha),\, \1 \mid \1 \mid \1)}", swap] & U^{|\gamma| + j}_{\omega \circledcirc_b \alpha}\big(\Gamma, (\Gamma'')_{<j} \mid \Delta_1, \widehat U', \Delta_2, \Delta'', \Delta_3 \mid (\Gamma'')_{>j}\big) \arrow[d, "{\epsilon^U_{|\gamma| + j}\ (\beta,\, \omega \circledcirc_b \alpha)}"] \\
				U^{1 + j}_\omega\big(F^{|\delta_1| + |\delta'| + |\delta_2| + 1}_{\beta \circledcirc_a \alpha}(\Delta_1, \Delta', \Delta_2 \mid \Gamma', \Gamma \mid \Delta_3), (\Gamma'')_{<j} \mid \Delta'' \mid (\Gamma'')_{>j}\big) \arrow[r, "{\eta^U_{\to j}\ (\omega,\, \beta \circledcirc_a \alpha)}", swap] & U^{|\gamma'| + |\gamma| + j}_\chi\big(\Gamma', \Gamma, (\Gamma'')_{<j} \mid \Delta_1, \Delta', \Delta_2, \Delta'', \Delta_3 \mid (\Gamma'')_{>j}\big)
			\end{tikzcd}
			}
		\end{center}
		\item \textbf{Disjoint Invariance, input configuration.} Dually, let
		$$\otimes(\theta_1, a, \theta_2, b, \theta_3) \xto{\omega} \oplus_{m''}\xi, \qquad \otimes_n\gamma \xto{\beta} \oplus(\delta_1, a), \qquad \otimes_{n_2}\gamma' \xto{\beta'} \oplus(b, \delta'_2)$$
        be arbitrary. Let
\[
\Theta_1, \Theta_2, \Theta_3, \Xi, \Gamma, \Gamma', \Delta_1, \Delta'_2
\]
 be arbitrary lists of objects over $\theta_1, \theta_2, \theta_3, \xi, \gamma, \gamma', \delta_1, \delta'_2$, respectively. Write $\widehat F := F^{|\delta_1| + 1}_\beta(\Delta_1 \mid \Gamma \mid \,)$ and $\widehat F' := F^{1}_{\beta'}(\, \mid \Gamma' \mid \Delta'_2)$. Write $c'_b := |\theta_1| + |\theta_2| + 2$ for the position of the input $b$ of $\omega$. Write
\[
\chi:=(\omega\circledcirc_a\beta)\circledcirc_b\beta'=(\omega\circledcirc_b\beta')\circledcirc_a\beta
\]
 for the common composite of the two cut orders, unambiguous by \cref{lem:poly-circledcirc-interchange}. The following eight mirror squares commute, one per zone of $\chi$.

		($\delta_1$\textit{-zone, $1 \leq i \leq |\delta_1|$):}
		\begin{center}
			\adjustbox{scale=0.6}{
			\begin{tikzcd}[row sep=huge, column sep=large]
				F^i_\chi\big((\Delta_1)_{<i} \mid \Theta_1, \Gamma, \Theta_2, \Gamma', \Theta_3 \mid (\Delta_1)_{>i}, \Xi, \Delta'_2\big) \arrow[r, "{\epsilon^F_i\ (\omega \circledcirc_a \beta,\, \beta')}"] \arrow[d, "{\eta^F_{i \to}\ (\omega \circledcirc_b \beta',\, \beta)}", swap] & F^i_{\omega \circledcirc_a \beta}\big((\Delta_1)_{<i} \mid \Theta_1, \Gamma, \Theta_2, \widehat F', \Theta_3 \mid (\Delta_1)_{>i}, \Xi\big) \arrow[d, "{\eta^F_{i \to}\ (\omega,\, \beta)}"] \\
				F^i_\beta\big((\Delta_1)_{<i} \mid \Gamma \mid (\Delta_1)_{>i}, U^{|\theta_1| + 1}_{\omega \circledcirc_b \beta'}(\Theta_1 \mid \Xi, \Delta'_2 \mid \Theta_2, \Gamma', \Theta_3)\big) \arrow[r, "{F^i_\beta(\1 \mid \1 \mid \1,\, \eta^U_{(|\theta_1| + 1) \to}\ (\omega, \beta'))}", swap] & F^i_\beta\big((\Delta_1)_{<i} \mid \Gamma \mid (\Delta_1)_{>i}, U^{|\theta_1| + 1}_\omega(\Theta_1 \mid \Xi \mid \Theta_2, \widehat F', \Theta_3)\big)
			\end{tikzcd}
			}
		\end{center}
		($\xi$\textit{-zone, $1 \leq i \leq |\xi|$):}
		\begin{center}
			\adjustbox{scale=0.6}{
			\begin{tikzcd}[row sep=huge, column sep=large]
				F^{|\delta_1| + i}_\chi\big(\Delta_1, \Xi_{<i} \mid \Theta_1, \Gamma, \Theta_2, \Gamma', \Theta_3 \mid \Xi_{>i}, \Delta'_2\big) \arrow[r, "{\epsilon^F_{|\delta_1| + i}\ (\omega \circledcirc_a \beta,\, \beta')}"] \arrow[d, "{\epsilon^F_i\ (\omega \circledcirc_b \beta',\, \beta)}", swap] & F^{|\delta_1| + i}_{\omega \circledcirc_a \beta}\big(\Delta_1, \Xi_{<i} \mid \Theta_1, \Gamma, \Theta_2, \widehat F', \Theta_3 \mid \Xi_{>i}\big) \arrow[d, "{\epsilon^F_i\ (\omega,\, \beta)}"] \\
				F^i_{\omega \circledcirc_b \beta'}\big(\Xi_{<i} \mid \Theta_1, \widehat F, \Theta_2, \Gamma', \Theta_3 \mid \Xi_{>i}, \Delta'_2\big) \arrow[r, "{\epsilon^F_i\ (\omega,\, \beta')}", swap] & F^i_\omega\big(\Xi_{<i} \mid \Theta_1, \widehat F, \Theta_2, \widehat F', \Theta_3 \mid \Xi_{>i}\big)
			\end{tikzcd}
			}
		\end{center}
		($\delta'_2$\textit{-zone, $1 \leq i \leq |\delta'_2|$):}
		\begin{center}
			\adjustbox{scale=0.6}{
			\begin{tikzcd}[row sep=huge, column sep=large]
				F^{|\delta_1| + |\xi| + i}_\chi\big(\Delta_1, \Xi, (\Delta'_2)_{<i} \mid \Theta_1, \Gamma, \Theta_2, \Gamma', \Theta_3 \mid (\Delta'_2)_{>i}\big) \arrow[r, "{\eta^F_{\to i}\ (\omega \circledcirc_a \beta,\, \beta')}"] \arrow[d, "{\epsilon^F_{|\xi| + i}\ (\omega \circledcirc_b \beta',\, \beta)}", swap] & F^{1 + i}_{\beta'}\big(U^{|\theta_1| + |\gamma| + |\theta_2| + 1}_{\omega \circledcirc_a \beta}(\Theta_1, \Gamma, \Theta_2 \mid \Delta_1, \Xi \mid \Theta_3), (\Delta'_2)_{<i} \mid \Gamma' \mid (\Delta'_2)_{>i}\big) \arrow[d, "{F^{1 + i}_{\beta'}(\eta^U_{\to (|\theta_2| + 1)}\ (\omega, \beta),\, \1 \mid \1 \mid \1)}"] \\
				F^{|\xi| + i}_{\omega \circledcirc_b \beta'}\big(\Xi, (\Delta'_2)_{<i} \mid \Theta_1, \widehat F, \Theta_2, \Gamma', \Theta_3 \mid (\Delta'_2)_{>i}\big) \arrow[r, "{\eta^F_{\to i}\ (\omega,\, \beta')}", swap] & F^{1 + i}_{\beta'}\big(U^{c'_b}_\omega(\Theta_1, \widehat F, \Theta_2 \mid \Xi \mid \Theta_3), (\Delta'_2)_{<i} \mid \Gamma' \mid (\Delta'_2)_{>i}\big)
			\end{tikzcd}
			}
		\end{center}
		($\theta_1$\textit{-zone, $1 \leq j \leq |\theta_1|$):}
		\begin{center}
			\adjustbox{scale=0.6}{
			\begin{tikzcd}[row sep=huge, column sep=large]
				U^j_\omega\big((\Theta_1)_{<j} \mid \Xi \mid (\Theta_1)_{>j}, \widehat F, \Theta_2, \widehat F', \Theta_3\big) \arrow[r, "{\eta^U_{j \to}\ (\omega,\, \beta)}"] \arrow[d, "{\eta^U_{j \to}\ (\omega,\, \beta')}", swap] & U^j_{\omega \circledcirc_a \beta}\big((\Theta_1)_{<j} \mid \Delta_1, \Xi \mid (\Theta_1)_{>j}, \Gamma, \Theta_2, \widehat F', \Theta_3\big) \arrow[d, "{\eta^U_{j \to}\ (\omega \circledcirc_a \beta,\, \beta')}"] \\
				U^j_{\omega \circledcirc_b \beta'}\big((\Theta_1)_{<j} \mid \Xi, \Delta'_2 \mid (\Theta_1)_{>j}, \widehat F, \Theta_2, \Gamma', \Theta_3\big) \arrow[r, "{\eta^U_{j \to}\ (\omega \circledcirc_b \beta',\, \beta)}", swap] & U^j_\chi\big((\Theta_1)_{<j} \mid \Delta_1, \Xi, \Delta'_2 \mid (\Theta_1)_{>j}, \Gamma, \Theta_2, \Gamma', \Theta_3\big)
			\end{tikzcd}
			}
		\end{center}
		($\gamma$\textit{-zone, $1 \leq j \leq |\gamma|$):}
		\begin{center}
			\adjustbox{scale=0.6}{
			\begin{tikzcd}[row sep=huge, column sep=large]
				U^j_\beta\big(\Gamma_{<j} \mid \Delta_1, U^{|\theta_1| + 1}_\omega(\Theta_1 \mid \Xi \mid \Theta_2, \widehat F', \Theta_3) \mid \Gamma_{>j}\big) \arrow[r, "{\epsilon^U_j\ (\omega,\, \beta)}"] \arrow[d, "{U^j_\beta(\1 \mid \1,\, \eta^U_{(|\theta_1| + 1) \to}\ (\omega, \beta') \mid \1)}", swap] & U^{|\theta_1| + j}_{\omega \circledcirc_a \beta}\big(\Theta_1, \Gamma_{<j} \mid \Delta_1, \Xi \mid \Gamma_{>j}, \Theta_2, \widehat F', \Theta_3\big) \arrow[d, "{\eta^U_{(|\theta_1| + j) \to}\ (\omega \circledcirc_a \beta,\, \beta')}"] \\
				U^j_\beta\big(\Gamma_{<j} \mid \Delta_1, U^{|\theta_1| + 1}_{\omega \circledcirc_b \beta'}(\Theta_1 \mid \Xi, \Delta'_2 \mid \Theta_2, \Gamma', \Theta_3) \mid \Gamma_{>j}\big) \arrow[r, "{\epsilon^U_j\ (\omega \circledcirc_b \beta',\, \beta)}", swap] & U^{|\theta_1| + j}_\chi\big(\Theta_1, \Gamma_{<j} \mid \Delta_1, \Xi, \Delta'_2 \mid \Gamma_{>j}, \Theta_2, \Gamma', \Theta_3\big)
			\end{tikzcd}
			}
		\end{center}
		($\theta_2$\textit{-zone, $1 \leq j \leq |\theta_2|$):}
		\begin{center}
			\adjustbox{scale=0.6}{
			\begin{tikzcd}[row sep=huge, column sep=large]
				U^{|\theta_1| + 1 + j}_\omega\big(\Theta_1, \widehat F, (\Theta_2)_{<j} \mid \Xi \mid (\Theta_2)_{>j}, \widehat F', \Theta_3\big) \arrow[r, "{\eta^U_{\to j}\ (\omega,\, \beta)}"] \arrow[d, "{\eta^U_{(|\theta_1| + 1 + j) \to}\ (\omega,\, \beta')}", swap] & U^{|\theta_1| + |\gamma| + j}_{\omega \circledcirc_a \beta}\big(\Theta_1, \Gamma, (\Theta_2)_{<j} \mid \Delta_1, \Xi \mid (\Theta_2)_{>j}, \widehat F', \Theta_3\big) \arrow[d, "{\eta^U_{(|\theta_1| + |\gamma| + j) \to}\ (\omega \circledcirc_a \beta,\, \beta')}"] \\
				U^{|\theta_1| + 1 + j}_{\omega \circledcirc_b \beta'}\big(\Theta_1, \widehat F, (\Theta_2)_{<j} \mid \Xi, \Delta'_2 \mid (\Theta_2)_{>j}, \Gamma', \Theta_3\big) \arrow[r, "{\eta^U_{\to j}\ (\omega \circledcirc_b \beta',\, \beta)}", swap] & U^{|\theta_1| + |\gamma| + j}_\chi\big(\Theta_1, \Gamma, (\Theta_2)_{<j} \mid \Delta_1, \Xi, \Delta'_2 \mid (\Theta_2)_{>j}, \Gamma', \Theta_3\big)
			\end{tikzcd}
			}
		\end{center}
		($\gamma'$\textit{-zone, $1 \leq j \leq |\gamma'|$):}
		\begin{center}
			\adjustbox{scale=0.6}{
			\begin{tikzcd}[row sep=huge, column sep=large]
				U^j_{\beta'}\big((\Gamma')_{<j} \mid U^{c'_b}_\omega(\Theta_1, \widehat F, \Theta_2 \mid \Xi \mid \Theta_3), \Delta'_2 \mid (\Gamma')_{>j}\big) \arrow[r, "{\epsilon^U_j\ (\omega,\, \beta')}"] \arrow[d, "{U^j_{\beta'}(\1 \mid \eta^U_{\to (|\theta_2| + 1)}\ (\omega, \beta),\, \1 \mid \1)}", swap] & U^{|\theta_1| + 1 + |\theta_2| + j}_{\omega \circledcirc_b \beta'}\big(\Theta_1, \widehat F, \Theta_2, (\Gamma')_{<j} \mid \Xi, \Delta'_2 \mid (\Gamma')_{>j}, \Theta_3\big) \arrow[d, "{\eta^U_{\to (|\theta_2| + j)}\ (\omega \circledcirc_b \beta',\, \beta)}"] \\
				U^j_{\beta'}\big((\Gamma')_{<j} \mid U^{|\theta_1| + |\gamma| + |\theta_2| + 1}_{\omega \circledcirc_a \beta}(\Theta_1, \Gamma, \Theta_2 \mid \Delta_1, \Xi \mid \Theta_3), \Delta'_2 \mid (\Gamma')_{>j}\big) \arrow[r, "{\epsilon^U_j\ (\omega \circledcirc_a \beta,\, \beta')}", swap] & U^{|\theta_1| + |\gamma| + |\theta_2| + j}_\chi\big(\Theta_1, \Gamma, \Theta_2, (\Gamma')_{<j} \mid \Delta_1, \Xi, \Delta'_2 \mid (\Gamma')_{>j}, \Theta_3\big)
			\end{tikzcd}
			}
		\end{center}
		($\theta_3$\textit{-zone, $1 \leq j \leq |\theta_3|$):}
		\begin{center}
			\adjustbox{scale=0.6}{
			\begin{tikzcd}[row sep=huge, column sep=large]
				U^{c'_b + j}_\omega\big(\Theta_1, \widehat F, \Theta_2, \widehat F', (\Theta_3)_{<j} \mid \Xi \mid (\Theta_3)_{>j}\big) \arrow[r, "{\eta^U_{\to (|\theta_2| + 1 + j)}\ (\omega,\, \beta)}"] \arrow[d, "{\eta^U_{\to j}\ (\omega,\, \beta')}", swap] & U^{|\theta_1| + |\gamma| + |\theta_2| + 1 + j}_{\omega \circledcirc_a \beta}\big(\Theta_1, \Gamma, \Theta_2, \widehat F', (\Theta_3)_{<j} \mid \Delta_1, \Xi \mid (\Theta_3)_{>j}\big) \arrow[d, "{\eta^U_{\to j}\ (\omega \circledcirc_a \beta,\, \beta')}"] \\
				U^{|\theta_1| + 1 + |\theta_2| + |\gamma'| + j}_{\omega \circledcirc_b \beta'}\big(\Theta_1, \widehat F, \Theta_2, \Gamma', (\Theta_3)_{<j} \mid \Xi, \Delta'_2 \mid (\Theta_3)_{>j}\big) \arrow[r, "{\eta^U_{\to (|\theta_2| + |\gamma'| + j)}\ (\omega \circledcirc_b \beta',\, \beta)}", swap] & U^{|\theta_1| + |\gamma| + |\theta_2| + |\gamma'| + j}_\chi\big(\Theta_1, \Gamma, \Theta_2, \Gamma', (\Theta_3)_{<j} \mid \Delta_1, \Xi, \Delta'_2 \mid (\Theta_3)_{>j}\big)
			\end{tikzcd}
			}
		\end{center}
		\item \textbf{Correspondence axioms.} Two axioms say how representations behave: one for identities, one for cuts.
		\begin{itemize}
			\item \textbf{(Correspondence--unit.)} The identity $\iota^a\circ\iota_a$ of $a$ has two positions, $1^+$ and $1^-$. For every $a\in\mc M$ and $X\in\mc T_a$, the identity of $F^1_{\iota^a\circ\iota_a}(X)$ generates a representation of $\iota^a\circ\iota_a$ at $(X;X)$. This representation is the one prescribed by $\tau$:
			$$\big\langle \1_{F^1_{\iota^a \circ \iota_a}(X)} \big\rangle^{1^+}_{1^-} \ = \ \tau^U_a\big(F^1_{\iota^a \circ \iota_a}(X)\big) \circ \big(\tau^F_a(X)\big)^{-1} \ : \ X \to U^1_{\iota^a \circ \iota_a}\big(F^1_{\iota^a \circ \iota_a}(X)\big).$$
			By naturality of $\varphi$, this determines $\varphi^{1^+,1^-}_{\iota^a \circ \iota_a}$ on every element.

			\item \textbf{(Correspondence--decomposition.)} For every cut configuration and every element $u:\widehat F\to \widehat U$ of its cut set, the cut reading $\kappa\langle u\rangle$ of \cref{def:cut-reading} is a representation of $\beta\circledcirc\alpha$. Equivalently,
			$$\varphi^{p,q}_{\beta\circledcirc\alpha}\circ K_p=K_q \qquad \text{for all positions } p,q \text{ of } \beta\circledcirc\alpha.$$
		\end{itemize}
	\end{itemize}
\end{definition}

\subsection{The Underlying Planar Polycategory}\label{subsec:underlying-planar-polycategory}

We construct the underlying planar polycategory $\mc T_p$ of a planar poly-indexed category $(\mc T, \mc M, \pi)$, together with a planar polyfunctor $\pi_p : \fib[T_p]{M_p}$. \Cref{thm:planar-cartesianness} analyzes the cartesianness of its canonical lifting candidates, including the exceptional nullary case.

\subsubsection*{Polymorphisms, Identities, and Cut}

\begin{definition}\label{def:underlying-planar-polycategory}
	The polycategory $\mc T_p$ is defined as follows.
	\begin{itemize}
		\item \textbf{Objects:} an object $A_p$ of $\mc T_p$ consists of an object $A \in \mc T$.
		\item \textbf{Polymorphisms:} for lists $\Gamma_p := (\Gamma_1)_p, \ldots, (\Gamma_n)_p$ and $\Delta_p := (\Delta_1)_p, \ldots, (\Delta_m)_p$, a polymorphism $\Gamma_p \xto[(g_1, \ldots, g_n)]{(f_1, \ldots, f_m)} \Delta_p$ consists of
		\begin{itemize}
			\item a morphism $\otimes_n \pi\Gamma \xto{\alpha} \oplus_m \pi\Delta$ in $\mc M$, i.e.\ a polymap $\pi\Gamma \to \pi\Delta$ of $\mc M_p$;
			\item a representation of $\alpha$ at $(\Gamma;\Delta)$, in the sense fixed in \cref{def:planar-poly-indexed-category}. This representation consists of one vertical morphism at every position of $\alpha$, i.e.\ elements
\[
\begin{gathered}
f_i \in \mathrm{Rep}^{i^+}_\alpha(\Gamma; \Delta) = \mc T_{\pi\Delta_i}\big(F^i_{\alpha}(\Delta_{<i} \mid \Gamma \mid \Delta_{>i}),\, \Delta_i\big)
			\\ \text{and}\quad
			g_j \in \mathrm{Rep}^{j^-}_\alpha(\Gamma; \Delta) = \mc T_{\pi\Gamma_j}\big(\Gamma_j,\, U^j_\alpha(\Gamma_{<j} \mid \Delta \mid \Gamma_{>j})\big).
\end{gathered}
\]
			These elements all present the same map. For all positions $p,q$ of $\alpha$, the component at $q$ is the image of the component at $p$ under $\varphi^{p,q}_\alpha$.
		\end{itemize}
		We abbreviate a polymorphism by $f_p$, its base by $\pi_p f_p := \alpha_p$, and its components by $f_i$ and $g_j$. We call a component at a position a \textit{reading} of $f_p$.
	\end{itemize}
\end{definition}

\phantomsection\label{anchor:representation-coordinate-system}
\textit{Remark:} When $\alpha$ has at least one position, a polymorphism over $\alpha_p$ is determined by a single element of the representation set $\mathrm{Rep}^p_\alpha(\Gamma;\Delta)$. Choosing a position $p$ is somewhat the same as choosing a coordinate system for the representation. It is possible to recover the representative map at any other position using the maps $\varphi$.

\begin{itemize}
	\item \textbf{Identity polymorphisms:} the identity of $A_p$, over the identity $\iota^{\pi A} \circ \iota_{\pi A}$ of $(\pi A)_p$, consists of the two components
	$$\tau^F_{\pi A}(A) : F^1_{\iota^{\pi A} \circ \iota_{\pi A}}(A) \to A \quadt{and} \tau^U_{\pi A}(A) : A \to U^1_{\iota^{\pi A} \circ \iota_{\pi A}}(A).$$
	\item \textbf{Cut:} Let $f_p : \Gamma_p \to (\Delta_1)_p, A_p, (\Delta_2)_p$ be a polymorphism over $\alpha$. Let $g_p : (\Gamma_1)_p, A_p, (\Gamma_2)_p \to \Delta_p$ be a polymorphism over $\beta$. Assume that the base pair $(\beta,\alpha)$ is planar. Write their components as $(f_i; g_j)$ and $(f'_i; g'_j)$, respectively. The cut positions are
	$$c := |\delta_1| + 1 \text{ (output of } \alpha\text{)}, \qquad c' := |\gamma_1| + 1 \text{ (input of } \beta\text{)}.$$
	The cut $g_p \circ_{A_p} f_p : (\Gamma_1)_p, \Gamma_p, (\Gamma_2)_p \to (\Delta_1)_p, \Delta_p, (\Delta_2)_p$ is the polymorphism over $\beta \circledcirc \alpha$ with components
	\begin{align*}
		f^{\beta \circledcirc \alpha}_i &:= f_i \circ F^{i}_\alpha\big(\1_{(\Delta_1)_{<i}} \mid \1_{\Gamma} \mid \1_{(\Delta_1)_{>i}}, g'_{c'}, \1_{\Delta_2}\big) \circ \eta^F_{i \to} && 1 \leq i \leq |\delta_1| \\
		f^{\beta \circledcirc \alpha}_{|\delta_1| + i} &:= f'_i \circ F^i_\beta\big(\1_{\Delta_{<i}} \mid \1_{\Gamma_1}, f_{c}, \1_{\Gamma_2} \mid \1_{\Delta_{>i}}\big) \circ \epsilon^F_i && 1 \leq i \leq |\delta| \\
		f^{\beta \circledcirc \alpha}_{|\delta_1| + |\delta| + i} &:= f_{c + i} \circ F^{c + i}_\alpha\big(\1_{\Delta_1}, g'_{c'}, \1_{(\Delta_2)_{<i}} \mid \1_{\Gamma} \mid \1_{(\Delta_2)_{>i}}\big) \circ \eta^F_{\to i} && 1 \leq i \leq |\delta_2| \\
		g^{\beta \circledcirc \alpha}_j &:= \eta^U_{j \to} \circ U^j_\beta\big(\1_{(\Gamma_1)_{<j}} \mid \1_\Delta \mid \1_{(\Gamma_1)_{>j}}, f_{c}, \1_{\Gamma_2}\big) \circ g'_j && 1 \leq j \leq |\gamma_1| \\
		g^{\beta \circledcirc \alpha}_{|\gamma_1| + j} &:= \epsilon^U_j \circ U^j_\alpha\big(\1_{\Gamma_{<j}} \mid \1_{\Delta_1}, g'_{c'}, \1_{\Delta_2} \mid \1_{\Gamma_{>j}}\big) \circ g_j && 1 \leq j \leq |\gamma| \\
		g^{\beta \circledcirc \alpha}_{|\gamma_1| + |\gamma| + j} &:= \eta^U_{\to j} \circ U^{c' + j}_\beta\big(\1_{\Gamma_1}, f_{c}, \1_{(\Gamma_2)_{<j}} \mid \1_\Delta \mid \1_{(\Gamma_2)_{>j}}\big) \circ g'_{c' + j} && 1 \leq j \leq |\gamma_2|.
	\end{align*}
\end{itemize}

We now explain how this cut is constructed. The six families are the six zones of the composite. Recall that the components of $f_p$ and $g_p$ are
$$F^i_\alpha\big((\Delta_1, A, \Delta_2)_{<i} \mid \Gamma \mid (\Delta_1, A, \Delta_2)_{>i}\big) \xto{f_i} (\Delta_1, A, \Delta_2)_i \quadt{and} \Gamma_j \xto{g_j} U^j_\alpha(\Gamma_{<j} \mid \Delta_1, A, \Delta_2 \mid \Gamma_{>j})$$
$$F^i_\beta\big(\Delta_{<i} \mid \Gamma_1, A, \Gamma_2 \mid \Delta_{>i}\big) \xto{f'_i} \Delta_i \quadt{and} (\Gamma_1, A, \Gamma_2)_j \xto{g'_j} U^j_\beta\big((\Gamma_1, A, \Gamma_2)_{<j} \mid \Delta \mid (\Gamma_1, A, \Gamma_2)_{>j}\big).$$
In particular, $f_c : F^c_\alpha(\Delta_1 \mid \Gamma \mid \Delta_2) \to A$ and $g'_{c'} : A \to U^{c'}_\beta(\Gamma_1 \mid \Delta \mid \Gamma_2)$ are the components at the cut positions. Recall that we use the concatenation notation such as $(\Delta_1, A, \Delta_2)_{<i} = (\Delta_1)_{<i}$ for $i \leq |\delta_1|$.
\begin{itemize}
	\item For $1 \leq i \leq |\delta_1|$, we are looking at the first contravariant zone of the composite. We need to produce a vertical morphism $F^i_{\beta \circledcirc \alpha}\big((\Delta_1)_{<i} \mid \Gamma_1, \Gamma, \Gamma_2 \mid (\Delta_1)_{>i}, \Delta, \Delta_2\big) \to (\Delta_1)_i$. We form the composite
\begin{align*}
&F^i_{\beta \circledcirc \alpha}\big((\Delta_1)_{<i} \mid \Gamma_1, \Gamma, \Gamma_2 \mid (\Delta_1)_{>i}, \Delta, \Delta_2\big)
\\
&\quad \xto{\eta^F_{i \to}} F^{i}_\alpha\big((\Delta_1)_{<i} \mid \Gamma \mid (\Delta_1)_{>i}, U^{c'}_\beta(\Gamma_1 \mid \Delta \mid \Gamma_2), \Delta_2\big)
\\
&\quad \xto{F^{i}_\alpha(\1 \mid \1 \mid \1, g'_{c'}, \1)} F^{i}_\alpha\big((\Delta_1)_{<i} \mid \Gamma \mid (\Delta_1)_{>i}, A, \Delta_2\big)
\\
&\quad \xto{f_i} (\Delta_1)_i ,
\end{align*}
	where the middle map plugs $g'_{c'}$ into the contravariant slot over $a$.
	\item For $1 \leq i \leq |\delta|$, we are looking at the covariant zone. We form
\begin{align*}
&F^{|\delta_1| + i}_{\beta \circledcirc \alpha}\big(\Delta_1, \Delta_{<i} \mid \Gamma_1, \Gamma, \Gamma_2 \mid \Delta_{>i}, \Delta_2\big)
\\
&\quad \xto{\epsilon^F_i} F^i_\beta\big(\Delta_{<i} \mid \Gamma_1, F^{c}_\alpha(\Delta_1 \mid \Gamma \mid \Delta_2), \Gamma_2 \mid \Delta_{>i}\big)
\\
&\quad \xto{F^i_\beta(\1 \mid \1, f_{c}, \1 \mid \1)} F^i_\beta\big(\Delta_{<i} \mid \Gamma_1, A, \Gamma_2 \mid \Delta_{>i}\big)
\\
&\quad \xto{f'_i} \Delta_i .
\end{align*}
	\item For $1 \leq i \leq |\delta_2|$, in the second contravariant zone, we form
\begin{align*}
&F^{|\delta_1| + |\delta| + i}_{\beta \circledcirc \alpha}\big(\Delta_1, \Delta, (\Delta_2)_{<i} \mid \Gamma_1, \Gamma, \Gamma_2 \mid (\Delta_2)_{>i}\big)
\\
&\quad \xto{\eta^F_{\to i}} F^{c + i}_\alpha\big(\Delta_1, U^{c'}_\beta(\Gamma_1 \mid \Delta \mid \Gamma_2), (\Delta_2)_{<i} \mid \Gamma \mid (\Delta_2)_{>i}\big)
\\
&\quad \xto{F^{c + i}_\alpha(\1, g'_{c'}, \1 \mid \1 \mid \1)} F^{c + i}_\alpha\big(\Delta_1, A, (\Delta_2)_{<i} \mid \Gamma \mid (\Delta_2)_{>i}\big)
\\
&\quad \xto{f_{c + i}} (\Delta_2)_i ,
\end{align*}
	noting that $\Delta_1, A, (\Delta_2)_{<i} = (\Delta_1, A, \Delta_2)_{< c + i}$ and that the codomain of $f_{c+i}$ is $(\Delta_1, A, \Delta_2)_{c + i} = (\Delta_2)_i$ as expected.
	\item For $1 \leq j \leq |\gamma_1|$, in the first contravariant input zone, we form
\begin{align*}
&(\Gamma_1)_j
\\
&\quad \xto{g'_j} U^j_\beta\big((\Gamma_1)_{<j} \mid \Delta \mid (\Gamma_1)_{>j}, A, \Gamma_2\big)
\\
&\quad \xto{U^j_\beta(\1 \mid \1 \mid \1, f_{c}, \1)} U^j_\beta\big((\Gamma_1)_{<j} \mid \Delta \mid (\Gamma_1)_{>j}, F^{c}_\alpha(\Delta_1 \mid \Gamma \mid \Delta_2), \Gamma_2\big)
\\
&\quad \xto{\eta^U_{j \to}} U^j_{\beta \circledcirc \alpha}\big((\Gamma_1)_{<j} \mid \Delta_1, \Delta, \Delta_2 \mid (\Gamma_1)_{>j}, \Gamma, \Gamma_2\big) ,
\end{align*}
	where the middle map plugs $f_c$ into a contravariant slot.
	\item For $1 \leq j \leq |\gamma|$, in the covariant input zone, we form
\begin{align*}
&\Gamma_j
\\
&\quad \xto{g_j} U^j_\alpha\big(\Gamma_{<j} \mid \Delta_1, A, \Delta_2 \mid \Gamma_{>j}\big)
\\
&\quad \xto{U^j_\alpha(\1 \mid \1, g'_{c'}, \1 \mid \1)} U^j_\alpha\big(\Gamma_{<j} \mid \Delta_1, U^{c'}_\beta(\Gamma_1 \mid \Delta \mid \Gamma_2), \Delta_2 \mid \Gamma_{>j}\big)
\\
&\quad \xto{\epsilon^U_j} U^{|\gamma_1| + j}_{\beta \circledcirc \alpha}\big(\Gamma_1, \Gamma_{<j} \mid \Delta_1, \Delta, \Delta_2 \mid \Gamma_{>j}, \Gamma_2\big) .
\end{align*}
	\item For $1 \leq j \leq |\gamma_2|$, in the second contravariant input zone, we form
\begin{align*}
&(\Gamma_2)_j
\\
&\quad \xto{g'_{c' + j}} U^{c' + j}_\beta\big(\Gamma_1, A, (\Gamma_2)_{<j} \mid \Delta \mid (\Gamma_2)_{>j}\big)
\\
&\quad \xto{U^{c' + j}_\beta(\1, f_{c}, \1 \mid \1 \mid \1)} U^{c' + j}_\beta\big(\Gamma_1, F^{c}_\alpha(\Delta_1 \mid \Gamma \mid \Delta_2), (\Gamma_2)_{<j} \mid \Delta \mid (\Gamma_2)_{>j}\big)
\\
&\quad \xto{\eta^U_{\to j}} U^{|\gamma_1| + |\gamma| + j}_{\beta \circledcirc \alpha}\big(\Gamma_1, \Gamma, (\Gamma_2)_{<j} \mid \Delta_1, \Delta, \Delta_2 \mid (\Gamma_2)_{>j}\big) ,
\end{align*}
	noting again that $(\Gamma_1, A, \Gamma_2)_{c' + j} = (\Gamma_2)_j$.
\end{itemize}
Every component is a composite of vertical morphisms, hence vertical, as required. This concludes the specification of the cut; we now verify that it is well defined.

\subsubsection*{Well-Definedness and Projection}

\begin{proposition}[well-definedness]\label{prop:tp-cut-well-defined}
	The identity family of $A_p$ and the cut family are representations.
\end{proposition}
\textit{Proof:}

\textbf{Identity.} We must verify that the family of morphisms for the identity is a representation. This means that its two morphisms present the same map. Thus, we must check that $\varphi^{1^+, 1^-}_{\iota^{\pi A} \circ \iota_{\pi A}}(\tau^F_{\pi A}(A)) = \tau^U_{\pi A}(A)$. This is the required change-of-representative equation. Let $a := \pi A$. Factor $\tau^F_a(A)$ as $\tau^F_a(A) = \tau^F_a(A) \circ \1_{F^1(A)}$. The following diagram expresses the naturality of $\varphi^{1^+,1^-}$:
\begin{center}
	\adjustbox{scale=0.8}{
	\begin{tikzcd}[row sep=large, column sep=huge]
		\mathrm{Rep}^{1^+}_{\iota^a \circ \iota_a}\big(A; F^1(A)\big) \arrow[r, "\varphi^{1^+,1^-}"] \arrow[d, "\tau^F_a(A) \circ (\--)", swap] & \mathrm{Rep}^{1^-}_{\iota^a \circ \iota_a}\big(A; F^1(A)\big) \arrow[d, "{U^1_{\iota^a \circ \iota_a}\big(\tau^F_a(A)\big) \circ (\--)}"] \\
		\mathrm{Rep}^{1^+}_{\iota^a \circ \iota_a}\big(A; A\big) \arrow[r, "\varphi^{1^+,1^-}", swap] & \mathrm{Rep}^{1^-}_{\iota^a \circ \iota_a}\big(A; A\big)
	\end{tikzcd}
	}
\end{center}
The correspondence--unit axiom then gives
$$\varphi^{1^+,1^-}\big(\tau^F_a(A)\big) = U^1_{\iota^a \circ \iota_a}\big(\tau^F_a(A)\big) \circ \varphi^{1^+,1^-}\big(\1_{F^1(A)}\big) = U^1_{\iota^a \circ \iota_a}\big(\tau^F_a(A)\big) \circ \tau^U_a\big(F^1(A)\big) \circ \big(\tau^F_a(A)\big)^{-1}.$$
\noindent The following square expresses naturality of $\tau^U_a$ at $\tau^F_a(A) : F^1(A) \to A$:
\begin{center}
\adjustbox{scale=0.8}{
\begin{tikzcd}[row sep=large, column sep=huge]
F^1(A) \arrow[r, "{\tau^U_a\big(F^1(A)\big)}"] \arrow[d, "\tau^F_a(A)", swap] & U^1_{\iota^a \circ \iota_a}\big(F^1(A)\big) \arrow[d, "{U^1_{\iota^a \circ \iota_a}\big(\tau^F_a(A)\big)}"] \\
A \arrow[r, "\tau^U_a(A)", swap] & U^1_{\iota^a \circ \iota_a}(A)
\end{tikzcd}
}
\end{center}
It gives $U^1_{\iota^a \circ \iota_a}(\tau^F_a(A)) \circ \tau^U_a(F^1(A)) = \tau^U_a(A) \circ \tau^F_a(A)$.

\noindent Substituting in the equation and cancelling with $(\tau^F_a(A))^{-1}$ leads to the result.

\textbf{Cut.} Let
$u \ := \ g'_{c'} \circ f_c \ : \ \widehat F \longrightarrow \widehat U$
be the composition of the two cut maps $f_c : \widehat F \to A$ and $g'_{c'} : A \to \widehat U$. It suffices to show that
$$g_p \circ_{A_p} f_p \ = \ \kappa\langle u\rangle ,$$
where $\kappa\langle u\rangle$ is defined in \cref{def:cut-reading}. Using the correspondence--decomposition axiom, we show that each component of the cut equals the corresponding component of $\kappa\langle u\rangle$.

Define the maps
$$g\langle f \rangle_{i^+} = f'_i \circ F^i_\beta(\1 \mid \1, f_c, \1 \mid \1), \qquad
g\langle f \rangle_{j^-} = U^j_\beta(\ldots, f_c, \ldots) \circ g'_j,$$
and, at the cut input itself, $g\langle f \rangle_{c'^-} = g'_{c'} \circ f_c = u$. Each $\mathrm{Rep}^r_\beta(\Gamma_1,-,\Gamma_2;\Delta)$ is functorial in the object of the cut slot, and $\varphi_\beta$ is natural. Hence $g\langle f\rangle$ is again a representation:
\begin{equation*}
\varphi^{r, s}_\beta\big(g\langle f \rangle_r\big) = g\langle f \rangle_s \quad \text{for all positions } r, s \text{ of } \beta. \tag{$*$}
\end{equation*}
Similarly, define $f\langle g \rangle_{i^+} = f_i \circ F^i_\alpha(\ldots, g'_{c'}, \ldots)$ and $f\langle g \rangle_{j^-} = U^j_\alpha(\ldots, g'_{c'}, \ldots) \circ g_j$. At the cut output, set $f\langle g \rangle_{c^+} = g'_{c'} \circ f_c = u$. Again, these maps satisfy
\begin{equation*}
\varphi^{r, s}_\alpha\big(f\langle g \rangle_r\big) = f\langle g \rangle_s. \tag{$**$}
\end{equation*}
The two families live at the lists $(\Gamma_1,\widehat F,\Gamma_2;\ \Delta)$ and $(\Gamma;\ \Delta_1,\widehat U,\Delta_2)$ of \cref{def:cut-configuration}. Both take the value $u$ at their cut position. Taking $r := c'^-$ in ($*$) and $r := c^+$ in ($**$), we identify them as the representations generated by $u$:
$$g\langle f\rangle_s = \varphi^{c'^-, s}_\beta(u) = (u_\beta)_s \quadt{and} f\langle g\rangle_s = \varphi^{c^+, s}_\alpha(u) = (u_\alpha)_s .$$
By construction, at a position $p$ of $\beta \circledcirc \alpha$ we have
$$\big(g_p \circ_{A_p} f_p\big)_p = \kappa_p\big(g\langle f\rangle_{\bar p}\big) \ \text{ if } \mathsf{fac}(p) = \beta, \qquad \big(g_p \circ_{A_p} f_p\big)_p = \kappa_p\big(f\langle g\rangle_{\bar p}\big) \ \text{ if } \mathsf{fac}(p) = \alpha.$$
Here $\kappa_p$ and $\bar p$ are as in \cref{def:coordinate-change,def:inherited-position}. Thus,
$$\big(g_p \circ_{A_p} f_p\big)_p = \kappa_p\Big(\big(u_{\mathsf{fac}(p)}\big)_{\bar p}\Big) = K_p(u) \qquad \text{for every position } p \text{ of } \beta \circledcirc \alpha.$$
This is the claimed equality $g_p \circ_{A_p} f_p = \kappa\langle u\rangle$. \hfill $\square$

\begin{proposition}\label{prop:pip-planar-polyfunctor}
	The assignments $A_p \mapsto (\pi A)_p$ and $f_p \mapsto \alpha_p$ define a functor of planar polycategories $\pi_p : \fib[T_p]{M_p}$.
\end{proposition}
\textit{Proof:} The identity of $A_p$ lies over $\iota^{\pi A} \circ \iota_{\pi A}$. By definition, this is the identity of $(\pi A)_p$ in $\mc M_p$. The cut of a polymorphism over $\alpha$ into one over $\beta$ lies over $\beta \circledcirc \alpha = \beta_p \circ_{a} \alpha_p$, again by definition of $\mc M_p$. \hfill $\square$

\subsubsection*{The Polycategory Laws}

\begin{theorem}\label{thm:tp-planar-polycategory}
	$\mc T_p$ is a planar polycategory, and $\pi_p : \fib[T_p]{M_p}$ is a functor of planar polycategories.
\end{theorem}

\textit{Proof:} Two polymorphisms over the same base are equal exactly when their components agree at every position. Hence each law is verified componentwise. The bases agree in every case by the corresponding one of \cref{lem:poly-circledcirc-units,lem:poly-circledcirc-associativity,lem:poly-circledcirc-interchange}. We treat the identity laws, then associativity, then the interchange laws.

\textit{Remark:} Since both sides of each law are representations, they are already equal as soon as their components agree at a \emph{single} position, by the \hyperref[anchor:representation-coordinate-system]{coordinate-system remark}. The case analyses below nevertheless verify every position. Any single zone may be the only non-empty one, so all cases are needed for full generality. The redundancy in the overlapping configurations is a useful check on the axioms.

\textbf{Identity laws.} Let $f_p : \Gamma_p \to (\Delta_1)_p, A_p, (\Delta_2)_p$ be a polymorphism over $\alpha$ with components $(f_i; g_j)$. Set $c := |\delta_1| + 1$ and write $a := \pi A$. By \cref{lem:poly-circledcirc-units}, the cut $\1_{A_p} \circ_{A_p} f_p$ lies over $(\iota^a \circ \iota_a) \circledcirc \alpha = \alpha$. Its components are given by the defining formulas with outer components $f'_1 = \tau^F_a(A)$ and $g'_1 = \tau^U_a(A)$:
\begin{align*}
&f_i \circ F^i_\alpha\big(\1 \mid \1 \mid \1, \tau^U_a(A), \1\big) \circ \eta^F_{i \to} \\
&\quad = f_i \\
&\quad \qquad \text{for } i \leq |\delta_1| \text{, by Unit Invariance (left)}, \\
&\quad \tau^F_a(A) \circ F^1_{\iota^a \circ \iota_a}\big(f_{c}\big) \circ \epsilon^F_1 \\
&\quad = f_{c} \circ \tau^F_a\big(F^{c}_\alpha(\Delta_1 \mid \Gamma \mid \Delta_2)\big) \circ \epsilon^F_1 \\
&\quad = f_{c} \\
&\quad \qquad \text{at the cut position}, \\
&\quad f_{c + i} \circ F^{c + i}_\alpha\big(\1, \tau^U_a(A), \1 \mid \1 \mid \1\big) \circ \eta^F_{\to i} \\
&\quad = f_{c+i} \\
&\quad \qquad \text{for } i \leq |\delta_2| \text{, by Unit Invariance (left)}, \\
&\quad \epsilon^U_j \circ U^j_\alpha\big(\1 \mid \1, \tau^U_a(A), \1 \mid \1\big) \circ g_j \\
&\quad = g_j \\
&\quad \qquad \text{for } j \leq |\gamma| \text{, by Unit Invariance (left)},
\end{align*}
where the second line first uses naturality of $\tau^F_a$ at the vertical morphism $f_c$ and then the last equation of the left unit family. Hence $\1_{A_p} \circ_{A_p} f_p = f_p$. Dually, let $g_p : (\Gamma_1)_p, A_p, (\Gamma_2)_p \to \Delta_p$ be a polymorphism over $\beta$ with components $(f'_i; g'_j)$. Set $c' := |\gamma_1| + 1$. The cut $g_p \circ_{A_p} \1_{A_p}$ lies over $\beta \circledcirc (\iota^a \circ \iota_a) = \beta$. Its components are
\begin{align*}
&f'_i \circ F^i_\beta\big(\1 \mid \1, \tau^F_a(A), \1 \mid \1\big) \circ \epsilon^F_i \\
&\quad = f'_i \\
&\quad \qquad \text{for } i \leq |\delta| \text{, by Unit Invariance (right)}, \\
&\quad \eta^U_{j \to} \circ U^j_\beta\big(\1 \mid \1 \mid \1, \tau^F_a(A), \1\big) \circ g'_j \\
&\quad = g'_j \\
&\quad \qquad \text{for } j \leq |\gamma_1| \text{, by Unit Invariance (right)}, \\
&\quad \epsilon^U_1 \circ U^1_{\iota^a \circ \iota_a}\big(g'_{c'}\big) \circ \tau^U_a(A) \\
&\quad = \epsilon^U_1 \circ \tau^U_a\big(U^{c'}_\beta(\Gamma_1 \mid \Delta \mid \Gamma_2)\big) \circ g'_{c'} \\
&\quad = g'_{c'} \\
&\quad \qquad \text{at the cut position}, \\
&\quad \eta^U_{\to j} \circ U^{c' + j}_\beta\big(\1, \tau^F_a(A), \1 \mid \1 \mid \1\big) \circ g'_{c' + j} \\
&\quad = g'_{c' + j} \\
&\quad \qquad \text{for } j \leq |\gamma_2| \text{, by Unit Invariance (right)},
\end{align*}
using naturality of $\tau^U_a$ at $g'_{c'}$ in the third line. Hence $g_p \circ_{A_p} \1_{A_p} = g_p$.

\textbf{Associativity.} Consider three polymorphisms forming an admissible iterated configuration:
\begin{align*}
	f_p &: \Gamma_p \to (\Delta_1)_p, A_p, (\Delta_2)_p && \text{over } \alpha : \otimes_n\gamma \to \oplus(\delta_1, a, \delta_2), && \text{components } (f_i; g_j), \\
	g_p &: (\Gamma_1)_p, A_p, (\Gamma_2)_p \to (\Lambda_1)_p, B_p, (\Lambda_2)_p && \text{over } \beta : \otimes(\gamma_1, a, \gamma_2) \to \oplus(\lambda_1, b, \lambda_2), && \text{components } (f'_i; g'_j), \\
	h_p &: (\Theta_1)_p, B_p, (\Theta_2)_p \to \Xi_p && \text{over } \omega : \otimes(\theta_1, b, \theta_2) \to \oplus\xi, && \text{components } (f''_i; g''_j).
\end{align*}
Set $c_\alpha := |\delta_1| + 1$ for the output $a$ of $\alpha$. Set $c'_\beta := |\gamma_1| + 1$ and $c_\beta := |\lambda_1| + 1$ for the input $a$ and output $b$ of $\beta$, respectively. Set $c'_\omega := |\theta_1| + 1$ for the input $b$ of $\omega$. We must show
$$(h_p \circ_{B_p} g_p) \circ_{A_p} f_p = h_p \circ_{B_p} (g_p \circ_{A_p} f_p).$$
Both sides lie over $(\omega \circledcirc \beta) \circledcirc \alpha = \omega \circledcirc (\beta \circledcirc \alpha)$ by \cref{lem:poly-circledcirc-associativity}. We first record the components of the two intermediate cuts, directly from the defining formulas. The cut $g_p \circ_{A_p} f_p$, over $\beta \circledcirc \alpha$, has components
\begin{align*}
	f^{\beta\alpha}_i &= f_i \circ F^i_\alpha(\1 \mid \1 \mid \1, g'_{c'_\beta}, \1) \circ \eta^F_{i \to} && i \leq |\delta_1| \\
	f^{\beta\alpha}_{|\delta_1| + i} &= f'_i \circ F^i_\beta(\1 \mid \1, f_{c_\alpha}, \1 \mid \1) \circ \epsilon^F_i && i \leq |\lambda_1| + 1 + |\lambda_2| \\
	f^{\beta\alpha}_{|\delta_1| + |\lambda_1| + 1 + |\lambda_2| + i} &= f_{c_\alpha + i} \circ F^{c_\alpha + i}_\alpha(\1, g'_{c'_\beta}, \1 \mid \1 \mid \1) \circ \eta^F_{\to i} && i \leq |\delta_2| \\
	g^{\beta\alpha}_j &= \eta^U_{j \to} \circ U^j_\beta(\1 \mid \1 \mid \1, f_{c_\alpha}, \1) \circ g'_j && j \leq |\gamma_1| \\
	g^{\beta\alpha}_{|\gamma_1| + j} &= \epsilon^U_j \circ U^j_\alpha(\1 \mid \1, g'_{c'_\beta}, \1 \mid \1) \circ g_j && j \leq |\gamma| \\
	g^{\beta\alpha}_{|\gamma_1| + |\gamma| + j} &= \eta^U_{\to j} \circ U^{c'_\beta + j}_\beta(\1, f_{c_\alpha}, \1 \mid \1 \mid \1) \circ g'_{c'_\beta + j} && j \leq |\gamma_2|,
\end{align*}
and the cut $h_p \circ_{B_p} g_p$, over $\omega \circledcirc \beta$, has components
\begin{align*}
	f^{\omega\beta}_i &= f'_i \circ F^i_\beta(\1 \mid \1 \mid \1, g''_{c'_\omega}, \1) \circ \eta^F_{i \to} && i \leq |\lambda_1| \\
	f^{\omega\beta}_{|\lambda_1| + i} &= f''_i \circ F^i_\omega(\1 \mid \1, f'_{c_\beta}, \1 \mid \1) \circ \epsilon^F_i && i \leq |\xi| \\
	f^{\omega\beta}_{|\lambda_1| + |\xi| + i} &= f'_{c_\beta + i} \circ F^{c_\beta + i}_\beta(\1, g''_{c'_\omega}, \1 \mid \1 \mid \1) \circ \eta^F_{\to i} && i \leq |\lambda_2| \\
	g^{\omega\beta}_j &= \eta^U_{j \to} \circ U^j_\omega(\1 \mid \1 \mid \1, f'_{c_\beta}, \1) \circ g''_j && j \leq |\theta_1| \\
	g^{\omega\beta}_{|\theta_1| + j} &= \epsilon^U_j \circ U^j_\beta(\1 \mid \1, g''_{c'_\omega}, \1 \mid \1) \circ g'_j && j \leq |\gamma_1| + 1 + |\gamma_2| \\
	g^{\omega\beta}_{|\theta_1| + |\gamma_1| + 1 + |\gamma_2| + j} &= \eta^U_{\to j} \circ U^{c'_\omega + j}_\omega(\1, f'_{c_\beta}, \1 \mid \1 \mid \1) \circ g''_{c'_\omega + j} && j \leq |\theta_2|.
\end{align*}
In particular, the components at the new cut positions are
$$f^{\beta\alpha}_{|\delta_1| + c_\beta} = f'_{c_\beta} \circ F^{c_\beta}_\beta(\1 \mid \1, f_{c_\alpha}, \1 \mid \1) \circ \epsilon^F_{c_\beta}[\beta,\alpha] \quadt{and} g^{\omega\beta}_{|\theta_1| + c'_\beta} = \epsilon^U_{c'_\beta}[\omega,\beta] \circ U^{c'_\beta}_\beta(\1 \mid \1, g''_{c'_\omega}, \1 \mid \1) \circ g'_{c'_\beta}.$$
The total composites are computed by applying the defining formulas once more. On the left-hand side, $f_p$ is cut into $h_p \circ g_p$ at $a$. On the right-hand side, $g_p \circ f_p$ is cut into $h_p$ at $b$. We verify the equality of the components zone by zone. The composite has ten zones: $(\delta_1, \lambda_1, \xi, \lambda_2, \delta_2)$ on the output side and $(\theta_1, \gamma_1, \gamma, \gamma_2, \theta_2)$ on the input side.

Three moves recur in every zone, and we fix their names here. The \textit{action} of a vertical map $v : P \to Q$ of $\mc T$ in a slot of a representative functor is the value of that functor at $v$ and identities elsewhere. In a covariant slot, it points from the $P$-argument to the $Q$-argument. In a contravariant one, it points the other way. By functoriality, two actions in disjoint slots commute, and successive actions in the same slot merge into the action of their composite. Finally, crossing $v$ past a decomposition map $D$ of \cref{def:planar-poly-indexed-category} means invoking the naturality of $D$ in the slot occupied by $v$. The two vertical actions of such a naturality square need not have the same shape. The reason is that the object of a cut slot occurs on one side of $D$ wrapped in the representative functor of the other factor. Every instance of the three moves is displayed in full below. We write
$$\widehat F_\alpha := F^{c_\alpha}_\alpha(\Delta_1 \mid \Gamma \mid \Delta_2) \quadt{and} \widehat U_\omega := U^{c'_\omega}_\omega(\Theta_1 \mid \Xi \mid \Theta_2).$$
The only two vertical morphisms crossed in the whole argument are
$$v_a := f_{c_\alpha} : \widehat F_\alpha \to A \quadt{and} v_b := g''_{c'_\omega} : B \to \widehat U_\omega.$$
The first is always crossed past a decomposition map of the pair $[\omega,\beta]$ in the $a$-slot. The second is always crossed past one of the pair $[\beta,\alpha]$ in the $b$-slot. We display the instance of the square at each use.
{\small
\begin{itemize}
	\item \textbf{($\delta_1$-zone, $1 \leq i \leq |\delta_1|$.)} The two candidate components are
	\begin{align*}
&\text{left} \\
&\quad = f_i \circ F^i_\alpha\big(\1 \mid \1 \mid \1, g^{\omega\beta}_{|\theta_1| + c'_\beta}, \1\big) \circ \eta^F_{i \to}[\omega \circledcirc \beta,\alpha], \\
&\quad \qquad \text{right} = f^{\beta\alpha}_i \circ F^i_{\beta \circledcirc \alpha}\big(\1 \mid \1 \mid \1, g''_{c'_\omega}, \1\big) \circ \eta^F_{i \to}[\omega,\beta \circledcirc \alpha].
\end{align*}
	Substituting the definition of $f^{\beta\alpha}_i$ into the right-hand side places the action of $v_b$ immediately to the right of $\eta^F_{i \to}[\beta,\alpha]$. On the target side, the $b$-slot of $\eta^F_{i \to}[\beta,\alpha]$ occurs inside $U^{c'_\beta}_\beta$. We cross $v_b$ there in the following square:
	\begin{center}
		\adjustbox{max width=\linewidth}{
		\begin{tikzcd}[row sep=large, column sep=huge]
			F^i_{\beta \circledcirc \alpha}\big((\Delta_1)_{<i} \mid \Gamma_1, \Gamma, \Gamma_2 \mid (\Delta_1)_{>i}, \Lambda_1, \widehat U_\omega, \Lambda_2, \Delta_2\big) \arrow[r, "{\eta^F_{i \to}[\beta,\alpha]}"] \arrow[d, "{F^i_{\beta \circledcirc \alpha}(\1 \mid \1 \mid \1, g''_{c'_\omega}, \1)}", swap] & F^i_\alpha\big((\Delta_1)_{<i} \mid \Gamma \mid (\Delta_1)_{>i}, U^{c'_\beta}_\beta(\Gamma_1 \mid \Lambda_1, \widehat U_\omega, \Lambda_2 \mid \Gamma_2), \Delta_2\big) \arrow[d, "{F^i_\alpha(\1 \mid \1 \mid \1, U^{c'_\beta}_\beta(\1 \mid \1, g''_{c'_\omega}, \1 \mid \1), \1)}"] \\
			F^i_{\beta \circledcirc \alpha}\big((\Delta_1)_{<i} \mid \Gamma_1, \Gamma, \Gamma_2 \mid (\Delta_1)_{>i}, \Lambda_1, B, \Lambda_2, \Delta_2\big) \arrow[r, "{\eta^F_{i \to}[\beta,\alpha]}", swap] & F^i_\alpha\big((\Delta_1)_{<i} \mid \Gamma \mid (\Delta_1)_{>i}, U^{c'_\beta}_\beta(\Gamma_1 \mid \Lambda_1, B, \Lambda_2 \mid \Gamma_2), \Delta_2\big)
		\end{tikzcd}
		}
	\end{center}
	This gives
	$$\text{right} = f_i \circ F^i_\alpha\big(\1 \mid \1 \mid \1, g'_{c'_\beta}, \1\big) \circ F^i_\alpha\big(\1 \mid \1 \mid \1, U^{c'_\beta}_\beta(\1 \mid \1, g''_{c'_\omega}, \1 \mid \1), \1\big) \circ \eta^F_{i \to}[\beta,\alpha] \circ \eta^F_{i \to}[\omega,\beta \circledcirc \alpha].$$
	The contravariant pentagon ($\delta_1$-zone) gives $\eta^F_{i \to}[\beta,\alpha] \circ \eta^F_{i \to}[\omega,\beta \circledcirc \alpha] = F^i_\alpha(\1 \mid \1 \mid \1, \epsilon^U_{c'_\beta}[\omega,\beta], \1) \circ \eta^F_{i \to}[\omega \circledcirc \beta,\alpha]$. We then merge the three actions in the contravariant slot by functoriality of $F^i_\alpha$:
	\begin{center}
		\adjustbox{max width=\linewidth}{
		\begin{tikzcd}[row sep=large]
			F^i_\alpha\big((\Delta_1)_{<i} \mid \Gamma \mid (\Delta_1)_{>i}, U^{|\theta_1| + c'_\beta}_{\omega \circledcirc \beta}(\Theta_1, \Gamma_1 \mid \Lambda_1, \Xi, \Lambda_2 \mid \Gamma_2, \Theta_2), \Delta_2\big) \arrow[d, "{F^i_\alpha(\1 \mid \1 \mid \1, \epsilon^U_{c'_\beta}[\omega,\beta], \1)}"] \arrow[ddd, bend right=45, "{F^i_\alpha(\1 \mid \1 \mid \1, g^{\omega\beta}_{|\theta_1| + c'_\beta}, \1)}"'] \\
			F^i_\alpha\big((\Delta_1)_{<i} \mid \Gamma \mid (\Delta_1)_{>i}, U^{c'_\beta}_\beta(\Gamma_1 \mid \Lambda_1, \widehat U_\omega, \Lambda_2 \mid \Gamma_2), \Delta_2\big) \arrow[d, "{F^i_\alpha(\1 \mid \1 \mid \1, U^{c'_\beta}_\beta(\1 \mid \1, g''_{c'_\omega}, \1 \mid \1), \1)}"] \\
			F^i_\alpha\big((\Delta_1)_{<i} \mid \Gamma \mid (\Delta_1)_{>i}, U^{c'_\beta}_\beta(\Gamma_1 \mid \Lambda_1, B, \Lambda_2 \mid \Gamma_2), \Delta_2\big) \arrow[d, "{F^i_\alpha(\1 \mid \1 \mid \1, g'_{c'_\beta}, \1)}"] \\
			F^i_\alpha\big((\Delta_1)_{<i} \mid \Gamma \mid (\Delta_1)_{>i}, A, \Delta_2\big)
		\end{tikzcd}
		}
	\end{center}
	Thus,
	\[
\begin{aligned}
&\text{right} \\
&\quad = f_i \circ F^i_\alpha\big(\1 \mid \1 \mid \1,\ \epsilon^U_{c'_\beta}[\omega,\beta] \circ U^{c'_\beta}_\beta(\1 \mid \1, g''_{c'_\omega}, \1 \mid \1) \circ g'_{c'_\beta},\ \1\big) \circ \eta^F_{i \to}[\omega \circledcirc \beta,\alpha] \\
&\quad = f_i \circ F^i_\alpha\big(\1 \mid \1 \mid \1, g^{\omega\beta}_{|\theta_1| + c'_\beta}, \1\big) \circ \eta^F_{i \to}[\omega \circledcirc \beta,\alpha] \\
&\quad = \text{left}.
\end{aligned}
\]
	\item \textbf{($\lambda_1$-zone, $1 \leq j \leq |\lambda_1|$.)} Here
	\begin{align*}
&\text{left} \\
&\quad = f^{\omega\beta}_j \circ F^j_{\omega \circledcirc \beta}\big(\1 \mid \1, f_{c_\alpha}, \1 \mid \1\big) \circ \epsilon^F_j[\omega \circledcirc \beta,\alpha], \\
&\quad \qquad \text{right} = f^{\beta\alpha}_{|\delta_1| + j} \circ F^{|\delta_1| + j}_{\beta \circledcirc \alpha}\big(\1 \mid \1 \mid \1, g''_{c'_\omega}, \1\big) \circ \eta^F_{(|\delta_1| + j) \to}[\omega,\beta \circledcirc \alpha].
\end{align*}
	We first substitute $f^{\beta\alpha}_{|\delta_1| + j} = f'_j \circ F^j_\beta(\1 \mid \1, f_{c_\alpha}, \1 \mid \1) \circ \epsilon^F_j[\beta,\alpha]$. We then cross $v_b$ past $\epsilon^F_j[\beta,\alpha]$, where the $b$-slot is an ordinary output argument on both sides:
	\begin{center}
		\adjustbox{max width=\linewidth}{
		\begin{tikzcd}[row sep=large, column sep=huge]
			F^{|\delta_1| + j}_{\beta \circledcirc \alpha}\big(\Delta_1, (\Lambda_1)_{<j} \mid \Gamma_1, \Gamma, \Gamma_2 \mid (\Lambda_1)_{>j}, \widehat U_\omega, \Lambda_2, \Delta_2\big) \arrow[r, "{\epsilon^F_j[\beta,\alpha]}"] \arrow[d, "{F^{|\delta_1| + j}_{\beta \circledcirc \alpha}(\1 \mid \1 \mid \1, g''_{c'_\omega}, \1)}", swap] & F^j_\beta\big((\Lambda_1)_{<j} \mid \Gamma_1, \widehat F_\alpha, \Gamma_2 \mid (\Lambda_1)_{>j}, \widehat U_\omega, \Lambda_2\big) \arrow[d, "{F^j_\beta(\1 \mid \1 \mid \1, g''_{c'_\omega}, \1)}"] \\
			F^{|\delta_1| + j}_{\beta \circledcirc \alpha}\big(\Delta_1, (\Lambda_1)_{<j} \mid \Gamma_1, \Gamma, \Gamma_2 \mid (\Lambda_1)_{>j}, B, \Lambda_2, \Delta_2\big) \arrow[r, "{\epsilon^F_j[\beta,\alpha]}", swap] & F^j_\beta\big((\Lambda_1)_{<j} \mid \Gamma_1, \widehat F_\alpha, \Gamma_2 \mid (\Lambda_1)_{>j}, B, \Lambda_2\big)
		\end{tikzcd}
		}
	\end{center}
	The mixed square ($\lambda_1$-zone) gives $\epsilon^F_j[\beta,\alpha] \circ \eta^F_{(|\delta_1| + j) \to}[\omega,\beta \circledcirc \alpha] = \eta^F_{j \to}[\omega,\beta] \circ \epsilon^F_j[\omega \circledcirc \beta,\alpha]$. Applying this identity, we obtain
	$$\text{right} = f'_j \circ F^j_\beta\big(\1 \mid \1, f_{c_\alpha}, \1 \mid \1\big) \circ F^j_\beta\big(\1 \mid \1, \1, \1 \mid \1, g''_{c'_\omega}, \1\big) \circ \eta^F_{j \to}[\omega,\beta] \circ \epsilon^F_j[\omega \circledcirc \beta,\alpha].$$
	The actions of $v_a$ and $v_b$ occupy disjoint slots of $F^j_\beta$ and hence commute:
	\begin{center}
		\adjustbox{max width=\linewidth}{
		\begin{tikzcd}[row sep=large, column sep=huge]
			F^j_\beta\big((\Lambda_1)_{<j} \mid \Gamma_1, \widehat F_\alpha, \Gamma_2 \mid (\Lambda_1)_{>j}, \widehat U_\omega, \Lambda_2\big) \arrow[r, "{F^j_\beta(\1 \mid \1, f_{c_\alpha}, \1 \mid \1)}"] \arrow[d, "{F^j_\beta(\1 \mid \1 \mid \1, g''_{c'_\omega}, \1)}", swap] & F^j_\beta\big((\Lambda_1)_{<j} \mid \Gamma_1, A, \Gamma_2 \mid (\Lambda_1)_{>j}, \widehat U_\omega, \Lambda_2\big) \arrow[d, "{F^j_\beta(\1 \mid \1 \mid \1, g''_{c'_\omega}, \1)}"] \\
			F^j_\beta\big((\Lambda_1)_{<j} \mid \Gamma_1, \widehat F_\alpha, \Gamma_2 \mid (\Lambda_1)_{>j}, B, \Lambda_2\big) \arrow[r, "{F^j_\beta(\1 \mid \1, f_{c_\alpha}, \1 \mid \1)}", swap] & F^j_\beta\big((\Lambda_1)_{<j} \mid \Gamma_1, A, \Gamma_2 \mid (\Lambda_1)_{>j}, B, \Lambda_2\big)
		\end{tikzcd}
		}
	\end{center}
	This brings the action of $v_a$ next to $\eta^F_{j \to}[\omega,\beta]$. We then cross $v_a$ past $\eta^F_{j \to}[\omega,\beta]$. Its $a$-slot is a covariant input argument on both sides:
	\begin{center}
		\adjustbox{max width=\linewidth}{
		\begin{tikzcd}[row sep=large, column sep=huge]
			F^j_{\omega \circledcirc \beta}\big((\Lambda_1)_{<j} \mid \Theta_1, \Gamma_1, \widehat F_\alpha, \Gamma_2, \Theta_2 \mid (\Lambda_1)_{>j}, \Xi, \Lambda_2\big) \arrow[r, "{\eta^F_{j \to}[\omega,\beta]}"] \arrow[d, "{F^j_{\omega \circledcirc \beta}(\1 \mid \1, f_{c_\alpha}, \1 \mid \1)}", swap] & F^j_\beta\big((\Lambda_1)_{<j} \mid \Gamma_1, \widehat F_\alpha, \Gamma_2 \mid (\Lambda_1)_{>j}, \widehat U_\omega, \Lambda_2\big) \arrow[d, "{F^j_\beta(\1 \mid \1, f_{c_\alpha}, \1 \mid \1)}"] \\
			F^j_{\omega \circledcirc \beta}\big((\Lambda_1)_{<j} \mid \Theta_1, \Gamma_1, A, \Gamma_2, \Theta_2 \mid (\Lambda_1)_{>j}, \Xi, \Lambda_2\big) \arrow[r, "{\eta^F_{j \to}[\omega,\beta]}", swap] & F^j_\beta\big((\Lambda_1)_{<j} \mid \Gamma_1, A, \Gamma_2 \mid (\Lambda_1)_{>j}, \widehat U_\omega, \Lambda_2\big)
		\end{tikzcd}
		}
	\end{center}
	Recognising $f^{\omega\beta}_j = f'_j \circ F^j_\beta(\1 \mid \1 \mid \1, g''_{c'_\omega}, \1) \circ \eta^F_{j \to}[\omega,\beta]$, we obtain
	$$\text{right} = f^{\omega\beta}_j \circ F^j_{\omega \circledcirc \beta}\big(\1 \mid \1, f_{c_\alpha}, \1 \mid \1\big) \circ \epsilon^F_j[\omega \circledcirc \beta,\alpha] = \text{left}.$$
	\item \textbf{($\xi$-zone, $1 \leq i \leq |\xi|$.)} Here
	\begin{align*}
&\text{left} \\
&\quad = f^{\omega\beta}_{|\lambda_1| + i} \circ F^{|\lambda_1| + i}_{\omega \circledcirc \beta}\big(\1 \mid \1, f_{c_\alpha}, \1 \mid \1\big) \circ \epsilon^F_{|\lambda_1| + i}[\omega \circledcirc \beta,\alpha], \\
&\quad \qquad \text{right} = f''_i \circ F^i_\omega\big(\1 \mid \1, f^{\beta\alpha}_{|\delta_1| + c_\beta}, \1 \mid \1\big) \circ \epsilon^F_i[\omega,\beta \circledcirc \alpha].
\end{align*}
	We substitute the displayed formula for $f^{\beta\alpha}_{|\delta_1| + c_\beta}$. By functoriality of $F^i_\omega$, the covariant action splits into the following three factors:
	\begin{center}
		\adjustbox{max width=\linewidth}{
		\begin{tikzcd}[row sep=large]
			F^i_\omega\big(\Xi_{<i} \mid \Theta_1, F^{|\delta_1| + c_\beta}_{\beta \circledcirc \alpha}(\Delta_1, \Lambda_1 \mid \Gamma_1, \Gamma, \Gamma_2 \mid \Lambda_2, \Delta_2), \Theta_2 \mid \Xi_{>i}\big) \arrow[d, "{F^i_\omega(\1 \mid \1, \epsilon^F_{c_\beta}[\beta,\alpha], \1 \mid \1)}"] \arrow[ddd, bend right=45, "{F^i_\omega(\1 \mid \1, f^{\beta\alpha}_{|\delta_1| + c_\beta}, \1 \mid \1)}"'] \\
			F^i_\omega\big(\Xi_{<i} \mid \Theta_1, F^{c_\beta}_\beta(\Lambda_1 \mid \Gamma_1, \widehat F_\alpha, \Gamma_2 \mid \Lambda_2), \Theta_2 \mid \Xi_{>i}\big) \arrow[d, "{F^i_\omega(\1 \mid \1, F^{c_\beta}_\beta(\1 \mid \1, f_{c_\alpha}, \1 \mid \1), \1 \mid \1)}"] \\
			F^i_\omega\big(\Xi_{<i} \mid \Theta_1, F^{c_\beta}_\beta(\Lambda_1 \mid \Gamma_1, A, \Gamma_2 \mid \Lambda_2), \Theta_2 \mid \Xi_{>i}\big) \arrow[d, "{F^i_\omega(\1 \mid \1, f'_{c_\beta}, \1 \mid \1)}"] \\
			F^i_\omega\big(\Xi_{<i} \mid \Theta_1, B, \Theta_2 \mid \Xi_{>i}\big)
		\end{tikzcd}
		}
	\end{center}
	This gives
	\[
\begin{aligned}
&\text{right} \\
&\quad  = f''_i \circ F^i_\omega\big(\1 \mid \1, f'_{c_\beta}, \1 \mid \1\big) \circ F^i_\omega\big(\1 \mid \1, F^{c_\beta}_\beta(\1 \mid \1, f_{c_\alpha}, \1 \mid \1), \1 \mid \1\big) \\
&\quad \circ F^i_\omega\big(\1 \mid \1, \epsilon^F_{c_\beta}[\beta,\alpha], \1 \mid \1\big) \circ \epsilon^F_i[\omega,\beta \circledcirc \alpha].
\end{aligned}
\]
	The covariant pentagon ($\xi$-zone) gives $F^i_\omega(\1 \mid \1, \epsilon^F_{c_\beta}[\beta,\alpha], \1 \mid \1) \circ \epsilon^F_i[\omega,\beta \circledcirc \alpha] = \epsilon^F_i[\omega,\beta] \circ \epsilon^F_{|\lambda_1| + i}[\omega \circledcirc \beta,\alpha]$. We then cross $v_a$ past $\epsilon^F_i[\omega,\beta]$. Its $a$-slot occurs inside $F^{c_\beta}_\beta$ on the target side:
	\begin{center}
		\adjustbox{max width=\linewidth}{
		\begin{tikzcd}[row sep=large, column sep=huge]
			F^{|\lambda_1| + i}_{\omega \circledcirc \beta}\big(\Lambda_1, \Xi_{<i} \mid \Theta_1, \Gamma_1, \widehat F_\alpha, \Gamma_2, \Theta_2 \mid \Xi_{>i}, \Lambda_2\big) \arrow[r, "{\epsilon^F_i[\omega,\beta]}"] \arrow[d, "{F^{|\lambda_1| + i}_{\omega \circledcirc \beta}(\1 \mid \1, f_{c_\alpha}, \1 \mid \1)}", swap] & F^i_\omega\big(\Xi_{<i} \mid \Theta_1, F^{c_\beta}_\beta(\Lambda_1 \mid \Gamma_1, \widehat F_\alpha, \Gamma_2 \mid \Lambda_2), \Theta_2 \mid \Xi_{>i}\big) \arrow[d, "{F^i_\omega(\1 \mid \1, F^{c_\beta}_\beta(\1 \mid \1, f_{c_\alpha}, \1 \mid \1), \1 \mid \1)}"] \\
			F^{|\lambda_1| + i}_{\omega \circledcirc \beta}\big(\Lambda_1, \Xi_{<i} \mid \Theta_1, \Gamma_1, A, \Gamma_2, \Theta_2 \mid \Xi_{>i}, \Lambda_2\big) \arrow[r, "{\epsilon^F_i[\omega,\beta]}", swap] & F^i_\omega\big(\Xi_{<i} \mid \Theta_1, F^{c_\beta}_\beta(\Lambda_1 \mid \Gamma_1, A, \Gamma_2 \mid \Lambda_2), \Theta_2 \mid \Xi_{>i}\big)
		\end{tikzcd}
		}
	\end{center}
	This gives
	$$\text{right} = f''_i \circ F^i_\omega\big(\1 \mid \1, f'_{c_\beta}, \1 \mid \1\big) \circ \epsilon^F_i[\omega,\beta] \circ F^{|\lambda_1| + i}_{\omega \circledcirc \beta}\big(\1 \mid \1, f_{c_\alpha}, \1 \mid \1\big) \circ \epsilon^F_{|\lambda_1| + i}[\omega \circledcirc \beta,\alpha] = \text{left}.$$
	\item \textbf{($\lambda_2$-zone, $1 \leq j \leq |\lambda_2|$.)} Here
	\begin{align*}
&\text{left} \\
&\quad = f^{\omega\beta}_{|\lambda_1| + |\xi| + j} \circ F^{|\lambda_1| + |\xi| + j}_{\omega \circledcirc \beta}\big(\1 \mid \1, f_{c_\alpha}, \1 \mid \1\big) \circ \epsilon^F_{|\lambda_1| + |\xi| + j}[\omega \circledcirc \beta,\alpha], \\
&\quad \qquad \text{right} = f^{\beta\alpha}_{|\delta_1| + c_\beta + j} \circ F^{|\delta_1| + c_\beta + j}_{\beta \circledcirc \alpha}\big(\1, g''_{c'_\omega}, \1 \mid \1 \mid \1\big) \circ \eta^F_{\to j}[\omega,\beta \circledcirc \alpha].
\end{align*}
	We first substitute $f^{\beta\alpha}_{|\delta_1| + c_\beta + j} = f'_{c_\beta + j} \circ F^{c_\beta + j}_\beta(\1 \mid \1, f_{c_\alpha}, \1 \mid \1) \circ \epsilon^F_{c_\beta + j}[\beta,\alpha]$. We then cross $v_b$ past $\epsilon^F_{c_\beta + j}[\beta,\alpha]$:
	\begin{center}
		\adjustbox{max width=\linewidth}{
		\begin{tikzcd}[row sep=large, column sep=huge]
			F^{|\delta_1| + c_\beta + j}_{\beta \circledcirc \alpha}\big(\Delta_1, \Lambda_1, \widehat U_\omega, (\Lambda_2)_{<j} \mid \Gamma_1, \Gamma, \Gamma_2 \mid (\Lambda_2)_{>j}, \Delta_2\big) \arrow[r, "{\epsilon^F_{c_\beta + j}[\beta,\alpha]}"] \arrow[d, "{F^{|\delta_1| + c_\beta + j}_{\beta \circledcirc \alpha}(\1, g''_{c'_\omega}, \1 \mid \1 \mid \1)}", swap] & F^{c_\beta + j}_\beta\big(\Lambda_1, \widehat U_\omega, (\Lambda_2)_{<j} \mid \Gamma_1, \widehat F_\alpha, \Gamma_2 \mid (\Lambda_2)_{>j}\big) \arrow[d, "{F^{c_\beta + j}_\beta(\1, g''_{c'_\omega}, \1 \mid \1 \mid \1)}"] \\
			F^{|\delta_1| + c_\beta + j}_{\beta \circledcirc \alpha}\big(\Delta_1, \Lambda_1, B, (\Lambda_2)_{<j} \mid \Gamma_1, \Gamma, \Gamma_2 \mid (\Lambda_2)_{>j}, \Delta_2\big) \arrow[r, "{\epsilon^F_{c_\beta + j}[\beta,\alpha]}", swap] & F^{c_\beta + j}_\beta\big(\Lambda_1, B, (\Lambda_2)_{<j} \mid \Gamma_1, \widehat F_\alpha, \Gamma_2 \mid (\Lambda_2)_{>j}\big)
		\end{tikzcd}
		}
	\end{center}
	The mixed square ($\lambda_2$-zone) gives $\epsilon^F_{c_\beta + j}[\beta,\alpha] \circ \eta^F_{\to j}[\omega,\beta \circledcirc \alpha] = \eta^F_{\to j}[\omega,\beta] \circ \epsilon^F_{|\lambda_1| + |\xi| + j}[\omega \circledcirc \beta,\alpha]$. The actions of $v_a$ and $v_b$ occupy disjoint slots of $F^{c_\beta + j}_\beta$ and hence commute:
	\begin{center}
		\adjustbox{max width=\linewidth}{
		\begin{tikzcd}[row sep=large, column sep=huge]
			F^{c_\beta + j}_\beta\big(\Lambda_1, \widehat U_\omega, (\Lambda_2)_{<j} \mid \Gamma_1, \widehat F_\alpha, \Gamma_2 \mid (\Lambda_2)_{>j}\big) \arrow[r, "{F^{c_\beta + j}_\beta(\1 \mid \1, f_{c_\alpha}, \1 \mid \1)}"] \arrow[d, "{F^{c_\beta + j}_\beta(\1, g''_{c'_\omega}, \1 \mid \1 \mid \1)}", swap] & F^{c_\beta + j}_\beta\big(\Lambda_1, \widehat U_\omega, (\Lambda_2)_{<j} \mid \Gamma_1, A, \Gamma_2 \mid (\Lambda_2)_{>j}\big) \arrow[d, "{F^{c_\beta + j}_\beta(\1, g''_{c'_\omega}, \1 \mid \1 \mid \1)}"] \\
			F^{c_\beta + j}_\beta\big(\Lambda_1, B, (\Lambda_2)_{<j} \mid \Gamma_1, \widehat F_\alpha, \Gamma_2 \mid (\Lambda_2)_{>j}\big) \arrow[r, "{F^{c_\beta + j}_\beta(\1 \mid \1, f_{c_\alpha}, \1 \mid \1)}", swap] & F^{c_\beta + j}_\beta\big(\Lambda_1, B, (\Lambda_2)_{<j} \mid \Gamma_1, A, \Gamma_2 \mid (\Lambda_2)_{>j}\big)
		\end{tikzcd}
		}
	\end{center}
	Finally, we cross $v_a$ past $\eta^F_{\to j}[\omega,\beta]$:
	\begin{center}
		\adjustbox{max width=\linewidth}{
		\begin{tikzcd}[row sep=large, column sep=huge]
			F^{|\lambda_1| + |\xi| + j}_{\omega \circledcirc \beta}\big(\Lambda_1, \Xi, (\Lambda_2)_{<j} \mid \Theta_1, \Gamma_1, \widehat F_\alpha, \Gamma_2, \Theta_2 \mid (\Lambda_2)_{>j}\big) \arrow[r, "{\eta^F_{\to j}[\omega,\beta]}"] \arrow[d, "{F^{|\lambda_1| + |\xi| + j}_{\omega \circledcirc \beta}(\1 \mid \1, f_{c_\alpha}, \1 \mid \1)}", swap] & F^{c_\beta + j}_\beta\big(\Lambda_1, \widehat U_\omega, (\Lambda_2)_{<j} \mid \Gamma_1, \widehat F_\alpha, \Gamma_2 \mid (\Lambda_2)_{>j}\big) \arrow[d, "{F^{c_\beta + j}_\beta(\1 \mid \1, f_{c_\alpha}, \1 \mid \1)}"] \\
			F^{|\lambda_1| + |\xi| + j}_{\omega \circledcirc \beta}\big(\Lambda_1, \Xi, (\Lambda_2)_{<j} \mid \Theta_1, \Gamma_1, A, \Gamma_2, \Theta_2 \mid (\Lambda_2)_{>j}\big) \arrow[r, "{\eta^F_{\to j}[\omega,\beta]}", swap] & F^{c_\beta + j}_\beta\big(\Lambda_1, \widehat U_\omega, (\Lambda_2)_{<j} \mid \Gamma_1, A, \Gamma_2 \mid (\Lambda_2)_{>j}\big)
		\end{tikzcd}
		}
	\end{center}
	After this crossing, we obtain
	\[
\begin{aligned}
&\text{right} \\
&\quad  = f'_{c_\beta + j} \circ F^{c_\beta + j}_\beta\big(\1, g''_{c'_\omega}, \1 \mid \1 \mid \1\big) \circ \eta^F_{\to j}[\omega,\beta] \\
&\quad \circ F^{|\lambda_1| + |\xi| + j}_{\omega \circledcirc \beta}\big(\1 \mid \1, f_{c_\alpha}, \1 \mid \1\big) \circ \epsilon^F_{|\lambda_1| + |\xi| + j}[\omega \circledcirc \beta,\alpha],
\end{aligned}
\]
	where the composite of the first three factors is $f^{\omega\beta}_{|\lambda_1| + |\xi| + j}$. Hence $\text{right}=\text{left}$.
	\item \textbf{($\delta_2$-zone, $1 \leq j \leq |\delta_2|$.)} Here
	\begin{align*}
&\text{left} \\
&\quad = f_{c_\alpha + j} \circ F^{c_\alpha + j}_\alpha\big(\1, g^{\omega\beta}_{|\theta_1| + c'_\beta}, \1 \mid \1 \mid \1\big) \circ \eta^F_{\to j}[\omega \circledcirc \beta,\alpha], \\
&\quad \qquad \text{right} = f^{\beta\alpha}_{|\delta_1| + c_\beta + |\lambda_2| + j} \circ F^{|\delta_1| + c_\beta + |\lambda_2| + j}_{\beta \circledcirc \alpha}\big(\1, g''_{c'_\omega}, \1 \mid \1 \mid \1\big) \circ \eta^F_{\to (|\lambda_2| + j)}[\omega,\beta \circledcirc \alpha].
\end{align*}
	We first substitute $f^{\beta\alpha}_{|\delta_1| + c_\beta + |\lambda_2| + j} = f_{c_\alpha + j} \circ F^{c_\alpha + j}_\alpha(\1, g'_{c'_\beta}, \1 \mid \1 \mid \1) \circ \eta^F_{\to j}[\beta,\alpha]$. We then cross $v_b$ past $\eta^F_{\to j}[\beta,\alpha]$. Its $b$-slot occurs inside $U^{c'_\beta}_\beta$ on the target side:
	\begin{center}
		\adjustbox{max width=\linewidth}{
		\begin{tikzcd}[row sep=large, column sep=huge]
			F^{|\delta_1| + c_\beta + |\lambda_2| + j}_{\beta \circledcirc \alpha}\big(\Delta_1, \Lambda_1, \widehat U_\omega, \Lambda_2, (\Delta_2)_{<j} \mid \Gamma_1, \Gamma, \Gamma_2 \mid (\Delta_2)_{>j}\big) \arrow[r, "{\eta^F_{\to j}[\beta,\alpha]}"] \arrow[d, "{F^{|\delta_1| + c_\beta + |\lambda_2| + j}_{\beta \circledcirc \alpha}(\1, g''_{c'_\omega}, \1 \mid \1 \mid \1)}", swap] & F^{c_\alpha + j}_\alpha\big(\Delta_1, U^{c'_\beta}_\beta(\Gamma_1 \mid \Lambda_1, \widehat U_\omega, \Lambda_2 \mid \Gamma_2), (\Delta_2)_{<j} \mid \Gamma \mid (\Delta_2)_{>j}\big) \arrow[d, "{F^{c_\alpha + j}_\alpha(\1, U^{c'_\beta}_\beta(\1 \mid \1, g''_{c'_\omega}, \1 \mid \1), \1 \mid \1 \mid \1)}"] \\
			F^{|\delta_1| + c_\beta + |\lambda_2| + j}_{\beta \circledcirc \alpha}\big(\Delta_1, \Lambda_1, B, \Lambda_2, (\Delta_2)_{<j} \mid \Gamma_1, \Gamma, \Gamma_2 \mid (\Delta_2)_{>j}\big) \arrow[r, "{\eta^F_{\to j}[\beta,\alpha]}", swap] & F^{c_\alpha + j}_\alpha\big(\Delta_1, U^{c'_\beta}_\beta(\Gamma_1 \mid \Lambda_1, B, \Lambda_2 \mid \Gamma_2), (\Delta_2)_{<j} \mid \Gamma \mid (\Delta_2)_{>j}\big)
		\end{tikzcd}
		}
	\end{center}
	The contravariant pentagon ($\delta_2$-zone) gives $\eta^F_{\to j}[\beta,\alpha] \circ \eta^F_{\to (|\lambda_2| + j)}[\omega,\beta \circledcirc \alpha] = F^{c_\alpha + j}_\alpha(\1, \epsilon^U_{c'_\beta}[\omega,\beta], \1 \mid \1 \mid \1) \circ \eta^F_{\to j}[\omega \circledcirc \beta,\alpha]$. We then merge the three actions in the contravariant slot:
	\begin{center}
		\adjustbox{max width=\linewidth}{
		\begin{tikzcd}[row sep=large]
			F^{c_\alpha + j}_\alpha\big(\Delta_1, U^{|\theta_1| + c'_\beta}_{\omega \circledcirc \beta}(\Theta_1, \Gamma_1 \mid \Lambda_1, \Xi, \Lambda_2 \mid \Gamma_2, \Theta_2), (\Delta_2)_{<j} \mid \Gamma \mid (\Delta_2)_{>j}\big) \arrow[d, "{F^{c_\alpha + j}_\alpha(\1, \epsilon^U_{c'_\beta}[\omega,\beta], \1 \mid \1 \mid \1)}"] \arrow[ddd, bend right=45, "{F^{c_\alpha + j}_\alpha(\1, g^{\omega\beta}_{|\theta_1| + c'_\beta}, \1 \mid \1 \mid \1)}"'] \\
			F^{c_\alpha + j}_\alpha\big(\Delta_1, U^{c'_\beta}_\beta(\Gamma_1 \mid \Lambda_1, \widehat U_\omega, \Lambda_2 \mid \Gamma_2), (\Delta_2)_{<j} \mid \Gamma \mid (\Delta_2)_{>j}\big) \arrow[d, "{F^{c_\alpha + j}_\alpha(\1, U^{c'_\beta}_\beta(\1 \mid \1, g''_{c'_\omega}, \1 \mid \1), \1 \mid \1 \mid \1)}"] \\
			F^{c_\alpha + j}_\alpha\big(\Delta_1, U^{c'_\beta}_\beta(\Gamma_1 \mid \Lambda_1, B, \Lambda_2 \mid \Gamma_2), (\Delta_2)_{<j} \mid \Gamma \mid (\Delta_2)_{>j}\big) \arrow[d, "{F^{c_\alpha + j}_\alpha(\1, g'_{c'_\beta}, \1 \mid \1 \mid \1)}"] \\
			F^{c_\alpha + j}_\alpha\big(\Delta_1, A, (\Delta_2)_{<j} \mid \Gamma \mid (\Delta_2)_{>j}\big)
		\end{tikzcd}
		}
	\end{center}
	Thus,
	$$\text{right} = f_{c_\alpha + j} \circ F^{c_\alpha + j}_\alpha\big(\1,\ \epsilon^U_{c'_\beta}[\omega,\beta] \circ U^{c'_\beta}_\beta(\1 \mid \1, g''_{c'_\omega}, \1 \mid \1) \circ g'_{c'_\beta},\ \1 \mid \1 \mid \1\big) \circ \eta^F_{\to j}[\omega \circledcirc \beta,\alpha] = \text{left}.$$
	\item \textbf{($\theta_1$-zone, $1 \leq j \leq |\theta_1|$.)} Here
	\begin{align*}
&\text{left} \\
&\quad = \eta^U_{j \to}[\omega \circledcirc \beta,\alpha] \circ U^j_{\omega \circledcirc \beta}\big(\1 \mid \1 \mid \1, f_{c_\alpha}, \1\big) \circ g^{\omega\beta}_j, \\
&\quad \qquad \text{right} = \eta^U_{j \to}[\omega,\beta \circledcirc \alpha] \circ U^j_\omega\big(\1 \mid \1 \mid \1, f^{\beta\alpha}_{|\delta_1| + c_\beta}, \1\big) \circ g''_j.
\end{align*}
	The contravariant slot reverses composites, so substituting the formula for $f^{\beta\alpha}_{|\delta_1| + c_\beta}$ splits the middle action of the right-hand side as
	\begin{center}
		\adjustbox{max width=\linewidth}{
		\begin{tikzcd}[row sep=large]
			U^j_\omega\big((\Theta_1)_{<j} \mid \Xi \mid (\Theta_1)_{>j}, B, \Theta_2\big) \arrow[d, "{U^j_\omega(\1 \mid \1 \mid \1, f'_{c_\beta}, \1)}"] \arrow[ddd, bend right=45, "{U^j_\omega(\1 \mid \1 \mid \1, f^{\beta\alpha}_{|\delta_1| + c_\beta}, \1)}"'] \\
			U^j_\omega\big((\Theta_1)_{<j} \mid \Xi \mid (\Theta_1)_{>j}, F^{c_\beta}_\beta(\Lambda_1 \mid \Gamma_1, A, \Gamma_2 \mid \Lambda_2), \Theta_2\big) \arrow[d, "{U^j_\omega(\1 \mid \1 \mid \1, F^{c_\beta}_\beta(\1 \mid \1, f_{c_\alpha}, \1 \mid \1), \1)}"] \\
			U^j_\omega\big((\Theta_1)_{<j} \mid \Xi \mid (\Theta_1)_{>j}, F^{c_\beta}_\beta(\Lambda_1 \mid \Gamma_1, \widehat F_\alpha, \Gamma_2 \mid \Lambda_2), \Theta_2\big) \arrow[d, "{U^j_\omega(\1 \mid \1 \mid \1, \epsilon^F_{c_\beta}[\beta,\alpha], \1)}"] \\
			U^j_\omega\big((\Theta_1)_{<j} \mid \Xi \mid (\Theta_1)_{>j}, F^{|\delta_1| + c_\beta}_{\beta \circledcirc \alpha}(\Delta_1, \Lambda_1 \mid \Gamma_1, \Gamma, \Gamma_2 \mid \Lambda_2, \Delta_2), \Theta_2\big)
		\end{tikzcd}
		}
	\end{center}
	The contravariant pentagon ($\theta_1$-zone) gives $\eta^U_{j \to}[\omega,\beta \circledcirc \alpha] \circ U^j_\omega(\1 \mid \1 \mid \1, \epsilon^F_{c_\beta}[\beta,\alpha], \1) = \eta^U_{j \to}[\omega \circledcirc \beta,\alpha] \circ \eta^U_{j \to}[\omega,\beta]$. We then cross $v_a$ past the inner $\eta^U_{j \to}[\omega,\beta]$. The $a$-slot of this map occurs inside $F^{c_\beta}_\beta$ on the source side and is contravariant on both sides:
	\begin{center}
		\adjustbox{max width=\linewidth}{
		\begin{tikzcd}[row sep=large, column sep=huge]
			U^j_\omega\big((\Theta_1)_{<j} \mid \Xi \mid (\Theta_1)_{>j}, F^{c_\beta}_\beta(\Lambda_1 \mid \Gamma_1, A, \Gamma_2 \mid \Lambda_2), \Theta_2\big) \arrow[r, "{\eta^U_{j \to}[\omega,\beta]}"] \arrow[d, "{U^j_\omega(\1 \mid \1 \mid \1, F^{c_\beta}_\beta(\1 \mid \1, f_{c_\alpha}, \1 \mid \1), \1)}", swap] & U^j_{\omega \circledcirc \beta}\big((\Theta_1)_{<j} \mid \Lambda_1, \Xi, \Lambda_2 \mid (\Theta_1)_{>j}, \Gamma_1, A, \Gamma_2, \Theta_2\big) \arrow[d, "{U^j_{\omega \circledcirc \beta}(\1 \mid \1 \mid \1, f_{c_\alpha}, \1)}"] \\
			U^j_\omega\big((\Theta_1)_{<j} \mid \Xi \mid (\Theta_1)_{>j}, F^{c_\beta}_\beta(\Lambda_1 \mid \Gamma_1, \widehat F_\alpha, \Gamma_2 \mid \Lambda_2), \Theta_2\big) \arrow[r, "{\eta^U_{j \to}[\omega,\beta]}", swap] & U^j_{\omega \circledcirc \beta}\big((\Theta_1)_{<j} \mid \Lambda_1, \Xi, \Lambda_2 \mid (\Theta_1)_{>j}, \Gamma_1, \widehat F_\alpha, \Gamma_2, \Theta_2\big)
		\end{tikzcd}
		}
	\end{center}
	After this crossing, we obtain
	$$\text{right} = \eta^U_{j \to}[\omega \circledcirc \beta,\alpha] \circ U^j_{\omega \circledcirc \beta}\big(\1 \mid \1 \mid \1, f_{c_\alpha}, \1\big) \circ \eta^U_{j \to}[\omega,\beta] \circ U^j_\omega\big(\1 \mid \1 \mid \1, f'_{c_\beta}, \1\big) \circ g''_j = \text{left},$$
	where the composite of the last three factors is $g^{\omega\beta}_j$.
	\item \textbf{($\gamma_1$-zone, $1 \leq j \leq |\gamma_1|$.)} Here
	\begin{align*}
&\text{left} \\
&\quad = \eta^U_{(|\theta_1| + j) \to}[\omega \circledcirc \beta,\alpha] \circ U^{|\theta_1| + j}_{\omega \circledcirc \beta}\big(\1 \mid \1 \mid \1, f_{c_\alpha}, \1\big) \circ g^{\omega\beta}_{|\theta_1| + j}, \\
&\quad \qquad \text{right} = \epsilon^U_j[\omega,\beta \circledcirc \alpha] \circ U^j_{\beta \circledcirc \alpha}\big(\1 \mid \1, g''_{c'_\omega}, \1 \mid \1\big) \circ g^{\beta\alpha}_j.
\end{align*}
	We first substitute $g^{\beta\alpha}_j = \eta^U_{j \to}[\beta,\alpha] \circ U^j_\beta(\1 \mid \1 \mid \1, f_{c_\alpha}, \1) \circ g'_j$. We then cross $v_b$ past $\eta^U_{j \to}[\beta,\alpha]$. Its $b$-slot is a covariant output argument on both sides:
	\begin{center}
		\adjustbox{max width=\linewidth}{
		\begin{tikzcd}[row sep=large, column sep=huge]
			U^j_\beta\big((\Gamma_1)_{<j} \mid \Lambda_1, B, \Lambda_2 \mid (\Gamma_1)_{>j}, \widehat F_\alpha, \Gamma_2\big) \arrow[r, "{\eta^U_{j \to}[\beta,\alpha]}"] \arrow[d, "{U^j_\beta(\1 \mid \1, g''_{c'_\omega}, \1 \mid \1)}", swap] & U^j_{\beta \circledcirc \alpha}\big((\Gamma_1)_{<j} \mid \Delta_1, \Lambda_1, B, \Lambda_2, \Delta_2 \mid (\Gamma_1)_{>j}, \Gamma, \Gamma_2\big) \arrow[d, "{U^j_{\beta \circledcirc \alpha}(\1 \mid \1, g''_{c'_\omega}, \1 \mid \1)}"] \\
			U^j_\beta\big((\Gamma_1)_{<j} \mid \Lambda_1, \widehat U_\omega, \Lambda_2 \mid (\Gamma_1)_{>j}, \widehat F_\alpha, \Gamma_2\big) \arrow[r, "{\eta^U_{j \to}[\beta,\alpha]}", swap] & U^j_{\beta \circledcirc \alpha}\big((\Gamma_1)_{<j} \mid \Delta_1, \Lambda_1, \widehat U_\omega, \Lambda_2, \Delta_2 \mid (\Gamma_1)_{>j}, \Gamma, \Gamma_2\big)
		\end{tikzcd}
		}
	\end{center}
	The mixed square ($\gamma_1$-zone) gives $\epsilon^U_j[\omega,\beta \circledcirc \alpha] \circ \eta^U_{j \to}[\beta,\alpha] = \eta^U_{(|\theta_1| + j) \to}[\omega \circledcirc \beta,\alpha] \circ \epsilon^U_j[\omega,\beta]$. Applying this identity, we obtain
	$$\text{right} = \eta^U_{(|\theta_1| + j) \to}[\omega \circledcirc \beta,\alpha] \circ \epsilon^U_j[\omega,\beta] \circ U^j_\beta\big(\1 \mid \1, g''_{c'_\omega}, \1 \mid \1, \1, \1\big) \circ U^j_\beta\big(\1 \mid \1 \mid \1, f_{c_\alpha}, \1\big) \circ g'_j.$$
	The actions of $v_a$ and $v_b$ occupy disjoint slots of $U^j_\beta$ and hence commute:
	\begin{center}
		\adjustbox{max width=\linewidth}{
		\begin{tikzcd}[row sep=large, column sep=huge]
			U^j_\beta\big((\Gamma_1)_{<j} \mid \Lambda_1, B, \Lambda_2 \mid (\Gamma_1)_{>j}, A, \Gamma_2\big) \arrow[r, "{U^j_\beta(\1 \mid \1, g''_{c'_\omega}, \1 \mid \1)}"] \arrow[d, "{U^j_\beta(\1 \mid \1 \mid \1, f_{c_\alpha}, \1)}", swap] & U^j_\beta\big((\Gamma_1)_{<j} \mid \Lambda_1, \widehat U_\omega, \Lambda_2 \mid (\Gamma_1)_{>j}, A, \Gamma_2\big) \arrow[d, "{U^j_\beta(\1 \mid \1 \mid \1, f_{c_\alpha}, \1)}"] \\
			U^j_\beta\big((\Gamma_1)_{<j} \mid \Lambda_1, B, \Lambda_2 \mid (\Gamma_1)_{>j}, \widehat F_\alpha, \Gamma_2\big) \arrow[r, "{U^j_\beta(\1 \mid \1, g''_{c'_\omega}, \1 \mid \1)}", swap] & U^j_\beta\big((\Gamma_1)_{<j} \mid \Lambda_1, \widehat U_\omega, \Lambda_2 \mid (\Gamma_1)_{>j}, \widehat F_\alpha, \Gamma_2\big)
		\end{tikzcd}
		}
	\end{center}
	This brings the action of $v_a$ next to $\epsilon^U_j[\omega,\beta]$. We then cross $v_a$ past $\epsilon^U_j[\omega,\beta]$. Its $a$-slot is a contravariant input argument on both sides:
	\begin{center}
		\adjustbox{max width=\linewidth}{
		\begin{tikzcd}[row sep=large, column sep=huge]
			U^j_\beta\big((\Gamma_1)_{<j} \mid \Lambda_1, \widehat U_\omega, \Lambda_2 \mid (\Gamma_1)_{>j}, A, \Gamma_2\big) \arrow[r, "{\epsilon^U_j[\omega,\beta]}"] \arrow[d, "{U^j_\beta(\1 \mid \1 \mid \1, f_{c_\alpha}, \1)}", swap] & U^{|\theta_1| + j}_{\omega \circledcirc \beta}\big(\Theta_1, (\Gamma_1)_{<j} \mid \Lambda_1, \Xi, \Lambda_2 \mid (\Gamma_1)_{>j}, A, \Gamma_2, \Theta_2\big) \arrow[d, "{U^{|\theta_1| + j}_{\omega \circledcirc \beta}(\1 \mid \1 \mid \1, f_{c_\alpha}, \1)}"] \\
			U^j_\beta\big((\Gamma_1)_{<j} \mid \Lambda_1, \widehat U_\omega, \Lambda_2 \mid (\Gamma_1)_{>j}, \widehat F_\alpha, \Gamma_2\big) \arrow[r, "{\epsilon^U_j[\omega,\beta]}", swap] & U^{|\theta_1| + j}_{\omega \circledcirc \beta}\big(\Theta_1, (\Gamma_1)_{<j} \mid \Lambda_1, \Xi, \Lambda_2 \mid (\Gamma_1)_{>j}, \widehat F_\alpha, \Gamma_2, \Theta_2\big)
		\end{tikzcd}
		}
	\end{center}
	Recognising $g^{\omega\beta}_{|\theta_1| + j} = \epsilon^U_j[\omega,\beta] \circ U^j_\beta(\1 \mid \1, g''_{c'_\omega}, \1 \mid \1) \circ g'_j$, we obtain $\text{right}=\text{left}$.
	\item \textbf{($\gamma$-zone, $1 \leq j \leq |\gamma|$.)} Here
	\begin{align*}
&\text{left} \\
&\quad = \epsilon^U_j[\omega \circledcirc \beta,\alpha] \circ U^j_\alpha\big(\1 \mid \1, g^{\omega\beta}_{|\theta_1| + c'_\beta}, \1 \mid \1\big) \circ g_j, \\
&\quad \qquad \text{right} = \epsilon^U_{|\gamma_1| + j}[\omega,\beta \circledcirc \alpha] \circ U^{|\gamma_1| + j}_{\beta \circledcirc \alpha}\big(\1 \mid \1, g''_{c'_\omega}, \1 \mid \1\big) \circ g^{\beta\alpha}_{|\gamma_1| + j}.
\end{align*}
	We first substitute $g^{\beta\alpha}_{|\gamma_1| + j} = \epsilon^U_j[\beta,\alpha] \circ U^j_\alpha(\1 \mid \1, g'_{c'_\beta}, \1 \mid \1) \circ g_j$. We then cross $v_b$ past $\epsilon^U_j[\beta,\alpha]$. Its $b$-slot occurs inside $U^{c'_\beta}_\beta$ on the source side:
	\begin{center}
		\adjustbox{max width=\linewidth}{
		\begin{tikzcd}[row sep=large, column sep=huge]
			U^j_\alpha\big(\Gamma_{<j} \mid \Delta_1, U^{c'_\beta}_\beta(\Gamma_1 \mid \Lambda_1, B, \Lambda_2 \mid \Gamma_2), \Delta_2 \mid \Gamma_{>j}\big) \arrow[r, "{\epsilon^U_j[\beta,\alpha]}"] \arrow[d, "{U^j_\alpha(\1 \mid \1, U^{c'_\beta}_\beta(\1 \mid \1, g''_{c'_\omega}, \1 \mid \1), \1 \mid \1)}", swap] & U^{|\gamma_1| + j}_{\beta \circledcirc \alpha}\big(\Gamma_1, \Gamma_{<j} \mid \Delta_1, \Lambda_1, B, \Lambda_2, \Delta_2 \mid \Gamma_{>j}, \Gamma_2\big) \arrow[d, "{U^{|\gamma_1| + j}_{\beta \circledcirc \alpha}(\1 \mid \1, g''_{c'_\omega}, \1 \mid \1)}"] \\
			U^j_\alpha\big(\Gamma_{<j} \mid \Delta_1, U^{c'_\beta}_\beta(\Gamma_1 \mid \Lambda_1, \widehat U_\omega, \Lambda_2 \mid \Gamma_2), \Delta_2 \mid \Gamma_{>j}\big) \arrow[r, "{\epsilon^U_j[\beta,\alpha]}", swap] & U^{|\gamma_1| + j}_{\beta \circledcirc \alpha}\big(\Gamma_1, \Gamma_{<j} \mid \Delta_1, \Lambda_1, \widehat U_\omega, \Lambda_2, \Delta_2 \mid \Gamma_{>j}, \Gamma_2\big)
		\end{tikzcd}
		}
	\end{center}
	The covariant pentagon ($\gamma$-zone) gives $\epsilon^U_{|\gamma_1| + j}[\omega,\beta \circledcirc \alpha] \circ \epsilon^U_j[\beta,\alpha] = \epsilon^U_j[\omega \circledcirc \beta,\alpha] \circ U^j_\alpha(\1 \mid \1, \epsilon^U_{c'_\beta}[\omega,\beta], \1 \mid \1)$. We then merge the three actions in the covariant slot:
	\begin{center}
		\adjustbox{max width=\linewidth}{
		\begin{tikzcd}[row sep=large]
			U^j_\alpha\big(\Gamma_{<j} \mid \Delta_1, A, \Delta_2 \mid \Gamma_{>j}\big) \arrow[d, "{U^j_\alpha(\1 \mid \1, g'_{c'_\beta}, \1 \mid \1)}"] \arrow[ddd, bend right=45, "{U^j_\alpha(\1 \mid \1, g^{\omega\beta}_{|\theta_1| + c'_\beta}, \1 \mid \1)}"'] \\
			U^j_\alpha\big(\Gamma_{<j} \mid \Delta_1, U^{c'_\beta}_\beta(\Gamma_1 \mid \Lambda_1, B, \Lambda_2 \mid \Gamma_2), \Delta_2 \mid \Gamma_{>j}\big) \arrow[d, "{U^j_\alpha(\1 \mid \1, U^{c'_\beta}_\beta(\1 \mid \1, g''_{c'_\omega}, \1 \mid \1), \1 \mid \1)}"] \\
			U^j_\alpha\big(\Gamma_{<j} \mid \Delta_1, U^{c'_\beta}_\beta(\Gamma_1 \mid \Lambda_1, \widehat U_\omega, \Lambda_2 \mid \Gamma_2), \Delta_2 \mid \Gamma_{>j}\big) \arrow[d, "{U^j_\alpha(\1 \mid \1, \epsilon^U_{c'_\beta}[\omega,\beta], \1 \mid \1)}"] \\
			U^j_\alpha\big(\Gamma_{<j} \mid \Delta_1, U^{|\theta_1| + c'_\beta}_{\omega \circledcirc \beta}(\Theta_1, \Gamma_1 \mid \Lambda_1, \Xi, \Lambda_2 \mid \Gamma_2, \Theta_2), \Delta_2 \mid \Gamma_{>j}\big)
		\end{tikzcd}
		}
	\end{center}
	Thus,
	$$\text{right} = \epsilon^U_j[\omega \circledcirc \beta,\alpha] \circ U^j_\alpha\big(\1 \mid \1,\ \epsilon^U_{c'_\beta}[\omega,\beta] \circ U^{c'_\beta}_\beta(\1 \mid \1, g''_{c'_\omega}, \1 \mid \1) \circ g'_{c'_\beta},\ \1 \mid \1\big) \circ g_j = \text{left}.$$
	\item \textbf{($\gamma_2$-zone, $1 \leq j \leq |\gamma_2|$.)} Here, at position $|\theta_1| + |\gamma_1| + |\gamma| + j$,
	\begin{align*}
&\text{left} \\
&\quad = \eta^U_{\to j}[\omega \circledcirc \beta,\alpha] \circ U^{|\theta_1| + c'_\beta + j}_{\omega \circledcirc \beta}\big(\1, f_{c_\alpha}, \1 \mid \1 \mid \1\big) \circ g^{\omega\beta}_{|\theta_1| + c'_\beta + j}, \\
&\quad \qquad \text{right} = \epsilon^U_{|\gamma_1| + |\gamma| + j}[\omega,\beta \circledcirc \alpha] \circ U^{|\gamma_1| + |\gamma| + j}_{\beta \circledcirc \alpha}\big(\1 \mid \1, g''_{c'_\omega}, \1 \mid \1\big) \circ g^{\beta\alpha}_{|\gamma_1| + |\gamma| + j}.
\end{align*}
	We first substitute $g^{\beta\alpha}_{|\gamma_1| + |\gamma| + j} = \eta^U_{\to j}[\beta,\alpha] \circ U^{c'_\beta + j}_\beta(\1, f_{c_\alpha}, \1 \mid \1 \mid \1) \circ g'_{c'_\beta + j}$. We then cross $v_b$ past $\eta^U_{\to j}[\beta,\alpha]$:
	\begin{center}
		\adjustbox{max width=\linewidth}{
		\begin{tikzcd}[row sep=large, column sep=huge]
			U^{c'_\beta + j}_\beta\big(\Gamma_1, \widehat F_\alpha, (\Gamma_2)_{<j} \mid \Lambda_1, B, \Lambda_2 \mid (\Gamma_2)_{>j}\big) \arrow[r, "{\eta^U_{\to j}[\beta,\alpha]}"] \arrow[d, "{U^{c'_\beta + j}_\beta(\1 \mid \1, g''_{c'_\omega}, \1 \mid \1)}", swap] & U^{|\gamma_1| + |\gamma| + j}_{\beta \circledcirc \alpha}\big(\Gamma_1, \Gamma, (\Gamma_2)_{<j} \mid \Delta_1, \Lambda_1, B, \Lambda_2, \Delta_2 \mid (\Gamma_2)_{>j}\big) \arrow[d, "{U^{|\gamma_1| + |\gamma| + j}_{\beta \circledcirc \alpha}(\1 \mid \1, g''_{c'_\omega}, \1 \mid \1)}"] \\
			U^{c'_\beta + j}_\beta\big(\Gamma_1, \widehat F_\alpha, (\Gamma_2)_{<j} \mid \Lambda_1, \widehat U_\omega, \Lambda_2 \mid (\Gamma_2)_{>j}\big) \arrow[r, "{\eta^U_{\to j}[\beta,\alpha]}", swap] & U^{|\gamma_1| + |\gamma| + j}_{\beta \circledcirc \alpha}\big(\Gamma_1, \Gamma, (\Gamma_2)_{<j} \mid \Delta_1, \Lambda_1, \widehat U_\omega, \Lambda_2, \Delta_2 \mid (\Gamma_2)_{>j}\big)
		\end{tikzcd}
		}
	\end{center}
	The mixed square ($\gamma_2$-zone) gives $\epsilon^U_{|\gamma_1| + |\gamma| + j}[\omega,\beta \circledcirc \alpha] \circ \eta^U_{\to j}[\beta,\alpha] = \eta^U_{\to j}[\omega \circledcirc \beta,\alpha] \circ \epsilon^U_{c'_\beta + j}[\omega,\beta]$. The actions of $v_a$ and $v_b$ occupy disjoint slots of $U^{c'_\beta + j}_\beta$ and hence commute:
	\begin{center}
		\adjustbox{max width=\linewidth}{
		\begin{tikzcd}[row sep=large, column sep=huge]
			U^{c'_\beta + j}_\beta\big(\Gamma_1, A, (\Gamma_2)_{<j} \mid \Lambda_1, B, \Lambda_2 \mid (\Gamma_2)_{>j}\big) \arrow[r, "{U^{c'_\beta + j}_\beta(\1 \mid \1, g''_{c'_\omega}, \1 \mid \1)}"] \arrow[d, "{U^{c'_\beta + j}_\beta(\1, f_{c_\alpha}, \1 \mid \1 \mid \1)}", swap] & U^{c'_\beta + j}_\beta\big(\Gamma_1, A, (\Gamma_2)_{<j} \mid \Lambda_1, \widehat U_\omega, \Lambda_2 \mid (\Gamma_2)_{>j}\big) \arrow[d, "{U^{c'_\beta + j}_\beta(\1, f_{c_\alpha}, \1 \mid \1 \mid \1)}"] \\
			U^{c'_\beta + j}_\beta\big(\Gamma_1, \widehat F_\alpha, (\Gamma_2)_{<j} \mid \Lambda_1, B, \Lambda_2 \mid (\Gamma_2)_{>j}\big) \arrow[r, "{U^{c'_\beta + j}_\beta(\1 \mid \1, g''_{c'_\omega}, \1 \mid \1)}", swap] & U^{c'_\beta + j}_\beta\big(\Gamma_1, \widehat F_\alpha, (\Gamma_2)_{<j} \mid \Lambda_1, \widehat U_\omega, \Lambda_2 \mid (\Gamma_2)_{>j}\big)
		\end{tikzcd}
		}
	\end{center}
	We then cross $v_a$ past $\epsilon^U_{c'_\beta + j}[\omega,\beta]$:
	\begin{center}
		\adjustbox{max width=\linewidth}{
		\begin{tikzcd}[row sep=large, column sep=huge]
			U^{c'_\beta + j}_\beta\big(\Gamma_1, A, (\Gamma_2)_{<j} \mid \Lambda_1, \widehat U_\omega, \Lambda_2 \mid (\Gamma_2)_{>j}\big) \arrow[r, "{\epsilon^U_{c'_\beta + j}[\omega,\beta]}"] \arrow[d, "{U^{c'_\beta + j}_\beta(\1, f_{c_\alpha}, \1 \mid \1 \mid \1)}", swap] & U^{|\theta_1| + c'_\beta + j}_{\omega \circledcirc \beta}\big(\Theta_1, \Gamma_1, A, (\Gamma_2)_{<j} \mid \Lambda_1, \Xi, \Lambda_2 \mid (\Gamma_2)_{>j}, \Theta_2\big) \arrow[d, "{U^{|\theta_1| + c'_\beta + j}_{\omega \circledcirc \beta}(\1, f_{c_\alpha}, \1 \mid \1 \mid \1)}"] \\
			U^{c'_\beta + j}_\beta\big(\Gamma_1, \widehat F_\alpha, (\Gamma_2)_{<j} \mid \Lambda_1, \widehat U_\omega, \Lambda_2 \mid (\Gamma_2)_{>j}\big) \arrow[r, "{\epsilon^U_{c'_\beta + j}[\omega,\beta]}", swap] & U^{|\theta_1| + c'_\beta + j}_{\omega \circledcirc \beta}\big(\Theta_1, \Gamma_1, \widehat F_\alpha, (\Gamma_2)_{<j} \mid \Lambda_1, \Xi, \Lambda_2 \mid (\Gamma_2)_{>j}, \Theta_2\big)
		\end{tikzcd}
		}
	\end{center}
	After this crossing, we obtain
	\[
\begin{aligned}
&\text{right} \\
&\quad  = \eta^U_{\to j}[\omega \circledcirc \beta,\alpha] \circ U^{|\theta_1| + c'_\beta + j}_{\omega \circledcirc \beta}\big(\1, f_{c_\alpha}, \1 \mid \1 \mid \1\big) \circ \epsilon^U_{c'_\beta + j}[\omega,\beta] \\
&\quad \circ U^{c'_\beta + j}_\beta\big(\1 \mid \1, g''_{c'_\omega}, \1 \mid \1\big) \circ g'_{c'_\beta + j} \\
&\quad = \text{left},
\end{aligned}
\]
	where the composite of the last three factors is $g^{\omega\beta}_{|\theta_1| + c'_\beta + j}$.
	\item \textbf{($\theta_2$-zone, $1 \leq j \leq |\theta_2|$.)} Here, at position $|\theta_1| + |\gamma_1| + |\gamma| + |\gamma_2| + j$,
	\begin{align*}
&\text{left} \\
&\quad  = \eta^U_{\to (|\gamma_2| + j)}[\omega \circledcirc \beta,\alpha] \circ U^{|\theta_1| + c'_\beta + |\gamma_2| + j}_{\omega \circledcirc \beta}\big(\1, f_{c_\alpha}, \1 \mid \1 \mid \1\big) \\
&\quad \circ g^{\omega\beta}_{|\theta_1| + c'_\beta + |\gamma_2| + j}, \\
&\quad \qquad \text{right} = \eta^U_{\to j}[\omega,\beta \circledcirc \alpha] \circ U^{c'_\omega + j}_\omega\big(\1, f^{\beta\alpha}_{|\delta_1| + c_\beta}, \1 \mid \1 \mid \1\big) \circ g''_{c'_\omega + j}.
\end{align*}
	As in the $\theta_1$-zone, the contravariant slot splits the middle action of the right-hand side into the following three factors:
	\begin{center}
		\adjustbox{max width=\linewidth}{
		\begin{tikzcd}[row sep=large]
			U^{c'_\omega + j}_\omega\big(\Theta_1, B, (\Theta_2)_{<j} \mid \Xi \mid (\Theta_2)_{>j}\big) \arrow[d, "{U^{c'_\omega + j}_\omega(\1, f'_{c_\beta}, \1 \mid \1 \mid \1)}"] \arrow[ddd, bend right=45, "{U^{c'_\omega + j}_\omega(\1, f^{\beta\alpha}_{|\delta_1| + c_\beta}, \1 \mid \1 \mid \1)}"'] \\
			U^{c'_\omega + j}_\omega\big(\Theta_1, F^{c_\beta}_\beta(\Lambda_1 \mid \Gamma_1, A, \Gamma_2 \mid \Lambda_2), (\Theta_2)_{<j} \mid \Xi \mid (\Theta_2)_{>j}\big) \arrow[d, "{U^{c'_\omega + j}_\omega(\1, F^{c_\beta}_\beta(\1 \mid \1, f_{c_\alpha}, \1 \mid \1), \1 \mid \1 \mid \1)}"] \\
			U^{c'_\omega + j}_\omega\big(\Theta_1, F^{c_\beta}_\beta(\Lambda_1 \mid \Gamma_1, \widehat F_\alpha, \Gamma_2 \mid \Lambda_2), (\Theta_2)_{<j} \mid \Xi \mid (\Theta_2)_{>j}\big) \arrow[d, "{U^{c'_\omega + j}_\omega(\1, \epsilon^F_{c_\beta}[\beta,\alpha], \1 \mid \1 \mid \1)}"] \\
			U^{c'_\omega + j}_\omega\big(\Theta_1, F^{|\delta_1| + c_\beta}_{\beta \circledcirc \alpha}(\Delta_1, \Lambda_1 \mid \Gamma_1, \Gamma, \Gamma_2 \mid \Lambda_2, \Delta_2), (\Theta_2)_{<j} \mid \Xi \mid (\Theta_2)_{>j}\big)
		\end{tikzcd}
		}
	\end{center}
    The contravariant pentagon ($\theta_2$-zone) gives
\[
\begin{aligned}
&\eta^U_{\to (|\gamma_2| + j)}[\omega \circledcirc \beta,\alpha] \circ \eta^U_{\to j}[\omega,\beta]
\\
&\quad = \eta^U_{\to j}[\omega,\beta \circledcirc \alpha] \circ U^{c'_\omega + j}_\omega(\1, \epsilon^F_{c_\beta}[\beta,\alpha], \1 \mid \1 \mid \1).
\end{aligned}
\]
This identity absorbs the $\epsilon^F_{c_\beta}[\beta,\alpha]$-factor. We then cross $v_a$ past the inner $\eta^U_{\to j}[\omega,\beta]$:
	\begin{center}
		\adjustbox{max width=\linewidth}{
		\begin{tikzcd}[row sep=large, column sep=huge]
			U^{c'_\omega + j}_\omega\big(\Theta_1, F^{c_\beta}_\beta(\Lambda_1 \mid \Gamma_1, A, \Gamma_2 \mid \Lambda_2), (\Theta_2)_{<j} \mid \Xi \mid (\Theta_2)_{>j}\big) \arrow[r, "{\eta^U_{\to j}[\omega,\beta]}"] \arrow[d, "{U^{c'_\omega + j}_\omega(\1, F^{c_\beta}_\beta(\1 \mid \1, f_{c_\alpha}, \1 \mid \1), \1 \mid \1 \mid \1)}", swap] & U^{|\theta_1| + c'_\beta + |\gamma_2| + j}_{\omega \circledcirc \beta}\big(\Theta_1, \Gamma_1, A, \Gamma_2, (\Theta_2)_{<j} \mid \Lambda_1, \Xi, \Lambda_2 \mid (\Theta_2)_{>j}\big) \arrow[d, "{U^{|\theta_1| + c'_\beta + |\gamma_2| + j}_{\omega \circledcirc \beta}(\1, f_{c_\alpha}, \1 \mid \1 \mid \1)}"] \\
			U^{c'_\omega + j}_\omega\big(\Theta_1, F^{c_\beta}_\beta(\Lambda_1 \mid \Gamma_1, \widehat F_\alpha, \Gamma_2 \mid \Lambda_2), (\Theta_2)_{<j} \mid \Xi \mid (\Theta_2)_{>j}\big) \arrow[r, "{\eta^U_{\to j}[\omega,\beta]}", swap] & U^{|\theta_1| + c'_\beta + |\gamma_2| + j}_{\omega \circledcirc \beta}\big(\Theta_1, \Gamma_1, \widehat F_\alpha, \Gamma_2, (\Theta_2)_{<j} \mid \Lambda_1, \Xi, \Lambda_2 \mid (\Theta_2)_{>j}\big)
		\end{tikzcd}
		}
	\end{center}
	This leaves
	\[
\begin{aligned}
&\text{right} \\
&\quad  = \eta^U_{\to (|\gamma_2| + j)}[\omega \circledcirc \beta,\alpha] \circ U^{|\theta_1| + c'_\beta + |\gamma_2| + j}_{\omega \circledcirc \beta}\big(\1, f_{c_\alpha}, \1 \mid \1 \mid \1\big) \circ \eta^U_{\to j}[\omega,\beta] \\
&\quad \circ U^{c'_\omega + j}_\omega\big(\1, f'_{c_\beta}, \1 \mid \1 \mid \1\big) \circ g''_{c'_\omega + j} \\
&\quad = \text{left}.
\end{aligned}
\]
\end{itemize}
}
This finishes the proof of associativity.

\textbf{Interchange laws.} We prove the two laws in the planar configurations of \cref{lem:poly-circledcirc-interchange}.

\textit{Output interchange.} Consider polymorphisms
\[
\begin{gathered}
f_p : \Gamma_p \to (\Delta_1)_p, A_p, (\Delta_2)_p, B_p, (\Delta_3)_p \text{ over } \alpha, \\
g_p : \Gamma'_p, A_p \to \Delta'_p \text{ over } \beta, \\
h_p : B_p, \Gamma''_p \to \Delta''_p \text{ over } \omega,
\end{gathered}
\]
with components $(f_i;g_j)$, $(f'_i;g'_j)$ and $(f''_i;g''_j)$. Write
$$c_a:=|\delta_1|+1, \qquad c_b:=|\delta_1|+|\delta_2|+2,$$
and assume $\Delta_1=()$ or $\Gamma'=()$, and $\Delta_3=()$ or $\Gamma''=()$. We must show that
$$h_p\circ_{B_p}(g_p\circ_{A_p}f_p)=g_p\circ_{A_p}(h_p\circ_{B_p}f_p).$$
Note that both sides lie over
$\chi:=\omega\circledcirc_b(\beta\circledcirc_a\alpha)=\beta\circledcirc_a(\omega\circledcirc_b\alpha).$ We adopt the shorthands 
$\widehat U':=U^{|\gamma'|+1}_\beta(\Gamma'\mid\Delta'\mid\,)$ and $\widehat U'':=U^1_\omega(\,\mid\Delta''\mid\Gamma''),$
and we write 
$$p_a:=f_{c_a}, \qquad q_a:=g'_{|\gamma'|+1}, \qquad p_b:=f_{c_b}, \qquad q_b:=g''_1.$$
Thus $q_a:A\to\widehat U'$ and $q_b:B\to\widehat U''$. Let 
$$\bar p_a:=p_a\circ F^{c_a}_\alpha(q_b):F^{c_a}_\alpha(\Delta_1\mid\Gamma\mid\Delta_2,\widehat U'',\Delta_3)\to A,$$
$$\bar p_b:=p_b\circ F^{c_b}_\alpha(q_a):F^{c_b}_\alpha(\Delta_1,\widehat U',\Delta_2\mid\Gamma\mid\Delta_3)\to B.$$
The two components used by the second cuts are
\begin{align*}
	(h_p\circ_{B_p}f_p)_{c_a}&=\bar p_a\circ\eta^F_{c_a\to}[\omega,\alpha], \\
	(g_p\circ_{A_p}f_p)_{|\delta_1|+|\delta'|+|\delta_2|+1}&=\bar p_b\circ\eta^F_{\to(|\delta_2|+1)}[\beta,\alpha].
\end{align*}
We now compute the two cuts at every position. 

{
\begin{itemize}
	\item ($\delta_1$-zone, $1\leq i\leq|\delta_1|$.) The left-hand component follows the upper and right sides of the square below. The right-hand component follows the left and lower sides. The last two maps are the same.
	\begin{center}
		\adjustbox{max width=\linewidth}{
		\begin{tikzcd}[row sep=huge, column sep=large]
			F^i_\chi\big((\Delta_1)_{<i}\mid\Gamma',\Gamma,\Gamma''\mid(\Delta_1)_{>i},\Delta',\Delta_2,\Delta'',\Delta_3\big) \arrow[r, "{\eta^F_{i\to}[\omega,\beta\circledcirc_a\alpha]}"] \arrow[d, "{\eta^F_{i\to}[\beta,\omega\circledcirc_b\alpha]}", swap] & F^i_{\beta\circledcirc_a\alpha}\big((\Delta_1)_{<i}\mid\Gamma',\Gamma\mid(\Delta_1)_{>i},\Delta',\Delta_2,\widehat U'',\Delta_3\big) \arrow[d, "{\eta^F_{i\to}[\beta,\alpha]}"] \\
			F^i_{\omega\circledcirc_b\alpha}\big((\Delta_1)_{<i}\mid\Gamma,\Gamma''\mid(\Delta_1)_{>i},\widehat U',\Delta_2,\Delta'',\Delta_3\big) \arrow[r, "{\eta^F_{i\to}[\omega,\alpha]}", swap] & F^i_\alpha\big((\Delta_1)_{<i}\mid\Gamma\mid(\Delta_1)_{>i},\widehat U',\Delta_2,\widehat U'',\Delta_3\big) \arrow[r, "{F^i_\alpha(q_a,q_b)}"] & F^i_\alpha\big((\Delta_1)_{<i}\mid\Gamma\mid(\Delta_1)_{>i},A,\Delta_2,B,\Delta_3\big) \arrow[r, "f_i"] & (\Delta_1)_i
		\end{tikzcd}
		}
	\end{center}
	The square commutes by Disjoint Invariance in the $\delta_1$-zone. Hence the two components are equal.

	\item ($\delta'$-zone, $1\leq i\leq|\delta'|$.) Naturality at $q_b$ and then at $p_a$ gives the following diagram. The left-hand component follows the upper and right sides. The right-hand component follows the left and lower sides.
	\begin{center}
		\adjustbox{max width=\linewidth}{
		\begin{tikzcd}[row sep=huge, column sep=large]
			F^{|\delta_1|+i}_\chi\big(\Delta_1,(\Delta')_{<i}\mid\Gamma',\Gamma,\Gamma''\mid(\Delta')_{>i},\Delta_2,\Delta'',\Delta_3\big) \arrow[r, "{\eta^F_{(|\delta_1|+i)\to}[\omega,\beta\circledcirc_a\alpha]}"] \arrow[d, "{\epsilon^F_i[\beta,\omega\circledcirc_b\alpha]}", swap] & F^{|\delta_1|+i}_{\beta\circledcirc_a\alpha}\big(\Delta_1,(\Delta')_{<i}\mid\Gamma',\Gamma\mid(\Delta')_{>i},\Delta_2,\widehat U'',\Delta_3\big) \arrow[d, "{\epsilon^F_i[\beta,\alpha]}"] \\
			F^i_\beta\big((\Delta')_{<i}\mid\Gamma',F^{c_a}_{\omega\circledcirc_b\alpha}(\Delta_1\mid\Gamma,\Gamma''\mid\Delta_2,\Delta'',\Delta_3)\mid(\Delta')_{>i}\big) \arrow[r, "{F^i_\beta(\eta^F_{c_a\to}[\omega,\alpha])}", swap] & F^i_\beta\big((\Delta')_{<i}\mid\Gamma',F^{c_a}_\alpha(\Delta_1\mid\Gamma\mid\Delta_2,\widehat U'',\Delta_3)\mid(\Delta')_{>i}\big) \arrow[r, "{F^i_\beta(\bar p_a)}"] & F^i_\beta\big((\Delta')_{<i}\mid\Gamma',A\mid(\Delta')_{>i}\big) \arrow[r, "f'_i"] & (\Delta')_i
		\end{tikzcd}
		}
	\end{center}
	The square is Disjoint Invariance in the $\delta'$-zone. Thus the components agree.

	\item ($\delta_2$-zone, $1\leq i\leq|\delta_2|$.) Naturality in the $a$- and $b$-slots gives
	\begin{center}
		\adjustbox{max width=\linewidth}{
		\begin{tikzcd}[row sep=huge, column sep=large]
			F^{|\delta_1|+|\delta'|+i}_\chi\big(\Delta_1,\Delta',(\Delta_2)_{<i}\mid\Gamma',\Gamma,\Gamma''\mid(\Delta_2)_{>i},\Delta'',\Delta_3\big) \arrow[r, "{\eta^F_{(|\delta_1|+|\delta'|+i)\to}[\omega,\beta\circledcirc_a\alpha]}"] \arrow[d, "{\eta^F_{\to i}[\beta,\omega\circledcirc_b\alpha]}", swap] & F^{|\delta_1|+|\delta'|+i}_{\beta\circledcirc_a\alpha}\big(\Delta_1,\Delta',(\Delta_2)_{<i}\mid\Gamma',\Gamma\mid(\Delta_2)_{>i},\widehat U'',\Delta_3\big) \arrow[d, "{\eta^F_{\to i}[\beta,\alpha]}"] \\
			F^{|\delta_1|+1+i}_{\omega\circledcirc_b\alpha}\big(\Delta_1,\widehat U',(\Delta_2)_{<i}\mid\Gamma,\Gamma''\mid(\Delta_2)_{>i},\Delta'',\Delta_3\big) \arrow[r, "{\eta^F_{(|\delta_1|+1+i)\to}[\omega,\alpha]}", swap] & F^{c_a+i}_\alpha\big(\Delta_1,\widehat U',(\Delta_2)_{<i}\mid\Gamma\mid(\Delta_2)_{>i},\widehat U'',\Delta_3\big) \arrow[r, "{F^{c_a+i}_\alpha(q_a,q_b)}"] & F^{c_a+i}_\alpha\big(\Delta_1,A,(\Delta_2)_{<i}\mid\Gamma\mid(\Delta_2)_{>i},B,\Delta_3\big) \arrow[r, "f_{c_a+i}"] & (\Delta_2)_i
		\end{tikzcd}
		}
	\end{center}
	The left-hand component follows the upper and right sides. The right-hand component follows the left and lower sides. The square is the $\delta_2$-zone axiom, so they are equal.

	\item ($\delta''$-zone, $1\leq i\leq|\delta''|$.) Naturality at $q_a$ and then at $p_b$ gives
	\begin{center}
		\adjustbox{max width=\linewidth}{
		\begin{tikzcd}[row sep=huge, column sep=large]
			F^{|\delta_1|+|\delta'|+|\delta_2|+i}_\chi\big(\Delta_1,\Delta',\Delta_2,(\Delta'')_{<i}\mid\Gamma',\Gamma,\Gamma''\mid(\Delta'')_{>i},\Delta_3\big) \arrow[r, "{\epsilon^F_i[\omega,\beta\circledcirc_a\alpha]}"] \arrow[d, "{\eta^F_{\to(|\delta_2|+i)}[\beta,\omega\circledcirc_b\alpha]}", swap] & F^i_\omega\big((\Delta'')_{<i}\mid F^{|\delta_1|+|\delta'|+|\delta_2|+1}_{\beta\circledcirc_a\alpha}(\Delta_1,\Delta',\Delta_2\mid\Gamma',\Gamma\mid\Delta_3),\Gamma''\mid(\Delta'')_{>i}\big) \arrow[d, "{F^i_\omega(\eta^F_{\to(|\delta_2|+1)}[\beta,\alpha])}"] \\
			F^{|\delta_1|+1+|\delta_2|+i}_{\omega\circledcirc_b\alpha}\big(\Delta_1,\widehat U',\Delta_2,(\Delta'')_{<i}\mid\Gamma,\Gamma''\mid(\Delta'')_{>i},\Delta_3\big) \arrow[r, "{\epsilon^F_i[\omega,\alpha]}", swap] & F^i_\omega\big((\Delta'')_{<i}\mid F^{c_b}_\alpha(\Delta_1,\widehat U',\Delta_2\mid\Gamma\mid\Delta_3),\Gamma''\mid(\Delta'')_{>i}\big) \arrow[r, "{F^i_\omega(\bar p_b)}"] & F^i_\omega\big((\Delta'')_{<i}\mid B,\Gamma''\mid(\Delta'')_{>i}\big) \arrow[r, "f''_i"] & (\Delta'')_i
		\end{tikzcd}
		}
	\end{center}
	The left-hand component follows the upper and right sides. The right-hand component follows the left and lower sides. Disjoint Invariance in the $\delta''$-zone gives equality.

	\item ($\delta_3$-zone, $1\leq i\leq|\delta_3|$.) Naturality in both marked output slots gives
	\begin{center}
		\adjustbox{max width=\linewidth}{
		\begin{tikzcd}[row sep=huge, column sep=large]
			F^{|\delta_1|+|\delta'|+|\delta_2|+|\delta''|+i}_\chi\big(\Delta_1,\Delta',\Delta_2,\Delta'',(\Delta_3)_{<i}\mid\Gamma',\Gamma,\Gamma''\mid(\Delta_3)_{>i}\big) \arrow[r, "{\eta^F_{\to i}[\omega,\beta\circledcirc_a\alpha]}"] \arrow[d, "{\eta^F_{\to(|\delta_2|+|\delta''|+i)}[\beta,\omega\circledcirc_b\alpha]}", swap] & F^{|\delta_1|+|\delta'|+|\delta_2|+1+i}_{\beta\circledcirc_a\alpha}\big(\Delta_1,\Delta',\Delta_2,\widehat U'',(\Delta_3)_{<i}\mid\Gamma',\Gamma\mid(\Delta_3)_{>i}\big) \arrow[d, "{\eta^F_{\to(|\delta_2|+1+i)}[\beta,\alpha]}"] \\
			F^{|\delta_1|+1+|\delta_2|+|\delta''|+i}_{\omega\circledcirc_b\alpha}\big(\Delta_1,\widehat U',\Delta_2,\Delta'',(\Delta_3)_{<i}\mid\Gamma,\Gamma''\mid(\Delta_3)_{>i}\big) \arrow[r, "{\eta^F_{\to i}[\omega,\alpha]}", swap] & F^{c_b+i}_\alpha\big(\Delta_1,\widehat U',\Delta_2,\widehat U'',(\Delta_3)_{<i}\mid\Gamma\mid(\Delta_3)_{>i}\big) \arrow[r, "{F^{c_b+i}_\alpha(q_a,q_b)}"] & F^{c_b+i}_\alpha\big(\Delta_1,A,\Delta_2,B,(\Delta_3)_{<i}\mid\Gamma\mid(\Delta_3)_{>i}\big) \arrow[r, "f_{c_b+i}"] & (\Delta_3)_i
		\end{tikzcd}
		}
	\end{center}
	The two cut orders give the two routes through the $\delta_3$-zone square. They therefore agree.

	\item ($\gamma'$-zone, $1\leq j\leq|\gamma'|$.) Naturality at $q_b$ and $p_a$ gives
	\begin{center}
		\adjustbox{max width=\linewidth}{
		\begin{tikzcd}[row sep=huge, column sep=large]
			(\Gamma')_j \arrow[r, "g'_j"] & U^j_\beta\big((\Gamma')_{<j}\mid\Delta'\mid(\Gamma')_{>j},A\big) \arrow[r, "{U^j_\beta(\bar p_a)}"] & U^j_\beta\big((\Gamma')_{<j}\mid\Delta'\mid(\Gamma')_{>j},F^{c_a}_\alpha(\Delta_1\mid\Gamma\mid\Delta_2,\widehat U'',\Delta_3)\big) \arrow[r, "{\eta^U_{j\to}[\beta,\alpha]}"] \arrow[d, "{U^j_\beta(\eta^F_{c_a\to}[\omega,\alpha])}", swap] & U^j_{\beta\circledcirc_a\alpha}\big((\Gamma')_{<j}\mid\Delta_1,\Delta',\Delta_2,\widehat U'',\Delta_3\mid(\Gamma')_{>j},\Gamma\big) \arrow[d, "{\epsilon^U_j[\omega,\beta\circledcirc_a\alpha]}"] \\
			& & U^j_\beta\big((\Gamma')_{<j}\mid\Delta'\mid(\Gamma')_{>j},F^{c_a}_{\omega\circledcirc_b\alpha}(\Delta_1\mid\Gamma,\Gamma''\mid\Delta_2,\Delta'',\Delta_3)\big) \arrow[r, "{\eta^U_{j\to}[\beta,\omega\circledcirc_b\alpha]}", swap] & U^j_\chi\big((\Gamma')_{<j}\mid\Delta_1,\Delta',\Delta_2,\Delta'',\Delta_3\mid(\Gamma')_{>j},\Gamma,\Gamma''\big)
		\end{tikzcd}
		}
	\end{center}
	The left-hand component follows the upper and right sides, while the right-hand component follows the left and lower sides. The $\gamma'$-zone square therefore gives equality.
	\item ($\gamma$-zone, $1\leq j\leq|\gamma|$.) Naturality in the $a$- and $b$-slots gives
	\begin{center}
		\adjustbox{max width=\linewidth}{
		\begin{tikzcd}[row sep=huge, column sep=large]
			\Gamma_j \arrow[r, "g_j"] & U^j_\alpha\big(\Gamma_{<j}\mid\Delta_1,A,\Delta_2,B,\Delta_3\mid\Gamma_{>j}\big) \arrow[r, "{U^j_\alpha(q_a,q_b)}"] & U^j_\alpha\big(\Gamma_{<j}\mid\Delta_1,\widehat U',\Delta_2,\widehat U'',\Delta_3\mid\Gamma_{>j}\big) \arrow[r, "{\epsilon^U_j[\beta,\alpha]}"] \arrow[d, "{\epsilon^U_j[\omega,\alpha]}", swap] & U^{|\gamma'|+j}_{\beta\circledcirc_a\alpha}\big(\Gamma',\Gamma_{<j}\mid\Delta_1,\Delta',\Delta_2,\widehat U'',\Delta_3\mid\Gamma_{>j}\big) \arrow[d, "{\epsilon^U_{|\gamma'|+j}[\omega,\beta\circledcirc_a\alpha]}"] \\
			& & U^j_{\omega\circledcirc_b\alpha}\big(\Gamma_{<j}\mid\Delta_1,\widehat U',\Delta_2,\Delta'',\Delta_3\mid\Gamma_{>j},\Gamma''\big) \arrow[r, "{\epsilon^U_j[\beta,\omega\circledcirc_b\alpha]}", swap] & U^{|\gamma'|+j}_\chi\big(\Gamma',\Gamma_{<j}\mid\Delta_1,\Delta',\Delta_2,\Delta'',\Delta_3\mid\Gamma_{>j},\Gamma''\big)
		\end{tikzcd}
		}
	\end{center}
	The two components follow the two routes through the $\gamma$-zone square, in the same order as above. They are equal.

	\item ($\gamma''$-zone, $1\leq j\leq|\gamma''|$.) Naturality at $q_a$ and $p_b$ gives
	\begin{center}
		\adjustbox{max width=\linewidth}{
		\begin{tikzcd}[row sep=huge, column sep=large]
			(\Gamma'')_j \arrow[r, "g''_{1+j}"] & U^{1+j}_\omega\big(B,(\Gamma'')_{<j}\mid\Delta''\mid(\Gamma'')_{>j}\big) \arrow[r, "{U^{1+j}_\omega(\bar p_b)}"] & U^{1+j}_\omega\big(F^{c_b}_\alpha(\Delta_1,\widehat U',\Delta_2\mid\Gamma\mid\Delta_3),(\Gamma'')_{<j}\mid\Delta''\mid(\Gamma'')_{>j}\big) \arrow[r, "{\eta^U_{\to j}[\omega,\alpha]}"] \arrow[d, "{U^{1+j}_\omega(\eta^F_{\to(|\delta_2|+1)}[\beta,\alpha])}", swap] & U^{|\gamma|+j}_{\omega\circledcirc_b\alpha}\big(\Gamma,(\Gamma'')_{<j}\mid\Delta_1,\widehat U',\Delta_2,\Delta'',\Delta_3\mid(\Gamma'')_{>j}\big) \arrow[d, "{\epsilon^U_{|\gamma|+j}[\beta,\omega\circledcirc_b\alpha]}"] \\
			& & U^{1+j}_\omega\big(F^{|\delta_1|+|\delta'|+|\delta_2|+1}_{\beta\circledcirc_a\alpha}(\Delta_1,\Delta',\Delta_2\mid\Gamma',\Gamma\mid\Delta_3),(\Gamma'')_{<j}\mid\Delta''\mid(\Gamma'')_{>j}\big) \arrow[r, "{\eta^U_{\to j}[\omega,\beta\circledcirc_a\alpha]}", swap] & U^{|\gamma'|+|\gamma|+j}_\chi\big(\Gamma',\Gamma,(\Gamma'')_{<j}\mid\Delta_1,\Delta',\Delta_2,\Delta'',\Delta_3\mid(\Gamma'')_{>j}\big)
		\end{tikzcd}
		}
	\end{center}
	Here the left-hand component follows the left and lower sides. The right-hand component follows the upper and right sides. The $\gamma''$-zone square again gives equality.
\end{itemize}
}
This finishes all the cases of the output interchange.

\textit{Input interchange.} Consider the maps
\[
\begin{gathered}
w_p:(\Theta_1)_p,A_p,(\Theta_2)_p,B_p,(\Theta_3)_p\to\Xi_p \text{ over }\omega, \\
u_p:\Gamma_p\to(\Delta_1)_p,A_p \text{ over }\beta, \\
v_p:\Gamma'_p\to B_p,(\Delta'_2)_p \text{ over }\beta',
\end{gathered}
\]
with components $(f^w_i;g^w_j)$, $(u_i;u^-_j)$ and $(v_i;v^-_j)$. Write
$a_\omega:=|\theta_1|+1$, $b_\omega:=|\theta_1|+|\theta_2|+2$, and $c_\beta:=|\delta_1|+1$ for the cut positions. We assume that $\Delta_1=()$ or $\Theta_1=()$, and $\Delta'_2=()$ or $\Theta_3=()$. We must show
$$(w_p\circ_{A_p}u_p)\circ_{B_p}v_p=(w_p\circ_{B_p}v_p)\circ_{A_p}u_p.$$
We use the shorthands
$\omega_a:=\omega\circledcirc_a\beta$ and $\omega_b:=\omega\circledcirc_b\beta'$. We write
$\chi:=\omega_a\circledcirc_b\beta'=\omega_b\circledcirc_a\beta$ for the composition involved.
By \cref{lem:poly-circledcirc-interchange}, both sides of the required equality of compositions lie over $\chi$.

Let
$$\widehat F:=F^{c_\beta}_\beta(\Delta_1\mid\Gamma\mid\,), \qquad
\widehat F':=F^1_{\beta'}(\,\mid\Gamma'\mid\Delta'_2),$$
$$p_a:=u_{c_\beta}, \qquad q_a:=g^w_{a_\omega}, \qquad p_b:=v_1, \qquad q_b:=g^w_{b_\omega}.$$
Thus $p_a:\widehat F\to A$ and $p_b:\widehat F'\to B$. Let
$Q_a:=U^{a_\omega}_\omega(\Theta_1\mid\Xi\mid\Theta_2,\widehat F',\Theta_3),$ and $Q_b:=U^{b_\omega}_\omega(\Theta_1,\widehat F,\Theta_2\mid\Xi\mid\Theta_3).$
We have 
$$\bar q_a:=U^{a_\omega}_\omega(p_b)\circ q_a:A\to Q_a, \qquad
\bar q_b:=U^{b_\omega}_\omega(p_a)\circ q_b:B\to Q_b.$$
The components used by the second cuts are
\begin{align*}
	(w_p\circ_{B_p}v_p)_{a_\omega}&=\eta^U_{a_\omega\to}[\omega,\beta']\circ\bar q_a, \\
	(w_p\circ_{A_p}u_p)_{|\theta_1|+|\gamma|+|\theta_2|+1}&=\eta^U_{\to(|\theta_2|+1)}[\omega,\beta]\circ\bar q_b.
\end{align*}
As in the preceding case, we compute each cut.

{
\begin{itemize}
	\item ($\delta_1$-zone, $1\leq i\leq|\delta_1|$.) The left-hand component follows the upper and right sides. The right-hand component follows the left and lower sides.
	\begin{center}
		\adjustbox{max width=\linewidth}{
		\begin{tikzcd}[row sep=huge, column sep=large]
			F^i_\chi\big((\Delta_1)_{<i}\mid\Theta_1,\Gamma,\Theta_2,\Gamma',\Theta_3\mid(\Delta_1)_{>i},\Xi,\Delta'_2\big) \arrow[r, "{\epsilon^F_i[\omega_a,\beta']}"] \arrow[d, "{\eta^F_{i\to}[\omega_b,\beta]}", swap] & F^i_{\omega_a}\big((\Delta_1)_{<i}\mid\Theta_1,\Gamma,\Theta_2,\widehat F',\Theta_3\mid(\Delta_1)_{>i},\Xi\big) \arrow[d, "{\eta^F_{i\to}[\omega,\beta]}"] \\
			F^i_\beta\big((\Delta_1)_{<i}\mid\Gamma\mid(\Delta_1)_{>i},U^{a_\omega}_{\omega_b}(\Theta_1\mid\Xi,\Delta'_2\mid\Theta_2,\Gamma',\Theta_3)\big) \arrow[r, "{F^i_\beta(\eta^U_{a_\omega\to}[\omega,\beta'])}", swap] & F^i_\beta\big((\Delta_1)_{<i}\mid\Gamma\mid(\Delta_1)_{>i},Q_a\big) \arrow[r, "{F^i_\beta(\bar q_a)}"] & F^i_\beta\big((\Delta_1)_{<i}\mid\Gamma\mid(\Delta_1)_{>i},A\big) \arrow[r, "u_i"] & (\Delta_1)_i
		\end{tikzcd}
		}
	\end{center}
	The square is Disjoint Invariance in the input $\delta_1$-zone. Hence the components agree.

	\item ($\xi$-zone, $1\leq i\leq|\xi|$.) Naturality in the $a$- and $b$-slots gives
	\begin{center}
		\adjustbox{max width=\linewidth}{
		\begin{tikzcd}[row sep=huge, column sep=large]
			F^{|\delta_1|+i}_\chi\big(\Delta_1,\Xi_{<i}\mid\Theta_1,\Gamma,\Theta_2,\Gamma',\Theta_3\mid\Xi_{>i},\Delta'_2\big) \arrow[r, "{\epsilon^F_{|\delta_1|+i}[\omega_a,\beta']}"] \arrow[d, "{\epsilon^F_i[\omega_b,\beta]}", swap] & F^{|\delta_1|+i}_{\omega_a}\big(\Delta_1,\Xi_{<i}\mid\Theta_1,\Gamma,\Theta_2,\widehat F',\Theta_3\mid\Xi_{>i}\big) \arrow[d, "{\epsilon^F_i[\omega,\beta]}"] \\
			F^i_{\omega_b}\big(\Xi_{<i}\mid\Theta_1,\widehat F,\Theta_2,\Gamma',\Theta_3\mid\Xi_{>i},\Delta'_2\big) \arrow[r, "{\epsilon^F_i[\omega,\beta']}", swap] & F^i_\omega\big(\Xi_{<i}\mid\Theta_1,\widehat F,\Theta_2,\widehat F',\Theta_3\mid\Xi_{>i}\big) \arrow[r, "{F^i_\omega(p_a,p_b)}"] & F^i_\omega\big(\Xi_{<i}\mid\Theta_1,A,\Theta_2,B,\Theta_3\mid\Xi_{>i}\big) \arrow[r, "f^w_i"] & \Xi_i
		\end{tikzcd}
		}
	\end{center}
	The two cut orders give the two routes through the input $\xi$-zone square. They are equal.

	\item ($\delta'_2$-zone, $1\leq i\leq|\delta'_2|$.) Naturality at $p_a$ and $q_b$ gives
	\begin{center}
		\adjustbox{max width=\linewidth}{
		\begin{tikzcd}[row sep=huge, column sep=large]
			F^{|\delta_1|+|\xi|+i}_\chi\big(\Delta_1,\Xi,(\Delta'_2)_{<i}\mid\Theta_1,\Gamma,\Theta_2,\Gamma',\Theta_3\mid(\Delta'_2)_{>i}\big) \arrow[r, "{\eta^F_{\to i}[\omega_a,\beta']}"] \arrow[d, "{\epsilon^F_{|\xi|+i}[\omega_b,\beta]}", swap] & F^{1+i}_{\beta'}\big(U^{|\theta_1|+|\gamma|+|\theta_2|+1}_{\omega_a}(\Theta_1,\Gamma,\Theta_2\mid\Delta_1,\Xi\mid\Theta_3),(\Delta'_2)_{<i}\mid\Gamma'\mid(\Delta'_2)_{>i}\big) \arrow[d, "{F^{1+i}_{\beta'}(\eta^U_{\to(|\theta_2|+1)}[\omega,\beta])}"] \\
			F^{|\xi|+i}_{\omega_b}\big(\Xi,(\Delta'_2)_{<i}\mid\Theta_1,\widehat F,\Theta_2,\Gamma',\Theta_3\mid(\Delta'_2)_{>i}\big) \arrow[r, "{\eta^F_{\to i}[\omega,\beta']}", swap] & F^{1+i}_{\beta'}\big(Q_b,(\Delta'_2)_{<i}\mid\Gamma'\mid(\Delta'_2)_{>i}\big) \arrow[r, "{F^{1+i}_{\beta'}(\bar q_b)}"] & F^{1+i}_{\beta'}\big(B,(\Delta'_2)_{<i}\mid\Gamma'\mid(\Delta'_2)_{>i}\big) \arrow[r, "v_{1+i}"] & (\Delta'_2)_i
		\end{tikzcd}
		}
	\end{center}
	The left-hand component follows the upper and right sides. The right-hand component follows the left and lower sides. The input $\delta'_2$-zone axiom proves equality.

	\item ($\theta_1$-zone, $1\leq j\leq|\theta_1|$.) Naturality in both marked input slots gives
	\begin{center}
		\adjustbox{max width=\linewidth}{
		\begin{tikzcd}[row sep=huge, column sep=large]
			(\Theta_1)_j \arrow[r, "g^w_j"] & U^j_\omega\big((\Theta_1)_{<j}\mid\Xi\mid(\Theta_1)_{>j},A,\Theta_2,B,\Theta_3\big) \arrow[r, "{U^j_\omega(p_a,p_b)}"] & U^j_\omega\big((\Theta_1)_{<j}\mid\Xi\mid(\Theta_1)_{>j},\widehat F,\Theta_2,\widehat F',\Theta_3\big) \arrow[r, "{\eta^U_{j\to}[\omega,\beta]}"] \arrow[d, "{\eta^U_{j\to}[\omega,\beta']}", swap] & U^j_{\omega_a}\big((\Theta_1)_{<j}\mid\Delta_1,\Xi\mid(\Theta_1)_{>j},\Gamma,\Theta_2,\widehat F',\Theta_3\big) \arrow[d, "{\eta^U_{j\to}[\omega_a,\beta']}"] \\
			& & U^j_{\omega_b}\big((\Theta_1)_{<j}\mid\Xi,\Delta'_2\mid(\Theta_1)_{>j},\widehat F,\Theta_2,\Gamma',\Theta_3\big) \arrow[r, "{\eta^U_{j\to}[\omega_b,\beta]}", swap] & U^j_\chi\big((\Theta_1)_{<j}\mid\Delta_1,\Xi,\Delta'_2\mid(\Theta_1)_{>j},\Gamma,\Theta_2,\Gamma',\Theta_3\big)
		\end{tikzcd}
		}
	\end{center}
	The left-hand component follows the upper and right sides. The right-hand component follows the left and lower sides. The input $\theta_1$-zone square gives equality.

	\item ($\gamma$-zone, $1\leq j\leq|\gamma|$.) Naturality at $p_b$ and $q_a$ gives
	\begin{center}
		\adjustbox{max width=\linewidth}{
		\begin{tikzcd}[row sep=huge, column sep=large]
			\Gamma_j \arrow[r, "u^-_j"] & U^j_\beta\big(\Gamma_{<j}\mid\Delta_1,A\mid\Gamma_{>j}\big) \arrow[r, "{U^j_\beta(\bar q_a)}"] & U^j_\beta\big(\Gamma_{<j}\mid\Delta_1,Q_a\mid\Gamma_{>j}\big) \arrow[r, "{\epsilon^U_j[\omega,\beta]}"] \arrow[d, "{U^j_\beta(\eta^U_{a_\omega\to}[\omega,\beta'])}", swap] & U^{|\theta_1|+j}_{\omega_a}\big(\Theta_1,\Gamma_{<j}\mid\Delta_1,\Xi\mid\Gamma_{>j},\Theta_2,\widehat F',\Theta_3\big) \arrow[d, "{\eta^U_{(|\theta_1|+j)\to}[\omega_a,\beta']}"] \\
			& & U^j_\beta\big(\Gamma_{<j}\mid\Delta_1,U^{a_\omega}_{\omega_b}(\Theta_1\mid\Xi,\Delta'_2\mid\Theta_2,\Gamma',\Theta_3)\mid\Gamma_{>j}\big) \arrow[r, "{\epsilon^U_j[\omega_b,\beta]}", swap] & U^{|\theta_1|+j}_\chi\big(\Theta_1,\Gamma_{<j}\mid\Delta_1,\Xi,\Delta'_2\mid\Gamma_{>j},\Theta_2,\Gamma',\Theta_3\big)
		\end{tikzcd}
		}
	\end{center}
	The two components are the two routes through the input $\gamma$-zone square. Thus they agree.

	\item ($\theta_2$-zone, $1\leq j\leq|\theta_2|$.) Naturality in the $a$- and $b$-slots gives
	\begin{center}
		\adjustbox{max width=\linewidth}{
		\begin{tikzcd}[row sep=huge, column sep=large]
			(\Theta_2)_j \arrow[r, "g^w_{a_\omega+j}"] & U^{a_\omega+j}_\omega\big(\Theta_1,A,(\Theta_2)_{<j}\mid\Xi\mid(\Theta_2)_{>j},B,\Theta_3\big) \arrow[r, "{U^{a_\omega+j}_\omega(p_a,p_b)}"] & U^{a_\omega+j}_\omega\big(\Theta_1,\widehat F,(\Theta_2)_{<j}\mid\Xi\mid(\Theta_2)_{>j},\widehat F',\Theta_3\big) \arrow[r, "{\eta^U_{\to j}[\omega,\beta]}"] \arrow[d, "{\eta^U_{(a_\omega+j)\to}[\omega,\beta']}", swap] & U^{|\theta_1|+|\gamma|+j}_{\omega_a}\big(\Theta_1,\Gamma,(\Theta_2)_{<j}\mid\Delta_1,\Xi\mid(\Theta_2)_{>j},\widehat F',\Theta_3\big) \arrow[d, "{\eta^U_{(|\theta_1|+|\gamma|+j)\to}[\omega_a,\beta']}"] \\
			& & U^{a_\omega+j}_{\omega_b}\big(\Theta_1,\widehat F,(\Theta_2)_{<j}\mid\Xi,\Delta'_2\mid(\Theta_2)_{>j},\Gamma',\Theta_3\big) \arrow[r, "{\eta^U_{\to j}[\omega_b,\beta]}", swap] & U^{|\theta_1|+|\gamma|+j}_\chi\big(\Theta_1,\Gamma,(\Theta_2)_{<j}\mid\Delta_1,\Xi,\Delta'_2\mid(\Theta_2)_{>j},\Gamma',\Theta_3\big)
		\end{tikzcd}
		}
	\end{center}
	The left-hand component follows the upper and right sides. The right-hand component follows the left and lower sides. The input $\theta_2$-zone axiom gives equality.

	\item ($\gamma'$-zone, $1\leq j\leq|\gamma'|$.) Naturality at $p_a$ and $q_b$ gives
	\begin{center}
		\adjustbox{max width=\linewidth}{
		\begin{tikzcd}[row sep=huge, column sep=large]
			(\Gamma')_j \arrow[r, "v^-_j"] & U^j_{\beta'}\big((\Gamma')_{<j}\mid B,\Delta'_2\mid(\Gamma')_{>j}\big) \arrow[r, "{U^j_{\beta'}(\bar q_b)}"] & U^j_{\beta'}\big((\Gamma')_{<j}\mid Q_b,\Delta'_2\mid(\Gamma')_{>j}\big) \arrow[r, "{\epsilon^U_j[\omega,\beta']}"] \arrow[d, "{U^j_{\beta'}(\eta^U_{\to(|\theta_2|+1)}[\omega,\beta])}", swap] & U^{|\theta_1|+1+|\theta_2|+j}_{\omega_b}\big(\Theta_1,\widehat F,\Theta_2,(\Gamma')_{<j}\mid\Xi,\Delta'_2\mid(\Gamma')_{>j},\Theta_3\big) \arrow[d, "{\eta^U_{\to(|\theta_2|+j)}[\omega_b,\beta]}"] \\
			& & U^j_{\beta'}\big((\Gamma')_{<j}\mid U^{|\theta_1|+|\gamma|+|\theta_2|+1}_{\omega_a}(\Theta_1,\Gamma,\Theta_2\mid\Delta_1,\Xi\mid\Theta_3),\Delta'_2\mid(\Gamma')_{>j}\big) \arrow[r, "{\epsilon^U_j[\omega_a,\beta']}", swap] & U^{|\theta_1|+|\gamma|+|\theta_2|+j}_\chi\big(\Theta_1,\Gamma,\Theta_2,(\Gamma')_{<j}\mid\Delta_1,\Xi,\Delta'_2\mid(\Gamma')_{>j},\Theta_3\big)
		\end{tikzcd}
		}
	\end{center}
	Here the left-hand component follows the left and lower sides. The right-hand component follows the upper and right sides. The input $\gamma'$-zone square proves equality.

	\item ($\theta_3$-zone, $1\leq j\leq|\theta_3|$.) Naturality in both marked input slots gives
	\begin{center}
		\adjustbox{max width=\linewidth}{
		\begin{tikzcd}[row sep=huge, column sep=large]
			(\Theta_3)_j \arrow[r, "g^w_{b_\omega+j}"] & U^{b_\omega+j}_\omega\big(\Theta_1,A,\Theta_2,B,(\Theta_3)_{<j}\mid\Xi\mid(\Theta_3)_{>j}\big) \arrow[r, "{U^{b_\omega+j}_\omega(p_a,p_b)}"] & U^{b_\omega+j}_\omega\big(\Theta_1,\widehat F,\Theta_2,\widehat F',(\Theta_3)_{<j}\mid\Xi\mid(\Theta_3)_{>j}\big) \arrow[r, "{\eta^U_{\to(|\theta_2|+1+j)}[\omega,\beta]}"] \arrow[d, "{\eta^U_{\to j}[\omega,\beta']}", swap] & U^{|\theta_1|+|\gamma|+|\theta_2|+1+j}_{\omega_a}\big(\Theta_1,\Gamma,\Theta_2,\widehat F',(\Theta_3)_{<j}\mid\Delta_1,\Xi\mid(\Theta_3)_{>j}\big) \arrow[d, "{\eta^U_{\to j}[\omega_a,\beta']}"] \\
			& & U^{|\theta_1|+1+|\theta_2|+|\gamma'|+j}_{\omega_b}\big(\Theta_1,\widehat F,\Theta_2,\Gamma',(\Theta_3)_{<j}\mid\Xi,\Delta'_2\mid(\Theta_3)_{>j}\big) \arrow[r, "{\eta^U_{\to(|\theta_2|+|\gamma'|+j)}[\omega_b,\beta]}", swap] & U^{|\theta_1|+|\gamma|+|\theta_2|+|\gamma'|+j}_\chi\big(\Theta_1,\Gamma,\Theta_2,\Gamma',(\Theta_3)_{<j}\mid\Delta_1,\Xi,\Delta'_2\mid(\Theta_3)_{>j}\big)
		\end{tikzcd}
		}
	\end{center}
	The left-hand component follows the upper and right sides. The right-hand component follows the left and lower sides. The input $\theta_3$-zone axiom gives equality.
\end{itemize}
}
This finishes the proof of both interchange laws.\hfill $\square$

\subsection{Canonical Liftings }\label{subsec:canonical-planar-liftings}

We now construct the lifting of a base map. 
\begin{definition}[lifting candidates]\label{def:lifting-candidates}
	For the lists $\Gamma, \Delta_1, \Delta_2$, set $c := |\delta_1| + 1$. The \textit{opcartesian lift of $\otimes_n\gamma \xto{\alpha} \oplus_m(\delta_1, b, \delta_2)$ at $c$} is the polymorphism
	$$\Phi = \Phi^c_\alpha(\Delta_1 \mid \Gamma \mid \Delta_2) \ : \ \Gamma_p \to (\Delta_1)_p, \widehat F_p, (\Delta_2)_p, \qquad \widehat F := F^c_\alpha(\Delta_1 \mid \Gamma \mid \Delta_2).$$
	It lies over $\alpha$ and is generated at the position $c^+$ by the identity of $\widehat F$. That is, $\Phi := \langle \1_{\widehat F}\rangle^{c^+}$.

	Dually, consider $\otimes_{n'}(\gamma_1, a, \gamma_2) \xto{\beta} \oplus_{m'}\delta$ with input position $c' := |\gamma_1| + 1$. For the lists $\Gamma_1, \Gamma_2, \Delta$, the \textit{cartesian lift of $\beta$ at $c'$} is the polymorphism
	$$\Upsilon = \Upsilon^{c'}_\beta(\Gamma_1 \mid \Delta \mid \Gamma_2) \ : \ (\Gamma_1)_p, \widehat U_p, (\Gamma_2)_p \to \Delta_p, \qquad \widehat U := U^{c'}_\beta(\Gamma_1 \mid \Delta \mid \Gamma_2).$$
	It lies over $\beta$ and is generated at $c'^-$ by the identity of $\widehat U$. That is, $\Upsilon := \langle \1_{\widehat U}\rangle^{c'^-}$. 
\end{definition}
\textit{Remark.} Both are representations by the change of representative conditions and therefore polymorphisms.

\begin{theorem}[planar cartesianness]\label{thm:planar-cartesianness}
	Let $\Phi : \Gamma_p \to (\Delta_1)_p, \widehat F_p, (\Delta_2)_p$ be a canonical output lifting from \cref{def:lifting-candidates}. Let $\otimes(\theta_1, b, \theta_2) \xto{\beta} \oplus\lambda$ be a morphism of $\mc M$. Let
	$$\Xi : (\Theta_1)_p, \Gamma_p, (\Theta_2)_p \to (\Delta_1)_p, \Lambda_p, (\Delta_2)_p$$
	be a polymorphism over $\beta \circledcirc \alpha$ with the same lists $\Delta_1, \Delta_2$ and $\Gamma$. Assume that $(\beta,\alpha)$ is planar.
	\begin{itemize}
  	\item If $\beta \circledcirc \alpha$ has at least one position, then there is a unique polymorphism $\Psi : (\Theta_1)_p, \widehat F_p, (\Theta_2)_p \to \Lambda_p$ over $\beta$ with $\Psi \circ_{\widehat F_p} \Phi = \Xi$.
  	\item If $\beta \circledcirc \alpha$ has no position, then $\alpha:()\to b$ and $\beta:b\to()$, and the set of such factorisations is $\mc T_b(F^1_\alpha(),U^1_\beta())$.
	\end{itemize}
	The dual statements hold for the canonical input liftings. Consequently, $\pi_p$ is a bifibration of planar polycategories whenever $\mc T_b(F^1_\alpha(),U^1_\beta())$ is a singleton for every pair $\alpha:()\to b$, $\beta:b\to()$. 
\end{theorem}

\textit{Proof:} Let the components of $\Xi$ be $(\Xi_p)_p$, where $p$ ranges over the positions of $\beta \circledcirc \alpha$. The zones of this map are $(\delta_1, \lambda, \delta_2)$ on the output side and $(\theta_1, \gamma, \theta_2)$ on the input side. The components of $\Phi$ are $\Phi_{c^+} = \1_{\widehat F}$ and $\Phi_q = \varphi^{c^+, q}_\alpha(\1_{\widehat F})$. This follows because $\Phi$ maps to the representative $\widehat F_p$. The representations of $\beta\circledcirc\alpha$, of $\alpha$ and of $\beta$ are taken respectively at the contexts
$$(\Theta_1,\Gamma,\Theta_2;\ \Delta_1,\Lambda,\Delta_2), \qquad (\Gamma;\ \Delta_1,\widehat U,\Delta_2), \qquad (\Theta_1,\widehat F,\Theta_2;\ \Lambda).$$
Here $\widehat U:=U^{c'}_\beta(\Theta_1\mid\Lambda\mid\Theta_2)$ is the representative object of the configuration for the map $\alpha$. We use the coordinate changes $\kappa_p$ and the bijections $K_p$ of \cref{def:coordinate-change,def:cut-reading} at these lists. We first assume that $\beta\circledcirc\alpha$ has at least one position.

\textbf{Construction of $\Psi$.} We impose the requirement $\Psi \circ_{\widehat F_p} \Phi = \Xi$ at the $\beta$-side zones. This determines the components of $\Psi$ at every position of $\beta$ other than the cut input $c'^- := (|\theta_1| + 1)^-$. Since $\Phi_{c^+} = \1_{\widehat F}$, the defining formulas of the cut give
$$(\Psi \circ_{\widehat F_p} \Phi)_{|\delta_1| + i} = \Psi_{i^+} \circ \epsilon^F_i, \qquad (\Psi \circ_{\widehat F_p} \Phi)_{j^-} = \eta^U_{j \to} \circ \Psi_{j^-}, \qquad (\Psi \circ_{\widehat F_p} \Phi)_{(\text{$\theta_2$-zone})} = \eta^U_{\to j} \circ \Psi_{(c' + j)^-}.$$
We therefore set
$$\Psi_{i^+} := \Xi_{(|\delta_1| + i)^+} \circ (\epsilon^F_i)^{-1}, \qquad \Psi_{j^-} := (\eta^U_{j \to})^{-1} \circ \Xi_{j^-} \ (j \leq |\theta_1|), \qquad \Psi_{(c' + j)^-} := (\eta^U_{\to j})^{-1} \circ \Xi_{(\text{$\theta_2$-zone } j)}.$$
Thus $\Psi_r := \kappa_{r'}^{-1}(\Xi_{r'})$ for every position $r \neq c'^-$ of $\beta$. Here $r'$ is the corresponding position of the composite, and $\kappa_{r'}$ is the coordinate change of \cref{def:coordinate-change}. For the component at the cut input, we distinguish two cases:
\begin{itemize}
	\item if $\beta$ has a position $r \neq c'^-$, set
	$$\Psi_{c'^-} := \varphi^{r, c'^-}_\beta(\Psi_r) = K_{r'}^{-1}(\Xi_{r'}).$$
	The second description holds because $K_{r'} = \kappa_{r'}\circ\varphi^{c'^-, r}_\beta$ at a $\beta$-side position. Thus the change of representative conditions give $K_{r'}\big(\varphi^{r, c'^-}_\beta(\Psi_r)\big) = \kappa_{r'}(\Psi_r) = \Xi_{r'}$. To verify independence, let $s$ be another such position. The correspondence--decomposition axiom, together with the fact that $\Xi$ is a representation, gives
	$$K_{s'}\big(K_{r'}^{-1}(\Xi_{r'})\big) = \varphi^{r', s'}_{\beta \circledcirc \alpha}\big(\Xi_{r'}\big) = \Xi_{s'} = K_{s'}\big(K_{s'}^{-1}(\Xi_{s'})\big).$$
	Since $K_{s'}$ is a bijection, the displayed equality proves independence. The same identification shows that $\Psi_r = \varphi^{c'^-, r}_\beta(\Psi_{c'^-})$ for every $r \neq c'^-$. That is, $\Psi = \langle \Psi_{c'^-}\rangle^{c'^-}$. Hence $\Psi$ is a representation;
	\item if $c'^-$ is the only position of $\beta$ (i.e.\ $\theta_1, \theta_2, \lambda$ are all empty), pick a position $q$ of $\beta \circledcirc \alpha$ and set $\Psi_{c'^-} := K_q^{-1}(\Xi_q)$. Here
	$$K_q \ : \ \mc T_{b}\big(\widehat F, \widehat U\big) \xto{\ \varphi^{c^+, \bar q}_\alpha\ } \mathrm{Rep}^{\bar q}_\alpha(\Gamma;\ \Delta_1,\widehat U,\Delta_2) \xto{\ \kappa_q\ } \mathrm{Rep}^{q}_{\beta \circledcirc \alpha}(\Theta_1,\Gamma,\Theta_2;\ \Delta_1,\Lambda,\Delta_2)$$
	is the bijection of \cref{def:cut-reading} at a position in $\alpha$. Recall from \cref{def:inherited-position} that the domain is the cut set $\mathrm{Rep}^{c^+}_\alpha(\Gamma;\ \Delta_1,\widehat U,\Delta_2)=\mc T_b(\widehat F,\widehat U)$. As in the previous case, the result is independent of $q$. Thus $\Psi$ has a single reading. Hence it is a representation.
\end{itemize}

We now show that $\Psi \circ_{\widehat F_p} \Phi = \Xi$. At the $\beta$-side zones, this holds by construction. Put $u:=\Psi_{c'^-}:\widehat F\to\widehat U$. At the three kinds of $\alpha$-side positions, the defining formulas of the cut give
\begin{align*}
	(\Psi \circ_{\widehat F_p} \Phi)_{i^+}
	&= \Phi_{i^+}\circ F^i_\alpha\big(\1_{(\Delta_1)_{<i}}\mid\1_\Gamma\mid\1_{(\Delta_1)_{>i}},u,\1_{\Delta_2}\big)\circ\eta^F_{i\to}, && 1\leq i\leq|\delta_1|,\\
	(\Psi \circ_{\widehat F_p} \Phi)_{(|\delta_1|+|\lambda|+i)^+}
	&= \Phi_{(c+i)^+}\circ F^{c+i}_\alpha\big(\1_{\Delta_1},u,\1_{(\Delta_2)_{<i}}\mid\1_\Gamma\mid\1_{(\Delta_2)_{>i}}\big)\circ\eta^F_{\to i}, && 1\leq i\leq|\delta_2|,\\
	(\Psi \circ_{\widehat F_p} \Phi)_{(|\theta_1|+j)^-}
	&= \epsilon^U_j\circ U^j_\alpha\big(\1_{\Gamma_{<j}}\mid\1_{\Delta_1},u,\1_{\Delta_2}\mid\1_{\Gamma_{>j}}\big)\circ\Phi_{j^-}, && 1\leq j\leq|\gamma|.
\end{align*}
Now, for every non-cut position $r$ of $\alpha$, let
$$\rho_r(u):\mathrm{Rep}^r_\alpha(\Gamma;\Delta_1,\widehat F,\Delta_2)\longrightarrow\mathrm{Rep}^r_\alpha(\Gamma;\Delta_1,\widehat U,\Delta_2)$$
be the map induced by $u$ in the $c$-th output object. Explicitly, its three possible forms are
\begin{align*}
	\rho_{i^+}(u)(h)
	&:=h\circ F^i_\alpha\big(\1_{(\Delta_1)_{<i}}\mid\1_\Gamma\mid\1_{(\Delta_1)_{>i}},u,\1_{\Delta_2}\big), && 1\leq i\leq|\delta_1|,\\
	\rho_{(c+i)^+}(u)(h)
	&:=h\circ F^{c+i}_\alpha\big(\1_{\Delta_1},u,\1_{(\Delta_2)_{<i}}\mid\1_\Gamma\mid\1_{(\Delta_2)_{>i}}\big), && 1\leq i\leq|\delta_2|,\\
	\rho_{j^-}(u)(h)
	&:=U^j_\alpha\big(\1_{\Gamma_{<j}}\mid\1_{\Delta_1},u,\1_{\Delta_2}\mid\1_{\Gamma_{>j}}\big)\circ h, && 1\leq j\leq|\gamma|.
\end{align*}
Naturality of $\varphi^{c^+,r}_\alpha$ in that object makes the following diagram commute:
\begin{center}
	\adjustbox{max width=\linewidth}{
	\begin{tikzcd}[row sep=large, column sep=huge]
		\mathrm{Rep}^{c^+}_\alpha(\Gamma;\Delta_1,\widehat F,\Delta_2) \arrow[r, "{\varphi^{c^+,r}_\alpha}"] \arrow[d, "{u\circ(\--)}", swap] & \mathrm{Rep}^{r}_\alpha(\Gamma;\Delta_1,\widehat F,\Delta_2) \arrow[d, "{\rho_r(u)}"] \\
		\mathrm{Rep}^{c^+}_\alpha(\Gamma;\Delta_1,\widehat U,\Delta_2) \arrow[r, "{\varphi^{c^+,r}_\alpha}", swap] & \mathrm{Rep}^{r}_\alpha(\Gamma;\Delta_1,\widehat U,\Delta_2).
	\end{tikzcd}
	}
\end{center}
Since $\Phi_r=\varphi^{c^+,r}_\alpha(\1_{\widehat F})$, applying this square to $\1_{\widehat F}$ gives 
$$\rho_r(u)(\Phi_r)=\varphi^{c^+,r}_\alpha(u\circ\1_{\widehat F})=\varphi^{c^+,r}_\alpha(u).$$
Unpacking $\rho_r(u)$ gives respectively
\begin{align*}
	\Phi_{i^+}\circ F^i_\alpha\big(\1_{(\Delta_1)_{<i}}\mid\1_\Gamma\mid\1_{(\Delta_1)_{>i}},u,\1_{\Delta_2}\big)
	&=\varphi^{c^+,i^+}_\alpha(u),\\
	\Phi_{(c+i)^+}\circ F^{c+i}_\alpha\big(\1_{\Delta_1},u,\1_{(\Delta_2)_{<i}}\mid\1_\Gamma\mid\1_{(\Delta_2)_{>i}}\big)
	&=\varphi^{c^+,(c+i)^+}_\alpha(u),\\
	U^j_\alpha\big(\1_{\Gamma_{<j}}\mid\1_{\Delta_1},u,\1_{\Delta_2}\mid\1_{\Gamma_{>j}}\big)\circ\Phi_{j^-}
	&=\varphi^{c^+,j^-}_\alpha(u).
\end{align*}
Comparing these identities with the three corresponding definitions of $\kappa_q$ in \cref{def:coordinate-change}, we obtain, uniformly for every $\alpha$-side position $q$,
$$(\Psi \circ_{\widehat F_p} \Phi)_q = \kappa_q\big(\varphi^{c^+, \bar q}_\alpha(u)\big) = K_q(u)=K_q(\Psi_{c'^-}).$$
If $\beta$ has a position $r \neq c'^-$, then $\Psi_{c'^-} = K_{r'}^{-1}(\Xi_{r'})$. The correspondence--decomposition axiom, together with the fact that $\Xi$ is a representation, gives
$$K_q(\Psi_{c'^-}) = \varphi^{r', q}_{\beta \circledcirc \alpha}\big(\Xi_{r'}\big) = \Xi_q.$$
If $c'^-$ is the only position of $\beta$, then $K_q(\Psi_{c'^-}) = \Xi_q$ by the definition of $\Psi_{c'^-}$ and its independence of $q$.

\textbf{Uniqueness.} Let $\Psi'$ be any polymorphism over $\beta$ with $\Psi' \circ_{\widehat F_p} \Phi = \Xi$. At the $\beta$-side zones, the formulas give $\kappa_{r'}(\Psi'_r) = \Xi_{r'}$ for every position $r \neq c'^-$ of $\beta$. Since the coordinate changes are bijections, $\Psi'_r = \Psi_r$ there.
For the cut component, first suppose that $\beta$ has a position $r \neq c'^-$. Since $\Psi'$ is a representation, $\Psi'_{c'^-} = \varphi^{r, c'^-}_\beta(\Psi'_r) = \varphi^{r, c'^-}_\beta(\Psi_r) = \Psi_{c'^-}$. Otherwise, the computation of the previous paragraph applied to $\Psi'$ gives $K_q(\Psi'_{c'^-}) = \Xi_q = K_q(\Psi_{c'^-})$. Since $K_q$ is a bijection, $\Psi'_{c'^-} = \Psi_{c'^-}$. This proves uniqueness. \hfill $\square$

\endgroup{}

\subfilebibliography

\endgroup{}
\end{mainf}

\begin{bibliof}
\bibliography{references}
\end{bibliof}

\end{document}